\documentclass[pdflatex,sn-basic]{sn-jnl}% Basic Springer Nature Reference Style/Chemistry Reference Style
\usepackage{graphicx}%
\usepackage{multirow}%
\usepackage{amsmath,amssymb,amsfonts}%
\usepackage{amsthm}%
\usepackage{mathrsfs}%
\usepackage[title]{appendix}%
\usepackage{xcolor}%
\usepackage{textcomp}%
\usepackage{manyfoot}%
\usepackage{booktabs}%
\usepackage{algorithm}%
\usepackage{algorithmicx}%
\usepackage{algpseudocode}%
\usepackage{listings}%
\usepackage{aas_macros}
\usepackage{rotating}
\newcommand{\fnum}{erg cm$^{-2}$ s$^{-1}$}
\newcommand{\flam}{erg cm$^{-2}$ s$^{-1}$ \AA$^{-1}$}
\newcommand{\kms}{km s$^{-1}$}
\newcommand\arcsec{\mbox{$^{\prime\prime}$}}% 
\newcommand{\nodata}{...}
\definecolor{bred}{rgb}{.93 , .40, 0.47 }

\begin{document}

\title[Stellar flares]{Stellar flares}


\author[1,2,3]{\fnm{Adam F.} \sur{Kowalski}}\email{adam.f.kowalski@colorado.edu}

\affil[1]{\orgdiv{Department of Astrophysical and Planetary Sciences}, \orgname{University of Colorado}, \orgaddress{\street{2000 Colorado Ave}, \city{Boulder}, \postcode{80305}, \state{CO}, \country{United States}}}

\affil[2]{\orgdiv{National Solar Observatory}, \orgname{University of Colorado}, \orgaddress{\street{3665 Discovery Drive}, \city{Boulder}, \postcode{80303}, \state{CO}, \country{United States}}}

\affil[3]{\orgdiv{Laboratory for Atmospheric and Space Sciences}, \orgname{University of Colorado}, \orgaddress{\street{3665 Discovery Drive}, \city{Boulder}, \postcode{80303}, \state{CO}, \country{United States}}}


\abstract{Magnetic storms on stars manifest as remarkable, randomly occurring changes of the luminosity over durations that are tiny in comparison to the normal evolution of stars.   These stellar flares are bursts of electromagnetic radiation from X-ray to radio wavelengths, and they occur on most stars with outer convection zones.  They are analogous to the events on the Sun known as solar flares, which impact our everyday life and modern technological society.  Stellar flares, however, can attain much greater energies than those on the Sun.  Despite this, we think that these phenomena are rather similar in origin to solar flares, which result from a catastrophic conversion of latent magnetic field energy into atmospheric heating within a region that is relatively small in comparison to normal stellar sizes.   We review the last several decades of stellar flare research.  We summarize multi-wavelength observational results and the associated thermal and nonthermal processes in flaring stellar atmospheres.  Static and hydrodynamic models are reviewed with an emphasis on recent progress in radiation-hydrodynamics and the physical diagnostics in flare spectra.  Thanks to their effects on the space weather of exoplanetary systems (and thus in our search for life elsewhere in the universe) and their preponderance in \emph{Kepler} mission data, white-light stellar flares have re-emerged in the last decade as a widely-impactful area of study within astrophysics.  Yet, there is still much we do not understand, both empirically and theoretically, about the spectrum of flare radiation, its origin, and its time evolution.  We conclude with several big-picture questions that are fundamental in our pursuit toward a greater understanding of these enigmatic stellar phonemena and, by extension, those on the Sun.}

\keywords{Solar-stellar connection}



\maketitle

\clearpage
\setcounter{tocdepth}{3} % TOC subsubsections
\tableofcontents

\section{Introduction}\label{sec1}
Stars produce energetic bursts of electromagnetic radiation that follows a sudden magnetic energy release into their atmospheres.   These electromagnetic bursts are called flares, which occur on a very wide range of timescales, from seconds to days.  Stellar flares are observed in all spectral windows, from the high-energy X-rays through the long wavelength radio waves.  However, all wavelength regimes do not respond equally in energy and simultaneously in time.  Some wavelengths (e.g., microwave) are dominated by nonthermal radiation, while others (optical, low-energy X-rays) are thought to result from primarily thermal radiative processes.  The flare energy -- both thermal and nonthermal -- ultimately originates in the  energy that is released from magnetic fields in stellar coronae.  These magnetic fields, in turn, originate from turbulent convective energy transport and shear in rotating stellar envelopes, beneath the visible photosphere.   The Sun is the best studied flare star due to its proximity to Earth and a fleet of multi-messenger instruments that continuously provide spatially resolved observations.

From Earth, solar and stellar flares are not detectable by the unaided human eye.  Stars like the Sun, and other active stars like Algol that are visible to the naked eye, have too much background glare for our eyes to sense the flare brightness increases.   
In all-sky surveys that are sensitive to the apparent magnitudes of low-mass stars (fainter than a Johnson $V$-band magnitude of $\sim9$), however, stellar flares are the most dramatic source of variability.  By \emph{most dramatic}, we mean that flares produce  the largest changes in their apparent magnitudes per unit time when compared against nearly all other astrophysical phenomena\footnote{Perspective provided by Z. Ivezic (priv.\ communication, 2010).}.

In data from the upcoming Vera C. Rubin Observatory's ten-year Legacy Survey of Space and Time (LSST), stellar flares will constitute a major source of variability \citep{Hawley2016}.  Further, many of the largest events that are observed serendipitously (and those from very faint but populous low-mass stars) will be bona-fide transients, whereby the changes in brightness occur from sources that have not been detected in quiescence.  Stellar flares are much shorter in duration and much less luminous than extra-galactic transient phenomena, such as supernovae, tidal disruption events, and optical counterparts to neutron star mergers.  Most stellar flares can in principle be readily distinguished from these events given enough empirical information, despite some striking spectral similarities \citep{Chang2020} in the optical spectra of kilonovae \citep{Shappee2017}.  Due to their proximity to the solar system, the largest stellar flare events can trigger gamma ray observatories; these events are known as ``superflares'' in X-rays.  Stellar superflares are factors of $\approx 10^2-10^4$ more energetic and luminous in comparison to the largest solar flares, which have bolometric energies of $E_{\rm{bol}} \approx 3-6 \times 10^{32}$ erg \citep{Woods2004, Woods2006, Cliver2022, Hayakawa2023}.  Such superflares were likely common when the Sun was very young and rotating much more rapidly than today \citep{Maehara2012}.  Understanding the physical processes in stellar flares thus provides insight into the heliospheric conditions in the early history of our solar system \citep{Ribas2005}.

It is timely for a review of recent observations and models of stellar flares.  The next sunspot cycle maximum is approaching in late 2024 \citep{Upton2023}\footnote{See also the recent assessments in \citet{Gopalswamy2023} and \citet{McIntosh2023} and the continuously updated web resource, \url{https://helioforecast.space/solarcycle}.}, and with it a deluge of flares and eruptions from the Sun.  New and upcoming solar observatories, such as the Daniel K. Inouye Solar Observatory \citep{Rimmele2020}, the Expanded Owens Solar Valley Array \citep{Gary2018}, and the Interface Region Imaging Spectrograph \citep{DePontieu2014}, will provide new avenues for solar-stellar comparisons.  The immaculate precision of \emph{Kepler} \citep{Koch2010}, K2 \citep{Howell2014}, and the Transiting Exoplanet Survey Satellite \citep[TESS;][]{Ricker2015} observations has recently facilitated a resurgence in the study of optical broadband stellar flares.  These missions provide access to stellar magnetic activity over long time baselines previously not feasible to observe from ground-based campaigns, and over a much larger variety of stars.  Models of energy transport, atmospheric response, and emission line broadening have increased in accuracy and sophistication over a large range of heating parameters.  As the community looks to next-generation modeling paths, analysis methods, and observational capabilities (such as new space missions and ground-based instruments), synthesizing recent findings and outstanding problems could help to steer these into productive directions.  Additionally, stellar flare radiation is now considered an important factor in assessing exoplanet habitability and photochemistry \citep{Shields2016, Segura2018}, and general interest within the solar and astrophysics communities has grown over the last decade.

 \subsection{Overview of this review}
There have been several reviews on stellar flares prior to \emph{ca.} 1990 \citep{Pettersen1989, Byrne1989, Haisch1991}. The current review supplements these with results from the past three decades.
This review features many results from flare studies of low-mass, M-dwarf (dM / MV) stars. Because of their low background glare from non-flaring regions, large contributions to Galactic populations, and inherently high flare rates, the M dwarfs tend to be the most commonly studied flare stars.  
Flares from active binary systems have also been very well-observed across the electromagnetic spectrum.  Results from studies of post main-sequence single stars and young solar-type stars are covered as well.  Surprisingly, certain wavelength regimes in M-dwarf flares have been more thoroughly observed than solar flares.   Solar flares are not reviewed in any detail here, but a general overview is provided, since the interpretation of and modeling approach to stellar flares are dependent on knowledge of the particle acceleration, magnetic fields,  and spatial resolution from the study of the Sun.

The primary purpose of this review is to serve as a compendium of references and to facilitate research in multi-wavelength observations and models of stellar flares.  The target audience consists of beginning PhD students or interested scientists in other areas of astrophysics or solar physics.
Every current topic in the study of stellar flares is not included here:  for example, I leave all results from the TESS mission to the next iteration of this \emph{Living Review}.   Unfortunately, many important results and references cannot be discussed extensively in order to keep this review a reasonable length, while also allowing growth with future addenda.  For the same reasons, every important result from the references cited herein regrettably cannot be included at this time. 
There are several comprehensive introductions to sub-topics in the stellar flare literature; these are indicated throughout instead of being repeated.  Broader applications of flare research beyond stellar astrophysics (e.g., exoplanet habitability and exoplanet atmospheric photochemistry) are outside the scope of this review and are not covered.   The study of stellar coronal mass ejections and their associated energetic particles is a vast and newly emerging field; these topics are only very briefly alluded to. \emph{ This review focuses on the electromagnetic response of flaring stellar atmospheres and detailed modeling of the associated physical processes. }

This review is organized as follows.  We begin with a brief tour of the flare star next door, Proxima Centauri, which is about the same age as the Sun but is otherwise a much different star (Sect.~\ref{sec:proxcen}). Then we take a detour for a brief introduction to solar flare terminology and phenomenology (Sect.~\ref{sec:solar}).   A general overview of all known types of stars that tend to flare is summarized in Sect.~\ref{sec:overview}.  The bulk of the review contains a synthesis of stellar flare observations across the electromagnetic spectrum with a focus on the near-ultraviolet (NUV) and optical response, which has not been the subject of most other reviews.  Observational studies can be separated into two general topics:  flare rates, which are primarily diagnosed through single-band photometry (Sect.~\ref{sec:rates}), and multi-wavelength analyses, which are primarily accomplished through spectroscopy (Sect.~\ref{sec:spectra}).  The final third of the review summarizes stellar flare modeling approaches (slab, semi-empirical, and radiative-hydrodynamic), beginning in Sect.~\ref{sec:modelsoverview}.  We go into detail in several areas within stellar flare modeling, focusing on key elements of radiation-hydrodynamics (Sect.~\ref{sec:rhdphysics}) and the interpretation of chromospheric spectral line broadening in flares (Sect.~\ref{sec:chromlines}).  A comprehensive analysis of an ``ideal'' multi-wavelength data set is discussed in Sect.~\ref{sec:multilam}, and inferences of the stellar flare geometries are reviewed in Sect.~\ref{sec:geom}.  We conclude in Sect.~\ref{sec:conclusions} with six questions that we think are currently central in stellar flare research (excluding questions directly related to exoplanets).  Throughout, we use the \emph{cgs} (Gaussian) units system except for wavelengths in \AA\ or in cases that are otherwise noted.  In the appendices, we include a visual guide to common filters used in optical studies of stellar flares (Appendix \ref{sec:filters}), some additional observational results from the literature are merged (Appendix \ref{sec:color_color_appendix}), several common slab modeling practices and assumptions are reviewed (Appendix \ref{sec:slabs}), and the basic microphysical processes that symmetrically broaden hydrogen lines in flares is summarized (Appendix \ref{sec:stark}).

\section{Our flare star neighbor: Proxima Centauri} \label{sec:proxcen}
The nearest star outside the solar system is Proxima Centauri, which is $\approx $1.3 parsecs, or 268,000 astronomical units (au), from the Earth, and is a low-mass ($\approx 0.12$ M$_{\rm{Sun}}$) M dwarf star of spectral type M5.5Ve (dM5.5e) \citep{PMSU2}.  The hydrogen Balmer $\alpha$ line (6562.81~{\AA}) is in emission (e)\footnote{A spectral line ``in emission'' means that it rises above the local continuum.   A spectral line ``in absorption'' means that  the line spectrum decreases below the local continuum or pseudo-continuum flux. } in quiescence outside of detectable flaring events, which occur very frequently.  This star is a relatively old ($\approx 5-6$ Gyr) main sequence star, but it is still very magnetically active, consistent with the long magnetic activity lifetimes of low mass stars of its spectral type \citep{West2008}.  We describe some of the flare properties of this star from a night of monitoring during an observing run at La Silla Observatory in 2010, while also introducing basic point-source photometry measurements and terminology that are common to almost all analyses of flare star data.    

\begin{figure*}
  \includegraphics[width=1.0\textwidth]{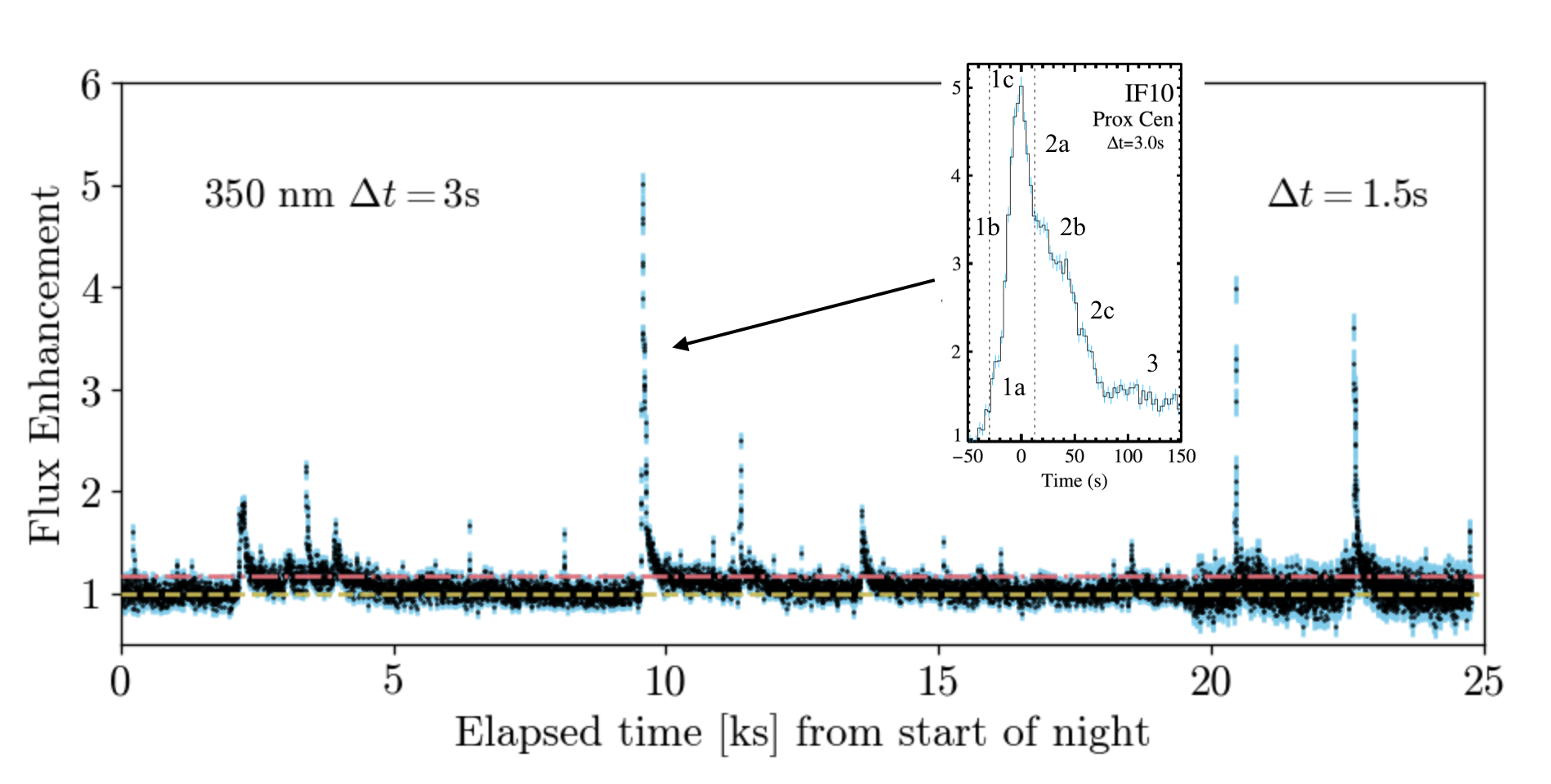} % published_data/ultracam_analysis.ipynb, .key on mac
\caption{A light curve of Proxima Centauri in a narrowband (filter FWHM of 100 \AA) centered in the near-ultraviolet, $U$-band wavelength regime.   These data are from \cite{Kowalski2016} and can be obtained from the Zenodo repository at \url{https://doi.org/10.5281/zenodo.45878}.  The horizontal dashed lines indicate the mean flux enhancement (1.0) and $3\sigma$ above the mean (where $\sigma$ is the standard deviation outside of flaring).  The inset shows the flare peaking at 2010 May 25 01:31:02 UT, which is referred to as the ``IF10'' event in \cite{Kowalski2016}.  The inset shows time relative to peak.  }
\label{fig:proxcen}      
\end{figure*}

 A NUV light curve of Prox Cen  \citep{Kowalski2016} is shown in Fig.~\ref{fig:proxcen}.   The photometry was measured through a custom filter with a full-width-at-half-maximum (FWHM) of 100 \AA\ centered at $\lambda_{\rm{cen}} = 3500$ \AA.  This is a wavelength range over which the star has a very faint luminosity during non-flaring (``quiescent'', denoted by $q$) intervals.   In less than seven hours of continuous 3~s exposures (and 1.5~s exposures at the end of the night), the star produced at least 16 flares at $>$3$\sigma$ (1$\sigma \approx 6$\%) above the mean, which is indicated by a flux enhancement equal to 1.0 in the light curve.  In quiescence, the star is very ``red'' since it has a very cool photospheric temperature ($< 3000$ K), but the flare light causes the star's total flux at Earth to become much ``bluer''.  In addition to the time-integrated flare energy (fluence), the peak flux enhancement is sometimes used to characterize the size of a flare.  The peak magnitude change of a flare is related to the peak flux enhancement by

\begin{equation}   \label{eq:magchange}
 \Delta \mathrm{mag}_{\mathrm{peak}} =-2.5\ \mathrm{log}_{10}\ \left( C_{\mathrm{rel}} (t_{\mathrm{peak}})/C_{\mathrm{rel},q} \right) = -2.5\ \mathrm{log}_{10} \big( I_{f}(t_{\rm{peak}})+1 \big)
\end{equation}
 
\noindent where $C_{\rm{rel}}$ represents the measured counts per exposure through a bandpass and is calculated relative to the counts in that same exposure from a nearby comparison (non-variable) star or set of stars.  The count flux enhancement, $C_{\rm{rel}} (t)/C_{\rm{rel},q}$, is normalized to a time interval over which the flare star is not clearly varying, usually taken right before a flare.  
The count flux enhancement has traditionally been denoted as $I_f + 1$ \citep{Gershberg1972}.  Because the star is spatially unresolved, the values of $I_f(t) + 1$ during a flare consist of the flux from the non-flaring regions on a flare star in addition to the flaring source;  if the flare source has an optical thickness $\tau \gtrsim 1$ at wavelengths in this bandpass, then some or all of the pre-flare flux from the flare area may be diminished  at time $t$ \citep{Hawley1995}.

None of the flares in Fig.~\ref{fig:proxcen} are all that energetic.  The  flare with the largest peak flux enhancement of five would correspond to an energy of $3\times10^{29}$ erg in the Johnson $U$ bandpass, which is traditionally the optimal broadband filter \citep[$\lambda_{\rm{peak}} \sim 3700$ \AA, a full-width-at-half maximum of $\sim 700$ \AA;][]{Moffett1974, Bessel2013}\footnote{Hereafter, following the convention established in \cite{Kowalski2019HST}, we refer to the wavelength regime from the atmospheric cutoff at $\approx 3200$ \AA\ to the violet at $4000$ \AA\ as the $U$ band, in order to distinguish it from the near-ultraviolet (NUV) band at $\lesssim 3200$ \AA, which can only be observed from space.}  for stellar flare observations at ground-based observatories.  This energy is nearly four times larger than the average $U$-band flare energy from Proxima Centauri \citep{Walker1981} and a factor of 100 smaller than the largest event on the Sun that has been observed at similar wavelengths \citep{Neidig1994}.  Multi-wavelength scaling relations \citep{HP91,Kretzschmar2011,Schrijver2012,Osten2015,Namekata2017,Cliver2022} facilitate comparison to the standard classification of solar flares according to their peak flux at Earth over the $\lambda = 1-8$ \AA\ band, as measured by the X-ray Sensor (XRS) on the Geostationary  Operational Environmental Satellite (GOES).  The A1.0, B1.0, C1.0, M1.0, and X1.0-class flares correspond to a range of peak X-ray irradiances logarithmically spanning $10^{-8}$ to $10^{-4}$ W m$^{-2}$, while those at $\ge 10^{-3}$ W m$^{-2}$ are sometimes clipped \citep{Hudson2023} and are given  designations beginning with X10.0.
The scaling relations imply that the large event on Proxima Centauri would correspond to approximately a GOES $1-8$ \AA\ C-class flare if it had occurred on the Sun.  However, Prox Cen  occasionally erupts with energies that are comparable to, or even much greater than, typical GOES X-class flares  from the Sun \citep{Gudel2002, Howard2018}.  Prox Cen also hosts a near Earth-mass planet at $\sim0.05$ au \citep{Escude2016}, and unlike our relatively safe distance from solar flares at 1 au, equivalent flare energies from this star would bathe the surrounding planet in $400x$ the flux of high-energy radiations.  Higher-mass, M3-M4 stars are known to emit even higher-energy flares than Prox Cen.  For example, the most luminous event known resulted in a remarkably fast $\Delta t \approx 35$~s rise to a peak flux change of $\Delta V = - 5$ mags ($L_{V,\rm{peak}} \approx 1.7 \times 10^{32}$ erg s$^{-1}$) from a young M4$+$M4 binary DG CVn \citep{Cabellero2015}, for which the estimated values of the $U$-band energy and GOES class are $4.7 \times 10^{34}$ erg and X600,000, respectively  \citep{Osten2016, Youngblood2017}. \cite{Pettersen2016} describes a remarkable EV Lac flare with a peak magnitude change in the $U$-band of $-7.2$ mags ($L_{U,\rm{peak}}  = 4.6 \times 10^{31}$ erg s$^{-1}$) and a $U$-band energy of $7.23 \times 10^{33}$ erg.  \cite{Howard2019} highlight several flares with extreme amplitudes  ($\Delta g^{\prime} \le -3$ mag) and energies ($E \approx 10^{35} - 10^{36}$ erg) on M1-M4 stars in their optical survey. %, which is more impulsive than the YZ CMi ``megaflare'' with $E_U \approx 2 \times 10^{34}$ erg and 
% Delta V = - 5mag is for two stars.
% V two stars is 12.02 or 12.05
%; julia> 10^(12.77 / -2.5) * 3.63e-9 * 190.0 * 836.0 * 4.0 * 3.1415 * (18.0 * 3.086e18)^2
%;1.743413531622151e32

The most common qualitative description of a flare light curve is a ``FRED'':  a fast-rise, exponential-decay.  In Fig.~\ref{fig:proxcen}, the flares exhibit a  simple FRED shape at low time-resolution;  however, at high-time resolution, much more variation in the temporal morphology is apparent, including two periods of the rise phase (1a, 1b), an extended peak (1c), two intervals of fast decay (2a, 2c), a stall between these two intervals (2b), and a gradual decay phase (3). These features are clear in some other events too \citep[e.g., Figure 2 of][which is reproduced in the middle, right of Fig.~\ref{fig:lcexamples} here]{Kowalski2013}, though the respective phases may have different durations, relative amplitudes, and integrated energies.   High-time resolution observations of stellar flares were actually routine in the 1970s and 1980s using photometers, which have been superseded by high efficiency CCDs, culminating in the unprecedented precision from \emph{Kepler}, K2, and TESS.   The short timescale variations in stellar flares have actually long been recognized.  The seminal study of \cite{Bopp1973} aptly summarized their findings as follows:   ``As the time resolution of observations has improved, the great complexity of the
flare phenomenon has been revealed. The classical definition of a flare (i.e., rapid rise
to maximum followed by a slower quasi-exponential decay) appears to be a gross oversimplification of the complex structures observed''.

\section{Solar Flares and the Standard Flare Model} \label{sec:solar}

Only the very largest solar flare events could be detected in the optical if the Sun was at a similar distance as other stars.  Recently, Kepler has provided the precision to detect a signal of $0.0170$\%$- 0.0270$\%  in optical enhancements that have been observed in Sun-as-a-star data \citep{Woods2004, Kretzschmar2011, Moore2014} and from slowly rotating G dwarfs \citep{Maehara2012, Notsu2019, Okamoto2021}.   Nonetheless, it is generally thought that flares from other stars originate from the same or similar processes as solar flares.  This is supported by the empirical Neupert effect \citep{Neupert1968}, which was first reported in stellar flares in  \cite{Hawley1995} and \cite{Gudel1996} (Sect.~\ref{sec:neupert}).  In this section, we briefly review the standard solar flare model paradigm, whose phenomenology (e.g., footpoints, loops) and fundamental physical processes are widely adopted in the interpretation and analyses of stellar flares -- either through direct application or through some sort of physical scaling to higher densities, magnetic fields, accelerated particle fluxes, etc...  
 This section synthesizes decades of observations and theoretical work, and it presents our own (rather highly simplified) viewpoint of the entire process.  Thus we give a non-exhaustive reference list.   For more extensive reviews of solar flares, see \cite{Svestka1976}, \cite{Hudson2007}, \cite{Hudson2011}, \cite{Benz2017},  \cite{Hudson2021}, the entries within the \emph{Space Science Reviews} Volume \citep{Dennis2011}, the books by \cite{Aschwanden2004Book} and \cite{Tandberg2009}, and the review by \cite{Shibata2011}.  For modern, comprehensive reviews of the observations and modeling of solar flares, see \cite{Fletcher2011}, \cite{Reeves2022}, \cite{Kerr2022}, and \cite{Kerr2023}.

\begin{figure*}
\begin{center}
 % \includegraphics[width=1.0\textwidth]{../sdo_image.png}
   \includegraphics[width=0.9\textwidth]{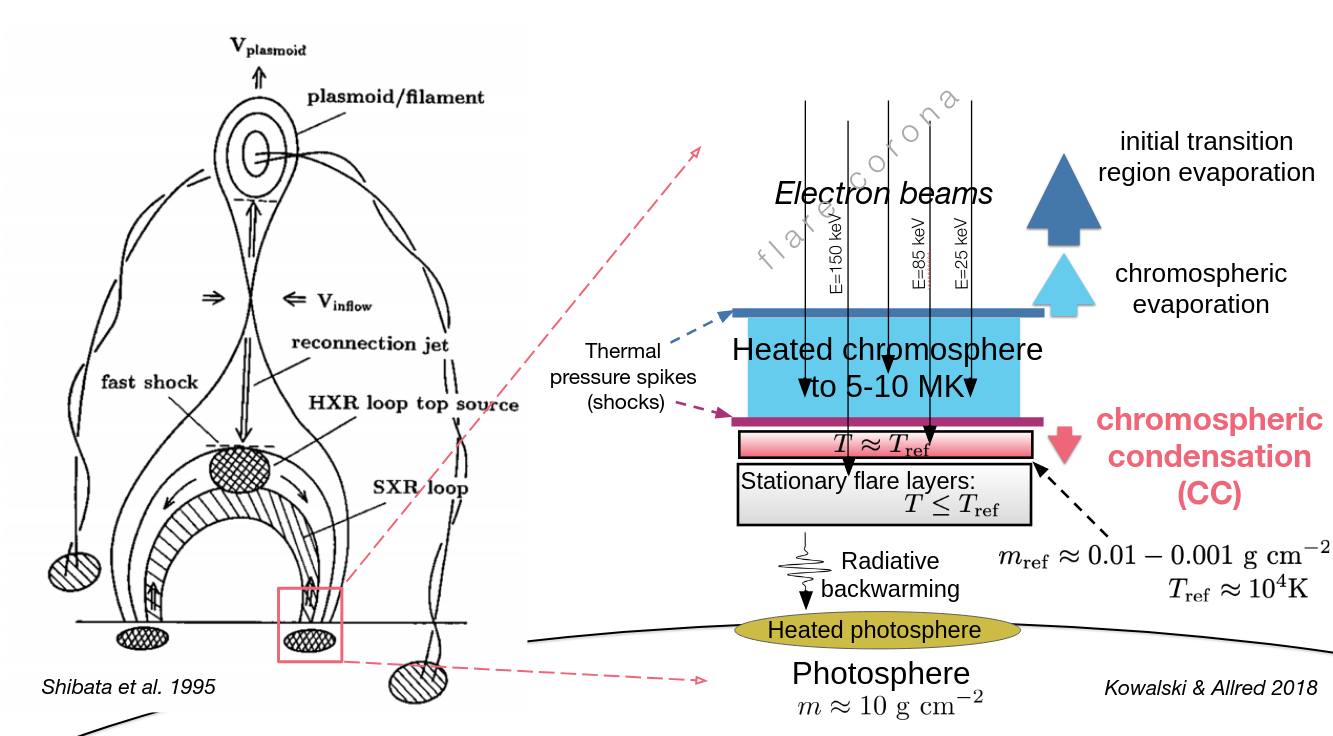}
\caption{(\emph{Left}) Illustration of several salient features and processes in the standard model of solar flares, reproduced from \citet{Shibata1995} with permission.  (\emph{Right)} Generalization of the chromospheric evaporation and condensation phenomena in radiative-hydrodynamic models of high-flux electron-beam heating at the footpoints of newly-reconnected flare loops \citep[adapted from][]{KA18}. Here, we also include an illustration of the radiative backwarming of the photosphere, which is predicted by some 1D  models \citep{Allred2006} of flare chromospheres that are optically thin at Balmer and Paschen continuum wavelengths \citep[see][for the expected geometry of 3D radiative backwarming of the photosphere]{Fisher2012}.  }
\label{fig:standardcartoon}     
\end{center}
\end{figure*}

In the standard flare paradigm, magnetic potential energy  is transferred to the atmosphere, which responds by radiating away this energy as the flare.  There are generally two categories of solar flares \citep{Thalmann2019, Kazachenko2023}.  Mass and magnetic field are ejected away from the Sun during eruptive flares.  In contrast, magnetized plasma is not ejected in confined, or compact, flares   or the eruption may be undetectable.    The collective process of the mass eruption and the electromagnetic flare is called a ``solar eruptive event''  (SEE).  The left panel of Figure \ref{fig:standardcartoon}
illustrates the geometry of eruptive magnetic field above compact flare loops, with reconnection of magnetic fields in between \citep{Shibata1995}.  This framework underlies most modern generalizations of flare-productive magnetic field topologies \citep[e.g., Figure 1 of ][]{Kazachenko2022}, which we now describe in further detail.

An SEE begins in active regions where there are colliding, non-conjugate\footnote{Meaning that they are not N and S poles that are connected by the same field lines.} sunspots of opposite polarity \citep{Chintz2019, Toriumi2019, Rempel2023}.  The  demarcation that separates the bulk of the north and south polarities in an active region is termed the polarity inversion line (PIL).  At the PIL, mixed negative and positive magnetic polarities appear as ``salt-and-pepper'' patterns in magnetograms where there is newly emerging magnetic flux.  Ongoing magnetic flux emergence adds to a twisted, cool filament (flux rope) that is parallel to the PIL, and it is thus driven to an instability \citep{Lin2001}.  The start of the SEE occurs when this filament  becomes unstable and erupts through overarching magnetic field into interplanetary space, developing into a coronal mass ejection (CME).  The CME produces a shock that accelerates a large number of solar energetic particles \citep[SEPs;][]{Reames2021} away from the Sun;  these particles may reach Earth within $\approx 10$ minutes of the X-ray flare, peak about $\approx 6-12$ hours after, and persist for days.   Another class of SEPs are prompt or impulsive-type SEPs, which originate from the flare site.  The arrival of the CME itself within $\approx 18 - 50$ hours of the flare  disturbs the terrestrial magnetic field (if the orientation of the CME magnetic field is opposite that of the Earth's magnetic field upon arrival), induces DC electric fields and currents in the ground, and increases the particle flux into the poles and radiation belts.  The ground-induced currents can damage transformers if power is not diverted away from high-load regions.  

%;

Returning to the Sun, the solar magnetic field lines that initially arch over a filament current channel \citep[which is relatively low-lying in the corona;][]{Sun2012, Rempel2023} are ``stretched''  during the eruption.   Oppositely-directed magnetic field lines pinch together in a ``current sheet'' in the wake of the erupting filament (a simplified analogy to magnetic tension and retraction is often made to a rubber band building up elastic tension and releasing it by snapping back).  The fields undergo magnetic reconnection at many so-called X-points, and the rapid speed of this process is thought to be facilitated by the tearing mode/plasmoid instability and MHD turbulence.   The magnetic potential energy in the stretched field line  is released as they shorten and retract \citep[Section 4 of][]{Longcope2018}.   

The fundamental physics of the conversion of magnetic energy into particle kinetic energy is described in the overview by \cite{Dahlin2020}.  As the retraction of field proceeds to a more relaxed state, an Alfven-ish-speed outflow (``jet''/``exhaust'') is directed from the reconnection region in the direction toward the stellar surface.   Slow-mode MHD (``Petschek'') shocks \citep[e.g.,][]{Longcope2016} and parallel electric fields \citep{Egedal2015} form in the just-reconnected, bent field lines, enhancing the temperature and density in the reconnection outflow.  Thus, the magnetic potential energy initially stored in the contorted magnetic field is converted into kinetic energy of bulk flows, the thermal energy of the plasma, and the kinetic energy of particles out of the ambient/thermal distribution.    %   Alfven wave pulses \citep{Fletcher2008} may generate a significant Poynting flux into the lower atmosphere as well.  

In large solar flares, a sizable fraction of the magnetic energy that is released seems to go into accelerating ambient particles (electrons, protons, ions) to mildly relativistic energies \citep{Lin2003, Emslie2012, WarmuthMann2016, Warmuth2020}.  The nonthermal (``beam'') distribution is a power-law extending from $\sim 10$ keV to tens of MeV for electrons and $\sim 1 - 1000$ MeV for protons.   Recent observations have shown that the Sun is a prolific particle accelerator:   $\approx 1 - 100$\% of the pre-flare coronal electrons are accelerated \citep{White2003, Kundu2009, Krucker2010, Oka2013, Krucker2014, Fleishman2022, Kontar2023}.  The inferred nonthermal particle flux density into the lower atmosphere can attain a correspondingly extreme range,  $\sim 10^{12} - 10^{13}$ \fnum, at some locations in solar flares \citep{McClymont1986, Canfield1991, Wulser1992, Neidig1993, Krucker2011}.

Determining how and where the bulk of nonthermal particles is accelerated in the solar atmosphere is actually a very active research topic \citep[for an overview, see][]{Zharkova2011}.  Generally speaking, there are three classes of particle acceleration that may operate in the solar atmosphere \citep[][and see also \citealt{Knuth2020} and \citealt{Chen2020} for succinct overviews of the more modern classifications]{Aschwanden2004Book}:  first-order Fermi and shock (including shock-drift and diffusive shock), stochastic (``AC'', or resonant interaction with a spectrum of magnetic turbulence), and  electric field (``DC'') acceleration.  Many theoretical frameworks roughly fall under one of these types in one way or another, and all may operate to some degree over the volume and time-evolution of a flare.

In and around the volume of reconnection and its outflow jets, there have been exciting advances in producing power-laws from multiple first-order Fermi reflections (through curvature-drift motions of particles) among volume-filling plasmoids, which are contracting and coalescing magnetic ``islands'' \citep{Shibata2001, Drake2006, Oka2010,  Karpen2012,Drake2013, Dahlin2014, Guidoni2016}.   These models have recently been extended to 3D \citep{Dahlin2015} and expanded to much larger physical sizes \citep{Drake2019,Arnold2021} than particle-in-cell simulations, demonstrating efficient acceleration of high-energy electrons into power-law tails that extend over many orders of magnitude.   The Alfvenic-speed reconnection exhaust may develop turbulence, and stochastic acceleration within turbulence  \citep{Miller1996, Petrosian2004, Petrosian2012, Bian2012} in the presence of Coulomb collisions \citep{Bian2014} is thought to be one of the possible acceleration mechanisms\footnote{Stochastic acceleration in gyro-resonant interaction with Whistler-modes of an inertial Alfven-wave turbulent spectrum in the lower atmosphere has also been suggested \citep{Fletcher2008}.}.   Sub-Driecer \citep[e.g.,][]{Holman1985, Benka1992, Alaoui2021} electric fields have also been investigated in accelerating a small fraction of the ambient particles, and super-Driecer electric fields (accelerating the entire ambient population) that result from the observed decay of magnetic field over a large loop-like volume have also been discussed \citep{Fleishman2022}, possibly related to a type of hybrid particle acceleration involving electric fields and turbulent reconnection \citep{Vlahos2019}.    Despite the  apparent lack of consensus among these impressive modeling and theoretical directions, a commonality is that  a significant amount of energy is transferred to the atmosphere through shortening of magnetic field lines (facilitated by reconnection), and particles are efficiently accelerated over a large volume.   For reviews of magnetic reconnection and particle acceleration, see also \cite{Uzdensky2007, Cassak2008, Priest2014, Cassak2017, Pontin2022, Arber2015, Lazarian2020, Nishikawa2021, Ji2022}.

We should mention that acceleration at a coronal termination shock where the reconnection outflow/exhaust collides with the lower-lying (previously reconnected) magnetic loops could also be important \citep[e.g.,][]{Somov1997}, and plasmoid interaction with a fast-mode MHD shock at the collision interface could contribute to the accelerated particle flux into the lower atmosphere \citep{Shibata2001}.  Acceleration based on terrestrial aurorae or magnetotail sub-storm processes are also considered \citep{Birn2017, Haerendel2018}, and some models argue for a mechanism where the bulk of electron acceleration takes place close to \citep{Simnett1990, Tsiklauri2017}  or within \citep{Fletcher2008} the chromosphere.  Whatever the case or cases may be, acceleration models should attempt to reconcile one way \citep[e.g.,][ and see also Section 3.6 in \citealt{Aschwanden1996Masuda}]{Brown1998} or another \citep{Chen2012} with energy-dependent time delays of hard X-rays \citep{Aschwanden1995, Aschwanden1996, Aschwanden1996Masuda, Aschwanden1996Conf, Qiu2012, Alyz2019, Knuth2020} and the general agreement between inferred path lengths and direct imaging measurements of flare loop sizes.  Other critical tests are matching the properties of the white-light continuum radiation \citep{Fletcher2007}, reproducing the energy content in accelerated protons \citep[e.g., $E \approx 7 \times 10^{32}$ erg above a proton kinetic energy of 30 MeV;][]{Murphy1997}, and generating strong nonthermal electron radio signatures at coronal heights \citep{Chen2020}.

In order to consider the rest of the flare process, let us assume that the nonthermal particle beams are injected into a region around the apex of a just-reconnected-and-retracted coronal loop. Initially, the particles are injected in relatively low-lying magnetic fields ($\lesssim 10$ Mm), resulting in the onset of the flare impulsive phase.  Over time, larger (and weaker) magnetic fields reconnect and energize particle beams.  The particle beams are injected into the newly reconnected loops with a distribution of pitch angles with respect to the magnetic field orientation.  The $E \gtrsim 100$ keV electrons in the beams produce gyrosynchrotron radiation as they spiral (with Larmor radii of $\sim$ cm) in the magnetic fields.  If the magnetic flux density [G] increases along the beam path (e.g., if the cross-sectional area of the loop correspondingly converges into the lower atmosphere), then the particles with small pitch angles with respect to the magnetic field direction will freely stream (``precipitate'') into the chromosphere.  Those with larger initial pitch angles will gradually and adiabatically increase their pitch angle, which is a result of Faraday's Law of Induction in the frame of the gyrating electrons.  These particles can eventually reflect, or ``mirror'', at the footpoints (which could occur in the transition region, high chromosphere, low chromosphere, or low corona)\footnote{The particles can also mirror as the loops retract after they reconnect;  see \cite{Aschwanden2004}, and they may also exhibit non-adiabatic/chaotic orbits \citep{Egedal2015}.}.   After some time, the mirrored particles too will scatter off enough ambient particles in the corona and precipitate out of the magnetic ``trap'' into the chromosphere.   This is the so-called ``trap$+$precipitation'' paradigm \citep{Melrose1976}.  The highest energy particles take the longest  to scatter and thus remain trapped  for a time proportional to $\frac{E^{1.5}}{n_e}$ where $E$ is the kinetic energy of the nonthermal electrons and $n_e$ is the ambient/thermal electron density that the beam heats (under the dilute beam assumption: $n_{\rm{e}} \gg n_{\rm{e-beam}}$).  These times are rather short ($< 10$~s) for typical conditions and energies.  In solar flares, proton/ion beams can certainly contribute to atmospheric heating \citep{Ondrej2, Allred2020}, nuclear excitation, a gamma-ray continuum spectrum, the 2.2 MeV neutron-capture Deuterium-formation line, the 511 keV positron-annihilation line, and neutrons that are directly detected at Earth \citep{Murphy1997,Vilmer2011}.   For the sake of brevity, we further consider the effects of only accelerated electrons.

When the electron beam particles stream into the chromosphere, they encounter a wall of dense, partially ionized gas to which they rapidly lose their energy through Coulomb collisions with ambient electrons.  At the same time, free-free radiative transitions in collisions with ambient protons occur in the chromosphere, producing hard X-ray ($E \gtrsim 25$ keV), nonthermal bremsstrahlung radiation that exhibits a power-law distribution.  Hard X-ray light curves define the impulsive phase in solar flares.    The standard model of hard X-ray emissivity that accounts for Coulomb energy loss as the electrons radiate bremsstrahlung is called the collisional (cold) thick target model \cite[CTTM;][for reviews, see \citealt{Brown2003, Holman2011, Kontar2011}]{Brown1971}.  Electron beam power-laws that are inferred from the CTTM and injected into radiative-hydrodynamic simulations are typically characterized by power-law indices of $\delta \approx 5$, a low-energy cutoff of $E_c \approx 15-25$ keV, and energy flux densities in the range of $\approx 10^{10}$ to $\approx 10^{11}$ \fnum\ \citep{Carlsson2023}.  
Often, the hard X-ray light curves consist of gradual variations with superimposed short bursts of $\mathcal{O}(0.5-5)$~s.  In some interpretations, these timescales are sensibly related to the processes of  prompt and delayed precipitation (described in the previous paragraph) of nonthermal electrons.  

The relative displacement of the accelerated electrons from the slower ions in the corona results in a ``return current electric field'' \citep[e.g.,][]{Oord1990, Siversky2009};  this electric field drains energy from the beam and transfers it as a drifting velocity component to the ambient Maxwellian distribution of electrons.  The drift is towards the loop apex.  In steady state, the flux densities ($J$; el s$^{-1}$ cm$^2$) of the beam and return current drifting electrons are equal and opposite, thus preventing pulsar-strength, $B \approx 10^9$ G, magnetic fields from forming in the solar atmosphere.  The resistivity of the background results in ``Joule heating'' of the coronal plasma \citep[$\approx \eta  e^2  J_{\rm{beam}}^2$;][]{Holman2012, Allred2020, Alaoui2017}.  For non-dilute beams, the density of the electron beam is comparable to the ambient electron density, and interactions of plasma wave turbulence and parallel electric fields (e.g., Langmuir waves, electrostatic double layers) with the beam are additionally expected.

The nonthermal electron beam kinetic energy is thermalized in the low corona and mid-to-upper chromosphere; see right side of Figure \ref{fig:standardcartoon}.  
Initially hydrogen primarily provides the radiative cooling that balances the chromospheric temperature increase.
As the chromosphere increases from $T \approx 8000$ K to $T \approx 30,000$ K, hydrogen becomes fully ionized, and then helium I and helium II take over the radiative cooling in the range of $T  \approx 30,000 - 80,000$ K.  Thus, it is thought that non-equilibrium rates of helium and detailed radiative transfer are important for calculating accurate evolutionary states of the atmosphere.  After helium is fully ionized, oxygen, carbon, and neon regulate the cooling at $T \approx 10^5$ K, which is the peak of the optically thin cooling curve \citep[e.g.,][]{Cox1969, RTV78, Dere1997, Dorfi1998}.   If enough heat is deposited to fully ionize these elements, then a temperature runaway, or explosion, ensues because the optically thin cooling decreases as a function of temperature up to $T \approx 20$ MK.  Thermal conduction transports energy away from the region of maximum beam deposition.  Thus, the temperature ``bubble'' expands, resembling a one-dimensional blast wave (which is more or less confined to the magnetic field direction) in an exponential atmosphere with pervasive beam heating.  The increased pressures and their gradients result in a multitude of mass motions.   Additionally, an increase in temperature of the very low corona (through thermal heat conduction) can also drive mass motions on top of the beam-generated mass motions.  The sources and sinks of the equation for energy conservation  are discussed further in Sect.~\ref{sec:rhdphysics}.

The impulsively-generated  upflows and downflows have been studied numerically as part of the standard solar flare paradigm for many decades \citep{Livshits1981, Fisher1985V, Fisher1985VI, Fisher1985VII, Fisher1989}.  They are respectively known as explosive chromospheric evaporation and chromospheric condensation (Figure \ref{fig:standardcartoon}, right)\footnote{The terms ``condensation'' and ``evaporation'' are used even though they are misnomers because there are no better words in the English language for the phase transitions from neutral gas to plasma and vice-versa.  ``Ionization'' and ``recombination'' could be appropriate but these are reserved for the element-dependent, microscopic atomic processes rather than the macroscopic flows and density changes. }.  The condensations and evaporations manifest as redshifts and blueshifts in cool and hot lines, respectively, with a transition temperature that has been constrained in solar flares \citep{Milligan2009}.   The high-speed upflows fill the loops with $T \approx 10-30$ MK plasma, while the higher density downflows accrue/accrete mass over time and radiatively cool through $T  \approx 10,000$ K  (i.e., this is a non-adiabatic process).  The chromosphere is compressed as in a snow plow effect while also experiencing continuous energy deposition directly by the beam.  The condensations are a result of shock phenomena because they increase in density up to 10x the ambient density (whereas sound waves would ostensibly smooth out the density perturbations).  The closest physical process to the temperature bubble and condensation that we have found \citep[e.g.,][]{Kowalski2022} is that of a thermal wave \citep[Section 6.6, pp 671-672 of ][and see also \citealt{Atzeni2004} for similarities to supersonic ablative heating waves in laboratory implosion experiments]{ZeldovichRazier}, which are described as a ``second-kind temperature wave'' in \citet{Livshits1981}.

The representative properties of the condensation and evaporation flows at any time in the evolution follow from a simple momentum balance equation:
 
 \begin{equation} \label{eq:momentum}
\left( v \rho \Delta z \right)_{\rm{cond}} \approx - \left( v \rho \Delta z \right)_{\rm{evap}},
\end{equation}
 
 \noindent where $\Delta z$ is a physical depth range over which gas with a mass density of $\rho$ has a bulk velocity of $v$.  
 
Here, we ignore the downward momentum of proton and electron beam particles \citep{Ichimoto1984, Allred2015}.
 Typical values of the condensation are $\Delta z \approx 30$ km \cite[which scales inversely with surface gravity;][]{KA18} and $v \approx -50$ km s$^{-1}$.  Typical evaporation flows are fully ionized and have $n_e \approx 10^{11}$ cm$^{-3}$ ($\rho \approx 3 \times 10^{-13}$ g cm$^{-3}$) with flow speeds of $v \approx 100-500$ km s$^{-1}$.  The density of the condensation increases over time, which is compensated by an increase in the $\Delta z$ of the evaporation as chromospheric/transition region mass ablates into and fills the newly-reconnected magnetic loop.  After the flows fill up a $\Delta z \approx 10$ Mm loop, then Equation \ref{eq:momentum} implies  a gas density of $10^{-9}$ g cm$^{-3}$ in the condensation, in agreement with numerical simulations\footnote{In reality, the densities and velocities have large gradients in both the condensation and evaporation, so these representative, average values are used only to elicit some basic insight.}.   The coronal loops thus emit in soft X-rays, while the footpoints emit in optical and UV radiation.    The condensation manifests as red-shifts in broad, chromospheric lines (Fe II, Mg II, H$\alpha$, Si II) that exhibit a red-wing asymmetry  (RWA).  The RWA evolves in velocity and intensity over $\approx 30$~s \citep{Ichimoto1984, Zarro1989, Wulser1992, Graham2020}.  

 %  gas density associated with an electron density of $10^{11}$ cm$^{-3}$ ($\rho \approx 3 \times 10^{-13}$ g cm$^{-3}$) implies
 
 \cite{Fisher1989} calculated the analytic relationship 
 between the beam energy flux deposited above the flare transition region ($F_{\rm{evap}}$), the peak downflow (condensation) speed  ($v_{\rm{peak}}$) just below the flare transition region, and the pre-flare chromospheric mass density ($\rho_{\rm{chrom}}$) at which the flare transition region forms:
 
 \begin{equation} \label{eq:fevap}
 v_{\rm{peak}} \approx 0.6 \big( \frac{F_{\rm{evap}}}{\rho_{\rm{chrom}}} \big)^{1/3}
 \end{equation}
 
 %%\vspace{1mm}

\noindent Typical upflows of $100 - 500$ km s$^{-1}$ imply durations up to $\approx 30-90$s to fill a flare loop, at which time the confined, field-aligned flows from conjugate footpoints collide.
Superhot (lower volume and lower density) $T \approx 30-50$ MK sources can be produced above the looptops of the cooler, $T \approx 20$ MK, plasma that is thought to be due to chromospheric evaporation;  these are thought to be due to heating from slow-mode (Petschek) shocks in reconnection, as demonstrated in recent multi-dimensional MHD simulations \citep[e.g.,][]{Longcope2016}.
However, some of such sources that are consistent with ultrahot $100-200$ MK thermal spectra that appear early on are generally more consistent with nonthermal power-law spectra.

\begin{figure*}
\begin{center}
 % \includegraphics[width=1.0\textwidth]{../sdo_image.png}
   \includegraphics[width=0.60\textwidth]{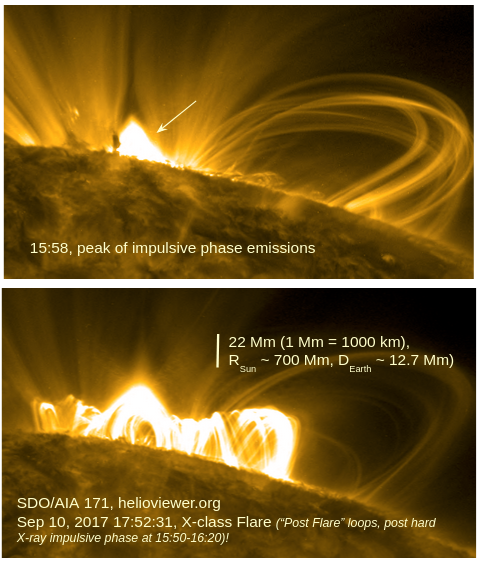}
\caption{SDO/AIA images around the EUV wavelength of $171$ \AA\ \cite{Lemen2012}, obtained from \url{Helioviewer.org}, showing  thermal emission from flare loops at two times during a large X-class flare on the solar limb.  The top panel shows the compact, low-lying thermal flare arcade (indicated by the arrow) during what is usually the peak of the nonthermal hard X-ray (not shown) luminosity from the footpoints.  Later in the flare, the bright arcade volume exhibits an expansion horizontally and vertically -- the expansion is, however, apparent in the sense that newly reconnected, larger magnetic loops participate in the flare as time progresses.  The large arcades have traditionally been called ``post-flare'' loops, which shine brightly much longer after the end of the main hard X-ray impulsive phase (they are perhaps more accurately called ``post-nonthermal-X-ray-footpoint'' loops).  It is generally thought that the flare loops in these images are the result of magnetic structures that have cooled down  from much higher temperatures to $T \approx 1$ MK \citep[e.g.][]{AschwandenAlexander2001}.  Note that a cusp-like (fork-shaped) dark region just above the bright loops in the top panel is where many models and observations suggest that the bulk of magnetic reconnection and particle acceleration occur \citep[see][]{Longcope2018, Chen2020}.  For illustrative comparisons of flare and active region loops, see \cite{Gudel2004Rev}. }
\label{fig:sdo}     
\end{center}
\end{figure*}

In the impulsive phase, bright sources are observed from the footpoints (Figure \ref{fig:standardcartoon}, right) of flare loops in cooler emission lines and in the UV, optical, and infrared continuum radiation.   A variety of source morphologies appear and could be affected by temporal and spatial resolution. However, one can typically separate the geometry into elongated ``ribbons'' and compact, brighter circular ``kernels'' \citep[see, for example, the remarkable H$\alpha$ images in ][]{Kawate2016}.
Within a relatively short time, the footpoints light up quasi-sequentially along the PIL, forming flare ribbons \citep[e.g.,][]{Qiu2017, Kazachenko2022}.  One ribbon typically develops in plage\footnote{Plage is typically considered non-spot, but highly magnetized, locations of the non-flaring active region Sun that is bright in the core of H$\alpha$}, and another ribbon develops in a penumbral or umbral region of sunspot.
As the flare progresses, the distance between the two ribbons increases, which is the so-called perpendicular apparent motion.  This is thought to be a signature of the reconnection occurring higher and higher in the corona, thus releasing energy from larger and larger loops, which is a hallmark of the famous ``CSHKP'' model of two-ribbon flares \citep{Carmichael1964, Sturrock1966, Hirayama1974, Kopp1976}.   As a result, the brightest emission is generally seen at the ``leading bright edge'' of the ribbons as they spread apart (Figure \ref{fig:standardcartoon} left).  It is thought that excitation from above, e.g. through particle beams, changes location rather than through cross-field thermal heat transport / diffusion in the low atmosphere.  However, radiative backwarming from the X-ray, EUV, and FUV lines may play an important role in heating the surrounding atmosphere away from the brightest kernels \citep[such as in the so-called ``core-halo'' morphology; ][]{Neidig1993, Allred2006, Isobe2007, Fisher2012, Namekata2022}.  After the excitation front passes, the impulsively-generated chromospheric condensation continues to propagate even in the absence of electron beam heating, but it eventually reaches pressure equilibrium with the lower atmosphere.  The loops cool from tens of MK to 1 MK, first primarily through thermal conduction,  then primarily through radiation, resulting in the appearance of the famous ``post-flare'' loops  as seen in filtergrams such as AIA 171 \AA\  (which are, really, post-impulsive phase loops; see Fig.~\ref{fig:sdo} and \citealt{Liu2013}).  In cooler lines such as H$\alpha$ or He II 304, the very late phase of each flare loop is sometimes associated with the phenomena of coronal rain, which is a second phase of ``condensation''.   In this phase, dense material that was chromospheric before the flare drains out of the corona.  New flare loops are continuously formed through the gradual decay phase of GOES soft X-ray solar flares \citep{Warren2006}, but the details of the partition among various possible heating sources in the low atmosphere (thermal heat flux\footnote{Thermal conductivity during the flare cools the corona and heats the top of the chromosphere.}, electron beams, proton beams, direct heating by waves) remain poorly understood, especially during stellar flares.  We note that the origin of the long cooling times of post-flare loops on the Sun is still a very active area of research \citep[e.g.,][]{Ryan2013, LiuQiu2013, Qiu2016, Bian2018, Zhu2018, Kerr2020, Reep2022, Allred2022, Ashfield2023Alf}.

Several notable and well-studied solar flares are the SOL1992-Jan-13T17:25 M2.0 flare \citep[``the Masuda flare'';][]{Masuda1994}, the SOL2000-Jul-14T10:00 X5.7 flare \citep[``the Bastille Day flare''\footnote{Referred to colloquially as ``the slinky flare'' within the stellar community.}, e.g.,][]{Qiu2010}, the SOL2003-Oct-28T11:30 X17 flare \citep[one among the ``the Halloween storms'', e.g.,][]{Woods2004}, the SOL2002-Jul-23T00:30 X4.8 \citep[ ``double power-law'';][]{Holman2003} flare, the SOL2014-Mar-29T17:48 X1.0 flare \citep[NASA's ``best observed X-class flare''; e.g.,][]{Kleint2016}, the SOL2014-Sep-10T17:45 X1.6 flare \citep[e.g.,][]{Graham2015}, and the SOL2017-Sep-10T16:06 X8.2 flare \citep[e.g.,][]{Chen2020}.

\section{ A Survey of Flare Stars} \label{sec:overview}
\begin{figure*}
\begin{center}
   \includegraphics[width=0.60\textwidth]{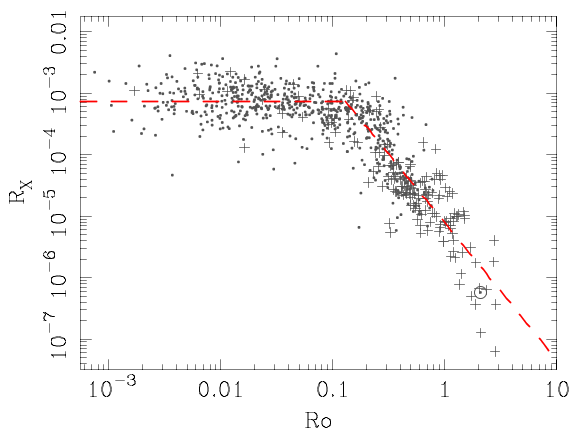}
\caption{ Quiescent $R_X = L_X / L_{\rm{bol}}$ \emph{vs.} the Rossby number for a large sample of stars, reproduced  from \citet{Wright2011} with permission.  $R_X$ is ``saturated'' at $\approx 10^{-3}$ for stars with Rossby numbers $\lesssim 0.1$, and $R_X$ decreases according to a power-law for lower-activity stars.  }
\label{fig:saturatedstars}     
\end{center}
\end{figure*}

 The characteristics that lead to flaring in stellar atmospheres are generally thought to be some combination of the following:  rapid rotation, an outer convective zone, and disorganized surface magnetic fields.  The ratio of stellar rotation to the convective turnover time\footnote{This is derived from models or theory \citep[e.g., see][and references therein]{Magaudda2020}.  One qualitative reasoning for M stars being such prolific flare stars early in their lives ($\lesssim 500$ Myr) is that for stars born with a large amount of angular momentum and rapid rotation, the low-luminosity of these stars facilities slow energy escape through convection and thus long convective turnover times.  Very low-luminosity main sequence stars, like Proxima Centauri, thus, may maintain high rates of flare activity at relatively long rotation periods (and old ages) because this ratio stays at a small value for a long time. } is called the Rossby number, or the inverse Coriolis number, which may be indicative of how much internal shear, differential rotation, and turbulent amplification of kinetic energy into magnetic energy contribute to the dynamo.  The dynamo may operate in the shear flows around the tachocline interface, and/or throughout the turbulent convection zone, and/or in the near-surface shear layers.   Stars with small Rossby numbers are found in the so-called ``saturated regime'' of quiescent magnetic activity (outside of large flares).  The saturated regime is empirically characterized by nearly constant luminosity ratios of $L_X/L_{\rm{bol}} \approx 10^{-3} $ (Figure \ref{fig:saturatedstars}) and $L_{H\alpha}/L_{bol} \approx 10^{-4}$ \citep[e.g.,][]{Pizzolato2003, Wright2011, Reiners2014,  West2015,Wright2016, Newton2017, Brun2017,Wright2018}, where $L_{\rm{bol}}$ is the quiescent bolometric luminosity, and the numerator is a luminosity \citep[e.g., the soft X-ray luminosity, $L_X$;][]{Schmitt2004, Wright2011} that is a proxy for magnetic heating in the corona  or chromosphere\footnote{Saturated regimes are also seen in many other quiescent magnetic activity indicators, such as Ca II \citep{CaIIKsatur1, CaIIKsatur2} and the photospheric magnetic field \citep{Vidotto2014, Reiners2022}.}.   There is a rather large spread about the saturated value from star to star, and the Sun varies from $L_X/L_{\rm{bol}} \approx 5 \times 10^{-8}$ to $\approx 5 \times 10^{-7}$ over its 11 year sunspot cycle \citep{Judge2003, Ayres2015Alpha}.  For comparison, the least active M dwarfs have ratios of $10^{-6}$ \citep{Wright2016, France2020, Brown2023, Engle2023}.  At very small Rossby numbers, a super-saturated regime may occur, while stellar activity proxies follow a power-law with increasing Rossby numbers $\gtrsim 0.1$, relatively independent of spectral type\footnote{The Sun's Rossby number is often discussed as being around a value of two, but it is not widely accepted whether it is even meaningful to characterize the Sun with a single number; further, the Rossby number itself is a model-dependent parameter.  }  For an introduction to dynamo theory in the stars and the Sun, see \cite{Brandenburg2005}, \cite{Brandenburg2005B}, \cite{Charb2013}, \cite{Brun2017}, and \cite{Schrijver2019}. For a review of the literature, see \cite{Brun2017} and \cite{Kapyla2023}.  Much recent progress has been made in modeling stellar dynamos in low-mass stars just below \citep{Bice1, Bice2} and above \citep{Browning2008, Brown2020, Kaplya2021} the fully convective transition, and with especially high-resolution simulations of the solar magnetic field and differential rotation \citep{Hotta2022}.

\begin{figure*}
\begin{center}
  \includegraphics[width=0.95\textwidth]{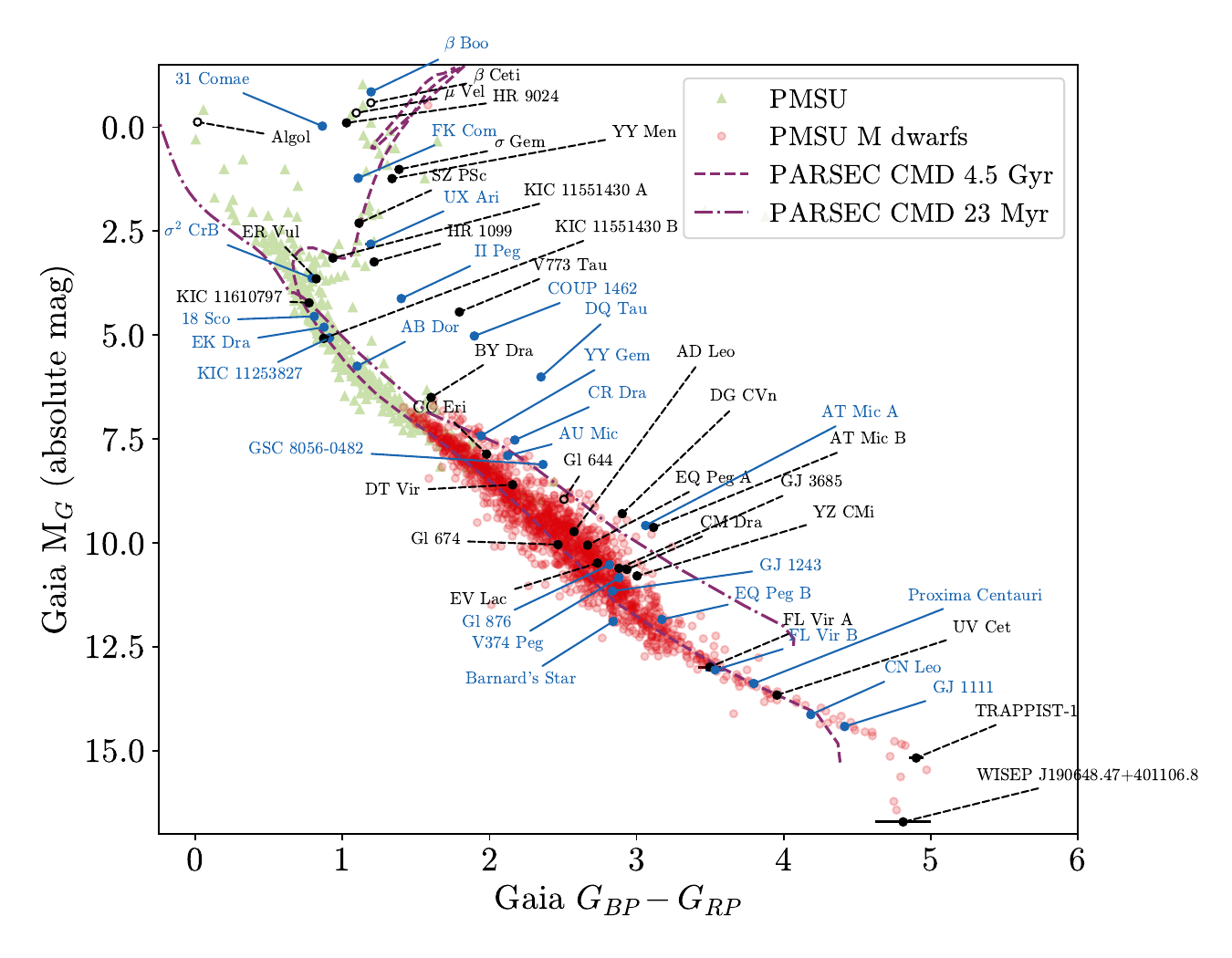}
\caption{Gaia color-magnitude diagram of absolute magnitude vs. color for notable flare stars (outside of flaring).  Note that corrections for extinction and reddening from interstellar dust are not applied because most of these stars are in close proximity (within $\sim 20$ pc) of Earth.  The linecolors and linestyles for the annotations are arbitrary.  Open circles signify that the data were transformed from Hipparcos magnitudes and/or parallaxes.  Several solar-metallicity isochrones from the PARSEC stellar evolutionary models are shown for reference. \label{fig:hrdiagram} }
      % Give a unique label
\end{center}
\end{figure*}

Magnetic fields buoyantly rise through the stellar photosphere;  once in the corona, they undergo destabilization and reconnection to release energy into the atmosphere to produce flares.  
Flaring occurs at different times through stellar evolution as convective envelopes come and go with changes in internal structure, core fusion processes, and external factors such as tidal locking and synchronous rotation.  An overview of the myriad of flare stars is best summarized in a color-magnitude diagram (CMD).
Prolific and/or representative flare stars are shown on a Gaia CMD in Fig.~\ref{fig:hrdiagram}.  The data were obtained from Gaia Data Release 2 \citep{GaiaDR2}, except for a few cases that are only available through the Early Data Release 3 or through common filter transformations \citep{Hipparcos2007} for some of the brightest stars.  The background data are the Gaia photometry for a collection of stars compiled for the Palomar/Michigan State University spectroscopic survey \citep[PMSU;][]{PMSU1,PMSU2,PMSU3,PMSU4,NLDS}, downloaded from Neill Reid's website\footnote{\url{https://www.stsci.edu/~inr/cmd.html}}.  Many of these stars are discussed in the recent 10 pc Gaia sample \citep{Gaia10pc} and the CNS5 catalog \citep{CNS5}.  Isochrones from the PARSEC v1.2S \citep{PARSEC1} stellar evolution models are shown for solar age and metallicity and the age of the $\beta$ Pic moving group \citep{Mamajek2014}.   Characteristics\footnote{We have tried to update properties with some ``standard'' values as best as possible within the time frame of writing this review but we have certainly missed some notable flare stars.  Spectral types of low-mass stars are determined to only about $\pm$0.5 spectral types using broadband flux spectral-typing facilities \citep{Covey2007}.  For example, \cite{Davison2015} quote a spectral type of M4.0Ve for YZ CMi, whereas spectral typing with the Hammer facility can result in determinations closer to M5 (and similar differences result for the M0/M1 star AU Mic).  \cite{Gaia10pc} discuss the evolutionary status of YZ CMi, a notorious flare star, as a candidate pre-main sequence star with an age of around 24 Myr.  Gyrochronology suggests an age range of $810 \pm^{60}_{56}$ Myr \citep{Engle2023}; the large intrinsic age spread at a given rotation period, especially at young ages for low-mass stars, is widely-recognized \citep[e.g.,][]{Irwin2011, EngleGuinan2023}.  There is a correspondingly large range of radius estimates for this star found in the literature \citep{Morin2008, Baroch2020}.  Aside from coordinates, parallaxes, and proper motions, one should use caution for some fundamental stellar parameters (binarity, surface gravity, evolutionary status) obtained from large catalog databases such as SIMBAD and Gaia.  For example, the Gaia DR3 archive returns an inaccurate surface gravity of log $g$/(cm s$^{-2})= 4.0$ for the notorious flare star AD Leo (as discussed by \cite{Hawley1999} and \cite{Davison2015}, surface gravities determined from spectra of active low mass stars are unreliable).  } of selected flare stars in the CMD are summarized in Tables \ref{table:flarestars1} - \ref{table:flarestars6}, and in some cases a reference to an example of a flare study featuring the respective star.

\clearpage
\begin{sidewaystable}
%%\sidewaystablefn%
\begin{center}
\begin{minipage}{\textheight}
\caption{Table of common and notable flare stars.}
\label{table:flarestars1}
\begin{tabular*}{\textheight}{@{\extracolsep{\fill}}lccccccl@{\extracolsep{\fill}}}
\toprule%
%& \multicolumn{3}{@{}c@{}}{Element 1\footnotemark[1]}& \multicolumn{3}{@{}c@{}}{Element\footnotemark[2]} \\\cmidrule{2-4}\cmidrule{5-7}%
Name & SpT & Class &  $d$  & Gaia $G$ & $U$ & RA, Dec & Ref \\
 &  &  & (pc) & (app mag) & (app mag) & (J2000) &  \\
\midrule
\hline
dMe / UV Cet & & & & & & & \\
\hline
YY Gem (Gl 278C) & M0+M0 & BY Dra/EB & 15.0984 & 8.3 & 11.6 & 07 34 37, +31 52 10 & \cite{Doyle1990B},\cite{Stelzer2002} \\
DT Vir (Gl 494) & M1 & dMe & 11.5132 & 8.91 & 12.3 & 13 00 46, +12 22 32 &  \\
AU Mic (Gl 803)$^a$ & M0/M1 & (d)Me & 9.7248 & 7.84 & 11.3  & 20 45 09, -31 20 27 & \cite{Cully1994}, \cite{Tristan2023}  \\
CR Dra (GJ 9552)$^b$ & M1  & dMe & 20.2626 & 9.06 & 12.0 & 16 17 05, +55 16 08 & \cite{Welsh2006}\\ %  , Res.\ by \cite{Tamazian2008} 
SDSS J211827.00 & M1 & dMe/S82 & 711.9443 & 17.01 & \nodata & 21 18 27, +00 02 48 & \cite{Kowalski2009} \\
\hspace{2mm}  +000248.5$^c$ &   &   &   &   &   &   &   \\
SDSS J005903.60- & M1 &  dMe/S82 & 123.2745 & 13.89 & \nodata & 00 59 03, -01 13 31 & \cite{Kowalski2009} \\
\hspace{2mm} 011331.1 & &  &  &  &  & &  \\
Gl 411 & M2 & dMe & 2.55 & \nodata & 10.1 & 11 03 20, +35 58 11 & \cite{Audard2000} \\ 
VV Lyn (Gl 277A)  & M2.5 & dMe & 11.86 & \nodata & \nodata & 07 31 57, +36 13 09 &  \\
%V1054 Oph (Gl 644AB) & M3 & dMe/ &     &         &         &                     &   \\
AD Leo (Gl 388) & M3 & dMe & 4.966 & 8.21 & 11.9 & 10 19 36, +19 52 12 & \cite{HP91}  \\
GQ And (Gl 15B) & M3.5 & dMe & 3.5614 & 9.68 & \nodata & 00 18 25, +44 01 38 &  \\
EQ Peg A (Gl 896A) & M3.5 & dMe & 6.2614 & 9.04 & 13.2 & 23 31 52, +19 56 14 &  \cite{Liefke2008}, \cite{Kowalski2016} \\
EV Lac (Gl 873) & M3.5 & dMe & 5.0502 & 9.00 & 13.0 & 22 46 49, +44 20 02 & \cite{Osten2005} \\
GJ 3685 & M4 & dMe & 18.7988 & 11.98 & \nodata & 11 47 40, +00 15 20 &  \cite{Robinson2005} \\
GJ 1243 & M4 & dMe & 11.9787 & 11.55 & 15.5 & 19 51 09, +46 29 00 &  \cite{Hawley2014} \\
DG CVn (GJ 3789) & M4+M4 & (d)Me & 18.2857 & 10.61 & \nodata & 13 31 46, +29 16 36 &\cite{Osten2016} \\ %    0.\arcsec 2 sep., 
BL Lyn (Gl 277B) & M4 & dMe & 12.001 & 10.47 & \nodata & 07 31 57, +36 13 47 &  \\
V374 Peg (GJ 4247) & M4 & dMe & 9.1041 & 10.63 & \nodata & 22 01 13, +28 18 24 &  \cite{Vida2016}  \\
YZ CMi (Gl 285) & M4.5 & dMe & 5.9874 & 9.68 & 13.8 & 07 44 40, +03 33 08 &  \cite{Kowalski2010} \\
EQ Peg B (Gl 896B) & M4.5 & dMe & 6.2477 & 10.82 & $\approx$15 & 23 31 52, +19 56 14 & \cite{Robrade2004}, \cite{Mathioudakis2006} \\
AT Mic A & M4.5 & (d)Me & 9.8816 & 9.56 & \nodata & 20 41 51, -32 26 06 &  \cite{Garcia2002} \\
AT Mic B & M4.5 & (d)Me & 9.8312 & 9.59 & \nodata & 20 41 51, -32 26 10 &  \cite{Garcia2002}\\
\bottomrule
\end{tabular*}  
\footnotetext{The Gaia $G$ and Johnson/Bessel $U$-band magnitudes are apparent magnitudes, and the Gaia $G$ mags are not adjusted for galactic dust extinction.  Note that many of the stars have large proper motions,
and truncated J2000 coordinates are given here.  ``EB'' indicates an eclipsing binary. ``S82'' refers to a SDSS 82 flare star with very large magnitude enhancements during the observed events. $^a$The reported $U-$band range in the literature is 11.2--11.4 \citep{Leggett1992, Tristan2023}.  $^b$CR Dra has a combined spectral type of M1, but the system consists of components with orbital parameters consistent with a M0 and an M3 with angular separations between $0.\arcsec 06-0.\arcsec 16$ \citep{Tamazian2008}. $^c$Produced the largest flare from an early-type M dwarf in the SDSS Stripe 82 flare star sample.  Other notable early/mid-type dMe flare stars are the $U=11.7$ M1.5e star, FK Aqr \citep{Byrne1990, Kashyap2002} and the  $U=11.6$  M3e spectroscopic binary, Gl 644AB (V1054 Oph) \citep{Doyle644, Math644, Crespo644}. }
\end{minipage}
\end{center}
\end{sidewaystable}
\clearpage

\begin{sidewaystable}
%%\sidewaystablefn%
\begin{center}
\begin{minipage}{\textheight}
\caption{Continued ... }
\label{table:flarestars2}
\begin{tabular*}{\textheight}{@{\extracolsep{\fill}}lccccccl@{\extracolsep{\fill}}}
\toprule%
%& \multicolumn{3}{@{}c@{}}{Element 1\footnotemark[1]}& \multicolumn{3}{@{}c@{}}{Element\footnotemark[2]} \\\cmidrule{2-4}\cmidrule{5-7}%
Name & SpT & Class &  $d$  & Gaia $G$ & $U$ & RA, Dec & Ref \\
 &  &  & (pc) & (app mag) & (app mag) & (J2000) &  \\
\midrule
\hline
V639 Her (Gl 669B) & M4.5 & dMe & 10.7566 & 11.46 & \nodata & 17 19 52, +26 30 02 &  \\
AF PSc (GJ 1285) & M4.5 & dMe & 34.957 & 12.81 & \nodata & 23 31 44, -02 44 39  &  \cite{Welsh2006}  \\
CM Dra (Gl 630.1) & M4.5+M4.5 & dMe/EB & 14.85 & 11.5 & \nodata & 16 34 20, +57 09 44  & \cite{Morales2009, Lacy1977} \\
GJ 1245A & M5 & dMe & 4.6919 & 11.47 & \nodata & 19 53 54, +44 24 51  & \cite{Lurie2015} \\
GJ 1245B & M5 & dMe & 4.6614 & 11.92 & \nodata & 19 53 55, +44 24 54  &  \cite{Lurie2015}  \\
FL Vir (Wolf 424 AB/Gl 473AB) & M5+M7 & dMe & 4.40 & 11.24 & 15.4 & 12 33 17, +09 01 15  &   \\
%FL Vir / Wolf 424 A & M5+M7 & dMe & 4.4747 & 11.24 & 15.4 & 12 33 17, +09 01 15  &   \\
%FL Vir / Wolf 424 B & M5+M7 & dMe & 4.3268 & 11.24 & 15.4 & 12 33 17, +09 01 15  & \\
CU Cnc$^d$ (GJ 2069) & M3.5+M3.5 & dMe/EB & 16.6012 & 10.55 & \nodata & 08 31 37, +19 23 39  &  \\
SDSS J025951.71  & M5 & dMe/S82 & 106.8607 & 15.87 & \nodata & 02 59 51, +00 46 19  &  \cite{Kowalski2009} \\
\hspace{2mm} +004619.1$^e$ &   &   &  &   &  &    &   \\
Prox Cen (Gl 551) & M5.5 & dMe & 1.3012 & 8.95 & 14.3 & 14 29 42, -62 40 46  & \cite{Gudel2002} \\
CX Cnc (GJ 1111) & M6 & dMe & 3.5805 & 12.19 & 19.0 & 08 29 49, +26 46 33  & \cite{Hilton2011} \\
UV Ceti$^f$ (Gl 65B) & M6 & dMe & 2.687 & 10.81 & \nodata & 01 39 01, -17 57 01  & \cite{Eason1992} \\
CN Leo (Gl 406) & M5.5-6 & dMe & 2.4086 & 11.04 & 17.03 & 10 56 28, +07 00 52  &   \cite{Schmitt2008} \\
SDSS J001309.33 & M6 & dMe/S82 & 48.2787 & 15.93 & \nodata & 00 13 09, -00 25 51  & \cite{Kowalski2009} \\
\hspace{2mm} $-$002552.0$^g$ &  &   &   &   &   &  &    \\
\bottomrule
\end{tabular*}
\footnotetext{$^d$Stellar parameters extensively reviewed and discussed in \cite{Feiden2013}; $^e$Most flares in the SDSS Stripe 82 flare sample.  $^f$2.\arcsec 3 from Gl 65A (BL Ceti), which is a dM5.5e BY Dra system. ``EB'' indicates an eclipsing binary. ``S82'' refers to a SDSS 82 flare star with very large magnitude enhancements during the observed events.  $^g$Largest magnitude enhancement in the SDSS Stripe 82 flare sample. } 
\end{minipage}
\end{center}
\end{sidewaystable}
\clearpage

\begin{sidewaystable}
%%\sidewaystablefn%
\begin{center}
\begin{minipage}{\textheight}
\caption{Continued ... }
\label{table:flarestars3}
%\caption{Tables which are too long to fit, should be written using the ``sidewaystable'' environment as shown here}
\begin{tabular*}{\textheight}{@{\extracolsep{\fill}}lccccccl@{\extracolsep{\fill}}}
\toprule%
%& \multicolumn{3}{@{}c@{}}{Element 1\footnotemark[1]}& \multicolumn{3}{@{}c@{}}{Element\footnotemark[2]} \\\cmidrule{2-4}\cmidrule{5-7}%
Name & SpT & Class &  $d$  & Gaia $G$ & $U$ & RA, Dec & Ref \\
 &  &  & (pc) & (app mag) & (app mag) & (J2000) &  \\
\midrule
\hline
Ultracool dMe / VLM$^h$ & & & & & & & \\
\hline
VB8 (Gl 644C) & M7 & dMe & 6.5014 & 13.84 & \nodata & 16 55 35, -08 23 40  &  \\
VB10 (Gl 752B) & M8 & dMe & 5.9185 & 14.32 & \nodata & 19 16 57, +05 09 01  & \cite{Kanodia2022} \\
TRAPPIST-1 & M8 & dMe & 12.4299 & 15.65 & \nodata & 23 06 29, -05 02 29 &  \cite{Vida2017} \\
Gaia DR2 87125610- & M8 & \nodata & 75.5971 & 19.28 & \nodata & 02 21 16, +19 40 20  &  \cite{Schmidt2014} \\
\hspace{2mm} 622839424 &   &   &   &   &   &    &    \\
WISEP J190648.47 & L1 & \nodata & 16.7867 & 17.84 & \nodata & 19 06 48, +40 11 08  & \cite{Gizis2013} \\
\hspace{2mm} +401106.8 &   &  &   &   &  &    &  \\
\hline
Optically inactive$^i$ dM & & & & & & & \\
\hline
%Gl 676A & M0 & dM  & 16.0346 & 8.87 & \nodata & 17 30 11, -51 38 13 &  \\
%Gl 15A & M1 & dM & 3.5626 & 7.22 & \nodata & 00 18 22, +44 01 22  &  \\
%Gl 667C & M1.5 & dM& 7.2455 & 9.38 & \nodata & 17 18 58, -34 59 48  &  \\
Gl 832$^j$ & M1.5 & dM & 4.9651 & 7.74 & 11.5 & 21 33 33, -49 00 32 & \\
%Gl 649 & M1.5 & dM & 10.3827 & 8.82 & \nodata & 16 58 08, +25 44 38  &  \\
%Gl 526 & M2 & dM & 5.4353 & 7.61 & \nodata & 13 45 43, +14 53 29  &  \\
%Gl 176 & M2.5 & dM & 9.485 & 9.00 & \nodata & 04 42 55, +18 57 29 &  \\
Gl 674 & M2.5 & dM & 4.5496 & 8.33 & 12.0 & 17 28 39, -46 53 42  & \cite{Froning2019}  \\
%Gl 163 & M3.5 & dM& 15.1351 & 10.68 & \nodata & 04 09 15, -53 22 25  &  \\
%Gl 849 & M3.5 & dM & 8.8028 & 9.22 & \nodata & 22 09 40, -04 38 26  &  \\
%Gl 729 & M3.5 & dM & 2.9751 & 9.13 & \nodata & 18 49 49, -23 50 10  &  \\
%Gl 436 & M3.5 & dM & 9.756 & 9.57 & \nodata & 11 42 11, +26 42 23  &  \\
Barnard's star (Gl 699) & M4 & dM & 1.8266 & 8.2 & 12.5 & 17 57 48, +04 41 36  & \cite{France2020},\cite{Paulson2006} \\
%%Gl 408 & M4 & dM & 6.7509 & 8.97 & \nodata & 11 00 04, +22 49 58  &  \\
%GJ 1132 & M4 & dM & 12.6176 & 12.14 & \nodata & 10 14 51,  -47 09 24  &  \\
%L 980-5 & M4 & dM & 13.4112 & 11.84 & \nodata & 14 21 15, -01 07 19  &  \\
%LP 756-18 & M4.5 & dM & 12.4639 & 12.92 & \nodata & 20 55 37,  -14 03 54  &  \\
%GJ 1214 & M4.5 & dM & 14.6487 & 13.01 & \nodata & 17 15 18,  +04 57 50 &  \\
%LHS 2686 & M5 & dM & 12.1901 & 12.82 & \nodata & 13 10 12  +47 45 18  &  \\
IL Aqr (Gl 876) & M4/5 & dM & 4.6758 & 8.88 & 12.9 & 22 53 16, -14 15 49  &  \cite{France2016} \\
%Gl 581 & M5 & dM & 6.2992 & 9.41 & \nodata & 15 19 26,  -07 43 20  &  \\
%LHS 2090 & M6.5 & dM(?) & 6.3792 & 13.43 & \nodata & 09 00 23,  +21 50 04  &  \\
\hline
YSO's and Pre-Main Sequence & & & & & & & \\
\hline
GSC 8056-0482$^k$ & M2 & \nodata & 38.8466 & 11.06 & \nodata & 02 36 51,  -52 03 03 & \cite{Loyd2018Hazmat} \\
V773 Tau & \nodata & T Tau & 128.1264 & 9.98 & \nodata & 04 14 12,  +28 12 12  & \cite{Massi2006} \\
DQ Tau$^l$ & \nodata & T Tau & 197.4475 & 12.49 & $\approx 15.5$ & 04 46 53, +17 00 00 & \cite{Getman2011, Tofflemire2017} \\
COUP 1462 & \nodata & YSO & 376.405 & 12.9 & \nodata & 05 35 31,  -05 15 33 & \cite{Getman2008} \\
\bottomrule
\end{tabular*}
\footnotetext{$^h$Other notable VLMs/brown-dwarfs that have produced spectacular flares are 2MASSW J0149 \citep{Liebert1999}, LP 412-3 \citep{Schmidt2007},  LP 944-20 \citep{Rutledge2000}, and V780Tau \citep{V780Tau}.  $^i$The optically inactive dM stars were mostly taken from the MUSCLES \citep{France2016, Loyd2018} and Mega-MUSCLES \citep{Wilson2021Mega, Froning2022} HST Treasury programs;  see also \cite{Brown2023}. $^j$Quiescent chromosphere recently modeled in \cite{Fontenla2016}. $^k$Colloquially nicknamed the ``Hazflare'' star.  $^l$See \cite{Tofflemire2017}:  this PMS star exhibits a large amount of $U$-band variability due to accretion phenomena.  }
\end{minipage}
\end{center}
\end{sidewaystable}
\clearpage

\begin{sidewaystable}
%%\sidewaystablefn%
\begin{center}
\begin{minipage}{\textheight}
%\caption{Tables which are too long to fit, should be written using the ``sidewaystable'' environment as shown here}
\caption{Continued ... }
\label{table:flarestars4}
\begin{tabular*}{\textheight}{@{\extracolsep{\fill}}lccccccl@{\extracolsep{\fill}}}
\toprule%
%& \multicolumn{3}{@{}c@{}}{Element 1\footnotemark[1]}& \multicolumn{3}{@{}c@{}}{Element\footnotemark[2]} \\\cmidrule{2-4}\cmidrule{5-7}%
Name & SpT & Class &  $d$  & Gaia $G$ & $U$ & RA, Dec & Ref \\
 &  &  & (pc) & (app mag) & (app mag) & (J2000) &  \\
\midrule
\hline
Active Binary & & & & & & & \\
\hline
HR 1099 (V711 Tau) & G5IV/K1IV & RS CVn & 29.6272 & 5.6 & \nodata & 03 36 47,  +00 35 15  & \cite{Osten2004}  \\
II Peg & K2-3V-IV/?? & RS CVn & 39.363 & 7.1 & \nodata & 23 55 04,  +28 38 01  &  \cite{Osten2007} \\
$\sigma$ Gem & K1III/?? & RS CVn & 37.0194 & 3.86 & \nodata & 07 43 18,  +28 53 00  &  \cite{BrownBrown2006} \\
ER Vul & G0V+G5V & RS CVn & 50.7822 & 7.18 & \nodata & 21 02 25,  +27 48 26 &  \\
$\sigma^2$ CrB & F6V+G0V & RS CVn & 22.658 & 5.41 & \nodata & 16 14 40,  +33 51 31  &  \cite{Osten2003} \\
V824 Ara & G5IV+K0V-IV & RS CVn & 30.5063 & 6.45 & \nodata & 17 17 25,  -66 57 03  &  \\
V815 Her & G5V+M1-2V & RS CVn & 32.396 & 7.32 & \nodata & 18 08 16,  +29 41 28  &  \\
EI Eri & G5IV/?? & RS CVn & 55.068 & 6.95 & \nodata & 04 09 40, -07 53 34  &  \\
AR Lac & G2IV & RS CVn & 42.6744 & 5.89 & \nodata & 22 08 40,  +45 44 32  &  \\
BH CVn & F2IV/K2IV & RS CVn & 46.1482 & 4.73 & \nodata & 13 34 47,  +37 10 56  &  \\
CF Tuc & G0V/K4IV & RS CVn & 87.9555 & 7.45 & \nodata & 00 53 07,  -74 39 05  &  \\
UX Ari & K3-4V-IV/?? & RS CVn & 50.4723 & 6.33 & \nodata & 03 26 35,  +28 42 54  &  \cite{Gudel1999} \\
VY Ari & K3-4-IV/?? & RS CVn & 41.2999 & 6.61 & \nodata & 02 48 43,  +31 06 54  &  \\
AR Psc & K2V/?? & RS CVn & 45.9586 & 7.03 & \nodata & 01 22 56, +07 25 09 &  \\
Lambda And & G8IV-III/?? & RS CVn & 24.1695 & 3.26 & \nodata & 23 37 33,  +46 27 29 &  \\
SZ PSc & K1IV+F8V & RS CVn/EB & 90.1098 & 7.08 & \nodata & 23 13 23,  +02 40 31  & \cite{Karmakar2023}  \\
CC Eri & K7V+M4V & RS CVn/BY Dra & 11.5373 & 8.17 & \nodata & 02 34 22,  -43 47 46  &  \cite{Karmakar2017} \\
ER Vul & G0V+G5V & RS CVn & 50.7822 & 7.18 & \nodata & 21 02 25,  +27 48 26 &  \\
EI Eri & G5IV+dM & RS CVn & 55.068 & 6.95 & \nodata & 04 09 40,  -07 53 34  &  \\
KIC 11551430 A$^{m}$ & \nodata & \nodata & 324.4975 & 10.7 & \nodata & 19 09 40, +49 30 05  &  \cite{Maehara2015}  \\
KIC 11551430 B & \nodata & \nodata & 321.2161 & 12.62 & \nodata & 19 09 40, +49 30 05  & \cite{Maehara2015}  \\
\bottomrule
\end{tabular*}
\footnotetext{$^m$AB system colloquially referred to as ``Pearl'' (D. Soderblom, priv. communication 2012).  The spectral types for the RS CVn systems were mostly reproduced from the tables in \cite{Osten1999}.}
\end{minipage}
\end{center}
\end{sidewaystable}
\clearpage

\begin{sidewaystable}
%%\sidewaystablefn%
\begin{center}
\begin{minipage}{\textheight}
%\caption{Tables which are too long to fit, should be written using the ``sidewaystable'' environment as shown here}
\caption{Continued ... }
\label{table:flarestars5}
\begin{tabular*}{\textheight}{@{\extracolsep{\fill}}lccccccl@{\extracolsep{\fill}}}
\toprule%
%& \multicolumn{3}{@{}c@{}}{Element 1\footnotemark[1]}& \multicolumn{3}{@{}c@{}}{Element\footnotemark[2]} \\\cmidrule{2-4}\cmidrule{5-7}%
Name & SpT & Class &  $d$  & Gaia $G$ & $U$ & RA, Dec & Ref \\
 &  &  & (pc) & (app mag) & (app mag) & (J2000) &  \\
\midrule
\hline
Single Evolved / Active Giants & & & & & & & \\
\hline
HR 9024 & G1III & Gap & 139.485 & 5.62 & \nodata & 23 49 40, +36 25 31  &  \cite{Testa2008} \\
FK Com & G5III & Gap & 216.9119 & 7.91 & \nodata & 13 30 46, +24 13 57  &  \\
v Peg & F8III & Gap & 2705.5292 & 9.84 & \nodata & 22 01 02, +06 07 11  &  \\
31 Comae & G0III & Gap & 87.0072 & 4.67 & \nodata & 12 51 41,  +27 32 26  &  \\
EK Eri & G8IV & Gap & 64.1692 & 6.07 & \nodata & 04 20 38, -06 14 45  &  \\
$\beta$ Ceti & K0III & Clump & 29.53 & $\approx 1.77$ & 3.90 & 00 43 35, -17 59 11  & \cite{Ayres2001} \\
Gamma Tau & G9.5III & Clump & 44.202 & 3.29 & \nodata & 04 19 47, +15 37 39  &  \\
$\theta^1$ Tau & G9III & Clump & 46.689 & 3.49 & \nodata & 04 28 34, +15 57 43  &  \\
$\beta$ Boo/ Nekkar & G8 III & Pre-RG?$^n$ & 62.7702 & 3.14 & 5.2 & 15 01 56, +40 23 26  & \citep{Heunsch1995} \\
YY Men & rapid & FK Com & 216.6981 & 7.92 & \nodata & 04 58 17, -75 16 37  &  \\
$\mu$ Vel & G6III(+ dF) & Pre-RG & 35.7812 & $\approx 2.35$ & 4.16 & 10 46 46, -49 25 12  & \cite{Ayres1999} \\
\hline
Dwarf K stars$^{o}$  & & & & & & & \\
\hline
AB Dor & K0V & Ke & 15.3093 & 6.67 & \nodata & 05 28 44, -65 26 55 & \citep{Maggio2000} \\
BY Dra & Ke+Ke & BY Dra & 16.5108 & 7.59 & \nodata & 18 33 55, +51 43 08  &  \\
$\epsilon$ Eri & K2V & \nodata & 3.2029 & 3.37 & \nodata & 03 32 55, -09 27 29  &  \\
\bottomrule
\end{tabular*}
\footnotetext{Pre-RG are pre-red giants, i.e., sub-giants about to ascend the red giant branch; $^{n}$\citet{Koning2006}, Jim Kaler's star database \url{http://stars.astro.illinois.edu/sow/sowlist.html}.  $^o$Other notable active K (Ke) stars are \citep{NLDS} PZ Mon (K2e), DK Leo (K7e), EQ Vir (K5e), and 61 Cyg AB (K5e$+$K7e). }
\end{minipage}
\end{center}
\end{sidewaystable}
\clearpage

\begin{sidewaystable}
%%\sidewaystablefn%
\begin{center}
\begin{minipage}{\textheight}
%\caption{Tables which are too long to fit, should be written using the ``sidewaystable'' environment as shown here}
\caption{Continued ... }
\label{table:flarestars6}
\begin{tabular*}{\textheight}{@{\extracolsep{\fill}}lccccccl@{\extracolsep{\fill}}}
\toprule%
%& \multicolumn{3}{@{}c@{}}{Element 1\footnotemark[1]}& \multicolumn{3}{@{}c@{}}{Element\footnotemark[2]} \\\cmidrule{2-4}\cmidrule{5-7}%
Name & SpT & Class &  $d$  & Gaia $G$ & $U$ & RA, Dec & Ref \\
 &  &  & (pc) & (app mag) & (app mag) & (J2000) &  \\
\midrule
\hline
Young Solar$^p$ & & & & & & & \\
\hline
EK Dra & G1.5V & solar & 34.4502 & 7.5 & \nodata & 14 39 00, +64 17 29  &  \cite{Ayres2015} \\
%$\kappa$ Ceti & G5 & solar &  &  &  &   &  \\
%HD 30495 & G2.5 & solar & 13.2413 & 5.29 & \nodata & 04 47 36, -16 56 04  &  \\
%18 Sco & G2 & solar & 14.1308 & 5.3 & \nodata & 16 15 37.2703 -08 22 09.981 &  \\
KIC 11610797 & G & solar & 279.3635 & 11.46 & \nodata & 19 27 36, +49 40 14  & \cite{Maehara2015} \\
% The following FIVE are the GKs from our sample!
%KIC 10745663 & dG2 & solar & 1546.4778 & 14.39 & \nodata & 19 47 53, +48 04 10  &  \\
%KIC 10646889 & dG1 & solar & 731.5621 & 13.61 & \nodata & 18 50 17, +47 54 15  &  \\
%KIC 11568624 & K5 & giant & 306.9281 & 13.68 & \nodata & 19 45 03, +49 34 57  &  \\
%KIC 7936309 & K & ?? & 87.8492 & 11.03 & \nodata & 18 41 39, +43 45 53  &  \\
%KIC 4043389 & G & solar & 27.7248 & 11.37 & \nodata & 19 09 31, +39 10 51  &  \\
KIC 11253827 & G & solar & 225.1926 & 11.83 & \nodata & 19 44 31, +48 58 38  &  \cite{Maehara2015}  \\
KIC 10422252 & G & solar &  &  &  &  &  \cite{Shibayama2013}  \\
%Feige 34 & sdO & std & 227.3302 & 11.11 & \nodata & 10 39 36.7377 +43 06 09.209 &  \\
%BD+284211 & sdO & std & 113.6079 & 10.46 & \nodata & 21 51 11.0220 +28 51 50.366 &  \\
%Feige 67 & sdO & std & 306.6078 & 11.77 & \nodata & 12 41 51.7901 +17 31 19.750 &  \\
%PG1708+602 & sdO & std & 1016.0683 & 13.71 & \nodata & 17 09 15.8898 +60 10 11.009 &  \\
%G191-B2B & WD & std & 52.9254 & 11.74 & \nodata & 05 05 30.6182 +52 49 51.920 &  \\
%Feige 110 & sdO & std & 268.327 & 11.79 & \nodata & 23 19 58.3995 -05 09 56.170 &  \\
%HD 85512 & K6 & dMM & 11.2847 & 7.19 & \nodata & 09 51 07.0520 -43 30 10.021 &  \\
%HD 40307 & K2.5 & dMM & 12.9412 & 6.84 & \nodata & 05 54 04.2409 -60 01 24.491 &  \\
%HD 97658 & K1 & dMM & 21.5752 & 7.51 & \nodata & 11 14 33.1613 +25 42 37.391 &  \\
KIC 12023498 & \nodata & \nodata & 530.1268 & 12.12 & \nodata & 19 47 02, +50 27 18  &  \cite{Brasseur2019} \\
KIC 8164653 & \nodata & \nodata & 3816.1876 & 14.32 & \nodata & 19 25 36, +44 05 10  &  \cite{Brasseur2019}\\
KIC 8803106 & \nodata & \nodata & 867.4893 & 13.79 & \nodata & 18 57 37, +45 05 13  &  \cite{Brasseur2019} \\
\bottomrule
\end{tabular*}
\footnotetext{$^p$Several other well-known young G-type flare stars are $\kappa$ Ceti  and 47 Cas \citep{Audard2000}.   }
\end{minipage}
\end{center}
\end{sidewaystable}
\clearpage

Stellar classification according to binarity, evolutionary status, and spectral type is important for understanding the origin and release of magnetic energy.
We describe general classifications of the types of stars that flare across the CMD (Fig.~\ref{fig:hrdiagram}).  In the following ``early'' spectral types refer to those of hotter effective temperatures (e.g., M0-M2),
and ``later'' spectral types refer to the cooler effective temperatures (e.g., M7-M9);  this jargon has no direct correspondence to time or age.  Mid-type M dwarfs span the fully convective transition and roughly correspond to stars with spectral types $\approx $ M3-M6, though there is over an order of magnitude range of bolometric luminosities just within this sub-grouping.
 
 \begin{itemize}
     \item  \textbf{RS CVn}.  Though rapid rotation is often associated with stellar youth, tidal locking of binaries can lead to prolonged magnetic activity and enhanced flare rates for billions of years \citep{Osten2012} in systems that would not normally produce energetic events.   RS CVn systems are synchronously rotating, detached binaries consisting of an early-type main sequence star and an evolved, cooler G/K sub-giant or giant companion.  In RS CVn systems, the flares are thought to originate from the cooler component.   The cool star component has a large convection zone, which combined with very rapid rotation for its size, leads to large flares with rise and decay times on the order of days.   The flares of RS CVns are among the highest energy ($10^{36}-10^{37}$ erg) and longest-lasting stellar flares.  Flare rates from the EUV and X-rays indicate about one flare per day \citep{Osten1999}, but optical enhancements are relatively rare due to the enormous background glare from the non-flaring source; peak $U$-band changes (Equation \ref{eq:magchange}) are typically much less than a magnitude for energies of $E_U>10^{33}$ erg \citep[rates of $\approx 0.17$ hour$^{-1}$;][]{Mathioudakis1992}; the largest flares increase the $U$-band flux by a factor of $\approx 2$ \citep{Doyle1991}.  The best-studied RS CVn system is arguably HR 1099 (V711 Tau), which consists of a primary K1 IV (sub-giant) and a G5 V (main sequence) star with orbital separation of $\approx 3R_{\rm{Sun}}$ and a 2.8 day synchronized orbital and rotation period  \citep{Fekel1983}.   The most extensive multi-wavelength study of the flaring and quiescence of RS CVns is presented in \cite{Osten1999, Osten2000, Osten2002, Osten2003, Osten2004}.  A large catalog of the non-flaring properties of these stars is contained in \cite{Strass1993}.

%%\vspace{2mm} 

     \item \textbf{Algol-type flare stars} are semi-detached binaries with a history of significant mass transfer from a cool evolved sub-giant onto a main sequence B-type star.  Algol (B8V + G8III/K1IV/K2III) consists of eclipsing binary (EB) stars separated by 14.14$R_{\rm{Sun}}$, whose components have radii of 2.9 and 3.5$R_{\rm{Sun}}$ with a 2.87-day orbital period.  Algol produces giant flares on occasion \citep{Oord1989B, Algol4}, and serendipitous eclipses of the flares constrain the geometries and locations of the events on the cooler, less-massive subgiant component, Algol B \citep{Algol1, Algol2}, which is also the source of the coronal radio emission \citep{Lestrade1993}.    Algol B is one of the few stars besides the Sun whose magnetosphere has been spatially resolved in radio observations \citep{Petersen2010}.   Detailed comparisons of Algol-type and RS CVn systems have been presented in \cite{Singh1996, Sarna1998, Drake2003}. They are otherwise quite similar, except the total optical (non-flaring) light  from the system is dominated by the main sequence B star in Algol-like systems, whereas RS CVns consist of two cool stars (spectral type G or later) with one subgiant or giant star that dominates the total system light\footnote{In Algols, the components have evolved such that the subgiant is no longer the more massive star due to mass transfer, whereas in RS CVn systems, the more evolved star is the more massive component.}.  The different locations on the CMD (Fig.~\ref{fig:hrdiagram}) are clear.   

%%\vspace{2mm} 

     \item \textbf{BY Dra}-type stars are, here, informally referred to as closely orbiting (EB or non-EB), detached binaries consisting of two main-sequence, typically late K or early M, stars.  Arguably, the best known BY Dra flare star is the eclipsing 0.8 day period binary YY Gem, which consists of nearly identical, and probably rotationally synchronized, dM0e stars with radii of $\approx 0.6R_{\rm{Sun}}$ and masses of $\approx 0.6M_{\rm{Sun}}$ separated by 3.83 $R_{\rm{Sun}}$ \citep{Torres2002, Feiden2013, Butler2015, YYGem19}.  Other 
     eclipsing BY Dra flare star systems are CM Dra and CU Cnc.
     BY Dra itself is composed of two active main-sequence K stars 
     with semi-major axes of $7.4-8.4 \times R_{\rm{Sun}}$ and pseudo-synchronous rotation with a period of 5.9 d \citep{BYDra2012}.  Formally, BY Dra is a term that describes the class of magnetically active spotted stars that exhibit periodic variation in their optical light curves, due to starspots and rotational modulation \citep[General Catalogue of Variable Stars: Version GCVS 5.1;][]{VarCat}\footnote{http://www.sai.msu.su/gcvs/gcvs/vartype.htm}); however, the BY Dra classification more uniquely describes \citep[e.g.,][]{NLDS} detached active binary systems that consist of main sequence late-type (K or M) stars that orbit very closely together\footnote{In contrast to UV Ceti-type binary flare stars like EQ Peg AB whose components orbit each other at much larger separations of $4-5$ au.  At these orbital separations, the active M dwarf stars are considered ``non-interacting'' binaries because it is not that plausible that  interacting magnetospheres, given their radio sizes \citep{TopkaMarsh1982}, are the source of the flares.}, like BY Dra itself.
    
     %%\vspace{2mm} 

     \item \textbf{Pre-Main Sequence} (PMS) stars, such as Young stellar objects (YSO's) and T Tauri stars exhibit flaring emission \citep[e.g.,][]{Flaccomio2018} in addition to the emission from accretion: the flow, shocks, and hotspots at the stellar surface all generate optical, UV, IR, and X-ray continuum and emission line radiation \citep[e.g.][]{Herczeg2008} that can be both steady-state and transient.  The optical, NUV, and FUV flares on T Tauri stars tend to be very energetic, $\gtrsim 10^{35}$ erg \citep{Tofflemire2017, Hinton2022, Getman2023}.   \cite{Tofflemire2017} conducted a large optical photometric monitoring campaign on the T Tauri binary star DQ Tau and compared the color differences between accretion and flare variations. \cite{Getman2011} performed an extensive study of X-ray giant flares in the DQ Tau system.  The Chandra Orion Ultradeep Project \citep[COUP;][]{Getman2005} detected many X-ray superflares from YSOs;  the flares produced X-ray hardness ratios consistent with coronal temperatures of $T >10^8$ K \citep{Getman2008}.  M-type stars take longer to contract onto the main sequence after losing their gaseous disks and are thus not actively accreting, but may exhibit inflated radii for tens of Myr (e.g., AU Mic) as they are still contracting.  These are still PMS stars but are also often considered as dwarf, UV Ceti-type stars.

%%\vspace{2mm} 
       
\item  \textbf{UV Ceti}-type stars are low-mass, M-type main sequence (dwarf) flare stars that exhibit frequent flaring and H$\alpha$ in emission in quiescence \citep{GershbergAtlas}. There is a high correspondence between the dMe status and high flare rates \citep{Kowalski2009}. These stars are traditionally  denoted as ``dMe'' or ``MVe'' stars, they exhibit optical rotational modulation (with periods typically in the $0.2 - 5$ day range, but see \citealt{West2015} and \citealt{Newton2017}), and many but not all are in the saturated regime ($L_x/L_{\rm{bol}} \approx 10^{-3}$) of quiescent X-ray luminosity.  Many of the UV Cet/dMe-type stars are in the early stages of their main sequence evolution and are possibly only several hundred Myr or less in age.  UV Ceti itself is a triple star system with two dM5.5e stars (A and B components) separated by 5.3 au \citep{Benz1998}; UV Ceti B is a binary BY Dra-type system (using our adopted definition above).   This category includes some non-accreting PMS stars (AU Mic, AT Mic).  A catalog of UV Ceti stars and their quiescent properties was compiled by \cite{GershbergAtlas} and is provided on VizieR\footnote{\url{http://vizier.u-strasbg.fr/viz-bin/VizieR?-source=J/A+AS/139/555}}.  Though the M-dwarfs are the most populous stars in the Galaxy \citep[e.g.,][]{NLDS}, none are visible to the naked eye ($V \lesssim 6$) from Earth due to their low quiescent luminosities.

%%\hspace{2mm} 

\item  \textbf{Very Low-Mass (VLM) Stars, Ultracool dwarfs, Brown dwarfs.}  The spectral types M7 through early L span the edge of the core-hydrogen burning regime (i.e., the star/brown-dwarf boundary).  The exoplanet host star TRAPPIST-1 is now a well-known, M8 flare star, due to its system of planets in or around the traditional habitable zone \citep{Gillon2017, Agol2021}. VB 10 is another prolific VLM flare star \citep{Herbig1956, Linsky1995, Fleming2000, Kanodia2022}.  The G{\''u}del-Benz relationship \citep{GudelBenz} is a correlation between the quiescent radio and soft X-ray fluxes that holds over many orders of magnitudes but breaks down in the VLM regime (with the VLMs being radio overluminous compared to the X-rays).   Nonetheless, flares from VLMs and L-type stars have shown dramatic optical continuum enhancements and broad and bright H$\alpha$ lines \citep{Liebert1999, Schmidt2007, Gizis2013} that otherwise resemble the very energetic flares from the earlier M dwarf spectral types, as noted  by \cite{NLDS}.  White-light flare rates and magnitude variations have been rather well-characterized recently in time-domain surveys \citep[e.g., ASAS-SN, Kepler, K2, NGTS;][]{Schmidt2014, Schmidt2016, Gizis2017,Vida2017, Paudel2018, Paudel2019,  Jackman2019Ldwarf, Paudel2020}. The L-dwarf flare luminosities are remarkably super-bolometric and exceed magnitude changes of $-9$ in the $V$-band \citep{Schmidt2014}.   X-ray superflares have been reported from L dwarfs \citep[e.g.,][]{DeLuca2020} as well.  
 
     %%\vspace{2mm} 

   \item \textbf{Weakly active and ``inactive'' M dwarfs} (dM or MV) stars are ``optically inactive'' stars without H$\alpha$ in emission during quiescence.  The inactive dM's also flare on occasion \citep{Paulson2006, Kowalski2009}.  These are thought to be descendants of MVe stars, which are inferred to hold on to their high levels of activity for several billion years \citep{Hawley2000, West2008, Kiman2021}.  After several Gyr on the main sequence, the M-dwarf stars maintain a lower level of chromospheric emission in Ca II H \& K (with sometimes H$\alpha$ showing a deeper absorption profile), Mg II h \& k emission lines, transition region emission (e.g., C II), and faint X-rays.  One notable example of a low-activity M dwarf is the $\approx 10$ Gyr-old, slowly rotating ($P \approx 150$ d) star Gl 699 \citep[Barnard's star; ][]{WH09, Fontenla2016, Toledo2019, France2020}.  Optical flares are rare \citep{Hilton2011, Hawley2014} on optically inactive stars, which do not exhibit much if any rotational modulation in \emph{Kepler}.  The visibility (contrast) of flares at shorter wavelengths is much greater, however, and thus the FUV and X-ray flares are rather prolific \citep{Loyd2014,  France2016,  Loyd2018, Froning2019, France2020, Brown2023}.
     
     %%\vspace{2mm} 

     \item \textbf{Solar-type stars}
	Flares from solar-type, G-dwarf stars have been observed rather infrequently, and mostly through serendipitous means in the FUV \citep{Ayres1994}, X-ray \citep {Getman2008, Pye2015}, and optical \citep{Schaeffer2000, Maehara2012}.  The detectable flares above the spatially integrated, background luminosity are very energetic, and in almost all cases they exceed broadband optical energies of $E > 10^{33}$ erg.  Detailed flare rates in the EUV have been studied for the rapidly rotating, young, single G-type stars, EK Dra, 47 Cas, and $\kappa$ Ceti \citep{Audard2000}.
   
The \emph{Kepler} and K2 missions have transformed our knowledge about the white-light flare rates of G-type stars.  The literature has been summarized in detail by \cite{Cliver2022} and \cite{NotsuBook}, and we refer the reader to these works.  Here, we provide only a brief synopsis.  After the discovery of G-dwarf white-light superflares in Kepler \citep{Maehara2012}, \cite{Shibayama2013} and \cite{Notsu2013B} performed a comprehensive flare rate analysis of the 30-minute cadence data (with some comparisons to the 1-minute cadence data), and \cite{Candel2014} extended analyses to lower mass K and M stars \citep[see also][for analyses of 30-minute cadence data of M star flares]{Walkowicz2011}.  \cite{Yang2018} also compares short and long cadence \emph{Kepler} data.  \cite{Notsu2013}, \cite{Nogami2014}, and \cite{Karoff2016} began the detailed spectroscopic characterization of the chromospheric and photospheric properties of the solar-type superflare stars reported in \cite{Maehara2012} and \cite{Shibayama2013}.
  \cite{Notsu2015A} and \cite{Notsu2015B} presented  spectroscopic follow-up of 46 of the flare stars with 30-minute cadence data in \citet{Shibayama2013}.  \cite{Maehara2015} analyzed   1-minute cadence data of 23 solar-type superflare stars.  The most prolific flare star (KIC 11551430; Table \ref{table:flarestars4}) was confirmed as a visual and spectroscopic binary in \cite{Notsu2019}, who presented  spectroscopic follow-up of 18 of these 23 stars, leveraging Gaia DR2 data as well to constrain evolutionary statuses on the subgiant branch or main sequence.  \cite{Okamoto2021} re-analyzed all solar-type \emph{Kepler} data and presented up-to-date flare rates.  As a result of the group's detailed follow-up, the calculated superflare occurrence rates on Sun-like stars ($T_{\rm{eff}} = 5600-6000$ K, $P_{\rm{rot}} = 20-40$ d) has decreased by $\approx$ an order of magnitude since \cite{Shibayama2013}.  These refined constraints provide interesting comparisons to extrapolated, multi-wavelength scaling relations from smaller solar flare energies and from large solar energetic particle events  \citep[see also, e.g., Figure 33 of][]{Usoskin2023LRSP}.
\cite{Wu2015}   presented flare rates and power-law indices for 77 individual G stars that produce superflares, including KIC 10422252 which is the prolific flare star that is discussed in \cite{Shibayama2013}.
  \cite{Davenport2016}, \cite{VanD}, and \cite{Yang2019} present \emph{Kepler} flare catalogs, where the latter includes all spectral types including A-type stars (see below) and only long-cadence data.  \cite{Lawson2019} discussed source contamination in the \cite{Davenport2016} catalog in the higher mass stellar range.  Recently, \cite{Aschwanden2021} analyzed the power-law distributions of the \emph{Kepler} long-cadence data flare catalog of \cite{Yang2019} and tested against self-organized criticality theory \citep{Aschwanden2014}.

     \item \textbf{Single Active Giants} \citep{Simon1989} are classified into one of two general categories \citep[see][]{Ayres1998}.  The first type is
     helium core burning (post Helium flash) red clump active giants like $\beta$ Ceti.    Giant flare events with rise times of $\sim1$ day are observed from these stars:  an impressive EUV light curve of $\beta$ Ceti is shown in Fig.~\ref{fig:BetaCeti} from \cite{Ayres2001}.  The active clump stars are rotating slower than evolutionary models predict, yet they have detectable magnetic  fields (unlike the rest of this stellar population).  The origin of the magnetic activity in these stars is not well understood, but it has been hypothesized that swallowing a companion planet or brown dwarf increases the shear in the stellar interior \citep{Siess1999}, resulting in ``magnetic rejuvenation'' in old age.  An alternative idea is that these stars, like some other active  giant stars \citep{Stepien1993}, are descendants of Ap stars with strong fossil fields \citep{Tsvetkova2013}.  The second type  of active giants occur in the Hertzsprung Gap, which is a short-lived phase of stellar evolution after rapidly rotating A and B-type stars leave the main sequence before rotationally breaking down and ascending the red giant branch.    White-light flares have been widely reported in \emph{Kepler} data of subgiants and giants \citep{Notsu2019, Okamoto2021, Katsova2018, Kovari2020, Olah2022}, and some exceed $10^{38}$ erg from KIC 2852961, whose binary status is not yet clear.

     \item \textbf{B5-F5} single stars have convective cores and radiative exteriors and are not expected to produce complex surface magnetic fields and flares.    B stars and Ap stars are magnetic but the magnetism is thought to be a remnant of fossil fields, which are not disorganized enough to facilitate reconnection and impulsive energy release.     
The F2 star HR 120 \citep{Mullan2000} is a surprisingly early-type flare star that has a detailed EUV flare rate presented in \cite{Audard2000}.  White-light superflares from A-type stars have been reported in \emph{Kepler} data \citep{Balona2012, Balona2019}, and spectroscopic follow-up  by \cite{Pedersen2017} found evidence for binarity in most of these sources. If originating from the companion, \cite{Balona2012} argues that the energy requirements are yet rather large and unreasonable \citep[see also][]{Mullan2009}.  The A7 star Altair is magnetically active, and its dynamo may operate locally in an equatorial zone that is cool enough for convection to occur due to the star's very fast rotation \citep{Altair}.  But to our knowledge, this ``backyard star'' is not known to produce white-light superflares.  Algol-like systems consist of a B- or A-type star, but the flares originate from the cooler, evolved star.

 \end{itemize}

\begin{figure*}
  \includegraphics[width=1.0\textwidth]{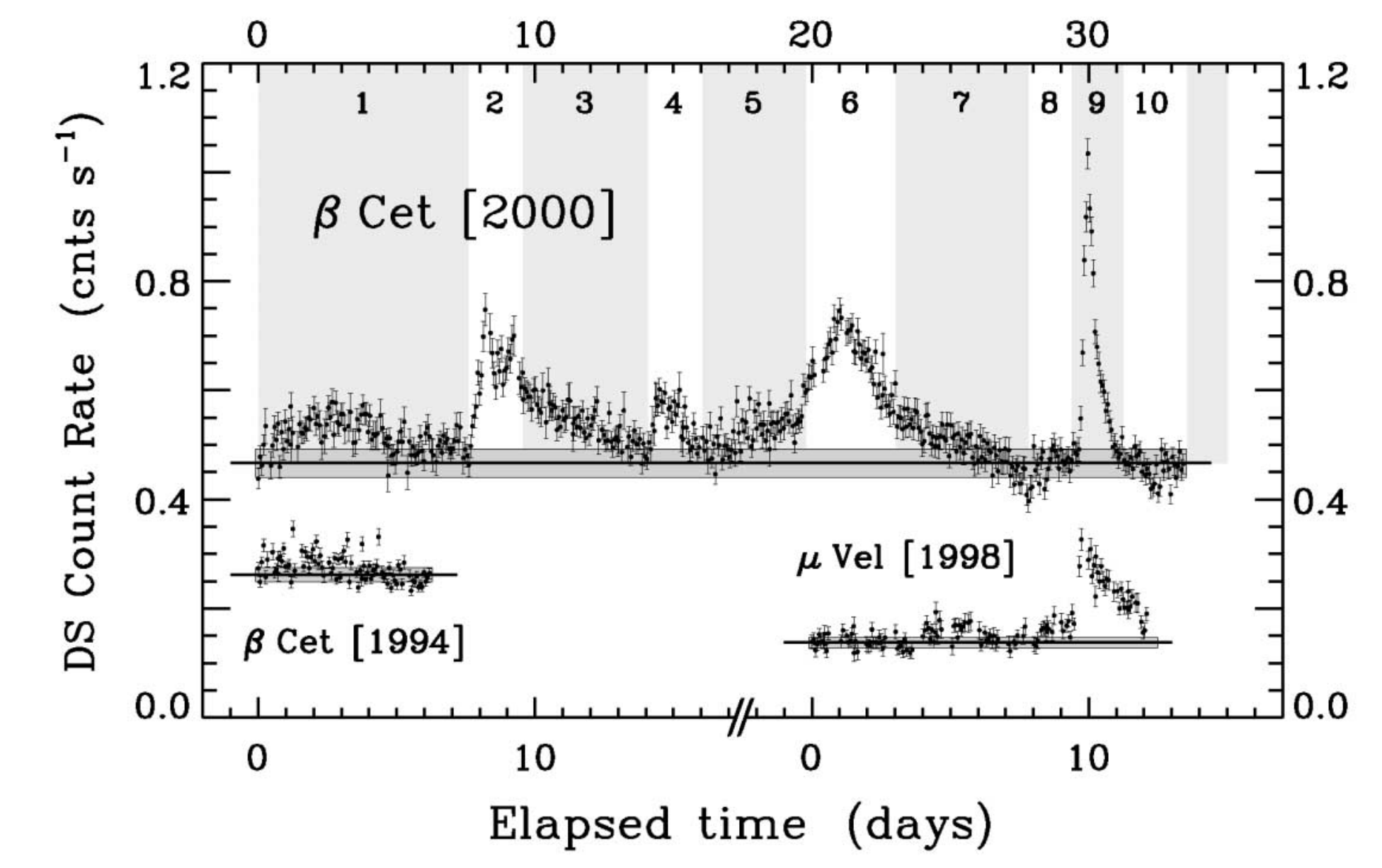}
\caption{A small fraction of single, red clump giant stars like $\beta$ Ceti exhibit continuous variability that consists of week-long, giant EUV ($60-180$ \AA) flares. A  flare on $\mu$ Velorum (also shown) is even more unexpected in its evolutionary stage. Figure reproduced from \cite{Ayres2001} with permission. }
\label{fig:BetaCeti}       % Give a unique label
\end{figure*}

\section{Flare Rates and Flare Frequency Distributions} \label{sec:rates}

The rates at which stars flare have been measured through  continuous ground-based observations, continuous space-based observations, and coarse sampling with serendipitous detections.  The coarse sampling may result in just one data point per flare; thus an accurate characterization of the flaring source (Sect.~\ref{sec:overview}) is often desired through follow-up spectroscopy.  The study of flare rates is important for a wide variety of wider applications, including coronal heating physics \citep{Hudson1991}, characterizing serendipitous variability in time-domain and exoplanet surveys \citep{Becker2004, Welsh2007, Hilton2010, Berger2013, Gezari2013, Hawley2016, Fuhrmeister2018, Mondrik2019, Howard2019, Chang2020, Jackman2021NGTS, Rodriguez2020, Koller2021, Webb2021, Wu2022}, and studying the effects on exoplanet environments \citep[e.g.,][]{Howard2018, Tilley2019}.  Much knowledge of stellar flares from low-mass stars is a result of monitoring a handful of nearby dMe stars, which reliably flare at a high rate.  Large etendue\footnote{Defined as a survey telescope's primary mirror diameter multiplied by the field of view \citep{Ivezic2019}.}  time-domain capabilities of wide-field surveys such as the Sloan Digital Sky Survey, 2MASS, and ASAS-SN have been leveraged to characterize flare rates of inactive and early-type M dwarfs \cite{Kowalski2009}, at infrared wavelengths \citep{Davenport2012},  and of ultracool dwarfs \citep{Schmidt2014, Schmidt2016, Schmidt2019}. 

\subsection{Basic Methods and Early Studies}  \label{sec:earlyffd}

\cite{Hilton2010} distinguish among \emph{flare rates}, \emph{flare duty cycles} (flaring fractions), and \emph{flare frequency distributions}.  Here, we focus on flare frequency distributions (FFDs), which refer to the energy dependence of the flare rates from continuous monitoring.  Flaring fractions in sparsely sampled data are discussed in Section \ref{sec:age}.

The seminal study of \cite{Lacy1976} presented an extensive flare rate analysis of several nearby active M dwarf (dMe) stars from hundreds of hours of ground-based monitoring in typical Johnson bandpasses.  The high-time resolution data and observational details, along with some fundamental measurements such as equivalent durations, timescales, and peak amplitudes, can be found in \cite{Moffett1974}.  In this section, we first present several basic measurements that are derived from flare light curves. The most fundamental calculation is the equivalent duration of a flare through a bandpass.  The equivalent duration of a flare is the time that a star must spend in quiescence to produce the same energy as in the flare.  It is a  proxy of the flare energy, which is spectral type and luminosity class dependent, but can be readily obtained from relative photometry.  The equivalent duration is \citep{Gershberg1972}:

 \begin{equation} \label{eq:eqdur}
ED = \int_{t_{\rm{start}}}^{t_{\rm{end}}} {\frac{I(t)- I_q}{I_q}} \,dt = \int_{t_{\rm{start}}}^{t_{\rm{end}}} {I_f(t)} \,dt
\end{equation}

%%\vspace{1mm}

\noindent which has units of time and requires establishing flare start and end times.  A small equivalent duration can result from a small flare energy and/or a large quiescent count flux ($I_q$) at Earth.   Note that the symbol for intensity, $I$, is traditionally used, though this quantity is actually the spatially unresolved instrumental ($I$) count flux from the star\footnote{Some recent studies use $\delta$ instead of ``ED'' to refer to equivalent duration;  later in this review, we follow the standard convention and use $\delta$ to refer to the power-law index of accelerated particles. The equivalent duration is referred to as the \emph{photometric equivalent width} in several studies of flare rates from \emph{Kepler} data. }.   The equivalent duration is multiplied by quiescent luminosity through this bandpass to obtain the absolute flare energy.\footnote{The quiescent luminosity can be obtained from the literature for a handful of flare stars, it can be calculated from an apparent magnitude of the star assuming a bandpass zeropoint flux, or it can be synthesized from spectrophotometry of a quiescent spectrum -- all assuming that the quiescent flux of the star does not largely vary among observing dates.  Alternatively, $L_q$ can be calculated directly using standard star fields during the observations;  this is a time-consuming process that requires taking time to point to the standard star fields over the course of the night, potentially missing awesome flares!   Only with \emph{Kepler} and other high-precision ground-based datasets \citep[e.g.][]{Hebb2007} have we seen that the values of $I_q$ for typical flare stars can vary by several percent due to gradual stellar rotational modulation.  }

The integrand of the equivalent duration is known as the ``flare visibility'' or ``flare contrast'' and is denoted as $I_f(t)$.   The flux enhancement relative to the flare star's quiescence is known as $I_f(t) + 1$ (Sect.~\ref{sec:proxcen}). The numerator in the integrand is the ``flare excess'', which can be denoted using a prime symbol to indicate a change.  If the average flux is calculated from a continuum or pseudo-continuum region of a spectrum, then the flare-only flux is denoted as C$\lambda_c^{\prime}$, where $\lambda_c$ is a representative wavelength within the spectral window.   The $I$ quantity is formally the count flux of the flare star normalized to the count flux from a nearby non-variable star ($I(t) = C_{\mathrm{rel}} (t)$ in Eq. \ref{eq:magchange}) in the same exposure and same aperture radius.  For a photon-counting instrument, such as a CCD, the theoretical value of $I$ is 

\begin{equation}
  I(t) = \frac{\int{T(\lambda) f_{\lambda} (\lambda, t) \lambda \, d\lambda}}{\int{T(\lambda) f_{\lambda, 0} (\lambda) \lambda \, d\lambda}}
\end{equation}

%%\vspace{1mm}

\noindent where $T(\lambda)$ is the total system response or effective area, including filter, atmospheric transmission (if applicable), and the detector gain; $f_{\lambda,0}$ is the flux spectrum of a comparison (non-variable) star or stars, and  $f_{\lambda}$ is the flux spectrum of the target star in units\footnote{Throughout the review, we refer to $f_{\lambda}$ as the ``flux'', following traditional shorthand in the observational astronomy community.  Formally, this quantity is the specific/monochromatic/spectral flux density at Earth.} in \flam. 
The ED is then multiplied
by the bandpass ($T$) quiescent luminosity ($L_{q,T}$) to find the bandpass-integrated flare energy, $E_{T} = ED \times L_{q,T}$. 
For known and well-characterized bandpasses, these can be determined using zeropoint fluxes, $f_{\rm{ZP}}$ \citep{Willmer2018} and published apparent magnitudes\footnote{Using the results of others who have done the painstaking work of observing standard star fields and deriving atmospheric extinction corrections.} , $m_T$, at low-levels of flare activity:

\begin{equation}
m_T = -2.5 \, \log_{10} \frac{\langle f_{q,\lambda} \rangle_T}{f_{\mathrm{ZP}, T}} 
\end{equation}

%%\vspace{1mm}

\noindent where the zeropoint flux in the Vega mag system is determined by the numerical integration of a bandpass over the spectrum of the A0 V spectrophotometric standard star, Vega.
If a flux-calibrated quiescent spectrum of the flare star is available, numerical integration over the total system response, $T(\lambda)$, including bandpass, alternatively yields the following: 

\begin{equation} \label{eq:spectrophot}
L_{q,T} = \langle f_{q,\lambda} \rangle_T \, \Delta \lambda\, 4 \pi d^2  = \frac{\int{T(\lambda) f_{q,\lambda} (\lambda) \lambda} \,d\lambda}{\int{T (\lambda) \lambda} \,d\lambda} \Delta \lambda\, 4 \pi d^2,
\end{equation} 

%%\vspace{1mm}

\noindent where $\langle {f_{q,\lambda}} \rangle_T $ is the system-weighted flux \citep{Sirianni2005}, $f_{q,\lambda}(\lambda$) is the quiescent stellar spectrum at Earth in units of \flam, and $d$ is the distance to the star. The width of the bandpass, $\Delta \lambda$, is usually taken to be the FWHM or the effective width, which vary from $\Delta \lambda  \approx  600$ \AA\ to $4000$ \AA, for the SDSS $u$-band and \emph{Kepler} bands, respectively.   In recent years, quasi\footnote{Usually not accounting for very short wavelength radiation at EUV and X-ray energies.}-bolometric conversions from white-light filters have assumed a blackbody function for all flares, and all phases of flares \citep[e.g.,][]{Shibayama2013}.  Other studies assume constant spectral energy distributions, with energies reported over the limits of the bandpass \citep[e.g.,][]{Hawley2014}.   Several quiescent luminosities used for M dwarfs are given in Table 16 by \cite{Moffett1974}, but it should be noted that stars, especially active M dwarfs, have a rather large spread of $U$-band magnitudes reported in the literature (e.g., reported quiescent apparent $U$-band magnitudes for the flare star YZ CMi range between $13.70 - 13.85$).  Total system response characteristics through $U$ bandpasses are also inherently uncertain \citep[e.g.][]{MA2006} and to some extent, intractable.  Thus, non-negligible systematic uncertainties in absolute energies are to be expected from a combination of these various issues \citep[see][for empirical comparisons of flare energies in SDSS $u$ and Johnson $U$ from two different telescopes]{Hilton2011}. In Appendix \ref{sec:filters}, we show several of the broadband and narrowband filters that are commonly employed in optical stellar flare studies.

FFDs are presented as either downward cumulative, $Q(>E)$, or differential, $n(E)$, distributions in log$_{10}$-log$_{10}$ space \citep[see][for comparisons of these methods]{Audard2000}.  The most common practice is to fit a power-law (pareto) model to the unbinned, downward cumulative FFD:

\begin{equation}
  Q(>E) = N \big(\frac{E}{E_0} \big)^{\beta}
  \end{equation}

\noindent with power-law index, $\beta$, and $N$ total flares  greater than or equal to a fiducial low-energy limit, $E_0$ (e.g., the detection completeness limit) observed within a monitoring duration of $\Delta t$ (hr).  The usual convention is that $\beta$ is negative.  The differential FFD (\# of flares per unit energy) is then 

\begin{equation} \label{eq:ffdpowerlaw}
  n(E) = - \frac{dQ}{dE} = N \frac{\alpha - 1}{E_0} \big(\frac{E}{E_0} \big)^{-\alpha}
  \end{equation}

\noindent where $\beta = 1 - \alpha$ (again, the convention is that $\beta$ is negative and $\alpha$ is positive).  Note that if  $A \int_{E_0}^{\infty} n(E)dE = 1$ is solved for $A$, the converted differential distribution $A\ n(E)$  is a  probability density function (PDF): given a flare occurred, $A\ n(E)dE$  is the probability for a flare to have an energy between $E$ and $E+dE$. 

   Convenient analytic forms exist for the power-law index and its statistical uncertainty from a direct (i.e., unbinned) maximum likelihood (ML) analysis  \citep[][Appendix B]{WJ,Clauset2009}:

   \begin{equation}
     \hat{\beta}_{\text{ML}} = \frac{N}{\sum_{i=1}^N ln \frac{E_i}{E_0}}
     \end{equation}

\noindent from which the result follows that $\sigma_{\hat{\beta}_{\text{ML}}} \approx \hat{\beta}_{\text{ML}} / \sqrt{N}$.

Alternatively, weighted least-squares fits of the power-law index (slope) and intercept of a line in log-log space has been sometimes employed in stellar flare studies.  For example, \cite{Lacy1976} express the power-law for each flare star in their sample
of active M dwarfs as

\begin{equation} \label{eq:ffd}
\log_{10} \nu(>E)  = a + \beta \, \log_{10} \, E 
\end{equation}
 
 \noindent where $\nu(>E) = Q(>E)/\Delta t$ is the number of flares per hour greater than or equal to energy $E$ (expressed in erg) through an optical bandpass $T$.   (To increase statistical confidence, flares that are observed in several bandpasses were multiply-counted using empirical bandpass energy conversions with the effective observing time adjusted accordingly.) 
Of course, the counts in even well-populated cumulative distributions have statistical uncertainties that are asymmetric in logarithmic space, and the values in cumulative distributions are non-independent.  Nevertheless, this method is employed as a sufficient approximation in some cases given the statistics that are possible with  the relatively low number of flares in ground-based observing campaigns.

 Power-laws are typically fit to FFDs over an intermediate range of energy that is carefully chosen \citep[e.g.,][]{Silverberg2016} to exclude the high-energy and low-energy ends, which may have  incomplete sampling that can affect the fits.   Note that \cite{Clauset2009} present more sophisticated methods for uncertainty estimation for power-laws and have been employed in some more recent flare rate studies \citep{Medina2020}.  The Markov Chain Monte Carlo approach is described in \cite{Davenport2016Prox}, and \cite{Davenport2019} use an appropriate uncertainty analysis for the counts \citep{Gehrels1986}.   Modifications to the basic power-law form (Eq. \ref{eq:ffd}) and completeness functions have been used to make adjustments to the the low-energy end \citep{Rosner1978, Aschwanden2015, Davenport2012, Medina2020, Okamoto2021} and high-energy tail \citep{Aschwanden2021} of FFDs.    For reference, solar flare frequency distributions \citep{Cliver2022} exhibit power-law indices in (nonthermal) hard X-ray energy of $\approx 1.5$ and peak luminosity of $\approx 1.7$ \citep{Crosby1993}, whereas the thermal soft X-ray peak flux exhibits a larger power-law index of 2.0 \citep{Veronig2002Xrays, AschwandenFreeland2012, Hudson2023}.

 We summarize the general range of values of $\alpha$ calculated from FFDs in the X-ray and the optical regimes.   
Ground-based optical and space-based XEUV ($0.01-10$ keV, which corresponds to $1.24-1240$ \AA, and where only EUV data, such as DS count rates, are available, a two-temperature, $T = 6+23$ MK, optically thin free-free model extrapolates to the full XEUV range to calculate energies) flare rates have generally given values from $\alpha \approx 1.4 - 2.2$ with statistical uncertainties of $\gtrsim 0.1$ \citep{Audard2000}.  RS CVn systems exhibit power-law indices in the EUV
on the lower end of this distribution \citep{Osten1999}.  \cite{Wu2015} summarize the power-laws from 77 G-type stars.
\cite{Hawley2014} and \cite{Illin2021} summarize values of $\alpha$ for low-mass stars in the optical, and \cite{Loyd2018} present power-laws for low-mass stars of different activity levels in the FUV.
 A useful document that summarizes flare rates from low mass M stars of various spectral types is Osten (2016)\footnote{\url{https://www.stsci.edu/files/live/sites/www/files/home/hst/instrumentation/stis/documentation/instrument-science-reports/_documents/2017_02.pdf}.}.
 Among the active M dwarfs,  the higher luminosity stars (in quiescence) exhibit flatter power-laws \citep{Pettersen1984}, and thus the sum of the energy in all flares is dominated by the occurrence of the highest-energy flares \citep{Lacy1976}.
Alternatives to the power-law functional form (Eq \ref{eq:ffdpowerlaw}) are summarized in \cite{Aschwanden2011}, \cite{Rosner1978}, and \cite{Sakurai2022}. 

A few stars have had power-law indices measured independently several times; these stars show some interesting systematic differences in the reported power-law properties.
In the M dwarf study of \cite{Lacy1976}, the famous flare star AD Leo appeared to be an outlier from the others.  Due to the relatively short observing time in the \cite{Lacy1976}, its FFD was recalculated after more data were collected.  \cite{Pettersen1984} found a value of $\alpha = 1.62 \pm 0.09$ from 85 $U$-band flare events (with 115 individual flare peaks) in 111 hr of monitoring.  The most energetic flare was $10^{33}$ erg.   The $U$-band FFD of Proxima Centauri has been quantified by \cite{Walker1981} for flare energies between $5\times 10^{27}$ erg and $10^{30}$ erg, giving a similar value of $\alpha \approx 1.7$ to AD Leo but flatter than other stars of the same quiescent luminosity, such as CN Leo.  Recently, \cite{Davenport2016Prox} doubled the number of flares in the statistics for Prox Cen using white-light optical data from the MOST satellite and calculated a similar power-law slope to \cite{Walker1981}.  \cite{Howard2018} found a steeper power-law for Proxima Centauri from the Evryscope survey and investigated the effect on the power-law index with and without an extreme superflare outlier.

 \cite{Lacy1976} and \cite{Pettersen1984} found trends in power-law indices with effective temperature or quiescent $U$-band luminosity:  larger luminosity (earlier-type) main-sequence stars exhibiting flatter power-laws (i.e., smaller $\alpha $, $\mid \beta \mid$).  However, the study of \cite{Pettersen1984} did not include the results of \cite{Walker1981} for the interesting case of Proxima Centauri, which shows a flatter power-law than expected;   evidence for and against Proxima Centauri being less magnetically active than others of the same sub-type, such as CN Leo, are summarized in \cite{Reiners2008}.   \cite{Audard2000} suggested that $\alpha$ derived from the XEUV may exhibit an opposite trend with spectral type, though this is reported with low confidence.  They find an increasing rate of $E > 10^{32}$ erg flares corresponds with greater  quiescent X-ray luminosity.

 Other interesting quantities that have been calculated from flare star monitoring are the energy per flare (average flare energy) and the radiated flare energy per unit time (average luminosity due to flaring).  These quantities are usually calculated through the $U$ band.  They correlate with the non-flaring, bandpass quiescent luminosity \citep{Lacy1976} or non-flaring, bolometric luminosity of the star over several orders of magnitude.   It should be noted that the number of detectable flares is anti-correlated with the quiescent luminosity, an effect largely due to brighter stellar background that raises the detection limit \citep[e.g.,][]{Davenport2012}.  Nonetheless, including many small, unobserved flares in the total energy calculation does not change the trends.  From modern space-based data, it seems that these relationships generally extend to active binaries, which comprise much higher luminosity stellar components \citep{Osten2012}.    An extrapolation of these relationships to the Sun, however, dramatically over-predicts the flare energy release rate in optical solar flares by many orders of magnitude, as first noted by \cite{Lacy1976}.   From \emph{Kepler} data (Sect.~\ref{sec:kepler}), a somewhat similar quantity of $L_f$  vs.\ $ L_{\rm{Kp}}$ has been constructed by \cite{Lurie2015} and is making for powerful comparisons for normalized flaring efficiencies among stars of different spectral types \citep[e.g.,][]{Davenport2019}.  \citet{Davenport2016} investigates evidence for saturation in the relative flare luminosity given by $L_f/L_{\rm{Kp}}$ at small Rossby numbers, analogous to the quiescent $L_X/L_{\rm{bol}}$ saturation (Section \ref{sec:overview}).

 There has been a severe paucity of low-activity M dwarf flare rates until \cite{Hawley2014} presented white-light 1-minute cadence data from \emph{Kepler} and some previously unpublished ground-based $U$-band data from the PhD work within \cite{Hilton2011}.  These studies revealed that stars that are optically inactive in quiescence (Sect.~\ref{sec:overview})  produce broadband optical flares, a result that had been quantified in sparsely sampled data from the Sloan Digital Sky Survey \citep{Kowalski2009} and from serendipitous spectral detections \citep{Paulson2006}.   \cite{Loyd2018} compared FUV flare energies among stars with comparable quiescent bolometric luminosities and found that FFDs are shifted according to the quiescent flux in the FUV.

The different apparent behaviors at the high-energy ends of stellar flare FFDs are still largely unexplained.  Does the power-law that is well-fit to a middle range of energies continue on indefinitely, or does it turn over at the high-energy end and follow a steeper power-law index \citep{Doyle1990, Osten2012} or another type of roll-over \citep[e.g.,][]{Osten2004, Dal2020, Aschwanden2021, Cliver2022, Sakurai2022} for some stars?  What physical parameters determine the upper limits of flare energies (and peak luminosities) that are possible on a given star?  Extending the power-law that is fit to the FFD of YZ CMi predicts a $U$-band flare to occur about once per month with an energy that is a factor of $5000 - 10,000$ larger than its mean flare energy \citep{Lacy1976, Kowalski2010}.  \citet{Dal2011A} discuss an upper limit to the energy released in dMe flares is approached as the flares become longer in total duration.  \citet{Shibata2013} and \citet{Aulanier2013} present theoretical grounds and semi-empirical relations for upper limits of flare energies.  Empirical constraints from very long monitoring times have only been possible recently using Kepler data.

 %  , which then should have been better quantified with the long-duration monitoring provided by the Kepler mission (Section \ref{sec:kepler}).      

\subsection{White-Light Flare Rates \& \emph{Kepler}} \label{sec:kepler}

The release of \emph{Kepler} data \citep{Borucki2010} has transformed knowledge of flare rate statistics in several important ways.  First, the high precision and long monitoring times facilitate detection of relatively rare white-light flares from older, less active low-mass stars, from much larger luminosity stars such as solar-type stars, and from higher mass K and early-type M dwarf stars, and evolved stars.  Second, it provided many more hours of monitoring per star:  ground-based monitoring efforts resulted in $\approx 30-100$ hours per star, while the nominal baseline for Kepler was two contiguous months of observations per star.  The flare statistics correspondingly improved, but the high precision also revealed comparable amounts of background flux variation due to starspots and rotation.  Nonetheless, the \emph{Kepler} field began as a relatively poorly known star field (due to its relatively faint magnitude limits), and systematic contamination of the flare sources from binaries, subgiants, and other non-flare sources was an issue until targeted, ground-based spectroscopic (and asteroseismic) analyses caught up and Gaia DR2 parallaxes became available.  Additionally, the vast majority of data from the \emph{Kepler} mission is at 30-minute cadence, and most superflares span just two to four data points.  The 1-minute cadence data were made possible for select stars through Guest Observer programs \citep[e.g.,][]{Hawley2014}.

\begin{figure*}
\begin{center}
  \includegraphics[width=0.75\textwidth]{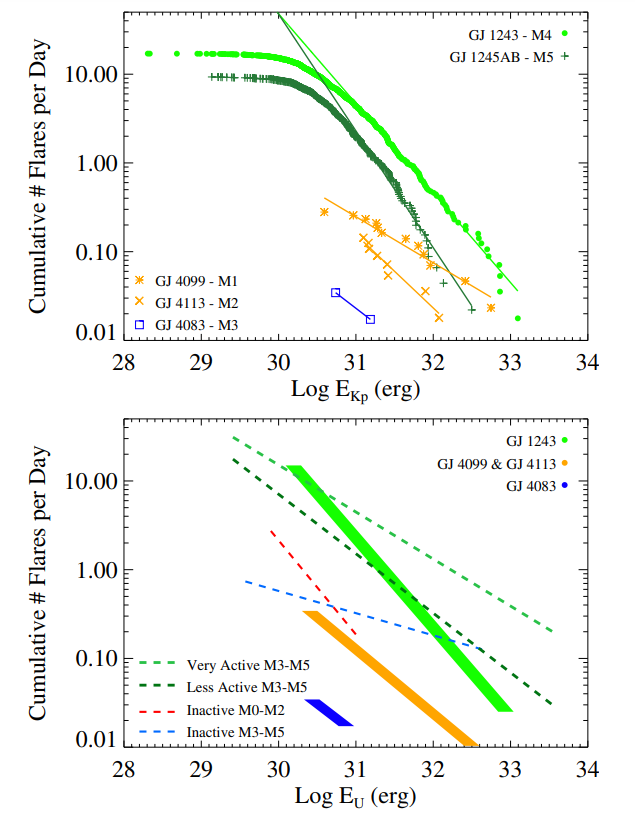} 
\caption{ Flare frequency distributions of several main-sequence M stars in the Kepler band (top) and in the $U$-band (bottom).  Figure reproduced from \cite{Hawley2014} with permission.  Note the spread in flare rates in the bottom panel among the very active and less active stars of the same spectral types. }
\label{fig:hawley2014}       % Give a unique label
\end{center}
\end{figure*}

The dM4e star GJ 1243 has been observed for 11 months at 1-minute cadence with \emph{Kepler}, making it the longest observed flare star in the sky besides the Sun.   The FFDs and flare properties of this star have been studied extensively by \cite{Ramsay2013, Hawley2014, Davenport2014, Davenport2016, Silverberg2016}.  From the database of over 6000 flares, \cite{Silverberg2016} obtained a value of $\alpha = 2.008 \pm 0.002$, and  \cite{Davenport2020} inferred $\alpha = 1.942 \pm 0.001$.  From the first two months of data, \cite{Hawley2014} found $\alpha = 2.01$.  They also noted that the turnover in the FFD at the low-energy end, which has often been attributed to a detection limit effect, may be astrophysical at some level since  flares of these energies should be readily detectable above the precision of \emph{Kepler}.  
The FFD of GJ 1243 is shown in Fig.~\ref{fig:hawley2014}(top) in comparison to several other M dwarfs spanning different spectral types and levels of quiescent magnetic activity (dM v. dMe). Comparable diagrams with compilations of FFDs of many individual stars are shown in \cite{Shakh1989}, \cite{Ramsay2013}, and \cite{Dal2020}.  \cite{Lurie2015} presented a detailed analysis of the GJ 1243 AB system, where they separated the \emph{Kepler} light into the two stellar components and found values of $\alpha \approx 2$ for both stars, but a higher flare rate occurs on the slower rotator.  
The bottom panel of Fig.~\ref{fig:hawley2014} shows the \emph{Kepler} energies converted to energies in the $U$-bandpass.  The flare rates of  ``less-active'' M3--M5 stars are closer to the FFD of GJ 1243,  whereas the very-active M3-M5 stars include the dMe stars from \cite{Lacy1976}, such as YZ CMi, EV Lac, and AD Leo.  A continuum of flare rates is clearly evident as the vertical offsets of the power-laws, while different power-law slopes may be a result of vast differences in statistics from ground and space-based monitoring \citep{Hawley2014}.

The spread in flare rates among active M-dwarfs of similar spectral types (Fig.~\ref{fig:hawley2014}) contrasts with the findings of \cite{Kowalski2009}, who found that the average M4Ve-M5Ve flare frequency distribution from sparsely sampled data of field stars in the Sloan Digital Sky Survey (SDSS) Stripe 82 is sensibly in line with that of the well-known flare star YZ CMi (which is one of those among the ``more active'' stars in Fig.~\ref{fig:hawley2014}). \cite{Hilton2011} began to investigate the chance alignment of FFDs in \cite{Kowalski2009} using Monte Carlo sampling of simulated flares at the SDSS $riuzg$ cadence.  The resolution of this issue (which is a non-trivial adjustment to the statistical rate calculation; E. Hilton, priv. communication 2013) will be important for accurately mapping flare rates of the Galaxy using sparsely sampled data from the LSST \citep{Ivezic2019}.

The immaculate flare statistics of \emph{Kepler} facilitated new timing analyses of flare occurrence and correlation.
The flare occurrence of GJ 1243 was investigated as a function of the phase of the longer-timescale, flux modulation that exhibits a $\approx 1$\% amplitude and is attributed to stellar rotation. No significant correlation was found at either maximum or minimum flux levels.  The relationship between flare occurrence and phase was further investigated in detail with larger samples of flares on GJ 1243 \citep{Silverberg2016} and for a large sample of M dwarfs  \citep{Lurie2015, Doyle2018}.   \citet{Morris2018} reported a larger flare occurrence when the very low-mass (M8V) star TRAPPIST-1 is slightly brighter than average and when it goes through the increasing flux phase in the Kepler light curve.  \cite{Dal2011B} investigated the  occurrence of different types of flares with phase from a large sample of ground-based observations of flares.

  \cite{Hawley2014} finds that the intervals between successive flares (``waiting times'') on an active M dwarf (GJ 1243) over many rotation periods is consistent with a single exponential probability distribution, $p(\Delta t_{\rm{next}}; \tau_0)$ $\propto \frac{1}{\tau_0} e^{-\Delta t_{\rm{next}} / \tau_0 }$ (where $p$ is the probability density and $\tau_0$ is the average interval between consecutive flares).  Equivalently, the number of flares occurring in fixed time intervals, $\Delta t$,  is Poisson-distributed, $p(N_{\rm{flares}}$; $r, \Delta t) = (r \Delta t)^{N_{\rm{flares}}}  e^{-r \Delta t} / (N_{\rm{flares}}!)$ \citep[e.g.,][]{Lacy1976}.  Stellar flares from active stars thus occur independently and continuously at a constant rate, $r = 1/\tau_0$ (note that some studies use $\lambda$ instead of $r$ for the rate), except possibly on some short time-intervals ($\Delta t \lesssim 30$ min) within complex events \citep{Hawley2014,Davenport2014}. Correlations between waiting time and flare energy, as may be expected from magnetic energy buildup and release \citep[e.g.,][]{Rosner1978} in a single region on the star, are not clearly consistent with the stellar data \citep{Hawley2014}.  Efforts to understand the flare occurrence as a function of light-curve phase are ongoing in the era of TESS data \citep{Ikuta2023}. % using maximum likelihood parameter estimation and chi-squred hypothesis testing \citep{LME76}.
% Wheatland 2010 shows double exponential for solar flare rates.
%  Also Mullan for energy buildup and release.
% equal probability per unit time.
% p($\Delta t_{\rm{next}}; \lambda t_{\rm{elapsed}}$)
% p(Nflare; $ \lambda \Delta t$)

\cite{Davenport2014} constructed a canonical flare template from the normalized, classical (single-peaked) flares of GJ 1243 in short-cadence \emph{Kepler} data.  They modeled the decay phase with a sum of two exponentials and the rise phase with a polynomial.  The template reflects a remarkable self-similarity in many flares on one-minute integrations.  Further analysis in this work demonstrated that many complex flares can be decomposed into linear superpositions of  several templates with adjustable parameters.  They then injected synthetic flares over  a long time baseline and simulated flare statistics, concluding that the number of complex flares observed in the data is not fully consistent with a chance superposition of many uncorrelated classical flares.  This is tantalizing, quantitative evidence for a solar-like ``sympathetic'' \citep[e.g.,][]{Lynch2016} flaring property of stars.

The rates of flares in the high-energy regime have been investigated in detail with Kepler data.  \citet{Silverberg2016} discuss a deviation from a single power-law between $E_{\rm{Kp}} = 10^{32.5} - 10^{34}$ erg flares in the GJ 1243 sample.  (In the following, flare energies now refer to the energies integrated over a $T=9000$ K blackbody function, as reported in the respective studies, rather than an energy over a particular bandpass). The upper limit flare energy for white-light flares has been pretty well established from the long monitoring times from \emph{Kepler} to be $E_{\rm{max}} \approx 10^{37}$ erg for subgiants, $E_{\rm{max}} \approx 10^{36}$ erg for rapidly rotating G-dwarfs, and  $E_{\rm{max}} \approx 10^{35}$ erg for slower rotating G dwarfs \citep{Notsu2019, Okamoto2021}.  For G, K, and M stars in the Kepler field, \citet{Candel2014} discuss a correlation between a star's superflare ($E> 5\times  10^{34}$ erg) rate and its light-curve amplitude modulation due to starspots.  The ground-based Evryscope network has also provided new insights into $g^{\prime}$-band superflare rates:  \citet{Howard2019} find that the rate of $E \ge 10^{33}$ erg superflares  (by extrapolating the best-fit power-laws down over $0.5 - 1.5$ orders of magnitude in energy for the K5-M2 stars) decrease from the averaged active K5-K7 stars to the averaged active M4 stars.  Similar to many previous studies (Section \ref{sec:earlyffd}), they report a larger average flare energy for earlier-type stars, which they attribute to the ``size of the stellar convective region''. \citet{Pettersen1989} review and analyze the time-averaged luminosity due to flaring (and quiescent magnetic activity proxies)  in a sample of dKe and dMe stars, and they correlate it with the convective envelope volume.  Recent stellar evolution models of dM flare stars in eclipsing binary systems \citep{Feiden2013, Feiden2014} and dK/dG stars \citep{Matt2011} also show that early type stars, which have larger mean flare energies, have larger convective zone volumes.

\subsection{Timing on Long Scales:  Flare Rates and Stellar Age}\label{sec:age}
By characterizing the FFDs across many types of stars in different evolutionary stages, astronomers begin to piece together how flare rates change over billions of years.  With age-dating techniques such as open cluster membership / association and gyrochronology, quantitative evolutionary tracks in the flare history of stars are possible.
 At the youngest stellar ages, high flare rates and energies have been reported from PMS stars in nearby open clusters, such as Orion, in the X-ray and optical \citep{Getman2005, Jackman2020Age}.  Signatures of accretion (e.g., in T Tauri systems) indicate a very young age range for some flare stars \citep[e.g.,][]{Tofflemire2017}.  Membership among moving groups, such as the well-known $\beta$ Pic moving group (with the flare star members AU Mic and AT Mic AB) probe ages around $\sim$ 20 Myr for stars that are no longer actively accreting gaseous material from a circumstellar disk.   At older ages ($\gtrsim 100$ Myr), flare rates and properties have been reported in open clusters, such as the Pleiades \citep{Stelzer2000, Illin2}.  At the oldest ages, globular cluster and the Galactic bulge membership facilitates calibration of the flare-rate--age relationship, but such stars are located at very large distances. As stars spin down due to angular momentum loss over their main sequence evolution, flare rates are expected to correspondingly decrease.  This correlation is supported by detections of stellar superflares from main sequence, rapidly rotating (periods $<$ 5 d) G-type stars.   However, binarity plays an important role in maintaining flaring among a population stars to old ages, compared to the Galactic disk populations.  This may occur through a co-evolutionary process in M dwarf-white dwarf binaries  \citep{Morgan2016}  or rotational tidal synchronization in RS CVn-like systems \citep{Osten2012}.  
Binarity may also directly affect flare occurrences and energies \citep[e.g.,][and see references in Section \ref{sec:alma}]{Doyle1990, Gao2008, Adams2011}, but it is in general a difficult property to assess without adaptive optics \citep[e.g.,][]{Clarke2018} and high-resolution spectroscopy \citep[e.g.,][]{Notsu2013}, which are observational techniques that are limited to nearby stars.

Following the statistical inference of quiescent magnetic activity lifetime analyses of \cite{West2008}, the vertical distances above or below the Galactic plane have been used as proxies of stellar age for M dwarf stars in the field.  The basic idea is that over time, stars experience repeated gravitational scatterings in the midplane of the Galaxy and are consequently more likely to be found at larger vertical displacements.  Large time-domain surveys, such as the SDSS and soon the LSST, allow characterization of flare rates in multiple populations of stars in sparsely sampled data sets.   The stellar flaring fraction and the flare fraction (duty cycle) are two  quantities that are useful in mapping the flare rate of a galaxy and the star's vertical distance from the galactic plane. A \emph{stellar flaring fraction} is the fraction of stars that  produce some number of flares, while \emph{flaring fraction}, or duty cycle, is the fraction of sparsely sampled epochs that flare.  UV and optical surveys have been leveraged to map these quantities with increasing vertical distances \citep{Welsh2007, Kowalski2009, Hilton2010, Walkowicz2011}.  There is suggestive evidence that flaring fraction from sparsely sampled data indicates that flaring occurrence among a population decreases faster with stellar age than other measures of quiescent magnetic activity \citep{Kowalski2009, Hilton2010}.   One may be tempted to speculate that this is one possible source for the spread in flare rates among the mid-type dMe stars in Fig.~\ref{fig:hawley2014}, but more investigation is needed.  The effects of solar-like activity cycles \citep{Crosby1993} might also sensibly contribute to the variation of flare rates within a population of stars of the same spectral type \citep[but see][]{Pettersen1984, Davenport2020}.

Most recently, the  improved methods and data samples for flare statistics, asteroseismology, gyrochronology, and open cluster calibrations using \emph{Kepler} and K2 data have facilitated 
 much more quantitative characterization of the flare rate as a function of stellar age \citep[e.g.,][]{Illin2021, Illin2}.  \cite{Davenport2019} used \emph{Kepler} field stars and gyrochronology \citep{Mamajek2008} to present the first stellar flaring fraction as a function of age and $g-i$ color, which maps to spectral type \citep{Covey2007}. \cite{Davenport2019} used MCMC non-linear least squares to constrain $\alpha(t)$ in an equation similar to Eq. \ref{eq:ffd} over a range of stellar masses, from G to M dwarfs, and ages older than $t \ge 10$ Myr.  They found that $\alpha(t)$ remains approximately constant, flare rates indeed decrease over time for the M dwarfs, but that the flare rate decreases much less than for the G dwarfs.  \cite{Notsu2019} and \cite{Okamoto2021} characterized the evolution of differential flare frequency of G-type stars in the \emph{Kepler} field by splitting up into different rotational period bins from $<5$ d to $20-40$ d.  They further divided the $< 5$ d sample into smaller bins and found that flare frequency is approximately constant for periods $<$ 3 d, for which the corresponding ages are $\lesssim 500$ Myr but are not as well constrained from gyrochronology calibrations.  \citet{Johnstone2021} combine several state-of-the-art techniques to calculate the quiescent $L_X$ as a function of age and mass, and then they leverage a scaling relation (discussed in Sect.~\ref{sec:earlyffd}) from \citet{Audard2000} to map the rate of high-energy XEUV flares on GKM stars over ages spanning 2 Myr to 5 Gyr.

\section{Light Curve Analyses} \label{sec:lc}

In addition to integrated flare energy in a particular bandpass (Sect.~\ref{sec:rates}), a number of other strictly empirical quantities can be calculated from single-bandpass data with a moderately fast cadence.  A useful decay timescale measurement is the time it takes for a light curve to reach $1/e$ of its maximum flux \citep[e.g.,][]{Maehara2015, Namekata2017}, which prevents ambiguities in determining the bona-fide end of a flare. This measure also avoids model-dependent uncertainties in fitting an exponential function to flares with events in the decay phase, which may also show lengthening timescales \citep{Osten2005, Davenport2014}.  Here, we summarize a few relationships between decay times and other quantities.
The total rise and total decay times of flares are not well-correlated \citep{Moffett1974,Lacy1976, Hawley2014}, which possibly suggests some loss of ``memory'' about the details of the rise phase after the peak.  Integrated energy and total duration exhibit a tighter correlation.  Larger peak-flare amplitudes typically correspond to longer decay times and larger energy flares, but there is significant scatter.    \cite{Namekata2017} find that the $1/e$ decay duration scales with total white-light energy to the $1/3$ power and magnetic field to the $-5/3$ power for flares on the Sun, rapidly rotating G stars \citep[see also][]{Maehara2015}, and M stars \citep[see also][]{Howard2019}.  \citet{Kovari2020} found that the $1/3$ power scaling between duration and energy also extends to a late-type giant.  For similar-duration ($\approx 5-10$ min) solar flares and white-light flares on solar-type stars, the energies of the latter are several orders of magnitude larger \citep[\emph{cf} Figures 8-9 of ][]{Namekata2017}.  

\subsection{Classification of Flares by Optical Broadband Evolution} \label{sec:flare_classification}

\begin{figure*}
\begin{center}
   \includegraphics[width=0.75\textwidth]{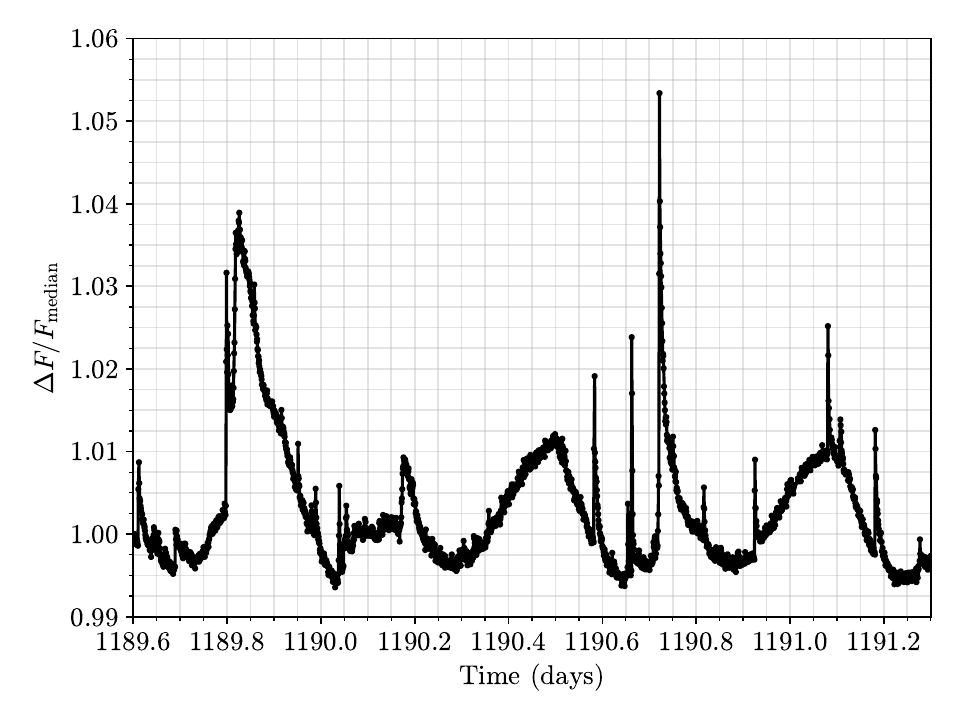}
  \includegraphics[width=0.75\textwidth]{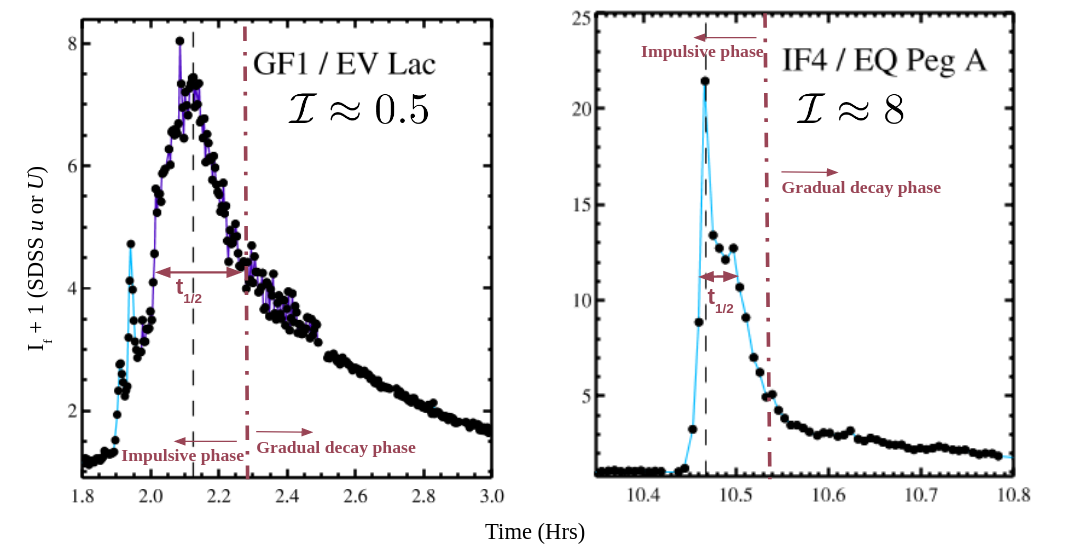} 
  \includegraphics[width=0.75\textwidth]{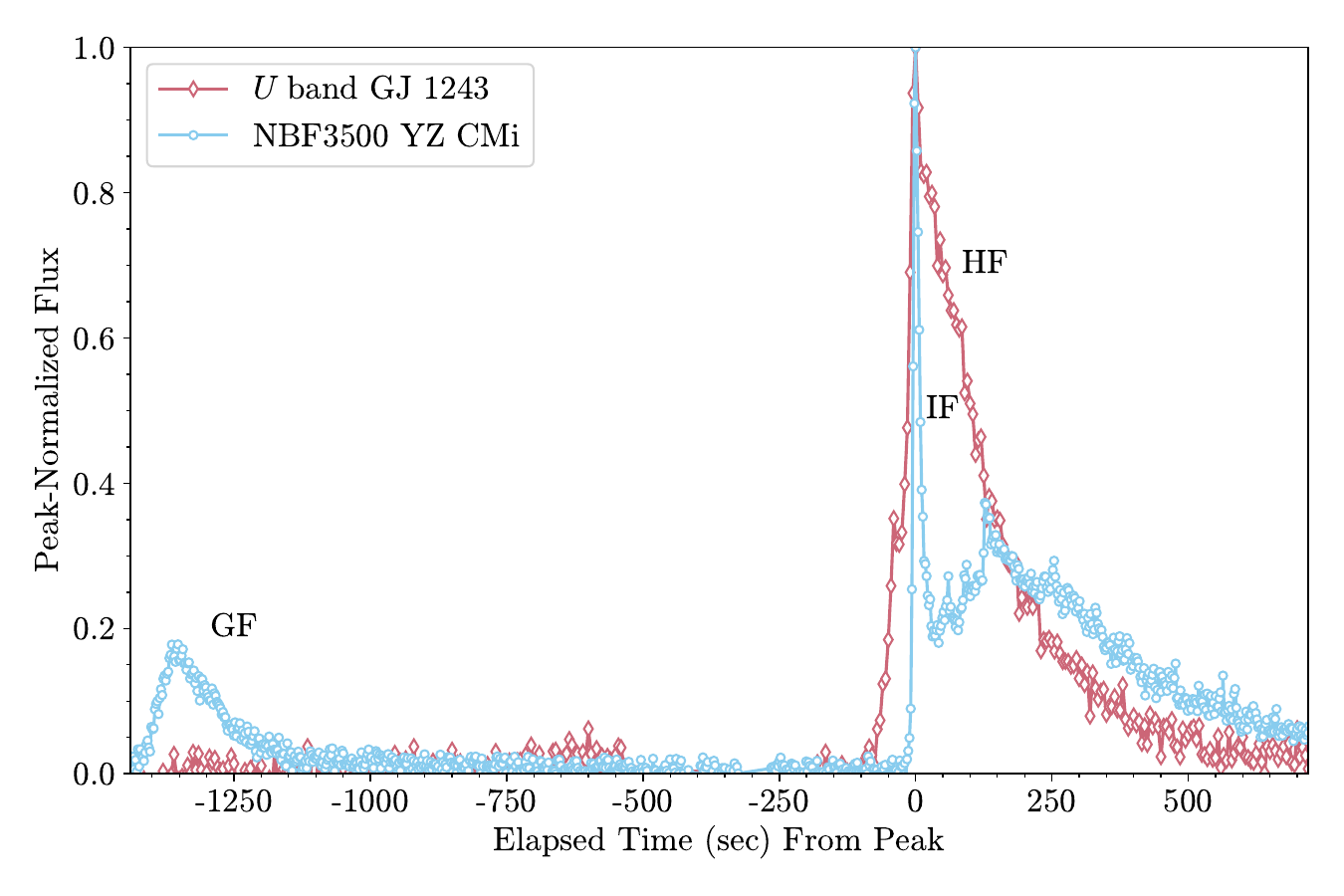} 
\caption{ Representative high-cadence optical light curves of M Ve flares. (Top) A randomly chosen  section of the 1-minute cadence Kepler light curve of GJ 1243 from \cite{Davenport2014} showing a variety of classical and complex flares superimposed on the modulation, which is due to starspots \citep{DavenportSpots}.  The data were obtained from \url{https://github.com/jradavenport/GJ1243-Flares}.  Note, there was a single flare event (not shown) that was observed simultaneously in both Kepler and in the $U$-band \citep{Hawley2014}.  For reference, this flare had a relative peak contrast of $\Delta F / F_{\rm{median}} = 0.014$ in the Kepler band, a peak contrast of $I_{f} = 0.9$ in the $U$ band, and a total $U$-band energy of $10^{31}$ erg.  (Middle left) The impulsive and gradual decay \emph{phases} of a high-energy gradual flare (GF) event on EV Lac;  reproduced with permission from \cite{Kowalski2013}.  The impulsive phase and gradual decay phases are indicated.  At the peak (vertical black dashed line), GF-type events like this one have large Balmer jumps, which are still much smaller than predicted by optically thin hydrogen recombination theory alone (e.g., Sect.~\ref{sec:peakphase}).  The $t_{1/2} = 13$ min is the FWHM of the light curve, and the $U$-band impulsiveness index, $\mathcal{I} = I_f / t_{1/2}$, is 0.5 for this flare.   (Middle right)  An example high-energy IF event ($t_{1/2} = 2.5$ min, $\mathcal{I} = 8.3$) on EQ Peg A, reproduced with permission from \citet{Kowalski2013}.   (Bottom) Examples of lower energy flares within the flare IF/HF/GF classification scheme based on the impulsiveness in the $U$-band or a narrow-band filter (NBF) with $\lambda_c = 3500$\AA.  The HF and IF events have similar peak contrasts ($I_f = 2.7 - 3.7$) but differ significantly in the values of $t_{1/2}$ (14~s and 120~s).  The IF and GF events are referred to as IF4 and GF2 in \citet{Kowalski2016};  the HF event is referred as the HST-2 flare in \citet{Kowalski2019HST}.  Note, the flare-only impulsive-phase spectrum of the IF event is shown in comparison to common photometry filters in Appendix \ref{sec:filters}.    }
\label{fig:lcexamples}      
\end{center}
\end{figure*}

Several classification schemes of flares have been developed based on optical broadband durations and light curve shapes (``morphologies'').  \cite{Bopp1973} and \cite{Moffett1974} established some of the basic descriptive terminology of flare classification, such as \emph{complex}, \emph{spike}, \emph{slow}, \emph{typical}, or \emph{multipeak} (a type of complex flare).   More recently, a condensed scheme denotes flares as either \emph{classical} or \emph{complex} \citep{Davenport2014};  several classical and complex events are evident throughout a small section of the Kepler data of GJ 1243 in Figure \ref{fig:lcexamples}(top) panel.  \cite{Kowalski2019HST} argued that flares with several events that comprise the rise phase can appear as relatively simple, ``classical'' flares if degraded to 1-minute time-integrations.   A classification of solar flares into ``gradual'', ``impulsive'', and ``secondary'' was proposed by \cite{Cliver1986}  based on extremely high-time resolution, hard X-ray data.  Hard X-ray data is rarely available for stellar flares, and impulsive and gradual classifications have thus been determined from broadband $U$- and optical light curves. 

A common classification utilizes a measure of the impulsiveness (also referred to as ``impulse'' or ``impulsivity''), of a flare.  
 \cite{HP91} quantified the impulsiveness as the fraction of energy emitted in the impulsive phase, which ends after the fast-decay phase(s) in broadband optical radiation.  They found that about 70\% of the energy was emitted in the impulsive phase during two\footnote{The light curve of the smaller event is shown in the review by \citet{Byrne1989}.} flares that are separated by $\sim 2$ orders of magnitude in total energy.  These were both considered very impulsive events. \cite{Kowalski2011ASPC} classified three events as ``gradual'', ``impulsive'', and ``traditional'' in ultra-high-cadence (0.16~s) photometry data.   \cite{Kowalski2013} calculated the impulsiveness index ($\mathcal{I}$) as the peak value of  $I_f$ divided by the FWHM ($t_{1/2}$) of the broadband ($U$) light curve to classify events into ``impulsive flares (IF)'', ``hybrid flares'' (HF), and ``gradual flares'' (GF).  Examples of high-energy, $E_U > 10^{32}$ erg, GF and IF events are shown in Figure \ref{fig:lcexamples}(middle).  Lower-energy and lower-amplitude IF, HF, and GF events are shown on the same axes in the bottom panel of Figure \ref{fig:lcexamples}. \cite{Dal2010} divide a large sample of dMe flares into fast and slow flares and suggest the differences are related to the position on the stellar disk.  This hypothesis could be verified with spectral properties that are thought to change with position on the solar disk \citep{Neidig1993Limb}. 
 
 The \citet{Kowalski2013} IF/HF/GF impulsiveness index is primarily a diagnostic of the broadband evolution during the impulsive \emph{phase} of a flare.  Thus, it is smaller for flares with a gradual decay phase that starts at a flux level that is a significant fraction of the peak flux.  Among the IF-type events, there is a large range, $30-70$\%, in the percentage of the total energy that is emitted in the impulsive phase.  Several spectral quantities (Balmer jump ratios, Balmer line-to-continuum ratios; Sect.~\ref{sec:wlcontinuum}) at the time of peak broadband flux correlate with this index;  one possible interpretation  is that the average impulsive-phase heating rate varies among events.  If the heating rate is attributed to the flux of electron beams, then the inter-flare variation of these optical spectral quantities suggests \citep{Kowalski2016} an important connection between large-scale flare evolution and electron acceleration, which is sometimes discussed in the context of nonthermal hard X-ray properties of solar flares \citep{McClymont1986, Holman2011}.    This expands upon the foundational paradigm that stellar flare scaling relations (e.g., H$\gamma$ \emph{v.} $U$-band energies), which extend over many orders of magnitude, mean that differences from flare to flare primarily result from the flaring surface area \citep[e.g.,][]{Butler1988, HP91}.    The empirical differences in IF/HF/GF spectra thus eventually motivated the semi-empirical modeling work in \citet{Kowalski2022Frontiers}, which suggests that differences among flares further result from scaling the relative areas of higher and lower concentrations of particle beams in the stellar atmosphere.  Recent analyses of H$\alpha$ line profiles from different locations within solar flare ribbons have begun to quantitatively explore the implications for spatially unresolved stellar data \citep{Namekata2022}.  

%  with a narrow range of injected heating fluxes.  See Kowalski 2022 frontiers, could be scaling of kernel and 

The fraction of flares within each of the IF, HF, and GF types is not yet well-established.    The reported numbers of flares that are IF/HF/GF \citep{Kowalski2013} are biased toward the IF type because the larger signal-to-noise at peak facilitates spectral characterization.  Note that rigorous distinctions cannot be made between an HF and GF classification for certain events \citep{Kowalski2019HST}.   It is also not clear how many IF events produce larger Balmer jumps \citep{Kowalski2016} than expected according to the original \cite{Kowalski2013} IF/HF/GF classification scheme. GF events in the decay phase of highly-impulsive events also have shown some of the smallest Balmer jumps  (in absorption). %IF-type events can show a large range of fraction of energy emitted in the impulsive phase (30-70\%).

Nonetheless, it is rather clear that the most gradual-type events in the light-curve evolution exhibit the largest Balmer jumps ratios and ratios of H$\gamma$ / C4170$^{\prime}$ at the peak time in broadband optical flux, whereas the most impulsive-type events show the smallest Balmer jumps in emission and the largest fraction of optical energy that is radiated in the continuum (Sect.~\ref{sec:lower}).  This is possibly an interesting connection to the Type I and Type II white-light flare classification on the Sun (Sect.~\ref{sec:wlcontinuum}), but the stellar events that would, ostensibly, be most analogous to Type II white-light flares in the spectral properties are actually the stellar events that are most impulsive.

We should clarify that an impulsive-type flare exhibits an impulsive phase and a distinct gradual decay phase, while a gradual-type flare can show a gradual decay phase that is clearly distinct from its impulsive phase.  Figure~\ref{fig:lcexamples} (middle left) shows an energetic, gradual-type flare event that exhibits several faster impulsive events superimposed on a more gradual rise phase;  the impulsive phase of this GF event ends around $t = 2.3$ hrs and transitions to a more gradually decaying phase.

\subsection{Quasi-Periodic Pulsations} \label{sec:qpp}

High-time resolution data of stellar flares facilitate searches for periodicity over the light curve evolution.  
Some events have shown fluctuations in the broadband flux having periods with a modulated amplitude or an evolving period over a cycle or two;  these are called quasi-periodic pulsations (QPPs).  This topic is of significant interest within the solar and stellar communities \citep{Vivering2023}, but an exhaustive discussion and reference list are outside the scope of this review.  I summarize a few selected results.  For detailed discussions and recent reviews, see \cite{Broomhall2021} and \cite {Zimovets2021}.

QPPs in stellar flares are reported on a variety of timescales, from 10~s to 30 minutes, and they occur in the impulsive and gradual phases.  
 \cite{Mathioudakis2006} report a 10~s period during the $u$-band flare in the peak phase using a wavelet analysis \citep{Torrence1998} and discuss different interpretations, including one in which the reconnection and particle acceleration are modulated through MHD oscillations in a nearby large loop \citep{Nakariakov2006}.
\cite{Anfin2013} detect a damped 32 minute oscillation during the decay phase of a $U$-band megaflare \citep{Kowalski2010} on YZ CMi using Lomb-Scargle and autocorrelation analyses of the detrended light curve.   \citet{Doyle2022QPP} discuss shorter, $1-4$ minute, QPPs in two giant ($E \approx 5-10 \times 10^{33}$ erg) flares observed on YZ CMi with $0.25 -0.6$~s sampling.  
The AFINO method \citep{Inglis2015, Inglis2016} is another powerful technique for evaluating the statistical significance of QPPs.  \cite{Inglis2015} apply the AFINO method to the stellar event in \cite{Anfin2013} and \citet{Kowalski2010} using a Fourier power-spectrum analysis that includes the component of the flare signal that is detrended in the wavelet technique.  This approach accounts for the fact that some types of power-spectra may naturally give rise to types of bursty light curve behavior in the temporal domain.

Stellar flare QPPs are reported in data from the optical to the X-ray regimes on a wide variety of stars.
 \citet{Cho2016} investigate a tight correlation between the damping times and periods of QPPs in a sample of X-ray stellar flares.  White-light QPPs in \emph{Kepler} data are presented in \cite{Balona2015}.  \cite{Mathioudakis2003} report remarkable oscillations with a 240~s period during the peak phase of a $U$-band flare on the RS CVn II Peg, and \cite{MitraKraev2005B} discuss several physical origins in the damped 750~s period in the X-ray flare on the lower-mass system AT Mic.   For an analysis of multi-band data during the decay phase of an X-ray flare on the young G star EK Dra, see \cite{Broomhall2019}.  QPPs of 320~s and 660~s periods were reported over the peak phase of a remarkably energetic flare on a PMS M3 star \citep{Jackman2019QPP}.

\section{Multi-Wavelength Spectral Observations} \label{sec:spectra}

Multi-wavelength observations of stellar flares are grouped into categories according to the regions of the atmosphere producing the electromagnetic response.  A predominantly thermal response in the optical and near-ultraviolet originates from the cool, $T \approx 10^4$ K, dense lower atmosphere (Sect.~\ref{sec:lower} -- \ref{sec:wlcontinuum}), while the far-ultraviolet emission lines probe the rapid response of transition region temperatures around $T \approx 10^5$ K (Section \ref{sec:FUV}).  Nonthermal radiation  originates from  accelerated particles gyrating in coronal magnetic fields (Sect.~\ref{sec:radioobs} -- \ref{sec:alma}), and a thermal response results from the hot, $T \gtrsim 10^7$ K, tenuous upper atmosphere (Sect.~\ref{sec:upper}).   The temporal correlations (Sect.~\ref{sec:neupert}) among these radiative responses justify a solar-like modeling paradigm with nonthermal electron beams and chromospheric evaporation and condensation processes (Sects.~\ref{sec:solar} and \ref{sec:models}).

\subsection{NUV and Optical:  The Thermal Radiative Response of the Footpoints to Impulsive Nonthermal Heating from Above} \label{sec:lower}
The optical and NUV response is one of the most enigmatic and energetic aspects of stellar flares.   This phenomenon is known for its dramatic impulsive phase, but it also exhibits a long-duration gradual decay phase that can extend for many hours after a bright peak flux phase.  Long-duration continuum flare radiation may persist after relatively small peaks as well \citep[see the GF1 event in \citealt{Kowalski2013} reproduced in the middle left panel of Fig.~\ref{fig:lcexamples} and the event studied in][]{Hawley1995}.  The optical and NUV radiation is thought to be thermal radiation that originates from photospheric and/or chromospheric heights.  
For reference, Table \ref{table:magchanges} summarizes several of the largest broadband optical flux enhancements in well-studied MVe flares with multi-band photometry data that include the $U$- or SDSS $u$-band. % The distribution of peak-flux enhancements in the $U$-band from \citet{Moffett1974} are shown in Fig.~\ref{fig:} (for YZ CMi).  

\begin{sidewaystable}
%%\sidewaystablefn%
%%\begin{center}
\caption{Flux enhancements at peak phase for several especially large MVe events with simultaneous multi-band photometry.  The flux enhancement values  ($I_f + 1$; Sect.~\ref{sec:proxcen}) are reported relative to the pre-flare or quiescence in each band.  The flux enhancements are related to magnitude changes through Equation \ref{eq:magchange}.}
\label{table:magchanges}
\begin{tabular*}{\textheight}{@{\extracolsep{\fill}}lcccccccccccccc@{\extracolsep{\fill}}}
%\begin{tabular*}{lccccccccccccccc}
\toprule%
(row) Star   & SpT   & $E_{U/u}$ (erg)                   & $U$  & $u$   &$B$ & $g$ & $V$ & $r$   & $R$ & $I$  & $i$ & 3500$^{\dagger}$ & 4170$^{\dagger}$ & 6010$^{\dagger}$  \\
\hline
(1a) YZ CMi & M4.5Ve &     $\approx 4\times10^{33}$     &      &       &    &     &      &      &     &      &     & 30.9 & 7.6 & 1.7 \\
(1b) YZ CMi & M4.5Ve &     $\approx 4\times10^{33}$     &      &       &    &     &      &      &     &      &     & 106 & 28.7 & 3.8 \\
(2) YZ CMi & M4.5Ve &     $1.85\times10^{33}$  &      &  78.2 &    & 6.3 &      &  2.6 &     &      & 1.35 &     &      &      \\
(3) AU Mic & M1(V)e    &   $1.6 \times 10^{33}$   &  3.4 &       &    &     & 1.13 &      &     &      &     &     &      &      \\
(4) AD Leo & M3Ve   &   $6.5 \times 10^{33}$   &  70  &       & 11 &     &  3.4 &      & 1.6 &      &     &     &      &       \\
(5) EV Lac & M3.5Ve  &   $3.9 \times 10^{32}$  &  40  &  40   &    & 4   &    &       &     &      &     &     &      &       \\
(6) EV Lac & M3.5Ve  &   $6.2 \times 10^{32}$  &  8.0 &  7.4  &    &  1.4 &    & 1.18  &     &      &     &     &      &       \\
(7) SDSS J001309 & M6Ve & \nodata              &      &  158  &    &  16.9 &   &  8.6   &    &      &  2.0&     &      &        \\
 % (7) AD Leo & M3.5Ve &  $> 10^{33}$ erg           &  
%Moffett flare  &             &     &     &       &     &     &     &     &    &     &     &      \\
%SDSS  &             &     &     &       &     &     &     &     &     &     &     &      \\
\bottomrule
\end{tabular*}
\footnotetext{(1a) and (1b): The maxima over the IF1 and IF3 flares, respectively, from \cite{Kowalski2016}; the combined, estimated $U$-band energy is given because they are part of the same ``Ultraflare'' event. (2): The IF3 event from \cite{Kowalski2013}; (3): Flare \# 23 from \cite{Tristan2023}; (4): The Great Flare from \cite{HP91}; (5): The biggest flare from \cite{Schmidt2012} and the most impulsive event (although labeled as IF10) in the sample of \cite{Kowalski2013};  (6): The GF1 event from \cite{Kowalski2013} that is shown in the middle left panel of Figure \ref{fig:lcexamples}.  $^{\dagger}$The 3500, 4170, and 6010 columns refer to the central wavelengths of the ULTRACAM narrow-band filters (NBF). (7) SDSS J001309.33-002552 from the SDSS Stripe 82 flare star sample \citep{Kowalski2009} and in Table \ref{table:flarestars1}.}
%%\end{center}
\end{sidewaystable}

\subsubsection{Overview of Emission Line Properties} \label{sec:emissionlinelc}
\begin{figure*}
\begin{center}
  \includegraphics[width=0.75\textwidth]{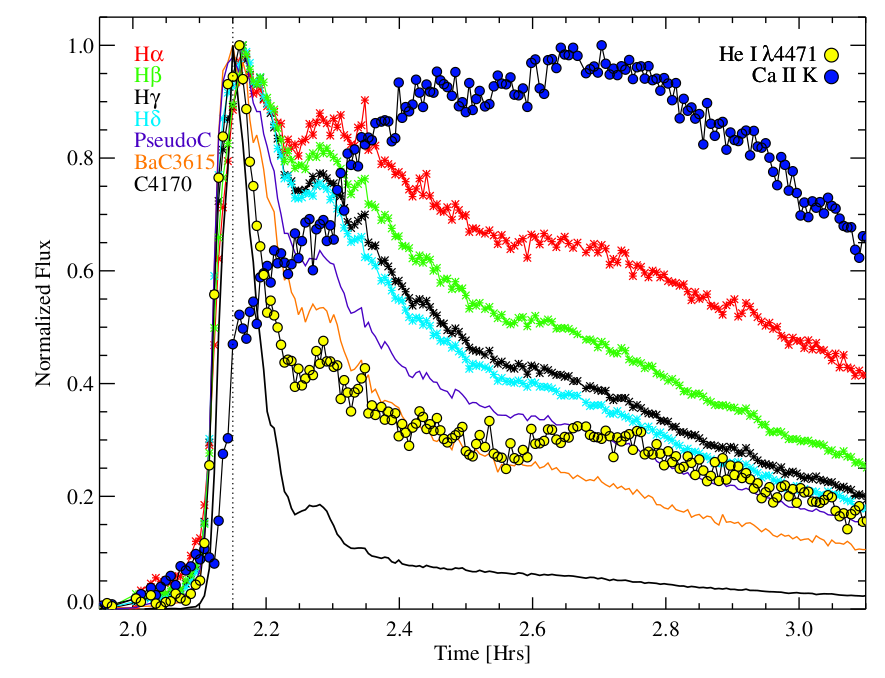} 
\caption{ Representative time-evolution of optical spectral quantities over a giant flare on the M4.5Ve star, YZ CMi.  The peak optical broadband magnitude enhancements are listed in row (2) of Table \ref{table:magchanges}.  The emission line fluxes are integrated over wavelength, and all light curves are normalized to their respective peak fluxes. The C4170$^{\prime}$ light curve is a proxy for the flare-only blue continuum flux averaged over $\lambda=4155-4185$ \AA\ and is the fastest to decay after its peak flux is attained.  The H$\alpha$ and Ca II K emission lines are the slowest. Figure reproduced from \cite{Kowalski2013} with permission. }
\label{fig:lcsummary}       % Give a unique label
\end{center}
\end{figure*}

An overview of the temporal response of various emission lines and continuum fluxes in the $U$ band and optical regimes throughout a high-energy flare ($E_u > 10^{33}$ erg; Table \ref{table:magchanges}) is shown in Fig.~\ref{fig:lcsummary}.  The temporal sequence in the figure is representative of many  stellar flares, although quantitative differences among events occur \citep{Houdebine1991, Houdebine2003, Kowalski2013, Kowalski2019HST}.  The line fluxes respond rapidly in the impulsive phase, except that Ca II K (and H) peak well into the gradual decay phase of the optical flare continuum flux.  The late peak of Ca II K is a remarkable phenomenon in stellar flares \citep{HP91, Houdebine1993A, Houdebine2003, Garcia2002, Kowalski2013}.   The blue continuum flux is the fastest to rise to its maximum and then begin its decay, followed by the highest order Balmer lines \citep{HP91}.  The Balmer H$\alpha$ line is the slowest to decay relative to its peak \citep[see also][]{Namekata2020}. In some flares, the Balmer line peaks occur significantly after the peak of the continuum, which may be empirically related to secondary events \citep[e.g.,][and see also the IF4/F2 event in \citealt{Kowalski2016}]{HP91, Garcia2002}.
 The neutral helium lines exhibit a very wide range of decay timescales:  The blue He I lines decay quickly \citep{HP91}, but He I 10830 \AA\  (studied in a different MVe flare) exhibits one of the most gradually evolving light curves \citep{Schmidt2012}. 

Quantitative assessments of the time-scales of emission line and continuum fluxes have been investigated with a time-decrement \citep{Kowalski2013, Kowalski2019HST}.  
 The time-decrement is defined as the empirical trend of the full-width-at-half-maximum ($t_{1/2}$) of each light curve, plotted as a function of the wavelength of the transition.  Are the relative decay timescales in Fig.~\ref{fig:lcsummary} consistent with a simple model consisting of a slowly decreasing average temperature over the flare footpoints on the star \citep{Gurzadian1984, Houdebine1991}?   This explanation would be analogous to the cooling thermal loop sequence in solar flares \citep{AschwandenAlexander2001} that is attributed to the temporal offsets in the maximum of optically thin, XEUV light curves.   The average temperature properties of the flare region are undoubtedly important in understanding stellar flare light curve evolution in Fig.~\ref{fig:lcsummary}.  However, the ratios of the Balmer line fluxes (which is called the Balmer flux decrement, or just ``Balmer decrement'') and line-to-continuum flux ratios in the observations  suggest that chromospheric optical depths \citep{Kunkel1970, Drake1980}, vertical and transverse spatial inhomogeneities \citep{Kowalski2015, Kowalski2022Frontiers}, and non-LTE radiative transfer \citep{HF92, Allred2006} are probably also critical to interpret the sustained chromospheric emission line fluxes (and the relative evolution of the optical continuum fluxes) in stellar flares.   In summary, there are not yet any comprehensive physical models that explain the absolute and relative timescales of the various optical spectral variations in Fig.~\ref{fig:lcsummary}.

\subsection{Spectral Properties of the Optical and NUV Continuum Radiation} \label{sec:wlcontinuum}
The continuum fluxes through the NUV and optical are clearly much more impulsive than the emission lines in Fig.~\ref{fig:lcsummary} (see also \cite{HP91, Garcia2002, Kowalski2019HST}).
The observed flare continuum radiation from the NUV through the optical is collectively referred to as the white-light\footnote{The $\lambda = 2000-4000$ \AA\  ultraviolet spectral region can be roughly divided into the NUV ($\lambda = 2000-3200$ \AA), and the $U$- (or SDSS $u$-) band ($\lambda=3200-4000$ \AA). The UV-C covers $\lambda = 2000-2800$ \AA , while the FUV typically refers to wavelengths between $\lambda=1000-2000$ \AA .   The blue-optical refers to $\lambda=4000-4800$ \AA , the green-optical to $\lambda=4800-5800$ \AA , and the red-optical to $\lambda=5800 - 7000$ \AA .   Sometimes the $U$-band is considered part of the optical because it can be observed from the ground.  The $\lambda=1800-2400$ \AA\ range is sometimes referred to as the mid-UV.  % 

\hspace{5mm} In reference to stellar flare radiation, white-light has traditionally meant ``detected in broadband optical filters''; before the immaculate precision provided by Kepler, this unambiguously referred to flare continuum radiation.  In solar physics, a common anecdotal use of ``white-light flare''  refers to an enhancement of photospheric continuum radiation and thus the excitation of the solar atmosphere in very deep layers.  Adding to the ambiguity, there may be two types of white-light solar flares, Type I and Type II, which are classified according to their optical and Balmer jump spectral properties \citep[see, e.g.,][]{Ondrej2}.   The introduction of  \cite{Kowalski2019IRIS} provides further discussion about the ambiguities in terminology surrounding the use of ``white-light'' in solar physics and astrophysics.  Some researchers colloquially refer to the white-light (``flare'' implied) as an increase in  the optical continuum radiation only (excluding shorter wavelength $U$-band and NUV continuum radiation).  Note that bona-fide ``white light'' is defined according to the CIE D65 white-light standard illuminant, which is noon sunlight.   See \cite{Cranmer2021} for human color vision synthesis of various stellar colors.  Note, we have synthesized the ``true'' colors in Fig.~\ref{fig:scales} from representative spectra of each source that is illustrated.}.  In this review, we adopt the empirical definition (without regard to its origin in the stellar atmosphere) that a white-light flare is a broad-wavelength increase in the observed NUV, $U$-band, and optical continuum stellar flare radiation that sometimes extends into the FUV and NIR. Thus, the white light would contribute a majority of the integrated flux in optical broadband filters (e.g., $UBVR$, Kepler, TESS).  

 In this section, we review observational properties of optical and NUV flare continuum radiation, as revealed by spectroscopic measurements. We separately discuss peak/impulsive phase spectra (Sect.~\ref{sec:peakphase}) and gradual decay phase spectra (Sect.~\ref{sec:decayphase}).    In Sect.~\ref{sec:colorimetry} and Sect.~\ref{sec:models}, we review some closely-related results from analyses of colors in broadband photometry (``colorimetry''). The M-dwarf flare spectral observations, combined with detailed modeling, suggest that there are unexpected amounts of heating over a large (deep) column mass density (hereafter, just ``column mass'' or $m_c$; g cm$^{-2}$).   Whether the $\lambda = 1300-9000$ \AA\ continuum radiation in stellar flares is caused by heating the photosphere to bona-fide incandescence (i.e., isotropic blackbody radiation or a blackbody-like spectral intensity) is an open question that is discussed in the next section and further in  Sect.~\ref{sec:models}.  The data discussed in this review come mostly from dMe flares, which produce large contrasts against their non-flaring photospheric radiative fluxes at blue and optical wavelengths.  This property facilitates isolating ``flare-only'' spectra, which can be readily compared to models without the ambiguities of subtraction artifacts.

\subsubsection{Rise and Peak Phase} \label{sec:peakphase}
\begin{figure*}
\begin{center}
  \includegraphics[width=0.8\textwidth]{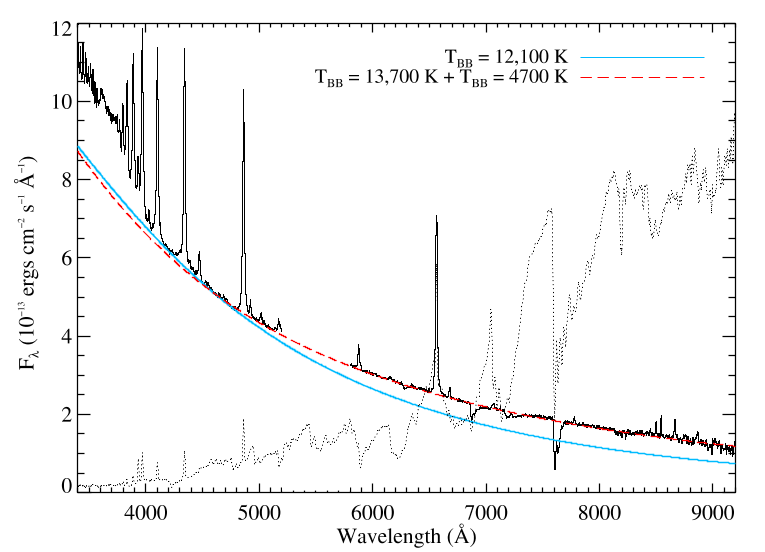}
\caption{ Flare spectrum at the peak of the large, impulsive-type (IF) event on YZ CMi in Fig.~\ref{fig:lcsummary}.  Several blackbody fits to optical ($\lambda \ge 4000$ \AA) continuum regions are shown.  The flare spectrum (black) has the pre-flare (dotted) spectrum subtracted.  Figure reproduced from \cite{Kowalski2013} with permission. }  % R = 590 spectrum. }
\label{fig:yzcmi_peak}      
\end{center}
\end{figure*}

A commonly reported empirical property of the optical, $\lambda \ge 4000$ \AA, flare-only continuum radiation in the spectra of the impulsive phase of stellar flares is a color temperature matching a $T \approx 8,000-14,000$ K blackbody \citep{Mochnacki1980, HP91, Katsova1991, Paulson2006, Kowalski2013, Gizis2013, Lalitha2013, Kowalski2016}.  The blackbody color temperatures might be purely phenomenological, in which case the color temperature values are parametrizations of the spectral shapes rather than analogs to in situ ``thermometer readings'' (Appendix \ref{sec:Tbb}).  A temperature   value is simply easier to relate to than continuum flux ratios, which are unique to specific spectral windows for a given color temperature.   The continuum radiation that is consistent with $T \approx 10^4$ K optical blackbody color temperatures is thus referred to as ``hot, blackbody-like''.   As we discuss further in Sect.~\ref{sec:models},  there are possible physical explanations from optically thick hydrogen recombination radiation.    A peak-phase spectrum is shown in Fig.~\ref{fig:yzcmi_peak}, which was taken at the peak of the event whose evolution is showcased in Fig.~\ref{fig:lcsummary}.   Single-component blackbody function models in the range of $T \approx 9000 - 12,000$ K can be fit to this flare spectrum  at $\lambda \ge 4000$ \AA\  outside of the major emission lines \citep{Kowalski2013}.  Multi-component blackbody curves with higher and lower temperatures are fit to the entire $\lambda \gtrsim 4150$ \AA\ wavelength range and are compared to a single blackbody fit to the blue-optical, $\lambda = 4000 - 4800$ \AA , range in Fig.~\ref{fig:yzcmi_peak}.

The broadband photometry observations from \citet{HP91} established the 9500 K blackbody hypothesis for the continuum in an energetic $E \sim 10^{34}$ erg flare from the M3Ve star AD Leo (see Sect.~\ref{sec:models} for additional discussion), which was verified in \cite{Kowalski2013} by fitting a blackbody temperature of $T \approx 11,600$ K to the optical range in the spectra of this flare.  Further spectral investigation at resolving powers of $R \sim 500-1000$ in other events has revealed that a $T_{\rm{BB}} \approx 8500 - 14,000$ K blackbody color temperature in the blue-optical, $\lambda = 4000-4800$ \AA , is  not a unique property of extremely energetic, large amplitude flares: for example, similar color temperatures were calculated in the impulsive phase  flare spectrum (Appendix \ref{sec:filters}) of the IF event in Fig.~\ref{fig:lcexamples}(bottom), which has an energy of $E_U\approx 1.6 \times 10^{31}$ erg \citep{Kowalski2016}.  Hot blackbody-like continuum radiation in the blue-optical regime tends to be more prominent, or more often reported, in the rise and peak phases of impulsive-type (IF; Sect.~\ref{sec:flare_classification}) events.   The flare event in Fig.~\ref{fig:yzcmi_peak} is a highly impulsive event.  We refer the reader also to the discussions in \citet{Kowalski2016} and \citet{Kowalski2023} pertaining to the variation in the optical blackbody color temperatures on $\Delta t= 3$~s cadences in the large, IF-type events listed in rows (1a) and (1b) in Table \ref{table:magchanges}.

\begin{figure*}
  \begin{center}
  \includegraphics[width=1.0\textwidth]{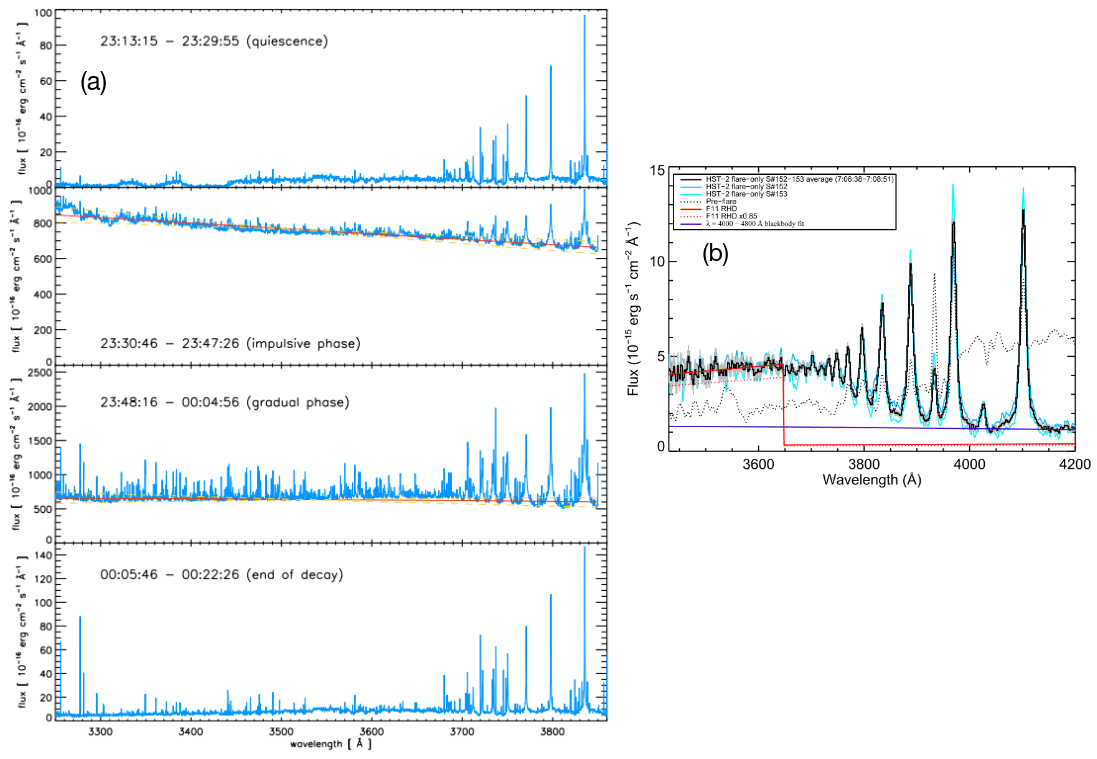}
\caption{ \textbf{(a)} VLT/UVES echelle ($R\sim 40,000$) spectral observations at $\lambda < 3860$ \AA\ that nearly extend to the terrestrial atmospheric cutoff.  These spectra show the evolution of a giant white-light flare on the M6Ve star CN Leo, reproduced from \cite{Fuhrmeister2008} with permission.  A $T \sim 11,300$ K blackbody is fit in the impulsive phase (2nd from top), and the gradual phase spectrum is fit with a  $T \sim 9100$ K blackbody in the third panel.   A catalog of the emission lines in the spectrum in the middle panel, extending to $\lambda = 3060$ \AA , is available through VizieR.   \textbf{(b)} Other MVe flares in the impulsive phase show a large Balmer jump and a flatter $U$-band continuum spectrum that decreases into the NUV at shorter wavelengths than shown here \citep{Kowalski2019HST}  are not satisfactorily explained by a $T \approx 9000$ K blackbody fit to the blue-optical continuum wavelengths.  Here, the two blue-colored spectra correspond to the rise/peak phase and early gradual decay phase, respectively, which are averaged into the black spectrum. The red-colored spectrum is an optically thin hydrogen recombination continuum model from a RHD simulation that invokes a solar-type electron beam heating function; the model is scaled to the observed spectrum at $\lambda \approx 3615$ \AA.  The spectra have a low-resolving power, $R \approx 600$, which is clearly sufficient for robust wavelength-integrated emission line fluxes.  Figure reproduced with permission from \citet{Kowalski2019HST}. }  % R = 590 spectrum. }
\label{fig:Fuhrmeister2008}
\end{center}
\end{figure*}

In  echelle flare spectra with high-resolving power, \citet{Schmitt2008} and \citet{Fuhrmeister2008} have shown that the continuum radiation within the $U$ band at $\lambda = 3250 - 3860$ \AA\  also exhibits hot blackbody color temperatures of $\approx 11,300$ K in the impulsive phase (Fig.~\ref{fig:Fuhrmeister2008}a).  At lower resolving power, similarly blue spectral trends at wavelengths shortward of the Balmer limit have been noted in \citet{Kowalski2013} for a large sample of flares.  Though spectral observations within the $U$-band benefit from increased flare contrast, especially afforded by M dwarfs, accurate measurements are generally very difficult due to the systematic uncertainties of the flux calibration near the terrestrial atmospheric limit.

Spectral observations with high signal-to-noise (S/N) at $\lambda > 3600$ \AA\ reveal a very important clue about the origin of the white-light continuum radiation and its hot blackbody-like property.  An extrapolation of a blackbody function that is fit to $\lambda = 4000-4800$ \AA\ (thus giving $T_{\rm{BB}}$), or an extrapolation of a blackbody that is fit to the continuum flux ratio, $f^{\prime}_{4170 \rm{\AA}} / f^{\prime}_{6010 \rm{\AA}} = $C4170$^{\prime}/$C6010$^{\prime}$ (thus giving $T_{\rm{FcolorR}}$), fails to account for excess continuum flux at wavelengths shorter than $\lambda \lesssim 3700$ \AA.  In other words, there is a Balmer jump between the blue-optical continuum flux and continuum flux at shorter wavelengths within the $U$ band.  An isothermal blackbody function thus does not comprehensively explain the panchromatic continuum flux properties.  A Balmer jump was tentatively noted in \citet{Kunkel1970} and \citet{Moffett1974}, and it was clearly evident and extensively characterized in impulsive-phase flare spectra using modern instrumentation and high-time resolution in \citet{Kowalski2013, Kowalski2016}.  An impulsive phase spectrum during a $E _U \approx 10^{31}$ erg, gradual/hybrid-type flare on GJ 1243 is shown in Fig.~\ref{fig:Fuhrmeister2008}(b) over a broader spectral range than in panel (a).  This is one of two flares analyzed in \cite{Kowalski2019HST} that produced similar decreases in the continuum flux into the NUV range at $\lambda < 3200$ \AA, as constrained by Hubble Space Telescope data (not shown here).  The Balmer jump in this event is the largest that has been detected spectroscopically over the impulsive phase of an M dwarf flare.  In flares with very prominent hot blackbody-like continuum radiation at optical wavelengths, such as the Great Flare of AD Leo and the flare in Fig.~\ref{fig:yzcmi_peak}, the Balmer jumps are evident as smaller flux excesses above the blackbody extrapolations to $\lambda \lesssim 3700$ \AA.  %  Flares that are completely dominated by a single are very rare.

\begin{figure*}
\begin{center}
   \includegraphics[width=1.0\textwidth]{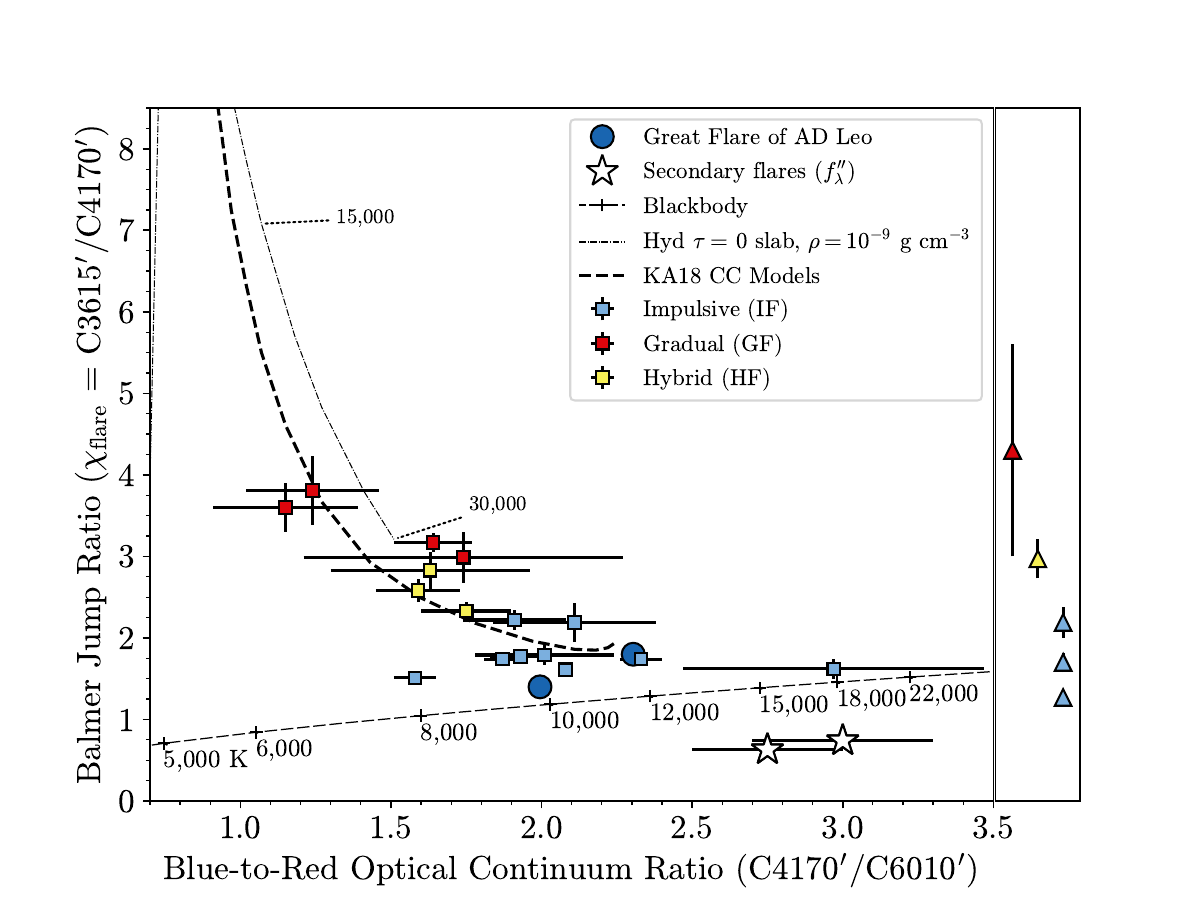} 
\caption{Compilation of impulsive-phase continuum flux ratios (color-color diagrams) from spectra \citep{Kowalski2013, Kowalski2016, Kowalski2019HST}.  The y-axis is a measure of the Balmer jump ratio, and the flux ratios on the x-axis correspond to optical color temperatures.   Note that the $x$ and $y$ uncertainties are correlated, but only marginal error bars are shown.  The prime symbols indicate that a pre-flare flux has been subtracted. The optical continuum colors from the Great Flare of AD Leo \citep{HP91} are estimated from extrapolations of blackbody curve fits over the $\lambda = 3800 - 4440$ \AA\ range \citep{Kowalski2013}.   The wavelength range of C4170$^{\prime}$ can be seen in the  right panel of  Fig.~\ref{fig:Fuhrmeister2008}.  The flux  ratio values at the peaks of flares indicate optical depths between blackbody radiation and optically thin hydrogen recombination radiation.  The general trend is well-reproduced by approximations \citep[KA18 CC models;][]{KA18} to the dynamic atmospheres in RHD simulations (Sect.~\ref{sec:models}).  In the right margin plot, flares with Balmer jump measurements only are shown. The flares are color-coded to their $U$-band impulsiveness (IF/HF/GF) classification.  Other flare-peak flux ratios from narrowband ULTRACAM photometry \citep{Kowalski2016} are summarized in Appendix \ref{sec:color_color_appendix}. The color temperatures corresponding to the blue-to-red flux ratios in this figure are typically $\gtrsim 1500$ K less than that obtained from spectral fitting over the $\lambda = 4000 - 4800$ \AA\ range.  All data in this figure are provided in supplementary online material hosted on Zenodo. }
\label{fig:colorcolor}      
\end{center}
\end{figure*}

Impulsive-phase continuum flux ratios, which are ``colors'' if they are converted to magnitude differences, are calculated from flare spectra in \cite{Kowalski2013}, \cite{Kowalski2016}, and \cite{Kowalski2019HST} and are shown in Fig.~\ref{fig:colorcolor}.   Flux ratios have been calculated from isothermal static, slab models for $\tau(\lambda) \ggg 1$ (blackbody) and $\tau = 0$ (continuous hydrogen LTE emissivity; Appendix \ref{sec:slabs}) and are shown in Fig.~\ref{fig:colorcolor} for comparison.  The flare peak ``color-color'' diagrams from the spectra demonstrate that the Balmer jump strengths ($f_{3615}^{\prime}$/$f_{4170}^{\prime} =$C3615$^{\prime}/$C4170$^{\prime}$) imply that moderate-to-large continuum optical depths develop in stellar flare atmospheres.  Further quantitative understanding about the observed offsets from the blackbody line in Fig.~\ref{fig:colorcolor} requires detailed RHD model spectra, which are represented in the figure by the thick dashed line.  These RHD model calculations include wavelength-dependent opacities and optical depths over column mass/depth-dependent temperatures and electron densities (Sect.~\ref{sec:models}), which are not included in blackbody or other simple slab models.

 In Appendix \ref{sec:color_color_appendix}, we discuss comparable color-color diagrams using narrow-band ($\Delta \lambda = 50-100$ \AA) filters around $\lambda= 3500$ \AA, $4170$ \AA, and $6010$ \AA, which were designed specifically to avoid emission lines in flares and to measure the Balmer jump with the ULTRACAM instrument \citep{Dhillon2007}.  These narrow-band filters provide high-cadence ($\Delta t \approx 0.3 - 3$~s), simultaneous constraints of continuum variations while avoiding degeneracies in model fits to broadband filters (Sect.~\ref{sec:colorimetry}).  ULTRACAM count-flux ratios from a large sample of MVe flares  were analyzed  and converted into flare-only continuum flux ratios in \cite{Kowalski2016}.  The flare color from relative photometry is the ratio of two values of 

\begin{equation} \label{eq:coloreq}
I_f(t)  \langle f_{q,\lambda} \rangle_T
  \end{equation}

  \noindent in the notation of Equation \ref{eq:spectrophot}.
Several flares were  observed simultaneously with spectra over the impulsive to gradual decay phases, which facilitated unambiguous interpretation of the filter ratios (\emph{cf} Figure 3 in \citealt{Kowalski2016}).  A relationship between color temperatures fit to blue-optical spectra and to blue-to-red optical flux ratios is characterized by an offset of $\Delta T \approx 1900$ K (see Table 5 in \citealt{Kowalski2016}), in line with the various blackbody fits to the peak flare spectrum in Fig.~\ref{fig:yzcmi_peak}.

\subsubsection{The Decay Phase and The Evolution of Optical Spectral Properties} \label{sec:decayphase}
The $T \approx 10,000$ K blackbody-like  optical color temperatures do not last for long during MVe flares \citep[e.g.,][]{Kowalski2013, Kowalski2023}.
After the peak of a broadband (e.g., $U$-band) light curve, a gradual decay phase almost always follows a shorter period of faster decay.  A characteristic blue-optical, blackbody color temperature value at the start of the gradual decay phase is $T_{\rm{BB}} \approx 8000$ K \citep{Kowalski2013}. For equal $\lambda \approx 3615$ \AA\ flare-only ﬂux levels during the fast rise and fast decay phases, the fast decay phase is more than $\Delta T \approx 1000$ K cooler.  \cite{Mochnacki1980} observed similar trends for optical color temperatures using continuum data during
several MVe ﬂares at $\lambda= 4200 -6900$ \AA\ (e.g., see their Figure 3).  At red-optical wavelengths, the continuum becomes very flat in the gradual decay phase, with even cooler temperatures.  This was referred to as a red ``conundruum'' in \cite{Kowalski2013};  the cooler blackbody fits were linked to the increasing fraction of energy that was previously reported in the redder bands of photometry in the decay phase of the Great Flare of AD Leo \citep[see Figure 10 and accompanying discussion in ][]{HP91} and retrospectively to the deviations from   blackbody fits to the photometry of that flare (bottom right panel of Figure 11 in \citealt{HF92}).   See Figure 31 of \cite{Kowalski2013} for examples of the stellar flare gradual phase continuum spectra that exhibit cool $T \lesssim 5000$ K blackbody color temperatures in large dMe events. In smaller flares, it is difficult to accurately subtract the preflare spectrum at $\lambda \gtrsim 7500$ \AA\ \citep{Kowalski2019HST} and thus determine the NIR continuum properties.

It is not clear how to explain such persistent flat optical continua with relatively moderate-sized Balmer jump ratios;  optically thin (bound-free) Paschen continua and photospheric backwarming predict cooler color temperatures but also much larger Balmer jumps, which are not observed -- hence the conundrum.  A related phenomenon during the impulsive phase may be the cause of the lower color temperatures that are calculated over a very wide optical wavelength range (such as those on the x-axis in Fig.~\ref{fig:colorcolor}) that spans the blue-optical and red-optical regimes.  On the other hand, \cite{Fuhrmeister2008} report hotter blackbody temperatures over the red-wavelengths (not shown) in  the flare in Fig.~\ref{fig:Fuhrmeister2008}a.   Clearly, further characterization of the red-optical and near-IR continuum properties is warranted.  There may be important contributions to \emph{Kepler} flare energetics in the gradual decay phase  \citep{Hawley2014}. 

\begin{figure*}
\begin{center}
  \includegraphics[width=0.75\textwidth]{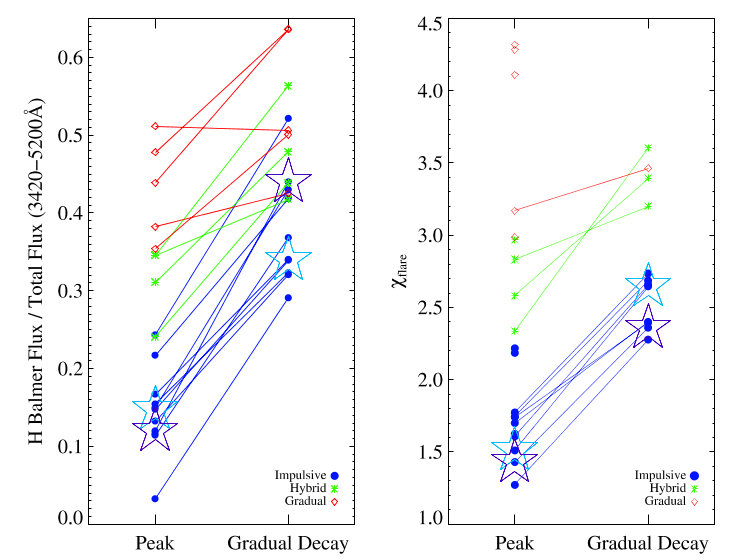} 
\caption{ Energy of the  hydrogen Balmer component vs. the optical continuum  energy evolution in a large sample of MVe flares (reproduced from \citealt{Kowalski2013} with permission).   The Balmer jump ratios ($\chi_{\rm{flare}}$) increase into the gradual decay phase and are also generally correlated with the IF/HF/GF classification scheme.  The hydrogen Balmer component includes an estimate for the Balmer continuum flux above an optical continuum model extrapolation to $\lambda < 3650$ \AA.  Several more flares with comparable spectroscopy \citep{Kowalski2016, Kowalski2019HST} since this result have been consistent with these trends.   }
\label{fig:energypartition}       % Give a unique label
\end{center}
\end{figure*}

At shorter wavelengths within the $U$-band, the Balmer jump ratio increases in the gradual decay phase of a flare.  The fraction of the wavelength-integrated energy in emission lines (primarily the hydrogen Balmer series in the optical and within the $U$-band wavelength range) also becomes larger.  The energy partition evolution is shown in Fig.~\ref{fig:energypartition} for a large sample of dMe flares.  It was found that the impulsive flare (IF) events were more continuum-dominated than the gradual flare (GF)-type events in both the impulsive phases and gradual phases of each event.  Thus, an empirical connection among the broadband impulsiveness, fraction of energy in the Balmer component, Balmer jump strength (right panel of Fig.~\ref{fig:energypartition}), and Balmer line-to-continuum ratios (namely, the H$\gamma$/C4170$^{\prime}$ ratios -- see Appendix \ref{sec:color_color_appendix}) was established from simultaneous spectral and broadband photometry observations \citep[see also][]{Kowalski2016, Kowalski2019HST}.

 Stellar flares in the NUV generally exhibit a rapid time evolution \citep{Brasseur2019}.
 However, the spectral properties in the NUV ($\lambda = 2000-3200$ \AA) are not nearly as well-observed as in the optical, in either impulsive- or gradual-type flares. The NUV spectral observations and model predictions are compiled and reviewed in \cite{Brasseur2023}.   \cite{Kowalski2019HST} investigated the NUV properties in two HF-type events (that show some optical spectral properties that are more in line with other GF-type events) with shorter-wavelength spectra ($\lambda = 2440 - 2840$ \AA) from the Hubble Space Telescope / Cosmic Origins Spectrograph.  They concluded that a Balmer continuum enhancement upon a blackbody-like optical continuum component is necessary to account for the continuum flux down to at least $\lambda \approx 2500$ \AA.  International Ultraviolet Explorer (IUE) FUV spectra covering $\lambda = 1900-3100$ \AA\ have been reported in the decaying phases of two large events on MVe stars \citep{HP91, Robinson1995}. For a review of the timing of the IUE/NUV and IUE/FUV spectral integrations within the AD Leo Great Flare, see \cite{Kowalski2022Frontiers}.  Of the few (relatively small) flares with NUV spectra that have been analyzed through their peak and decay phases, the fractions of energy in the continuum were calculated in the large range of $\approx 50-90$\% \citep{Hawley2007, Kowalski2019HST}, which is much larger than in quiescence (35\%).  
 The evolution of the $\lambda \approx 3300-3800$ \AA\ continuum-to-line energy fractionation is also noticeable after the peak phase in the M5.5/6Ve event in Fig.~\ref{fig:Fuhrmeister2008}(a).

\subsubsection{Empirically-determined, Broadband Radiated Energy Budgets Relative to the $U$ band} \label{sec:energies}
Empirical radiative energy fractionation from the X-ray to the radio provides model-independent\footnote{In some flare models of radiative backwarming \citep[e.g.,][]{Neidig1993}, the energy in one band is reprocessed energy from other bands.} characterization of stellar flares.  
The energy in the optical and NUV continuum provides a large contribution to the fractionation.  To date, \cite{Osten2015} is the most comprehensive analysis of  the multi-wavelength energy budget in archival observations of stellar flares across a wide variety of spectral types and quiescent activity levels.  At $\lambda= 1200-8000$ \AA, 96\% of the peak flare luminosity can originate from the continuum in the impulsive phase \citep{HP91}.  In the gradual decay phase, the continuum accounts for as much as 83\% when the emission lines remain highly elevated for longer durations (e.g., Fig.~\ref{fig:lcsummary}).  The energy radiated in soft X-rays (0.04 - 2 keV) is estimated to be   $\sim 11$x the H$\gamma$ energy \citep{Butler1988} and to be comparable to that in  the $U$-band, which is roughly 1/6 of the integrated $1200-8000$ \AA\ white-light energy \citep{HP91, Hawley1995}, and a factor of 1/10 of the bolometric flare energy \citep{Osten2015}. \cite{HP91} constrain $E_U = 0.64 E_{2000-3260\mathrm{\AA}}$, however the bandpass-averaged energy [erg $\rm{\AA}^{-1}$] in the $U$-band is reported as a factor of 1.2 greater.  Several estimates of the $U$-band to Kepler-band (Kp) energy ratios have been empirically determined to be $E_U = (0.4 - 0.65) E_{\rm{Kp}}$ \citep{Hawley2014}.  \cite{Lacy1976} determine $E_U = 1.2 E_B = 1.8 E_V$.  Line-integrated flare energies in H$\alpha$ are typically factors of $\gtrsim 0.04-0.08$ smaller than the $U$-band energies, and the H$\gamma$ energies are factors of $\approx 0.08$ of the $U$-band energies over several orders of magnitude in total flare energy \citep[see the large compilation in Figure 4.2 in][]{Kowalski2012}.

The relative energy in soft X-rays ($E \approx 0.1-5$ keV) is difficult to compare among flares.  This difficulty arises due to different temporal \citep[e.g.,][]{Hawley1995} and wavelength integration limits (e.g., by including a significant portion of the EUV range) among datasets.   The X-ray and optical thermal radiation occur on much different time-scales (Sect.~\ref{sec:neupert}), and a limiting assumption \citep[e.g.][]{Audard1999, Audard2000} is sometimes that the broad-band flare energy is accurately accounted for with a scaled 2-temperature quiescent coronal model \citep{Gudel1997EK}.  Nonetheless, many interesting empirical constraints on the X-ray to optical flare energies are reported in the literature.  The ratio of the soft X-ray (and/or EUV) to the $U$-band energy can take on a large range of values, from essentially no detectable response in the soft X-rays  within a $\pm 10-20$ minute window \citep{Doyle1988, Osten2005}), down to $1/3$ \citep{Hawley1995}, to as large as a factor of $\gtrsim 10$\ \citep{Gudel2002, Osten2016}.  \cite{Namekata2020} study the X-ray properties (Sect.~\ref{sec:multilam}) of several X-ray flares with NICER spectra.  The $E = 0.5-10$ keV soft X-ray energies fall in the range of $0.3 - 1\times 10^{32}$ erg, but the event with complete multi-wavelength coverage in their study showed an H$\alpha$ response without a detectable $g^{\prime}$-band (white-light) response.   In general, using the energy fractionation relation of $E_U= 1/3 E_{0.01-10 \rm{keV}} = 1/3 E_{\rm{XEUV}}$ in \cite{Osten2015}, one obtains consistent  cumulative flare rates in the XEUV and in the $U$-band \citep[e.g., $\sim 1$ per day greater than $10^{32}$ erg in $U$][]{Pettersen1984} corresponding to $1$ flare per day with $E_{\rm{XEUV}} \gtrsim 3 \times 10^{32}$ erg \citep{Audard2000} for non-simultaneous flares on AD Leo.  \cite{Tristan2023} calculate the energy ratios of six simultaneous $U$-band and soft X-ray ($E=0.2-12$ keV) flares to be $E_{\rm{SXR}} / E_U \approx 1.5$.  \cite{Schmitt2019} extend the comprehensive, multi-wavelength analysis of a $E \approx 10^{34}$ erg flare on the K0Ve star AB Dor from \cite{Lalitha2013}; they also report comparable energies released in soft X-rays and in an NUV bandpass.  This is consistent with a paradigm that a majority \citep{Osten2015, Kuznetsov2021} of the total bolometric response in many (but not all) white-light stellar flares occurs from energy deposition in the dense, chromospheric footpoints (Fig.~\ref{fig:standardcartoon}(right)).

 The peak radio (3.6 cm) specific luminosity (erg s$^{-1}$ Hz$^{-1}$) is $\sim 1/4$ of the $U$-band, but the bandpass-integrated energy is a factor of $\mathcal{O} (10^4)$ smaller \citep{Osten2005}.    New broadband observations have suggested much larger energy releases than expected\footnote{TESS results are not included in this iteration of the review, but we should note briefly the discrepancies with models are also discussed in \cite{Howard2020} and \cite{Jackman2023}.} in the NUV regime relative to the optical \citep{Brasseur2023}, while others are rather well-matched by current RHD models that invoke large heating rates \citep{Osten2016, Kowalski2019HST}.  Studies of high-cadence, simultaneous Galex/FUV and Galex/NUV photometry have reported much larger peak luminosities in the broadband FUV than in the broadband NUV \citep{Robinson2005} (but see the discussion of bright Galex sources in Sect.~\ref{sec:FUV}).  In the NIR, \citet{Davenport2012} present model predictions for the magnitude enhancements and energies relative to the $U$ band.  \cite{Tofflemire2012} used high-cadence, high-precision monitoring of $\Delta U = -1.5$ mag flares on mid-type M dwarfs to constrain the broadband NIR response to $|\Delta_{\rm{NIR}}| \lesssim 5-12$ milli-magnitudes.

\subsubsection{Optical Broadband Flare Colorimetry} \label{sec:colorimetry}

Optical flare colorimetry is the study of the temporal tracks of broadband flare colors or absolute fluxes.  For example, the $U-B$ vs. $B-V$ colors of flares on the dM3.5e star EV Lac were compared to regimes of optically thick and optically thin hydrogen spectra in \cite{Zhilyaev2007}.  They conclude that the flare continuum becomes optically thick (and more blackbody-like) in the peak phase of flares, but in the decay phase, the optical depths decrease.  \cite{Hawley2003} find a best-fit blackbody temperature of $T \sim $8500 K by fitting to the fluxes in $UBVR$ filters for a sample of eight moderate-sized flares on AD Leo.   Colorimetry with the traditional Johnson/Bessell bandpasses is a powerful and convenient method to test a wide variety of flare continuum models \citep{Kunkel1970, Mullan1976, MullanTarter1977, Cram1982, Doyle1989, HF92, Maas2022, Namekata2022, Brasseur2023}, and similar methods using broadband filters in the NUV and FUV have inferred a wide range of even hotter blackbody temperatures, $T \approx 20,000 - 50,000$ K \citep{Robinson2005, Getman2023}.

However, there are several important assumptions in broadband colorimetry that warrant spectroscopic verification or calibration \citep[e.g.,][]{Kowalski2016}.  We discuss a few here. First, a correction to the contribution from emission lines in the optical bandpasses must be considered \citep{HP91, HF92}.  The $V$ band is relatively free of major flare emission lines, but the contributions from the Balmer lines vary from flare to flare  in the $B$ band (e.g., Fig.~\ref{fig:energypartition}, Fig.~\ref{fig:filters} in Appendix \ref{sec:filters}).  The $U$ band integrates over the Balmer jump, a pseudo-continuum of blended Balmer lines, and to a lesser extent Ca II K and H emission.  \cite{Allred2006} demonstrated that a RHD model spectrum with a large Balmer jump and a discontinuity at $\lambda=3646$ \AA\ exhibits colors that are consistent with a $T\sim 9000$ K blackbody when convolved with the $UBVR$ filters -- but a blackbody function has no Balmer discontinuity or jump!  \cite{Kowalski2019HST} convolved an observed flare spectrum with broadband filters and showed that very large blackbody temperatures $\approx 15,000 - 22,000$ K could be inferred if the Balmer jump and high-order Balmer lines are not included in a model for the $U$-band and/or a constraint on the NUV continuum flux is not available (see their Figure 16).  They showed that the full optical continuum regime may be flatter than a hot blackbody fit to the blue-optical wavelengths in such flares.  The analysis of \cite{Hawley2003} included a bona-fide continuum flux constraint from HST spectra in the FUV, which may have otherwise   resulted in  much hotter peak blackbody temperatures, as in \cite{Zhilyaev2007}.

The results from broadband flare colorimetry have largely motivated spectroscopic investigations \citep{Giampapa1982, HP91, Kowalski2013} of flares and detailed modeling \citep{HF92}, which will be discussed in Sect.~\ref{sec:models}.

\subsection{FUV Observations: The Rapidly Evolving Flare Transition Region (TR)}  \label{sec:FUV}
The FUV continuum and emission line response may be a complex mixture of thermal, nonthermal, and nonlocal radiative processes in the flare transition region ($T \gtrsim 50,000$ K) and the deeper atmosphere at $T \lesssim 10,000$ K \citep{HP91, Phillips1992, Hawley2003, Ayres2015, Froning2019, Kowalski2022Frontiers, ADAS1, ADAS2}.  Many 
questions remain about the origin of the spectral and temporal properties in stellar flares in this wavelength regime.  A few observational properties are summarized in this section.

\citet{Hawley2003} presented a comprehensive, high-resolving power ($R=70,000$) study of eight moderate-sized M dwarf flares in the FUV with contemporaneous optical spectroscopy and broadband photometry.   The energy budgets and timing properties in the brightest spectral lines and continua were compared in detail.  In the biggest flare, the peak line fluxes in C II$\lambda$1335, Si IV$\lambda$1394, C IV$\lambda$1548, and C III$\lambda$1176 were calculated in the ratio of $\approx$ 25:35:50:100, respectively (note that the lineshifts in this study are discussed in Sect.~\ref{sec:chromlines}).  The study also presented a novel timing analysis of the light curves of Si IV, C IV, N V, and the FUV continuum region (averaged over $\lambda \approx 1266-1295, 1420-1452, 1675-1710$ \AA)   compared to the $U$-band by fitting $b$ and $m$ in the relation  $\log_{10}$($F_{X}$) $= b + m$ $\log_{10}$($F_U$).  For C IV 1551 and Si IV 1403, the slopes are rather consistent with a value of $\approx 1$, but the value of $m_{\rm{FUVcont}} \approx 1.7$ indicates a significantly more rapid evolution of the FUV continuum compared to the $U$-band, an effect that has also been noted in a much larger event on AD Leo \citep{HP91}.  The power-law relationship between the FUV continuum and the $U$ band is not yet explained by physical models.  A linear scaling among neutral silicon FUV continua, C IV, and C II luminosities among a sample of dMe and RS CVn flares was interpreted in \cite{Phillips1992} to be consistent with radiative backheating of the temperature minimum region.

\begin{figure*}[htbp!]
\begin{center}
    \includegraphics[width=0.75\textwidth]{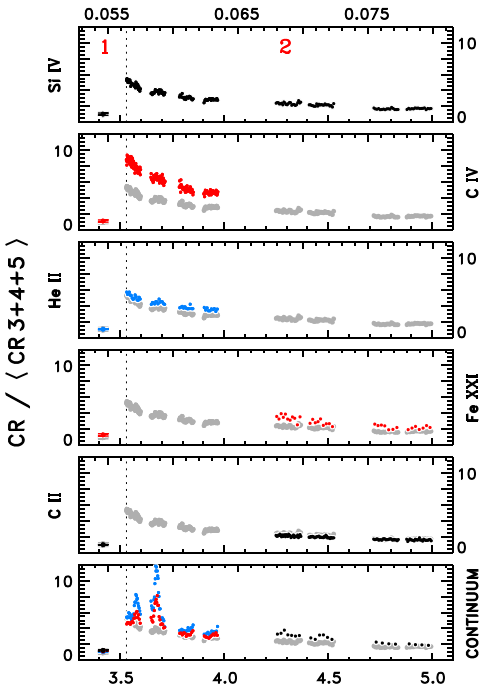} 
\caption{ Figure 7 from \cite{Ayres2015}, reproduced with permission, showing various FUV spectral quantities from HST/COS spectra during the decay phase of a large flare on the young solar analog, EK Dra.  The bottom $x$-axis is elapsed time (hr) and top $x$-axis is rotational phase.   Note the bursts that only occur in the FUV continuum ($\lambda = 1435$ $\pm$ 25 \AA , 1610 $\pm$ 25 \AA ).  Later in the decay phase, shorter FUV wavelengths were observed by HST/COS, providing constraints on the continuum at  $\lambda = 1344.5 \pm 6$ \AA\ and $\lambda = 1381.0 \pm 8.0$ \AA , the forbidden coronal line [Fe XXI]$\lambda$1354 ($T_{\rm{form}} \approx 10^7$ K), and C II$\lambda$1335.     }
\label{fig:ayres2015}      
\end{center}
\end{figure*}

Remarkable M-star FUV flare spectra were analyzed in \cite{HP91}, \cite{Redfield2002}, \cite{Loyd2018Hazmat}, \cite{Froning2019}, \cite{MacGregor2021}, and \cite{Feinstein2022}.   In nearly all cases, the continuum spectra are either very flat \citep[in units of  \flam; see also the spectra compiled in][]{Butler1981, Bromage1986, Byrne1990, Phillips1992} or an isothermal blackbody continuum fit to the  regions between emission lines results in $T \gtrsim 15,000$ K, up to $T \approx 30,000- 40,000$ K.   Of these, \cite{HP91} and \cite{MacGregor2021} had simultaneous multi-wavelength observations that were analyzed in detail, and \cite{Froning2019} and \cite{Loyd2018Hazmat} discuss redshifted line asymmetries in the FUV.  \cite{Ayres2015} discuss interesting FUV continuum bursts during the gradual decay phase of a Si IV light curve in an energetic event from the young solar analog EK Dra (Fig.~\ref{fig:ayres2015});  similar responses during the FUV bursts do not show up in higher temperature lines.  These continuum-only bursts were interpreted as possible stellar analogs of ``Type II'' white-light solar flares \citep{Fang1995}.

 Broadband comparisons of FUV and NUV responses during flares have been possible due to the time-tagged photon capability of the Galex mission.  \cite{Robinson2005} analyzed a giant flare in both Galex bands, and \cite{Welsh2006} showed similar  trends in other flares: very large ratios of the FUV to NUV fluxes that are far larger than expected from any known continuum model constrained by optical spectroscopy (Sect.~\ref{sec:wlcontinuum}).  It has been pointed out that non-linearity corrections are important considerations at these large count rates in Galex \citep{Morrissey2007,Fleming2022}.  More recently, the \texttt{gPhoton} tool \citep{Million2016} has been employed to study events in the NUV without such concerns at large peak count rates, revealing remarkably fast time variations on a wide variety of stars  \citep{Brasseur2019}.  Simultaneous constraints from Kepler and Galex observations favor very large flux ratios that are, again, far larger than RHD models of optically thick continuum radiation from  high-energy electron-beam heating models \citep{Brasseur2023}.

Potential signatures of proton beams are found in both the FUV and in the gamma rays.  \cite{Loh2017} and \cite{Song2020} searched for gamma-ray stellar flares, which result from the higher energy ($E \gtrsim 30$ MeV / nucleon) ions and protons interacting with the stellar chromosphere.  Lower-energy protons ($E \ll 1$ MeV) in the beams are expected to undergo charge-exchange with the neutral hydrogen in the chromosphere before it explodes in temperature.  The nonthermal hydrogen beams then emit red-shifted Lyman $\alpha$ photons as a captured electron falls to the ground state.  This is called the Orrall-Zirker effect \citep{OZ, Kerr2023}.  \citep{Woodgate1992} reported a broad satellite component that was highly redshifted from Ly$\alpha$ and was attributed to low-energy proton beams in the early phase of a stellar flare in AU Mic.  Follow-up observations with high-time resolution spectroscopy of the Ly $\alpha$ red and blue wings have not resulted in similar detections \citep{Robinson1993, Robinson2001, Feinstein2022}.

\subsection{Centimeter (Microwave/Radio) Observations:  Nonthermal Gyrosynchrotron Radiation from Mildly Relativistic Electrons}  \label{sec:radioobs}
The radio/microwave regime at cm wavelengths (GHz frequencies) is a probe of nonthermal emission from accelerated electrons 
in flare coronae.  Lower frequencies, $\lesssim 2$ GHz, tend to be highly circularly polarized, shorter in duration, and narrowband, indicating plasma or electron-cyclotron maser emission,  while higher frequencies are largely unpolarized, broadband, and longer in duration.  The unpolarized higher frequency radiation is consistent with gyrosynchrotron radiation from a power-law distribution of mildly relativistic electrons trapped in the magnetic fields of the flare region.  The peak frequency of the gyrosynchrotron spectrum determines the demarcation at which the spectrum transitions from optically thin (at frequencies higher than the peak frequency) to optically thick (at frequencies lower than the peak frequency).  For a homogeneously emitting source, the power-law index of accelerated electrons is directly related to the power-law index of the optically thin gyrosynchrotron flux spectrum at Earth;
see \cite{Dulk1985} and Sect.~\ref{sec:multilam} for more discussion and references.

\begin{figure*}
  \includegraphics[width=1.0\textwidth]{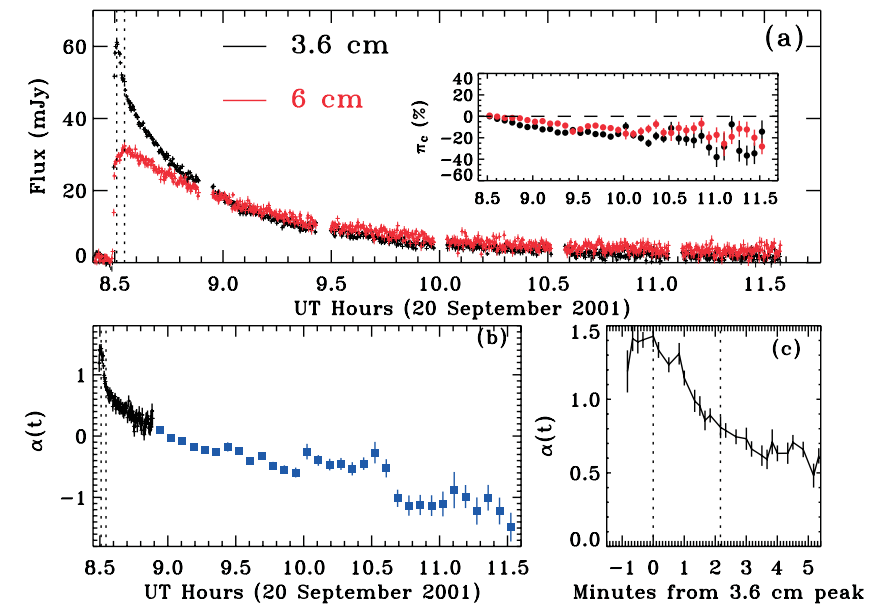} 
\caption{  A remarkable radio flare on the dM3.5e EV Lac from \cite{Osten2005} with the spectral indices calculated from observations at the two observed frequencies.  Reproduced with permission.  }
\label{fig:Osten2005}     
\end{figure*}

Radio stellar flares have typically been observed in two bands, such as 6 cm (4.9 GHz) and 3.6 cm (8.4 GHz), which are close to or at lower frequencies than typical peak frequencies ($\approx 10$ GHz) \citep[note that some solar flares have peak frequencies constrained to at least $\gtrsim 30$ GHz;][]{White2003}.  A large multi-wavelength flare from the dM3.5e star EV Lac was studied in \cite{Osten2005} and a light curve is shown Fig.~\ref{fig:Osten2005}.  The peak radio luminosity at 8.4 GHz is $1.9 \times 10^{15}$ erg s$^{-1}$ Hz$^{-1}$, but the $U$-band (not reproduced here; $L_{U,\rm{peak}} \approx 10^{30}$ erg s$^{-1}$, $E_U \approx 5 \times 10^{31}$ erg) peaks about 54~s earlier and decays much more rapidly.  The spectral index ($\alpha; S_{\nu} \propto \nu^{\alpha}$) evolution is discussed in \cite{Osten2005} and indicates rapidly varying optical depths at these frequencies over the flare; in the decay phase, the higher frequency likely becomes optically thin, while in the peak phase,  a  reasonable range of expected power-law indices for the nonthermal electrons cannot explain the values of $\alpha$ for a homogeneous optically thick source.  The lengthening decay timescales were interpreted in terms of the ``trap+precipitation'' model (Section \ref{sec:solar}). Source parameters were derived during the long decay, indicating very large loops on comparable scales to the stellar radius with small trapped nonthermal electron densities ($10^4 - 10^6$ cm$^{-3}$), and large magnetic fields (100 G) increasing over time and decreasing in source size as the higher frequency becomes more optically thin (R. Osten, priv. communication 2011).

\begin{figure*}
\begin{center}
  \includegraphics[width=0.75\textwidth]{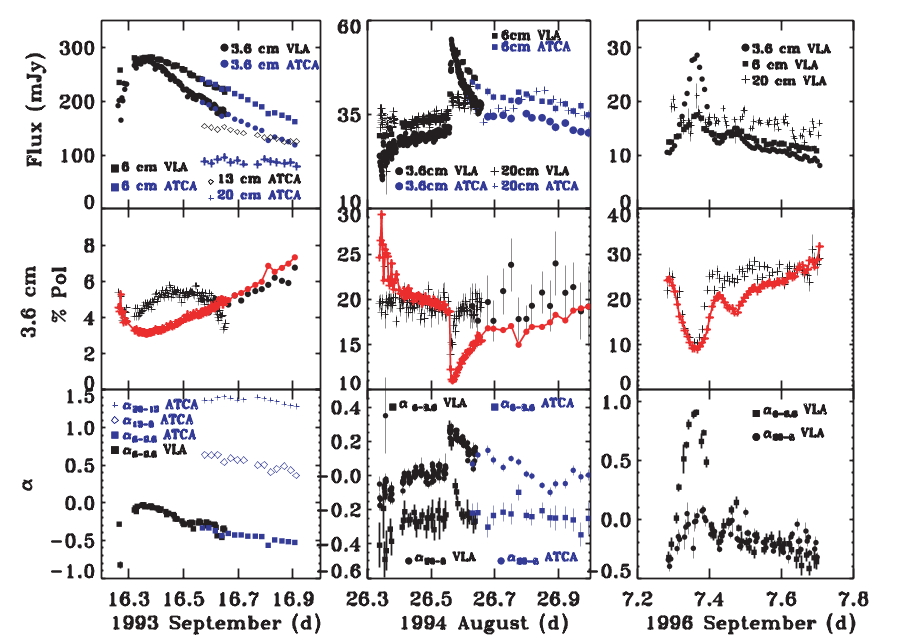} 
\caption{ Flux (top row), polarization (middle row), and spectral indices (bottom row) during three gyrosynchrotron radio flares from the RS CVn HR 1099, reproduced from \cite{Osten2004} with permission.  }
\label{fig:Osten2004_Figure5}       % Give a unique label
\end{center}
\end{figure*}

Fig.~\ref{fig:Osten2004_Figure5} shows three radio flares from the comprehensive multi-wavelength study of the RS CVn system HR 1099 \citep{Osten2004}.  The data show low polarization during the flares and a similar spectral index evolution patterns among these long-duration flares.  \cite{Gudel1996} and \cite{Smith2005} present radio analyses of stellar flares that will be discussed below (Sect.~\ref{sec:neupert}, Sect.~\ref{sec:multilam}) in multi-wavelength contexts.

The emission at lower frequencies results from a plethora of possible phenomena. Decimetric flares\footnote{\cite{Cliver2022} review extreme solar decimetric events and physical processes.} can be very luminous and highly circularly polarized indicating coherent emission from beam-plasma instabilities (``plasma emission'') or electron cyclotron maser \citep[][see references and discussion in \citealt{Bastian1990, Gudel1996, Osten2005}]{Lang1983,Bastian1990B, Osten2008, Villadsen2019}.  Coherent bursts can occur at higher frequencies too, and a remarkable example is the 100\% left-hand circularly polarized 30 mJy peak-flux, 3-minute burst at 5 GHz discussed in \cite{Smith2005} from the dM3e AD Leo.  Assuming plasma emission, they derive a coronal electron density of $\approx 3\times 10^{11}$ cm$^{-3}$; assuming electron cyclotron, they derive a magnetic field of $B \approx 1.8$ kG.  Between 1--2 GHz, stellar analogies to Type III decimetric bursts\footnote{The ``plasma emission'' in this case is produced by excitation of Langmuir waves and conversion to electromagnetic radiation as electron beams travel over the density gradient of the corona.} have been reported in very high time-resolution data from  Arecibo \citep{Osten2006}.  A surprising lack of Type II radio bursts has been constrained in the context of extensive ground-based monitoring of optical flaring \citep{Crosley2018A}.  Many recent numerical, theoretical, and observational efforts have sought to explain the lack of Type II radio bursts from other stars \citep[][see also \citealt{Cliver2022B}]{AlvaradoGomez2018, Mullan2019, AlvaradoGomez2020, Wood2021}.  
%crosley:  230-470 MHz 

\subsection{Millimeter Observations from ALMA:  Nonthermal Radiation with Linear Polarization} \label{sec:alma}
Bright nonthermal flare radiation occurs at much higher frequencies around $\nu \approx 230$ GHz ($\lambda \approx 1.3$ mm) too.  Millimeter flares have been reported in ALMA data from AU Mic and Proxima Centauri \citep{MacGregor2018, MacGregor2020}. Particularly remarkable is  the extremely impulsive flare from Proxima Centauri with multi-wavelength observations \citep{MacGregor2021}.  The dMe mm flares are short in duration ($2-30$~s), have rather symmetric (Gaussian-like) impulsive-phase time profiles, and exhibit only a very weak gradually decaying tail.  They note that these properties are rather similar to the FUV continuum response in the same flares (see Sect.~\ref{sec:FUV}). The spectral indices, $\alpha$ ($S_{\nu} \propto \nu^{\alpha}$), are negative with a significant amount of linear polarization that ``flip-flops'' between negative and positive values during the peak phase.   

The millimeter flare emission is interpreted as either an optically thin extension to the gyrosynchrotron radiation component that peaks around $\nu = 10$ GHz (see the previous section), or as optically thin synchrotron radiation that undergoes some type of depolarization to the observed levels of $\pm 20$\% \citep[see][]{MacGregor2020}. The radiation sources are relativistic or ultra-relativistic electrons, but the magnetic field strengths are largely different between these hypotheses.   Being in the optically thin regime of either interpretation, these observations have facilitated important inferences of the power-law indices of the accelerated electrons in the range of $\delta_{r} \approx 2.8-5.2$, indicating rather hard distributions of accelerated particles.  At much higher frequencies, \cite{Beskin2017} reported short (FWHM durations of $\lesssim 1$~s) synchrotron bursts in the optical $U$-band superimposed on a giant, unpolarized dMe event from UV Ceti.  Only lower limits on the linear polarization were possible, but the analyses of the energetics of the bursts favored very hard, $\delta < 3.4$, power-laws of ultra-relativistic electrons.

Flares at $\nu \approx 100$ GHz ($\lambda \approx 3$ mm) have been reported from PMS stars, RS CVn's, and eccentric binaries with colliding magnetospheres \citep{Phillips1996, Bower2003, Salter2008, Salter2010, BrownBrown2006, Adams2011}.  In the case of the remarkable periastron event in the binary T Tauri V773 A \citep{Massi2006}, the flare was interpreted as synchrotron radiation.  There is a notable observation of an RS CVn event at $\nu \le 100$ GHz showing a break and a rise toward higher frequencies  \citep{Beasley1998}:  i.e., an opposite spectral index to the dMe events reported in ALMA, and one that is more similar to some reports in solar flares \citep{Kaufmann2004}.   Colliding magnetospheres at periastron are also attributed to some flare-like optical and NUV variability in the PMS DQ Tau system \citep{Salter2010, Getman2023}, while the optical variability properties at periastron at other times follow the expectations of accretion alone \citep{Mathieu1997, Tofflemire2017}.

\subsection{Soft X-rays and EUV Observations: Loops Filled with Evaporated, Previously-Chromospheric/TR, $T>10^7$ K Thermal Plasma} \label{sec:upper}

\begin{figure*}
\begin{center}
  \includegraphics[width=1.0\textwidth]{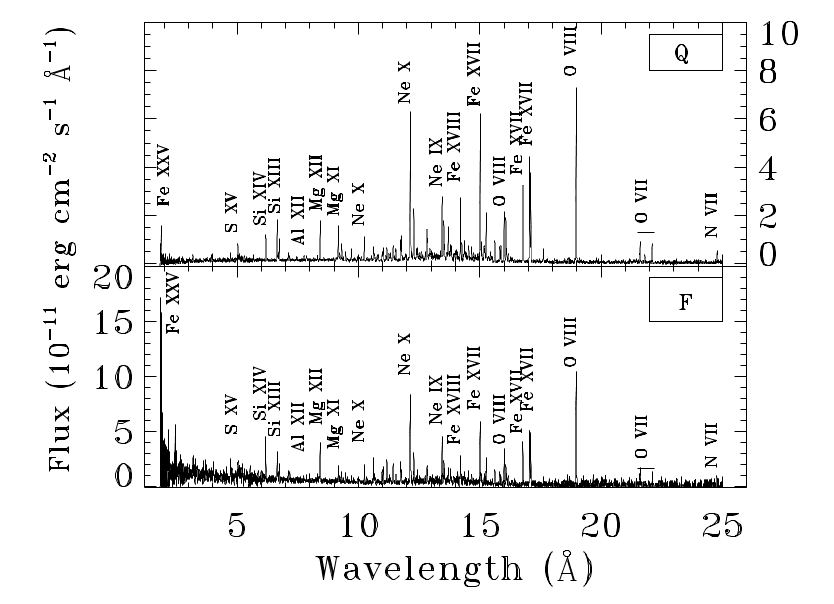} 
\caption{ X-ray spectra from Chandra during quiescence (top) and flaring (bottom) in the active binary $\sigma^2$ Coronae Borealis, reproduced from \cite{Osten2003} with permission.  The authors note the enhanced continuum flux increasing towards shorter wavelengths in the flare spectrum.  \label{fig:Osten2003_1}}  % Give a unique label
\end{center}
\end{figure*}

Stellar flares have been studied in detail in soft X-rays ($E \approx 0.1 - 10$ keV) for many decades through XMM-Newton, Chandra, Swift, and NICER spectroscopy and photometry.   X-ray flare spectra are a composite of many optically thin, highly-ionized emission lines and free-free bremsstrahlung continua that originate from the flaring coronal volume.  The bulk of the radiation is thought to originate in the chromospheric gas that has been heated to millions of K and ablated into the magnetic loops, which confine the flows (Sect.~\ref{sec:solar}).  One of the highest signal-to-noise flare spectra during a stellar flare, which occurred in an RS CVn system, is showcased in Fig.~\ref{fig:Osten2003_1}.  

An extensive review of stellar X-ray flares and analysis methods (prior to and including 2004) are provided by \cite{Gudel2004Rev} and \cite{Gudel2006}.   A general review of X-ray spectroscopy of stars is contained in \cite{Gudel2009Rev}.  Since 2004, there have been many important X-ray flare studies \citep{Osten2005, Robrade2005, Stelzer2006, Fuhrmeister2007, Nordon2007, Nordon2008, Pandey2008, Huenemoerder2010,Liefke2010, Fuhrmeister2011, Pandey2012, Lalitha2013, Pye2015, Namekata2020, Paudel2021, Karmakar2022}.  In this section, we focus on a few key results from the last several decades pertaining to abundance changes and the stellar flare Neupert effect.  Several other results are included throughout this review in other contexts where they are especially complementary (e.g., multi-wavelength analyses).

\subsection{The Neupert Effect in Stellar Flares} \label{sec:neupert}
The backbone of the solar-stellar flare connection is the Neupert effect \citep{Neupert1968}.  This was originally established in three solar flares as a correlation  between the cumulative time-integral of the nonthermal microwave emission around 2.7 GHz  and the thermal soft X-ray flux at 1.87 \AA, attributed to the Fe XXV line emission (1s$^2$--1s2p, which is sensitive to $T \approx 30$ MK).  It has since been the subject of many follow-up studies in solar flares using either the hard X-rays or the derivative of the GOES 1--8 \AA\ soft X-rays as a proxy for the nonthermal particle heating in the chromosphere \citep[e.g.,][]{Dennis1993, McTiernan1999, Veronig2002, Veronig2005, Warmuth2009}.  Alternatively, the empirical Neupert effect is expressed as a relationship between integral of the hard X-ray flux and the luminosity of the GOES soft X-rays.  The relationship between the impulsive, early nonthermal radiation and the gradual, later-peaking thermal flux is thought to be linked by the physics of impulsive nonthermal particle propagation and bombardment of the chromosphere, rapid evaporation/ablation of $T \approx 10-50$ MK chromospheric mass into the coronal loops, and the eventual cooling to pre-flare coronal temperatures (Sect.~\ref{sec:solar}).

\begin{figure*}
  \includegraphics[width=1.0\textwidth]{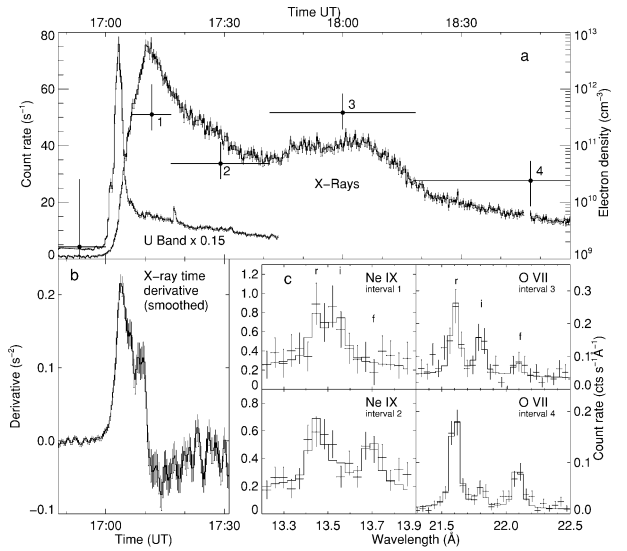} 
\caption{ A remarkable example of the Neupert effect in a stellar flare in Proxima Centauri; figure reproduced with permission, from \cite{Gudel2002}.  The $U$-band light curve is a (thermal) proxy for the nonthermal, impulsive energy deposition into the lower atmosphere.  This flare was subsequently studied and modeled in \cite{Gudel2004} and \cite{Reale2004}. }
\label{fig:Gudel2002}     
\end{figure*}

The Neupert effect is also reported in multi-wavelength observations of stellar flares.  Because stellar flare hard X-ray ($E \gg 10$ keV) emission is almost always too faint to detect, and radio observations are relatively difficult to plan and secure, 
the (thermal) response in the $U$-band is utilized as a proxy.  This approach is justified by the close temporal and spatial correspondence between hard X-ray and white-light emission on the Sun \citep[e.g.,][]{Kane1985, Neidig1989,Hudson1992,NeidigKane1993,Fletcher2007,Krucker2011,Krucker2015,Kleint2016}.   A remarkable example of the stellar Neupert effect is shown for a flare from Proxima Centauri \citep{Gudel2002} in Fig.~\ref{fig:Gudel2002}.  

 A stellar flare Neupert effect was first reported in \cite{Hawley1995} using the correlation between the extreme-ultraviolet (65--190 \AA) flux and the integrated $U$-band energy during a long-duration, 3.5-hour rise phase of a flare from the dM3e AD Leo.  This was followed shortly thereafter in \cite{Gudel1996} using a correlation between the $E=0.5 - 2$ keV soft X-rays and 3.6 cm radio emission during a flare on the dM5.5e UV Ceti A.  The Neupert effect has subsequently been reported in stellar flares from RS CVn systems \citep{Gudel2002b, Osten2004}, dMe stars \citep{Gudel2002, Gudel2004, Mitra2005, Wargelin2008, Fuhrmeister2011, Cabellero2015, Tristan2023}, dK-stars \citep{Lalitha2013}, T Tauri stars \citep{Audard2007}.   There are also notable exceptions to the expected multi-wavelength relations from the Neupert effect in stellar flares \citep{Doyle1988, Osten2005};  see these papers for discussions of possible explanations, which include deep heating from ultra-relativistic electrons or MeV protons.  On the Sun, a clear violation of the Neupert effect in a ``late impulsive [hard X-ray] burst'' \citep[adopting the terminology from][]{Xia2021} has been modeled with a large, low-energy cutoff $E \gtrsim 100$ keV in the nonthermal electron beam \citep{Warmuth2009}, which would reasonably generate small amounts of collisional heating within and chromospheric evaporation into the corona \citep[however, the explanation is still being investigated;][]{Alaoui2017}.

The Neupert effect in Flare D in \cite{Gudel1996} was investigated in detail and was compared to a similar multi-wavelength relationship in a solar flare from \cite{Dennis1993}.  Intriguingly, the microwave response of flare D returned to pre-flare levels well before the maximum of the soft X-ray light curve, which is not in strict agreement with the expected Neupert effect.
It was shown that similar deviations in the solar flare could be explained by a temperature sensitivity on the Neupert effect, such that radiation from hotter plasma more strongly followed the expectations \citep[see also][]{McTiernan1999}.  \cite{Gudel1996} thus separately derived a \emph{generalized Neupert effect for the light curve} and a \emph{generalized Neupert effect}, following energy conservation over the flaring coronal volume, $V$:

\begin{equation} \label{eq:generaln}
\frac{d}{dt} \Big( 3 n_e k_B T V \Big) = \alpha f_{\rm{radio}}(t) - n_e^2 V \psi(T)
\end{equation}

\noindent where $n_e$ is the ambient/thermal electron density of the heated plasma at a temperature $T$, $k_B$ is Boltzmann's constant, $f_{\rm{radio}}$ is the flare radio flux as a function of time $t$, $\alpha$ is a proportionality constant, and $\psi(T)$ is the optically thin radiative loss function in units of erg cm$^{3}$ s$^{-1}$.   The left hand side of Equation \ref{eq:generaln} is the change of the thermal energy of the corona due to chromospheric mass evaporation, and the right hand side is the balance between radio/ microwave flux at Earth (assumed to be proportional to the kinetic energy of nonthermal electrons causing the chromospheric evaporation) and the optically thin radiative cooling of the coronal volume \citep[assuming conductive cooling is overall small; see also][]{Fisher1990}.  Thus, accounting for the total plasma energetics and the bolometric luminosity evolution more naturally explains the apparent deviations from the Neupert effect assessed in a specific X-ray bandpass.
An additional finding from the solar-stellar flare comparison was that the M dwarf flare was ``radio-overluminous'', possibly suggesting a greater efficiency of acceleration of microwave-emitting ($E \gtrsim 100$ keV) electrons (Sect.~\ref{sec:radioobs}) and/or a lower efficiency of chromospheric evaporation, among several other possibilities \citep[see][]{Gudel1996}.

The long rise times of Ca II K and H in dMe flares are suggestive of a Neupert-like effect, which could provide insight into their origin \citep[e.g., through XEUV radiative backwarming;][]{HF92}.  \cite{Kowalski2013} investigated the relationship between the time-integral of the blue continuum flux around $\lambda = 4170$ \AA\ and the Ca II K line flux for a large sample of dMe flares.   They found a Neupert-like relationship for a wide variety of flare types, both impulsive and gradual-type events, but none of these observations were complemented with X-ray data.

\subsection{Abundance Changes}
Impulsive heating in the chromosphere evaporates gas into the coronal loops, but all chemical elements are not  evaporated equally.  This results in relative abundance changes in the hot, coronal flare plasma such that elements with low first ionization potential (FIP) of $< 10$ eV, like Fe, Si, and Mg, become preferentially enhanced above their photospheric values. The fractionation may be explained by ponderomotive forces \citep{Laming2004} that selectively act on singly-ionized species in the upper pre-flare chromosphere.

\begin{figure*}
\begin{center}
  \includegraphics[width=0.5\textwidth]{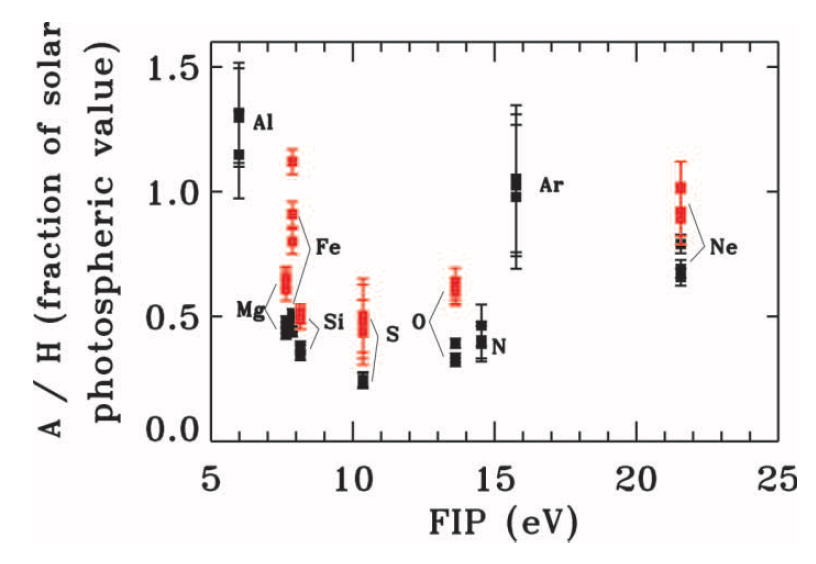} 
\caption{ Abundances relative to hydrogen \emph{vs.} the first ionization potential (FIP) in quiescent (black) and flaring (red) intervals in Chandra spectra of the RS CVn $\sigma^2$ Coronae Borealis.  Figure reproduced from \cite{Osten2003} with permission.   They note that there is a slight enhancement of Mg, Si, O, and Ne during the flare, and Fe increases by nearly twice.  \label{fig:fipeffect}       }
\end{center}
\end{figure*}

Optically thin, coronal equilibrium modeling of X-ray and EUV spectra has been used to search for the so-called FIP effect (or its inverse, the IFIP effect) in stellar flares.   Detections of changes in low-FIP abundances have been suggestive but not highly statistically significant \citep{Gudel1999}, as have relative changes in high-FIP abundances such as Ne \citep{Osten2005}.  An example of an abundance analysis of an RS CVn flare in Fig.~\ref{fig:Osten2003_1} is shown in Fig.~\ref{fig:fipeffect} where an abundance change in Fe was seen, but other expected FIP dependencies were not.   Recently, \cite{Paudel2021} searched  for the (I)FIP effect in individual flares from the dM3.5e star EV Lac;  they discuss the importance of field geometry in the context of the Laming theory.  A homogeneous Sun-as-a-star (i.e., stellar) EUV spectral analysis of a large sample of solar flare Fe lines revealed coronal abundances that were remarkably close to photospheric, thus suggesting that chromospheric heating and evaporation occur deeper in the atmosphere than the standard models of electron beam heating and elemental fractionation \citep{Warren2014}, but \cite{Laming2021} briefly contests the interpretation.  Nonetheless, these solar results provide tantalizing connections to the deep heating rates (Sect.~\ref{sec:models}) that are inferred and hydrogen line blueshifts (Sect.~\ref{sec:chromlines}) that are reported from optical observations of M dwarf flares. 
 
\section{Atmosphere Modeling of Stellar Flares:  Overview of Results} \label{sec:modelsoverview}

The physics of stellar flares is a complicated (and therefore interesting!) intersection of particle and nuclear physics, radiative transfer, plasma processes, gas dynamics and shock phenomena.  All of the physical processes are inter-dependent and occur in a highly magnetized, coupled atmosphere, which is probably also highly turbulent.  Flares are transient non-equilibrium departures of the stellar luminosity, and all layers of the atmosphere are thought to produce  a response to some degree. In this section, we review different stellar flare modeling techniques. We focus on efforts toward understanding chromospheric flare processes in the context of the response of the entire stellar atmosphere.  In contrast to purely magneto-hydrodynamic models, radiative models include detailed, time-dependent spectral predictions, which are critical for connecting to observations.

Stellar flare modeling can be sub-divided into several types of techniques.  Isothermal, static slab modeling and atmospheric modeling are the two most general categories.    
Atmospheric modeling can be further divided into a static semi-empirical approach \citep[e.g.,][wherein the atmospheric structure is adjusted by hand to match observations]{Cram1982}  and radiative-hydrodynamic (RHD) forward modeling, whereby the input heating function is data-constrained and is not fine-tuned by the modeler to match observations \citep[e.g.,][]{Allred2006}.  A hybrid, semi-empirical RHD approach is also possible, whereby the injected flare heating function is varied by hand over a large parameter space to find a time-dependent spectral response that matches the observations; thus, the atmospheric response and depth-dependent heating rates are calculated in a fully self-consistent manner  \citep[e.g.,][]{Kowalski2015}.  Though atmospheric modeling accounts for the vertical heterogeneity at a location in a flare source, additional techniques are required to model the lateral heterogeneity of the flare source and the source geometries (to be discussed further in Sect.~\ref{sec:geom}).

\subsection{Slab Models}

Uniform, static slab modeling techniques include the following:  fitting a Planck function to spectra or multi-band photometry \citep{Hawley1995, Hawley2003, Zhilyaev2007, Kowalski2013}, calculating LTE continuum spectra for a range of optical depths $\tau(\lambda) = [0,\infty)$ \citep{Kunkel1970, Eason1992}, and modeling the NLTE Balmer line decrements, optical depths, and continuum flux ratios \citep{Drake1980, Jev1998, Garcia2002, Morchenko2015}.  The pioneering study of \citet{Kunkel1970} found evidence for a prominent Balmer jump in their spectral observations of M dwarf flares.   Their models of optically thin hydrogen (emissivity) continuum calculations, however, were largely inconsistent with the colors of M dwarf flares.  They concluded that an additional moderately heated photospheric flare component could better account for observations (it was not until the radiative-hydrodynamic models of \citet{Allred2006} that essentially produced this scenario through a chromospheric hydrogen recombination spectrum and a radiatively backheated photospheric spectrum;  see below).  In Appendix \ref{sec:slabs}, we update the LTE hydrogen continuum emissivity calculations from \citet{Kunkel1970} with non-ideal occupational probabilities around the Balmer limit and extend them to higher temperatures and shorter wavelengths.  These are useful for verifying for oneself how poorly first-principle simplifications hold up against continuum observations of actual stellar flares\footnote{Note, the Balmer jump ratios and optical continuum flux ratios from these slab models are plotted in Figure \ref{fig:colorcolor}.}.  

The higher quality spectral observations of the Great Flare of AD Leo \citep{HP91} were found to (apparently) exhibit little to no Balmer jump in the flux, but there was unprecedented power in the optical continuum radiation and highly broadened hydrogen line wings.  The broadband colors from the FUV through the optical ($R$ band) were tested against all of the models available at the time \citep{HF92}, and a $T \approx 8500 - 9500$ K blackbody was clearly favored over a hydrogen recombination model and a hot $T = 10^7$ K free-free slab model interpretation;  the former model has a slope that is too red, and the latter interpretation is inconsistent with X-ray to optical energy fractionations \citep[see Section 4.3.2 of \citealt{HF92},][and Sect.~\ref{sec:energies} here]{Osten2015} and representative volume emission measures (Sect.~\ref{sec:multilam}).  The various interpretations (blackbody \emph{vs.}~ optically thin Balmer continuum) and reported characteristics of the properties at the Balmer limit eventually motivated additional spectral observations \citep{Kowalski2013}.  More immediately, they paved the way toward advances in the realism of models of stellar flares.  It became essential to consider the comprehensive atmospheric response that includes the complexities of depth-dependent heterogeneities of opacities, temperatures, and densities.  Then, self-consistent non-equilibrium calculations of velocity fields, mass advection, radiative cooling, and flare heating became possible.  We summarize these efforts next.
% Also 1e56 emission measures are required and wrong slopes in FUV ...

  \subsection{Atmospheric Models} \label{sec:models}
\begin{figure*}
\begin{center}
  \includegraphics[width=0.65\textwidth]{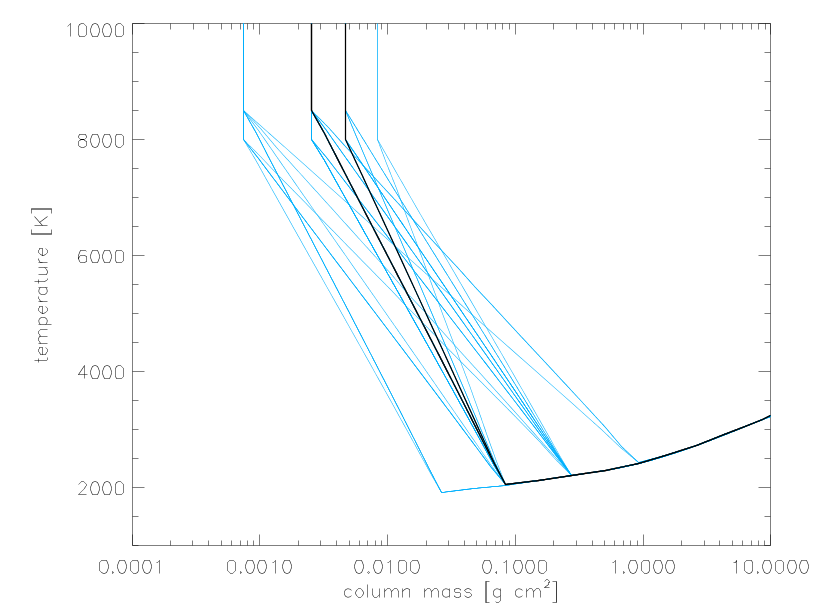} 
\caption{ Demonstration of semi-empirical temperature adjustments used to model the red- and blue-wavelength optical spectra of a giant flare from CN Leo.  Figure from \cite{Fuhrmeister2010} reproduced with permission from the author.  See also \cite{Fuhrmeister2011} and \cite{Fuhrmeister2005} for similar approaches for modeling flares from Proxima Centauri and the dM6 star AZ Cnc, respectively.  }
\label{fig:semi_empirical}
\end{center}
\end{figure*}

A traditional approach to modeling stellar flare spectra is to start with a hydrostatic equilibrium model atmosphere and semi-empirically adjust the temperature gradients of the chromosphere, photosphere, and the location of a transition region.  \cite{Cram1982} explored the results of this technique using six atmospheric variations, including two very extreme adjustments where the chromosphere and transition region are placed at large column mass depths.  They successfully reproduced a $T \approx 14,000$ K blackbody-like continuum spectral property in one of these models (their model \#5), which also had very broad hydrogen lines and deep central reversals.  Similar approaches have been widely adopted to model EUV, optical, and infrared stellar flare spectra \citep{Houdebine1992, Mauas1996, Jev1998B, Christian2003, Fuhrmeister2010, Schmidt2012}.   An example of this modeling technique from \cite{Fuhrmeister2010} is shown in Fig.~\ref{fig:semi_empirical}, where the chromospheric temperature adjustments result in satisfactory fits to the emission lines in the giant CN Leo flare (Fig.~\ref{fig:Fuhrmeister2008}(a)) from \cite{Fuhrmeister2008}.   Further, a depth-dependent filling factor and missing photospheric heating were inferred through this technique.

How to create these deep transition region locations and chromospheres self-consistently with a flare heating function, $Q_{\rm{flare}}$?   One early hypothesis \citep{MullanTarter1977} was that intense irradiation of soft, thermal X-rays from the flare corona penetrated to the low atmosphere, causing the optical emission lines and continuum in the gradual decay phase (whereas, the impulsive phase optical emission was assumed to be caused by particle beams, as was known to be consistent with hard X-ray and white-light timing in solar flares).   \cite{HF92} made a major advancement in stellar flare X-ray backheating models by calculating the full atmosphere, including the coronal structure.  The apex temperatures were set and all else were determined self-consistently through energy balance (X-ray heating and photoionization \emph{v.} radiative cooling) and hydrostatic equilibrium.  Although the models produced hydrogen lines that were too narrow, and the optical and NUV continuum radiation was too weak compared to the observations \citep{HP91}, a large amount of broad, Ca II K line flux was predicted (perhaps offering clues to understanding its very slow temporal response).  Updates to XEUV irradiation modeling (e.g., the atomic physics and geometrical assumptions therein)  were employed in evolving atmospheres \citep{HF94, Abbett1998, Allred2005, Allred2006, Allred2015} that  facilitated comparisons to the role of nonthermal electron heating.  The role of reprocessing much larger X-ray emissivities into lower atmosphere optical continuum radiation in superflares has been recently reconsidered in static atmosphere calculations  \citep{Nizamov2019}.  

Significant advances were made in modeling the temporal evolution of the atmospheric response in gas-dynamic simulations, starting in the 1980s \citep{Livshits1981, McClymont1983, Fisher1985V, Fisher1985VI, Fisher1985VII, Emslie1992}. An  accurate treatment of energy deposition in a partially ionized, nonuniform chromosphere due to an injected nonthermal electron distribution  was developed \citep{HF94} based on analytic formulae for Coulomb collisions \citep{Emslie1978, Emslie1981}.   With the inclusion of radiative-transfer, this effort resulted in a series of papers on radiative\footnote{The gas-dynamic models, of course, include radiative cooling in the energy equation, but it is  treated using optically thin, coronal equilibrium approximations.}-hydrodynamic modeling of solar flares \citep{Abbett1999, Allred2005} using a version of the RADYN code \citep{Carlsson1992B, Carlsson1995, Carlsson1997} that was configured for a larger surface gravity, a cooler photosphere, and a hotter and denser corona of an M dwarf in \citet{Allred2006}.   The RHD simulations are currently limited to one spatial dimension, which facilitates resolving short-timescale shock phenomena, non-equilibrium ionization rates of hydrogen and helium, and detailed radiative transport.

The Allred et al.\ stellar flare models studied the atmospheric response to electron beam heating inputs from the first widely studied solar flare (SOL2002-07-23T00:30 GOES class X4.8) of the RHESSI \citep{Lin2002} hard X-ray spectroscopic imaging era \citep{Dennis2022}.  A double-power law distribution with a low-energy cutoff of $E_c=37$ keV was obtained from the hard X-ray modeling of \cite{Holman2003}, and several injected fluxes were investigated.   These models reproduced the broad Balmer lines for the first time and bright optical and NUV continuum radiation.  However, the Balmer jump was even more conspicuous than in the energy equilibrium models with X-ray flare irradiation \citep{HF92} and semi-empirical models \citep{Fuhrmeister2010}.  The dominant mode of radiative backheating in the RHD models takes place through irradiation of the photosphere by the Paschen and Balmer continuum fluxes of optically thin thermal radiation in response to the nonthermal electron beam energy deposition in the mid-to-upper chromosphere (the beam electrons do not penetrate to the photosphere; see Sect.~\ref{sec:rhdphysics}).
The detailed RHD spectrum provided an alternative explanation to the $T\approx 9000$ K blackbody interpretation of broadband photometry reported in earlier observations \citep[e.g.,][]{Hawley2003}, and the general shape of the Balmer continuum in the $U$-band was later found to be a satisfactory model in the decay phase spectra of a megaflare from YZ CMi \citep{Kowalski2010}.  The lower flux model was run for several minutes in \cite{Allred2006}, and it well-reproduced several transition region flare  lines from \cite{Hawley2003}.  Thus, the Allred et al.\ models provided a comprehensive solar-flare modeling framework for interpreting many of the observed properties of M dwarf flares.

Yet, re-analyses of archival spectra and new flare spectra and narrowband ULTRACAM photometry still did not match the shape and relative strength of the optical continuum radiation predicted by the highest-beam flux model in the \cite{Allred2006} simulations.  Some important details of the line broadening and blending at the Balmer limit remained unexplained as well (Sect.~\ref{sec:edge}). 
The M dwarf gas-dynamic models of \citet{Livshits1981} produced a chromospheric condensation using a large nonthermal electron flux in an initially isothermal atmosphere and suggested that a large optical depth in the continuum could be produced well above the photosphere \citep[see also][]{Katsova1997}.  \cite{Gan1992} investigated chromospheric condensations as a source of white-light in solar flares.   Impulsive-phase chromospheric condensations have been thought to be important in generating emission line spectral features in solar flares, such as the red-wing asymmetries in H$\alpha$ \citep{Ichimoto1984, Kowalski2022, Namekata2022} and in more optically thin lines such as Fe II in IRIS NUV spectra \citep{Kowalski2017Broadening, Graham2020}.  However, stellar radiative-hydrodynamic models did not produce chromospheric condensations that were dense enough to explain the observed continuum radiation until recently when the effects of very large electron beam fluxes were explored in RADYN \citep{Kowalski2015} and the RH code \citep{Uitenbroek2001}.  (A realistic M dwarf atmosphere has a dense chromosphere and very efficient radiative cooling that can make the onset of complete ionization and explosive hydrodynamics a rather energetic threshold to attain, especially for a beam with a moderately large low-energy cutoff of $E_c = 37$ keV, as traditionally employed in the RADYN dMe flare models.)   The large electron beam fluxes were actually motivated by scaling the radiative backwarming fluxes from the \citet{Allred2006} models, but when the large beam fluxes were simulated, the lower chromosphere become too optically thick for the NUV and optical continuum radiation to penetrate to the pre-flare photosphere.  Instead, the $\tau(\lambda)$ surfaces shift to a dense chromospheric condensation and the beam-heated layers just below the condensation (Fig.~\ref{fig:standardcartoon}(right)).

\begin{figure*}
\begin{center}
  \includegraphics[width=0.65\textwidth]{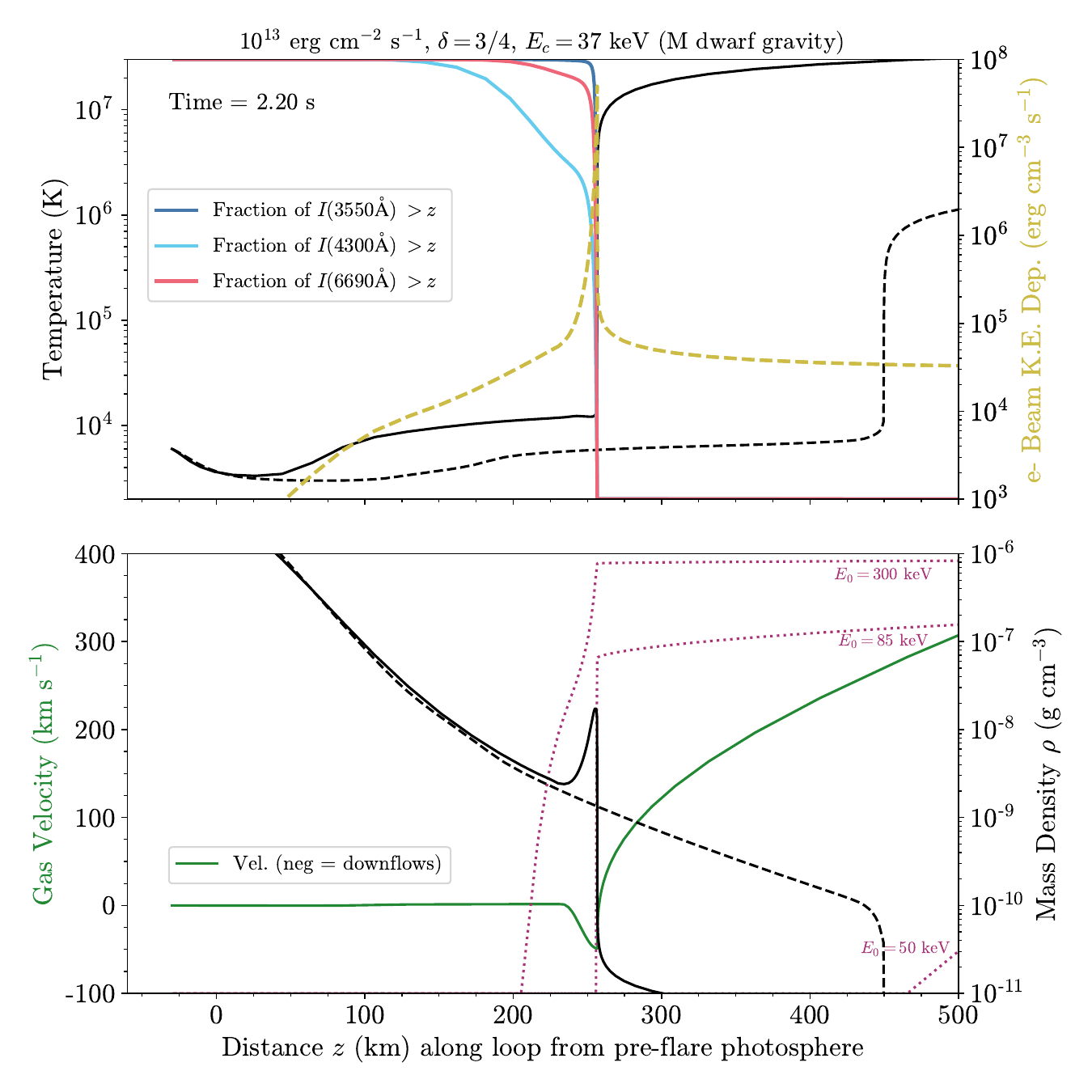} 
\caption{  RADYN model of the low flare atmosphere after 2.2~s of heating by a high-flux, F13 ($10^{13}$ erg cm$^{-2}$ s$^{-1}$), double power-law electron beam \citep{Kowalski2015}.  (top panel) The cumulative of the contribution function to the emergent intensity, $I(\lambda=3550\rm{\AA})$, shows that the chromospheric condensation becomes optically thick in the Balmer continuum.  The emergent blue-optical ($\lambda = 4300$ \AA) continuum intensity has larger contribution from the layers below the condensation, which is the large increase in gas density in the bottom panel.  (Bottom panel)  Several trajectories of nonthermal electrons are shown on top of the gas dynamic quantities (the initial kinetic energies are indicated, and the kinetic energy variations, $E(z)$, are normalized from 0 to 1 on a linear scale).  The initial temperatures and gas densities are shown as black dashed lines.  A movie is available online. }
\label{fig:CC_F13}     
\end{center}
\end{figure*}

A snapshot of the lower flare atmosphere in a RADYN model from \cite{Kowalski2015} is shown in Fig.~\ref{fig:CC_F13}.  The electron beam flux that is injected at the model loop apex is very large, $10^{13}$ erg cm$^{-2}$ s$^{-1}$ (which is hereafter referred to as a beam flux of ``F13'', whereas ``F12'' refers to an injected beam flux of $10^{12}$ erg cm$^{-2}$ s$^{-1}$, and so on).   By the time shown ($t=2.2$~s), the chromospheric condensation has cooled and accrued mass into a narrow $\Delta z \approx 15$ km region with a maximum electron density of $n_e \approx 5 \times 10^{15}$ cm$^{-3}$.  This was an exciting development because the model produced a small Balmer jump ratio and an optical color temperature that was in line with the impulsive phase spectral observations of dMe flares.  In these models, the emergent optical radiative flux spectrum is a result of thermal response to the nonthermal electron beam energy deposition. The hydrogen recombination rates are determined by the Maxwellian velocity distribution of recombining ambient electrons, which are rapidly equilibrated as the beam kinetic energy is transferred to them (Sect.~\ref{sec:rhdphysics}).  Non-thermal ionization rates of hydrogen by the beam electrons (and subsequent nonthermal recombination) \citep{Hudson1972, RC83, AH86, Fang1993}  are much less than the thermal rates in these atmospheres at all times except for the first small fraction of a second. Optical and NUV spectral calculations with dominant non-thermal ionization rates \citep{Zharkova1993} do not seem to predict the observed blue continuum properties and Balmer jump strengths (e.g., Fig.~\ref{fig:yzcmi_peak}).

The formation of the optical and NUV continuum radiation in a F13 electron beam model is illustrated in Fig.~\ref{fig:CC_F13}.  The cumulative contribution functions are shown for $\lambda = 3550$ \AA, $4300$ \AA, and $6690$ \AA.  The detailed analyses of emissivity sources and optical depths explain the emergent spectrum as Balmer and Paschen recombination radiation from optically thick layers \citep[see also][for further analyses of these models]{Kowalski2016, Kowalski2016B}.  The $\lambda = 3550$ \AA\ and $6690$ \AA\ radiative fluxes are very optically thick and escape from the top of the condensation;  the blue ($\sim 4300$ \AA) continuum flux is less optically thick and emerges from the condensation and deeper layers that have a lower electron density of $n_e \approx 10^{15}$ cm$^{-3}$ and a lower gas temperature.  The excitation by the highest energy electrons in the beam, and to a lesser extent, Balmer continuum backwarming, heat these deep layers and lead to an emergent spectrum that departs significantly from an optically thin hydrogen recombination spectrum from a homogeneous slab.  \cite{KA18}  reported self-similar patterns in chromospheric condensation models (using a basic HI\footnote{Human Intelligence} pattern recognition algorithm) and parameterized the RHD atmospheres at especially interesting times in their evolution.  The two main parameters are a reference column mass ($m_{\rm{ref}}$) and a reference temperature ($T_{\rm{ref}}$), which are indicated in the cartoon in Fig.~\ref{fig:standardcartoon}(right).  These parameterized models facilitate LTE estimates of the Balmer jump and optical continuum ratios over a large range of condensation column masses and chromospheric temperature gradients.  A prediction from \cite{KA18} is shown in Fig.~\ref{fig:colorcolor} for  $T_{\rm{ref}}=11,000$ K and a range of $m_{\rm{ref}}$.  The chromospheric condensation models can account for the trend over most of the  impulsive-phase, flare-only Balmer jump ratios and optical continuum flux ratios in spectral observations.

There has been follow-up work by Kowalski et al.\ on the modeling of chromospheric condensations to further evaluate their role as the origin of cooler line and continuum radiation in flares. Here, we summarize some of the challenges and successes with the hypothesis.  First, these model atmospheres evolve very quickly, and the emergent flux spectra around the Balmer jump rapidly change on short timescales.  Thus, an exposure-time averaged model spectrum does not necessarily exhibit properties that are consistent with the observational constraints.  An average condensation model that produces a continuum spectrum that resembles observations requires a very hard, $\delta \approx 3$, electron beam distribution \citep[e.g.,][]{Kowalski2016, Kowalski2017Broadening}.  In other cases, an average model spectrum has satisfactorily produced the broadband continuum shape and Balmer jump strength in two moderate-sized dMe flares with HST/COS observations \citep{Kowalski2019HST}, and success has been achieved in an even broader-wavelength comparisons to a superflare event \citep{Osten2016}.  There are also important theoretical issues with the huge current density and charge displacement due to an electron beam with a flux of $10^{13}$ \fnum\ that have not yet been addressed. 

\begin{figure*}
\begin{center}
  \includegraphics[width=0.65\textwidth]{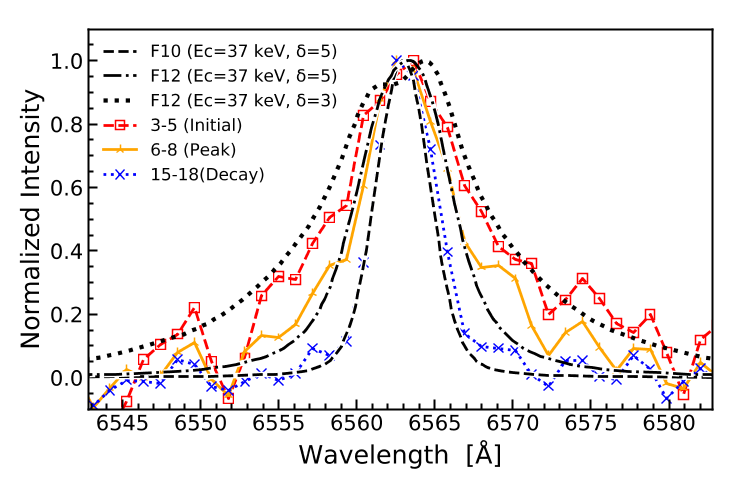} 
  \includegraphics[width=0.95\textwidth]{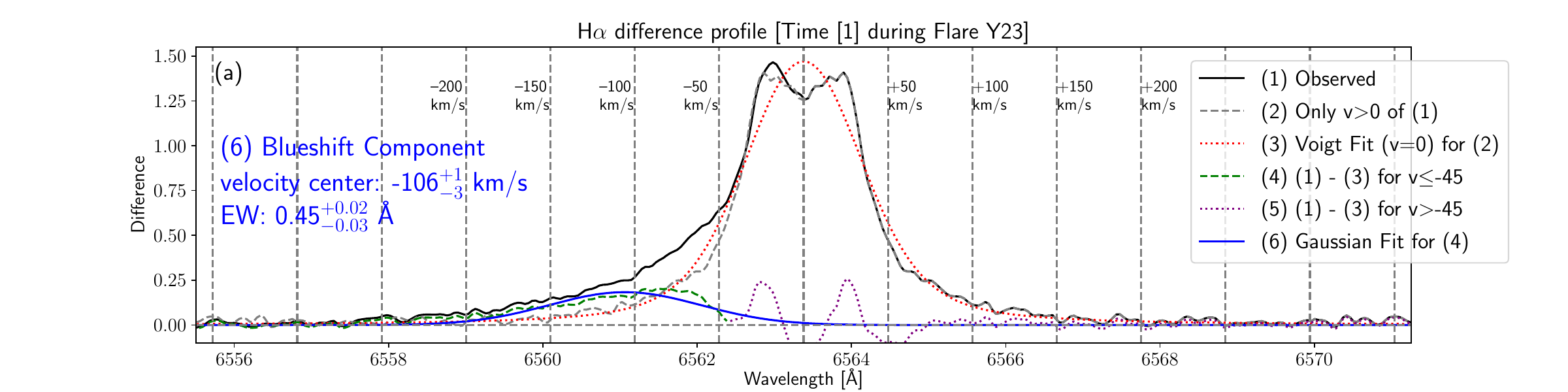} 
\caption{ (Top) Model spectra of H$\alpha$ that were calculated with the RADYN code and analyzed in \cite{Namekata2020}.  The larger electron beam heating rates reproduce the broad, symmetric Balmer lines in an AD Leo flare. Figure reproduced with permission from \cite{Namekata2020}. (Bottom) Echelle resolving power, flare-only spectra of the H$\alpha$ line during the decay of a large flare on YZ CMi.  This figure demonstrates the phenomenological Voigt profile $+$ Gaussian modeling that is used to isolate and characterize the asymmetries in the wings of the Balmer lines. In this spectrum, a Gaussian model is fit to the broad blue asymmetry;  reproduced from Fig. 42(a) in \citet{Notsu2023}, with permission. }
\label{fig:namekata}     
\end{center}
\end{figure*}

A further severe challenge has been encountered in dense chromospheric condensation modeling of dMe flares, as detailed in \cite{Kowalski2017Broadening} and \cite{Kowalski2022Frontiers}.  Updated hydrogen emission line broadening treatments (Sect.~\ref{sec:chromlines}) reveal that the optical depths and the ambient (thermal) charge densities in the model condensations are far too large to be consistent with the symmetric broadening in archival dMe flare observations. \cite{Namekata2020} modeled the broad H$\alpha$ emission lines in a $E > 10^{33}$ erg, $\Delta g \approx -1.35$ mag superflare from AD Leo. They used updated hydrogen broadening profiles in RADYN simulations and concluded that lower electron beam fluxes with hard power-law distributions produce optical depths and electron densities in the low atmosphere that explain the broadening.  The models with energy transport through only thermal conduction were largely discrepant (Fig.~\ref{fig:namekata}).

Similarly to \cite{Namekata2020}, the approaches in \cite{Kowalski2017Broadening}, \cite{Kowalski2022Frontiers}, and \citet{Brasseur2023} employ semi-empirical forward modeling with the RADYN and RH codes.  These studies use large low-energy cutoffs ($E_c \ge 85$ keV), hard power-law indices ($\delta = 3$), and high energy fluxes ($10^{12} - 10^{13}$ erg cm$^{-2}$ s$^{-1}$) to heat the low chromosphere at $\log_{10} m_c/[\rm{g\ cm}^{-2}] \gtrsim -2$ to gas temperatures hotter than $10,000$ K \citep[see also][]{Kowalski2023}.  This approach results in optically thick continuum radiation at NUV and optical wavelengths and broad Balmer wings in the early impulsive and peak phase.  However, an additional model component that is attributed to lateral spatial heterogeneity in the flare heating is required to simultaneously account for the narrow and broad components of the hydrogen Balmer emission line fluxes.  The slow-rising Ca II K emission line flux is not yet explained with this approach.  Large, low-energy electron beam cutoffs or some other energy transport mechanism \citep[e.g.,][]{Kontar2012} that similarly results in a large fraction of the beam heating over $\log_{10} m_c = -2.2$ to $-1.2$ \citep{Kowalski2023} rather than in the upper chromosphere  ($\log_{10} m_c \lesssim -3$) seems to be a promising avenue for explaining some of the more challenging M dwarf flare observations.  For example, a model with $2\times 10^{12}$ \fnum\ and  $E_c =500$ keV (i.e., a high-flux, fully relativistic electron beam) adequately reproduces the spectra of the secondary flare events in the decay phase of the YZ CMi megaflare \citep{Kowalski2010, Kowalski2013}.  These events produced an ``A star on an M star'', such that the flare-only spectra resembled the main sequence A0 V star, Vega.  The Balmer lines and Balmer continuum are ``in absorption''.  The colors from two secondary events are shown in Fig.~\ref{fig:colorcolor} as star symbols.  Alternative explanations are summarized in \cite{Anfin2013}.  \cite{Kowalski2013} claimed that some flares show a narrowing of the hydrogen Balmer lines in their impulsive-phase peaks that is consistent with a spatially unresolved component on the star that has the hydrogen Balmer lines in absorption.

Relatively little attention has been given to the coronal and late phase predictions in the recent RADYN models of stellar flares.  The RADYN models are generally viewed as  simulations of the short, pulsed beam-heating events that sequentially light up along a two ribbon flare arcade, as in typical solar flare geometries (see Sect.~\ref{sec:geom}).   Ostensibly, the very hot $T \gtrsim 50$ MK coronal temperatures produced within a short time are expected to eventually cool down and shine in emission lines that probe cooler temperatures \citep{HF92, AschwandenAlexander2001}.  However, it is imprudent to speculate further on these issues without X-ray spectral synthesis because of the vast range of electron densities over the hot temperatures in the model corona and transition region.

\begin{figure*}[ht!]
\begin{center}
  \includegraphics[width=0.4\textwidth]{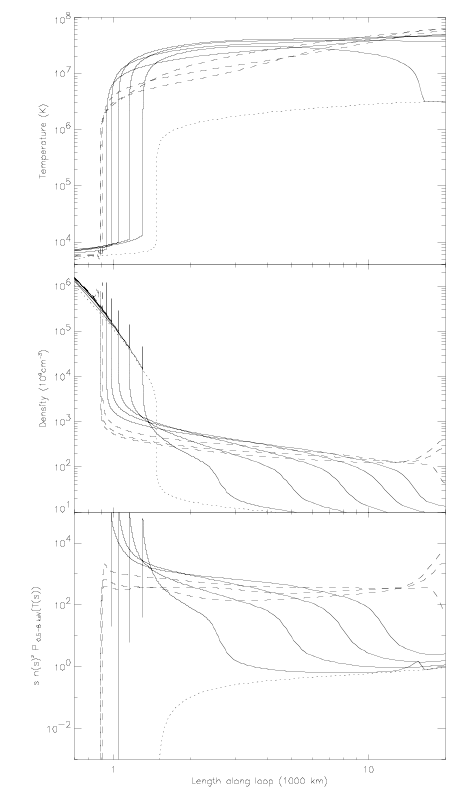} 
    \includegraphics[width=0.4\textwidth]{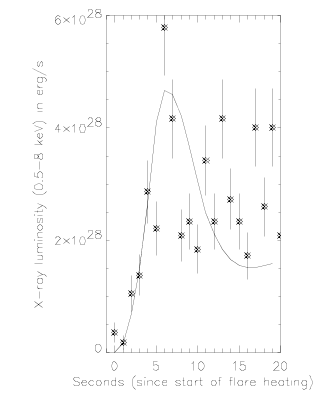} 
\caption{ (\emph{Left}) Hydrodynamic modeling of a coronal explosion on CN Leo.  The panel shows temperature, number density, and the emission measure weighted cooling function. The solid lines show the evolution at $t=0, 2, 4, 8, 10$~s when a flare heating function was applied, while the dashed lines correspond to $t=13, 16$, and $19$~s.   (\emph{Right}) Modeled X-ray light curve over the first 20~s of the model compared to XMM-Newton data of the early phase of the giant X-ray flare.  Figures reproduced from \cite{Schmitt2008} with permission.  }
\label{fig:schmitt2008}     
\end{center}
\end{figure*}

\cite{Schmitt2008}  used the Palermo-Harvard adaptive regridding code \citep{PM1, PM2} to model the hydrodynamic response of the giant X-ray CN Leo flare in Fig.~\ref{fig:Fuhrmeister2008}(a) over the first 20~s.  They assumed a Gaussian heating function centered in the low corona and extended into the chromosphere.  The atmospheric evolution is shown in Fig.~\ref{fig:schmitt2008}.  The flare transition region moves downward, and there is an increase in density  that resembles a chromospheric condensation, but low-temperature spectral synthesis was not the goal of these modeling efforts \citep[see][and Fig.~\ref{fig:semi_empirical} here]{Fuhrmeister2010}.  They successfully reproduced the X-ray luminosity evolution by multiplying the model by an area of $10^{19}$ cm$^{2}$.  The model X-ray luminosity is shown in the right panel of Fig.~\ref{fig:schmitt2008}.  They explain  the early phase X-ray light curve as a result of heating within the bottom parts of a flaring loop during the first 10~s, followed by a $v \sim 1000$ km s$^{-1}$ evaporation upflow that dominates the coronal emission after that.  Compressional heating at the apex occurs later on as well (see their dashed temperature profiles) and persists after the applied flare heating function  is turned off at 10~s.
The Palermo-Harvard code has modeled longer heating durations to simulate X-ray flares from the M dwarf Proxima Centauri \citep{Reale2004} and the active giant HR 9024 as well \citep{Testa2007, Argiroffi2019}. \cite{Reale2004} model the second X-ray flare in the decay of the event in Fig.~\ref{fig:Gudel2002} here as the first in a series of loops.  They develop a two-phase heating model whereby the total energy input consists of energy deposition near the loop apex and near the low-altitude coronal footpoints.

\section{Atmosphere Modeling of Stellar Flares: Overview of Physics and Methods} \label{sec:rhdphysics}

\subsection{An Introduction to Radiative-Hydrodynamic Flare Modeling} \label{sec:rhdbasics}

Radiative-hydrodynamic flare modeling consists of solving the coupled, non-linear
conservation equations for mass, momentum, energy, charge, radiative transfer, level populations, and grid motion.  The atmosphere begins in a steady-state, and flare heating is introduced through a term in the energy equation, $Q_{\rm{flare}}$.   In this section, I follow the general framework and modeling assumptions employed in the RADYN flare code, which has unique capabilities for stellar flare applications (e.g., radiative transfer and non-solar gravity).

A conservation/continuity equation for a volumetric quantity $w$ ($w$hatever cm$^{-3}$), which is a function of space and time, can be written (in laboratory frame coordinates) as

\begin{equation} \label{eq:continuity}
\frac{\partial w}{\partial t} + \nabla \cdot (w \mathbf{v}) = 0
\end{equation}

%%\vspace{1mm}
\noindent in three spatial dimensions (if the time-derivative is 0, then the equation leads to the steady-state solution), where $\mathbf{v}$ is the macroscopic gas velocity vector, $w\mathbf{v}$ is the flux\footnote{More precisely, one means the flux per unit area, or flux density (current density).} of the quantity in $w$, and we have excluded a diffusion term, $-D \nabla w$ \citep{Weinberg2021}, in the parentheses.
  The second term is the advection term which is a gradient of the flux (while in more than one spatial dimension, it is the divergence of the flux).  From here on, we discuss the 1D form of Eq.\ \ref{eq:continuity} and consider the spatial coordinate $z$ (``height'') along a hypothetical magnetic loop extending from below the photosphere to the loop apex in the corona.  At the loop apex, the $z$ direction is parallel to the stellar surface, and the surface gravity in the momentum equation decreases accordingly.  In 1D flare modeling, one  assumes that flows and energy transfer are aligned with the magnetic field, meaning that the magnetic pressure is assumed to be much larger than the gas pressure everywhere at all times (which may be appropriate for the center of a strong magnetic flux tube).  Note also that the typical 1D form of the continuity equation in Eq.\ \ref{eq:continuity} usually assumes a constant flux tube cross-sectional area; for a non-variable magnetic field along the flow direction, the adjustments that must be made to account for funneling effects are written in \cite{Emslie1992} and \cite{Reep2022}.

The conservation equations are solved simultaneously with an adaptive grid equation \citep{Dorfi1987} using a semi-implicit numerical scheme that allows for timesteps that are much longer than the Courant step, which is very small in the transition region during flares. Implicit radiation-hydrodynamics essentially consists of linearization and iteration (i.e., multi-dimensional Newton-Raphson) of the conservation equations with a five-point stencil monotonic upstream scheme \citep{vanLeer1977, vanLeer1979} to ensure numerical stability of advected quantities around steep gradients, such as shocks.  The spatial derivatives are written as finite differences and are centered in time between the current and next time (semi-implicitly) during the linearization of the Taylor-expansion about the iterative guesses.  We do not discuss the details here and instead direct the interested reader to \cite{Kneer1976}, \cite{McClymont1983}, \cite{Carlsson1998}, \cite{Abbett1998} for general overviews\footnote{The advantages of Newton-Raphson methods of solution are also leveraged in stellar evolution calculations, such as with the MESA code \citep{Paxton2011}.}.  We especially recommend the lucid and comprehensive lecture on the topic in \cite{Dorfi1998}.

The Eulerian (fixed space) forms of the conservation equations that are solved in RADYN flare modeling are written in  \cite{Allred2015}.  Here, I highlight the conservation of internal energy density, $e$, which is the first law of thermodynamics (i.e., $dU + PdV = Q$) connecting several of the critical ingredients in flare RHD:

\begin{equation}  \label{eq:firstlaw}
\frac{\partial e}{\partial t} + \frac{\partial (e v_z)}{\partial z} + (P + q) \frac{\partial v_z}{\partial z} =  -  \frac{\partial \left( F_c  + F_r \right) }{\partial z} + Q_{\rm{flare}} + \sum_n Q_n
\end{equation}

\noindent where $P$ is the thermal gas pressure from all species $s$ ($P = k_B T \sum_s n_s $) and enters into the $P dV$ work term, $v_z$ is the gas velocity along the loop axis ($z$ direction), $q$ is a viscous stress term added to aid in numerical stability, $F_c$ is the thermal conduction flux (temperature diffusion), $F_r$ is the radiative flux, and $Q_n$ can be any number of external heating terms, which can be flare-related (e.g., beam-generated electric fields) or non-radiative sources to keep the photosphere and corona hot in the pre-flare equilibrium state.  The form of thermal conduction is taken from \cite{Smith1980} and \cite{Fisher1985V}.   The internal energy density is the sum of thermal and excitation energy densities:

% flux limiting of conductive flux Spitzer 1962, Fisher1985 289, 414:
%  smith & Auer 1980: F_c -kappa_0 (1e-6 cgs) T^(5/2) dT/dz

\begin{equation} \label{eq:ienergy}
e = \frac{3}{2} k_B T  \sum_s n_s +  \sum_{s,i} n_{s,i} \chi_{s,i}
\end{equation}

\noindent where $n_{s,i}$ is the population level density of species $s$ and level $i$ and $\chi$ represents atomic excitation energies ($\chi=0$ for the ground state).  Equation \ref{eq:firstlaw} is derived by starting from the continuity equation (setting Eq. \ref{eq:continuity} equal to all $Q$ terms, instead of 0) for the energy density, $\varepsilon$, which is the usual \citep[e.g.,][]{Thorne2017}:

\begin{equation}
\varepsilon = \rho \left( \frac{1}{2} v_z^2 + e^{\prime} + \Phi \right)  
\end{equation} 

\noindent where $e^{\prime}$ is the internal energy per unit mass ($e = \rho e^{\prime}$), $\Phi$ is the gravitational potential energy, $g z$.  Magnetic energy density is neglected.  The energy flux is 

\begin{equation}
\mathfrak{F} = \rho \left( \frac{1}{2} v_z^2 + h + \Phi \right) v_z 
\end{equation}

\noindent where $h$ is the enthalpy per unit mass (``specific'' enthalpy; $h = e^{\prime} + \frac{P}{\rho}$).  After substituting the continuity equations for mass density and linear momentum density, some algebra and reduction, the form of Equation \ref{eq:firstlaw} is readily obtained\footnote{J.\ Allred, priv. communication 2017.}.

The equation of radiative transfer is solved simultaneously with the other equations through complete linearization and a low-order Feautrier method that is stable in the presence of steep gradients and shocks (M. Carlsson, priv. communication 2022).  The monochromatic specific radiative intensity, $I_{\nu}(z,\nu,\mu)$, at frequency $\nu$, at an angle $\theta = \cos^{-1}\mu$ from the atmospheric normal, at a height $z$, is assumed to be in steady-state ($\frac{\partial I_{\nu}(z,\nu,\mu)/c}{\partial t} = 0$ where $c$ is the speed of light; \emph{cf.} Section 11.2 of \citealt{Hubeny2014}) within each hydrodynamic time-step.  The solution to the intensity at each depth point is integrated over angle using a Gaussian quadrature numerical weighting \citep{Chandrasekhar1960}, and then it is integrated over frequency to give the net radiative flux.  The gradient in the energy equation (Eq. \ref{eq:firstlaw}) thus gives the radiative heating or cooling rate at each time-step in the flare simulation.  Due to this coupling, the time-history of the atomic level populations, ionization, rates, and radiative transfer are critical in the self-consistent energy balance, pressure gradients, and hydrodynamics in a flare simulation.

\subsection{The Flare Heating Term} \label{sec:qbeam}

The flare heating term, $Q_{\rm{flare}}$, in the energy equation (Eq. \ref{eq:firstlaw}) may be a sum from a large number of possible sources:  nonthermal electrons, nonthermal protons, Joule heating, shocks in the reconnection process, and Alfven waves.  Here, we assume the major source of heating in stellar flares is through nonthermal electrons, as is widely accepted in and borrowed from the solar flare analogy.  The radio gyrosynchrotron radiation from mildly relativistic electrons in stellar flares also supports this approach (Sect.~\ref{sec:multilam}).  The following is a synopsis of several relevant points from the  collisional energy loss theory of \cite{Emslie1978} and a unified model calculation presented in \cite{Allred2015}.  Additional formalism about the theory of energy loss on ionized species and neutrals is contained in \cite{Trubnikov1965}, \cite{Chambe1979}, \cite{Emslie1981}, \cite{Fang1993}, \cite{HF94}, \cite{Mott1949}, \cite{Snyder1949}.
We consider particle velocities with magnitude, $v$, and three directional components (whereas in the previous section, one-dimensional bulk velocities of the ambient/thermalized gas were denoted as $v_z$).

A nonthermal (beam) electron penetrating a plasma deposits heat thus affecting the internal energy of the plasma by increasing the temperature, excitation, and ionization of the plasma constituents.  A beam particle is assumed to begin with a kinetic energy that is much greater than the average kinetic energy of the free electrons in the background gas or plasma (thus, the background plasma is a ``cold target'');  the beam particle is then followed until it decreases to an energy that is close ($\sim 2.5 k_B T$) to the energy of the thermal distribution, at which point it joins the background/thermal pool of particles.   Energy loss from the beam electron occurs through Coulomb collisions, and the interactions with all particles in the plasma require an integral that, when evaluated (with somewhat artificially imposed distance limits), is known as a Coulomb logarithm \citep[e.g.,][]{Benz2002}.  The Coulomb logarithms give a sense of the relative importance of distant to nearby interactions with the beam particle.  A beam particle loses energy to ambient (thermal) electrons and protons, with an associated Coulomb logarithm\footnote{Most sources refer to the Coulomb logarithm as $\lambda = \ln \Lambda$, but we follow the abbreviated notation in \cite{Emslie1978}.}  $ \Lambda$.  The integral that describes collisions with ambient neutral atoms also has an associated Coulomb logarithm, $\Lambda^{\prime}$, for each neutral species.  For a typical beam electron,  values for $\Lambda$ and $\Lambda^{\prime}$ are $\approx 20$ and 8.5, respectively.  Including the ionization fraction of a hydrogen gas, $x$, the fraction of energy loss that goes into the neutrals is \citep{RC83} $\frac{1-x}{x \Lambda + (1-x)\Lambda^{\prime}}$.
For a 50\% ionized gas, the energy loss is about a factor of three more efficient on the ionized component (i.e., the ambient thermal electrons).  

The pitch angle of the beam particle changes through scattering, which can be elastic (for collisions with ambient protons, electrons, and neutrals) or inelastic (for collisions with neutrals).  The momentum changes in the direction parallel to the trajectory of the beam particle must include an additional Coulomb logarithm, $\Lambda^{\prime\prime}$, for elastic scattering off of neutrals (e.g., collisions with bound electrons that do not undergo atomic transitions). Distribution-averaged approximations to the Coulomb logarithms are given in \cite{Emslie1978} and \cite{HF94}; \cite{Allred2015, Allred2020} gives the energy-dependent, relativistic formulae.  Then, the total differential kinetic energy change per unit path length [keV km$^{-1}$] of an electron beam particle interacting with hydrogen and ambient electrons in a partially ionized gas is, following from  Eq. 24a of \cite{Emslie1978},

\begin{align} \label{eq:dedz}
\Big( \frac{dE}{dz} \Big)_{\rm{hyd}} & =  \Big( \frac{dE}{dz} \Big)_{\rm{hyd, I}} + \Big( \frac{dE}{dz} \Big)_{\rm{hyd, N}}  \\
  &    = -10^5 \frac{2 \pi e^4 }{\mu E \times a^2} \Big(x \Lambda_{ee}  + (1 - x) \Lambda^{\prime}  \Big) \gamma n_{\rm{hyd}}   
\end{align}

\noindent which is the sum of the ionized (I) and neutral (N) components of energy loss in a hydrogen gas.  We ignore the $\sim 10^3$ less energy loss on ambient protons, $E$ is the kinetic energy of the electron beam particle in keV, $a = 1.602 \times 10^{-9}$ keV erg$^{-1}$, $\mu$ specifies its pitch angle ($\mu =1$ for beamed along hypothetical magnetic field), $x$ is the ionization fraction of the hydrogen target, $n_{\rm{hyd}}$ is the number density of neutral hydrogen atoms plus ambient protons in the target, $\gamma$ is the relativistic gamma factor of the beam electron, and all of the rest of the units are in \emph{cgs}.  
As an example, if $x=0.75$, $n_{\rm{hyd}} = 1.7 \times 10^{13}$ cm$^{-3}$, $T \approx 10^4$ K, which are representative conditions at the top of a model M dwarf chromosphere, then  $\frac{dE}{dx} = -0.1$ keV km$^{-1}$ for $E=40$ keV.  Additional energy losses on helium and other gas constituents can be included.

 \begin{figure*}
  \includegraphics[width=1.0\textwidth]{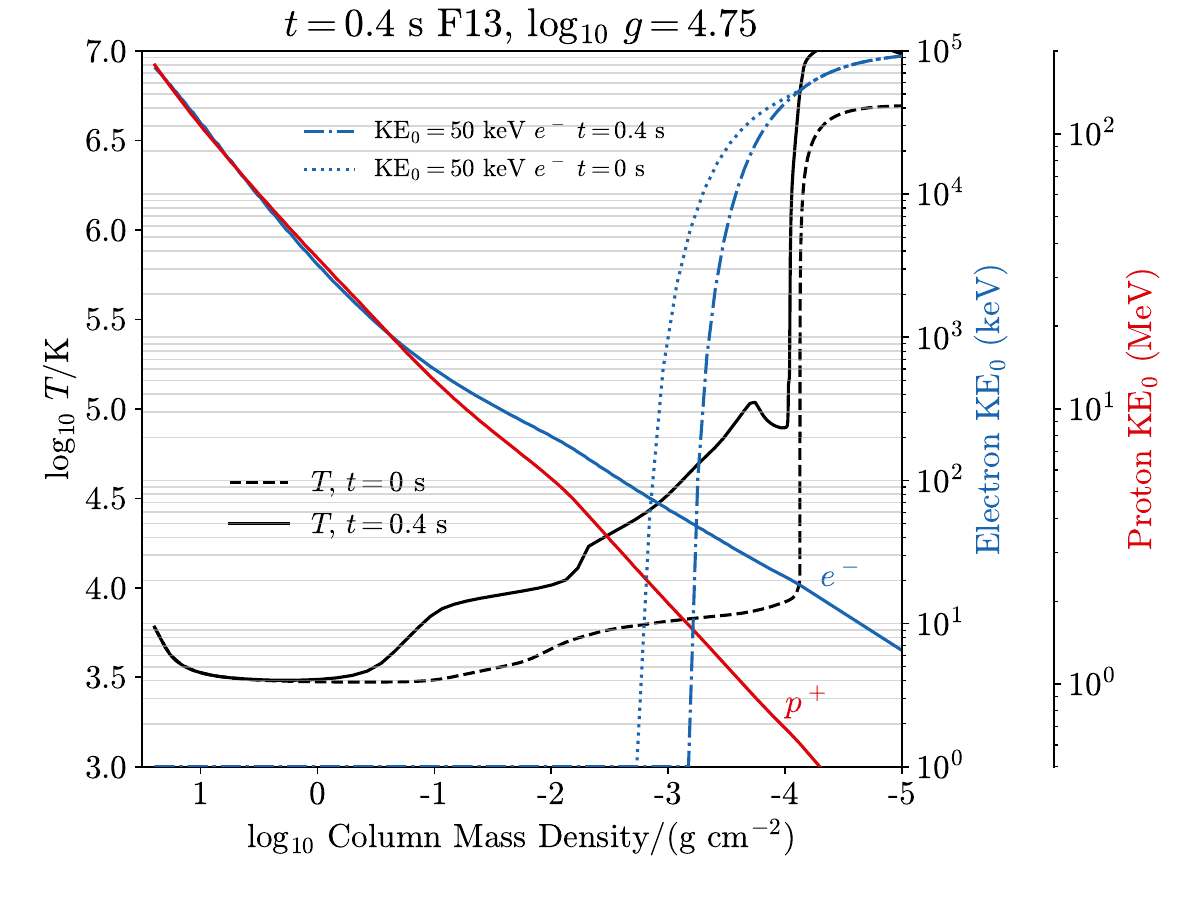} 
\caption{ A pre-flare and flare model (from the RADYN code) of the temperature distribution in the low atmosphere compared to Coulomb stopping depths of beam particles.  For reference the (pre-flare) photosphere and upper photosphere corresponds to $\log_{10} m_c > 1$ on this plot ($m_c$ is the column mass density in units of g cm$^{-2}$).  The collisional stopping depths of mono-energetic beam particles are shown on the right using Equation \ref{eq:dedz}.  The kinetic energy decrease of an electron with injected energy of 50 keV is shown for each state of the atmosphere; these two  curves are on a linear scale and normalized to 1.0 at the top of the loop (where they are injected) outside the range of the plot on the right.  An electron gets stopped at higher altitudes when the chromosphere has experienced a large amount of hydrogen ionization;  as the atmosphere further evolves, the lower energy electrons in the beam can be stopped in the evaporation flows (\emph{cf.} Fig.~\ref{fig:CC_F13}), and all the rest of the beam particles except for the highest energy electrons are stopped in the chromospheric condensation (Fig.~\ref{fig:standardcartoon}(right)).  Note, these ``cold-target'' stopping depth approximations for the protons are not accurate significantly below $\lesssim 1$ MeV at which the warm-target transport treatment becomes essential \citep{Allred2020}.   }
\label{fig:beamdepths}     
\end{figure*}

Assuming a fully ionized, uniform hydrogen plasma, Eq. \ref{eq:dedz} can be integrated from the electron's initial energy $E=E_0$ to $E=0$ to find the stopping length of an electron, $L_{\rm{stop}} \propto E_0^2 n_e^{-1}$, which defines an effective absorption cross section, $\sigma$,  according to $\sigma n_e L_{\rm{stop}} = 1$. Higher energy particles penetrate deeper because they have more energy to lose and because they spend less time losing energy per collision.  The energy loss rate per electron is, $\frac{dE}{dz}\frac{dz}{dt} = dE/dt = - 2\pi e^4 n_e v \Lambda/ E$ which can be integrated over a uniform plasma  to give a timescale for energy loss, $\tau \propto E^{1.5}/n_e$ at non-relativistic energies.  Note, this is the same proportionality in the estimate for scattering timescales out of a magnetic trap (Sect.~\ref{sec:solar}).

The heating from a power-law distribution of electrons and protons is analytically solved in \cite{Emslie1978} following \cite{Brown1973B} and \cite{Lin1976} in an integral from, $Q_{\rm{beam}}(m_c)$, which is a function of column mass 
over an atmosphere with arbitrary, constant ionization fraction.  \cite{HF94} extended the formulae to atmospheres with non-uniform ionization fractions.  The analytic formulae for $Q_{\rm{beam}}$ were used in RADYN simulations of stellar flares up through the models in \cite{Kowalski2015, Kowalski2016}.  

An alternative method is to numerically solve the distribution function, $f(z,\mu,E,t)$, of the nonthermal electrons \citep{Leach1981, McTiernan1990}, and evaluate the spatial gradient/divergence.  This is the Fokker-Planck solution  \citep{Rosenbluth1957}, which tracks the phase space evolution due to systematic decrease (drag) and diffusion of the kinetic energies, $E$, and of the cosines of the beam pitch angles, $\mu$.  Generally speaking, the Fokker-Planck solution is required when some or all of the following are important:
time dependence of the beam propagation, diffusion of pitch angle and energy, external forces (e.g., magnetic mirror, synchrotron losses), and beam feedback effects such as return currents and plasma waves \citep{Hamilton1987, Mauas1997, Zharkova2006, Allred2015}.   We refer the reader to the comprehensive description of the collisional Boltzmann transport equation in \cite{Trubnikov1965} and the application of the Fokker-Planck solution to flare modeling in \cite{Allred2020}. Other helpful references are \cite{Thorne2017} and \cite{BoydSanderson2003}.  In current RHD simulations, numerical limitations necessitate a steady-state solution ($\frac{\partial f}{\partial t} = 0$) at each hydrodynamic time-step, thus giving the flare heating rate as:

\begin{equation}
Q_{\rm{flare}}(z) = Q_{\rm{beam}}(z) = \frac{d }{dz} \int_{\mu} \int_{E} \mu v E f(z,\mu,E) dE d\mu
\end{equation}

\noindent which is the spatial gradient of the beam flux at height $z$ in the atmosphere.

\section{Models of Chromospheric \& Transition Region Flare Emission Line Broadening (\& Asymmetries)} \label{sec:chromlines}
The emission line properties (strength and shapes) during flares encode much of the evolution of the atmospheric physics, which may be quite dissimilar to the density, optical depth, and velocity regimes in the atmosphere before a flare\footnote{\emph{cf.} the vast differences in terrestrial atmospheric states and radiative transfer in a blue-emitting lightning channel and the clear blue sky.}.  The interpretation of flare emission lines, especially in the chromosphere, in general requires forward modeling with non-equilibrium physics.  Thus, there is still much we do not know about the origin of certain well-observed changes in the spectral line profile shapes during flares.  The efforts to make headway in this area have been stymied because, further, not all  flares show similar behavior in the same emission line (\emph{cf.} in solar flares, H$\alpha$ shows a rather wide range of empirical properties; \citealt{Canfield1990}).  Detailed line shape measurements are further obfuscated by the heterogeneity of the flare source and long exposure times, which are generally required of echelle observations.  

This section is divided into two parts. First, we categorize the line broadening mechanisms that are thought to be important in the interpretation of transition-region and chromospheric line widths and asymmetries during stellar flares (Sect.~\ref{sec:broadsources}).  Then, in Sect.~\ref{sec:hlines}, we summarize the symmetric pressure broadening of hydrogen lines in more detail.  Observations and modeling are reviewed together in this section.

\subsection{An Overview of Broadening Sources} \label{sec:broadsources}

There are many types of possible sources of chromospheric and transition region emission line broadening in stellar flare spectral observations.
Microscopic broadening occurs over scales that are much smaller than the optical depth, and each process is generally assumed to be statistically independent.  Thus, the total microscopic broadening enters the line opacity after a convolution of all the respective probability density functions.  Microscopic broadening is largely a source of symmetric broadening (to first order).  Macroscopic broadening
sources can affect the line opacity though the mean intensity, $J_{\nu}$,  but this term typically refers to the broadening from spatially unresolved, inhomogeneous sources in the plane transverse to the observer.  

\begin{figure*}
\begin{center}
  \includegraphics[width=0.65\textwidth]{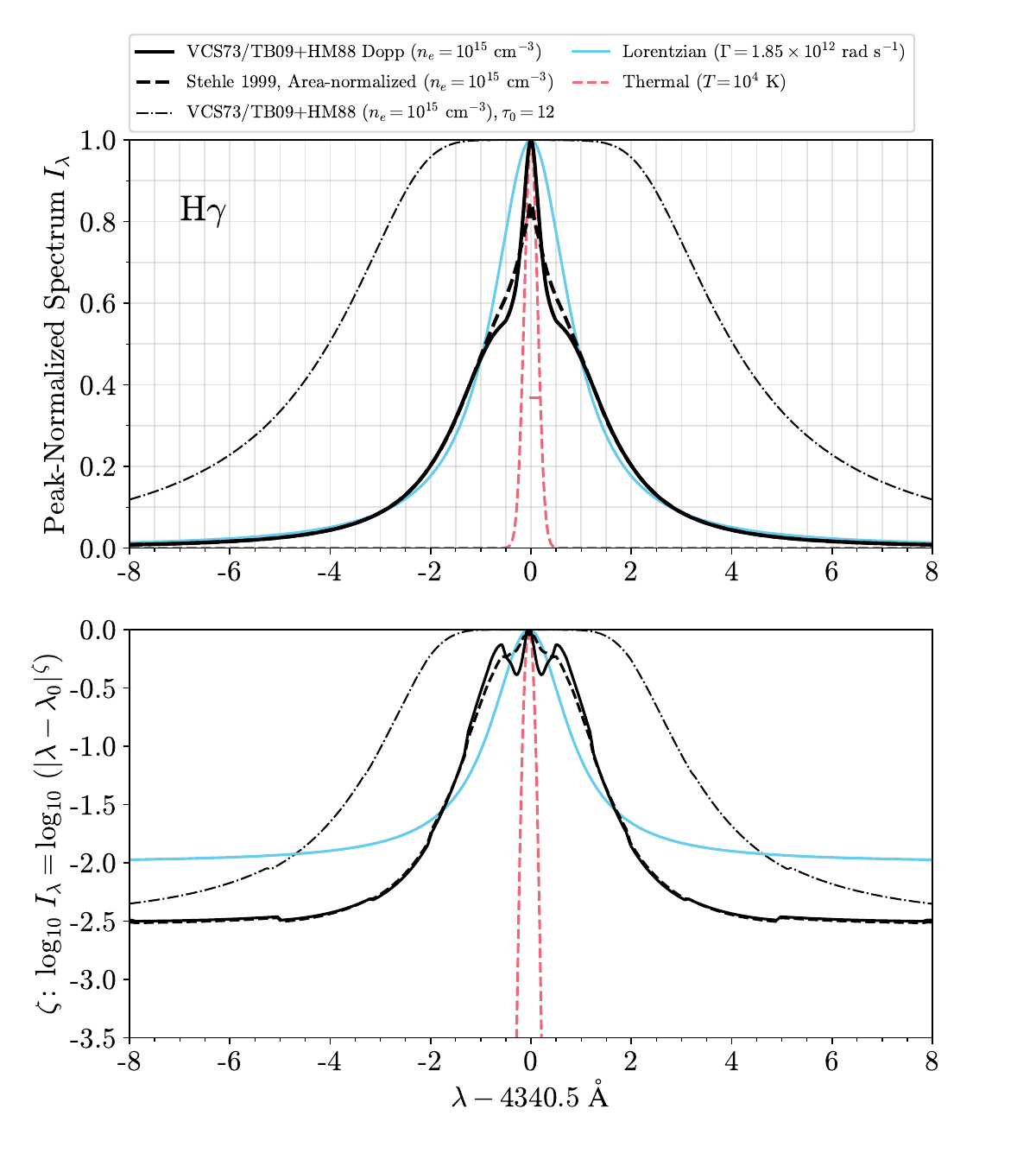} 
  \includegraphics[width=0.65\textwidth]{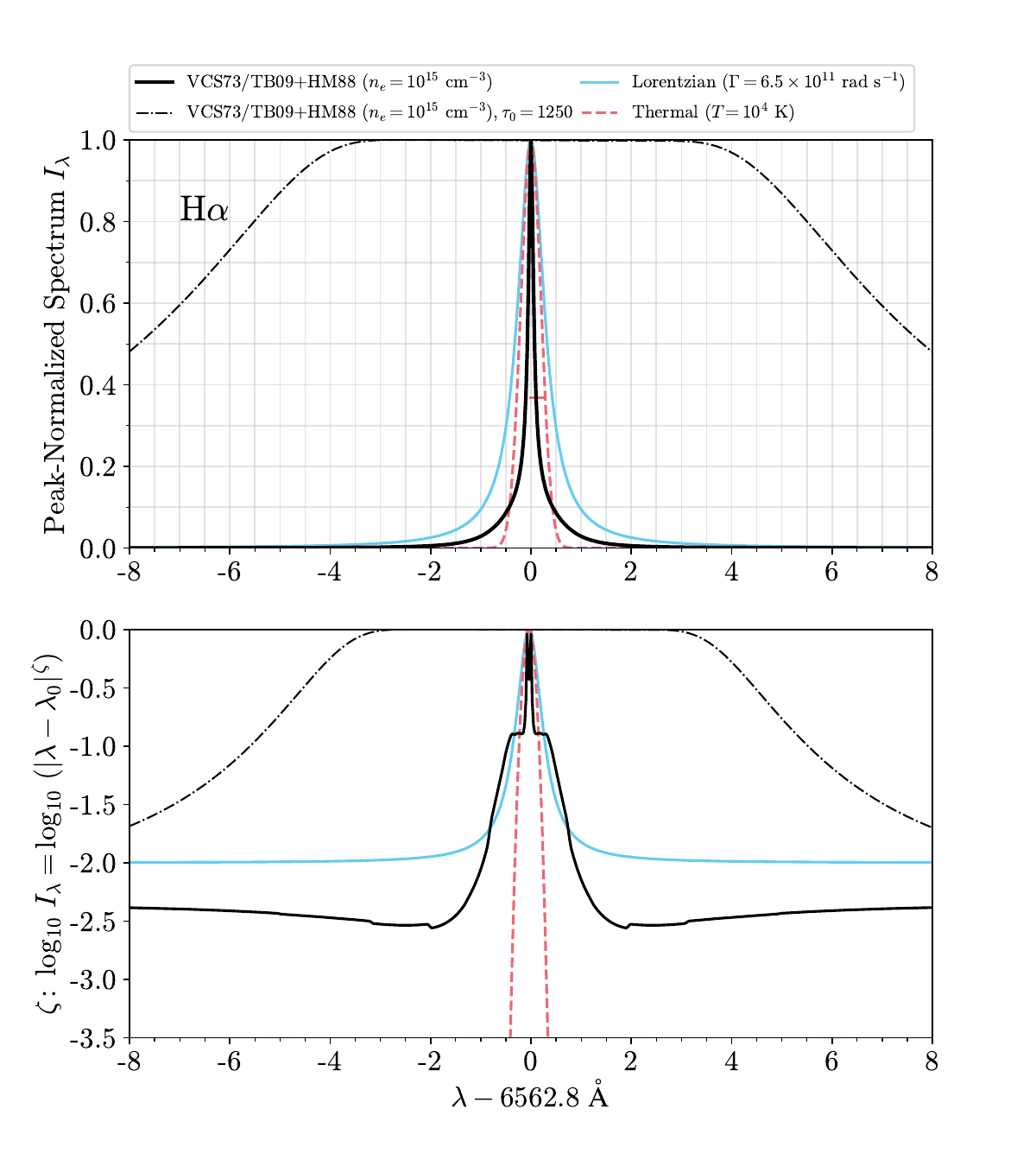} 
\caption{Optically thin spectral line profiles from the TB09$+$HM88 pressure broadening theory for Balmer H$\gamma$ (top two panels) and H$\alpha$ (bottom two panels) at $n_e \approx 10^{15}$ cm$^{-3}$. These are identical to the \citet{Vidal1973} (VCS73) calculations at this density, and they do not include thermal Doppler convolution in this figure.  The 2nd and 4th panels show the slopes, $\eta$, calculated for each of the profiles in 1st and 3rd panels.  Lorentzian profiles with scaled damping widths are overplotted for comparison.  The slope of the H$\alpha$ pressure broadening profile is between the Holstmark and Lorentzian because of the higher importance of collisional broadening from electrons for this transition. The higher order lines, including H$\gamma$, are more obviously dominated by the quasistatic (Holtsmark-distributed) perturbations \citep[\emph{cf.} Fig 3 of ][]{Vidal1971}, are less optically thick, are closer to LTE because of less scattering in this line, and are less affected by thermal broadening (the FWHM of H$\alpha$ is affected by thermal Doppler broadening even at such high densities;  $1/e$ Doppler widths are indicated by horizontal red lines).  The ``self-broadening'' of hydrogen is relatively less in H$\gamma$ and in the higher order lines, but it is important when the atmosphere is largely neutral \citep{Barklem2000}. The Balmer H$\gamma$ line is also less affected by thermal, $T=10^4$ K ion-dynamic effects \cite{Stehle1994}:  we show an area-normalized (rather than peak-normalized) profile from \cite{Stehle1999}, which includes additional Lorentzian damping near line-center from non-quasistatic components of the perturbing ionic field.   Also shown are emergent spectral intensities (dashed-dotted black), with Doppler broadening included, from slabs with a physical depth of $dl = 500$ m to demonstrate the large differences that result from curve-of-growth-like optical depth effects.  }
\label{fig:lineguides}       % Give a unique label
\end{center}
\end{figure*}

  Here, we summarize the most important  sources of micro- and macroscopic broadening that are relevant to stellar flare chromospheric, transition region, and photospheric lines.  We exclude rotational and Zeeman broadening, which affect the broadening of lines in active stars in their quiescent states.  Note that microscopic broadening sources are modeled with an ambient, thermal distribution of perturbers even when they are considered separately from thermal Doppler broadening. In Fig.~\ref{fig:lineguides} we show several profiles for the hydrogen Balmer $\gamma$ and $\alpha$ lines, which complement the descriptions throughout this section.  For each type of broadening, we state the probability distribution that most closely describes the perturbation magnitudes.

\begin{itemize}
\item \textbf{H I$+$p$^{+}$/e$^{-}$}: ``Stark broadening'', or electric pressure broadening of hydrogen, hydrogen-like ions, or Rydberg atoms by ambient charged particles (electrons, protons, ions) in the flare chromosphere, which is partially ionized by definition.  The perturbations are caused by both electron collisional ``damping'' with a Lorentzian probability distribution and quasi-static splitting of energy levels from ambient positive charged protons (and ions) with a Holtsmark-like\footnote{More accurately, this is a Hooper distribution, which accounts for plasma correlations and Debye screening \citep{Hooper1968, Nayfonov1999}.  A Hooper distribution approaches a Holtsmark distribution at large field strengths, which translate to the far wings of a spectral line \citep{Hooper1968B}.} probability distribution.   The splitting of energy levels is due to the first-order/linear Stark effect which is discussed in more detail in Appendix \ref{sec:stark}.  \emph{Optically thin emission line profiles from a homogeneous slab exhibit a Holtsmark power-law slope ($\log_{10} I \propto \log_{10} \mid \Delta \lambda \mid^{-5/2}$) in the far wings}, if non-ideal effects are not important.

 \item \textbf{LEPT}: \cite{Oks2016} discuss the role of broadening of hydrogen lines due to chromospheric plasma turbulence.  Power-law electron beams drive a cospatial drifting return current, whose speed may exceed the threshold for the ion-acoustic / current instability and therefore lead to the development of ion-acoustic turbulence in the chromosphere (E. Oks, priv. communication 2022).  The low-frequency electrostatic plasma turbulence (LEPT) is broadband, and it is quasi-static compared to radiative timescales.  LEPT is treated as a statistically independent broadening process to the ionic microfield broadening  and is therefore convolved with the Holtsmark or Hooper distribution of field strengths that enters into the nominal hydrogen line broadening (above). \cite{Oks2016} analyzed several archival dMe flare observations of the Balmer series.  Without detailed wing shapes, the values of FWHM/$\lambda_0$ from the data are degenerate with either electron densities of $> 10^{15}$ cm$^{-3}$ or \emph{rms} LEPT fields of $10-30$ kV cm$^{-1}$.  Lower electron densities combined with LEPT development in the chromosphere are favored by considering the Balmer flux decrements or  the last resolved Balmer line. \emph{Profiles dominated by LEPT have an exponential dependence in the far wings, $I \propto exp(-k \mid \Delta \lambda \mid^{\gamma}), \gamma \approx 1$.}

\item \textbf{non-H$+$e$^{-}$, non-H$+$H}: Collisional broadening of non-hydrogenic lines due to collisions with ambient electrons (the quadratic Stark effect) or with neutral hydrogen atoms \citep[van der Waals broadening or the more complete ABO theory;][]{Anstee1995, Barklem1998}.   The quadratic (second order) electron impact damping of resonance ion lines, such as Ca II H and K, causes a small wavelength shift and a broadening that is typically $10^3$ smaller than the linear effect in hydrogen.   The passage of H atoms is relatively fast and so this is a damping.  \textbf{non-H$+$non-H}:  Collisions among like atoms or ions results in resonance broadening.  This is an exchange and thus described as a lifetime broadening effect (Bohm 1951). \emph{Optically thin emission line profiles have a Lorentzian shape (or Voigt if a Doppler broadening is included).}

\item \textbf{HI$+$HI}: Collisional damping of hydrogen lines from transient perturbations from other hydrogen atoms.   van der Waals and resonance forces, collectively known as self-broadening, drop off rapidly with distance and so this interaction has a short duration. The neutral fraction of hydrogen is too small, even at $\sim 50$\%, in the flare chromosphere for self-broadening  to be significant in comparison to broadening by charged particles, unlike in the quiet Sun photosphere \citep{Barklem2000}.  \emph{Optically thin emission line profiles have a Lorentzian shape (Voigt if thermal Doppler included).}   

%\item PRD effects.

\item \textbf{Thermal Doppler broadening} is important for line opacity near $\lambda_o$ but decreases precipitously according to the Gaussian distribution.  Note that one Doppler width is $\approx$ 0.2 \AA\ for H$\gamma$ at $T=10,000$ K.   Thus thermal Doppler broadening cannot explain broad wing emission unless there are very high temperatures $T \ggg 10,000$ K.  \emph{Optically thin emission line profiles exhibit a Gaussian profile.}

\item \textbf{Microturbulence}\footnote{A succinct overview of microturbulence and macroturbulence in reference to quiet-Sun MHD models can be found in the discussion in \citet{Carlsson2016}.}, or nonthermal Gaussian Doppler broadening.    An \emph{ad hoc} microturbulence parameter is often used to account for missing broadening sources when all other known broadening sources have been exhausted.  This parameter enters into the opacity (extinction coefficient) through  a convolution of a Gaussian with the thermal Doppler core.   (Certain values of this parameter may propagate through decades of stellar models with little critical reconsideration of their physical meaning or necessity.)  For recent use in solar flare modeling of Fe II and Mg II lines, see \cite{Kowalski2017Broadening} and \cite{Zhu2019}, respectively.   Non-Gaussian microturbulence is discussed often in the interpretation of solar flare transition region lines \citep{Dudik2017}, but the densities are rather large and the optical depths are non-negligible even in transition region lines of quiescent state of active stars \citep[e.g.,][]{Mathioudakis1999}.  \cite{Osten2005} report nonthermal broadening velocities of several tens of km s$^{-1}$ in transition region flare lines with a temperature dependence, such that N V ($T>10^5$ K) does not exhibit excess broadening.

  \hspace{5mm} Does the microturbulence parameter represent bona-fide gas-dynamic turbulence (eddies, vortices, random gas motions) in the footpoints of a flare loop \citep[e.g.,][]{Bornmann1987}, and if so, through which fluid instabilities (e.g., Kelvin-Helmoltz, Rayleigh-Taylor) does the flare turbulence occur?   The turbulence hypothesis can be investigated by considering the energetics inferred through spectral lines and the emission line flaring area on the star (see Sect.~\ref{sec:geom}).  \cite{Hawley2007} reported very broad, symmetric wings of Mg II h and k emission that had a small, bulk blueshift during a NUV flare observed by the Hubble Space Telescope/STIS.  The wings would require electron densities as high as $10^{17}$ cm$^{-3}$ if due to collisional broadening.  A Gaussian fit to the broad wings gives the line-of-sight velocity component root-mean-square (rms), $\sigma_{\rm{los}}$, and the kinetic energy of the chromospheric turbulence is

\begin{equation} \label{eq:turb}
  KE = \frac{3}{2} \sigma_{\rm{los}}^2 \rho_{\rm{chrom}} \Delta z_{\rm{chrom}} A_{\rm{flare}}
\end{equation}

  \noindent where $A_{\rm{flare}}$ is the projected area on the star (Section \ref{sec:geom}) that produces the emission line flux at Earth. In this flare, the  turbulence would be highly supersonic (200 km s$^{-1}$) and its kinetic energy would be several orders 
of magnitude larger than the radiated NUV energy.  %This is contrary to the energetics in solar flares \citep{Antonucci1982}.

\item  \textbf{Opacity broadening, optical depth effects}:  It has long been known that simultaneous modeling of electron density and optical depth is required to interpret the hydrogen lines properly in stellar and solar flares \citep[e.g.,][]{Drake1980}.  Optical depth effects cause a type of macroscopic broadening in chromospheric flare lines in the emergent spectral lines from stellar atmospheres.  Sometimes this is loosely referred to as opacity broadening, which is described in \cite{Wood1996}, \cite{Rathore2015A}, and \cite{Hansteen2023}.  Essentially,  opacity broadening  follows from the Eddington-Barbier relation  for a depth-dependent source function over optically thick wavelengths of a line\footnote{In laboratory plasma physics, opacity broadening is also called ``self-absorption'', and it is usually described in the context of a depth-independent source function \citep[e.g.,][]{Yaakobi1977, Kunze2009}.  Stellar atmospheres, and their flaring states, are not homogeneous slabs, however.  }.  If the source function decreases in the outer layers a central reversal can form due to an absorbing layer for very optically thick lines; this is also sometimes called ``self-absorption''.    An optical depth broadening effect that is closely related to opacity broadening occurs in the enhancement of the  emergent intensity at wavelengths that correspond to the transition from a formation over very large optical depths (where there is opacity broadening, proper) to smaller optical depths in the far wings.  This is similar to the broadening in textbook illustrations of the curve-of-growth of a spectral line, whereby the Doppler core saturates at the local source function, and the Lorentzian or Holtsmark wings are very prominently in emission but are not close to saturation.  The core saturation and wing enhancements of H$\gamma$ and H$\alpha$ are demonstrated in Fig.~\ref{fig:lineguides} for a slab thickness of $dl = 500$ m and electron density $n_e = 10^{15}$ cm$^{-3}$.  An analysis of these effects in a heterogeneous, non-equilibrium model chromospheric condensation is presented in \citet{Kowalski2022}.  \cite{Namekata2020} analyze in detail the self-absorption and pressure broadening in stellar flare H$\alpha$ emission lines.

\item \textbf{Red-wing and blue-wing asymmetries} cause an increase in the breadth of a spectral line, especially if the line is spectrally and/or spatially unresolved.  In stellar flares, red- and blue-shifted components of H$\alpha$ have maximum line-of-sight velocity extents of $\approx 100-300$ km s$^{-1}$ from the rest wavelength \citep{Vida2019}.
  Reports of line asymmetries in the literature attribute the asymmetries to a variety of phenomena that involve chromospheric mass motions:  beam-generated impulsive phase chromospheric condensations, coronal rain (a ``coronal condensation'') from post-flare loop plasma draining and cooling to transition region and chromospheric temperatures, cool filament eruptions of plasma, absorbing downflows, and chromospheric evaporation of cool material into magnetic loops \citep[see][for an overview]{Wu2022}. 
  Many stellar flare observations, however, with echelle resolving power  have rather long exposure times ($t_{\rm{exp}} \approx 15$ minutes), and some are serendipitous single-exposure observations from exoplanet radial velocity monitoring. 
Additionally, optical depth effects and other microscopic broadening sources are important for interpretation.  Optically thin emission lines can exhibit a velocity ``smearing'' in the emergent spectra, simply due to an integration of the emissivity over the velocity gradient along the line-of-sight, and optically thick lines are complicated by opacity broadening \citep{Namekata2020, Wu2022}.
  
 There are many reports of line asymmetries in hydrogen, helium I, and metallic lines throughout the UV, optical, and NIR that have been interpreted with these  phenomena.  The spectral observations consist of a bright emission line component around the rest wavelength and a fainter asymmetric wing component, or more than one wing component.  Most modeling of these lines uses linear superpositions of Gaussians to characterize the shifts and widths of each component.  \cite{Wu2022}, \cite{Namekata2022B}, and \cite{Namizaki2023} use Voigt/Lorentzian profiles to isolate a Gaussian fit to a red-shifted components in large stellar flares.  The bottom panel of Fig.~\ref{fig:namekata} demonstrates this method, which effectively isolates a $\approx -100$ km s$^{-1}$ blueshifted component in the H$\alpha$ line flux during the decay phase of a white-light flare \citep{Notsu2023}.  \cite{Eason1992} and \cite{Fuhrmeister2008} report broad H$\alpha$ lines that are blueshifted during the rise and peak spectra during flares on UV Ceti and CN Leo, respectively. In the event on CN Leo, the blueshifts were -10 to -20 km s$^{-1}$ and the Gaussian FWHM values were 4.8 \AA;  the event on UV Ceti produced a similar width but a larger blueshift of -70 km/s.  \cite{Fuhrmeister2011} presents detailed line profiles at high-time resolution during flares on Proxima Centauri with VLT/UVES.   \cite{Vida2016} and \cite{Vida2019} discuss blueshifts observed across the Balmer series with some temporal resolution.  \cite{Notsu2023} present a large survey of echelle spectral observations of the H$\alpha$ and H$\beta$ lines in M dwarf flares;  the spectra had relatively short exposure times and were complemented with simultaneous optical broadband photometry, allowing the changes in the profile shapes to be characterized in the context of the impulsive and gradual phases of the white-light response (e.g., Fig.~\ref{fig:namekata}(bottom)).   \cite{Honda2018} report on high-time cadence monitoring of EV Lac and present persistent blue-wing asymmetries through the entire duration of a flare \citep[see also][]{Maehara2021}.  \cite{Namekata2021} discuss a transient blueshifted H$\alpha$ absorption feature, while \cite{Namekata2022B} analyze broad H$\alpha$ emission lines with red-shifted components, in flares on the young, solar-type flare star EK Dra, and \citet{Lalitha2013} analyze the redshift evolution of a broad H$\alpha$ line component in a flare on the K0Ve star AB Dor.  \cite{Gunn1994b} discuss blueshifted emission components in AT Mic. Of course, the location on the stellar disk is generally unconstrained in stellar flares, and the inferred velocities are not corrected for line-of-sight effects for field-confined flows, contrary to what is possible for solar flare observations \citep{Brosius2018}.

Flux-weighted line-center redshifts in Si IV and C IV during and after the impulsive phase of several dMe flares were reported in \cite{Hawley2003}.  Chromospheric-line, red-wing emission in H$\alpha$, Na I D, and He I 5876 \AA\ was described in \citet{Fuhrmeister2005}.  These observations were tested against scaling relations (e.g., Eq. \ref{eq:fevap}) from analytic models of impulsive-phase chromospheric condensations \citep{Fisher1989}.   Late-phase redshifts, on the other hand, are usually interpreted as condensations that are analogous to the draining of cool material from the post-flare loops on the Sun.  \cite{Wu2022} discuss evidence from the time-evolution of the broadening of the H$\alpha$ line that could favor beam-generated condensations in the late-phase instead (which we note would be more in line with the multi-thread RHD modeling of the continuum radiation in the decay phase of large flares;  Sect.~\ref{sec:geom}, Fig.~\ref{fig:megaflare_decay}).   \cite{Fuhrmeister2008} and \cite{Kanodia2022} report on red wing enhancements in He I 7065 \AA\ and 10830 \AA\ emission lines, respectively, in two flares. \cite{Fuhrmeister2020} show remarkably broad He I 10830 \AA\ wings during a flare on the dM3e star EV Lac;  correspondingly broad H$\alpha$ profiles were analyzed in \cite{Fuhrmeister2018} and consist of a spectrally unresolved red wing satellite component.  \cite{Loyd2018Hazmat}, \cite{Froning2019}, and \cite{France2020} report redshifts in other transition region lines, such as C II, from low-mass stellar flares.  \cite{Linsky1994} and \cite{Ayres2015} analyze the profile variations in FUV lines during flares on AU Mic and EK Dra, respectively.  The optical Balmer lines have been decomposed into ``separate'' broad and narrow components, which exhibit clearly different Doppler shifts in some flares \citep{Houdebine1992, Houdebine1993B, Fuhrmeister2008, Kowalski2022Frontiers}.  Extreme blueshifted  \citep{Houdebine1990} and redshifted \citep{Woodgate1992, Bookbinder1992}  emission has been reported out to several 1000 km s$^{-1}$ in chromospheric and transition region lines.   \cite{Koller2021} discuss blueshifts and redshifts in a large sample of galactic field stars at larger distances and over a wide range of spectral types.

\item \textbf{Macroturbulence} is a term that refers to a superposition of several Doppler-shifted profiles that originate from sequentially-ignited, neighboring flare loop footpoints, as in a solar arcade.  The superposition is intended to emulate spatially unresolved flows, which are termed `directed mass motions' in the literature.  The unresolved flows may be preferentially blueshifted, preferentially redshifted \citep{Rubio2017}, or equally blueshifted and redshifted \citep{Doyle1988}.
 To generate symmetric wings, the emergent radiation from downflows (e.g., a condensation) would have to precisely balance  the emergent radiation, and thus the optical depths, from the upflows (e.g., evaporation) in neighboring loops.  This precarious balance would have to be maintained over long durations, as for typical exposure times of $60-300$s.  \emph{The lineshapes may take on a variety of symmetric or asymmetric forms that depart from Holtsmark and Lorentzian wing profiles.}
  
\end{itemize}

\subsection{Symmetric Wing Broadening of H Lines} \label{sec:hlines}

The symmetric (or very nearly symmetric) wing broadening of hydrogen Balmer lines is a remarkable property of stellar flares. Full widths at 10\% maximum in emission line spectra are as large as $\approx 15-20$ \AA\ \citep{HP91, Namekata2020}.
The source of the symmetric broadening has been debated as a superposition of directed mass motions (macroturbulence), gas turbulence (as represented by a microturbulence parameter), and pressure broadening due to the Stark effect (Sect.~\ref{sec:broadsources}).  \cite{Doyle1988} and \cite{Eason1992} argue against the Stark effect in the large dMe flares that they analyze by comparing to the optically thin wing predictions of Holtsmark profiles and the \cite{Vidal1973}  profiles for electron densities of $n_e \approx 10^{15}$ cm$^{-3}$ \citep[see also][]{Houdebine1993A, Houdebine1993B, Gunn1994}.  The better fits with multiple Gaussian components in these studies support the hypothesis of some type\footnote{The solar flare H$\alpha$ profiles in \cite{Ichimoto1984} are invoked for justification, but these spectra are highly asymmetric to the red and are broadened out to $\Delta \lambda \sim 10$ \AA.} of supersonic laminar- or turbulent-emitting flows. The latter corresponds to a microturbulence parameter of $50$ km s$^{-1}$ around line center and $150-600$ \kms\ in the wings.  \cite{Montes1999} present similar two-component Gaussian fits to broad and narrow components in H$\alpha$ and H$\beta$ in flares from the K dwarf LQ Hya; the broad components show relative shifts with respect to the narrow components and are interpreted in terms of various types of unresolved mass motions.

The alternative hypothesis for the source of the broad wings of hydrogen lines is pressure (linear Stark) broadening from ambient charged particles, which imprints unique signatures in the observations.  Namely, the pressure broadening of hydrogen produces larger widths for larger $n_{j}$ within a series (e.g., within the Balmer, Paschen, or Lyman series)  in the absence of optical depth effects and heterogeneities (here, $j$ refers to the upper level of a transition).  However, the smoking gun signature is generally thought to be broader hydrogen emission line wings compared to nearby resonance lines of metallic ions, such as Mg II and Ca II, which experience a much smaller amount of broadening due to collisions with charged particles \citep{STARKB_CaIIA, STARKB_CaIIB}.  \cite{HP91} and \cite{Garcia2002} present much broader Balmer series lines in comparison to the Ca II K emission line \citep[see Fig.~4 of][]{Garcia2002} in the impulsive phases of two large dMe flares \citep[see also][]{Lalitha2013}.   

\begin{figure*}
\begin{center}
  \includegraphics[width=0.65\textwidth]{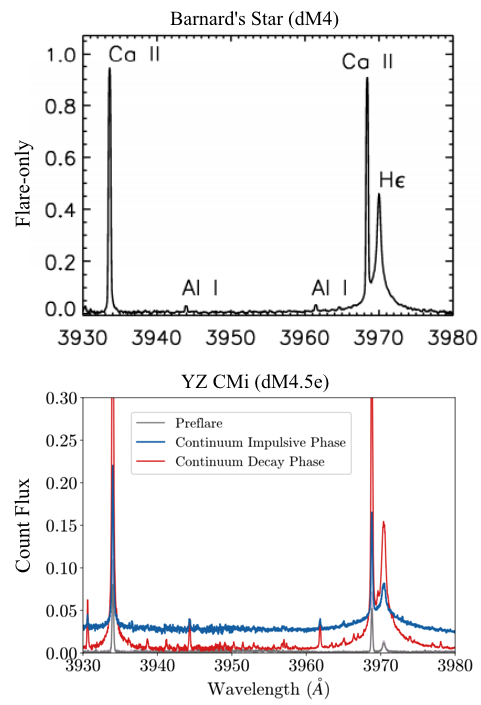}
\caption{ Dramatic broadening differences between hydrogen and metallic emission lines, such as the Ca II resonance lines, in flare spectra are considered evidence for the important role of pressure broadening of hydrogen.  The pressure broadening of hydrogen  constrains the charge densities (and optical depths) in the flare chromosphere.   This conclusion is supported by detailed spectral synthesis with, e.g., the RADYN, RH, MULTI codes.  The top figure shows these effects during a flare on the dM4 star Gl 699, reproduced from \citet{Paulson2006}, and the bottom panel shows the coarse temporal evolution of the broadening differences during a flare \citep{Vida2019} on the dM4.5e star YZ CMi. }
\label{fig:PaulsonVida}       % Give a unique label
\end{center}
\end{figure*}

We show two remarkable examples of the hydrogen broadening in stellar flare spectra in Fig.~\ref{fig:PaulsonVida}.  The spectral broadening and emission line shape differences among Ca II K, H, and Balmer H$\epsilon$ ($n_j=7 \rightarrow n_i=2$) are clearly evident in these echelle spectral observations.  The top panel shows a serendipitous flare spectrum ($t_{\rm{exp}} = 30$ min) from the inactive dM3.5-4 star Gl 6l 699 (Barnard's star) that was studied in detail and modeled with the RADYN and MULTI codes in \cite{Paulson2006}.  The bottom panel displays a sequence of spectra during a flare from the dM4.5e YZ CMi at the same times\footnote{We downloaded these data from the PolarBase public spectral archive \citep{PolarBase1, PolarBase2}.} as shown for the H$\gamma$ spectra in Fig. A.1, top panel, of \cite{Vida2019}. The inactive M-star flare shows some qualitative similarities to the line broadening in the spectrum of the decay phase of the continuum in the YZ CMi flare.   These similarities are remarkable given that the stars have very different ages ($> 7$ Gyr vs. $< 500$ Myr, respectively) and quiescent magnetic activity levels.  The impressive FUV flaring activity of Barnard's star has been reported recently in \cite{France2020}.
A detailed analysis of the H9 Balmer line was presented for the large flare on CN Leo in \cite{Fuhrmeister2008}.  Broadened Paschen lines have been reported in the decay phase of a large flare from the very low mass star VB10 \citep{Kanodia2022}.   Similar comparisons are, in principle, possible with the broadening of the high-order Paschen series and the Ca II infrared triplets  \citep{Neidig1984}, for which opacity broadening is perhaps less of a concern in comparisons to Ca II H and K.

\subsubsection{Recombination Edge Modeling} \label{sec:edge}

Accurate interpretation of stellar spectra from EUV through NIR wavelengths depends on modeling how the wings of the hydrogen lines overlap and merge near bound-free ionization limits \citep{Hubeny1994}.  Accurate modeling, which should lack any sharp edge feature, opens up diagnostic power from the rich spectral region at wavelengths just longward of a recombination limit.   For example, the Inglis-Teller relation is \citep{InglisTeller, Rutten2003}

\begin{equation}\label{eq:it}
  \log_{10} n_e = 23.2 - 7.5 \log_{10} n_{\rm{max}}^{\rm{Balmer}}
 \end{equation}

\noindent which has commonly been used to determine electron density from the  last discernible Balmer emission line near the Balmer recombination limit \citep{Kurochka1970, HP91}. Besides stellar flares, many types of astrophysical and laboratory phenomena  also lack discontinuities in spectra around the expected location of hydrogen recombination edges \citep{Herczeg2008, Esteban2004, Meigs2013, Shull2012}.  In solar flares, features such as a ``blue continuum bump'', a shift of the edge to redder wavelengths, or a completely flat continuum have been reported in observations \citep{Neidig1983, Donati1985}.  The complete lack of any type of edge feature, and a lack of a jump in flux across the expected edge location, is ostensibly an indication of blackbody radiation.

However, the lack of an edge is also expected from recombination theory that includes non-ideal, level dissolution.  
 The fundamental question pertaining to stellar flares is thus if there is enough instrumental blending of the wings and/or overlapping, linearly-superposed wing opacities to account for the lack of the observed discontinuities.  A remarkable example of stellar flare spectra with no edge feature is shown in Fig.~\ref{fig:Fuhrmeister2008}(a).  Instead, these spectra (and others) show a series of faint, broad hydrogen lines that fade into a continuum.  Other spectra, albeit at much lower resolving power, show some noisy features here, however \citep{HP91}.   Ostensibly, the conservation of oscillator strength density should result in a smooth transition of observed flux from a pseudo-continuum of a confluence of high-order, highly broadened and blended emission lines to the onset of recombination radiation. 
 A phenomenological description of the transfer of bound-bound opacity to bound-free opacity was developed by \cite{Dappen1987} using the occupational probability ($w_n$) formalism in the non-ideal equation-of-state of \cite{HM88}.  The ``dissolved-level'' bound-free Balmer continuum opacity longward of the Balmer edge has been recently incorporated into spectral models of stellar flares \citep{Kowalski2017Broadening, Kowalski2022Frontiers}.  However,  self-consistent modeling of the pressure broadening of the bound-bound Balmer lines \citep{Tremblay2009} was not also included in the first models \citep{Kowalski2015}.  This incorrectly supported the extremely dense chromospheric condensations in 1D RADYN models (Sect.~\ref{sec:models}) as realistic models of flare density enhancements at cool temperatures \citep{Kowalski2016}.  Otherwise, the dissolved-level continuum opacities have largely reconciled observations with models by explaining the lack of an edge around $\lambda = 3646 - 3700$ \AA\ and the range of observed pseudo-continuum properties in between the longer-wavelength, high-order Balmer lines at $\lambda = 3700 - 4000$ \AA.  The non-discontinuous, transition of dissolved-level opacities at $T \approx 10^{4}$ K have apparently reconciled the hydrogen recombination (Balmer jump) theory of the NUV with the blackbody-like theory of the optical continuum radiation.

 \begin{figure*}[h]
  \includegraphics[width=1.0\textwidth]{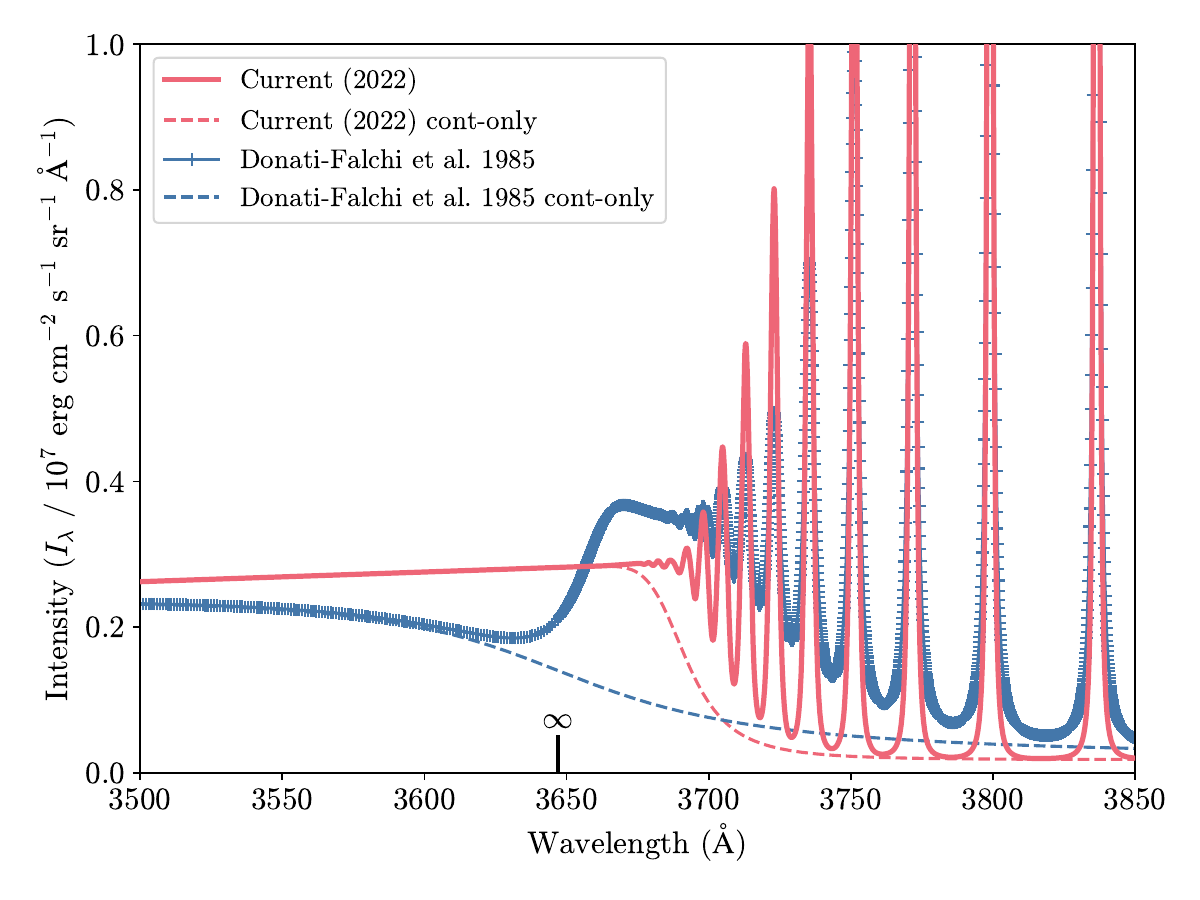}
\caption{ Direct comparisons of two methods \citep{Donati1985, Kowalski2017Broadening} for calculating the Balmer jump spectral region in solar and stellar flares.  The emergent LTE intensity is calculated from a slab with $n_e = 3.9 \times 10^{13}$ cm$^{-3}$.  The method from \cite{Donati1985} predicts a ``blue continuum bump'', whereas the current treatment with occupational probabilities predicts an extension of the Balmer continuum intensity to wavelengths longward of the ``ideal'' recombination limit (indicated by an $\infty$ symbol) at $\lambda = 3646$ \AA.  }
\label{fig:tb09df85}      
\end{figure*}

 The physical cause of the continuous, dissolved-level opacity that was parameterized in \cite{Dappen1987} for hydrogen is still not well understood.  Experiments with Rydberg atoms support the idea that Landau-Zener transitions \citep[e.g.,][]{Zwiebach2022} at avoided level crossings \citep{Zimmerman1979, Rubbmark1981, Pillet1984} facilitate a cascading ionization process.  In this picture, an electron is photo-excited to a high-lying level $n$ that is pressure-broadened  to overlap with the broader, next higher-energy level $n+1$, thus effectively ``dissolving'' the level $n$.  This description apparently explains the Inglis-Teller relationship through dynamic microfield ionization  \citep{HM88}.   There are actually many types of calculations \citep[][and see the review in Sections 86.2.2 - 86.2.3 in \citealt{Springer2006}, \citealt{Springer2006_Ch86}]{Littman1978, Damburg1979, Bergeman1984, Seidel1995, Benenti1999, FisherMaron2002, FisherMaron2003, Cho2022}\footnote{The dynamic microfield ionization hydrogen is assumed to be analogous to microwave field ionization of Rydberg atoms \citep{HM88}. The rather ambiguous term ``pressure ionization'' is usually reserved for the situation in which ground-state hydrogen atoms are so close together that electron degeneracy pressure causes the ionization (e.g., in the cores of stars and in white dwarfs).  }  that may contribute to  ``lowering the ionization potential'' or ``continuum lowering'' in a hydrogen plasma with microfield fluctuations and collisional perturbations\footnote{Nonetheless, one should note that Landau-Zener transitions have been invoked in microwave ionization experiments with Rydberg atoms in order to account for more rapid ionization than is predicted by the classical, hydrogenic saddle-point estimate of the lowering of the ionization potential;  see Fig.~\ref{fig:stark_effect} in Appendix \ref{sec:stark}.  }.

\begin{figure*}[h!]
  \includegraphics[width=1.0\textwidth]{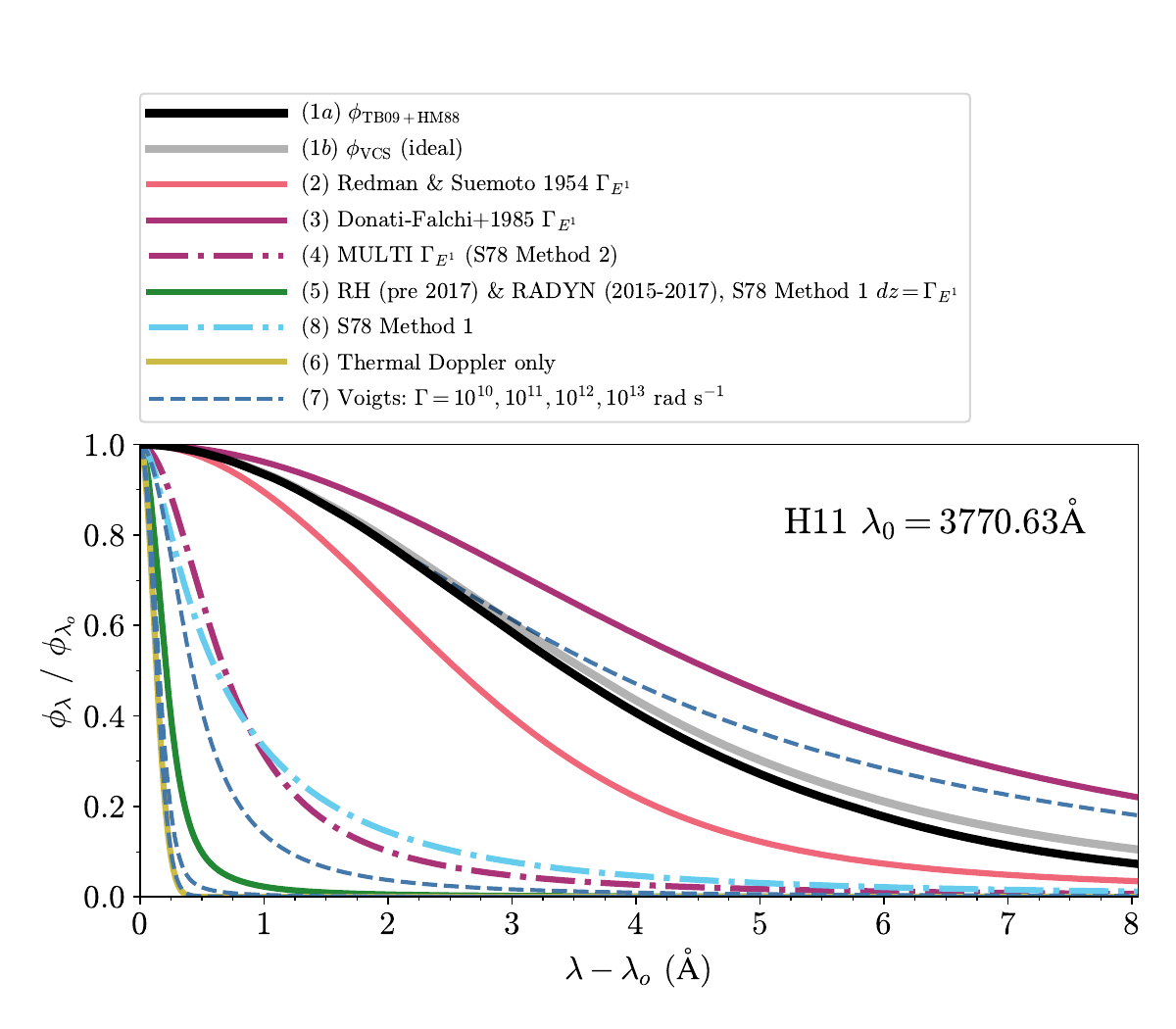}
\caption{A variety of approximations to opacity line profile functions for the Balmer series have been used in stellar flare modeling codes over many decades.  This figure summarizes the differences in models of the line profile function, $\phi_{\lambda}$, for the H11 Balmer line \citep[see also][for a similar figure for the H10 line]{JohnsKrull1997}.   The calculations use $n_e = n_p =  3.16 \times 10^{14}$ cm$^{-3}$, or about 10x larger than in Fig.~\ref{fig:tb09df85}, and thermal Doppler broadening at $T = 10^4$ K is included.  The ``ideal'' profiles, $\phi_{VCS}$, are the VCS73 calculations from \cite{Lemke1997}, and do not account for line narrowing due to occupational probability / level dissolution.  Several Voigt profiles are also plotted for a range of Lorentzian FWHM values, $\Gamma$.  Currently, the most accurate theory corresponds to the $\phi_{\rm{TB09+HM88}}$ (solid black) curve.  }
\label{fig:H11Methods}      
\end{figure*}

Due to level-dissolution in dense regions of flare atmospheres that are not very optically thick in the Balmer continuum, applying a simple relationship, such as the Inglis-Teller, that connects the last-visible hydrogen line to a single electron density may be misleading.   \cite{Kowalski2022} discuss a scenario wherein level-dissolution of hydrogen in a chromospheric condensation can generate an approximately grey opacity window at $\lambda \approx 3700 - 3750$ \AA.  Thus, the Balmer lines that are in emission in the emergent spectra actually originate from deeper layers with smaller ambient (thermal) electron densities and no apparent gas velocities.  Due to the heterogeneous vertical stratification of the flare atmosphere, the high-order hydrogen Balmer series in emergent intensity spectra can be as broad as or less broad than the low-order Balmer lines (H$\gamma$, H$\beta$, H$\alpha$), which are very optically thick in some of the more recent RHD flare models.  Some models of stellar flare spectra at the Balmer limit also incorporate spatial and temporal heterogeneity within the flare ribbons \citep{Kowalski2022Frontiers}.  These seek to explain the inconsistencies in large $n_{\rm{max}}^{\rm{Balmer}} \gtrsim 15$ (Eq. \ref{eq:it}) being observed (indicative of low electron densities, $n_e= 10^{13} - 10^{14}$ cm$^{-3}$) with the necessity to also produce the observed continuum properties from a location on the star with much larger electron density ($n_e \gtrsim 10^{15}$ cm$^{-3}$; $n_{\rm{max}}^{\rm{Balmer}} \approx 11$).  Note that in occupational probability theory, however, there is not one single maximum upper level for all atoms in a plasma.

Fig.~\ref{fig:tb09df85} directly compares two different modeling techniques within the Balmer jump spectral region.  The emergent intensity is calculated in LTE from a static slab with $dl = 400$ km, $T=10^4$ K, and $\rho=10^{-10}$ g cm$^{-3}$ ($n_e = 3.9 \times 10^{13}$ cm$^{-3}$).   \cite{Donati1985} modeled several solar flare spectra (with high resolving power) using Voigt profiles up to $n_{\rm{max}} = 2\times10^4 T^{0.25} n_e^{-0.25} \approx 65-85$; this is a saddle-point estimate with Debye-screening of the proton field \citep{Mihalas1978}.  This method convolves the Balmer edge with the Voigt function corresponding to the line with $n_{\rm{max}}$.  The ``current (2022)'' method uses the TB09$+$HM88 profiles \citep{Tremblay2009} and the dissolved level bound-free continuum opacity.  Many other various approximations for the pressure broadening of hydrogen have been used in models of solar and stellar flares.  The ambiguities were discussed extensively in \cite{JohnsKrull1997} and were resolved only in recent years.  In Fig.~\ref{fig:H11Methods}, we summarize the wide variety of broadening prescriptions for the Balmer H11 line, whose integrated flux is sensitive to the amount of level dissolution in the densities of flare atmosphere models \citep{Kowalski2015}.  For a description of each method presented in the figure, we refer the reader to \cite{Kowalski2022} where similar comparisons are shown for the Balmer H$\gamma$ line.

\section{A Comprehensive Multi-Wavelength Stellar Flare Dataset} \label{sec:multilam}

What fundamental physical parameters can one infer using a multi-wavelength dataset of a stellar (super)flare?  A comprehensive analysis of a series of two consecutive superflares from the binary dM4e$+$dM4e system DG CVn was presented in \cite{Osten2016}, who summarize the plasma and geometric properties that can be constrained in a flaring atmosphere.  This culminates an extensive history of stellar flare research, and we briefly review the salient points to the algorithm.

\subsection{Inferring Nonthermal Electron Properties from Stellar Data} \label{sec:xrayradio}
The nonthermal radiation from power-law electrons in stellar flares occurs as hard X-rays ($E \gtrsim 10$ keV) and as incoherent gyrosynchrotron radiation in the radio/microwaves at cm wavelengths. The radiative processes are thought to be similar to solar flares, in which the hard X-rays are emitted through collisional bremsstrahlung in the chromosphere, whereas the radio flux originates from predominantly higher energy electrons ($E \approx 100-300$ keV) trapped in magnetic fields; see the review in \cite{White2011}.  The distribution of electrons is parameterized in the hard X-rays and radio using two power-law indices, $\delta_x$ and $\delta_r$, respectively.  From hard X-rays, the particle flux distribution is

\begin{equation}  \label{eq:plxray}
    F(E) = F_o \left( \frac{E}{E_c} \right)^{-\delta_x}
\end{equation}

\noindent where $E$ is the electron kinetic energy in keV, $E_c$ is the low-energy cutoff, and $F_o$ is the differential (specific) electron flux density (el s$^{-1}$ cm$^{-2}$ keV$^{-1}$) around the cutoff energy.   Similarly, a nonthermal electron beam density distribution determined from the radio is written as 

\begin{equation} \label{eq:plradio}
    n(E) = n_o \left( \frac{E}{E_c} \right)^{-\delta_r}
\end{equation}

\noindent where $n_o$ is the differential number density (el cm$^{-3}$ keV$^{-1}$) of electrons around energy $E_c$.  If the same population of electrons is responsible for both X-rays and radio,  then $\delta_x \approx \delta_r$ is expected for electrons with velocity $v$ very close to the speed of light $c$; otherwise, $\delta_x \approx \delta_r - 0.5$.    The different forms (\emph{i.e.,}, flux density \emph{vs.} number density) of these equations is because the injected power-law flux distribution is  inferred through the collisional thick target model of the hard X-ray spectrum, whereas a power-law density distribution is determined through observations at optically thin microwave frequencies (see below).

Note that the integral of Eq. \ref{eq:plxray} multiplied by $E$ gives the total beam energy flux density (erg cm$^{-2}$ s$^{-1}$), which is an important input for flare modeling (e.g., F10, F13, etc...).
Integrating Eq. \ref{eq:plradio} over energy gives the total beam density, $N$, where $N$ is the quantity that is used in the gyrosynchrotron expressions in \cite{Dulk1985} if $E_c$ is set to $10$ keV. In the studies of \cite{Smith2005, Osten2016, Dulk1982, Dulk1985}, Eq. \ref{eq:plradio} is integrated from $E_c$ to $\infty$ to replace our $n_o$ with the prefactor $N(t) \frac{\delta_r-1}{E_c}$ where $N(t)$ is the total electron density (el cm$^{-3}$) above the cutoff (and the same is usually done for Eq. \ref{eq:plxray}).   Eq. \ref{eq:plxray} can be multiplied by a probability density function for the angular distribution, $M(\theta)$, to specify the directivity of particles with respect to the magnetic field \citep{Allred2020}.  A highly-collimated Gaussian distribution is usually chosen for $M(\theta)$.  Then, a ``1.5D'' solution to the Fokker-Planck equation (Sect.~\ref{sec:qbeam}) accounts for the changes in the pitch angles of the particles.

The best diagnostic of accelerated particles in stellar flares is optically thin frequencies in the mm or radio.  If the observations are in the radio at GHz frequencies, one must ensure that these frequencies are above the peak frequency ($\approx 10$ GHz) so as to be in the optically thin part of the gyrosynchrotron spectrum.  An expression from \cite{Dulk1985} converts from a power-law index in the optically thin flux spectrum, $S_{\nu} \propto \nu^{\alpha_r}$, to the power-law index in the electron distribution, $\delta_r$:

\begin{equation}
\alpha_r = 1.22 - 0.90 \delta_r
\end{equation}

\noindent The power-law index in the optically thick part of the spectrum is much less sensitive to the power-law index of nonthermal electrons.  

\begin{figure}
\begin{center}
  \includegraphics[width=0.65\textwidth]{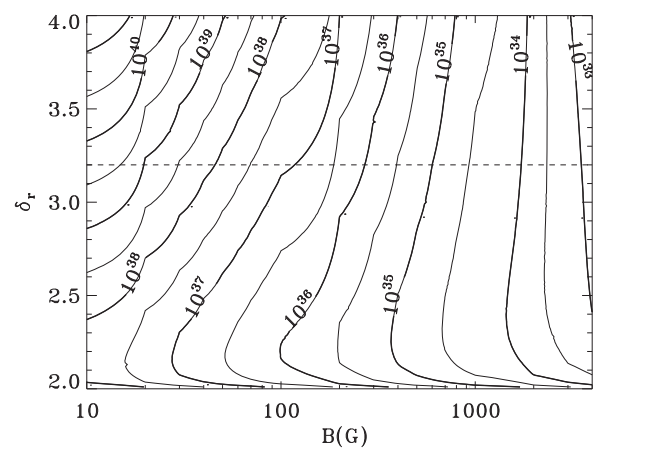} % published_data/
\caption{A range of possible nonthermal electron beam power-law indices and magnetic field strengths that can be constrained from gyrosynchrotron data of stellar flares; reproduced from Figure 6 in \cite{Osten2016} with permission.  See also \cite{Smith2005} and  \cite{MacGregor2021}.  Contours of constant electron beam kinetic energies add a third constraint.  Cross sectional footpoint areas from optical data can constrain the beam flux density for forward modeling inputs \citep{Osten2010, Osten2016}.    }
\label{fig:osten2016}       % Give a unique label
\end{center}
\end{figure}

Following \cite{Smith2005}, the kinetic energy of nonthermal particles can be constrained from the radio spectrum as follows.    The power per volume in nonthermal electrons above the assumed low-energy cutoff value, $E_c$, is given by Eq.\ 9 of \cite{Osten2016}.  In this equation, the total number of nonthermal electrons, $N(t) V(t)$, over the flaring volume $V(t)$ is an unknown, but it is directly proportional to the optically thin spectral radio luminosity, $L_{\nu}$.  By constraining the value of $\delta_r$ in this part of the spectrum and invoking some prior knowledge or assumption about the peak frequency, an integral over the entire radio spectrum can be calculated -- including the optically thick part -- to give the time-integrated energy of nonthermal electrons in the flare.  This leaves one unknown, the magnetic field, $B$, in the gyrosynchrotron emissivity formula for the optically thin part of the spectrum (see below for further constraints on the magnetic field, but it must be assumed that the $B$ derived from the radio originates from the same $B$ that is derived from the soft X-rays).  If the radio or mm data are available at only one frequency, and thus a spectral index cannot be determined from the data, then  contours of nonthermal energies and plausible ranges of magnetic field strengths and spectral indices are reported, as in Fig.~\ref{fig:osten2016}, which is reproduced from \cite{Osten2016} \citep[see also][where a spectral index was derived from mm observations]{MacGregor2021}.

X-ray and optical data can constrain the physical parameters of flares further.  From time-resolved X-ray spectra, one obtains the volume emission measure, $\mathcal{VEM}$.  The $\mathcal{VEM}$ is $\int n_e^2 dV$, which is the amount of optically thin emitting plasma at a fixed X-ray temperature, $T_X$.   Typical values of the $\mathcal{VEM}$ (in M-dwarf flares) are  $10^{51} - 10^{52}$ cm$^{-3}$.  In a $\log_{10}$ $T_X$ vs.\ $\log_{10}\ \sqrt{\mathcal{VEM}}$ diagram, the peak $T_X$ may clearly occur before the peak $\mathcal{VEM}$ \citep{Favata2000, AschwandenGudel2008, Liefke2010}.   Over the decay phase evolution of a flare, the slope in this diagram provides one of three inputs for the widely employed flare loop modeling of \cite{Reale1997}.  Combined with an exponential decay constant from the X-ray light curve and the maximum X-ray temperature, a semi-circular loop's half-length, $l$, can be inferred.  The loop sizes in giant X-ray flares are typically reported to be on the order of the stellar radius \citep[e.g.,][]{Favata2000, Osten2010, Lalitha2013, Osten2016}. Note, large $50 - 200$ Mm loop lengths are also inferred under some interpretations of QPPs (Section \ref{sec:qpp}) in optical broadband light curves \citep{Anfin2013, Doyle2022QPP}.

Optical and/or NUV data provide data to relax the geometrical assumption of a single, giant monolithic loop (Sect.~\ref{sec:geom}), while also providing constraints on the electron density  and the magnetic field strength in the flare corona.  The optical and NUV probe the footpoint radiation, from which  a dimensionless ratio of $R_{\rm{foot}} / (2l)$   can be calculated, where $R_{\rm{foot}}$ is an equivalent radius of the area of an optical flare footpoint:  $A_{\rm{flare,opt}}=2 \pi R_{\rm{foot}}^2$.  By combining with an alternative expression for the $\mathcal{VEM}$ in terms of these quantities and using the value of the $\mathcal{VEM}$ from X-ray spectral fitting, a representative value of the coronal $n_e$ is readily obtained that is  independent of the number of identical loops in a hypothetical stellar arcade \citep[Eq. 25 of][]{Osten2016}.
One also can assume that the magnetic pressure must be at least equal to the gas pressure of the evaporated MK plasma for it to be confined to the loops instead of cooling through expansion, as in a scenario analogous to a filament eruption or coronal mass ejection \citep{Cully1994}.  The magnetic field confinement condition places a lower-limit on the coronal magnetic field of the flaring loops.

 \cite{Mullan2006} homogeneously analyzed a large sample of many sizes of X-ray stellar flares from various spectral types and evolutionary stages using ``Haisch's simplified analysis'' \citep[HSA;][]{Haisch1983}, which involve scaling relations \citep[see also][]{Oord1989A} that infer coronal temperature, electron density, loop length, and confining magnetic field strength from an emission measure and light curve decay timescale.  Typical X-ray temperatures derived from multi-component fits to $E=0.5 - 10$ keV spectra are similar to those in solar flares, $T_X \approx 15-30$ MK \citep[e.g.,][]{Osten2005, Liefke2010, Namekata2020}.   At the low coronal temperature end, $T \approx 2-15$ MK, electron density variations, or lack thereof, are constrained with X-ray spectra of the $f$ (forbidden) and $i$ (intercombination) components of helium-like triplets   \citep[e.g., bottom right spectra in Fig.~\ref{fig:Gudel2002};][]{Gudel2002, Osten2003, Testa2004, Osten2005, Liefke2010}.  When significantly detected, the flare-enhanced electron densities are reported in the range of $10^{11}-10^{12}$ cm$^{-3}$ at these cooler coronal temperatures.  

\citet{Osten2016} summarize the physical parameters in three of the largest dMe flares that have been observed with X-ray spectra (\emph{cf.} their Table 5). The inferred temperatures are $T \approx 50$, $140$, $290$ MK;  the electron densities are 3 $\times 10^{11}$, 3 $\times 10^{12}$ 3 $\times 10^{11}$ cm$^{-3}$; the confining magnetic field flux densities are 230, 1100, 580 G, respectively; and $\mathcal{VEM}$ ranges between $10^{54} - 10^{55}$ cm$^{-3}$.  Generally similar numbers are found in the largest RS CVn flares, which produce integrated X-ray energies and luminosities one to two orders of magnitude larger \citep{Osten2007, Karmakar2023}.   YSO's also exhibit such extreme volume emission measures and temperature \citep[$40-80$ MK;][]{Vievering2019}.   These superhot temperatures correspond with X-ray luminosities that are close to or in excess of the bolometric luminosity of the star\footnote{Before the term ``superflare'' became widely associated with white-light $E \gtrsim 10^{33.5}$ erg superflares on G dwarfs in \emph{Kepler} data, \cite{Osten2007} proposed the term for flares that are super-bolometric in X-ray luminosity.  It is remarkable that so many types of stars -- from RS CVn's to tiny, red-dwarf stars -- can release more energy per second momentarily in their outer atmospheres than they do through nuclear fusion in their cores;  perspective provided by R. Osten, priv. communication, 2014.  }.
Note that in an X-ray superflare from the dM3.5e star EV Lac with comparable temperatures ($T_{X, \rm{peak}} \approx 70$ MK) that were derived from the hydrodynamic modeling analysis of \cite{Reale1997}, \cite{Favata2000} argue that accounting for the $E > 10^{34}$ erg of energy release  during the flare places further constraints on coronal magnetic fields  in the $\approx 4$ kG range before the flare.

The peak stellar flare values of $\mathcal{VEM}$ and $T_{X}$ generally compare well with scaling relations over many orders of magnitude, from superflares to solar flares, in the famous ``Hertzsprung-Russell (H-R) diagram of solar and stellar flares'' \citep{Yokoyama1998}.  The H-R diagram of solar and stellar flares was first theoretically explained using MHD scaling relations in \cite{Shibata1999}.  This diagram was investigated further in \cite{AschwandenGudel2008}, and the tightest correlation  was reported in the empirical relationship 

\begin{equation}  \label{eq:VEM}
\mathcal{VEM}(T_p) = 10^{50.8} \big( \frac{T_p}{\rm{10\ MK}} \big)^{4.5 \pm 0.4} [cm^{-3}]
\end{equation}

%%\vspace{1mm}

\noindent where $T_p$ is the peak flare X-ray temperature.  Their relation for decay time and peak temperature is, in principle, useful for inferring physical parameters if only the decay phase is caught in the observations \citep{Ayres2015};  however, the empirical scatter about the trends is an order of magnitude in both quantities.
\cite{Getman2008} and \cite{Osten2016}   suggest that the hottest ($T_X > 100$ MK) stellar flares exhibit a ``saturated''-like volume emission measure of $\approx 10^{55}$ cm$^{-3}$ and do not follow a solar extrapolation in the flare H-R diagram.   \cite{Shibata1999} include two additional parameters (either loop length or magnetic field and ambient thermal electron density) in their scaling relation analogue \citep[see also further alternatives summarized in][]{Heinzel2018} to Equation \ref{eq:VEM} to explain the apparent offset in volume emission measure from the solar extrapolation.     \cite{Howard2022EV} include a much smaller flare from an active M-dwarf flare in the context of stellar superflares and solar flares.  \cite{Osten2016} report on a stellar superflare (their ``F2'' event) with $T_X \approx 50$ MK that is within a factor of 2.5 in the predicted $\mathcal{VEM}$ given by Equation \ref{eq:VEM}.

Large stellar flares occasionally trigger \emph{Swift}'s Burst Alert Telescope (BAT trigger numbers 172969, 310125, 331592, 596958, 625898), which is intended to automatically slew to gamma-ray bursts and other extra-galactic transients\footnote{The \emph{Swift} project maintains a webpage of flare stars whose count rates are monitored;  \url{https://swift.gsfc.nasa.gov/results/transients/Flare_stars.html}.}.  The combination of X-ray data from the BAT and at softer X-ray energies from the \emph{Swift}'s XRT facilitates classical model hypothesis testing among thermal and nonthermal models out to $E \approx 100$ keV.  In superflares from II Peg and DG CVn, nonthermal bremsstrahlung models were found to be equally statistically robust fits as a superhot thermal model fit.  However, in the superflare event from DG CVn \citep{Osten2016} that the BAT spectra triggered on, the $T_X = 290$ MK thermal model was favored over a nonthermal model because the time-evolution of the X-rays was delayed with respect to the optical $V$-band data \citep[in line with the thermal, empirical Neupert effect;][]{Cabellero2015}, and a footpoint area from the optical gives (even by stellar flare standards) unrealistic collisional thick target electron beam energy and flux  into the lower atmosphere. The collisional relaxation times were estimated to be rather short compared to the (classical) thermal conduction times, thus further supporting the thermal interpretation.   Additional evidence for thermal interpretation of superflare X-ray emission from YSO's is discussed in \citet{Vievering2019}.

In soft X-ray flare spectra, features around the Fe 6.4 keV K$\alpha$ emission line and the 6.7 keV Fe XXV thermal emission line complex are also reported.  These features have been interpreted as either electron beam excitation \citep{Emslie1986} or non-LTE fluorescence of a larger-area  photospheric `halo' \citep{Osten2007, Drake2008,  Osten2010}.  However, there are some calibration issues in \emph{Swift} data around these features that have been reported  only recently \citep{Pagani2011}.  The recalibrations greatly diminish the 6.4 keV feature while enhancing the 6.7 keV complex.  Fe K$\alpha$ fluorescence was studied in a flare from an active giant star using a different observatory (Chandra) without instrumental problems and higher resolving power \citep{Testa2008}; see Sect.~\ref{sec:fekalpha}.

Table \ref{table:diagnostics} highlights a few important multi-wavelength diagnostics in stellar flares that have been discussed in this review.   Of course, the interpretations of these diagnostics are complicated by issues pertaining to heterogeneities of the unresolved nature of stellar flare sources (in time, and in all three spatial directions), and thus are generally  model-dependent.  In the final section of this review, we discuss assumptions for and inferences of stellar flare geometries and the associated implications for heterogeneity.

\clearpage
\begin{sidewaystable}
%%\sidewaystablefn%
\begin{center}
\begin{minipage}{\textheight}
\caption{A summary of flare diagnostics}\label{table:diagnostics}
\begin{tabular*}{\textheight}{@{\extracolsep{\fill}}ll@{\extracolsep{\fill}}}
  \toprule%
  Measurement/Data & Physical parameter of flare \\
 \hline
X-ray exponential decay constant, $T_{X,\rm{max}}$, slope in $T_X$ vs. $\mathcal{VEM}$ & semi-loop length $l$ \\
Slope at optically thin radio/mm frequencies & power-law index of NT electrons, $B$, $E_{\rm{NT,kinetic}}$ \\
Optical ($\lambda > 4000$ \AA) color temperature & Footpoint area \\
Optical or $U$-band light curve timing \emph{vs.} X-rays, impulsiveness index &  non-thermal vs. thermal interpretation of hard X-rays \\
Balmer jump ratio (or H$\gamma$/C4170$^{\prime}$) & Optical depth at $T \approx 10,000$ K \\
%High n balmer lines & chromospheric electron density, non-ideal effects \\
H$\epsilon$, H$\delta$, or H$\gamma$ vs. Ca II H or K & electric pressure (Stark) broadening vs. turbulent Doppler broadening \\
Fe K$\alpha$ spectral region & fluorescence vs. NT/beam excitation, viewing angle, halo geometry \\
$f/i$ ratios from X-ray spectra of helium-like triplets (e.g., O VII $\lambda \approx 22$\AA ) & thermal electron densities at $T \lesssim 15$ MK \\
\midrule
\hline
\\
\bottomrule
\end{tabular*}
\footnotetext{``NT'' is an abbreviation for nonthermal.}
\end{minipage}
\end{center}
\end{sidewaystable}
\clearpage

\section{Stellar Flare Geometries} \label{sec:geom}
Stellar flares are spatially unresolved point sources.
Yet, from spectral data, model fitting, and sufficient time-resolution, there are many properties of the spatial dimensions of flaring sources that can be inferred.  These reveal similarities to solar flare phenomenology at similar wavelengths and strongly reinforce the solar-stellar connectivity of flare physics.  In this section, we review the methods and results of stellar flare geometry calculations.

Optical and NUV observations constrain the chromospheric/photospheric flare areas.
A blackbody temperature, $T_{\rm{flare}}$, and the fraction (``filling factor'') of the visible stellar hemisphere that is flaring, $X_{\rm{flare}} = R_{\rm{flare}}^2 / R_{\rm{star}}^2$, can be fit to spectral data or multi-band photometry through the following equation \citep{Hawley1995, Hawley2003}:

 \begin{equation} \label{eq:bbx}
      f_{\lambda, \rm{Model}}^{\prime} = \Big(\pi B_{\lambda}(T_{\rm{flare}}) - S_{\lambda,q}\Big) X_{\rm{flare}} \frac{R^2_{\rm{star}}}{d^2} 
 \end{equation}

\noindent where $B_{\lambda}(T_{\rm{flare}})$ is the Planck function intensity model of the flare, $S_{\lambda,q}$ is the surface flux of the quiescent star, and  $f_{\lambda, \rm{Model}}^{\prime}$ is the `flare-only' model flux at Earth above the atmosphere that is to be fit to the observed, calibrated [\flam] flare-only flux, $f_{\lambda}^{\prime}$.  The ratio of flare-only fluxes at two wavelengths is a flare color and is the minimum information that is needed to solve for $T_{\rm{flare}}$ in Eq. \ref{eq:bbx} (through linearization and iteration or a lookup table). 
An $m$-component, linear superposition of RHD surface flux flare spectra,  $S_{\lambda, \mathrm{RHD}, m}$, or a multi-thermal blackbody model are typically required to reproduce the optical and NUV continuum observations over a wavelength range spanning more than $\Delta \lambda \approx 1000$ \AA .  Additional terms are thus included in Equation \ref{eq:bbx}. For an $m$-component RHD model superposition, the solution is a linear least squares fit to an observed flare-only spectrum, $f_{\lambda}^{\prime}$, resulting in the estimated parameters, $\hat{X}_{\mathrm{flare},m}$ \citep[see][]{Osten2016, Kowalski2017Broadening, Kowalski2022Frontiers}. 
 For further discussion, see Appendix \ref{sec:slabs}.

Constraints on $X_{\rm{flare}}$ from blackbody fitting to optical flare-only spectra result in inferences that a small fraction of the visible  stellar hemisphere flares for even the largest, $E \approx 10^{34}$ erg, dMe events.  For example, the broadband photometry in the Great Flare of AD Leo \citep{HP91} constrain $X_{\rm{flare}}$ to be 0.5\%  \citep{HF92} in the first impulsive peak. Fits to the flare-only, optical continuum in the rise/peak spectrum of this event are consistently small (0.2\%)  but with a correspondingly larger blue-optical blackbody color temperature of $T_{\rm{flare}} \approx T_{BB} = 11,600$ K \citep{Kowalski2013}.  An $m=2$ component RHD model of this spectrum gives a similar filling factor for a model with large electron-beam heating fluxes, which are combined with the second RHD model component with filling factors that are $\approx 2 - 10$x larger in order to account for most of the narrow emission line flux \citep[][see also \citealt{Cram1982, HF92, Kowalski2010}]{Kowalski2022Frontiers}.   The corresponding projected area at the star is $A_{\rm{flare}} = X_{\rm{flare}} \pi R_{\rm{star}}^2$, which may instead consist of a series of $K$ circular flare kernels.  Inferences in smaller dMe flares from broadband photometry \citep{Hawley2003} and optical spectroscopy \citep{Kowalski2016}\footnote{See Table 5 of \cite{Kowalski2016} for different inferences of optical color temperature and filling factors using different wavelength ranges over the optical.}  are consistent with very small areas $\approx 5 - 10\times 10^{16}$  cm$^{2}$, thus reinforcing the analogy to compact white-light kernels that are observed at the footpoints of impulsively heated loops in solar flares \citep[e.g.,][]{Fletcher2007}.    During the most energetic dMe events, however, peak-phase flare areas are inferred to be several orders of magnitude larger than solar flare white-light and NUV areas \citep{Krucker2011, Kowalski2017Mar29}.

\begin{figure*}
  \includegraphics[width=1.0\textwidth]{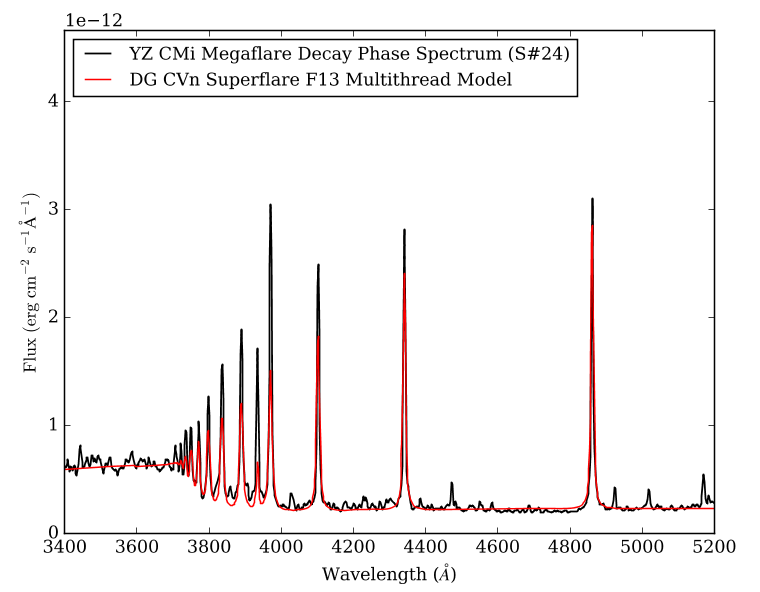} 
\caption{ Gradual-decay phase, flare-only spectrum of a megaflare on the MV4.5e star YZ CMi \citep{Kowalski2010}.   Multi-thread RHD modeling of another superflare on DG CVn \citep{Osten2016} adequately reproduces the Balmer jump, optical continuum shape, and some of the hydrogen Balmer emission line properties (but some important shortcomings remain).
Each RHD model component is calculated self-consistently, but the linear superposition of the  model components is semi-empirical; see discussion in \cite{Kowalski2017Broadening} and \cite{Kowalski2019IRIS}.  }
\label{fig:megaflare_decay}       % Give a unique label
\end{figure*}

In the gradual decay phase, and in the impulsive phase of gradual-type dMe flare events with large Balmer jump ratios, blackbody fits to broadband photometry may be misleading (Sect.~\ref{sec:colorimetry}).  Blackbody fits to the gradual phase continuum give a rather inconsistent picture of the post-peak footpoint areal evolution from event to event; perhaps, heterogeneous RHD model superpositions in the gradual decay phase are much more important for accurate areal inferences than in the $10^4$ K blackbody-like rise and peak phase.  A superposition of RHD model components in the gradual decay phase of a megaflare on the dM4.5e star YZ CMi is shown in Fig.~\ref{fig:megaflare_decay} from \cite{Kowalski2017Broadening}.  This model was previously developed to explain the NUV to $V$-band ratio in the DG CVn superflare event reported in \cite{Osten2016}, and it also adequately modeled the NUV continua and Balmer jumps in much smaller, gradual-type events \citep{Kowalski2019HST}. This ``DG CVn superflare multi-thread (F13) model'' consists of several RHD model spectra over the evolution of the atmosphere that is shown in Fig.\ref{fig:CC_F13}, thus superposing snapshots from a range of optical depths, temperatures, and densities (see Sect.~\ref{sec:arcade}) in order to explain the gradual decay phase.

    %The Megaflare 0.01\% and 25x.  
 
\begin{figure*}
  \includegraphics[width=1.0\textwidth]{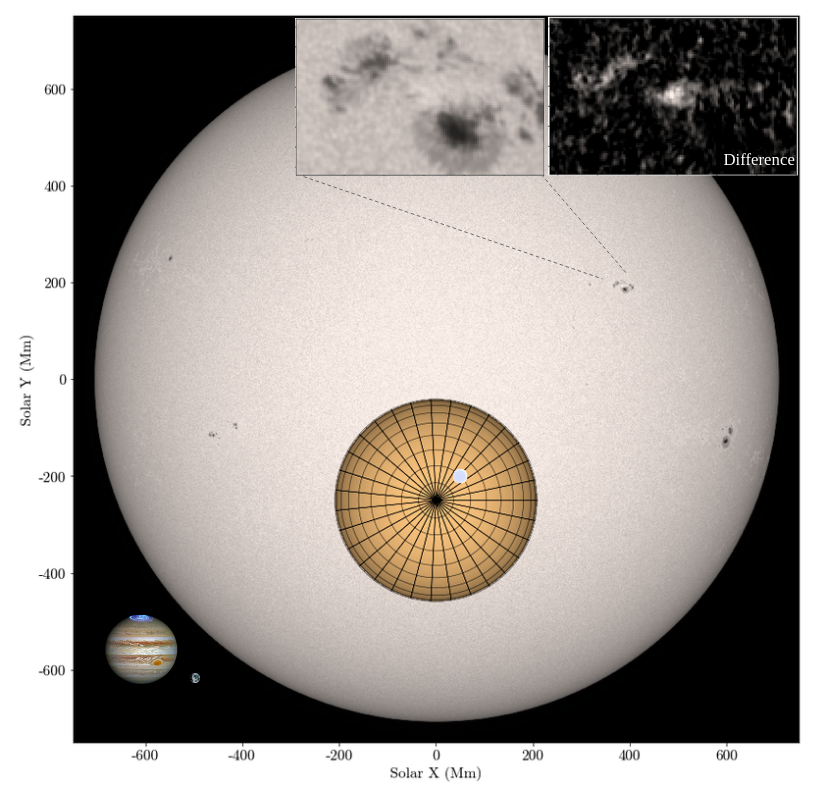} 
\caption{ Sizes to scale.  An image of the Sun at $\lambda = 6173$ \AA\ from the Helioseismic and Magnetic Imager \citep[HMI;][]{Scherrer2012} on SDO, a representative size of an active mid-type (MV3e-MV4.5e) M star ($R_{\rm{star}} = 0.3 R_{\rm{Sun}}$), and a circular flare area ($\approx 5 \times 10^{18}$ cm$^{2}$ ) that corresponds to about 0.5\% of the visible hemisphere of this hypothetical M star.     For comparison, Earth and Jupiter are shown to scale.  The insets each have a field-of-view of $\approx 10^{19}$ cm$^{2}$, and they show an image of the solar active region in the NW quadrant during the hard X-ray impulsive phase of the SOL2014-03-29T17:48 X1.0 flare (total intensity left, difference image right). For a detailed analysis of the HMI data of this solar flare, we refer the reader to \cite{Kleint2016}. Jupiter image credit: NASA, ESA, and J. Nichols;  Earth image credit:  GOES/NOAA/NESDIS.  }
\label{fig:scales}       % Give a unique label
 \end{figure*}

Various size scales in solar and stellar flares are shown in Fig.~\ref{fig:scales}.
 
 \subsection{A Dominant Loop or an Arcade?} \label{sec:arcade}
 There are two general approaches to modeling spatially unresolved flare light curves and spectra \citep{Liefke2010}.  The first approach is summarized in \cite{Karmakar2023}, and references therein, where it is most recently applied to an RS CVn X-ray superflare.  This modeling follows the analysis of \cite{Reale1997} and describes the rise and decay of a flare as result of a dominant monolithic loop (or a superposition of a few dominant loops that are heated and cooled simultaneously).   The heating and cooling timescales in the loop are on the order of tens of minutes to hours \citep[see also][]{Reale2004, Testa2007, Argiroffi2019}.  In \cite{Reale2004}, the initial X-ray event in the same flare in Fig.~\ref{fig:Gudel2002} was modeled with this method, and the subsequent X-ray flare in the event was modeled as an arcade of $\approx$ five loops that were sequentially heated in a nearby region.  Coincidentally, a solar analogy inspired a similar hypothetical explanation for the NUV and optical footpoint ignition during the decay phase of a different M dwarf flare in \citealt{Kowalski2017Broadening}.  Longitudinal MHD oscillations in a monolithic, giant (200 Mm) loop have been proposed as an explanation for the triggering of quasi-periodic, $U$-band, light-curve oscillations in the decay phase of this megaflare event \citep{Anfin2013}.  The comparisons of particularly dense snapshots \citep[e.g., $t=2.2$~s;][]{Kowalski2015} from RHD models to observations also inherently make the assumption that a single dominant kernel at any one time contributes to the bulk of the flux received at Earth \citep[but see][for recent tests of this assumption]{Namekata2022}.

 An alternative hypothesis is known as ``multi-thread modeling'', which was pioneered by \cite{Hori1997}, \cite{Reeves2002}, and \cite{Warren2006} to explain Sun-as-a-star  soft X-ray solar flare light curves.  These works addressed the large discrepancy between the observations of long decay times in GOES soft X-rays and the very short decay times in models of individual loops;  multithread techniques have further been used to model a variety of solar flare phenomena \citep[e.g.,][]{Reeves2007, Rubio2016, Reep2017}.  Elements of this approach and of observations of sequential footpoint brightenings in the hard X-ray and optical wavelengths in solar flare ribbons \citep[e.g.,][]{Kosovichev2001, Qiu2010} have inspired stellar flare RHD model superpositions  \citep{Osten2016, Kowalski2017Broadening} for more direct comparisons to spatially unresolved observations.   \cite{Getman2011} argue that the rise phase of a giant X-ray flare from the PMS star DQ Tau consists of a multi-threaded source;  see also \cite{Schmitt2008} in reference to the modeling approach showcased in  Fig.~\ref{fig:schmitt2008}.    \citet{Doyle2022QPP} 
 suggest that a series of tens of loops contribute to the geometrical dimensions, as inferred from their QPP (Section \ref{sec:qpp}) analysis, in two high-energy flares.   \cite{Mathioudakis2006} discuss the loop triggering scenario of \cite{Emslie1981Trigger} as an alternative hypothesis for short-duration QPPs in white-light.  \cite{Osten2005} hypothesize that an arcade of heterogeneous gyrosynchrotron-emitting loops may explain the microwave spectral indices in a large dMe flare.
 
 Multi-thread modeling approaches seek to be more representative of a solar flare arcade geometry, following arguments previously made in  stellar \citep{HF92} and solar modeling \citep{HF94}.  However, ad hoc geometrical assumptions and simplifications must be imposed due to the absence of appropriate (i.e., broad spectral coverage, high-time resolution, unsaturated data at comparable wavelengths) constraints from solar data.   Recent analyses of IRIS solar data at high-time resolution, such as the ``new area'' calculations of \cite{Graham2020}, and spatially resolved spectral line analyses \citep{Namekata2022}, may be able to provide clearer guidance to stellar models of sequential footpoint ignition.   Nonetheless, there may be important differences between solar and stellar flares that complicate a direct application of the thermal multi-thread modeling of \cite{Warren2006}:  shorter-during footpoint pulses \citep[e.g.,][]{Mathioudakis2006}, greater importance of the gradual-phase white-light continuum radiation \citep{HP91, Heinzel2018} and particle acceleration \citep{Osten2005}, and larger apparent loop height growth over the flare evolution \citep[$\tau_c \propto l^2$; e.g.,][]{Osten2005, Osten2016} all are examples of properties that distinguish stellar flares.  
 
 Novel stellar flare applications of the two-ribbon solar flare reconnection model of \cite{Kopp1984} are described in \cite{Poletto1988} and \citealt{Gudel1999} \citep[see also][]{Haisch1983}.  \cite{Gudel1999} model the $\Delta t \approx 8000$~s rising phase of a flare on the RS CVn UX Ari in this framework.  This flare exhibited   a remarkable peak X-ray luminosity of $>10^{32}$ erg s$^{-1}$, and they infer the evolution of the reconnection heights and electron densities (which evolve through $\approx 10^{11}$ cm$^{-3}$ at peak).  A long-duration decay phase is predicted by the model, and the paper discusses how the lack of nonthermal particles in the early rise phase is a  potential limitation of the original  assumptions.
 
 \subsection{Stellar Core-Halo } \label{sec:fekalpha}
\begin{figure*}
\begin{center}
  \includegraphics[width=0.5\textwidth]{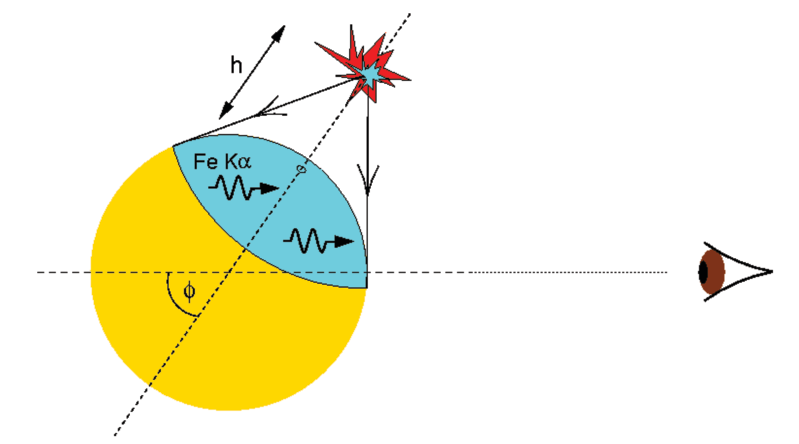} 
  \includegraphics[width=0.4\textwidth]{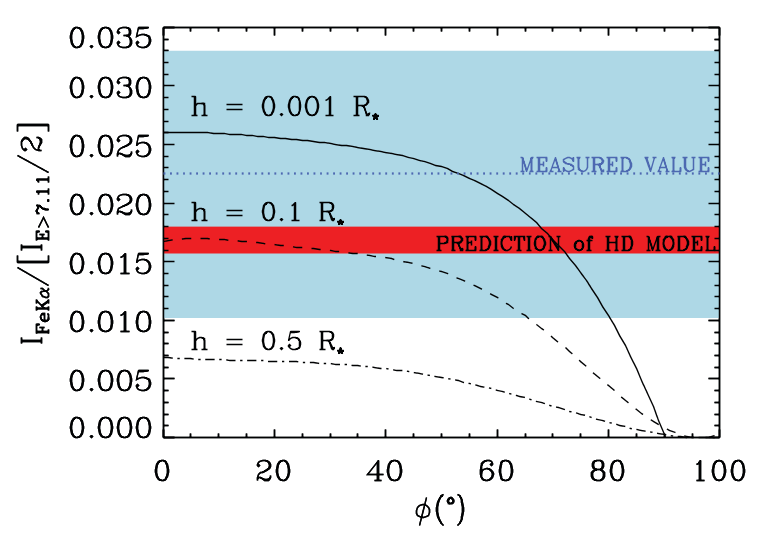} 
\caption{ Geometrical inference from spectral analysis of the 6.4 keV (1.94 \AA ) Fe K$\alpha$ emission line during a flare on the giant star HR 9024, reproduced from \cite{Testa2008} with permission.  The X-rays from hot flare loops fluoresce the neutral iron in the surrounding photosphere. The modeling leverages hydrodynamic modeling from \cite{Testa2007}.   \label{fig:testa1}   }
     % Give a unique label
\end{center}
\end{figure*}

Multi-dimensional geometrical analysis of flare sources is possible by combining hydrodynamic modeling of loop lengths and high resolution spectral observations at Fe K$\alpha$.  The technique is described in \cite{Testa2008}, who infer an X-ray source height that irradiates a large area of the photosphere, from which the Fe K$\alpha$ is observed at a viewing angle $\phi$.  The geometry is shown in Fig.~\ref{fig:testa1}. The combined constraints from the Palermo-Harvard code and the MOCASSIN 3D Monte Carlo radiation code  \citep{Erc03, Erc08, DE07} are reproduced from their paper on the right of Fig.~\ref{fig:testa1}.  Recently, Werner band H$_{2}$  flare emission has been reported in \cite{France2020} and interpreted as fluorescence due to O VI $\lambda = 1032$ \AA\ irradiation of the temperature minimum region ($T \le 1500$ K).

In 1D radiative-hydrodynamic modeling, XEUV backwarming of the photosphere is approximated using the treatments of \cite{GanFang1990}, \cite{HF92}, \cite{HF94} and \cite{Abbett1999} through exponential integrals and detailed radiative transfer \citep{Allred2015, Whalstrom1994}.  A multi-dimensional extension of UV and optical radiative backwarming is illustrated in a cartoon of Figure 4 of \cite{Fisher2012}, and the geometrical dilution of chromospheric radiative flux impinging on the photosphere is given by their Eq. 25.

The remarkable, state-of-the-art 3D MHD codes that have been recently developed with the primary goal to study quiet Sun magnetism are not ready to address some of the most pressing current problems in stellar flare physics.  Flare models require accurate treatments of non-equilibrium physics and detailed radiative transfer for comparisons to chromospheric observations.  An approximation to nonthermal particle heating can in principle be included as $Q_{\rm{flare}}$ in the energy equation in 3D simulations of small flares \citep{Frogner2020}.  The geometrical grid spacing in 3D MHD models, however, is immobile and too coarse (e.g., see Sect.~\ref{sec:models}) to resolve the gradients and contribution function features for the optically thick spectral lines and continua.   Section 4.6.1 of \cite{Bjorgen2019Flare} discusses the issues facing a 3D MHD solar flare simulation that predict chromospheric emission lines in response to thermally conducted fluxes.  The RADYN code (for example) was originally a ``quiet'' Sun code, but it had implemented the 1D adaptive grid scheme of \cite{Dorfi1987} and was flexible enough to simulate particle beam heating in a non-solar gravity.  These features proved fortuitous in applications to stellar flare models.  The stellar community is in urgent need of collaborative development of  multi-dimensional extensions of Carlsson's and Dorfi's works, such as \cite{Stokl2007}, in order to understand the comprehensive, radiative-hydrodynamic response of stellar atmospheres to the impulsive release of magnetic energy.

\section{Conclusions and Future Outlook} \label{sec:conclusions}

In conclusion, we summarize six, non-exhaustive big-picture questions that are (in our view) at the forefront of  stellar flare research.

\begin{itemize}

\item Stellar flares routinely attain energies and peak luminosities that are factors of $10^2-10^4$ greater than the largest solar flares.  How are the flare energies connected to photospheric magnetic flux densities/geometries,  particle acceleration properties, flaring footpoint areas, and energy transport processes in the atmospheres of other stars

\item Stellar superflares exceed the quiescent bolometric stellar luminosity.  The rate of superflares, however, is not 
characterized with high statistical significance.  What are the rates of events in the superflare regime for solar analogs and other low-mass cool stars as a function of stellar age?

\item  On the Sun, the brightness of chromospheric flaring sources is transient at any single location for (arguably) $\lesssim 10-20$~s.  A comprehensive, self-consistent model (e.g., within frameworks similar to \citealt{Reeves2005} and \citealt{Rempel2023} pertaining to solar flare soft X-rays) has not yet been developed for optical or radio stellar flare light curves that last for tens of minutes to hours. 
 What are the principal physical parameters that drive the observed temporal scales of stellar flare light curves of optically thick radiation in the rise and decay phases?

\item Blue asymmetries in Balmer lines are often interpreted as evidence of mass eruptions from a star.  Are there direct analogues \citep[e.g., as in Mg II;][]{Tei2018} on the Sun, and does chromospheric evaporation in stellar flares contribute to blueshifted spectral features in cool lines?

\item Turbulent mass motions and other non-equilibrium macroscale processes (e.g., UV and X-ray radiative backwarming) in stellar flares remain poorly understood in three spatial dimensions.  If gas-dynamic turbulence is important for explaining the observed spectral line shapes, what processes generate large Reynold's numbers in flare atmospheres and, more specifically, in flare chromospheres?

\item Optical and NUV stellar flare spectra are inconsistent with optically thin hydrogen recombination theory.  Electron-beam and thermally-conducted heating fluxes that are within our expectations of the standard solar flare model paradigm do not explain the observed strength and spectral shape of the continuum radiation at NUV and optical wavelengths in M-dwarf flares.  What is the source of stellar flare white-light continuum radiation, which is prominent in almost all of the \emph{Kepler} (and TESS) broadband flare data?

\end{itemize}

\backmatter

\bmhead{Supplementary information}
Additional data are provided in supplementary online material hosted on Zenodo 
at \url{https://doi.org/10.5281/zenodo.10641273}.

\bmhead{Acknowledgments}
I thank Dr. Yuta Notsu for providing helpful comments and corrections.  I gratefully acknowledge an anonymous referee for valuable comments and suggestions.
I am grateful to Dr. Suzanne Hawley for much guidance and many stimulating discussions about stellar flares over the years.  I thank Dr. Mihalas Mathioudakis and the solar physics group at Queen's University Belfast for many productive discussions. I thank Dr. Rachel Osten for many discussions about stellar flares, radio and X-ray observations, and about energy partitions.  I thank Dr. John Wisniewski for many conversations about near-ultraviolet stellar flares.  I thank Dr. Thomas Gomez for helpful discussions about line broadening theory, Dr. Eugene Oks for discussions about plasma line broadening, Dr. Jim Drake and Dr. Joel Dahlin for discussions about PIC simulations and related plasma physics topics, Dr. Joel Allred for extremely helpful guidance in radiative-hydrodynamics and assistance with flare modeling, Dr. Mats Carlsson with invaluable assistance with flare modeling, Dr. Han Uitenbroek for helpful conversations about stellar atmosphere codes, Dr. Lyndsay Fletcher, Dr. Gianna Cauzzi, Dr. Dana Longcope, Dr. Eduard Kontar, and Dr. Markus Aschwanden for stimulating discussions on solar flare topics, Dr. Gregory Fleishman and Dr. Stephen White for discussions about solar flare radio observations, Dr. Andy Inglis for helpful discussion about quasi-periodic pulsations in solar and stellar flares, Dr. Matthias Rempel and Cole Tamburri for conversations about 3D flare modeling, and Dr. Manuel G{\"u}del for discussions about the Neupert effect and stellar flares on Proxima Centauri.  I acknowledge the International Space Science Institute (ISSI) for my participation in several solar-stellar workshops that were organized and lead by Dr. Lyndsay Fletcher, Dr. Sven Wedemeyer, Dr. Louise Harra, Dr. Graham Kerr, Dr. Vanessa Polito, and Dr. Paola Testa.  Lastly, I thank all of the researchers in this field for developing such an engrossing area of study within astrophysics.

\begin{appendices}

    \section{Optical Filter Reference} \label{sec:filters}
We include several appendices in this Living Review.  This appendix, Appendix \ref{sec:filters}, serves as a quick-reference for optical broadband filter curves.   Appendix \ref{sec:color_color_appendix} shows supplementary spectral constraints from optical dMe flare data, Appendix  \ref{sec:slabs} reviews two continuum modeling approaches with slab geometries (blackbody fitting and optically thin hydrogen emissivity spectra) and the continuum modeling approach with parameterized chromospheric condensations \citep{KA18}, and Appendix \ref{sec:stark} reviews the salient physics (through a mostly qualitative exposition) of the pressure broadening of the spectral lines of hydrogen.

\begin{figure}[ht!]
\begin{center}
  \includegraphics[width=1.0\textwidth]{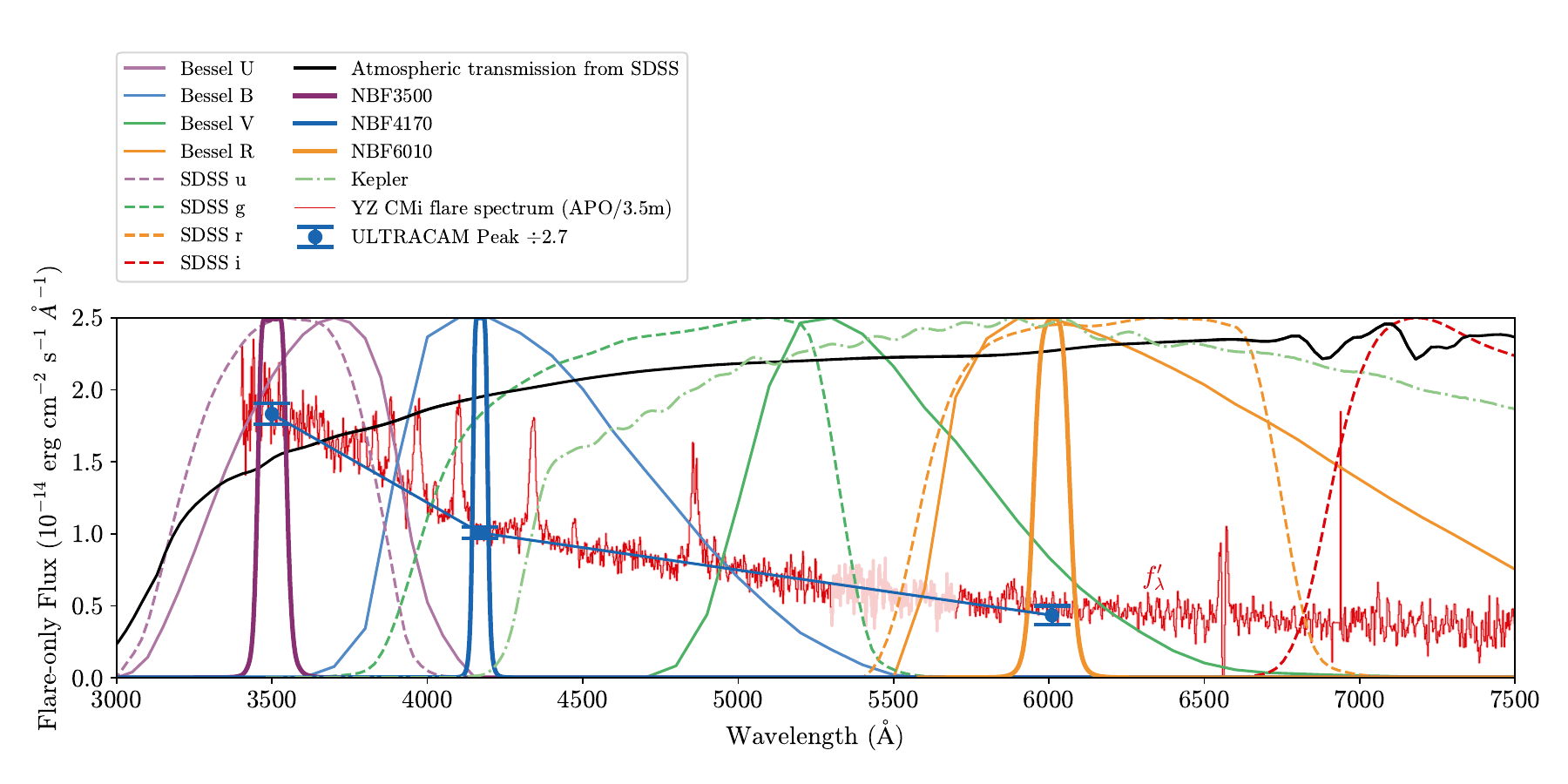}
  \end{center}
\caption{  Quantum efficiency curves of common optical filters, normalized to peak transmission. An example of a terrestrial atmospheric transmission curve from the Sloan Digital Sky Survey is shown for an airmass of 1.0 at the Apache Point Observatory.  One should note that the quantum efficiency of the CCD, telescope primary mirror, and other optical elements are not included. For reference, the impulsive phase flare spectrum from the IF4 event in \cite{Kowalski2016} is shown.  The wavelengths that are affected by the dichroic are in a lighter shade of red. The ULTRACAM photometry (NBF3500, NBF4170, NBF6010) at the peak of this flare were calibrated independently, and they are scaled to account for the factor of ten shorter integration time. The flare data were obtained from the Zenodo repository at:  \url{https://doi.org/10.5281/zenodo.45878}. }
\label{fig:filters}  
\end{figure}

    In Fig.~\ref{fig:filters}, we show the optical broadband and narrowband filters that are commonly used in stellar flare studies.  There are numerous databases that provide these, such as the SVO Filter Profile Service at \url{http://svo2.cab.inta-csic.es/theory/fps/}, and some are published alongside photometry calibration surveys \citep{Bessel2013}.  For comparison, we show the low-resolving power, flare-only spectrum \citep[from Figure 3 of][]{Kowalski2016} over the impulsive phase of the IF event from the bottom panel of Fig.~\ref{fig:lcexamples}.  The spectral shape from the calibrated ULTRACAM narrowband filters at the flare peak time is in agreement with the spectrum.

  \section{A Summary of Supplemental Model Constraints in the Optical} \label{sec:color_color_appendix}

\begin{figure}[htbp!]
\begin{center}
  \includegraphics[width=0.75\textwidth]{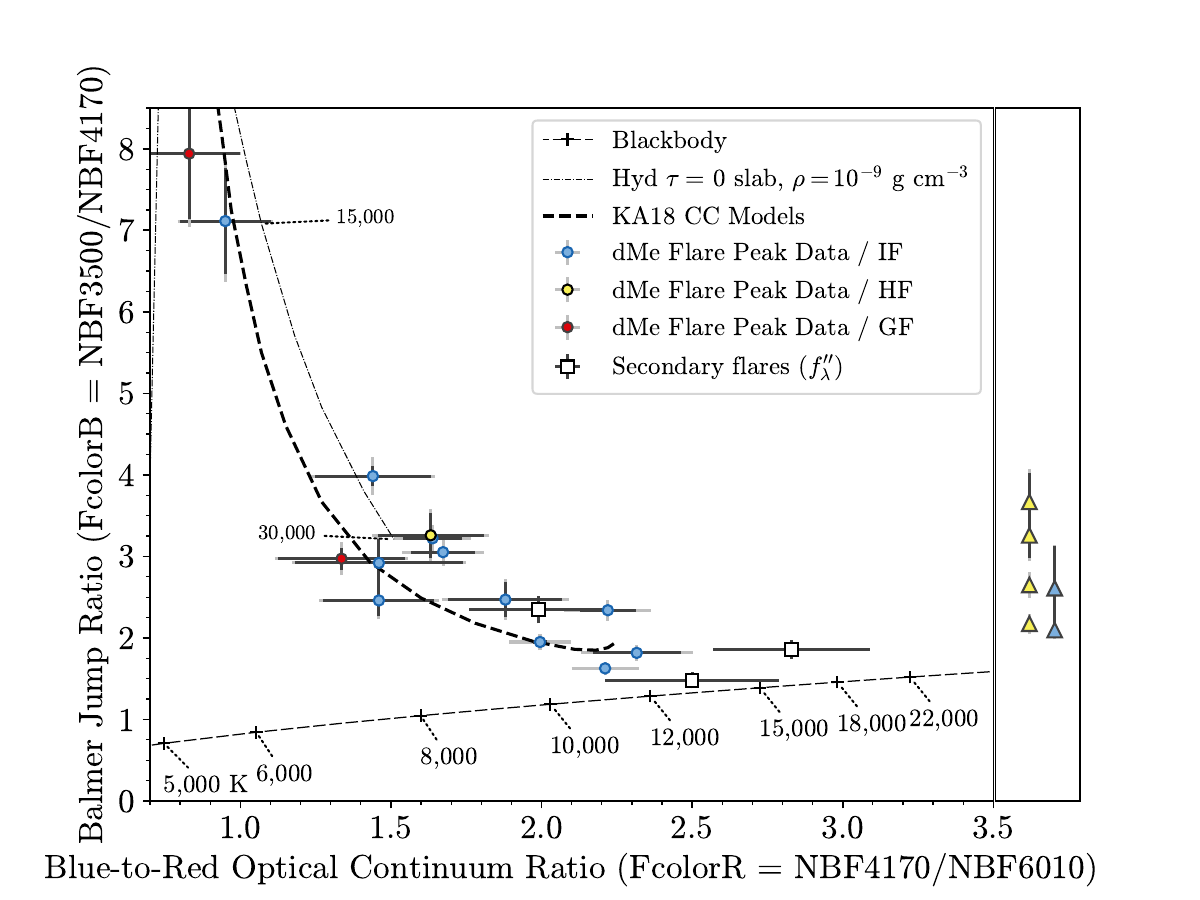}
    \includegraphics[width=0.75\textwidth]{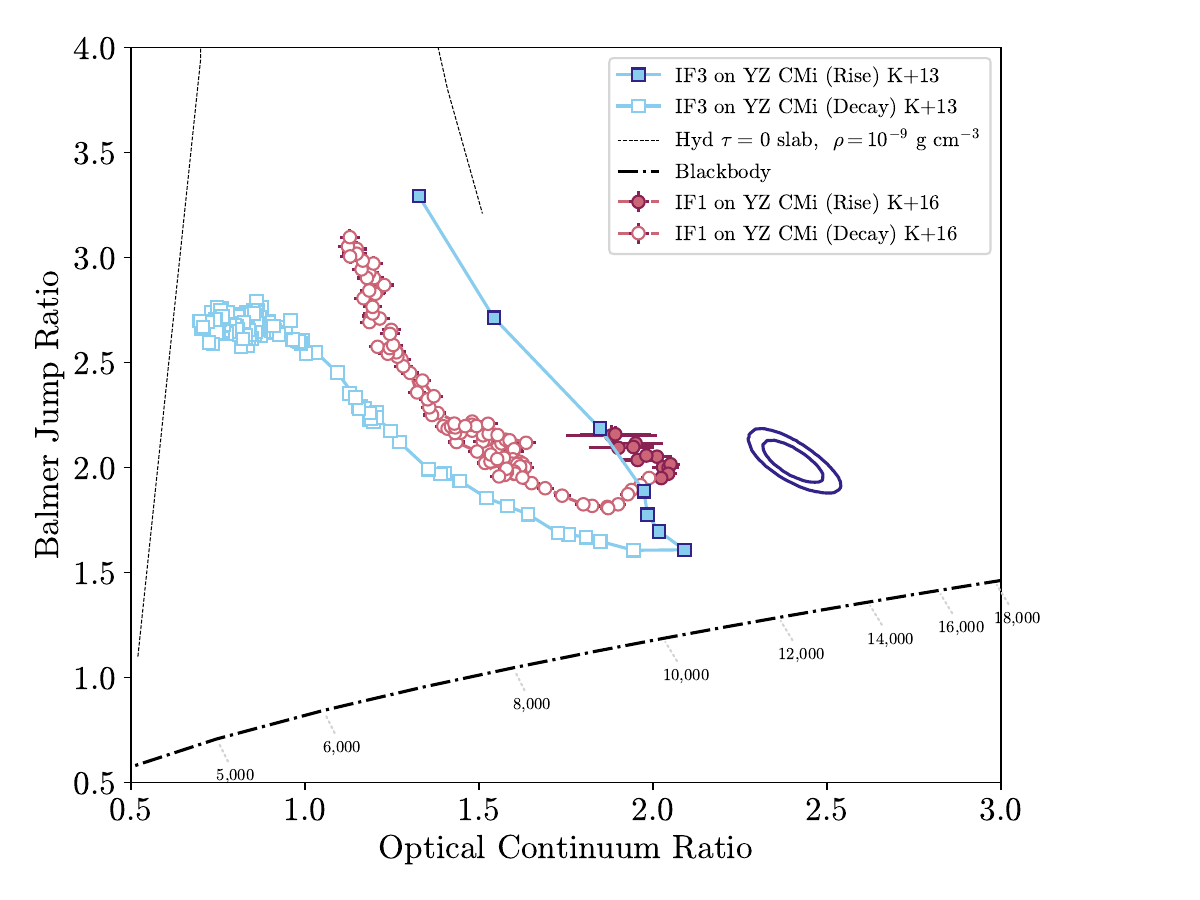}
        \includegraphics[width=0.75\textwidth]{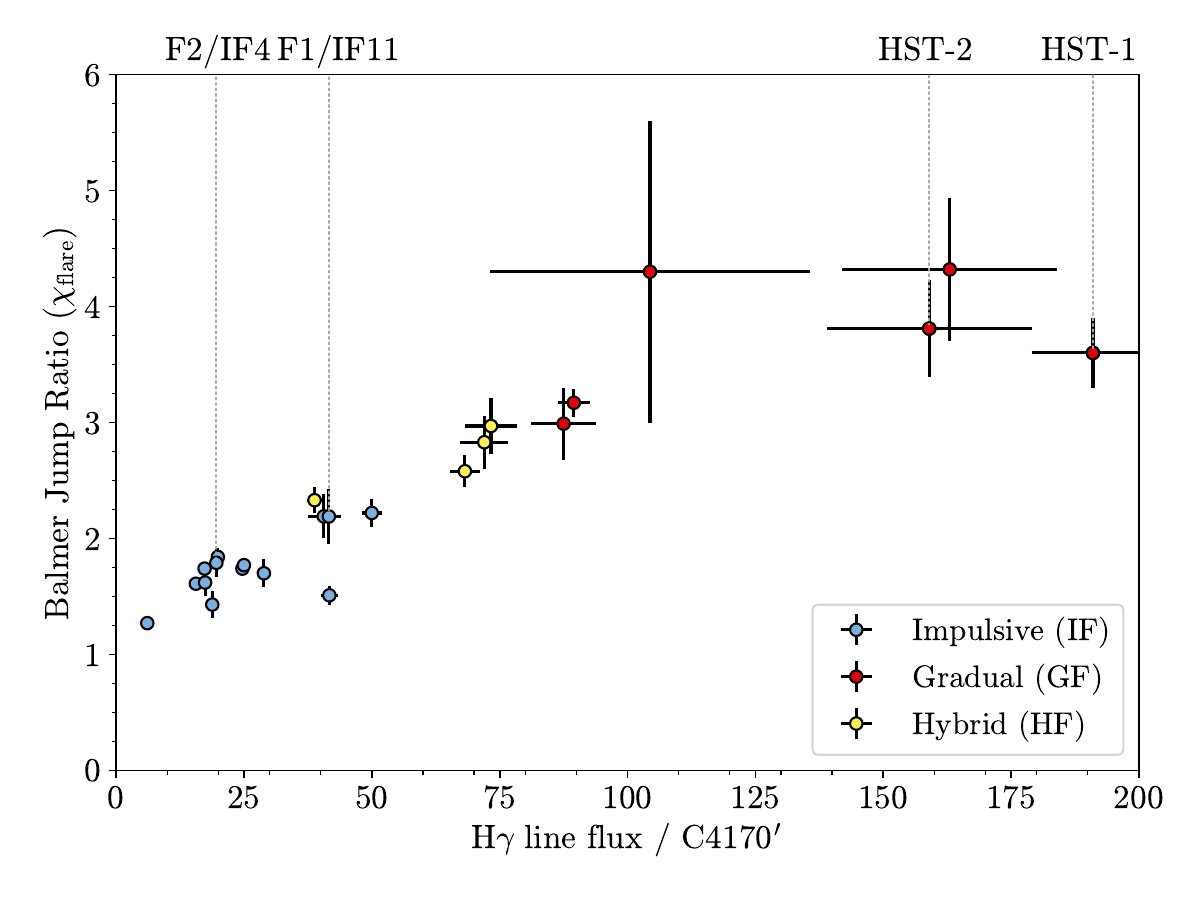}
  \end{center}
\caption{ (\emph{Top}) Flare-peak colors from ULTRACAM narrowband filter ratios;  similar to Fig.~\ref{fig:colorcolor} and see text for details.  (\emph{Middle}) Evolution of flare colors in two large flares on YZ CMi.  Time increases over the tracks in a clockwise direction.  These data have small statistical uncertainties (which are shown) but non-negligible systematic errors.  The uncertainties are correlated, which we show for representative 5\% systematic absolute color calibration uncertainties in both flare colors located at $(x,y) \approx  (2,2)$.  The error-ellipses indicate representative joint probabilities:  the outer is the joint 68\% confidence interval, while the inner translates to the marginal 68\% uncertainties.   (\emph{Bottom}) Flare-only, line-to-continuum flux ratios from the same peak spectra in Fig.~\ref{fig:colorcolor} in the main text.  }
\label{fig:moredata}  
\end{figure}

  In this appendix, we supplement the impulsive phase flare colors from spectra, shown in Fig.~\ref{fig:colorcolor}, with the peak colors calculated from narrow-band continuum filters.  Fig.~\ref{fig:moredata}(top) is a summary of the peak-phase ULTRACAM filter (Appendix \ref{sec:filters}) ratios from \cite{Kowalski2016}.  These flare colors were calculated at higher time resolution than from the spectra that were discussed in the main text (Sect.~\ref{sec:wlcontinuum}). Figure~\ref{fig:moredata}(middle) shows the evolution of the ULTRACAM filter ratios in the large IF1 event described in row (1a) of Table \ref{table:magchanges};  see also \citet{Kowalski2023}.  Average fluxes in the ULTRACAM narrowband filters are calculated from the spectra of the energetic IF3 event (also on YZ CMi) from \cite{Kowalski2013} and are shown for comparison in Fig.~\ref{fig:moredata}(middle).  Note, the peak flare spectrum for this event is showcased in Fig.~\ref{fig:yzcmi_peak}.     In the bottom panel, we show Balmer jump ratios against the ratios of the H$\gamma$ line flux to the flare-only continuum flux around $\lambda = 4170$ \AA\ at the time corresponding to the optical continuum peak flux. (The H$\gamma$ line flux is continuum-subtracted, integrated over the wavelengths of the line, and pre-flare subtracted; C4170$^{\prime}$ is the average the flare-only flux over $\lambda = 4155-4185$ \AA).  Fig.~\ref{fig:moredata}(bottom) combines the impulsive-phase spectral measurements from \cite{Kowalski2013} with the same quantities obtained in the four flares (indicated on the top axis) with low-resolving-power, optical spectra that were analyzed in \cite{Kowalski2016} and \cite{Kowalski2019HST}.  It is our hope that as more flares are observed, these figures will be updated to provide compelling and comprehensive constraints for future multi-dimensional flare models.  

\clearpage

\section{Slab Model Reference} \label{sec:slabs}

\subsection{Blackbody Fitting to Stellar Flare Optical Data}\label{sec:Tbb}

In this appendix, we summarize the assumptions and techniques employed by fitting isothermal blackbody functions to spectra, broadband photometry, and narrowband photometry of flares.  A blackbody ``color'' temperature refers to the  temperature of a blackbody that is fit to a ratio of two fluxes (e.g., a flare color calculated from a ratio of two flare-only continuum fluxes; Sect.~\ref{sec:peakphase}), where color is equivalently a difference of magnitudes in two bandpasses.  The term also refers to the temperature of a blackbody that is fit to the absolute value of more than two continuum flux values (outside of prominent emission lines) over a limited spectral range ($\Delta \lambda \approx 1000-2000$ \AA).  Generally, a color temperature contains rather limited information about the actual nature of the radiation, such as whether it is optically thick, thin, an intermediate optical thickness, or which atomic processes contribute to the emergent radiative flux.  A color temperature does not unambiguously constrain how the spectrum of continuum radiation extends to wavelengths outside of the range over which it is calculated.  Within the total wavelength span that a flare color is calculated (e.g., the ratio of flare fluxes in narrowband filters around 4170 \AA\ and 6010 \AA), there may be emission lines that affect the calculated value, and the actual spectrum between two wavelengths may be best represented by a more complex continuum model.  Therefore, flare color temperatures are \emph{equivalent} color temperatures of blackbody functions.

Other measures of temperature include gas temperature, radiation (or brightness) temperature, and effective temperature (in addition to the color temperatures of optically thin recombination and bremsstrahlung spectra).  None of these are possible to robustly measure directly from optical stellar flare observations and must be inferred from detailed modeling. In principle, flare effective temperatures could be measured, but there has not yet been a more complete broadband spectral observation of a flare than analyzed in \citealt{HP91}.  To demonstrate the ambiguities in these measures of temperature, it is interesting to note that for an optically thin hydrogen recombination and free-free spectrum calculated from LTE level populations for a slab with a  gas temperature of $T_{\rm{gas}}= 10,000$ K (Appendix \ref{sec:jlam}, Fig.~\ref{fig:kunkel}),  the radiation temperature at $\lambda = 4170$ \AA\ is $T_{\rm{rad}} = 3600$ K.  The emergent intensities are calculated  for a slab thickness of $dl = 0.1$ km to ensure that $\tau(\lambda) < 0.005$ across the optical regime.   The color temperature measured from the ratio of intensities at $\lambda = 4170$ \AA\ to 6010 \AA\ is $T_{\rm{FcolorR}}=5300$ K.  Thus, $T_{\rm{gas}} > T_{\rm{FcolorR}} > T_{\rm{rad}}$ in this example.

\begin{figure}
\begin{center}
  \includegraphics[width=0.8\textwidth]{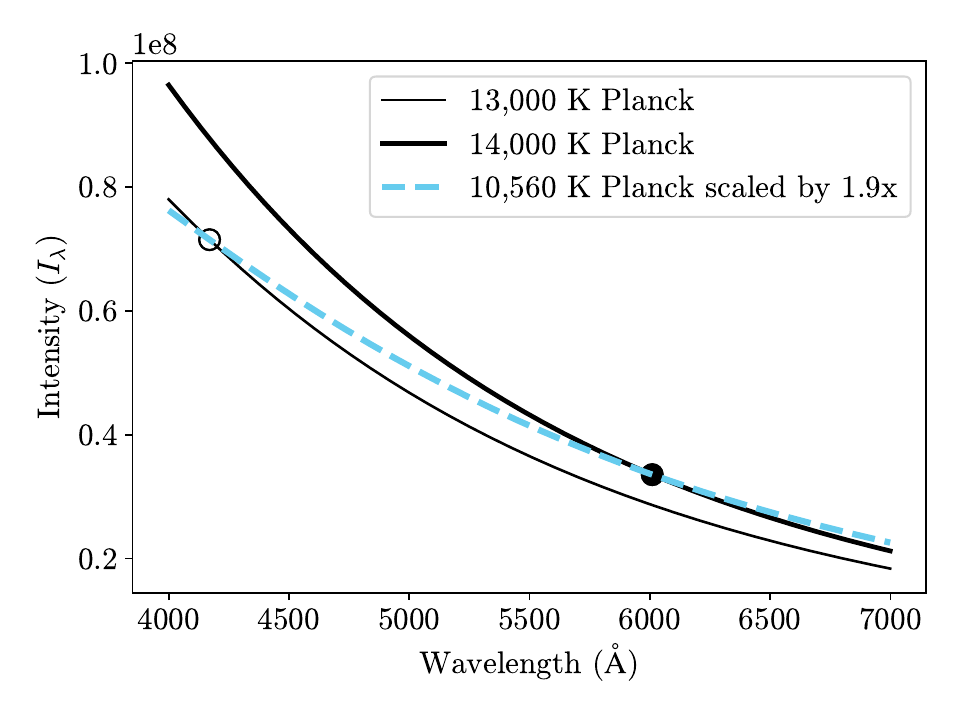}
  \end{center}
\caption{Demonstration showing the relationship between a color temperature and wavelength-dependent radiation (brightness) temperatures at $\lambda = 4170$ \AA\ and $\lambda = 6010$ \AA.  }
\label{fig:bbdemo}  
\end{figure}

Another interesting ambiguity in inferred gas temperatures occurs when the emergent continuum flux is formed over significant ($\tau \approx 1$) optical depth in dynamic, non-LTE, non-isothermal flare atmospheres.  In Fig.~\ref{fig:bbdemo}, we summarize the argument  in \cite{Kowalski2023}, which is also closely related to the analyses in the appendices of \cite{Kowalski2017Broadening}.  Two Planck functions with $T = 13,000$ K and $T = 14,000$ K are shown.  If wavelength-dependent, continuum optical depths are large over an inhomogeneous temperature structure, then one could possibly, for example, measure radiation temperatures of $T_{\rm{rad}} = 13,000$ K at $\lambda = 4170$ \AA\ and $T_{\rm{rad}} = 14,000$ K at $\lambda = 6010$ \AA.  The ratio of the corresponding emergent intensities gives a color temperature that is only $T_{\rm{FcolorR}} = 10,500$ K.  Thus, $T_{\rm{rad}} > T_{\rm{FcolorR}}$ in this example.  For semi-infinite atmospheres that exhibit a linear dependence between the optical depth and the (LTE or non-LTE) source function, the Eddington-Barbier approximation relates the radiation temperature to the gas temperature at the depth corresponding to $\tau(\lambda)=1$;  however, dynamic flare atmosphere calculations are usually much more complicated over the large physical depth ranges of continuum formation (e.g., in chromospheric condensations and the stationary layers below; \emph{cf.} Fig.~\ref{fig:standardcartoon}(right) and Fig.~\ref{fig:CC_F13}), which results in differences between the radiation temperature and the gas temperatures over which the emergent continuum intensity forms \citep[see][for details]{Kowalski2023}.

The temperatures and filling factors calculated from optical and NUV flare data are in general obtained through two methods:

\begin{enumerate} 

\item From a ratio of flare-only fluxes in two bandpasses, one can solve for the temperature and filling factor ($T_{\rm{flare}}$, $X_{\rm{flare}}$) of a Planck function following \citealt{Hawley1995} (i.e., by equating ratios of the RHS of Eq.~\ref{eq:bbx} in Sect.~\ref{sec:geom} to the observed flare-only ratio).  The ratio equation is a transcendental equation for $T_{\rm{flare}}$ and can be solved numerically using linearization and iteration (i.e., Newton-Raphson) or through a simple lookup table. The result for $T_{\rm{flare}}$ is used to infer $X_{\rm{flare}}$.

  As discussed in \cite{Kleint2016} and \cite{Castellanos2020}, by not accounting for the fraction of stellar flux ($X_{\rm{flare}}$) that no longer shines through with the quiescent brightness, such as in a blackbody flare \citep[see][]{Hawley1995} or a flare with a transient overlying absorbing layer, it is possible to calculate a color temperature of $\approx 8500$ K from flare-only irradiance observations of solar flares, as reported in \citet{Kretzschmar2011}.    In Fig.~\ref{fig:bbcalcs}, we demonstrate what happens to the calculated color temperature of an optically thick ($\tau \gg 1$) flare source if one does not use the proper equation in \cite{Hawley1995}.  For example, 
a 5800 K Planck function subtracted from a 7800 K Planck function at $\lambda_1=4170$ \AA\ and $\lambda_2 = 6010$ \AA\ gives a blackbody color temperature of 9000 K (\emph{c.f.}, where  the 9000 K solid line and the 7500 K dashed line approximately intersect).    
However, for certain conditions, $S_{\lambda, \rm{flare}} \gg S_{\lambda, q}  $ where $S_{\lambda,\rm{flare}}$ is the model flare surface flux spectrum, one may ignore the background subtraction errors (in which case, the problem must still be solved numerically or through a simple lookup table).  Using the actual quiet Sun photospheric intensity, which is larger than a Planck function at these wavelengths, a 7500 K Planck function is sufficient to produce a 9000 K blackbody shape in the subtraction residuals.  The surface fluxes of M dwarfs are much fainter than on G dwarfs.   M-dwarf surface fluxes are lower than blackbody functions given by their effective temperatures, and the errors indicated in Fig.~\ref{fig:bbcalcs} are consequently smaller.  For optically thick emission lines, similar issues arise if the model surface fluxes are comparable to the pre-flare surface fluxes (thus, it is fortuitous that large electron beam flux models produce spectra that are extremely bright compared to the pre-flare surface flux spectrum!).

\begin{figure}
\begin{center}
  \includegraphics[width=0.8\textwidth]{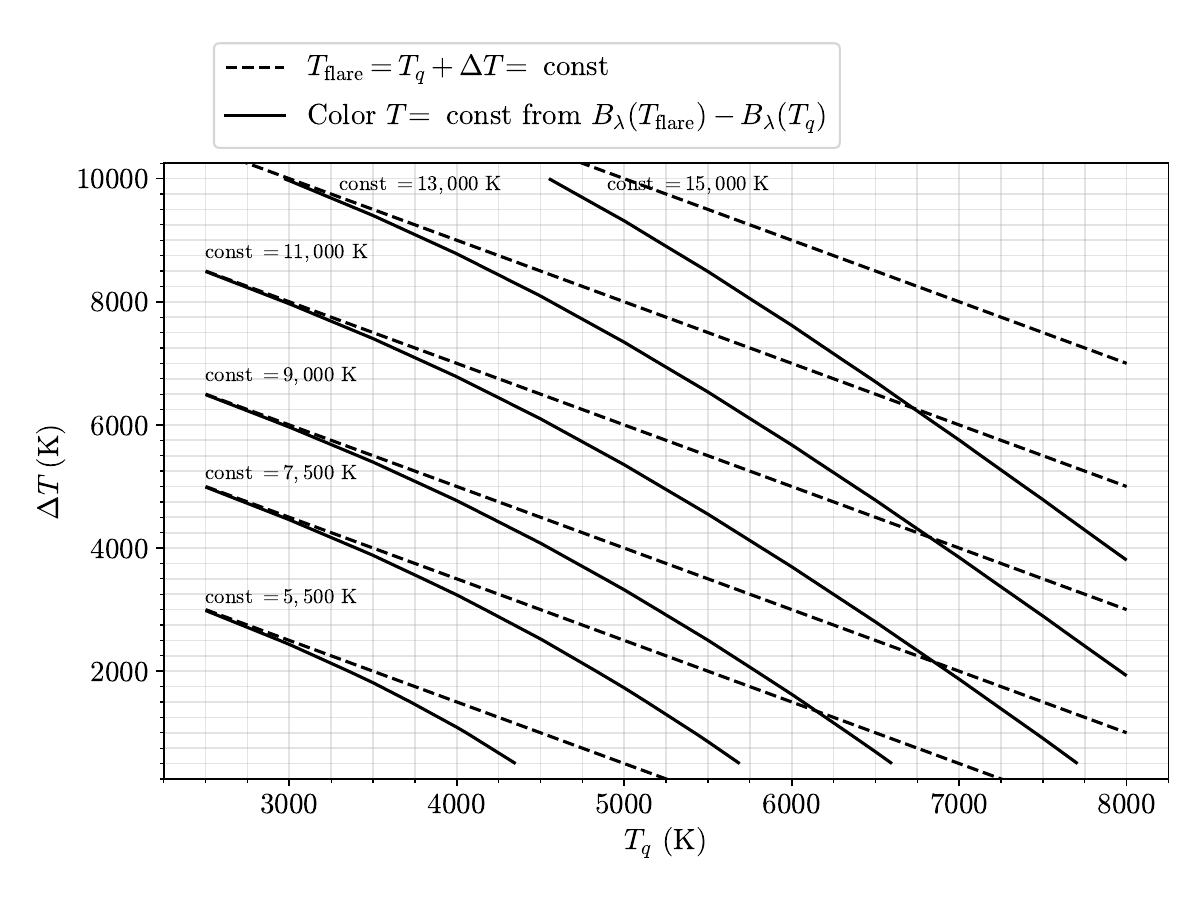}
  \end{center}
\caption{Lines of constant blackbody flare temperatures ($T_{\rm{flare}}$) and lines of constant blackbody color temperatures that are calculated after subtracting a quiescent blackbody surface flux spectrum, $S_{\lambda,q} = B_{\lambda}(T_q)$, from the flare model spectrum, $S_{\lambda, \mathrm{flare}} = B_{\lambda}(T_{\rm{flare}})$.   The color temperature calculations, here shown for $\lambda_1=4170$ \AA\ and $\lambda_2=6010$ \AA , from stellar flare data usually account for or consider these issues in M dwarf flares \citep{Hawley1995, Kowalski2016}, but hotter flare color temperatures in higher-mass stars could be a result of  ``oversubtraction'' effects.  }
\label{fig:bbcalcs}  
\end{figure}

\item  Second, a color temperature can be fit to a spectrum of observed fluxes.  A Planck function is typically fit to the excess (flare-only)  spectrum, or to flare-only multi-band photometry data \citep{Hawley2003}, with $n\lambda$ points ($n > 2$) and Gaussian uncertainties, thus deriving the maximum likelihood parameters $\hat{T}_{\rm{flare}}, \hat{X}_{\rm{flare}}$.  This can be accomplished with non-linear least squares through linearization and iteration for unconstrained parameter fits and with the Levenberg-Marquardt algorithm (e.g., IDL's \emph{mpcurvefit}) for constrained parameter fits.  The color temperatures (e.g., $T_{\rm{BB}}$) that were quoted from blackbody fitting in \cite{Kowalski2013} were obtained by fitting Planck functions directly to the observed flare-only spectral fluxes (i.e., assuming $B_{\lambda}(T_{\rm{flare}}) \gg S_{\lambda, q}$).  Also in that paper, the excess RHD model flux spectra were similarly compared to a Planck function temperature without subtracting a pre-flare from the Planck function.  In this approach, the color temperature is just a convenient parametrization for comparing models to flare-only spectral observations, since temperatures are easier to compare than color indices in different filters or at different wavelengths (as noted in the main text).   Whatever one does to the data, one should also do analogously to the model, whether the model is a blackbody or an RHD spectrum, irrespective of spectral type of the star (from RHD models of M dwarf flares that are optically thin in the chromospheric continuum radiation, it is actually desirable to isolate the chromospheric flare radiation by subtracting the pre-flare surface flux because the model photospheric spectrum, which does not have proper molecular opacities, otherwise shines through and obfuscates comparisons to observations).

\end{enumerate}

  The main purpose of subtracting a pre-flare spectrum from the flare observations is to remove the relatively large, wavelength-dependent flux from the area on the star that is not flaring so that the flare emission is more accurately characterized.  A method for determining an accurate subtraction of the preflare spectrum based on molecular bandpass veiling by the flare continuum is described in \cite{Kowalski2013}.  Alternatively, a very wide slit may be employed, but spectral purity and wavelength resolution are sacrificed;  see \cite{Kowalski2016}. 

In summary, the first method uses two filters and assumes that the flare spectrum is inherently blackbody radiation emitting at a single temperature $T_{\rm{flare}}$, and the second method is used for calculating color temperatures ($T_{\rm{BB}}, T_{\rm{FcolorR}}$) directly from a spectrum of flare-only fluxes assuming that the effects in Figure \ref{fig:bbdemo} are not important (although this assumption can readily be relaxed in the second method).

\subsection{Hydrogen Continuum Emissivity Models} \label{sec:jlam}

\begin{figure}
\begin{center}
  \includegraphics[width=0.8\textwidth]{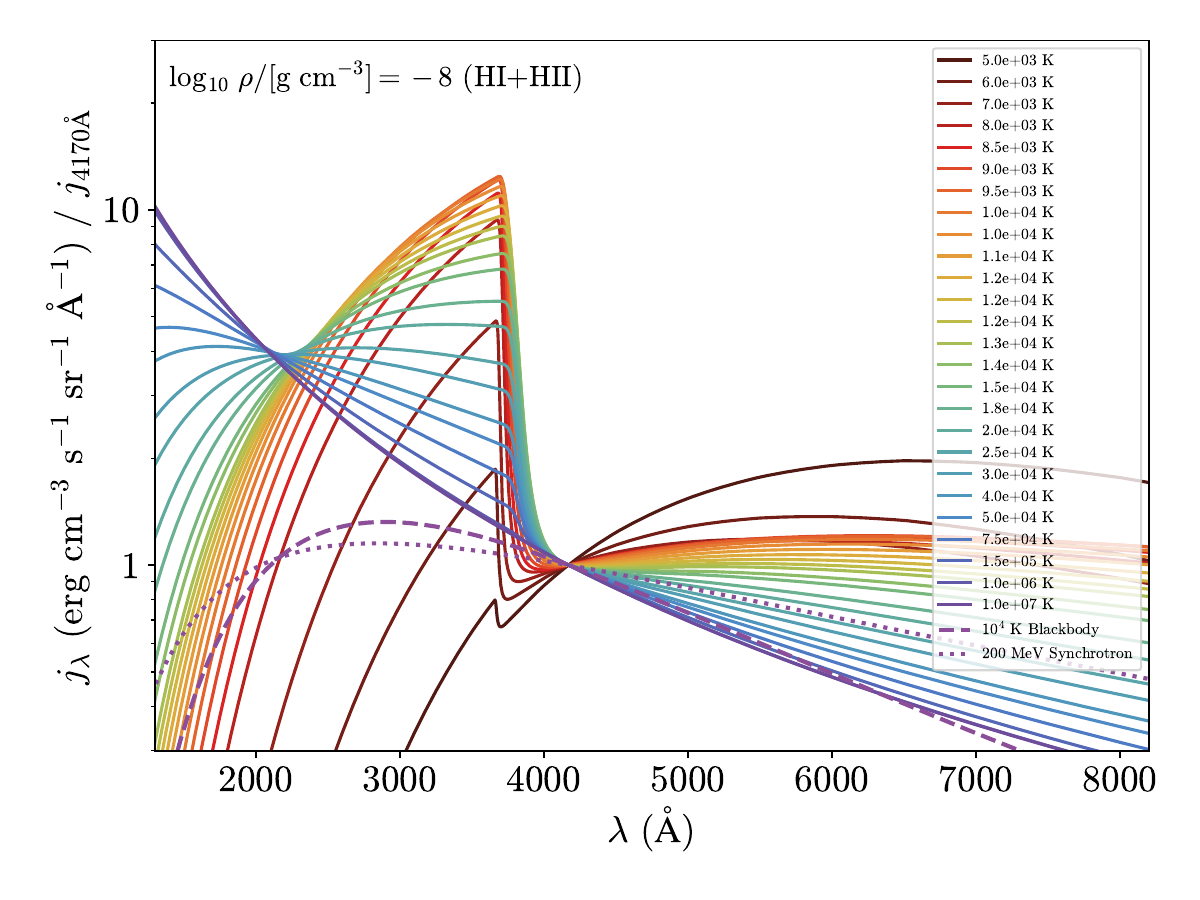}
\caption{Hydrogen LTE emissivity calculations with bound-free (recombination) emissivity from H I and H$^-$ and free-free (thermal bremsstrahlung) emissivity from H II.  The spectral shapes here are identical to the shapes of intensity spectra, $I_{\lambda}$, that emerge from optically thin, homogeneous slabs.  A blackbody function and a mono-energetic synchrotron spectrum is shown for comparison.  The smooth transitions from the Balmer jump to the optical wavelengths are caused by the ``dissolved level Balmer continuum'' \citep{Dappen1987}, which extends to redder wavelengths for larger electron densities. }
\label{fig:kunkel}  
\end{center}
\end{figure}

\cite{Kunkel1970} calculated a grid of optically thin, LTE hydrogen recombination models with free-free contributions.  These are often used as a starting place for comparisons to flare observations (e.g., Fig.~\ref{fig:colorcolor}).  The equations are not included in \cite{Kunkel1970} but were restated in \cite{Kowalski2012}.  Here, we update the calculations with free-free Gaunt factors from CHIANTI and include the dissolved-level Balmer continuum components (longward of the Balmer edge at $\lambda = 3646$ \AA) in the spectra.  These are shown in Fig.~\ref{fig:kunkel} for a range of temperatures and a gas density of $\rho = 10^{-8}$ g cm$^{-3}$.  The Balmer jump ratios are calculated from these spectra and are shown in Fig.~\ref{fig:bj_thin}.  All spectra are available as FITS tables for a range of gas densities in the supplemental online material hosted on Zenodo.

\begin{figure}
  \begin{center}
  \includegraphics[width=0.4\textwidth]{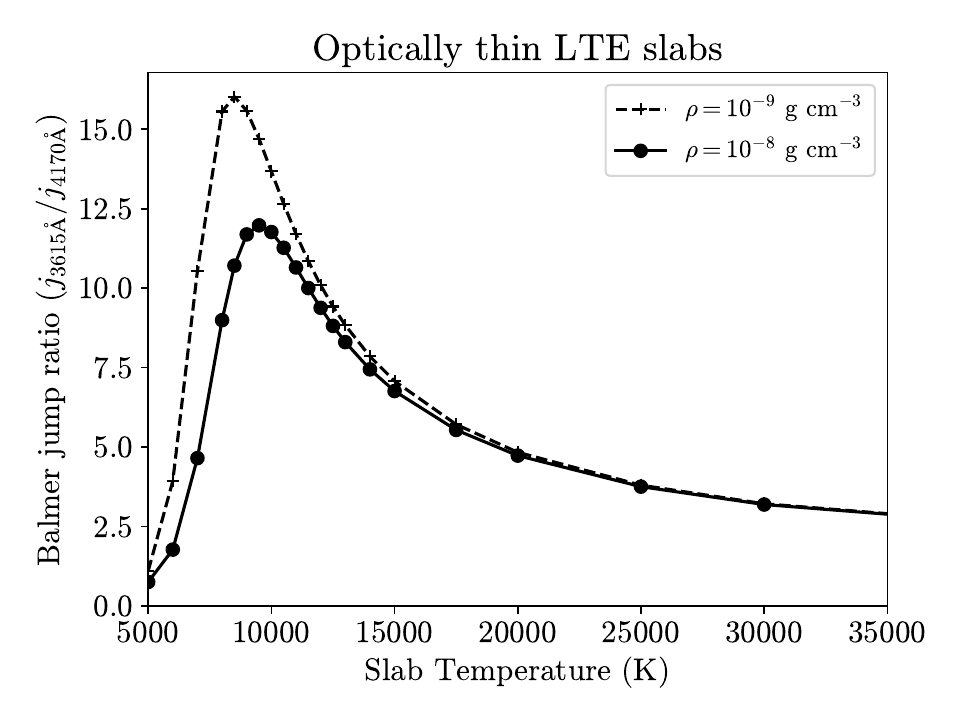}
\caption{Balmer jump ratios calculated from the LTE emissivity models in Fig.~\ref{fig:kunkel} and from the same calculations of the LTE emissivity in a hydrogen gas with a density of $\rho = 10^{-9}$ g cm$^{-3}$.}
\label{fig:bj_thin}
\end{center}
\end{figure}

\begin{figure}
\begin{center}
  \includegraphics[width=0.45\textwidth]{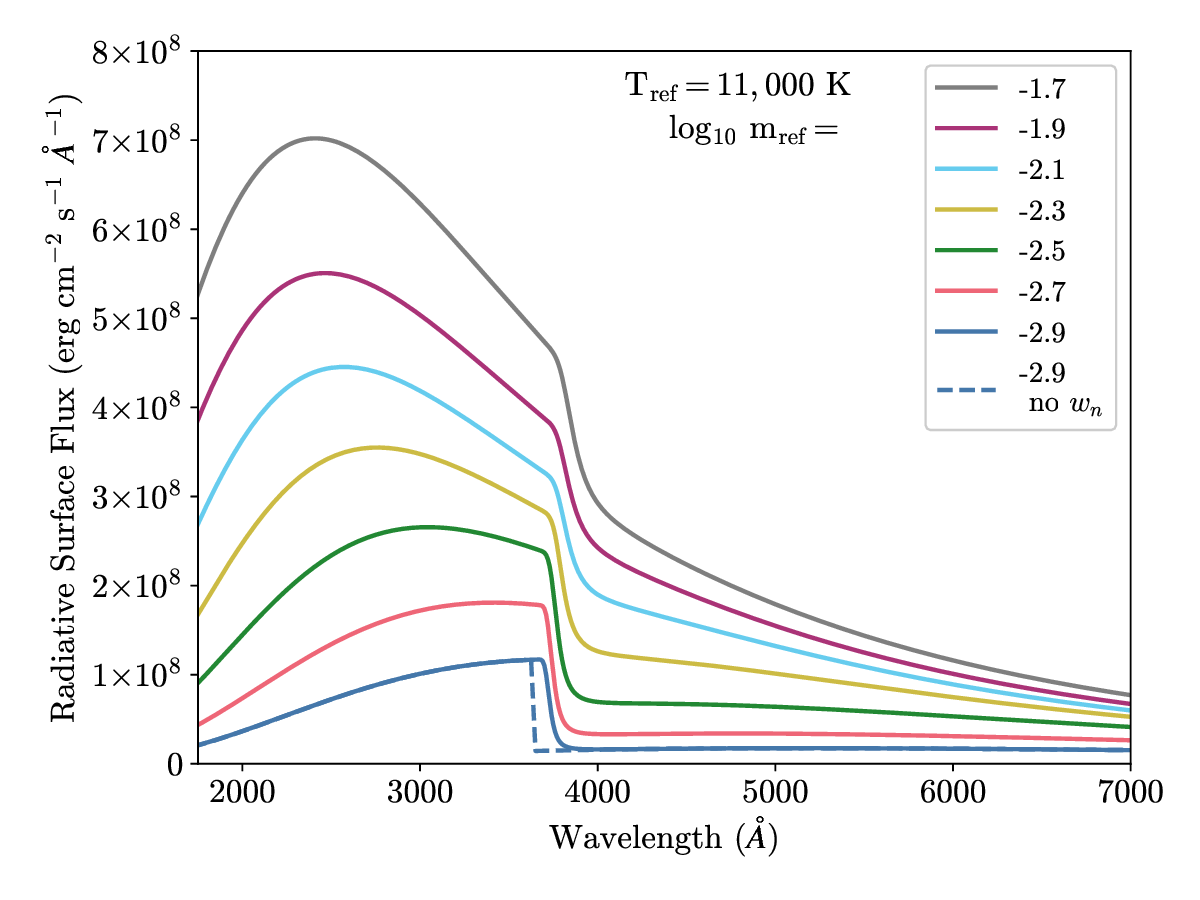}
  \includegraphics[width=0.45\textwidth]{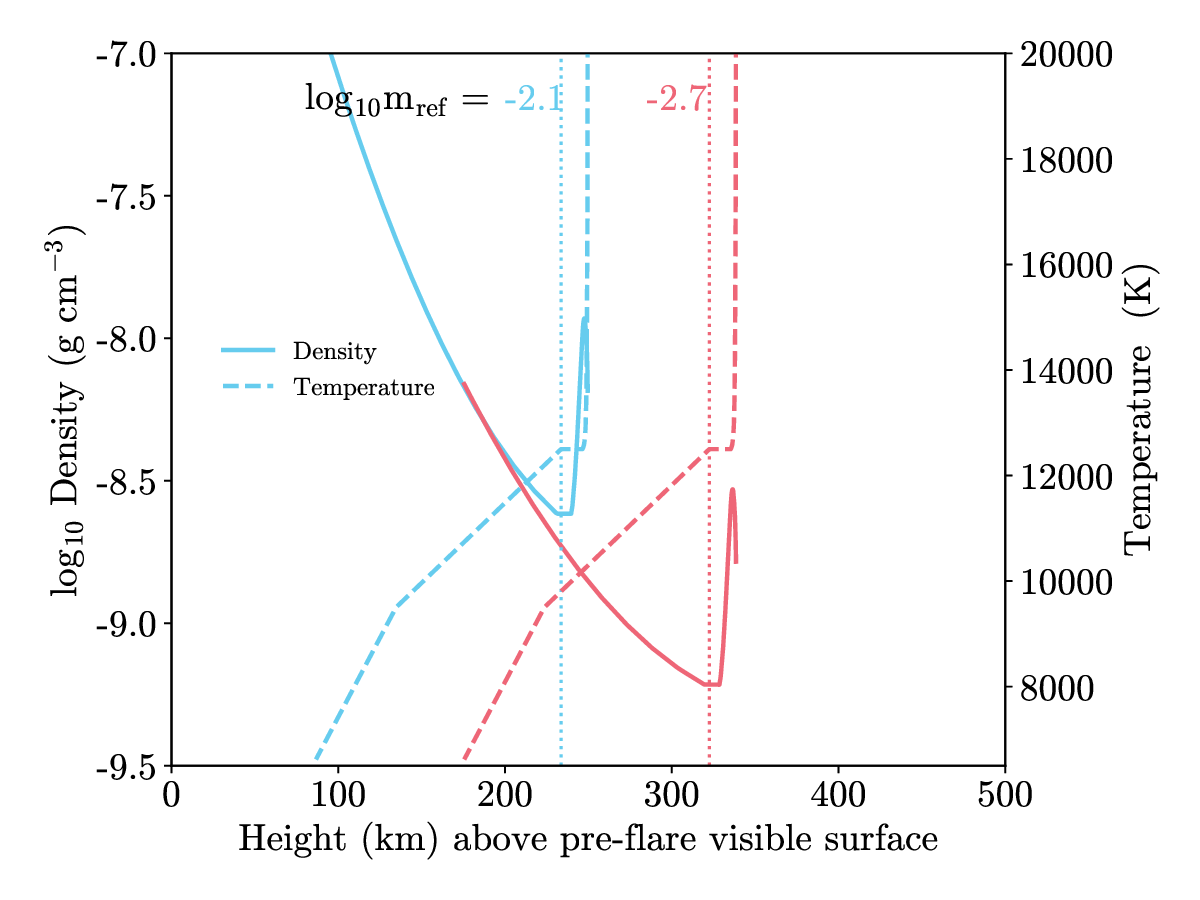}

  \end{center}
\caption{(Left) LTE continuum flux spectra calculated from the parameterized evolution of chromospheric condensations from RHD models.  (Right) Examples of two of the parameterized atmospheres.}
\label{fig:ccdemo}  
\end{figure}

At constant slab temperature, a large range of flare colors result from adjusting the continuum optical depths \citep{Morchenko2015}.  
In Fig.~\ref{fig:ccdemo}, we show the LTE continuum spectra from RHD model parameterizations \citep{KA18}; the colors of these spectra are calculated and are shown in Fig.~\ref{fig:colorcolor} as the ``KA18 CC Models''.  A range of reference column masses ($m_{\rm{ref}}$) were chosen to increase the continuum optical depths within the model chromospheric condensations.  Occupational probabilities ($w_n$) for the dissolved level Balmer continuum opacities are included as in Figure \ref{fig:kunkel}.  

\section{Pressure Broadening of Hydrogen Lines in Flares}  \label{sec:stark}

Electric fields split the degenerate energy levels of hydrogen atoms, Rydberg atoms, and hydrogen-like ions by an amount directly proportional to the field magnitude \citep[$\propto E^1$; e.g.,][]{Condon1963, BetheSalpeter, Gallagher2006, Goldman2006}, which is commonly known as the linear Stark, or Stark-Lo Surdo\footnote{This effect was discovered nearly simultaneously in experiments conducted by Johannes Stark and Antonino Lo Surdo in 1913 and 1914.  Additional ground-breaking experiments were conducted by Toshio Takamine and Noboru Kokubu.  The classical splitting of energy levels  by electric fields was predicted by Woldemar Voigt to be very small, and the large splitting in hydrogen that was subsequently observed by Stark and Lo Surdo was  one of the first major achievements of quantum theory, thanks to pioneering theoretical efforts of Bohr, Sommerfield, Epstein, K. Schwarzschild, Schrodinger, and Kramers. The relevant history is thoroughly summarized in \cite{Leone2004} and \cite{Longair2013}.  See \citet{Kleppner1981} for an excellent high-level overview.} effect.   The splitting increases for larger $n$, and $n=1$ experiences only second-order energy shifts, which are small.  Within a partially ionized gas, the electric fields from ambient, thermal charges (electrons and protons) within a Debye radius broaden spectral lines.  This explains the broad hydrogen absorption profiles in main sequence A stars and white dwarfs (combined with, of course, large optical depths in their photospheric layers).

  The symmetric, atomic energy-level shifts that are linearly proportional to an external, uniform electric field $\mathcal{E}$ are described by the quantum numbers $nqm$, according to the following absolute value (in \emph{cgs} units):
   
  \begin{equation} \label{eq:starkshifts}
 | \Delta E^{(1)} | =  | \frac{3 a_0 e}{2 Z} n q  \mathcal{E} | 
   \end{equation}
   
   %%\vspace{1mm}
 \noindent In Eq.~\ref{eq:starkshifts}, $n$ is the principal quantum number, $q$ is the difference of two parabolic quantum numbers, $k_1 - k_2$, that satisfy $n = k_1 + k_2 + \mid m \mid + 1$, $m = m_l$ is the magnetic quantum number, $a_0$ is the Bohr radius, $e$ is the electron charge, and $Z$ is the nuclear charge (where $Z=1$ for hydrogen). 
   
\begin{figure*}[ht!]
  \includegraphics[width=1.0\textwidth]{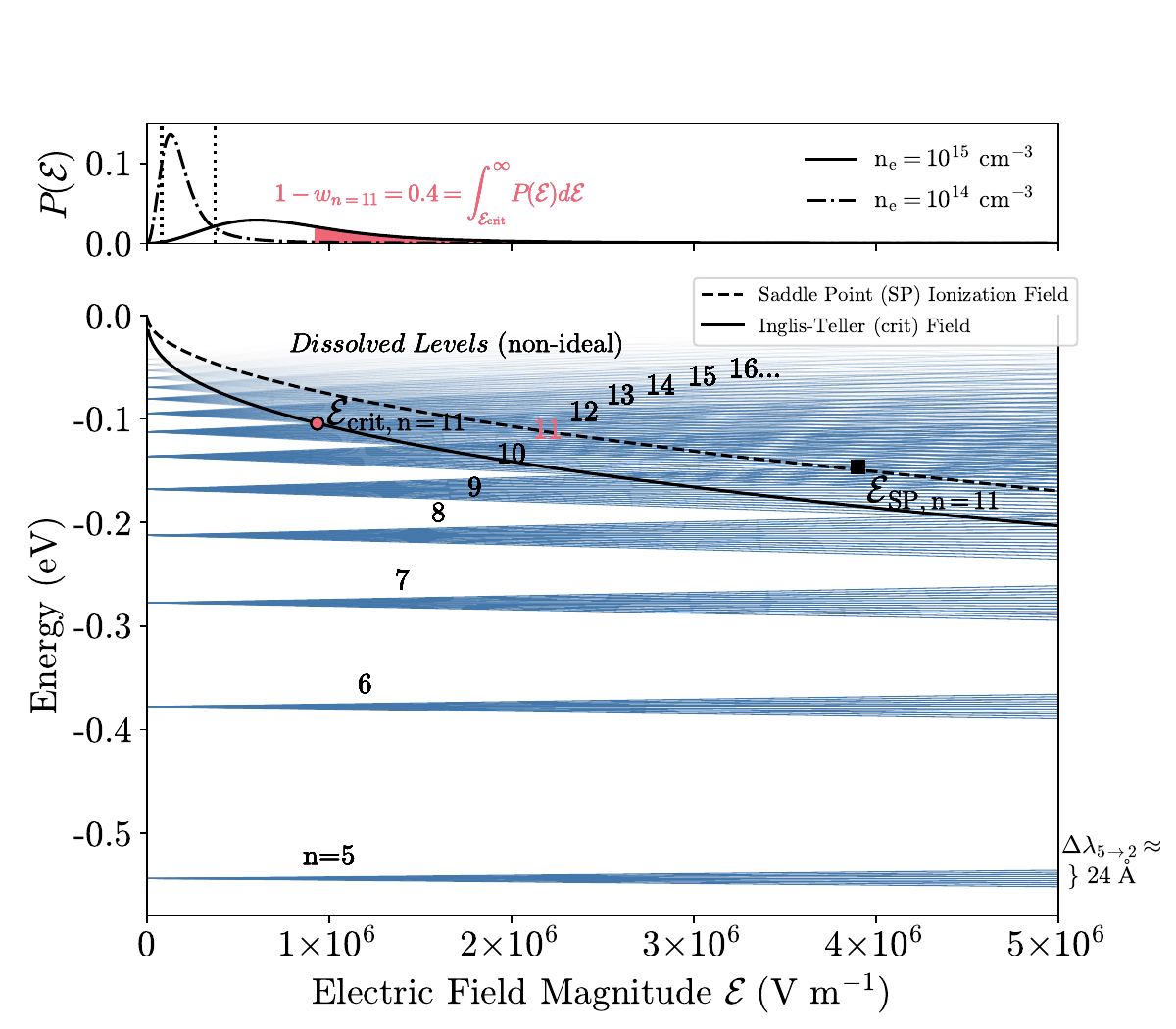}
\caption{Energy level splitting for hydrogen in a static external electric field as a function of the electric field (in SI units).  A principal level $n$ is split into roughly $2n-1$ (including the unshifted) sub-states. For even $n$, there is an unshifted component.  Every split sub-state ($k_1, k_2, m$) is a mix of all the $l$-states for that value of $n$.   The maximum splitting of the $n=5$ level shown here corresponds to  0.016 eV, which would correspond to a maximum wavelength displacement of $\Delta \lambda \approx 24$ \AA\ for the H$\gamma$ transition.  Only levels that can result in a dipole transition with $n=2$ have been plotted. The transitions with $\Delta m = 0$ are known as the $\pi$ components, and $\Delta m \pm 1$ are the $\sigma$ components.  The field strength that forms a ``saddle point'' (potential energy maximum) and destroys all levels $>n$ in hydrogen is shown as a function of $n$ \citep[if there is a saddle-point in the potential, then tunneling is allowed as well;][]{BetheSalpeter}.  The field ionization of hydrogen has been observed (see the photograph from \cite{Traubenberg1930} that is reproduced in \citealt{Montgomery2022}).  The Inglis-Teller relation corresponds to the critical field that causes the ``bluest'' sub-state of $n$ to intersect with the ``reddest'' sub-state of $n+1$, which occurs at a much smaller electric field for a given $n$ than the saddle point estimate (the ratio of the critical field to the saddle-point field is given by the factor known as $K_n$ in the literature). Holtsmark probability distributions of electric field magnitudes are shown in the top panel to illustrate an occupational probability ($w_n$) calculation for $n=11$ and $n_e = 10^{15}$ cm$^{-3}$.   A ``normal field'', $F_0$, is the electric field magnitude at an average interparticle separation; the critical fields for $n_e =10^{14}$ and $10^{15}$ cm$^{-3}$ are indicated by vertical dotted lines in the top panel.
 }
\label{fig:stark_effect}       
\end{figure*}

   Fig.~\ref{fig:stark_effect} shows the splitting of the energy levels of hydrogen in a static electric field.  The energy shifts were calculated using Equation \ref{eq:starkshifts} and the quadratic (second-order) and cubic (third-order) perturbation terms for hydrogen \citep{Condon1963, BetheSalpeter, Springer2006, Goldman2006}.  The second-order and third-order terms create asymmetric splittings but are only important for hydrogen at very high electric field strengths.  The maximum possible wavelength shift for an allowed Balmer transition ($n_{\rm{lower}} = 2$) and $\Delta m \pm1,0$ follows from Eq. \ref{eq:starkshifts} and is proportional to  \citep{vanDien1949}:
  
\begin{equation}
 \big( \frac{n_{j}^2}{n_{j}^2 - 4} \big)^2  ( n_{j} (n_{j} - 1) + 2 ) n_e^{2/3}
\end{equation}

   \noindent where $\big( \frac{n_j^2}{n_j^2 - 4} \big)^2 $ is the appropriate conversion from $dE$ to $d\lambda$, and $n_j$ is the upper level of the Balmer transition. This formula, which goes as $n_j^2$ in the limit of large $n_j$, is sometimes used as an equivalent Doppler width that is added in quadrature with the true Doppler width in a Voigt profile.  However, this approximation, among others \citep{Sutton1978}, predicts inaccurate scalings for the line shapes and widths within a series.  Thus, we recommend that these approximations not be used in place of proper line broadening calculations, which we briefly summarize next.

In a partially ionized gas, the electric microfield amplitude and direction vary from atom to atom and are not constant in time even at one atom because the charge distribution changes.  The microfield amplitude probability is described by a probability distribution function (e.g., in the top panel of Fig.~\ref{fig:stark_effect}).
A brief phenomenological summary of the microphysics of the broadening of hydrogen spectral lines follows.  See \cite{Barklem2016}, \cite{Gigosos2014}, \cite{Gomez2022}, \cite{Hubeny2014} for more comprehensive and rigorous reviews\footnote{My description herein combines elements from classical and quantum theory, while generally following the classical path ``unified theory'' framework of \cite{Vidal1970, Vidal1971, Vidal1973} (VCS73) that was extended to incorporate the \cite{HM88} model of non-ideal effects in the line profile shapes following \cite{Seaton1990} and \cite{Tremblay2009}. Note that some of the important approximations made by VCS73, and other groups at the time, are no longer necessary in newer, numerical simulations of the low-order Balmer lines \citep{Gomez2016, Gomez2017, Cho2022}. }.
The densities of perturbing, ambient (thermal) protons ($n_p$) and electrons ($n_e$) affect the total electric pressure broadening of  radiative (dipole) transitions in hydrogen differently.  The slow-moving protons and ions produce quasi-static \citep[meaning that the microfield changes slowly compared to radiative/orbital timescales; see][]{Griem1974, HM88} electric microfields that split the energy levels of hydrogen into sub-states (Fig.~\ref{fig:stark_effect}), which are superpositions of the various orbital angular momentum states, $l$, within each $n$ (``manifold'').  The microfield magnitude is given by a Hooper probability distribution \citep[which is a modified Holtsmark distribution that accounts for Debye screening and plasma correlations;][]{Nayfonov1999}.  The collisional perturbations from ambient electrons are more complicated.  Classically, the individual elastic electron impacts on each atom are randomly occurring and well-separated in time, thus causing a damping/Lorentzian profile function with broadening FWHM $\Gamma$ that is proportional to $n_e$.  However, on very short timescales, which correspond to large frequency displacements from line center (in the far line wings) in Fourier-conjugate space, the surrounding electrons are nearly still and thus produce a quasi-static (Holtsmark) broadening proportional to $n_e^{2/3}$.
  The transition in electron broadening from its impact to its quasistatic regimes occurs at wavelength shifts, of order $|\lambda - \lambda_{\rm{rest}}| \approx 20$ \AA\ (for the H$\gamma$ line). Thus, large optical depths can make the effects of the quasi-static limit of electrons important in the wings.  Additionally,  lifetime broadening occurs if there are inelastic effects, such as collisional transitions to states of different principle quantum number $n$ \citep[e.g.,][]{HM88, Bransden}, but these are generally small for well-separated levels (and are ignored in the ``no-quenching''  assumption in the VCS73 theory).

The energy-level perturbations from protons and from electron collisions are not independent random variables, and the line profiles thus cannot be modeled as convolutions of two probability distribution functions, such as a Lorentzian convolved with a Holtsmark (or Hooper) distribution.  This is because the electron collisions with hydrogen are transient (short duration) and non-adiabatic\footnote{\emph{cf.} the Adiabatic Theorem which is described in most textbooks on quantum mechanics.} \citep{Smith1969, Sobelman1995, Gigosos2014, Hubeny2014}:  they can cause transitions and redistribution among the microfield-split sub-states of the same $n$.
  Quantum calculations \citep[Eq 10 of][which is reproduced from \citealt{Vidal1970}]{Tremblay2009} that account for the perturbing electrons in both of their limiting regimes  are crucial.  If the dynamical aspects \citep{Stehle1994} of the ions are ignored, the normalized line profile functions are then the ionic microfield averages of the electronic broadening profiles calculated at each value of the quasi-static (QS), ionic microfield $\mathcal{E_{\rm{QS}}}$.  When the spacing between the upper levels $n_j$ becomes small and the broadening causes the states to have a significant probability of overlapping, level dissolution occurs and the bound-bound transitions become narrower \citep{Seaton1990, Tremblay2009} and less Holtsmark-like in the wings.  The extensions of the VCS73 profiles that account for the effects of level dissolution through occupational probabilities (Sect.~\ref{sec:edge}) are known as the TB09$+$HM88 profiles.   \cite{Cho2022} discuss comparisons of high-density analytic profiles to new numerical calculations without level dissolution effects \citep[see also][]{Stehle1993, Gomez2017}.  To our knowledge, the TB09$+$HM88 profiles are yet the only quantum calculations extended to large $n_{j}$ that also include a  framework, albeit phenomenological, for ionization-limit lowering consistently with the bound-bound opacities and line shapes.   This method is adequate for model spectra of the A0 V star Vega \citep[e.g.,][]{Kowalski2017Broadening}, and thus by extension for flare models, given that there is an unfortunate paucity of echelle observations  in this spectral region in stellar  \citep{Fuhrmeister2008, Fuhrmeister2011} and solar \citep{Donati1985} flares.

\end{appendices}


\phantomsection
\addcontentsline{toc}{section}{References}

\bibliography{main_v7.bib}% common bib file

\begin{thebibliography}{802}
\providecommand{\natexlab}[1]{#1}
\providecommand{\url}[1]{{#1}}
\providecommand{\urlprefix}{URL }
\providecommand{\doi}[1]{\url{https://doi.org/#1}}
\providecommand{\eprint}[2][]{\url{#2}}
 \bibcommenthead

\bibitem[{{Abbett}(1998)}]{Abbett1998}
{Abbett} WP (1998) {A Theoretical Investigation of Optical Emission in Solar
  Flares}. PhD thesis, Michigan State University

\bibitem[{{Abbett} and {Hawley}(1999)}]{Abbett1999}
{Abbett} WP, {Hawley} SL (1999) {Dynamic Models of Optical Emission in
  Impulsive Solar Flares}. \apj 521:906--919. \doi{10.1086/307576}

\bibitem[{{Aboudarham} and {Henoux}(1986)}]{AH86}
{Aboudarham} J, {Henoux} JC (1986) {Non-thermal excitation and ionization of
  hydrogen in solar flares. I. Effects on a flaring chromosphere.} \aap
  168(1-2):301--307

\bibitem[{{Adams} et~al(2011){Adams}, {Cai}, {Galli}, {Lizano}, and
  {Shu}}]{Adams2011}
{Adams} FC, {Cai} MJ, {Galli} D, et~al (2011) {Magnetic Interactions in
  Pre-main-sequence Binaries}. \apj 743(2):175.
  \doi{10.1088/0004-637X/743/2/175},
  {\href{https://arxiv.org/abs/1110.4562}{{arXiv:1110.4562}}} {[astro-ph.SR]}

\bibitem[{{Agol} et~al(2021){Agol}, {Dorn}, {Grimm}, {Turbet}, {Ducrot},
  {Delrez}, {Gillon}, {Demory}, {Burdanov}, {Barkaoui}, {Benkhaldoun},
  {Bolmont}, {Burgasser}, {Carey}, {de Wit}, {Fabrycky}, {Foreman-Mackey},
  {Haldemann}, {Hernandez}, {Ingalls}, {Jehin}, {Langford}, {Leconte},
  {Lederer}, {Luger}, {Malhotra}, {Meadows}, {Morris}, {Pozuelos}, {Queloz},
  {Raymond}, {Selsis}, {Sestovic}, {Triaud}, and {Van Grootel}}]{Agol2021}
{Agol} E, {Dorn} C, {Grimm} SL, et~al (2021) {Refining the Transit-timing and
  Photometric Analysis of TRAPPIST-1: Masses, Radii, Densities, Dynamics, and
  Ephemerides}. \planss 2(1):1. \doi{10.3847/PSJ/abd022},
  {\href{https://arxiv.org/abs/2010.01074}{{arXiv:2010.01074}}} {[astro-ph.EP]}

\bibitem[{{Alaoui} and {Holman}(2017)}]{Alaoui2017}
{Alaoui} M, {Holman} GD (2017) {Understanding Breaks in Flare X-Ray Spectra:
  Evaluation of a Cospatial Collisional Return-current Model}. \apj 851(2):78.
  \doi{10.3847/1538-4357/aa98de},
  {\href{https://arxiv.org/abs/1706.03897}{{arXiv:1706.03897}}} {[astro-ph.SR]}

\bibitem[{{Alaoui} et~al(2021){Alaoui}, {Holman}, {Allred}, and
  {Eufrasio}}]{Alaoui2021}
{Alaoui} M, {Holman} GD, {Allred} JC, et~al (2021) {Role of Suprathermal
  Runaway Electrons Returning to the Acceleration Region in Solar Flares}. \apj
  917(2):74. \doi{10.3847/1538-4357/ac0820},
  {\href{https://arxiv.org/abs/2103.13999}{{arXiv:2103.13999}}} {[astro-ph.SR]}

\bibitem[{{Allred} et~al(2005){Allred}, {Hawley}, {Abbett}, and
  {Carlsson}}]{Allred2005}
{Allred} JC, {Hawley} SL, {Abbett} WP, et~al (2005) {Radiative Hydrodynamic
  Models of the Optical and Ultraviolet Emission from Solar Flares}. \apj
  630:573--586. \doi{10.1086/431751},
  {\href{https://arxiv.org/abs/astro-ph/0507335}{{astro-ph/0507335}}}

\bibitem[{{Allred} et~al(2006){Allred}, {Hawley}, {Abbett}, and
  {Carlsson}}]{Allred2006}
{Allred} JC, {Hawley} SL, {Abbett} WP, et~al (2006) {Radiative Hydrodynamic
  Models of Optical and Ultraviolet Emission from M Dwarf Flares}. \apj
  644:484--496. \doi{10.1086/503314},
  {\href{https://arxiv.org/abs/astro-ph/0603195}{{astro-ph/0603195}}}

\bibitem[{{Allred} et~al(2015){Allred}, {Kowalski}, and
  {Carlsson}}]{Allred2015}
{Allred} JC, {Kowalski} AF, {Carlsson} M (2015) {A Unified Computational Model
  for Solar and Stellar Flares}. \apj 809:104.
  \doi{10.1088/0004-637X/809/1/104},
  {\href{https://arxiv.org/abs/1507.04375}{{arXiv:1507.04375}}} {[astro-ph.SR]}

\bibitem[{{Allred} et~al(2020){Allred}, {Alaoui}, {Kowalski}, and
  {Kerr}}]{Allred2020}
{Allred} JC, {Alaoui} M, {Kowalski} AF, et~al (2020) {Modeling the Transport of
  Nonthermal Particles in Flares Using Fokker-Planck Kinetic Theory}. \apj
  902(1):16. \doi{10.3847/1538-4357/abb239},
  {\href{https://arxiv.org/abs/2008.10671}{{arXiv:2008.10671}}} {[astro-ph.SR]}

\bibitem[{{Allred} et~al(2022){Allred}, {Kerr}, and {Gordon
  Emslie}}]{Allred2022}
{Allred} JC, {Kerr} GS, {Gordon Emslie} A (2022) {Solar Flare Heating with
  Turbulent Suppression of Thermal Conduction}. \apj 931(1):60.
  \doi{10.3847/1538-4357/ac69e8},
  {\href{https://arxiv.org/abs/2204.11684}{{arXiv:2204.11684}}} {[astro-ph.SR]}

\bibitem[{{Altyntsev} et~al(2019){Altyntsev}, {Meshalkina}, {Lysenko}, and
  {Fleishman}}]{Alyz2019}
{Altyntsev} AT, {Meshalkina} NS, {Lysenko} AL, et~al (2019) {Rapid Variability
  in the SOL2011-08-04 Flare: Implications for Electron Acceleration}. \apj
  883(1):38. \doi{10.3847/1538-4357/ab3808},
  {\href{https://arxiv.org/abs/1909.03593}{{arXiv:1909.03593}}} {[astro-ph.SR]}

\bibitem[{{Alvarado-G{\'o}mez} et~al(2018){Alvarado-G{\'o}mez}, {Drake},
  {Cohen}, {Moschou}, and {Garraffo}}]{AlvaradoGomez2018}
{Alvarado-G{\'o}mez} JD, {Drake} JJ, {Cohen} O, et~al (2018) {Suppression of
  Coronal Mass Ejections in Active Stars by an Overlying Large-scale Magnetic
  Field: A Numerical Study}. \apj 862(2):93. \doi{10.3847/1538-4357/aacb7f},
  {\href{https://arxiv.org/abs/1806.02828}{{arXiv:1806.02828}}} {[astro-ph.SR]}

\bibitem[{{Alvarado-G{\'o}mez} et~al(2020){Alvarado-G{\'o}mez}, {Drake},
  {Fraschetti}, {Garraffo}, {Cohen}, {Vocks}, {Poppenh{\"a}ger}, {Moschou},
  {Yadav}, and {Manchester}}]{AlvaradoGomez2020}
{Alvarado-G{\'o}mez} JD, {Drake} JJ, {Fraschetti} F, et~al (2020) {Tuning the
  Exospace Weather Radio for Stellar Coronal Mass Ejections}. \apj 895(1):47.
  \doi{10.3847/1538-4357/ab88a3},
  {\href{https://arxiv.org/abs/2004.05379}{{arXiv:2004.05379}}} {[astro-ph.SR]}

\bibitem[{{Anfinogentov} et~al(2013){Anfinogentov}, {Nakariakov},
  {Mathioudakis}, {Van Doorsselaere}, and {Kowalski}}]{Anfin2013}
{Anfinogentov} S, {Nakariakov} VM, {Mathioudakis} M, et~al (2013) {The Decaying
  Long-period Oscillation of a Stellar Megaflare}. \apj 773(2):156.
  \doi{10.1088/0004-637X/773/2/156}

\bibitem[{{Anglada-Escud{\'e}} et~al(2016){Anglada-Escud{\'e}}, {Amado},
  {Barnes}, {Berdi{\~n}as}, {Butler}, {Coleman}, {de La Cueva}, {Dreizler},
  {Endl}, {Giesers}, {Jeffers}, {Jenkins}, {Jones}, {Kiraga}, {K{\"u}rster},
  {L{\'o}pez-Gonz{\'a}lez}, {Marvin}, {Morales}, {Morin}, {Nelson}, {Ortiz},
  {Ofir}, {Paardekooper}, {Reiners}, {Rodr{\'\i}guez},
  {Rodr{\'\i}guez-L{\'o}pez}, {Sarmiento}, {Strachan}, {Tsapras}, {Tuomi}, and
  {Zechmeister}}]{Escude2016}
{Anglada-Escud{\'e}} G, {Amado} PJ, {Barnes} J, et~al (2016) {A terrestrial
  planet candidate in a temperate orbit around Proxima Centauri}. \nat
  536(7617):437--440. \doi{10.1038/nature19106},
  {\href{https://arxiv.org/abs/1609.03449}{{arXiv:1609.03449}}} {[astro-ph.EP]}

\bibitem[{{Anstee} and {O'Mara}(1995)}]{Anstee1995}
{Anstee} SD, {O'Mara} BJ (1995) {Width cross-sections for collisional
  broadening of s-p and p-s transitions by atomic hydrogen}. \mnras
  276(3):859--866. \doi{10.1093/mnras/276.3.859}

\bibitem[{{Arber} et~al(2015){Arber}, {Bennett}, {Brady}, {Lawrence-Douglas},
  {Ramsay}, {Sircombe}, {Gillies}, {Evans}, {Schmitz}, {Bell}, and
  {Ridgers}}]{Arber2015}
{Arber} TD, {Bennett} K, {Brady} CS, et~al (2015) {Contemporary
  particle-in-cell approach to laser-plasma modelling}. Plasma Physics and
  Controlled Fusion 57(11):113001. \doi{10.1088/0741-3335/57/11/113001}

\bibitem[{{Argiroffi} et~al(2019){Argiroffi}, {Reale}, {Drake}, {Ciaravella},
  {Testa}, {Bonito}, {Miceli}, {Orlando}, and {Peres}}]{Argiroffi2019}
{Argiroffi} C, {Reale} F, {Drake} JJ, et~al (2019) {A stellar flare-coronal
  mass ejection event revealed by X-ray plasma motions}. Nature Astronomy
  3:742--748. \doi{10.1038/s41550-019-0781-4},
  {\href{https://arxiv.org/abs/1905.11325}{{arXiv:1905.11325}}} {[astro-ph.SR]}

\bibitem[{{Arnold} et~al(2021){Arnold}, {Drake}, {Swisdak}, {Guo}, {Dahlin},
  {Chen}, {Fleishman}, {Glesener}, {Kontar}, {Phan}, and {Shen}}]{Arnold2021}
{Arnold} H, {Drake} JF, {Swisdak} M, et~al (2021) {Electron Acceleration during
  Macroscale Magnetic Reconnection}. \prl 126(13):135101.
  \doi{10.1103/PhysRevLett.126.135101},
  {\href{https://arxiv.org/abs/2011.01147}{{arXiv:2011.01147}}}
  {[physics.plasm-ph]}

\bibitem[{{Aschwanden}(1996)}]{Aschwanden1996Conf}
{Aschwanden} MJ (1996) {Hard X-ray timing}. In: {Ramaty} R, {Mandzhavidze} N,
  {Hua} XM (eds) High energy solar Physics, pp 300--310, \doi{10.1063/1.50965}

\bibitem[{{Aschwanden}(2004{\natexlab{a}})}]{Aschwanden2004Book}
{Aschwanden} MJ (2004{\natexlab{a}}) {Physics of the Solar Corona. An
  Introduction}

\bibitem[{{Aschwanden}(2004{\natexlab{b}})}]{Aschwanden2004}
{Aschwanden} MJ (2004{\natexlab{b}}) {Pulsed Particle Injection in a
  Reconnection-Driven Dynamic Trap Model in Solar Flares}. \apj
  608(1):554--561. \doi{10.1086/392494}

\bibitem[{{Aschwanden}(2011)}]{Aschwanden2011}
{Aschwanden} MJ (2011) {Self-Organized Criticality in Astrophysics}

\bibitem[{{Aschwanden}(2014)}]{Aschwanden2014}
{Aschwanden} MJ (2014) {A Macroscopic Description of a Generalized
  Self-organized Criticality System: Astrophysical Applications}. \apj
  782(1):54. \doi{10.1088/0004-637X/782/1/54},
  {\href{https://arxiv.org/abs/1310.4191}{{arXiv:1310.4191}}} {[astro-ph.SR]}

\bibitem[{{Aschwanden}(2015)}]{Aschwanden2015}
{Aschwanden} MJ (2015) {Magnetic Energy Dissipation during the 2014 March 29
  Solar Flare}. \apjl 804(1):L20. \doi{10.1088/2041-8205/804/1/L20},
  {\href{https://arxiv.org/abs/1504.03301}{{arXiv:1504.03301}}} {[astro-ph.SR]}

\bibitem[{{Aschwanden} and {Alexander}(2001)}]{AschwandenAlexander2001}
{Aschwanden} MJ, {Alexander} D (2001) {Flare Plasma Cooling from 30 MK down to
  1 MK modeled from Yohkoh, GOES, and TRACE observations during the Bastille
  Day Event (14 July 2000)}. \solphys 204:91--120.
  \doi{10.1023/A:1014257826116}

\bibitem[{{Aschwanden} and {Freeland}(2012)}]{AschwandenFreeland2012}
{Aschwanden} MJ, {Freeland} SL (2012) {Automated Solar Flare Statistics in Soft
  X-Rays over 37 Years of GOES Observations: The Invariance of Self-organized
  Criticality during Three Solar Cycles}. \apj 754(2):112.
  \doi{10.1088/0004-637X/754/2/112},
  {\href{https://arxiv.org/abs/1205.6712}{{arXiv:1205.6712}}} {[astro-ph.SR]}

\bibitem[{{Aschwanden} and {G{\"u}del}(2021)}]{Aschwanden2021}
{Aschwanden} MJ, {G{\"u}del} M (2021) {Self-organized Criticality in Stellar
  Flares}. \apj 910(1):41. \doi{10.3847/1538-4357/abdec7},
  {\href{https://arxiv.org/abs/2106.06490}{{arXiv:2106.06490}}} {[astro-ph.SR]}

\bibitem[{{Aschwanden} et~al(1995){Aschwanden}, {Schwartz}, and
  {Alt}}]{Aschwanden1995}
{Aschwanden} MJ, {Schwartz} RA, {Alt} DM (1995) {Electron Time-of-Flight
  Differences in Solar Flares}. \apj 447:923. \doi{10.1086/175930}

\bibitem[{{Aschwanden} et~al(1996{\natexlab{a}}){Aschwanden}, {Hudson},
  {Kosugi}, and {Schwartz}}]{Aschwanden1996Masuda}
{Aschwanden} MJ, {Hudson} H, {Kosugi} T, et~al (1996{\natexlab{a}}) {Electron
  Time-of-Flight Measurements during the Masuda Flare, 1992 January 13}. \apj
  464:985. \doi{10.1086/177386}

\bibitem[{{Aschwanden} et~al(1996{\natexlab{b}}){Aschwanden}, {Kosugi},
  {Hudson}, {Wills}, and {Schwartz}}]{Aschwanden1996}
{Aschwanden} MJ, {Kosugi} T, {Hudson} HS, et~al (1996{\natexlab{b}}) {The
  Scaling Law between Electron Time-of-Flight Distances and Loop Lengths in
  Solar Flares}. \apj 470:1198. \doi{10.1086/177943}

\bibitem[{{Aschwanden} et~al(2008){Aschwanden}, {Stern}, and
  {G{\"u}del}}]{AschwandenGudel2008}
{Aschwanden} MJ, {Stern} RA, {G{\"u}del} M (2008) {Scaling Laws of Solar and
  Stellar Flares}. \apj 672(1):659--673. \doi{10.1086/523926},
  {\href{https://arxiv.org/abs/0710.2563}{{arXiv:0710.2563}}} {[astro-ph]}

\bibitem[{{Ashfield} and {Longcope}(2023)}]{Ashfield2023Alf}
{Ashfield} W, {Longcope} D (2023) {A Model for Gradual-phase Heating Driven by
  MHD Turbulence in Solar Flares}. \apj 944(2):147.
  \doi{10.3847/1538-4357/acb1b2},
  {\href{https://arxiv.org/abs/2301.04592}{{arXiv:2301.04592}}} {[astro-ph.SR]}

\bibitem[{{Astudillo-Defru} et~al(2017){Astudillo-Defru}, {Delfosse},
  {Bonfils}, {Forveille}, {Lovis}, and {Rameau}}]{CaIIKsatur1}
{Astudillo-Defru} N, {Delfosse} X, {Bonfils} X, et~al (2017) {Magnetic activity
  in the HARPS M dwarf sample. The rotation-activity relationship for very
  low-mass stars through R'$_{HK}$}. \aap 600:A13.
  \doi{10.1051/0004-6361/201527078},
  {\href{https://arxiv.org/abs/1610.09007}{{arXiv:1610.09007}}} {[astro-ph.SR]}

\bibitem[{Atzeni and Meyer-ter Vehn(2004)}]{Atzeni2004}
Atzeni S, Meyer-ter Vehn J (2004) {The Physics of Inertial Fusion: BeamPlasma
  Interaction, Hydrodynamics, Hot Dense Matter}. Oxford University Press,
  \doi{10.1093/acprof:oso/9780198562641.001.0001},
  \urlprefix\url{https://doi.org/10.1093/acprof:oso/9780198562641.001.0001}

\bibitem[{{Audard} et~al(1999){Audard}, {G{\"u}del}, and {Guinan}}]{Audard1999}
{Audard} M, {G{\"u}del} M, {Guinan} EF (1999) {Implications from
  Extreme-Ultraviolet Observations for Coronal Heating of Active Stars}. \apjl
  513(1):L53--L56. \doi{10.1086/311907}

\bibitem[{{Audard} et~al(2000){Audard}, {G{\"u}del}, {Drake}, and
  {Kashyap}}]{Audard2000}
{Audard} M, {G{\"u}del} M, {Drake} JJ, et~al (2000) {Extreme-Ultraviolet Flare
  Activity in Late-Type Stars}. \apj 541(1):396--409. \doi{10.1086/309426},
  {\href{https://arxiv.org/abs/astro-ph/0005062}{{arXiv:astro-ph/0005062}}}
  {[astro-ph]}

\bibitem[{{Audard} et~al(2007){Audard}, {Briggs}, {Grosso}, {G{\"u}del},
  {Scelsi}, {Bouvier}, and {Telleschi}}]{Audard2007}
{Audard} M, {Briggs} KR, {Grosso} N, et~al (2007) {The XMM-Newton Optical
  Monitor survey of the Taurus molecular cloud}. \aap 468(2):379--390.
  \doi{10.1051/0004-6361:20066320},
  {\href{https://arxiv.org/abs/astro-ph/0611367}{{arXiv:astro-ph/0611367}}}
  {[astro-ph]}

\bibitem[{{Aulanier} et~al(2013){Aulanier}, {D{\'e}moulin}, {Schrijver},
  {Janvier}, {Pariat}, and {Schmieder}}]{Aulanier2013}
{Aulanier} G, {D{\'e}moulin} P, {Schrijver} CJ, et~al (2013) {The standard
  flare model in three dimensions. II. Upper limit on solar flare energy}. \aap
  549:A66. \doi{10.1051/0004-6361/201220406},
  {\href{https://arxiv.org/abs/1212.2086}{{arXiv:1212.2086}}} {[astro-ph.SR]}

\bibitem[{{Ayres}(2015{\natexlab{a}})}]{Ayres2015Alpha}
{Ayres} TR (2015{\natexlab{a}}) {The Far-Ultraviolet Ups and Downs of Alpha
  Centauri}. \aj 149(2):58. \doi{10.1088/0004-6256/149/2/58}

\bibitem[{{Ayres}(2015{\natexlab{b}})}]{Ayres2015}
{Ayres} TR (2015{\natexlab{b}}) {The Flare-ona of EK Draconis}. \aj 150:7.
  \doi{10.1088/0004-6256/150/1/7},
  {\href{https://arxiv.org/abs/1505.02320}{{arXiv:1505.02320}}} {[astro-ph.SR]}

\bibitem[{{Ayres} et~al(1994){Ayres}, {Stauffer}, {Simon}, {Stern},
  {Antiochos}, {Basri}, {Bookbinder}, {Brown}, {Doschek}, {Linsky}, {Ramsey},
  and {Walter}}]{Ayres1994}
{Ayres} TR, {Stauffer} JR, {Simon} T, et~al (1994) {A Far-Ultraviolet Flare on
  a Pleiades G Dwarf}. \apjl 420:L33. \doi{10.1086/187156}

\bibitem[{{Ayres} et~al(1998){Ayres}, {Simon}, {Stern}, {Drake}, {Wood}, and
  {Brown}}]{Ayres1998}
{Ayres} TR, {Simon} T, {Stern} RA, et~al (1998) {The Coronae of Moderate-Mass
  Giants in the Hertzsprung Gap and the Clump}. \apj 496(1):428--448.
  \doi{10.1086/305347}

\bibitem[{{Ayres} et~al(1999){Ayres}, {Osten}, and {Brown}}]{Ayres1999}
{Ayres} TR, {Osten} RA, {Brown} A (1999) {The Rise and Fall of
  {\ensuremath{\mu}} Velorum: A Remarkable Flare on a Yellow Giant Star
  Observed with the Extreme Ultraviolet Explorer}. \apj 526(1):445--450.
  \doi{10.1086/308001}

\bibitem[{{Ayres} et~al(2001){Ayres}, {Osten}, and {Brown}}]{Ayres2001}
{Ayres} TR, {Osten} RA, {Brown} A (2001) {3 Ms in the Life of
  {\ensuremath{\beta}} Ceti: Sustained Flare Activity on a Clump Giant Detected
  by the Extreme Ultraviolet Explorer}. \apjl 562(1):L83--L86.
  \doi{10.1086/337971}

\bibitem[{{Balona}(2012)}]{Balona2012}
{Balona} LA (2012) {Kepler observations of flaring in A-F type stars}. \mnras
  423(4):3420--3429. \doi{10.1111/j.1365-2966.2012.21135.x}

\bibitem[{{Balona}(2019)}]{Balona2019}
{Balona} LA (2019) {Evidence for spots on hot stars suggests major revision of
  stellar physics}. \mnras 490(2):2112--2116. \doi{10.1093/mnras/stz2808},
  {\href{https://arxiv.org/abs/1910.01584}{{arXiv:1910.01584}}} {[astro-ph.SR]}

\bibitem[{{Balona} et~al(2015){Balona}, {Broomhall}, {Kosovichev},
  {Nakariakov}, {Pugh}, and {Van Doorsselaere}}]{Balona2015}
{Balona} LA, {Broomhall} AM, {Kosovichev} A, et~al (2015) {Oscillations in
  stellar superflares}. \mnras 450(1):956--966. \doi{10.1093/mnras/stv661},
  {\href{https://arxiv.org/abs/1504.01491}{{arXiv:1504.01491}}} {[astro-ph.SR]}

\bibitem[{{Barklem}(2016)}]{Barklem2016}
{Barklem} PS (2016) {Accurate abundance analysis of late-type stars: advances
  in atomic physics}. \aapr 24(1):9. \doi{10.1007/s00159-016-0095-9},
  {\href{https://arxiv.org/abs/1604.07659}{{arXiv:1604.07659}}} {[astro-ph.SR]}

\bibitem[{{Barklem} and {O'Mara}(1998)}]{Barklem1998}
{Barklem} PS, {O'Mara} BJ (1998) {The broadening of strong lines of Ca\^+,
  Mg\^+ and Ba\^+ by collisions with neutral hydrogen atoms}. \mnras
  300(3):863--871. \doi{10.1046/j.1365-8711.1998.01942.x}

\bibitem[{{Barklem} et~al(2000){Barklem}, {Piskunov}, and
  {O'Mara}}]{Barklem2000}
{Barklem} PS, {Piskunov} N, {O'Mara} BJ (2000) {Self-broadening in Balmer line
  wing formation in stellar atmospheres}. \aap 363:1091--1105.
  {\href{https://arxiv.org/abs/astro-ph/0010022}{{arXiv:astro-ph/0010022}}}
  {[astro-ph]}

\bibitem[{{Baroch} et~al(2020){Baroch}, {Morales}, {Ribas}, {Herrero},
  {Rosich}, {Perger}, {Anglada-Escud{\'e}}, {Reiners}, {Caballero},
  {Quirrenbach}, {Amado}, {Jeffers}, {Cifuentes}, {Passegger}, {Schweitzer},
  {Lafarga}, {Bauer}, {B{\'e}jar}, {Colom{\'e}}, {Cort{\'e}s-Contreras},
  {Dreizler}, {Galad{\'\i}-Enr{\'\i}quez}, {Hatzes}, {Henning}, {Kaminski},
  {K{\"u}rster}, {Montes}, {Rodr{\'\i}guez-L{\'o}pez}, and
  {Zechmeister}}]{Baroch2020}
{Baroch} D, {Morales} JC, {Ribas} I, et~al (2020) {The CARMENES search for
  exoplanets around M dwarfs. Convective shift and starspot constraints from
  chromatic radial velocities}. \aap 641:A69.
  \doi{10.1051/0004-6361/202038213},
  {\href{https://arxiv.org/abs/2006.16608}{{arXiv:2006.16608}}} {[astro-ph.SR]}

\bibitem[{{Bastian}(1990)}]{Bastian1990}
{Bastian} TS (1990) {Radio Emission from Flare Stars}. \solphys
  130(1-2):265--294. \doi{10.1007/BF00156794}

\bibitem[{{Bastian} et~al(1990){Bastian}, {Bookbinder}, {Dulk}, and
  {Davis}}]{Bastian1990B}
{Bastian} TS, {Bookbinder} J, {Dulk} GA, et~al (1990) {Dynamic Spectra of Radio
  Bursts from Flare Stars}. \apj 353:265. \doi{10.1086/168613}

\bibitem[{{Beasley} and {Bastian}(1998)}]{Beasley1998}
{Beasley} AJ, {Bastian} TS (1998) {VLBA Imaging of UX Ari}. In: {Zensus} JA,
  {Taylor} GB, {Wrobel} JM (eds) IAU Colloq. 164: Radio Emission from Galactic
  and Extragalactic Compact Sources, p 321

\bibitem[{{Becker} et~al(2004){Becker}, {Wittman}, {Boeshaar}, {Clocchiatti},
  {Dell'Antonio}, {Frail}, {Halpern}, {Margoniner}, {Norman}, {Tyson}, and
  {Schommer}}]{Becker2004}
{Becker} AC, {Wittman} DM, {Boeshaar} PC, et~al (2004) {The Deep Lens Survey
  Transient Search. I. Short Timescale and Astrometric Variability}. \apj
  611(1):418--433. \doi{10.1086/421994},
  {\href{https://arxiv.org/abs/astro-ph/0404416}{{arXiv:astro-ph/0404416}}}
  {[astro-ph]}

\bibitem[{{Benenti} et~al(1999){Benenti}, {Casati}, and
  {Shepelyansky}}]{Benenti1999}
{Benenti} G, {Casati} G, {Shepelyansky} DL (1999) {Chaotic enhancement in
  microwave ionization of Rydberg atoms}. European Physical Journal D
  5(3):311--326. \doi{10.1007/s100530050261},
  {\href{https://arxiv.org/abs/cond-mat/9805216}{{arXiv:cond-mat/9805216}}}
  {[cond-mat]}

\bibitem[{{Benka} and {Holman}(1992)}]{Benka1992}
{Benka} SG, {Holman} GD (1992) {A Thermal/Nonthermal Model for Solar Microwave
  Bursts}. \apj 391:854. \doi{10.1086/171394}

\bibitem[{{Benz}(2002)}]{Benz2002}
{Benz} A (2002) {Plasma Astrophysics, second edition}, vol 279.
  \doi{10.1007/978-0-306-47719-5}

\bibitem[{{Benz}(2017)}]{Benz2017}
{Benz} AO (2017) {Flare Observations}. Living Reviews in Solar Physics 14(1):2.
  \doi{10.1007/s41116-016-0004-3}

\bibitem[{{Benz} et~al(1998){Benz}, {Conway}, and {Gudel}}]{Benz1998}
{Benz} AO, {Conway} J, {Gudel} M (1998) {First VLBI images of a main-sequence
  star}. \aap 331:596--600

\bibitem[{{Bergeman}(1984)}]{Bergeman1984}
{Bergeman} T (1984) {Relativistically enhanced ionization rates at Stark-effect
  level crossings in hydrogen}. \prl 52(19):1685--1688.
  \doi{10.1103/PhysRevLett.52.1685}

\bibitem[{{Berger} et~al(2013){Berger}, {Leibler}, {Chornock}, {Rest}, {Foley},
  {Soderberg}, {Price}, {Burgett}, {Chambers}, {Flewelling}, {Huber},
  {Magnier}, {Metcalfe}, {Stubbs}, and {Tonry}}]{Berger2013}
{Berger} E, {Leibler} CN, {Chornock} R, et~al (2013) {A Search for Fast Optical
  Transients in the Pan-STARRS1 Medium-Deep Survey: M-Dwarf Flares, Asteroids,
  Limits on Extragalactic Rates, and Implications for LSST}. \apj 779(1):18.
  \doi{10.1088/0004-637X/779/1/18},
  {\href{https://arxiv.org/abs/1307.5324}{{arXiv:1307.5324}}} {[astro-ph.HE]}

\bibitem[{{Beskin} et~al(2017){Beskin}, {Karpov}, {Plokhotnichenko},
  {Stepanov}, and {Tsap}}]{Beskin2017}
{Beskin} G, {Karpov} S, {Plokhotnichenko} V, et~al (2017) {Discovery of the
  Sub-second Linearly Polarized Spikes of Synchrotron Origin in the UV Ceti
  Giant Optical Flare}. \pasa 34:e010. \doi{10.1017/pasa.2017.3},
  {\href{https://arxiv.org/abs/1702.06660}{{arXiv:1702.06660}}} {[astro-ph.SR]}

\bibitem[{{Bessell} and {Murphy}(2012)}]{Bessel2013}
{Bessell} M, {Murphy} S (2012) {Spectrophotometric Libraries, Revised Photonic
  Passbands, and Zero Points for UBVRI, Hipparcos, and Tycho Photometry}. \pasp
  124:140. \doi{10.1086/664083},
  {\href{https://arxiv.org/abs/1112.2698}{{arXiv:1112.2698}}} {[astro-ph.SR]}

\bibitem[{{Bethe} and {Salpeter}(1957)}]{BetheSalpeter}
{Bethe} HA, {Salpeter} EE (1957) {Quantum Mechanics of One- and Two-Electron
  Atoms}

\bibitem[{{Betta} et~al(1997){Betta}, {Peres}, {Reale}, and {Serio}}]{PM2}
{Betta} R, {Peres} G, {Reale} F, et~al (1997) {An adaptive grid code for high
  resolution 1-D hydrodynamics of the solar and stellar transition region and
  corona}. \aaps 122:585--592. \doi{10.1051/aas:1997157}

\bibitem[{{Bian} et~al(2012){Bian}, {Emslie}, and {Kontar}}]{Bian2012}
{Bian} N, {Emslie} AG, {Kontar} EP (2012) {A Classification Scheme for
  Turbulent Acceleration Processes in Solar Flares}. \apj 754(2):103.
  \doi{10.1088/0004-637X/754/2/103},
  {\href{https://arxiv.org/abs/1206.0472}{{arXiv:1206.0472}}} {[astro-ph.SR]}

\bibitem[{{Bian} et~al(2018){Bian}, {Emslie}, {Horne}, and {Kontar}}]{Bian2018}
{Bian} N, {Emslie} AG, {Horne} D, et~al (2018) {Heating and Cooling of Coronal
  Loops with Turbulent Suppression of Parallel Heat Conduction}. \apj
  852(2):127. \doi{10.3847/1538-4357/aa9f29},
  {\href{https://arxiv.org/abs/1711.11388}{{arXiv:1711.11388}}} {[astro-ph.SR]}

\bibitem[{{Bian} et~al(2014){Bian}, {Emslie}, {Stackhouse}, and
  {Kontar}}]{Bian2014}
{Bian} NH, {Emslie} AG, {Stackhouse} DJ, et~al (2014) {The Formation of
  Kappa-distribution Accelerated Electron Populations in Solar Flares}. \apj
  796(2):142. \doi{10.1088/0004-637X/796/2/142},
  {\href{https://arxiv.org/abs/1410.0819}{{arXiv:1410.0819}}} {[astro-ph.SR]}

\bibitem[{{Bice} and {Toomre}(2020)}]{Bice1}
{Bice} CP, {Toomre} J (2020) {Probing the Influence of a Tachocline in
  Simulated M-dwarf Dynamos}. \apj 893(2):107. \doi{10.3847/1538-4357/ab8190},
  {\href{https://arxiv.org/abs/2001.05555}{{arXiv:2001.05555}}} {[astro-ph.SR]}

\bibitem[{{Bice} and {Toomre}(2022)}]{Bice2}
{Bice} CP, {Toomre} J (2022) {Longitudinally Modulated Dynamo Action in
  Simulated M-dwarf Stars}. \apj 928(1):51. \doi{10.3847/1538-4357/ac4be0},
  {\href{https://arxiv.org/abs/2202.02869}{{arXiv:2202.02869}}} {[astro-ph.SR]}

\bibitem[{{Birn} et~al(2017){Birn}, {Battaglia}, {Fletcher}, {Hesse}, and
  {Neukirch}}]{Birn2017}
{Birn} J, {Battaglia} M, {Fletcher} L, et~al (2017) {Can Substorm Particle
  Acceleration Be Applied to Solar Flares?} \apj 848(2):116.
  \doi{10.3847/1538-4357/aa8ad4}

\bibitem[{{Bj{\o}rgen} et~al(2019){Bj{\o}rgen}, {Leenaarts}, {Rempel},
  {Cheung}, {Danilovic}, {de la Cruz Rodr{\'\i}guez}, and
  {Sukhorukov}}]{Bjorgen2019Flare}
{Bj{\o}rgen} JP, {Leenaarts} J, {Rempel} M, et~al (2019) {Three-dimensional
  modeling of chromospheric spectral lines in a simulated active region}. \aap
  631:A33. \doi{10.1051/0004-6361/201834919},
  {\href{https://arxiv.org/abs/1906.01098}{{arXiv:1906.01098}}} {[astro-ph.SR]}

\bibitem[{{Bookbinder} et~al(1992){Bookbinder}, {Walter}, and
  {Brown}}]{Bookbinder1992}
{Bookbinder} JA, {Walter} FM, {Brown} A (1992) {HST Observations of AD Leo}.
  In: {Giampapa} MS, {Bookbinder} JA (eds) Cool Stars, Stellar Systems, and the
  Sun, p~27

\bibitem[{{Bopp} and {Moffett}(1973)}]{Bopp1973}
{Bopp} BW, {Moffett} TJ (1973) {High time resolution studies of UV Ceti.} \apj
  185:239. \doi{10.1086/152412}

\bibitem[{{Bornmann}(1987)}]{Bornmann1987}
{Bornmann} PL (1987) {Turbulence as a Contributor to Intermediate Energy
  Storage during Solar Flares}. \apj 313:449. \doi{10.1086/164984}

\bibitem[{{Borucki} et~al(2010){Borucki}, {Koch}, {Basri}, {Batalha}, {Brown},
  {Caldwell}, {Caldwell}, {Christensen-Dalsgaard}, {Cochran}, {DeVore},
  {Dunham}, {Dupree}, {Gautier}, {Geary}, {Gilliland}, {Gould}, {Howell},
  {Jenkins}, {Kondo}, {Latham}, {Marcy}, {Meibom}, {Kjeldsen}, {Lissauer},
  {Monet}, {Morrison}, {Sasselov}, {Tarter}, {Boss}, {Brownlee}, {Owen},
  {Buzasi}, {Charbonneau}, {Doyle}, {Fortney}, {Ford}, {Holman}, {Seager},
  {Steffen}, {Welsh}, {Rowe}, {Anderson}, {Buchhave}, {Ciardi}, {Walkowicz},
  {Sherry}, {Horch}, {Isaacson}, {Everett}, {Fischer}, {Torres}, {Johnson},
  {Endl}, {MacQueen}, {Bryson}, {Dotson}, {Haas}, {Kolodziejczak}, {Van Cleve},
  {Chandrasekaran}, {Twicken}, {Quintana}, {Clarke}, {Allen}, {Li}, {Wu},
  {Tenenbaum}, {Verner}, {Bruhweiler}, {Barnes}, and {Prsa}}]{Borucki2010}
{Borucki} WJ, {Koch} D, {Basri} G, et~al (2010) {Kepler Planet-Detection
  Mission: Introduction and First Results}. Science 327(5968):977.
  \doi{10.1126/science.1185402}

\bibitem[{{Boudreaux} et~al(2022){Boudreaux}, {Newton}, {Mondrik},
  {Charbonneau}, and {Irwin}}]{CaIIKsatur2}
{Boudreaux} EM, {Newton} ER, {Mondrik} N, et~al (2022) {The Ca II H and K
  Rotation-Activity Relation in 53 Mid-to-late-type M Dwarfs}. \apj 929(1):80.
  \doi{10.3847/1538-4357/ac5cbf},
  {\href{https://arxiv.org/abs/2203.04999}{{arXiv:2203.04999}}} {[astro-ph.SR]}

\bibitem[{{Bower} et~al(2003){Bower}, {Plambeck}, {Bolatto}, {McCrady},
  {Graham}, {de Pater}, {Liu}, and {Baganoff}}]{Bower2003}
{Bower} GC, {Plambeck} RL, {Bolatto} A, et~al (2003) {A Giant Outburst at
  Millimeter Wavelengths in the Orion Nebula}. \apj 598(2):1140--1150.
  \doi{10.1086/379101},
  {\href{https://arxiv.org/abs/astro-ph/0308277}{{arXiv:astro-ph/0308277}}}
  {[astro-ph]}

\bibitem[{{Boyd} and {Sanderson}(2003)}]{BoydSanderson2003}
{Boyd} TJM, {Sanderson} JJ (2003) {The Physics of Plasmas}

\bibitem[{{Brandenburg}(2005)}]{Brandenburg2005}
{Brandenburg} A (2005) {The Case for a Distributed Solar Dynamo Shaped by
  Near-Surface Shear}. \apj 625(1):539--547. \doi{10.1086/429584},
  {\href{https://arxiv.org/abs/astro-ph/0502275}{{arXiv:astro-ph/0502275}}}
  {[astro-ph]}

\bibitem[{{Brandenburg} and {Subramanian}(2005)}]{Brandenburg2005B}
{Brandenburg} A, {Subramanian} K (2005) {Astrophysical magnetic fields and
  nonlinear dynamo theory}. \physrep 417(1-4):1--209.
  \doi{10.1016/j.physrep.2005.06.005},
  {\href{https://arxiv.org/abs/astro-ph/0405052}{{arXiv:astro-ph/0405052}}}
  {[astro-ph]}

\bibitem[{Bransden and Joachain(2000)}]{Bransden}
Bransden BH, Joachain CJ (2000) Quantum Mechanics. Pearson, Prentice Hall,
  Harlow, England :

\bibitem[{{Brasseur} et~al(2019){Brasseur}, {Osten}, and
  {Fleming}}]{Brasseur2019}
{Brasseur} CE, {Osten} RA, {Fleming} SW (2019) {Short-duration Stellar Flares
  in GALEX Data}. \apj 883(1):88. \doi{10.3847/1538-4357/ab3df8},
  {\href{https://arxiv.org/abs/1908.08377}{{arXiv:1908.08377}}} {[astro-ph.SR]}

\bibitem[{{Brasseur} et~al(2023){Brasseur}, {Osten}, {Tristan}, and
  {Kowalski}}]{Brasseur2023}
{Brasseur} CE, {Osten} RA, {Tristan} II, et~al (2023) {Constraints on Stellar
  Flare Energy Ratios in the NUV and Optical from a Multiwavelength Study of
  GALEX and Kepler Flare Stars}. \apj 944(1):5. \doi{10.3847/1538-4357/acab59},
  {\href{https://arxiv.org/abs/2212.08696}{{arXiv:2212.08696}}} {[astro-ph.SR]}

\bibitem[{{Bressan} et~al(2012){Bressan}, {Marigo}, {Girardi}, {Salasnich},
  {Dal Cero}, {Rubele}, and {Nanni}}]{PARSEC1}
{Bressan} A, {Marigo} P, {Girardi} L, et~al (2012) {PARSEC: stellar tracks and
  isochrones with the PAdova and TRieste Stellar Evolution Code}. \mnras
  427(1):127--145. \doi{10.1111/j.1365-2966.2012.21948.x},
  {\href{https://arxiv.org/abs/1208.4498}{{arXiv:1208.4498}}} {[astro-ph.SR]}

\bibitem[{{Bromage} et~al(1986){Bromage}, {Phillips}, {Dufton}, and
  {Kingston}}]{Bromage1986}
{Bromage} GE, {Phillips} KJH, {Dufton} PL, et~al (1986) {Flares on dMe stars :
  IUE and optical observations of AT MIC and comparison of far-ultraviolet
  stellar and solar flares.} \mnras 220:1021--1046.
  \doi{10.1093/mnras/220.4.1021}

\bibitem[{{Broomhall} et~al(2019{\natexlab{a}}){Broomhall}, {Davenport},
  {Hayes}, {Inglis}, {Kolotkov}, {McLaughlin}, {Mehta}, {Nakariakov}, {Notsu},
  {Pascoe}, {Pugh}, and {Van Doorsselaere}}]{Broomhall2021}
{Broomhall} AM, {Davenport} JRA, {Hayes} LA, et~al (2019{\natexlab{a}}) {A
  Blueprint of State-of-the-art Techniques for Detecting Quasi-periodic
  Pulsations in Solar and Stellar Flares}. \apjs 244(2):44.
  \doi{10.3847/1538-4365/ab40b3},
  {\href{https://arxiv.org/abs/1910.08458}{{arXiv:1910.08458}}} {[astro-ph.SR]}

\bibitem[{{Broomhall} et~al(2019{\natexlab{b}}){Broomhall}, {Thomas}, {Pugh},
  {Pye}, and {Rosen}}]{Broomhall2019}
{Broomhall} AM, {Thomas} AEL, {Pugh} CE, et~al (2019{\natexlab{b}})
  {Multi-waveband detection of quasi-periodic pulsations in a stellar flare on
  EK Draconis observed by XMM-Newton}. \aap 629:A147.
  \doi{10.1051/0004-6361/201935653},
  {\href{https://arxiv.org/abs/1908.06033}{{arXiv:1908.06033}}} {[astro-ph.SR]}

\bibitem[{{Brosius} and {Inglis}(2018)}]{Brosius2018}
{Brosius} JW, {Inglis} AR (2018) {Localized Quasi-periodic Fluctuations in C
  II, Si IV, and Fe XXI Emission during Chromospheric Evaporation in a Flare
  Ribbon Observed by IRIS on 2017 September 9}. \apj 867(2):85.
  \doi{10.3847/1538-4357/aae5f5}

\bibitem[{{Brown} et~al(2023){Brown}, {Schneider}, {France}, {Froning},
  {Youngblood}, {J. Wilson}, {Loyd}, {Pineda}, {Duvvuri}, {Kowalski}, and
  {Berta-Thompson}}]{Brown2023}
{Brown} A, {Schneider} PC, {France} K, et~al (2023) {Coronal X-Ray Emission
  from Nearby, Low-mass, Exoplanet Host Stars Observed by the MUSCLES and
  Mega-MUSCLES HST Treasury Survey Projects}. \aj 165(5):195.
  \doi{10.3847/1538-3881/acc38a},
  {\href{https://arxiv.org/abs/2303.12929}{{arXiv:2303.12929}}} {[astro-ph.SR]}

\bibitem[{{Brown} et~al(2020){Brown}, {Oishi}, {Vasil}, {Lecoanet}, and
  {Burns}}]{Brown2020}
{Brown} BP, {Oishi} JS, {Vasil} GM, et~al (2020) {Single-hemisphere Dynamos in
  M-dwarf Stars}. \apjl 902(1):L3. \doi{10.3847/2041-8213/abb9a4},
  {\href{https://arxiv.org/abs/2008.02362}{{arXiv:2008.02362}}} {[astro-ph.SR]}

\bibitem[{{Brown}(1971)}]{Brown1971}
{Brown} JC (1971) {The Deduction of Energy Spectra of Non-Thermal Electrons in
  Flares from the Observed Dynamic Spectra of Hard X-Ray Bursts}. \solphys
  18(3):489--502. \doi{10.1007/BF00149070}

\bibitem[{{Brown}(1973)}]{Brown1973B}
{Brown} JC (1973) {The Temperature Structure of Chromospheric Flares Heated by
  Non-Thermal Electrons}. \solphys 31(1):143--169. \doi{10.1007/BF00156080}

\bibitem[{{Brown} et~al(1998){Brown}, {Conway}, and {Aschwanden}}]{Brown1998}
{Brown} JC, {Conway} AJ, {Aschwanden} MJ (1998) {The Electron Injection
  Function and Energy-dependent Delays in Thick-Target Hard X-Rays}. \apj
  509(2):911--917. \doi{10.1086/306522}

\bibitem[{{Brown} et~al(2003){Brown}, {Emslie}, and {Kontar}}]{Brown2003}
{Brown} JC, {Emslie} AG, {Kontar} EP (2003) {The Determination and Use of Mean
  Electron Flux Spectra in Solar Flares}. \apjl 595(2):L115--L117.
  \doi{10.1086/378169}

\bibitem[{{Brown} and {Brown}(2006)}]{BrownBrown2006}
{Brown} JM, {Brown} A (2006) {A Large Millimeter Flare on the RS CVn Binary
  {\ensuremath{\sigma}} Geminorum}. \apjl 638(1):L37--L40. \doi{10.1086/500929}

\bibitem[{{Browning}(2008)}]{Browning2008}
{Browning} MK (2008) {Simulations of Dynamo Action in Fully Convective Stars}.
  \apj 676(2):1262--1280. \doi{10.1086/527432},
  {\href{https://arxiv.org/abs/0712.1603}{{arXiv:0712.1603}}} {[astro-ph]}

\bibitem[{{Brun} and {Browning}(2017)}]{Brun2017}
{Brun} AS, {Browning} MK (2017) {Magnetism, dynamo action and the solar-stellar
  connection}. Living Reviews in Solar Physics 14(1):4.
  \doi{10.1007/s41116-017-0007-8}

\bibitem[{{Butler} et~al(1981){Butler}, {Byrne}, {Andrews}, and
  {Doyle}}]{Butler1981}
{Butler} CJ, {Byrne} PB, {Andrews} AD, et~al (1981) {Ultraviolet spectra of
  dwarf solar neighbourhood stars. I.} \mnras 197:815--827.
  \doi{10.1093/mnras/197.3.815}

\bibitem[{{Butler} et~al(1988){Butler}, {Rodono}, and {Foing}}]{Butler1988}
{Butler} CJ, {Rodono} M, {Foing} BH (1988) {A correlation between Balmer and
  soft X-ray emission from stellar andsolar flares.} \aap 206:L1--L4

\bibitem[{{Butler} et~al(2015){Butler}, {Erkan}, {Budding}, {Doyle}, {Foing},
  {Bromage}, {Kellett}, {Frueh}, {Huovelin}, {Brown}, and {Neff}}]{Butler2015}
{Butler} CJ, {Erkan} N, {Budding} E, et~al (2015) {A multiwavelength study of
  the M dwarf binary YY Geminorum}. \mnras 446(4):4205--4219.
  \doi{10.1093/mnras/stu2398},
  {\href{https://arxiv.org/abs/1501.03930}{{arXiv:1501.03930}}} {[astro-ph.SR]}

\bibitem[{{Byrne}(1989)}]{Byrne1989}
{Byrne} PB (1989) {Multi-Wavelength Observations of Stellar Flares}. \solphys
  121(1-2):61--74. \doi{10.1007/BF00161687}

\bibitem[{{Byrne} et~al(1990){Byrne}, {Butler}, and {Lyons}}]{Byrne1990}
{Byrne} PB, {Butler} CJ, {Lyons} MA (1990) {Activity in late-type stars. VI.
  Optical photometry and UV spectroscopy of the active dMe star, FK Aquarii in
  late 1983.} \aap 236:455

\bibitem[{{Caballero-Garc{\'\i}a} et~al(2015){Caballero-Garc{\'\i}a},
  {{\v{S}}imon}, {Jel{\'\i}nek}, {Castro-Tirado}, {Cwiek}, {Claret}, {Opiela},
  {{\.Z}arnecki}, {Gorosabel}, {Oates}, {Cunniffe}, {Jeong}, {Hudec},
  {Sokolov}, {Makarov}, {Tello}, {Lara-Gil}, {Kub{\'a}nek}, {Guziy}, {Bai},
  {Fan}, {Wang}, and {Park}}]{Cabellero2015}
{Caballero-Garc{\'\i}a} MD, {{\v{S}}imon} V, {Jel{\'\i}nek} M, et~al (2015)
  {Early optical follow-up of the nearby active star DG CVn during its 2014
  superflare}. \mnras 452(4):4195--4202. \doi{10.1093/mnras/stv1565},
  {\href{https://arxiv.org/abs/1507.03143}{{arXiv:1507.03143}}} {[astro-ph.HE]}

\bibitem[{{Candelaresi} et~al(2014){Candelaresi}, {Hillier}, {Maehara},
  {Brandenburg}, and {Shibata}}]{Candel2014}
{Candelaresi} S, {Hillier} A, {Maehara} H, et~al (2014) {Superflare Occurrence
  and Energies on G-, K-, and M-type Dwarfs}. \apj 792(1):67.
  \doi{10.1088/0004-637X/792/1/67},
  {\href{https://arxiv.org/abs/1405.1453}{{arXiv:1405.1453}}} {[astro-ph.SR]}

\bibitem[{{Canfield} et~al(1990){Canfield}, {Penn}, {Wulser}, and
  {Kiplinger}}]{Canfield1990}
{Canfield} RC, {Penn} MJ, {Wulser} JP, et~al (1990) {H alpha Spectra of Dynamic
  Chromospheric Processes in Five Well-observed X-Ray Flares}. \apj 363:318.
  \doi{10.1086/169345}

\bibitem[{{Canfield} et~al(1991){Canfield}, {Wulser}, {Zarro}, and
  {Dennis}}]{Canfield1991}
{Canfield} RC, {Wulser} JP, {Zarro} DM, et~al (1991) {A Study of Solar Flare
  Energy Transport Based on Coordinated H alpha and X-Ray Observations}. \apj
  367:671. \doi{10.1086/169663}

\bibitem[{{Carlsson}(1998)}]{Carlsson1998}
{Carlsson} M (1998) {Radiative Transfer and Radiation Hydrodynamics}. In:
  {Vial} JC, {Bocchialini} K, {Boumier} P (eds) Space Solar Physics:
  Theoretical and Observational Issues in the Context of the SOHO Mission, vol
  507. p 163, \doi{10.1007/BFb0106920}

\bibitem[{{Carlsson} and {Stein}(1992)}]{Carlsson1992B}
{Carlsson} M, {Stein} RF (1992) {Non-LTE Radiating Acoustic Shocks and CA II
  K2V Bright Points}. \apjl 397:L59. \doi{10.1086/186544}

\bibitem[{{Carlsson} and {Stein}(1995)}]{Carlsson1995}
{Carlsson} M, {Stein} RF (1995) {Does a Nonmagnetic Solar Chromosphere Exist?}
  \apjl 440:L29. \doi{10.1086/187753},
  {\href{https://arxiv.org/abs/astro-ph/9411036}{{arXiv:astro-ph/9411036}}}
  {[astro-ph]}

\bibitem[{{Carlsson} and {Stein}(1997)}]{Carlsson1997}
{Carlsson} M, {Stein} RF (1997) {Formation of Solar Calcium H and K Bright
  Grains}. \apj 481:500--514

\bibitem[{{Carlsson} et~al(2016){Carlsson}, {Hansteen}, {Gudiksen},
  {Leenaarts}, and {De Pontieu}}]{Carlsson2016}
{Carlsson} M, {Hansteen} VH, {Gudiksen} BV, et~al (2016) {A publicly available
  simulation of an enhanced network region of the Sun}. \aap 585:A4.
  \doi{10.1051/0004-6361/201527226},
  {\href{https://arxiv.org/abs/1510.07581}{{arXiv:1510.07581}}} {[astro-ph.SR]}

\bibitem[{{Carlsson} et~al(2023){Carlsson}, {Fletcher}, {Allred}, {Heinzel},
  {Ka{\v{s}}parov{\'a}}, {Kowalski}, {Mathioudakis}, {Reid}, and
  {Sim{\~o}es}}]{Carlsson2023}
{Carlsson} M, {Fletcher} L, {Allred} J, et~al (2023) {The F-CHROMA grid of 1D
  RADYN flare models}. \aap 673:A150. \doi{10.1051/0004-6361/202346087},
  {\href{https://arxiv.org/abs/2304.02618}{{arXiv:2304.02618}}} {[astro-ph.SR]}

\bibitem[{{Carmichael}(1964)}]{Carmichael1964}
{Carmichael} H (1964) {A Process for Flares}. In: NASA Special Publication,
  vol~50. p 451

\bibitem[{{Cassak} et~al(2008){Cassak}, {Mullan}, and {Shay}}]{Cassak2008}
{Cassak} PA, {Mullan} DJ, {Shay} MA (2008) {From Solar and Stellar Flares to
  Coronal Heating: Theory and Observations of How Magnetic Reconnection
  Regulates Coronal Conditions}. \apjl 676(1):L69. \doi{10.1086/587055},
  {\href{https://arxiv.org/abs/0710.3399}{{arXiv:0710.3399}}} {[astro-ph]}

\bibitem[{{Cassak} et~al(2017){Cassak}, {Liu}, and {Shay}}]{Cassak2017}
{Cassak} PA, {Liu} YH, {Shay} MA (2017) {A review of the 0.1 reconnection rate
  problem}. Journal of Plasma Physics 83(5):715830501.
  \doi{10.1017/S0022377817000666},
  {\href{https://arxiv.org/abs/1708.03449}{{arXiv:1708.03449}}}
  {[physics.plasm-ph]}

\bibitem[{{Castellanos Dur{\'a}n} and {Kleint}(2020)}]{Castellanos2020}
{Castellanos Dur{\'a}n} JS, {Kleint} L (2020) {The Statistical Relationship
  between White-light Emission and Photospheric Magnetic Field Changes in
  Flares}. \apj 904(2):96. \doi{10.3847/1538-4357/ab9c1e}

\bibitem[{{Chambe} and {Henoux}(1979)}]{Chambe1979}
{Chambe} G, {Henoux} JC (1979) {Direct excitation of hydrogen by photoelectron
  and accelerated electron collisions in solar chromospheric flares - Effects
  on the profile and polarization of LY alpha}. \aap 80(2):123--129

\bibitem[{{Chandrasekhar}(1960)}]{Chandrasekhar1960}
{Chandrasekhar} S (1960) {Radiative transfer}

\bibitem[{{Chang} et~al(2020){Chang}, {Wolf}, and {Onken}}]{Chang2020}
{Chang} SW, {Wolf} C, {Onken} CA (2020) {Photometric flaring fraction of M
  dwarf stars from the SkyMapper Southern Survey}. \mnras 491(1):39--50.
  \doi{10.1093/mnras/stz2898},
  {\href{https://arxiv.org/abs/1910.06478}{{arXiv:1910.06478}}} {[astro-ph.SR]}

\bibitem[{{Charbonneau}(2013)}]{Charb2013}
{Charbonneau} P (2013) {Solar and Stellar Dynamos}, Saas-Fee Advanced Course,
  vol~39. \doi{10.1007/978-3-642-32093-4}

\bibitem[{{Chen} et~al(2020){Chen}, {Shen}, {Gary}, {Reeves}, {Fleishman},
  {Yu}, {Guo}, {Krucker}, {Lin}, {Nita}, and {Kong}}]{Chen2020}
{Chen} B, {Shen} C, {Gary} DE, et~al (2020) {Measurement of magnetic field and
  relativistic electrons along a solar flare current sheet}. Nature Astronomy
  4:1140--1147. \doi{10.1038/s41550-020-1147-7},
  {\href{https://arxiv.org/abs/2005.12757}{{arXiv:2005.12757}}} {[astro-ph.SR]}

\bibitem[{{Cheng} et~al(2012){Cheng}, {Qiu}, {Ding}, and {Wang}}]{Chen2012}
{Cheng} JX, {Qiu} J, {Ding} MD, et~al (2012) {Solar flare hard X-ray spikes
  observed by RHESSI: a statistical study}. \aap 547:A73.
  \doi{10.1051/0004-6361/201118608},
  {\href{https://arxiv.org/abs/1210.7027}{{arXiv:1210.7027}}} {[astro-ph.SR]}

\bibitem[{{Chintzoglou} et~al(2019){Chintzoglou}, {Zhang}, {Cheung}, and
  {Kazachenko}}]{Chintz2019}
{Chintzoglou} G, {Zhang} J, {Cheung} MCM, et~al (2019) {The Origin of Major
  Solar Activity: Collisional Shearing between Nonconjugated Polarities of
  Multiple Bipoles Emerging within Active Regions}. \apj 871(1):67.
  \doi{10.3847/1538-4357/aaef30},
  {\href{https://arxiv.org/abs/1811.02186}{{arXiv:1811.02186}}} {[astro-ph.SR]}

\bibitem[{{Cho} et~al(2016){Cho}, {Cho}, {Nakariakov}, {Kim}, and
  {Kumar}}]{Cho2016}
{Cho} IH, {Cho} KS, {Nakariakov} VM, et~al (2016) {Comparison of Damped
  Oscillations in Solar and Stellar X-Ray flares}. \apj 830(2):110.
  \doi{10.3847/0004-637X/830/2/110}

\bibitem[{{Cho} et~al(2022){Cho}, {Gomez}, {Montgomery}, {Dunlap}, {Fitz Axen},
  {Hobbs}, {Hubeny}, and {Winget}}]{Cho2022}
{Cho} PB, {Gomez} TA, {Montgomery} MH, et~al (2022) {Simulation of
  Stark-broadened Hydrogen Balmer-line Shapes for DA White Dwarf Synthetic
  Spectra}. \apj 927(1):70. \doi{10.3847/1538-4357/ac4df3}

\bibitem[{{Christian} et~al(2003){Christian}, {Mathioudakis}, {Jevremovi{\'c}},
  {Dupuis}, {Vennes}, and {Kawka}}]{Christian2003}
{Christian} DJ, {Mathioudakis} M, {Jevremovi{\'c}} D, et~al (2003) {The
  Extreme-Ultraviolet Continuum of a Strong Stellar Flare}. \apjl
  593:L105--L108. \doi{10.1086/378217}

\bibitem[{{Clarke} et~al(2018){Clarke}, {Davenport}, {Covey}, and
  {Baranec}}]{Clarke2018}
{Clarke} RW, {Davenport} JRA, {Covey} KR, et~al (2018) {Flare Activity of Wide
  Binary Stars with Kepler}. \apj 853(1):59. \doi{10.3847/1538-4357/aaa0d3}

\bibitem[{{Clauset} et~al(2009){Clauset}, {Shalizi}, and
  {Newman}}]{Clauset2009}
{Clauset} A, {Shalizi} CR, {Newman} MEJ (2009) {Power-Law Distributions in
  Empirical Data}. SIAM Review 51(4):661--703. \doi{10.1137/070710111}

\bibitem[{{Cliver} et~al(1986){Cliver}, {Dennis}, {Kiplinger}, {Kane},
  {Neidig}, {Sheeley}, and {Koomen}}]{Cliver1986}
{Cliver} EW, {Dennis} BR, {Kiplinger} AL, et~al (1986) {Solar Gradual Hard
  X-Ray Bursts and Associated Phenomena}. \apj 305:920. \doi{10.1086/164306}

\bibitem[{{Cliver} et~al(2022{\natexlab{a}}){Cliver}, {P{\"o}tzi}, and
  {Veronig}}]{Cliver2022B}
{Cliver} EW, {P{\"o}tzi} W, {Veronig} AM (2022{\natexlab{a}}) {Large Sunspot
  Groups and Great Magnetic Storms: Magnetic Suppression of CMEs}. \apj
  938(2):136. \doi{10.3847/1538-4357/ac847d}

\bibitem[{{Cliver} et~al(2022{\natexlab{b}}){Cliver}, {Schrijver}, {Shibata},
  and {Usoskin}}]{Cliver2022}
{Cliver} EW, {Schrijver} CJ, {Shibata} K, et~al (2022{\natexlab{b}}) {Extreme
  solar events}. Living Reviews in Solar Physics 19(1):2.
  \doi{10.1007/s41116-022-00033-8},
  {\href{https://arxiv.org/abs/2205.09265}{{arXiv:2205.09265}}} {[astro-ph.SR]}

\bibitem[{{Condon} and {Shortley}(1963)}]{Condon1963}
{Condon} EU, {Shortley} GH (1963) {The theory of atomic spectra}

\bibitem[{{Covey} et~al(2007){Covey}, {Ivezi{\'c}}, {Schlegel}, {Finkbeiner},
  {Padmanabhan}, {Lupton}, {Ag{\"u}eros}, {Bochanski}, {Hawley}, {West},
  {Seth}, {Kimball}, {Gogarten}, {Claire}, {Haggard}, {Kaib}, {Schneider}, and
  {Sesar}}]{Covey2007}
{Covey} KR, {Ivezi{\'c}} {\v{Z}}, {Schlegel} D, et~al (2007) {Stellar SEDs from
  0.3 to 2.5 {\ensuremath{\mu}}m: Tracing the Stellar Locus and Searching for
  Color Outliers in the SDSS and 2MASS}. \aj 134(6):2398--2417.
  \doi{10.1086/522052}

\bibitem[{{Cox} and {Tucker}(1969)}]{Cox1969}
{Cox} DP, {Tucker} WH (1969) {Ionization Equilibrium and Radiative Cooling of a
  Low-Density Plasma}. \apj 157:1157. \doi{10.1086/150144}

\bibitem[{{Cram} and {Woods}(1982)}]{Cram1982}
{Cram} LE, {Woods} DT (1982) {Models for stellar flares}. \apj 257:269--275.
  \doi{10.1086/159985}

\bibitem[{{Cranmer}(2021)}]{Cranmer2021}
{Cranmer} SR (2021) {Brown Dwarfs are Violet: A New Calculation of Human-eye
  Colors of Main-sequence Stars and Substellar Objects}. Research Notes of the
  American Astronomical Society 5(9):201. \doi{10.3847/2515-5172/ac225c}

\bibitem[{{Crespo-Chac{\'o}n} et~al(2004){Crespo-Chac{\'o}n}, {Montes},
  {Fern{\'a}ndez-Figueroa}, {L{\'o}pez-Santiago}, {Garc{\'\i}a-Alvarez}, and
  {Foing}}]{Crespo644}
{Crespo-Chac{\'o}n} I, {Montes} D, {Fern{\'a}ndez-Figueroa} MJ, et~al (2004)
  {High Temporal Resolution Spectroscopic Observations of the Flare Star V1054
  Oph}. \apss 292(1):697--703. \doi{10.1023/B:ASTR.0000045077.58363.e4}

\bibitem[{{Crosby} et~al(1993){Crosby}, {Aschwanden}, and
  {Dennis}}]{Crosby1993}
{Crosby} NB, {Aschwanden} MJ, {Dennis} BR (1993) {Frequency distributions and
  correlations of solar X-ray flare parameters}. \solphys 143(2):275--299.
  \doi{10.1007/BF00646488}

\bibitem[{{Crosley} and {Osten}(2018)}]{Crosley2018A}
{Crosley} MK, {Osten} RA (2018) {Constraining Stellar Coronal Mass Ejections
  through Multi-wavelength Analysis of the Active M Dwarf EQ Peg}. \apj
  856(1):39. \doi{10.3847/1538-4357/aaaec2},
  {\href{https://arxiv.org/abs/1802.03440}{{arXiv:1802.03440}}} {[astro-ph.SR]}

\bibitem[{{Cully} et~al(1994){Cully}, {Fisher}, {Abbott}, and
  {Siegmund}}]{Cully1994}
{Cully} SL, {Fisher} GH, {Abbott} MJ, et~al (1994) {A coronal mass ejection
  model for the 1992 July 15 flare on AU Microscopii observed by the extreme
  ultraviolet explorer}. \apj 435:449--463. \doi{10.1086/174827}

\bibitem[{{Dahlin}(2020)}]{Dahlin2020}
{Dahlin} JT (2020) {Prospectus on electron acceleration via magnetic
  reconnection}. Physics of Plasmas 27(10):100601. \doi{10.1063/5.0019338}

\bibitem[{{Dahlin} et~al(2014){Dahlin}, {Drake}, and {Swisdak}}]{Dahlin2014}
{Dahlin} JT, {Drake} JF, {Swisdak} M (2014) {The mechanisms of electron heating
  and acceleration during magnetic reconnection}. Physics of Plasmas
  21(9):092304. \doi{10.1063/1.4894484},
  {\href{https://arxiv.org/abs/1406.0831}{{arXiv:1406.0831}}}
  {[physics.plasm-ph]}

\bibitem[{{Dahlin} et~al(2015){Dahlin}, {Drake}, and {Swisdak}}]{Dahlin2015}
{Dahlin} JT, {Drake} JF, {Swisdak} M (2015) {Electron acceleration in
  three-dimensional magnetic reconnection with a guide field}. Physics of
  Plasmas 22(10):100704. \doi{10.1063/1.4933212},
  {\href{https://arxiv.org/abs/1503.02218}{{arXiv:1503.02218}}}
  {[physics.plasm-ph]}

\bibitem[{{Dal}(2020)}]{Dal2020}
{Dal} HA (2020) {The flare cumulative frequencies of UV Ceti stars from
  different spectral types}. \mnras 495(4):4529--4541.
  \doi{10.1093/mnras/staa1484}

\bibitem[{{Dal} and {Evren}(2010)}]{Dal2010}
{Dal} HA, {Evren} S (2010) {A New Method for Classifying Flares of UV Ceti Type
  Stars: Differences Between Slow and Fast Flares}. \aj 140(2):483--489.
  \doi{10.1088/0004-6256/140/2/483},
  {\href{https://arxiv.org/abs/1206.5791}{{arXiv:1206.5791}}} {[astro-ph.SR]}

\bibitem[{{Dal} and {Evren}(2011{\natexlab{a}})}]{Dal2011B}
{Dal} HA, {Evren} S (2011{\natexlab{a}}) {Rotation Modulations and
  Distributions of the Flare Occurrence Rates on the Surface of Five UV Ceti
  Type Stars}. \pasj 63:427--447. \doi{10.1093/pasj/63.2.427},
  {\href{https://arxiv.org/abs/1206.5792}{{arXiv:1206.5792}}} {[astro-ph.SR]}

\bibitem[{{Dal} and {Evren}(2011{\natexlab{b}})}]{Dal2011A}
{Dal} HA, {Evren} S (2011{\natexlab{b}}) {Saturation Levels for White-light
  Flares of Flare Stars: Variation of Minimum Flare Duration for Saturation}.
  \aj 141(2):33. \doi{10.1088/0004-6256/141/2/33},
  {\href{https://arxiv.org/abs/1206.5793}{{arXiv:1206.5793}}} {[astro-ph.SR]}

\bibitem[{{Damburg} and {Kolosov}(1979)}]{Damburg1979}
{Damburg} RJ, {Kolosov} VV (1979) {A hydrogen atom in a uniform electric field.
  III}. Journal of Physics B Atomic Molecular Physics 12(16):2637--2643.
  \doi{10.1088/0022-3700/12/16/011}

\bibitem[{{Dappen} et~al(1987){Dappen}, {Anderson}, and {Mihalas}}]{Dappen1987}
{Dappen} W, {Anderson} L, {Mihalas} D (1987) {Statistical mechanics of
  partially ionized stellar plasma - The Planck-Larkin partition function,
  polarization shifts, and simulations of optical spectra}. \apj 319:195--206.
  \doi{10.1086/165446}

\bibitem[{{Davenport}(2016)}]{Davenport2016}
{Davenport} JRA (2016) {The Kepler Catalog of Stellar Flares}. \apj 829:23.
  \doi{10.3847/0004-637X/829/1/23},
  {\href{https://arxiv.org/abs/1607.03494}{{arXiv:1607.03494}}} {[astro-ph.SR]}

\bibitem[{{Davenport} et~al(2012){Davenport}, {Becker}, {Kowalski}, {Hawley},
  {Schmidt}, {Hilton}, {Sesar}, and {Cutri}}]{Davenport2012}
{Davenport} JRA, {Becker} AC, {Kowalski} AF, et~al (2012) {Multi-wavelength
  Characterization of Stellar Flares on Low-mass Stars Using SDSS and 2MASS
  Time-domain Surveys}. \apj 748(1):58. \doi{10.1088/0004-637X/748/1/58}

\bibitem[{{Davenport} et~al(2014){Davenport}, {Hawley}, {Hebb}, {Wisniewski},
  {Kowalski}, {Johnson}, {Malatesta}, {Peraza}, {Keil}, {Silverberg}, {Jansen},
  {Scheffler}, {Berdis}, {Larsen}, and {Hilton}}]{Davenport2014}
{Davenport} JRA, {Hawley} SL, {Hebb} L, et~al (2014) {Kepler Flares. II. The
  Temporal Morphology of White-light Flares on GJ 1243}. \apj 797:122.
  \doi{10.1088/0004-637X/797/2/122},
  {\href{https://arxiv.org/abs/1411.3723}{{arXiv:1411.3723}}} {[astro-ph.SR]}

\bibitem[{{Davenport} et~al(2015){Davenport}, {Hebb}, and
  {Hawley}}]{DavenportSpots}
{Davenport} JRA, {Hebb} L, {Hawley} SL (2015) {Detecting Differential Rotation
  and Starspot Evolution on the M Dwarf GJ 1243 with Kepler}. \apj 806(2):212.
  \doi{10.1088/0004-637X/806/2/212},
  {\href{https://arxiv.org/abs/1505.01524}{{arXiv:1505.01524}}} {[astro-ph.SR]}

\bibitem[{{Davenport} et~al(2016){Davenport}, {Kipping}, {Sasselov},
  {Matthews}, and {Cameron}}]{Davenport2016Prox}
{Davenport} JRA, {Kipping} DM, {Sasselov} D, et~al (2016) {MOST Observations of
  Our Nearest Neighbor: Flares on Proxima Centauri}. \apjl 829(2):L31.
  \doi{10.3847/2041-8205/829/2/L31},
  {\href{https://arxiv.org/abs/1608.06672}{{arXiv:1608.06672}}} {[astro-ph.SR]}

\bibitem[{{Davenport} et~al(2019){Davenport}, {Covey}, {Clarke}, {Boeck},
  {Cornet}, and {Hawley}}]{Davenport2019}
{Davenport} JRA, {Covey} KR, {Clarke} RW, et~al (2019) {The Evolution of Flare
  Activity with Stellar Age}. \apj 871(2):241. \doi{10.3847/1538-4357/aafb76}

\bibitem[{{Davenport} et~al(2020){Davenport}, {Mendoza}, and
  {Hawley}}]{Davenport2020}
{Davenport} JRA, {Mendoza} GT, {Hawley} SL (2020) {10 Years of Stellar Activity
  for GJ 1243}. \aj 160(1):36. \doi{10.3847/1538-3881/ab9536},
  {\href{https://arxiv.org/abs/2005.10281}{{arXiv:2005.10281}}} {[astro-ph.SR]}

\bibitem[{{Davison} et~al(2015){Davison}, {White}, {Henry}, {Riedel}, {Jao},
  {Bailey}, {Quinn}, {Cantrell}, {Subasavage}, and {Winters}}]{Davison2015}
{Davison} CL, {White} RJ, {Henry} TJ, et~al (2015) {A 3D Search for Companions
  to 12 Nearby M Dwarfs}. \aj 149(3):106. \doi{10.1088/0004-6256/149/3/106}

\bibitem[{{De Luca} et~al(2020){De Luca}, {Stelzer}, {Burgasser}, {Pizzocaro},
  {Ranalli}, {Raetz}, {Marelli}, {Novara}, {Vignali}, {Belfiore}, {Esposito},
  {Franzetti}, {Fumana}, {Gilli}, {Salvaterra}, and {Tiengo}}]{DeLuca2020}
{De Luca} A, {Stelzer} B, {Burgasser} AJ, et~al (2020) {EXTraS discovery of an
  X-ray superflare from an L dwarf}. \aap 634:L13.
  \doi{10.1051/0004-6361/201937163},
  {\href{https://arxiv.org/abs/2002.08078}{{arXiv:2002.08078}}} {[astro-ph.SR]}

\bibitem[{{De Pontieu} et~al(2014){De Pontieu}, {Title}, {Lemen}, {Kushner},
  {Akin}, {Allard}, {Berger}, {Boerner}, {Cheung}, {Chou}, {Drake}, {Duncan},
  {Freeland}, {Heyman}, {Hoffman}, {Hurlburt}, {Lindgren}, {Mathur}, {Rehse},
  {Sabolish}, {Seguin}, {Schrijver}, {Tarbell}, {W{\"u}lser}, {Wolfson},
  {Yanari}, {Mudge}, {Nguyen-Phuc}, {Timmons}, {van Bezooijen}, {Weingrod},
  {Brookner}, {Butcher}, {Dougherty}, {Eder}, {Knagenhjelm}, {Larsen},
  {Mansir}, {Phan}, {Boyle}, {Cheimets}, {DeLuca}, {Golub}, {Gates}, {Hertz},
  {McKillop}, {Park}, {Perry}, {Podgorski}, {Reeves}, {Saar}, {Testa}, {Tian},
  {Weber}, {Dunn}, {Eccles}, {Jaeggli}, {Kankelborg}, {Mashburn}, {Pust},
  {Springer}, {Carvalho}, {Kleint}, {Marmie}, {Mazmanian}, {Pereira}, {Sawyer},
  {Strong}, {Worden}, {Carlsson}, {Hansteen}, {Leenaarts}, {Wiesmann},
  {Aloise}, {Chu}, {Bush}, {Scherrer}, {Brekke}, {Martinez-Sykora}, {Lites},
  {McIntosh}, {Uitenbroek}, {Okamoto}, {Gummin}, {Auker}, {Jerram}, {Pool}, and
  {Waltham}}]{DePontieu2014}
{De Pontieu} B, {Title} AM, {Lemen} JR, et~al (2014) {The Interface Region
  Imaging Spectrograph (IRIS)}. \solphys 289:2733--2779.
  \doi{10.1007/s11207-014-0485-y},
  {\href{https://arxiv.org/abs/1401.2491}{{arXiv:1401.2491}}} {[astro-ph.SR]}

\bibitem[{{Dennis} and {Zarro}(1993)}]{Dennis1993}
{Dennis} BR, {Zarro} DM (1993) {The Neupert effect - What can it tell us about
  the impulsive and gradual phases of solar flares?} \solphys 146:177--190.
  \doi{10.1007/BF00662178}

\bibitem[{{Dennis} et~al(2011){Dennis}, {Emslie}, and {Hudson}}]{Dennis2011}
{Dennis} BR, {Emslie} AG, {Hudson} HS (2011) {Overview of the Volume}. \ssr
  159(1-4):3--17. \doi{10.1007/s11214-011-9802-z},
  {\href{https://arxiv.org/abs/1109.5831}{{arXiv:1109.5831}}} {[astro-ph.SR]}

\bibitem[{{Dennis} et~al(2022){Dennis}, {Shih}, {Hurford}, and
  {Saint-Hilaire}}]{Dennis2022}
{Dennis} BR, {Shih} AY, {Hurford} GJ, et~al (2022) {Ramaty High Energy Solar
  Spectroscopic Imager (RHESSI)}. arXiv e-prints arXiv:2206.00741.
  {\href{https://arxiv.org/abs/2206.00741}{{arXiv:2206.00741}}} {[astro-ph.SR]}

\bibitem[{{Dere} et~al(1997){Dere}, {Landi}, {Mason}, {Monsignori Fossi}, and
  {Young}}]{Dere1997}
{Dere} KP, {Landi} E, {Mason} HE, et~al (1997) {CHIANTI - an atomic database
  for emission lines}. \aaps 125:149--173. \doi{10.1051/aas:1997368}

\bibitem[{{Dhillon} et~al(2007){Dhillon}, {Marsh}, {Stevenson}, {Atkinson},
  {Kerry}, {Peacocke}, {Vick}, {Beard}, {Ives}, {Lunney}, {McLay}, {Tierney},
  {Kelly}, {Littlefair}, {Nicholson}, {Pashley}, {Harlaftis}, and
  {O'Brien}}]{Dhillon2007}
{Dhillon} VS, {Marsh} TR, {Stevenson} MJ, et~al (2007) {ULTRACAM: an ultrafast,
  triple-beam CCD camera for high-speed astrophysics}. \mnras 378(3):825--840.
  \doi{10.1111/j.1365-2966.2007.11881.x}

\bibitem[{{Dimitrijevi{\'c}} and {Sahal-Br{\'e}chot}(1992)}]{STARKB_CaIIA}
{Dimitrijevi{\'c}} MS, {Sahal-Br{\'e}chot} S (1992) {Stark broadening parameter
  tables for Ca II lines of astrophysical interest.} Bulletin Astronomique de
  Belgrade 145:83--99

\bibitem[{{Dimitrijevic} and {Sahal-Brechot}(1993)}]{STARKB_CaIIB}
{Dimitrijevic} MS, {Sahal-Brechot} S (1993) {Stark broadening of Ca II spectral
  lines}. \jqsrt 49(2):157--164. \doi{10.1016/0022-4073(93)90056-N}

\bibitem[{{Donati} et~al(1997){Donati}, {Semel}, {Carter}, {Rees}, and {Collier
  Cameron}}]{PolarBase1}
{Donati} JF, {Semel} M, {Carter} BD, et~al (1997) {Spectropolarimetric
  observations of active stars}. \mnras 291(4):658--682.
  \doi{10.1093/mnras/291.4.658}

\bibitem[{{Donati-Falchi} et~al(1985){Donati-Falchi}, {Falciani}, and
  {Smaldone}}]{Donati1985}
{Donati-Falchi} A, {Falciani} R, {Smaldone} LA (1985) {Analysis of the optical
  spectra of solar flares. IV - The 'blue' continuum of white light flares}.
  \aap 152:165--169

\bibitem[{{Dorfi}(1998)}]{Dorfi1998}
{Dorfi} EA (1998) {Radiation Hydrodynamics: Numerical Aspects and
  Applications}. In: {Steiner} O, {Gautschy} A (eds) Saas-Fee Advanced Course
  27: Computational Methods for Astrophysical Fluid Flow., p 263

\bibitem[{{Dorfi} and {Drury}(1987)}]{Dorfi1987}
{Dorfi} EA, {Drury} LO (1987) {Simple adaptive grids for 1-D initial value
  problems.} Journal of Computational Physics 69:175--195.
  \doi{10.1016/0021-9991(87)90161-6}

\bibitem[{{Doyle} and {Mathioudakis}(1990)}]{Doyle1990}
{Doyle} JG, {Mathioudakis} M (1990) {Flare activity and orbital rotation of YY
  Geminorum.} \aap 227:130--132

\bibitem[{{Doyle} et~al(1988{\natexlab{a}}){Doyle}, {Butler}, {Bryne}, and {van
  den Oord}}]{Doyle1988}
{Doyle} JG, {Butler} CJ, {Bryne} PB, et~al (1988{\natexlab{a}}) {Rotational
  modulation and flares on RS CVn and BY DRA systems}. \aap 193:229--247

\bibitem[{{Doyle} et~al(1988{\natexlab{b}}){Doyle}, {Butler}, {Callanan},
  {Tagliaferri}, {de La Reza}, {White}, {Torres}, and {Quast}}]{Doyle644}
{Doyle} JG, {Butler} CJ, {Callanan} PJ, et~al (1988{\natexlab{b}}) {Rotational
  modulation and flares on RS CVn and BY DRA systems. VIII. Simultaneous EXOSAT
  and H alpha observations of a flare on the dMe star GL 644 AB (Wolf 630) on
  24/25 August 1985.} \aap 191:79--86

\bibitem[{{Doyle} et~al(1989){Doyle}, {van den Oord}, and {Butler}}]{Doyle1989}
{Doyle} JG, {van den Oord} GHJ, {Butler} CJ (1989) {Optical flares from the
  dwarf M star V577 MON (Gliese 234 AB = Ross 614).} \aap 208:208--212

\bibitem[{{Doyle} et~al(1990{\natexlab{a}}){Doyle}, {Butler}, {van den Oord},
  and {Kiang}}]{Doyle1990B}
{Doyle} JG, {Butler} CJ, {van den Oord} GHJ, et~al (1990{\natexlab{a}}) {A
  periodicity in the flaring rate on the eclipsing binary YY Geminorum.} \aap
  232:83

\bibitem[{{Doyle} et~al(1990{\natexlab{b}}){Doyle}, {Butler}, {van den Oord},
  and {Kiang}}]{Doyle1991}
{Doyle} JG, {Butler} CJ, {van den Oord} GHJ, et~al (1990{\natexlab{b}}) {A
  periodicity in the flaring rate on the eclipsing binary YY Geminorum.} \aap
  232:83

\bibitem[{{Doyle} et~al(2012){Doyle}, {Giunta}, {Singh}, {Madjarska},
  {Summers}, {Kellett}, and {O'Mullane}}]{ADAS1}
{Doyle} JG, {Giunta} A, {Singh} A, et~al (2012) {The Diagnostic Potential of
  Transition Region Lines Undergoing Transient Ionization in Dynamic Events}.
  \solphys 280(1):111--124. \doi{10.1007/s11207-012-0025-6},
  {\href{https://arxiv.org/abs/1204.6598}{{arXiv:1204.6598}}} {[astro-ph.SR]}

\bibitem[{{Doyle} et~al(2013){Doyle}, {Giunta}, {Madjarska}, {Summers},
  {O'Mullane}, and {Singh}}]{ADAS2}
{Doyle} JG, {Giunta} A, {Madjarska} MS, et~al (2013) {Diagnosing transient
  ionization in dynamic events}. \aap 557:L9.
  \doi{10.1051/0004-6361/201321902},
  {\href{https://arxiv.org/abs/1307.8251}{{arXiv:1307.8251}}} {[astro-ph.SR]}

\bibitem[{{Doyle} et~al(2022){Doyle}, {Irawati}, {Kolotkov}, {Ramsay},
  {Nhalil}, {Dhillon}, {Marsh}, and {Yadav}}]{Doyle2022QPP}
{Doyle} JG, {Irawati} P, {Kolotkov} DY, et~al (2022) {Doubling of minute-long
  quasi-periodic pulsations from super-flares on a low-mass star}. \mnras
  514(4):5178--5182. \doi{10.1093/mnras/stac1695},
  {\href{https://arxiv.org/abs/2206.08070}{{arXiv:2206.08070}}} {[astro-ph.SR]}

\bibitem[{{Doyle} et~al(2018){Doyle}, {Ramsay}, {Doyle}, {Wu}, and
  {Scullion}}]{Doyle2018}
{Doyle} L, {Ramsay} G, {Doyle} JG, et~al (2018) {Investigating the rotational
  phase of stellar flares on M dwarfs using K2 short cadence data}. \mnras
  480(2):2153--2164. \doi{10.1093/mnras/sty1963},
  {\href{https://arxiv.org/abs/1807.08592}{{arXiv:1807.08592}}} {[astro-ph.SR]}

\bibitem[{{Drake}(2006)}]{Springer2006}
{Drake} GWF (2006) {Springer Handbook of Atomic, Molecular, and Optical
  Physics}. \doi{10.1007/978-0-387-26308-3}

\bibitem[{{Drake} et~al(2006){Drake}, {Swisdak}, {Che}, and {Shay}}]{Drake2006}
{Drake} JF, {Swisdak} M, {Che} H, et~al (2006) {Electron acceleration from
  contracting magnetic islands during reconnection}. \nat 443(7111):553--556.
  \doi{10.1038/nature05116}

\bibitem[{{Drake} et~al(2013){Drake}, {Swisdak}, and {Fermo}}]{Drake2013}
{Drake} JF, {Swisdak} M, {Fermo} R (2013) {The Power-law Spectra of Energetic
  Particles during Multi-island Magnetic Reconnection}. \apjl 763(1):L5.
  \doi{10.1088/2041-8205/763/1/L5},
  {\href{https://arxiv.org/abs/1210.4830}{{arXiv:1210.4830}}} {[astro-ph.SR]}

\bibitem[{{Drake} et~al(2019){Drake}, {Arnold}, {Swisdak}, and
  {Dahlin}}]{Drake2019}
{Drake} JF, {Arnold} H, {Swisdak} M, et~al (2019) {A computational model for
  exploring particle acceleration during reconnection in macroscale systems}.
  Physics of Plasmas 26(1):012901. \doi{10.1063/1.5058140},
  {\href{https://arxiv.org/abs/1809.04568}{{arXiv:1809.04568}}} {[astro-ph.SR]}

\bibitem[{{Drake}(2003)}]{Drake2003}
{Drake} JJ (2003) {From the Heart of the Ghoul: C and N Abundances in the
  Corona of Algol B}. \apj 594(1):496--509. \doi{10.1086/375837},
  {\href{https://arxiv.org/abs/astro-ph/0308230}{{arXiv:astro-ph/0308230}}}
  {[astro-ph]}

\bibitem[{{Drake} and {Ercolano}(2007)}]{DE07}
{Drake} JJ, {Ercolano} B (2007) {The Detectability of Neon Fluorescence and
  Measurement of the Solar Photospheric Neon Abundance}. \apjl
  665(2):L175--L178. \doi{10.1086/521190},
  {\href{https://arxiv.org/abs/0708.1022}{{arXiv:0708.1022}}} {[astro-ph]}

\bibitem[{{Drake} et~al(2008){Drake}, {Ercolano}, and {Swartz}}]{Drake2008}
{Drake} JJ, {Ercolano} B, {Swartz} DA (2008) {X-Ray-fluorescent Fe
  K{\ensuremath{\alpha}} Lines from Stellar Photospheres}. \apj
  678(1):385--393. \doi{10.1086/524976},
  {\href{https://arxiv.org/abs/0710.0621}{{arXiv:0710.0621}}} {[astro-ph]}

\bibitem[{{Drake} and {Ulrich}(1980)}]{Drake1980}
{Drake} SA, {Ulrich} RK (1980) {The emission-line spectrum from a slab of
  hydrogen at moderate to high densities}. \apjs 42:351--383.
  \doi{10.1086/190654}

\bibitem[{{Dud{\'\i}k} et~al(2017){Dud{\'\i}k}, {Polito},
  {Dzif{\v{c}}{\'a}kov{\'a}}, {Del Zanna}, and {Testa}}]{Dudik2017}
{Dud{\'\i}k} J, {Polito} V, {Dzif{\v{c}}{\'a}kov{\'a}} E, et~al (2017)
  {Non-Maxwellian Analysis of the Transition-region Line Profiles Observed by
  the Interface Region Imaging Spectrograph}. \apj 842(1):19.
  \doi{10.3847/1538-4357/aa71a8},
  {\href{https://arxiv.org/abs/1705.02104}{{arXiv:1705.02104}}} {[astro-ph.SR]}

\bibitem[{{Dulk}(1985)}]{Dulk1985}
{Dulk} GA (1985) {Radio emission from the sun and stars}. \araa 23:169--224.
  \doi{10.1146/annurev.aa.23.090185.001125}

\bibitem[{{Dulk} and {Marsh}(1982)}]{Dulk1982}
{Dulk} GA, {Marsh} KA (1982) {Simplified expressions for the gyrosynchrotron
  radiation from mildly relativistic, nonthermal and thermal electrons}. \apj
  259:350--358. \doi{10.1086/160171}

\bibitem[{{Eason} et~al(1992){Eason}, {Giampapa}, {Radick}, {Worden}, and
  {Hege}}]{Eason1992}
{Eason} ELE, {Giampapa} MS, {Radick} RR, et~al (1992) {Spectroscopic and
  photometric observations of a five-magnitude flare event on UV Ceti}. \aj
  104:1161--1173. \doi{10.1086/116305}

\bibitem[{{Egedal} et~al(2015){Egedal}, {Daughton}, {Le}, and
  {Borg}}]{Egedal2015}
{Egedal} J, {Daughton} W, {Le} A, et~al (2015) {Double layer electric fields
  aiding the production of energetic flat-top distributions and superthermal
  electrons within magnetic reconnection exhausts}. Physics of Plasmas
  22(10):101208. \doi{10.1063/1.4933055},
  {\href{https://arxiv.org/abs/1504.08045}{{arXiv:1504.08045}}}
  {[physics.plasm-ph]}

\bibitem[{{Emslie}(1978)}]{Emslie1978}
{Emslie} AG (1978) {The collisional interaction of a beam of charged particles
  with a hydrogen target of arbitrary ionization level}. \apj 224:241--246.
  \doi{10.1086/156371}

\bibitem[{{Emslie}(1981{\natexlab{a}})}]{Emslie1981}
{Emslie} AG (1981{\natexlab{a}}) {A comparison of the height distributions of
  solar flare hard X-rays in thick target and thermal models}. \apj
  245:711--720. \doi{10.1086/158846}

\bibitem[{{Emslie}(1981{\natexlab{b}})}]{Emslie1981Trigger}
{Emslie} AG (1981{\natexlab{b}}) {An Interacting Loop Model for Solar Flare
  Bursts}. \aplett 22:41

\bibitem[{{Emslie} et~al(1986){Emslie}, {Phillips}, and {Dennis}}]{Emslie1986}
{Emslie} AG, {Phillips} KJH, {Dennis} BR (1986) {The excitation of the iron
  K-alpha feature in solar flares}. \solphys 103:89--102

\bibitem[{{Emslie} et~al(1992){Emslie}, {Li}, and {Mariska}}]{Emslie1992}
{Emslie} AG, {Li} P, {Mariska} JT (1992) {Diagnostics of Electron-heated Solar
  Flare Models. III. Effects of Tapered Loop Geometry and Preheating}. \apj
  399:714. \doi{10.1086/171964}

\bibitem[{{Emslie} et~al(2012){Emslie}, {Dennis}, {Shih}, {Chamberlin},
  {Mewaldt}, {Moore}, {Share}, {Vourlidas}, and {Welsch}}]{Emslie2012}
{Emslie} AG, {Dennis} BR, {Shih} AY, et~al (2012) {Global Energetics of
  Thirty-eight Large Solar Eruptive Events}. \apj 759:71.
  \doi{10.1088/0004-637X/759/1/71},
  {\href{https://arxiv.org/abs/1209.2654}{{arXiv:1209.2654}}} {[astro-ph.SR]}

\bibitem[{{Engle}(2023)}]{Engle2023}
{Engle} SG (2023) {Living with a Red Dwarf: X-ray, UV, and Ca II Activity-Age
  Relationships of M Dwarfs}. arXiv e-prints arXiv:2310.04302.
  \doi{10.48550/arXiv.2310.04302},
  {\href{https://arxiv.org/abs/2310.04302}{{arXiv:2310.04302}}} {[astro-ph.SR]}

\bibitem[{{Engle} and {Guinan}(2023)}]{EngleGuinan2023}
{Engle} SG, {Guinan} EF (2023) {Living with a Red Dwarf: The Rotation-Age
  Relationships of M Dwarfs}. \apjl 954(2):L50. \doi{10.3847/2041-8213/acf472},
  {\href{https://arxiv.org/abs/2307.01136}{{arXiv:2307.01136}}} {[astro-ph.SR]}

\bibitem[{{Ercolano} et~al(2003){Ercolano}, {Barlow}, {Storey}, {Liu}, {Rauch},
  and {Werner}}]{Erc03}
{Ercolano} B, {Barlow} MJ, {Storey} PJ, et~al (2003) {Three-dimensional
  photoionization modelling of the hydrogen-deficient knots in the planetary
  nebula Abell 30}. \mnras 344(4):1145--1154.
  \doi{10.1046/j.1365-8711.2003.06892.x},
  {\href{https://arxiv.org/abs/astro-ph/0306230}{{arXiv:astro-ph/0306230}}}
  {[astro-ph]}

\bibitem[{{Ercolano} et~al(2008){Ercolano}, {Young}, {Drake}, and
  {Raymond}}]{Erc08}
{Ercolano} B, {Young} PR, {Drake} JJ, et~al (2008) {X-Ray Enabled MOCASSIN: A
  Three-dimensional Code for Photoionized Media}. \apjs 175(2):534--542.
  \doi{10.1086/524378},
  {\href{https://arxiv.org/abs/0710.2103}{{arXiv:0710.2103}}} {[astro-ph]}

\bibitem[{{Esteban} et~al(2004){Esteban}, {Peimbert}, {Garc{\'\i}a-Rojas},
  {Ruiz}, {Peimbert}, and {Rodr{\'\i}guez}}]{Esteban2004}
{Esteban} C, {Peimbert} M, {Garc{\'\i}a-Rojas} J, et~al (2004) {A reappraisal
  of the chemical composition of the Orion nebula based on Very Large Telescope
  echelle spectrophotometry}. \mnras 355(1):229--247.
  \doi{10.1111/j.1365-2966.2004.08313.x},
  {\href{https://arxiv.org/abs/astro-ph/0408249}{{arXiv:astro-ph/0408249}}}
  {[astro-ph]}

\bibitem[{{Fang} and {Ding}(1995)}]{Fang1995}
{Fang} C, {Ding} MD (1995) {On the spectral characteristics and atmospheric
  models of two types of white-light flares.} \aaps 110:99

\bibitem[{{Fang} et~al(1993){Fang}, {Henoux}, and {Gan}}]{Fang1993}
{Fang} C, {Henoux} JC, {Gan} WQ (1993) {Diagnostics of non-thermal processes in
  chromospheric flares. I. H alpha and CaII K line profiles of an atmosphere
  bombarded by hecta keV electrons.} \aap 274:917--922

\bibitem[{{Favata} and {Schmitt}(1999)}]{Algol4}
{Favata} F, {Schmitt} JHMM (1999) {Spectroscopic analysis of a super-hot giant
  flare observed on Algol by BeppoSAX on 30 August 1997}. \aap 350:900--916.
  \doi{10.48550/arXiv.astro-ph/9909041},
  {\href{https://arxiv.org/abs/astro-ph/9909041}{{arXiv:astro-ph/9909041}}}
  {[astro-ph]}

\bibitem[{{Favata} et~al(2000){Favata}, {Reale}, {Micela}, {Sciortino},
  {Maggio}, and {Matsumoto}}]{Favata2000}
{Favata} F, {Reale} F, {Micela} G, et~al (2000) {An extreme X-ray flare
  observed on EV Lac by ASCA in July 1998}. \aap 353:987--997.
  {\href{https://arxiv.org/abs/astro-ph/9909491}{{arXiv:astro-ph/9909491}}}
  {[astro-ph]}

\bibitem[{{Feiden} and {Chaboyer}(2013)}]{Feiden2013}
{Feiden} GA, {Chaboyer} B (2013) {Magnetic Inhibition of Convection and the
  Fundamental Properties of Low-mass Stars. I. Stars with a Radiative Core}.
  \apj 779(2):183. \doi{10.1088/0004-637X/779/2/183},
  {\href{https://arxiv.org/abs/1309.0033}{{arXiv:1309.0033}}} {[astro-ph.SR]}

\bibitem[{{Feiden} and {Chaboyer}(2014)}]{Feiden2014}
{Feiden} GA, {Chaboyer} B (2014) {Magnetic Inhibition of Convection and the
  Fundamental Properties of Low-mass Stars. II. Fully Convective Main-sequence
  Stars}. \apj 789(1):53. \doi{10.1088/0004-637X/789/1/53},
  {\href{https://arxiv.org/abs/1405.1767}{{arXiv:1405.1767}}} {[astro-ph.SR]}

\bibitem[{{Feinstein} et~al(2022){Feinstein}, {France}, {Youngblood},
  {Duvvuri}, {Teal}, {Cauley}, {Seligman}, {Gaidos}, {Kempton}, {Bean},
  {Diamond-Lowe}, {Newton}, {Ginzburg}, {Plavchan}, {Gao}, and
  {Schlichting}}]{Feinstein2022}
{Feinstein} AD, {France} K, {Youngblood} A, et~al (2022) {AU Microscopii in the
  Far-UV: Observations in Quiescence, during Flares, and Implications for AU
  Mic b and c}. \aj 164(3):110. \doi{10.3847/1538-3881/ac8107},
  {\href{https://arxiv.org/abs/2205.09606}{{arXiv:2205.09606}}} {[astro-ph.SR]}

\bibitem[{{Fekel}(1983)}]{Fekel1983}
{Fekel} JF.~C. (1983) {Spectroscopy of V711 Tauri (= HR 1099) : fundamental
  properties and evidence for starspots.} \apj 268:274--281.
  \doi{10.1086/160952}

\bibitem[{{Fisher} and {Maron}(2002)}]{FisherMaron2002}
{Fisher} DV, {Maron} Y (2002) {Effective statistical weights of bound states in
  plasmas}. European Physical Journal D 18(1):93--111.
  \doi{10.1140/e10053-002-0012-9}

\bibitem[{{Fisher} and {Maron}(2003)}]{FisherMaron2003}
{Fisher} DV, {Maron} Y (2003) {Characterization of electron states in dense
  plasmas and its use in atomic kinetics modeling}. \jqsrt 81:147--165.
  \doi{10.1016/S0022-4073(03)00068-2}

\bibitem[{{Fisher}(1989)}]{Fisher1989}
{Fisher} GH (1989) {Dynamics of flare-driven chromospheric condensations}. \apj
  346:1019--1029. \doi{10.1086/168084}

\bibitem[{{Fisher} and {Hawley}(1990)}]{Fisher1990}
{Fisher} GH, {Hawley} SL (1990) {An Equation for the Evolution of Solar and
  Stellar Flare Loops}. \apj 357:243. \doi{10.1086/168911}

\bibitem[{{Fisher} et~al(1985{\natexlab{a}}){Fisher}, {Canfield}, and
  {McClymont}}]{Fisher1985VII}
{Fisher} GH, {Canfield} RC, {McClymont} AN (1985{\natexlab{a}}) {Flare Loop
  Radiative Hydrodynamics - Part Seven - Dynamics of the Thick Target Heated
  Chromosphere}. \apj 289:434. \doi{10.1086/162903}

\bibitem[{{Fisher} et~al(1985{\natexlab{b}}){Fisher}, {Canfield}, and
  {McClymont}}]{Fisher1985VI}
{Fisher} GH, {Canfield} RC, {McClymont} AN (1985{\natexlab{b}}) {Flare Loop
  Radiative Hydrodynamics - Part Six - Chromospheric Evaporation due to Heating
  by Nonthermal Electrons}. \apj 289:425. \doi{10.1086/162902}

\bibitem[{{Fisher} et~al(1985{\natexlab{c}}){Fisher}, {Canfield}, and
  {McClymont}}]{Fisher1985V}
{Fisher} GH, {Canfield} RC, {McClymont} AN (1985{\natexlab{c}}) {Flare loop
  radiative hydrodynamics. V - Response to thick-target heating. VI -
  Chromospheric evaporation due to heating by nonthermal electrons. VII -
  Dynamics of the thick-target heated chromosphere}. \apj 289:414--441.
  \doi{10.1086/162901}

\bibitem[{{Fisher} et~al(2012){Fisher}, {Bercik}, {Welsch}, and
  {Hudson}}]{Fisher2012}
{Fisher} GH, {Bercik} DJ, {Welsch} BT, et~al (2012) {Global Forces in Eruptive
  Solar Flares: The Lorentz Force Acting on the Solar Atmosphere and the Solar
  Interior}. \solphys 277:59--76. \doi{10.1007/s11207-011-9907-2},
  {\href{https://arxiv.org/abs/1006.5247}{{arXiv:1006.5247}}} {[astro-ph.SR]}

\bibitem[{{Flaccomio} et~al(2018){Flaccomio}, {Micela}, {Sciortino}, {Cody},
  {Guarcello}, {Morales-Calder{\`o}n}, {Rebull}, and
  {Stauffer}}]{Flaccomio2018}
{Flaccomio} E, {Micela} G, {Sciortino} S, et~al (2018) {A multi-wavelength view
  of magnetic flaring from PMS stars}. \aap 620:A55.
  \doi{10.1051/0004-6361/201833308},
  {\href{https://arxiv.org/abs/1807.08525}{{arXiv:1807.08525}}} {[astro-ph.SR]}

\bibitem[{{Fleishman} et~al(2022){Fleishman}, {Nita}, {Chen}, {Yu}, and
  {Gary}}]{Fleishman2022}
{Fleishman} GD, {Nita} GM, {Chen} B, et~al (2022) {Solar flare accelerates
  nearly all electrons in a large coronal volume}. \nat 606(7915):674--677.
  \doi{10.1038/s41586-022-04728-8}

\bibitem[{{Fleming} et~al(2022){Fleming}, {Million}, {Osten}, {Kolotkov}, and
  {Brasseur}}]{Fleming2022}
{Fleming} SW, {Million} C, {Osten} RA, et~al (2022) {New Time-resolved,
  Multi-band Flares in the GJ 65 System with gPhoton}. \apj 928(1):8.
  \doi{10.3847/1538-4357/ac5037},
  {\href{https://arxiv.org/abs/2202.02861}{{arXiv:2202.02861}}} {[astro-ph.SR]}

\bibitem[{{Fleming} et~al(2000){Fleming}, {Giampapa}, and
  {Schmitt}}]{Fleming2000}
{Fleming} TA, {Giampapa} MS, {Schmitt} JHMM (2000) {An X-Ray Flare Detected on
  the M8 Dwarf VB 10}. \apj 533(1):372--377. \doi{10.1086/308657},
  {\href{https://arxiv.org/abs/astro-ph/0002065}{{arXiv:astro-ph/0002065}}}
  {[astro-ph]}

\bibitem[{{Fletcher} and {Hudson}(2008)}]{Fletcher2008}
{Fletcher} L, {Hudson} HS (2008) {Impulsive Phase Flare Energy Transport by
  Large-Scale Alfv{\'e}n Waves and the Electron Acceleration Problem}. \apj
  675:1645-1655. \doi{10.1086/527044},
  {\href{https://arxiv.org/abs/0712.3452}{{arXiv:0712.3452}}}

\bibitem[{{Fletcher} et~al(2007){Fletcher}, {Hannah}, {Hudson}, and
  {Metcalf}}]{Fletcher2007}
{Fletcher} L, {Hannah} IG, {Hudson} HS, et~al (2007) {A TRACE White Light and
  RHESSI Hard X-Ray Study of Flare Energetics}. \apj 656:1187--1196.
  \doi{10.1086/510446}

\bibitem[{{Fletcher} et~al(2011){Fletcher}, {Dennis}, {Hudson}, {Krucker},
  {Phillips}, {Veronig}, {Battaglia}, {Bone}, {Caspi}, {Chen}, {Gallagher},
  {Grigis}, {Ji}, {Liu}, {Milligan}, and {Temmer}}]{Fletcher2011}
{Fletcher} L, {Dennis} BR, {Hudson} HS, et~al (2011) {An Observational Overview
  of Solar Flares}. \ssr 159(1-4):19--106. \doi{10.1007/s11214-010-9701-8},
  {\href{https://arxiv.org/abs/1109.5932}{{arXiv:1109.5932}}} {[astro-ph.SR]}

\bibitem[{{Fontenla} et~al(2016){Fontenla}, {Linsky}, {Witbrod}, {France},
  {Buccino}, {Mauas}, {Vieytes}, and {Walkowicz}}]{Fontenla2016}
{Fontenla} JM, {Linsky} JL, {Witbrod} J, et~al (2016) {Semi-empirical Modeling
  of the Photosphere, Chromosphere, Transition Region, and Corona of the
  M-dwarf Host Star GJ 832}. \apj 830(2):154. \doi{10.3847/0004-637X/830/2/154}

\bibitem[{{France} et~al(2016){France}, {Loyd}, {Youngblood}, {Brown},
  {Schneider}, {Hawley}, {Froning}, {Linsky}, {Roberge}, {Buccino},
  {Davenport}, {Fontenla}, {Kaltenegger}, {Kowalski}, {Mauas}, {Miguel},
  {Redfield}, {Rugheimer}, {Tian}, {Vieytes}, {Walkowicz}, and
  {Weisenburger}}]{France2016}
{France} K, {Loyd} ROP, {Youngblood} A, et~al (2016) {The MUSCLES Treasury
  Survey. I. Motivation and Overview}. \apj 820(2):89.
  \doi{10.3847/0004-637X/820/2/89},
  {\href{https://arxiv.org/abs/1602.09142}{{arXiv:1602.09142}}} {[astro-ph.SR]}

\bibitem[{{France} et~al(2020){France}, {Duvvuri}, {Egan}, {Koskinen},
  {Wilson}, {Youngblood}, {Froning}, {Brown}, {Alvarado-G{\'o}mez},
  {Berta-Thompson}, {Drake}, {Garraffo}, {Kaltenegger}, {Kowalski}, {Linsky},
  {Loyd}, {Mauas}, {Miguel}, {Pineda}, {Rugheimer}, {Schneider}, {Tian}, and
  {Vieytes}}]{France2020}
{France} K, {Duvvuri} G, {Egan} H, et~al (2020) {The High-energy Radiation
  Environment around a 10 Gyr M Dwarf: Habitable at Last?} \aj 160(5):237.
  \doi{10.3847/1538-3881/abb465},
  {\href{https://arxiv.org/abs/2009.01259}{{arXiv:2009.01259}}} {[astro-ph.EP]}

\bibitem[{{Frogner} et~al(2020){Frogner}, {Gudiksen}, and
  {Bakke}}]{Frogner2020}
{Frogner} L, {Gudiksen} BV, {Bakke} H (2020) {Accelerated particle beams in a
  3D simulation of the quiet Sun}. \aap 643:A27.
  \doi{10.1051/0004-6361/202038529},
  {\href{https://arxiv.org/abs/2005.14483}{{arXiv:2005.14483}}} {[astro-ph.SR]}

\bibitem[{{Froning} et~al(2019){Froning}, {Kowalski}, {France}, {Loyd},
  {Schneider}, {Youngblood}, {Wilson}, {Brown}, {Berta-Thompson}, {Pineda},
  {Linsky}, {Rugheimer}, and {Miguel}}]{Froning2019}
{Froning} CS, {Kowalski} A, {France} K, et~al (2019) {A Hot Ultraviolet Flare
  on the M Dwarf Star GJ 674}. \apjl 871:L26. \doi{10.3847/2041-8213/aaffcd}

\bibitem[{{Froning} et~al(2022){Froning}, {Wilson}, {France}, {Youngblood},
  {Duvvuri}, {Brown}, {Mega-Muscles Collaboration}, and
  {Schneider}}]{Froning2022}
{Froning} CS, {Wilson} D, {France} K, et~al (2022) {The Mega-MUSCLES Treasury
  Survey}. In: Bulletin of the American Astronomical Society, p 102.23

\bibitem[{{Fuhrmeister} et~al(2005){Fuhrmeister}, {Schmitt}, and
  {Hauschildt}}]{Fuhrmeister2005}
{Fuhrmeister} B, {Schmitt} JHMM, {Hauschildt} PH (2005) {Detection of red line
  asymmetries in LHS 2034}. \aap 436(2):677--686.
  \doi{10.1051/0004-6361:20042518}

\bibitem[{{Fuhrmeister} et~al(2007){Fuhrmeister}, {Liefke}, and
  {Schmitt}}]{Fuhrmeister2007}
{Fuhrmeister} B, {Liefke} C, {Schmitt} JHMM (2007) {Simultaneous XMM-Newton and
  VLT/UVES observations of the flare star CN Leonis}. \aap 468(1):221--231.
  \doi{10.1051/0004-6361:20066229}

\bibitem[{{Fuhrmeister} et~al(2008){Fuhrmeister}, {Liefke}, {Schmitt}, and
  {Reiners}}]{Fuhrmeister2008}
{Fuhrmeister} B, {Liefke} C, {Schmitt} JHMM, et~al (2008) {Multiwavelength
  observations of a giant flare on CN Leonis. I. The chromosphere as seen in
  the optical spectra}. \aap 487(1):293--306.
  \doi{10.1051/0004-6361:200809379},
  {\href{https://arxiv.org/abs/0807.2025}{{arXiv:0807.2025}}} {[astro-ph]}

\bibitem[{{Fuhrmeister} et~al(2010){Fuhrmeister}, {Schmitt}, and
  {Hauschildt}}]{Fuhrmeister2010}
{Fuhrmeister} B, {Schmitt} JHMM, {Hauschildt} PH (2010) {Multi-wavelength
  observations of a giant flare on CN Leonis. II. Chromospheric modelling with
  PHOENIX}. \aap 511:A83. \doi{10.1051/0004-6361/200810224}

\bibitem[{{Fuhrmeister} et~al(2011){Fuhrmeister}, {Lalitha}, {Poppenhaeger},
  {Rudolf}, {Liefke}, {Reiners}, {Schmitt}, and {Ness}}]{Fuhrmeister2011}
{Fuhrmeister} B, {Lalitha} S, {Poppenhaeger} K, et~al (2011) {Multi-wavelength
  observations of Proxima Centauri}. \aap 534:A133.
  \doi{10.1051/0004-6361/201117447},
  {\href{https://arxiv.org/abs/1109.1130}{{arXiv:1109.1130}}} {[astro-ph.SR]}

\bibitem[{{Fuhrmeister} et~al(2018){Fuhrmeister}, {Czesla}, {Schmitt},
  {Jeffers}, {Caballero}, {Zechmeister}, {Reiners}, {Ribas}, {Amado},
  {Quirrenbach}, {B{\'e}jar}, {Galad{\'\i}-Enr{\'\i}quez}, {Guenther},
  {K{\"u}rster}, {Montes}, and {Seifert}}]{Fuhrmeister2018}
{Fuhrmeister} B, {Czesla} S, {Schmitt} JHMM, et~al (2018) {The CARMENES search
  for exoplanets around M dwarfs. Wing asymmetries of H{\ensuremath{\alpha}},
  Na I D, and He I lines}. \aap 615:A14. \doi{10.1051/0004-6361/201732204},
  {\href{https://arxiv.org/abs/1801.10372}{{arXiv:1801.10372}}} {[astro-ph.SR]}

\bibitem[{{Fuhrmeister} et~al(2020){Fuhrmeister}, {Czesla}, {Hildebrandt},
  {Nagel}, {Schmitt}, {Jeffers}, {Caballero}, {Hintz}, {Johnson},
  {Sch{\"o}fer}, {Zechmeister}, {Reiners}, {Ribas}, {Amado}, {Quirrenbach},
  {Nortmann}, {Bauer}, {B{\'e}jar}, {Cort{\'e}s-Contreras}, {Dreizler},
  {Galad{\'\i}-Enr{\'\i}quez}, {Hatzes}, {Kaminski}, {K{\"u}rster}, {Lafarga},
  and {Montes}}]{Fuhrmeister2020}
{Fuhrmeister} B, {Czesla} S, {Hildebrandt} L, et~al (2020) {The CARMENES search
  for exoplanets around M dwarfs. Variability of the He I line at 10 830
  {\r{A}}}. \aap 640:A52. \doi{10.1051/0004-6361/202038279},
  {\href{https://arxiv.org/abs/2006.09372}{{arXiv:2006.09372}}} {[astro-ph.SR]}

\bibitem[{{Gaia Collaboration} et~al(2018){Gaia Collaboration}, {Brown},
  {Vallenari}, {Prusti}, {de Bruijne}, {Babusiaux}, and
  {Bailer-Jones}}]{GaiaDR2}
{Gaia Collaboration}, {Brown} AGA, {Vallenari} A, et~al (2018) {Gaia Data
  Release 2. Summary of the contents and survey properties}. ArXiv e-prints
  {\href{https://arxiv.org/abs/1804.09365}{{arXiv:1804.09365}}}

\bibitem[{{Gallagher}(2006)}]{Gallagher2006}
{Gallagher} T (2006) {Rydberg Atoms}, p 235.
  \doi{10.1007/978-0-387-26308-3\_14}

\bibitem[{{Gan} and {Fang}(1990)}]{GanFang1990}
{Gan} WQ, {Fang} C (1990) {A Hydrodynamic Model of the Gradual Phase of the
  Solar Flare Loop}. \apj 358:328. \doi{10.1086/168989}

\bibitem[{{Gan} et~al(1992){Gan}, {Rieger}, {Zhang}, and {Fang}}]{Gan1992}
{Gan} WQ, {Rieger} E, {Zhang} HQ, et~al (1992) {The Role of Chromospheric
  Condensations in the Continuum Emission of White-Light Flares}. \apj 397:694.
  \doi{10.1086/171825}

\bibitem[{{Gao} et~al(2008){Gao}, {Chen}, {Ding}, and {Li}}]{Gao2008}
{Gao} DH, {Chen} PF, {Ding} MD, et~al (2008) {Simulations of the periodic
  flaring rate on YY Gem}. \mnras 384(4):1355--1362.
  \doi{10.1111/j.1365-2966.2007.12830.x},
  {\href{https://arxiv.org/abs/0712.2300}{{arXiv:0712.2300}}} {[astro-ph]}

\bibitem[{{Garc{\'{\i}}a-Alvarez} et~al(2002){Garc{\'{\i}}a-Alvarez},
  {Jevremovi{\'c}}, {Doyle}, and {Butler}}]{Garcia2002}
{Garc{\'{\i}}a-Alvarez} D, {Jevremovi{\'c}} D, {Doyle} JG, et~al (2002)
  {Observations and modelling of a large optical flare on AT Microscopii}. \aap
  383:548--557. \doi{10.1051/0004-6361:20011743},
  {\href{https://arxiv.org/abs/astro-ph/0112224}{{astro-ph/0112224}}}

\bibitem[{{Gary} et~al(2018){Gary}, {Chen}, {Dennis}, {Fleishman}, {Hurford},
  {Krucker}, {McTiernan}, {Nita}, {Shih}, {White}, and {Yu}}]{Gary2018}
{Gary} DE, {Chen} B, {Dennis} BR, et~al (2018) {Microwave and Hard X-Ray
  Observations of the 2017 September 10 Solar Limb Flare}. \apj 863(1):83.
  \doi{10.3847/1538-4357/aad0ef},
  {\href{https://arxiv.org/abs/1807.02498}{{arXiv:1807.02498}}} {[astro-ph.SR]}

\bibitem[{{Gehrels}(1986)}]{Gehrels1986}
{Gehrels} N (1986) {Confidence Limits for Small Numbers of Events in
  Astrophysical Data}. \apj 303:336. \doi{10.1086/164079}

\bibitem[{{Gershberg}(1972)}]{Gershberg1972}
{Gershberg} RE (1972) {Some results of the cooperative photometric observations
  of the UV Cet-type flare stars in the years 1967 71}. \apss 19:75--92.
  \doi{10.1007/BF00643168}

\bibitem[{{Gershberg} et~al(1999){Gershberg}, {Katsova}, {Lovkaya}, {Terebizh},
  and {Shakhovskaya}}]{GershbergAtlas}
{Gershberg} RE, {Katsova} MM, {Lovkaya} MN, et~al (1999) {Catalogue and
  bibliography of the UV Cet-type flare stars and related objects in the solar
  vicinity}. \aaps 139:555--558. \doi{10.1051/aas:1999407}

\bibitem[{{Getman} et~al(2005){Getman}, {Flaccomio}, {Broos}, {Grosso},
  {Tsujimoto}, {Townsley}, {Garmire}, {Kastner}, {Li}, {Harnden}, {Wolk},
  {Murray}, {Lada}, {Muench}, {McCaughrean}, {Meeus}, {Damiani}, {Micela},
  {Sciortino}, {Bally}, {Hillenbrand}, {Herbst}, {Preibisch}, and
  {Feigelson}}]{Getman2005}
{Getman} KV, {Flaccomio} E, {Broos} PS, et~al (2005) {Chandra Orion Ultradeep
  Project: Observations and Source Lists}. \apjs 160(2):319--352.
  \doi{10.1086/432092},
  {\href{https://arxiv.org/abs/astro-ph/0410136}{{arXiv:astro-ph/0410136}}}
  {[astro-ph]}

\bibitem[{{Getman} et~al(2008){Getman}, {Feigelson}, {Broos}, {Micela}, and
  {Garmire}}]{Getman2008}
{Getman} KV, {Feigelson} ED, {Broos} PS, et~al (2008) {X-Ray Flares in Orion
  Young Stars. I. Flare Characteristics}. \apj 688(1):418--436.
  \doi{10.1086/592033},
  {\href{https://arxiv.org/abs/0807.3005}{{arXiv:0807.3005}}} {[astro-ph]}

\bibitem[{{Getman} et~al(2011){Getman}, {Broos}, {Salter}, {Garmire}, and
  {Hogerheijde}}]{Getman2011}
{Getman} KV, {Broos} PS, {Salter} DM, et~al (2011) {The Young Binary DQ Tau: A
  Hunt for X-ray Emission from Colliding Magnetospheres}. \apj 730(1):6.
  \doi{10.1088/0004-637X/730/1/6},
  {\href{https://arxiv.org/abs/1101.4044}{{arXiv:1101.4044}}} {[astro-ph.SR]}

\bibitem[{{Getman} et~al(2023){Getman}, {K{\'o}sp{\'a}l}, {Arulanantham},
  {Semenov}, {Smirnov-Pinchukov}, and {van Terwisga}}]{Getman2023}
{Getman} KV, {K{\'o}sp{\'a}l} {\'A}, {Arulanantham} N, et~al (2023) {X-Ray,
  Near-ultraviolet, and Optical Flares Produced by Colliding Magnetospheres in
  the Young High-eccentricity Binary DQ Tau}. \apj 959(2):98.
  \doi{10.3847/1538-4357/ad054c},
  {\href{https://arxiv.org/abs/2310.12811}{{arXiv:2310.12811}}} {[astro-ph.SR]}

\bibitem[{{Gezari} et~al(2013){Gezari}, {Martin}, {Forster}, {Neill}, {Huber},
  {Heckman}, {Bianchi}, {Morrissey}, {Neff}, {Seibert}, {Schiminovich},
  {Wyder}, {Burgett}, {Chambers}, {Kaiser}, {Magnier}, {Price}, and
  {Tonry}}]{Gezari2013}
{Gezari} S, {Martin} DC, {Forster} K, et~al (2013) {The GALEX Time Domain
  Survey. I. Selection and Classification of Over a Thousand Ultraviolet
  Variable Sources}. \apj 766(1):60. \doi{10.1088/0004-637X/766/1/60},
  {\href{https://arxiv.org/abs/1302.1581}{{arXiv:1302.1581}}} {[astro-ph.CO]}

\bibitem[{{Giampapa}(1983)}]{Giampapa1982}
{Giampapa} MS (1983) {Results from optical and UV stellar flare spectroscopy}.
  In: {Byrne} PB, {Rodono} M (eds) IAU Colloq. 71: Activity in Red-Dwarf Stars,
  pp 223--233, \doi{10.1007/978-94-009-7157-8_26}

\bibitem[{{Gigosos}(2014)}]{Gigosos2014}
{Gigosos} MA (2014) {Stark broadening models for plasma diagnostics}. Journal
  of Physics D Applied Physics 47(34):343001.
  \doi{10.1088/0022-3727/47/34/343001}

\bibitem[{{Gillon} et~al(2017){Gillon}, {Triaud}, {Demory}, {Jehin}, {Agol},
  {Deck}, {Lederer}, {de Wit}, {Burdanov}, {Ingalls}, {Bolmont}, {Leconte},
  {Raymond}, {Selsis}, {Turbet}, {Barkaoui}, {Burgasser}, {Burleigh}, {Carey},
  {Chaushev}, {Copperwheat}, {Delrez}, {Fernandes}, {Holdsworth}, {Kotze}, {Van
  Grootel}, {Almleaky}, {Benkhaldoun}, {Magain}, and {Queloz}}]{Gillon2017}
{Gillon} M, {Triaud} AHMJ, {Demory} BO, et~al (2017) {Seven temperate
  terrestrial planets around the nearby ultracool dwarf star TRAPPIST-1}. \nat
  542(7642):456--460. \doi{10.1038/nature21360},
  {\href{https://arxiv.org/abs/1703.01424}{{arXiv:1703.01424}}} {[astro-ph.EP]}

\bibitem[{{Gizis} et~al(2002){Gizis}, {Reid}, and {Hawley}}]{PMSU3}
{Gizis} JE, {Reid} IN, {Hawley} SL (2002) {The Palomar/MSU Nearby Star
  Spectroscopic Survey. III. Chromospheric Activity, M Dwarf Ages, and the
  Local Star Formation History}. \aj 123(6):3356--3369. \doi{10.1086/340465},
  {\href{https://arxiv.org/abs/astro-ph/0203499}{{arXiv:astro-ph/0203499}}}
  {[astro-ph]}

\bibitem[{{Gizis} et~al(2013){Gizis}, {Burgasser}, {Berger}, {Williams},
  {Vrba}, {Cruz}, and {Metchev}}]{Gizis2013}
{Gizis} JE, {Burgasser} AJ, {Berger} E, et~al (2013) {Kepler Monitoring of an L
  Dwarf I. The Photometric Period and White Light Flares}. \apj 779:172.
  \doi{10.1088/0004-637X/779/2/172},
  {\href{https://arxiv.org/abs/1310.5940}{{arXiv:1310.5940}}} {[astro-ph.SR]}

\bibitem[{{Gizis} et~al(2017){Gizis}, {Paudel}, {Mullan}, {Schmidt},
  {Burgasser}, and {Williams}}]{Gizis2017}
{Gizis} JE, {Paudel} RR, {Mullan} D, et~al (2017) {K2 Ultracool Dwarfs Survey.
  II. The White Light Flare Rate of Young Brown Dwarfs}. \apj 845(1):33.
  \doi{10.3847/1538-4357/aa7da0},
  {\href{https://arxiv.org/abs/1703.08745}{{arXiv:1703.08745}}} {[astro-ph.SR]}

\bibitem[{{Goldman} and {Cassar}(2006)}]{Goldman2006}
{Goldman} S, {Cassar} M (2006) {Atoms in Strong Fields}, p 227.
  \doi{10.1007/978-0-387-26308-3\_13}

\bibitem[{{Golovin} et~al(2023){Golovin}, {Reffert}, {Just}, {Jordan}, {Vani},
  and {Jahrei{\ss}}}]{CNS5}
{Golovin} A, {Reffert} S, {Just} A, et~al (2023) {The Fifth Catalogue of Nearby
  Stars (CNS5)}. \aap 670:A19. \doi{10.1051/0004-6361/202244250}

\bibitem[{{Gomez}(2017)}]{Gomez2017}
{Gomez} TA (2017) {Improving calculations of the interaction between atoms and
  plasma particles and its effect on spectral line shapes}. PhD thesis,
  University of Texas, Austin

\bibitem[{{Gomez} et~al(2016){Gomez}, {Nagayama}, {Kilcrease}, {Montgomery},
  and {Winget}}]{Gomez2016}
{Gomez} TA, {Nagayama} T, {Kilcrease} DP, et~al (2016) {Effect of higher-order
  multipole moments on the Stark line shape}. \pra 94(2):022501.
  \doi{10.1103/PhysRevA.94.022501}

\bibitem[{{Gomez} et~al(2022){Gomez}, {Nagayama}, {Cho}, {Kilcrease}, {Fontes},
  and {Zammit}}]{Gomez2022}
{Gomez} TA, {Nagayama} T, {Cho} PB, et~al (2022) {Introduction to spectral line
  shape theory}. Journal of Physics B Atomic Molecular Physics 55(3):034002.
  \doi{10.1088/1361-6455/ac4f31}

\bibitem[{{Gopalswamy} et~al(2023){Gopalswamy}, {Michalek}, {Yashiro},
  {M{\"a}kel{\"a}}, {Akiyama}, and {Xie}}]{Gopalswamy2023}
{Gopalswamy} N, {Michalek} G, {Yashiro} S, et~al (2023) {What Do Halo CMEs Tell
  Us about Solar Cycle 25?} \apjl 952(1):L13. \doi{10.3847/2041-8213/acdde2},
  {\href{https://arxiv.org/abs/2306.06633}{{arXiv:2306.06633}}} {[astro-ph.SR]}

\bibitem[{{Graham} and {Cauzzi}(2015)}]{Graham2015}
{Graham} DR, {Cauzzi} G (2015) {Temporal Evolution of Multiple Evaporating
  Ribbon Sources in a Solar Flare}. \apjl 807:L22.
  \doi{10.1088/2041-8205/807/2/L22},
  {\href{https://arxiv.org/abs/1506.03465}{{arXiv:1506.03465}}} {[astro-ph.SR]}

\bibitem[{{Graham} et~al(2020){Graham}, {Cauzzi}, {Zangrilli}, {Kowalski},
  {Sim{\~o}es}, and {Allred}}]{Graham2020}
{Graham} DR, {Cauzzi} G, {Zangrilli} L, et~al (2020) {Spectral Signatures of
  Chromospheric Condensation in a Major Solar Flare}. \apj 895(1):6.
  \doi{10.3847/1538-4357/ab88ad},
  {\href{https://arxiv.org/abs/2004.05075}{{arXiv:2004.05075}}} {[astro-ph.SR]}

\bibitem[{{Griem}(1974)}]{Griem1974}
{Griem} HR (1974) {Spectral line broadening by plasmas}

\bibitem[{{G{\"u}del}(2004)}]{Gudel2004Rev}
{G{\"u}del} M (2004) {X-ray astronomy of stellar coronae}. \aapr
  12(2-3):71--237. \doi{10.1007/s00159-004-0023-2},
  {\href{https://arxiv.org/abs/astro-ph/0406661}{{arXiv:astro-ph/0406661}}}
  {[astro-ph]}

\bibitem[{{G\"udel}(2006)}]{Gudel2006}
{G\"udel} M (2006) {Physics of Stellar Coronae}. arXiv e-prints
  astro-ph/0609389. \doi{10.48550/arXiv.astro-ph/0609389},
  {\href{https://arxiv.org/abs/astro-ph/0609389}{{arXiv:astro-ph/0609389}}}
  {[astro-ph]}

\bibitem[{{G\"udel} and {Benz}(1993)}]{GudelBenz}
{G\"udel} M, {Benz} AO (1993) {X-Ray/Microwave Relation of Different Types of
  Active Stars}. \apjl 405:L63. \doi{10.1086/186766}

\bibitem[{{G{\"u}del} and {Naz{\'e}}(2009)}]{Gudel2009Rev}
{G{\"u}del} M, {Naz{\'e}} Y (2009) {X-ray spectroscopy of stars}. \aapr
  17(3):309--408. \doi{10.1007/s00159-009-0022-4},
  {\href{https://arxiv.org/abs/0904.3078}{{arXiv:0904.3078}}} {[astro-ph.SR]}

\bibitem[{{G\"udel} et~al(1996){G\"udel}, {Benz}, {Schmitt}, and
  {Skinner}}]{Gudel1996}
{G\"udel} M, {Benz} AO, {Schmitt} JHMM, et~al (1996) {The Neupert Effect in
  Active Stellar Coronae: Chromospheric Evaporation and Coronal Heating in the
  dMe Flare Star Binary UV Ceti}. \apj 471:1002. \doi{10.1086/178027}

\bibitem[{{G{\"u}del} et~al(1997){G{\"u}del}, {Guinan}, {Mewe}, {Kaastra}, and
  {Skinner}}]{Gudel1997EK}
{G{\"u}del} M, {Guinan} EF, {Mewe} R, et~al (1997) {A Determination of the
  Coronal Emission Measure Distribution in the Young Solar Analog EK Draconis
  from ASCA/EUVE Spectra}. \apj 479(1):416--426. \doi{10.1086/303859}

\bibitem[{{G{\"u}del} et~al(1999){G{\"u}del}, {Linsky}, {Brown}, and
  {Nagase}}]{Gudel1999}
{G{\"u}del} M, {Linsky} JL, {Brown} A, et~al (1999) {Flaring and Quiescent
  Coronae of UX Arietis: Results from ASCA and EUVE Campaigns}. \apj
  511(1):405--421. \doi{10.1086/306651}

\bibitem[{{G{\"u}del} et~al(2002{\natexlab{a}}){G{\"u}del}, {Audard},
  {Skinner}, and {Horvath}}]{Gudel2002}
{G{\"u}del} M, {Audard} M, {Skinner} SL, et~al (2002{\natexlab{a}}) {X-Ray
  Evidence for Flare Density Variations and Continual Chromospheric Evaporation
  in Proxima Centauri}. \apjl 580:L73--L76. \doi{10.1086/345404},
  {\href{https://arxiv.org/abs/astro-ph/0210190}{{astro-ph/0210190}}}

\bibitem[{{G{\"u}del} et~al(2002{\natexlab{b}}){G{\"u}del}, {Audard}, {Smith},
  {Behar}, {Beasley}, and {Mewe}}]{Gudel2002b}
{G{\"u}del} M, {Audard} M, {Smith} KW, et~al (2002{\natexlab{b}}) {Detection of
  the Neupert Effect in the Corona of an RS Canum Venaticorum Binary System by
  XMM-Newton and the Very Large Array}. \apj 577(1):371--376.
  \doi{10.1086/342122},
  {\href{https://arxiv.org/abs/astro-ph/0206080}{{arXiv:astro-ph/0206080}}}
  {[astro-ph]}

\bibitem[{{G{\"u}del} et~al(2004){G{\"u}del}, {Audard}, {Reale}, {Skinner}, and
  {Linsky}}]{Gudel2004}
{G{\"u}del} M, {Audard} M, {Reale} F, et~al (2004) {Flares from small to large:
  X-ray spectroscopy of Proxima Centauri with XMM-Newton}. \aap 416:713--732.
  \doi{10.1051/0004-6361:20031471},
  {\href{https://arxiv.org/abs/astro-ph/0312297}{{arXiv:astro-ph/0312297}}}
  {[astro-ph]}

\bibitem[{{Guidoni} et~al(2016){Guidoni}, {DeVore}, {Karpen}, and
  {Lynch}}]{Guidoni2016}
{Guidoni} SE, {DeVore} CR, {Karpen} JT, et~al (2016) {Magnetic-island
  Contraction and Particle Acceleration in Simulated Eruptive Solar Flares}.
  \apj 820(1):60. \doi{10.3847/0004-637X/820/1/60},
  {\href{https://arxiv.org/abs/1603.01309}{{arXiv:1603.01309}}} {[astro-ph.SR]}

\bibitem[{{Gunn} et~al(1994{\natexlab{a}}){Gunn}, {Doyle}, {Mathioudakis}, and
  {Avgoloupis}}]{Gunn1994}
{Gunn} AG, {Doyle} JG, {Mathioudakis} M, et~al (1994{\natexlab{a}}) {An optical
  flare on YZ Canis Minoris}. \aap 285:157--160

\bibitem[{{Gunn} et~al(1994{\natexlab{b}}){Gunn}, {Doyle}, {Mathioudakis},
  {Houdebine}, and {Avgoloupis}}]{Gunn1994b}
{Gunn} AG, {Doyle} JG, {Mathioudakis} M, et~al (1994{\natexlab{b}})
  {High-velocity evaporation during a flare on AT Microscopii}. \aap
  285:489--496

\bibitem[{{Gurzadian}(1984)}]{Gurzadian1984}
{Gurzadian} GA (1984) {Flare stars : postflare behaviour of emission lines.}
  \apss 106:1--34. \doi{10.1007/BF00653910}

\bibitem[{{Haerendel}(2018)}]{Haerendel2018}
{Haerendel} G (2018) {Reconnection Mediated by Magnetic Fractures and the Solar
  Flare}. \apj 855(2):95. \doi{10.3847/1538-4357/aab0a9}

\bibitem[{{Haisch} et~al(1991){Haisch}, {Strong}, and {Rodono}}]{Haisch1991}
{Haisch} B, {Strong} KT, {Rodono} M (1991) {Flares on the Sun and other stars.}
  \araa 29:275--324. \doi{10.1146/annurev.aa.29.090191.001423}

\bibitem[{{Haisch}(1983)}]{Haisch1983}
{Haisch} BM (1983) {X-ray observations of stellar flares}. In: {Byrne} PB,
  {Rodono} M (eds) IAU Colloq. 71: Activity in Red-Dwarf Stars, pp 255--268,
  \doi{10.1007/978-94-009-7157-8_32}

\bibitem[{{Hamilton} and {Petrosian}(1987)}]{Hamilton1987}
{Hamilton} RJ, {Petrosian} V (1987) {Generation of Plasma Waves by Thick-Target
  Electron Beams and the Expected Radiation Signature}. \apj 321:721.
  \doi{10.1086/165665}

\bibitem[{{Hansteen} et~al(2023){Hansteen}, {Martinez-Sykora}, {Carlsson}, {De
  Pontieu}, {Go{\v{s}}i{\'c}}, and {Bose}}]{Hansteen2023}
{Hansteen} VH, {Martinez-Sykora} J, {Carlsson} M, et~al (2023) {Numerical
  Simulations and Observations of Mg II in the Solar Chromosphere}. \apj
  944(2):131. \doi{10.3847/1538-4357/acb33c},
  {\href{https://arxiv.org/abs/2211.09277}{{arXiv:2211.09277}}} {[astro-ph.SR]}

\bibitem[{{Hawley} and {Fisher}(1992)}]{HF92}
{Hawley} SL, {Fisher} GH (1992) {X-ray-heated models of stellar flare
  atmospheres - Theory and comparison with observations}. \apjs 78:565--598.
  \doi{10.1086/191640}

\bibitem[{{Hawley} and {Fisher}(1994)}]{HF94}
{Hawley} SL, {Fisher} GH (1994) {Solar flare model atmospheres}. \apj
  426:387--403. \doi{10.1086/174075}

\bibitem[{{Hawley} and {Pettersen}(1991)}]{HP91}
{Hawley} SL, {Pettersen} BR (1991) {The great flare of 1985 April 12 on AD
  Leonis}. \apj 378:725--741. \doi{10.1086/170474}

\bibitem[{{Hawley} et~al(1995){Hawley}, {Fisher}, {Simon}, {Cully}, {Deustua},
  {Jablonski}, {Johns-Krull}, {Pettersen}, {Smith}, {Spiesman}, and
  {Valenti}}]{Hawley1995}
{Hawley} SL, {Fisher} GH, {Simon} T, et~al (1995) {Simultaneous
  Extreme-Ultraviolet Explorer and Optical Observations of AD Leonis: Evidence
  for Large Coronal Loops and the Neupert Effect in Stellar Flares}. \apj
  453:464. \doi{10.1086/176408}

\bibitem[{{Hawley} et~al(1996){Hawley}, {Gizis}, and {Reid}}]{PMSU2}
{Hawley} SL, {Gizis} JE, {Reid} IN (1996) {The Palomar/MSU Nearby Star
  Spectroscopic Survey.II.The Southern M Dwarfs and Investigation of Magnetic
  Activity}. \aj 112:2799. \doi{10.1086/118222}

\bibitem[{{Hawley} et~al(1999){Hawley}, {Tourtellot}, and {Reid}}]{Hawley1999}
{Hawley} SL, {Tourtellot} JG, {Reid} IN (1999) {Low-Mass Stars in Open
  Clusters. I. NGC 2516 and NGC 3680}. \aj 117(3):1341--1359.
  \doi{10.1086/300783}

\bibitem[{{Hawley} et~al(2000){Hawley}, {Reid}, and {Tourtellot}}]{Hawley2000}
{Hawley} SL, {Reid} IN, {Tourtellot} JG (2000) {Properties of M Dwarfs in
  Clusters and the Field}. In: {Rebolo} R, {Zapatero-Osorio} MR (eds) Very
  Low-Mass Stars and Brown Dwarfs, p 109

\bibitem[{{Hawley} et~al(2003){Hawley}, {Allred}, {Johns-Krull}, {Fisher},
  {Abbett}, {Alekseev}, {Avgoloupis}, {Deustua}, {Gunn}, {Seiradakis}, {Sirk},
  and {Valenti}}]{Hawley2003}
{Hawley} SL, {Allred} JC, {Johns-Krull} CM, et~al (2003) {Multiwavelength
  Observations of Flares on AD Leonis}. \apj 597(1):535--554.
  \doi{10.1086/378351}

\bibitem[{{Hawley} et~al(2007){Hawley}, {Walkowicz}, {Allred}, and
  {Valenti}}]{Hawley2007}
{Hawley} SL, {Walkowicz} LM, {Allred} JC, et~al (2007) {Near-Ultraviolet
  Spectra of Flares on YZ CMi}. \pasp 119(851):67--81. \doi{10.1086/510561},
  {\href{https://arxiv.org/abs/astro-ph/0611074}{{arXiv:astro-ph/0611074}}}
  {[astro-ph]}

\bibitem[{{Hawley} et~al(2014){Hawley}, {Davenport}, {Kowalski}, {Wisniewski},
  {Hebb}, {Deitrick}, and {Hilton}}]{Hawley2014}
{Hawley} SL, {Davenport} JRA, {Kowalski} AF, et~al (2014) {Kepler Flares. I.
  Active and Inactive M Dwarfs}. \apj 797:121.
  \doi{10.1088/0004-637X/797/2/121},
  {\href{https://arxiv.org/abs/1410.7779}{{arXiv:1410.7779}}} {[astro-ph.SR]}

\bibitem[{{Hawley} et~al(2016){Hawley}, {Angus}, {Buzasi}, {Davenport},
  {Giampapa}, {Kashyap}, and {Meibom}}]{Hawley2016}
{Hawley} SL, {Angus} R, {Buzasi} D, et~al (2016) {Maximizing Science in the Era
  of LSST, Stars Study Group Report: Rotation and Magnetic Activity in the
  Galactic Field Population and in Open Star Clusters}. arXiv e-prints
  arXiv:1607.04302.
  {\href{https://arxiv.org/abs/1607.04302}{{arXiv:1607.04302}}} {[astro-ph.IM]}

\bibitem[{{Hayakawa} et~al(2023{\natexlab{a}}){Hayakawa}, {Bechet}, {Clette},
  {Hudson}, {Maehara}, {Namekata}, and {Notsu}}]{Hayakawa2023}
{Hayakawa} H, {Bechet} S, {Clette} F, et~al (2023{\natexlab{a}}) {Magnitude
  Estimates for the Carrington Flare in 1859 September: As Seen from the
  Original Records}. \apjl 954(1):L3. \doi{10.3847/2041-8213/acd853}

\bibitem[{{Hayakawa} et~al(2023{\natexlab{b}}){Hayakawa}, {Notsu}, and
  {Ebihara}}]{NotsuBook}
{Hayakawa} H, {Notsu} Y, {Ebihara} Y (2023{\natexlab{b}}) {Explorations of
  Extreme Space Weather Events from Stellar Observations and Archival
  Investigations}. In: {Kusano} K (ed) Solar-Terrestrial Environmental
  Prediction. p 327--376, \doi{10.1007/978-981-19-7765-7_11}

\bibitem[{{Hebb} et~al(2007){Hebb}, {Petro}, {Ford}, {Ardila}, {Toledo},
  {Minniti}, {Golimowski}, and {Clampin}}]{Hebb2007}
{Hebb} L, {Petro} L, {Ford} HC, et~al (2007) {A search for planets transiting
  the M-dwarf debris disc host, AU Microscopii}. \mnras 379(1):63--72.
  \doi{10.1111/j.1365-2966.2007.11904.x}

\bibitem[{{Heinzel} and {Shibata}(2018)}]{Heinzel2018}
{Heinzel} P, {Shibata} K (2018) {Can Flare Loops Contribute to the White-light
  Emission of Stellar Superflares?} \apj 859(2):143.
  \doi{10.3847/1538-4357/aabe78},
  {\href{https://arxiv.org/abs/1804.09656}{{arXiv:1804.09656}}} {[astro-ph.SR]}

\bibitem[{{He{\l}miniak} et~al(2012){He{\l}miniak}, {Konacki}, {Muterspaugh},
  {Browne}, {Howard}, and {Kulkarni}}]{BYDra2012}
{He{\l}miniak} KG, {Konacki} M, {Muterspaugh} MW, et~al (2012) {New
  high-precision orbital and physical parameters of the double-lined low-mass
  spectroscopic binary BY Draconis}. \mnras 419(2):1285--1293.
  \doi{10.1111/j.1365-2966.2011.19785.x},
  {\href{https://arxiv.org/abs/1109.5059}{{arXiv:1109.5059}}} {[astro-ph.SR]}

\bibitem[{{Herbig}(1956)}]{Herbig1956}
{Herbig} GH (1956) {Observations of the Spectrum of the Companion to BD +
  4{\textdegree}4048}. \pasp 68(405):531. \doi{10.1086/126992}

\bibitem[{{Herczeg} and {Hillenbrand}(2008)}]{Herczeg2008}
{Herczeg} GJ, {Hillenbrand} LA (2008) {UV Excess Measures of Accretion onto
  Young Very Low Mass Stars and Brown Dwarfs}. \apj 681(1):594--625.
  \doi{10.1086/586728}

\bibitem[{{Hilton}(2011)}]{Hilton2011}
{Hilton} EJ (2011) {The Galactic M Dwarf Flare Rate}. PhD thesis, University of
  Washington, Seattle

\bibitem[{{Hilton} et~al(2010){Hilton}, {West}, {Hawley}, and
  {Kowalski}}]{Hilton2010}
{Hilton} EJ, {West} AA, {Hawley} SL, et~al (2010) {M Dwarf Flares from
  Time-resolved Sloan Digital Sky Survey Spectra}. \aj 140(5):1402--1413.
  \doi{10.1088/0004-6256/140/5/1402},
  {\href{https://arxiv.org/abs/1009.1158}{{arXiv:1009.1158}}} {[astro-ph.SR]}

\bibitem[{{Hinton} et~al(2022){Hinton}, {France}, {Batista}, {Serna},
  {Hern{\'a}ndez}, {G{\"u}nther}, {Kowalski}, and {Schneider}}]{Hinton2022}
{Hinton} PC, {France} K, {Batista} MG, et~al (2022) {Far-ultraviolet Flares on
  Accreting Protostars: Weak and Classical T Tauri Stellar Pair Analysis}. \apj
  939(2):82. \doi{10.3847/1538-4357/ac8f26}

\bibitem[{{Hirayama}(1974)}]{Hirayama1974}
{Hirayama} T (1974) {Theoretical Model of Flares and Prominences. I:
  Evaporating Flare Model}. \solphys 34(2):323--338. \doi{10.1007/BF00153671}

\bibitem[{{Holman}(1985)}]{Holman1985}
{Holman} GD (1985) {Acceleration of runaway electrons and Joule heating in
  solar flares}. \apj 293:584--594. \doi{10.1086/163263}

\bibitem[{{Holman}(2012)}]{Holman2012}
{Holman} GD (2012) {Understanding the Impact of Return-current Losses on the
  X-Ray Emission from Solar Flares}. \apj 745:52.
  \doi{10.1088/0004-637X/745/1/52}

\bibitem[{{Holman} et~al(2003){Holman}, {Sui}, {Schwartz}, and
  {Emslie}}]{Holman2003}
{Holman} GD, {Sui} L, {Schwartz} RA, et~al (2003) {Electron Bremsstrahlung Hard
  X-Ray Spectra, Electron Distributions, and Energetics in the 2002 July 23
  Solar Flare}. \apjl 595(2):L97--L101. \doi{10.1086/378488}

\bibitem[{{Holman} et~al(2011){Holman}, {Aschwanden}, {Aurass}, {Battaglia},
  {Grigis}, {Kontar}, {Liu}, {Saint-Hilaire}, and {Zharkova}}]{Holman2011}
{Holman} GD, {Aschwanden} MJ, {Aurass} H, et~al (2011) {Implications of X-ray
  Observations for Electron Acceleration and Propagation in Solar Flares}. \ssr
  159(1-4):107--166. \doi{10.1007/s11214-010-9680-9},
  {\href{https://arxiv.org/abs/1109.6496}{{arXiv:1109.6496}}} {[astro-ph.SR]}

\bibitem[{{Honda} et~al(2018){Honda}, {Notsu}, {Namekata}, {Notsu}, {Maehara},
  {Ikuta}, {Nogami}, and {Shibata}}]{Honda2018}
{Honda} S, {Notsu} Y, {Namekata} K, et~al (2018) {Time-resolved spectroscopic
  observations of an M-dwarf flare star EV Lacertae during a flare}. \pasj
  70:62. \doi{10.1093/pasj/psy055},
  {\href{https://arxiv.org/abs/1804.03771}{{arXiv:1804.03771}}} {[astro-ph.SR]}

\bibitem[{{Hooper}(1968{\natexlab{a}})}]{Hooper1968B}
{Hooper} CF (1968{\natexlab{a}}) {Asymptotic Electric Microfield Distributions
  in Low-Frequency Component Plasmas}. Physical Review 169(1):193--195.
  \doi{10.1103/PhysRev.169.193}

\bibitem[{{Hooper}(1968{\natexlab{b}})}]{Hooper1968}
{Hooper} CF (1968{\natexlab{b}}) {Low-Frequency Component Electric Microfield
  Distributions in Plasmas}. Physical Review 165(1):215--222.
  \doi{10.1103/PhysRev.165.215}

\bibitem[{{Hori} et~al(1997){Hori}, {Yokoyama}, {Kosugi}, and
  {Shibata}}]{Hori1997}
{Hori} K, {Yokoyama} T, {Kosugi} T, et~al (1997) {Pseudo-Two-dimensional
  Hydrodynamic Modeling of Solar Flare Loops}. \apj 489(1):426--441.
  \doi{10.1086/304754}

\bibitem[{{Hotta} et~al(2022){Hotta}, {Kusano}, and {Shimada}}]{Hotta2022}
{Hotta} H, {Kusano} K, {Shimada} R (2022) {Generation of Solar-like
  Differential Rotation}. \apj 933(2):199. \doi{10.3847/1538-4357/ac7395}

\bibitem[{{Houdebine}(1992)}]{Houdebine1992}
{Houdebine} ER (1992) {Investigating the spectroscopic signatures of stellar
  flares}. Irish Astronomical Journal 20:213--249

\bibitem[{{Houdebine}(2003)}]{Houdebine2003}
{Houdebine} ER (2003) {Dynamics of flares on late type dMe stars. IV.
  Constraints from spectrophotometry in the visible}. \aap 397:1019--1034.
  \doi{10.1051/0004-6361:20021537}

\bibitem[{{Houdebine} et~al(1990){Houdebine}, {Foing}, and
  {Rodono}}]{Houdebine1990}
{Houdebine} ER, {Foing} BH, {Rodono} M (1990) {Dynamics of flares on late-type
  dMe stars. I. Flare mass ejections and stellar evolution.} \aap 238:249

\bibitem[{{Houdebine} et~al(1991){Houdebine}, {Butler}, {Panagi}, {Rodono}, and
  {Foing}}]{Houdebine1991}
{Houdebine} ER, {Butler} CJ, {Panagi} PM, et~al (1991) {Cooling curves of
  stellar flare plasmas from time resolved optical spectroscopy.} \aaps 87:33

\bibitem[{{Houdebine} et~al(1993{\natexlab{a}}){Houdebine}, {Foing}, {Doyle},
  and {Rodono}}]{Houdebine1993A}
{Houdebine} ER, {Foing} BH, {Doyle} JG, et~al (1993{\natexlab{a}}) {Dynamics of
  flares on late type dMe stars. II. Mass motions and prominence oscillations
  during a flare on AD Leonis.} \aap 274:245--264

\bibitem[{{Houdebine} et~al(1993{\natexlab{b}}){Houdebine}, {Foing}, {Doyle},
  and {Rodono}}]{Houdebine1993B}
{Houdebine} ER, {Foing} BH, {Doyle} JG, et~al (1993{\natexlab{b}}) {Dynamics of
  flares on late-type dMe stars. III. Kinetic energy and mass momentum budget
  of a flare on AD Leonis.} \aap 278:109--128

\bibitem[{{Howard} et~al(2018){Howard}, {Tilley}, {Corbett}, {Youngblood},
  {Loyd}, {Ratzloff}, {Law}, {Fors}, {del Ser}, {Shkolnik}, {Ziegler}, {Goeke},
  {Pietraallo}, and {Haislip}}]{Howard2018}
{Howard} WS, {Tilley} MA, {Corbett} H, et~al (2018) {The First Naked-eye
  Superflare Detected from Proxima Centauri}. \apjl 860(2):L30.
  \doi{10.3847/2041-8213/aacaf3},
  {\href{https://arxiv.org/abs/1804.02001}{{arXiv:1804.02001}}} {[astro-ph.EP]}

\bibitem[{{Howard} et~al(2019){Howard}, {Corbett}, {Law}, {Ratzloff},
  {Glazier}, {Fors}, {del Ser}, and {Haislip}}]{Howard2019}
{Howard} WS, {Corbett} H, {Law} NM, et~al (2019) {EvryFlare. I. Long-term
  Evryscope Monitoring of Flares from the Cool Stars across Half the Southern
  Sky}. \apj 881(1):9. \doi{10.3847/1538-4357/ab2767},
  {\href{https://arxiv.org/abs/1904.10421}{{arXiv:1904.10421}}} {[astro-ph.SR]}

\bibitem[{{Howard} et~al(2020){Howard}, {Corbett}, {Law}, {Ratzloff},
  {Galliher}, {Glazier}, {Gonzalez}, {Vasquez Soto}, {Fors}, {del Ser}, and
  {Haislip}}]{Howard2020}
{Howard} WS, {Corbett} H, {Law} NM, et~al (2020) {EvryFlare. III. Temperature
  Evolution and Habitability Impacts of Dozens of Superflares Observed
  Simultaneously by Evryscope and TESS}. \apj 902(2):115.
  \doi{10.3847/1538-4357/abb5b4},
  {\href{https://arxiv.org/abs/2010.00604}{{arXiv:2010.00604}}} {[astro-ph.SR]}

\bibitem[{{Howard} et~al(2022){Howard}, {MacGregor}, {Osten}, {Forbrich},
  {Cranmer}, {Tristan}, {Weinberger}, {Youngblood}, {Barclay}, {Parke Loyd},
  {Shkolnik}, {Zic}, and {Wilner}}]{Howard2022EV}
{Howard} WS, {MacGregor} MA, {Osten} R, et~al (2022) {The Mouse That Squeaked:
  A Small Flare from Proxima Cen Observed in the Millimeter, Optical, and Soft
  X-Ray with Chandra and ALMA}. \apj 938(2):103. \doi{10.3847/1538-4357/ac9134}

\bibitem[{{Howell} et~al(2014){Howell}, {Sobeck}, {Haas}, {Still}, {Barclay},
  {Mullally}, {Troeltzsch}, {Aigrain}, {Bryson}, {Caldwell}, {Chaplin},
  {Cochran}, {Huber}, {Marcy}, {Miglio}, {Najita}, {Smith}, {Twicken}, and
  {Fortney}}]{Howell2014}
{Howell} SB, {Sobeck} C, {Haas} M, et~al (2014) {The K2 Mission:
  Characterization and Early Results}. \pasp 126(938):398.
  \doi{10.1086/676406},
  {\href{https://arxiv.org/abs/1402.5163}{{arXiv:1402.5163}}} {[astro-ph.IM]}

\bibitem[{{Hubeny} and {Mihalas}(2014)}]{Hubeny2014}
{Hubeny} I, {Mihalas} D (2014) {Theory of Stellar Atmospheres}

\bibitem[{{Hubeny} et~al(1994){Hubeny}, {Hummer}, and {Lanz}}]{Hubeny1994}
{Hubeny} I, {Hummer} DG, {Lanz} T (1994) {NLTE model stellar atmospheres with
  line blanketing near the series limits.} \aap 282:151--167

\bibitem[{{Hudson} et~al(2023){Hudson}, {Cliver}, {White}, {Machol}, {Peck},
  {Tolbert}, {Viereck}, and {Zarro}}]{Hudson2023}
{Hudson} H, {Cliver} E, {White} S, et~al (2023) {The greatest GOES soft X-ray
  flares: Saturation and recalibration over two Hale cycles}. arXiv e-prints
  arXiv:2310.11457. \doi{10.48550/arXiv.2310.11457},
  {\href{https://arxiv.org/abs/2310.11457}{{arXiv:2310.11457}}} {[astro-ph.SR]}

\bibitem[{{Hudson}(1972)}]{Hudson1972}
{Hudson} HS (1972) {Thick-Target Processes and White-Light Flares}. \solphys
  24(2):414--428. \doi{10.1007/BF00153384}

\bibitem[{{Hudson}(1991)}]{Hudson1991}
{Hudson} HS (1991) {Solar flares, microflares, nanoflares, and coronal
  heating}. \solphys 133(2):357--369. \doi{10.1007/BF00149894}

\bibitem[{{Hudson}(2007)}]{Hudson2007}
{Hudson} HS (2007) {Chromospheric Flares}. In: {Heinzel} P, {Dorotovi{\v{c}}}
  I, {Rutten} RJ (eds) The Physics of Chromospheric Plasmas, p 365,
  \doi{10.48550/arXiv.0704.0823}, \eprint{0704.0823}

\bibitem[{{Hudson}(2011)}]{Hudson2011}
{Hudson} HS (2011) {Global Properties of Solar Flares}. \ssr 158(1):5--41.
  \doi{10.1007/s11214-010-9721-4},
  {\href{https://arxiv.org/abs/1108.3490}{{arXiv:1108.3490}}} {[astro-ph.SR]}

\bibitem[{{Hudson}(2021)}]{Hudson2021}
{Hudson} HS (2021) {Carrington Events}. \araa 59:445--477.
  \doi{10.1146/annurev-astro-112420-023324}

\bibitem[{{Hudson} et~al(1992){Hudson}, {Acton}, {Hirayama}, and
  {Uchida}}]{Hudson1992}
{Hudson} HS, {Acton} LW, {Hirayama} T, et~al (1992) {White-Light Flares
  Observed by YOHKOH}. \pasj 44:L77--L81

\bibitem[{{Huenemoerder} et~al(2010){Huenemoerder}, {Schulz}, {Testa}, {Drake},
  {Osten}, and {Reale}}]{Huenemoerder2010}
{Huenemoerder} DP, {Schulz} NS, {Testa} P, et~al (2010) {X-ray Flares of EV
  Lac: Statistics, Spectra, and Diagnostics}. \apj 723(2):1558--1567.
  \doi{10.1088/0004-637X/723/2/1558},
  {\href{https://arxiv.org/abs/1006.2558}{{arXiv:1006.2558}}} {[astro-ph.HE]}

\bibitem[{{Huensch} and {Reimers}(1995)}]{Heunsch1995}
{Huensch} M, {Reimers} D (1995) {Detection of an X-ray flare on the
  low-activity G 8 III-type giant {\ensuremath{\beta}} Boo.} \aap 296:509

\bibitem[{{Hummer} and {Mihalas}(1988)}]{HM88}
{Hummer} DG, {Mihalas} D (1988) {The equation of state for stellar envelopes. I
  - an occupation probability formalism for the truncation of internal
  partition functions}. \apj 331:794--814. \doi{10.1086/166600}

\bibitem[{{Ichimoto} and {Kurokawa}(1984)}]{Ichimoto1984}
{Ichimoto} K, {Kurokawa} H (1984) {H-alpha red asymmetry of solar flares}.
  \solphys 93:105--121. \doi{10.1007/BF00156656}

\bibitem[{{Ikuta} et~al(2023){Ikuta}, {Namekata}, {Notsu}, {Maehara},
  {Okamoto}, {Honda}, {Nogami}, and {Shibata}}]{Ikuta2023}
{Ikuta} K, {Namekata} K, {Notsu} Y, et~al (2023) {Starspot Mapping with
  Adaptive Parallel Tempering. II. Application to TESS Data for M-dwarf Flare
  Stars AU Microscopii, YZ Canis Minoris, and EV Lacertae}. \apj 948(1):64.
  \doi{10.3847/1538-4357/acbd36},
  {\href{https://arxiv.org/abs/2302.09249}{{arXiv:2302.09249}}} {[astro-ph.SR]}

\bibitem[{{Ilin} et~al(2021{\natexlab{a}}){Ilin}, {Schmidt}, {Poppenh{\"a}ger},
  {Davenport}, {Kristiansen}, and {Omohundro}}]{Illin2021}
{Ilin} E, {Schmidt} SJ, {Poppenh{\"a}ger} K, et~al (2021{\natexlab{a}}) {Flares
  in open clusters with K2. II. Pleiades, Hyades, Praesepe, Ruprecht 147, and M
  67}. \aap 645:A42. \doi{10.1051/0004-6361/202039198},
  {\href{https://arxiv.org/abs/2010.05576}{{arXiv:2010.05576}}} {[astro-ph.SR]}

\bibitem[{{Ilin} et~al(2021{\natexlab{b}}){Ilin}, {Schmidt}, {Poppenh{\"a}ger},
  {Davenport}, {Kristiansen}, and {Omohundro}}]{Illin2}
{Ilin} E, {Schmidt} SJ, {Poppenh{\"a}ger} K, et~al (2021{\natexlab{b}}) {Flares
  in open clusters with K2. II. Pleiades, Hyades, Praesepe, Ruprecht 147, and M
  67}. \aap 645:A42. \doi{10.1051/0004-6361/202039198},
  {\href{https://arxiv.org/abs/2010.05576}{{arXiv:2010.05576}}} {[astro-ph.SR]}

\bibitem[{{Inglis} et~al(2015){Inglis}, {Ireland}, and
  {Dominique}}]{Inglis2015}
{Inglis} AR, {Ireland} J, {Dominique} M (2015) {Quasi-periodic Pulsations in
  Solar and Stellar Flares: Re-evaluating their Nature in the Context of
  Power-law Flare Fourier Spectra}. \apj 798(2):108.
  \doi{10.1088/0004-637X/798/2/108},
  {\href{https://arxiv.org/abs/1410.8162}{{arXiv:1410.8162}}} {[astro-ph.SR]}

\bibitem[{{Inglis} et~al(2016){Inglis}, {Ireland}, {Dennis}, {Hayes}, and
  {Gallagher}}]{Inglis2016}
{Inglis} AR, {Ireland} J, {Dennis} BR, et~al (2016) {A Large-scale Search for
  Evidence of Quasi-periodic Pulsations in Solar Flares}. \apj 833(2):284.
  \doi{10.3847/1538-4357/833/2/284},
  {\href{https://arxiv.org/abs/1610.07454}{{arXiv:1610.07454}}} {[astro-ph.SR]}

\bibitem[{{Inglis} and {Teller}(1939)}]{InglisTeller}
{Inglis} DR, {Teller} E (1939) {Ionic Depression of Series Limits in
  One-Electron Spectra.} \apj 90:439. \doi{10.1086/144118}

\bibitem[{{Irwin} et~al(2011){Irwin}, {Berta}, {Burke}, {Charbonneau},
  {Nutzman}, {West}, and {Falco}}]{Irwin2011}
{Irwin} J, {Berta} ZK, {Burke} CJ, et~al (2011) {On the Angular Momentum
  Evolution of Fully Convective Stars: Rotation Periods for Field M-dwarfs from
  the MEarth Transit Survey}. \apj 727(1):56. \doi{10.1088/0004-637X/727/1/56},
  {\href{https://arxiv.org/abs/1011.4909}{{arXiv:1011.4909}}} {[astro-ph.SR]}

\bibitem[{{Isobe} et~al(2007){Isobe}, {Kubo}, {Minoshima}, {Ichimoto},
  {Katsukawa}, {Tarbell}, {Tsuneta}, {Berger}, {Lites}, {Nagata}, {Shimizu},
  {Shine}, {Suematsu}, and {Title}}]{Isobe2007}
{Isobe} H, {Kubo} M, {Minoshima} T, et~al (2007) {Flare Ribbons Observed with
  G-band and FeI 6302{\AA}, Filters of the Solar Optical Telescope on Board
  Hinode}. \pasj 59:S807--S813. \doi{10.1093/pasj/59.sp3.S807},
  {\href{https://arxiv.org/abs/0711.3946}{{arXiv:0711.3946}}}

\bibitem[{{Ivezi{\'c}} et~al(2019){Ivezi{\'c}}, {Kahn}, {Tyson}, {Abel},
  {Acosta}, {Allsman}, {Alonso}, {AlSayyad}, {Anderson}, {Andrew}, {Angel},
  {Angeli}, {Ansari}, {Antilogus}, {Araujo}, {Armstrong}, {Arndt}, {Astier},
  {Aubourg}, {Auza}, {Axelrod}, {Bard}, {Barr}, {Barrau}, {Bartlett}, {Bauer},
  {Bauman}, {Baumont}, {Bechtol}, {Bechtol}, {Becker}, {Becla}, {Beldica},
  {Bellavia}, {Bianco}, {Biswas}, {Blanc}, {Blazek}, {Blandford}, {Bloom},
  {Bogart}, {Bond}, {Booth}, {Borgland}, {Borne}, {Bosch}, {Boutigny},
  {Brackett}, {Bradshaw}, {Brandt}, {Brown}, {Bullock}, {Burchat}, {Burke},
  {Cagnoli}, {Calabrese}, {Callahan}, {Callen}, {Carlin}, {Carlson},
  {Chandrasekharan}, {Charles-Emerson}, {Chesley}, {Cheu}, {Chiang}, {Chiang},
  {Chirino}, {Chow}, {Ciardi}, {Claver}, {Cohen-Tanugi}, {Cockrum}, {Coles},
  {Connolly}, {Cook}, {Cooray}, {Covey}, {Cribbs}, {Cui}, {Cutri}, {Daly},
  {Daniel}, {Daruich}, {Daubard}, {Daues}, {Dawson}, {Delgado}, {Dellapenna},
  {de Peyster}, {de Val-Borro}, {Digel}, {Doherty}, {Dubois},
  {Dubois-Felsmann}, {Durech}, {Economou}, {Eifler}, {Eracleous}, {Emmons},
  {Fausti Neto}, {Ferguson}, {Figueroa}, {Fisher-Levine}, {Focke}, {Foss},
  {Frank}, {Freemon}, {Gangler}, {Gawiser}, {Geary}, {Gee}, {Geha}, {Gessner},
  {Gibson}, {Gilmore}, {Glanzman}, {Glick}, {Goldina}, {Goldstein}, {Goodenow},
  {Graham}, {Gressler}, {Gris}, {Guy}, {Guyonnet}, {Haller}, {Harris},
  {Hascall}, {Haupt}, {Hernandez}, {Herrmann}, {Hileman}, {Hoblitt}, {Hodgson},
  {Hogan}, {Howard}, {Huang}, {Huffer}, {Ingraham}, {Innes}, {Jacoby}, {Jain},
  {Jammes}, {Jee}, {Jenness}, {Jernigan}, {Jevremovi{\'c}}, {Johns}, {Johnson},
  {Johnson}, {Jones}, {Juramy-Gilles}, {Juri{\'c}}, {Kalirai}, {Kallivayalil},
  {Kalmbach}, {Kantor}, {Karst}, {Kasliwal}, {Kelly}, {Kessler}, {Kinnison},
  {Kirkby}, {Knox}, {Kotov}, {Krabbendam}, {Krughoff}, {Kub{\'a}nek},
  {Kuczewski}, {Kulkarni}, {Ku}, {Kurita}, {Lage}, {Lambert}, {Lange},
  {Langton}, {Le Guillou}, {Levine}, {Liang}, {Lim}, {Lintott}, {Long},
  {Lopez}, {Lotz}, {Lupton}, {Lust}, {MacArthur}, {Mahabal}, {Mandelbaum},
  {Markiewicz}, {Marsh}, {Marshall}, {Marshall}, {May}, {McKercher}, {McQueen},
  {Meyers}, {Migliore}, {Miller}, {Mills}, {Miraval}, {Moeyens}, {Moolekamp},
  {Monet}, {Moniez}, {Monkewitz}, {Montgomery}, {Morrison}, {Mueller},
  {Muller}, {Mu{\~n}oz Arancibia}, {Neill}, {Newbry}, {Nief}, {Nomerotski},
  {Nordby}, {O'Connor}, {Oliver}, {Olivier}, {Olsen}, {O'Mullane}, {Ortiz},
  {Osier}, {Owen}, {Pain}, {Palecek}, {Parejko}, {Parsons}, {Pease},
  {Peterson}, {Peterson}, {Petravick}, {Libby Petrick}, {Petry},
  {Pierfederici}, {Pietrowicz}, {Pike}, {Pinto}, {Plante}, {Plate}, {Plutchak},
  {Price}, {Prouza}, {Radeka}, {Rajagopal}, {Rasmussen}, {Regnault}, {Reil},
  {Reiss}, {Reuter}, {Ridgway}, {Riot}, {Ritz}, {Robinson}, {Roby}, {Roodman},
  {Rosing}, {Roucelle}, {Rumore}, {Russo}, {Saha}, {Sassolas}, {Schalk},
  {Schellart}, {Schindler}, {Schmidt}, {Schneider}, {Schneider}, {Schoening},
  {Schumacher}, {Schwamb}, {Sebag}, {Selvy}, {Sembroski}, {Seppala}, {Serio},
  {Serrano}, {Shaw}, {Shipsey}, {Sick}, {Silvestri}, {Slater}, {Smith},
  {Smith}, {Sobhani}, {Soldahl}, {Storrie-Lombardi}, {Stover}, {Strauss},
  {Street}, {Stubbs}, {Sullivan}, {Sweeney}, {Swinbank}, {Szalay}, {Takacs},
  {Tether}, {Thaler}, {Thayer}, {Thomas}, {Thornton}, {Thukral}, {Tice},
  {Trilling}, {Turri}, {Van Berg}, {Vanden Berk}, {Vetter}, {Virieux},
  {Vucina}, {Wahl}, {Walkowicz}, {Walsh}, {Walter}, {Wang}, {Wang}, {Warner},
  {Wiecha}, {Willman}, {Winters}, {Wittman}, {Wolff}, {Wood-Vasey}, {Wu},
  {Xin}, {Yoachim}, and {Zhan}}]{Ivezic2019}
{Ivezi{\'c}} {\v{Z}}, {Kahn} SM, {Tyson} JA, et~al (2019) {LSST: From Science
  Drivers to Reference Design and Anticipated Data Products}. \apj 873(2):111.
  \doi{10.3847/1538-4357/ab042c}

\bibitem[{{Jackman} et~al(2019{\natexlab{a}}){Jackman}, {Wheatley}, {Bayliss},
  {Burleigh}, {Casewell}, {Eigm{\"u}ller}, {Goad}, {Pollacco}, {Raynard},
  {Watson}, and {West}}]{Jackman2019Ldwarf}
{Jackman} JAG, {Wheatley} PJ, {Bayliss} D, et~al (2019{\natexlab{a}})
  {Detection of a giant white-light flare on an L2.5 dwarf with the Next
  Generation Transit Survey}. \mnras 485(1):L136--L140.
  \doi{10.1093/mnrasl/slz039},
  {\href{https://arxiv.org/abs/1902.00900}{{arXiv:1902.00900}}} {[astro-ph.SR]}

\bibitem[{{Jackman} et~al(2019{\natexlab{b}}){Jackman}, {Wheatley}, {Pugh},
  {Kolotkov}, {Broomhall}, {Kennedy}, {Murphy}, {Raddi}, {Burleigh},
  {Casewell}, {Eigm{\"u}ller}, {Gillen}, {G{\"u}nther}, {Jenkins}, {Louden},
  {McCormac}, {Raynard}, {Poppenhaeger}, {Udry}, {Watson}, and
  {West}}]{Jackman2019QPP}
{Jackman} JAG, {Wheatley} PJ, {Pugh} CE, et~al (2019{\natexlab{b}}) {Detection
  of a giant flare displaying quasi-periodic pulsations from a
  pre-main-sequence M star by the Next Generation Transit Survey}. \mnras
  482(4):5553--5566. \doi{10.1093/mnras/sty3036},
  {\href{https://arxiv.org/abs/1811.02008}{{arXiv:1811.02008}}} {[astro-ph.SR]}

\bibitem[{{Jackman} et~al(2020){Jackman}, {Wheatley}, {Acton}, {Anderson},
  {Belardi}, {Burleigh}, {Casewell}, {Eigm{\"u}ller}, {Gill}, {Gillen}, {Goad},
  {Grange}, {Hodgkin}, {Jenkins}, {McCormac}, {Moyano}, {Queloz}, {Raynard},
  {Tilbrook}, {Watson}, and {West}}]{Jackman2020Age}
{Jackman} JAG, {Wheatley} PJ, {Acton} JS, et~al (2020) {NGTS clusters survey -
  II. White-light flares from the youngest stars in Orion}. \mnras
  497(1):809--817. \doi{10.1093/mnras/staa1971},
  {\href{https://arxiv.org/abs/2007.01553}{{arXiv:2007.01553}}} {[astro-ph.SR]}

\bibitem[{{Jackman} et~al(2021){Jackman}, {Wheatley}, {Acton}, {Anderson},
  {Bayliss}, {Briegal}, {Burleigh}, {Casewell}, {G{\"a}nsicke}, {Gill},
  {Gillen}, {Goad}, {G{\"u}nther}, {Henderson}, {Hodgkin}, {Jenkins}, {Pugh},
  {Queloz}, {Raynard}, {Tilbrook}, {Watson}, and {West}}]{Jackman2021NGTS}
{Jackman} JAG, {Wheatley} PJ, {Acton} JS, et~al (2021) {Stellar flares detected
  with the Next Generation Transit Survey}. \mnras 504(3):3246--3264.
  \doi{10.1093/mnras/stab979},
  {\href{https://arxiv.org/abs/2104.02648}{{arXiv:2104.02648}}} {[astro-ph.SR]}

\bibitem[{{Jackman} et~al(2023){Jackman}, {Shkolnik}, {Million}, {Fleming},
  {Richey-Yowell}, and {Loyd}}]{Jackman2023}
{Jackman} JAG, {Shkolnik} EL, {Million} C, et~al (2023) {Extending optical
  flare models to the UV: results from comparing of TESS and GALEX flare
  observations for M Dwarfs}. \mnras 519(3):3564--3583.
  \doi{10.1093/mnras/stac3135},
  {\href{https://arxiv.org/abs/2210.15688}{{arXiv:2210.15688}}} {[astro-ph.SR]}

\bibitem[{{Jevremovic} et~al(1998{\natexlab{a}}){Jevremovic}, {Butler},
  {Drake}, {O'Donoghue}, and {van Wyk}}]{Jev1998}
{Jevremovic} D, {Butler} CJ, {Drake} SA, et~al (1998{\natexlab{a}})
  {Ultraviolet and optical flares on GL 866}. \aap 338:1057--1065

\bibitem[{{Jevremovic} et~al(1998{\natexlab{b}}){Jevremovic}, {Houdebine}, and
  {Butler}}]{Jev1998B}
{Jevremovic} D, {Houdebine} ER, {Butler} CJ (1998{\natexlab{b}})
  {Semi-Empirical Models of Stellar Flares}. In: {Donahue} RA, {Bookbinder} JA
  (eds) Cool Stars, Stellar Systems, and the Sun, p 1500

\bibitem[{{Ji} et~al(2022){Ji}, {Daughton}, {Jara-Almonte}, {Le}, {Stanier},
  and {Yoo}}]{Ji2022}
{Ji} H, {Daughton} W, {Jara-Almonte} J, et~al (2022) {Magnetic reconnection in
  the era of exascale computing and multiscale experiments}. Nature Reviews
  Physics 4(4):263--282. \doi{10.1038/s42254-021-00419-x},
  {\href{https://arxiv.org/abs/2202.09004}{{arXiv:2202.09004}}}
  {[physics.plasm-ph]}

\bibitem[{{Johns-Krull} et~al(1997){Johns-Krull}, {Hawley}, {Basri}, and
  {Valenti}}]{JohnsKrull1997}
{Johns-Krull} CM, {Hawley} SL, {Basri} G, et~al (1997) {Hamilton Echelle
  Spectroscopy of the 1993 March 6 Solar Flare}. \apjs 112:221--243.
  \doi{10.1086/313030}

\bibitem[{{Johnstone} et~al(2021){Johnstone}, {Bartel}, and
  {G{\"u}del}}]{Johnstone2021}
{Johnstone} CP, {Bartel} M, {G{\"u}del} M (2021) {The active lives of stars: A
  complete description of the rotation and XUV evolution of F, G, K, and M
  dwarfs}. \aap 649:A96. \doi{10.1051/0004-6361/202038407},
  {\href{https://arxiv.org/abs/2009.07695}{{arXiv:2009.07695}}} {[astro-ph.SR]}

\bibitem[{{Judge} et~al(2003){Judge}, {Solomon}, and {Ayres}}]{Judge2003}
{Judge} PG, {Solomon} SC, {Ayres} TR (2003) {An Estimate of the Sun's
  ROSAT-PSPC X-Ray Luminosities Using SNOE-SXP Measurements}. \apj
  593(1):534--548. \doi{10.1086/376405}

\bibitem[{{Kane} et~al(1985){Kane}, {Love}, {Neidig}, and {Cliver}}]{Kane1985}
{Kane} SR, {Love} JJ, {Neidig} DF, et~al (1985) {Characteristics of the
  white-light source in the 1981 April 24 solar flare}. \apjl 290:L45--L48.
  \doi{10.1086/184439}

\bibitem[{{Kanodia} et~al(2022){Kanodia}, {Ramsey}, {Maney}, {Mahadevan},
  {Ca{\~n}as}, {Ninan}, {Monson}, {Kowalski}, {Goumas}, {Stefansson}, {Bender},
  {Cochran}, {Diddams}, {Fredrick}, {Halverson}, {Hearty}, {Janowiecki},
  {Metcalf}, {Odewahn}, {Robertson}, {Roy}, {Schwab}, and
  {Terrien}}]{Kanodia2022}
{Kanodia} S, {Ramsey} LW, {Maney} M, et~al (2022) {High-resolution
  Near-infrared Spectroscopy of a Flare around the Ultracool Dwarf vB 10}. \apj
  925(2):155. \doi{10.3847/1538-4357/ac3e61},
  {\href{https://arxiv.org/abs/2111.14647}{{arXiv:2111.14647}}} {[astro-ph.SR]}

\bibitem[{{K{\"a}pyl{\"a}}(2021)}]{Kaplya2021}
{K{\"a}pyl{\"a}} PJ (2021) {Star-in-a-box simulations of fully convective
  stars}. \aap 651:A66. \doi{10.1051/0004-6361/202040049},
  {\href{https://arxiv.org/abs/2012.01259}{{arXiv:2012.01259}}} {[astro-ph.SR]}

\bibitem[{{K{\"a}pyl{\"a}} et~al(2023){K{\"a}pyl{\"a}}, {Browning}, {Brun},
  {Guerrero}, and {Warnecke}}]{Kapyla2023}
{K{\"a}pyl{\"a}} PJ, {Browning} MK, {Brun} AS, et~al (2023) {Simulations of
  solar and stellar dynamos and their theoretical interpretation}. arXiv
  e-prints arXiv:2305.16790. \doi{10.48550/arXiv.2305.16790},
  {\href{https://arxiv.org/abs/2305.16790}{{arXiv:2305.16790}}} {[astro-ph.SR]}

\bibitem[{{Karmakar} et~al(2017){Karmakar}, {Pandey}, {Airapetian}, and
  {Misra}}]{Karmakar2017}
{Karmakar} S, {Pandey} JC, {Airapetian} VS, et~al (2017) {X-Ray Superflares on
  CC Eri}. \apj 840(2):102. \doi{10.3847/1538-4357/aa6cb0},
  {\href{https://arxiv.org/abs/1705.06930}{{arXiv:1705.06930}}} {[astro-ph.SR]}

\bibitem[{{Karmakar} et~al(2022){Karmakar}, {Naik}, {Pandey}, and
  {Savanov}}]{Karmakar2022}
{Karmakar} S, {Naik} S, {Pandey} JC, et~al (2022) {AstroSat observations of
  long-duration X-ray superflares on active M-dwarf binary EQ Peg}. \mnras
  509(3):3247--3257. \doi{10.1093/mnras/stab3099},
  {\href{https://arxiv.org/abs/2111.07527}{{arXiv:2111.07527}}} {[astro-ph.SR]}

\bibitem[{{Karmakar} et~al(2023){Karmakar}, {Naik}, {Pandey}, and
  {Savanov}}]{Karmakar2023}
{Karmakar} S, {Naik} S, {Pandey} JC, et~al (2023) {Swift and XMM-Newton
  observations of an RS CVn-type eclipsing binary SZ Psc: superflare and
  coronal properties}. \mnras 518(1):900--918. \doi{10.1093/mnras/stac2970},
  {\href{https://arxiv.org/abs/2210.07170}{{arXiv:2210.07170}}} {[astro-ph.SR]}

\bibitem[{{Karoff} et~al(2016){Karoff}, {Knudsen}, {De Cat}, {Bonanno},
  {Fogtmann-Schulz}, {Fu}, {Frasca}, {Inceoglu}, {Olsen}, {Zhang}, {Hou},
  {Wang}, {Shi}, and {Zhang}}]{Karoff2016}
{Karoff} C, {Knudsen} MF, {De Cat} P, et~al (2016) {Observational evidence for
  enhanced magnetic activity of superflare stars}. Nature Communications
  7:11058. \doi{10.1038/ncomms11058}

\bibitem[{{Karpen} et~al(2012){Karpen}, {Antiochos}, and {DeVore}}]{Karpen2012}
{Karpen} JT, {Antiochos} SK, {DeVore} CR (2012) {The Mechanisms for the Onset
  and Explosive Eruption of Coronal Mass Ejections and Eruptive Flares}. \apj
  760(1):81. \doi{10.1088/0004-637X/760/1/81}

\bibitem[{{Kashyap} et~al(2002){Kashyap}, {Drake}, {G{\"u}del}, and
  {Audard}}]{Kashyap2002}
{Kashyap} VL, {Drake} JJ, {G{\"u}del} M, et~al (2002) {Flare Heating in Stellar
  Coronae}. \apj 580(2):1118--1132. \doi{10.1086/343869},
  {\href{https://arxiv.org/abs/astro-ph/0208546}{{arXiv:astro-ph/0208546}}}
  {[astro-ph]}

\bibitem[{{Katsova} et~al(1991){Katsova}, {Livshits}, {Butler}, and
  {Doyle}}]{Katsova1991}
{Katsova} MM, {Livshits} MA, {Butler} CJ, et~al (1991) {A gas-dynamic model for
  a flare on YZ CMi : interpretation of high-temporal-resolution spectroscopic
  data.} \mnras 250:402. \doi{10.1093/mnras/250.2.402}

\bibitem[{{Katsova} et~al(1997){Katsova}, {Boiko}, and
  {Livshits}}]{Katsova1997}
{Katsova} MM, {Boiko} AY, {Livshits} MA (1997) {The gas-dynamic model of
  impulsive stellar flares.} \aap 321:549--556

\bibitem[{{Katsova} et~al(2018){Katsova}, {Kitchatinov}, {Moss}, {Ol{\'a}h},
  and {Sokoloff}}]{Katsova2018}
{Katsova} MM, {Kitchatinov} LL, {Moss} D, et~al (2018) {Superflares on Giant
  Stars}. Astronomy Reports 62(8):513--519. \doi{10.1134/S1063772918080036},
  {\href{https://arxiv.org/abs/1804.06315}{{arXiv:1804.06315}}} {[astro-ph.SR]}

\bibitem[{{Kaufmann} et~al(2004){Kaufmann}, {Raulin}, {de Castro}, {Levato},
  {Gary}, {Costa}, {Marun}, {Pereyra}, {Silva}, and {Correia}}]{Kaufmann2004}
{Kaufmann} P, {Raulin} JP, {de Castro} CGG, et~al (2004) {A New Solar Burst
  Spectral Component Emitting Only in the Terahertz Range}. \apjl
  603(2):L121--L124. \doi{10.1086/383186}

\bibitem[{{Kawate} et~al(2016){Kawate}, {Ishii}, {Nakatani}, {Ichimoto},
  {Asai}, {Morita}, and {Masuda}}]{Kawate2016}
{Kawate} T, {Ishii} TT, {Nakatani} Y, et~al (2016) {Temporal evolution and
  spatial distribution of white-light flare kernels in a solar flare}. ArXiv
  e-prints {\href{https://arxiv.org/abs/1610.04328}{{arXiv:1610.04328}}}
  {[astro-ph.SR]}

\bibitem[{{Kazachenko}(2023)}]{Kazachenko2023}
{Kazachenko} MD (2023) {A Database of Magnetic and Thermodynamic Properties of
  Confined and Eruptive Solar Flares}. \apj 958(2):104.
  \doi{10.3847/1538-4357/ad004e},
  {\href{https://arxiv.org/abs/2310.02878}{{arXiv:2310.02878}}} {[astro-ph.SR]}

\bibitem[{{Kazachenko} et~al(2022){Kazachenko}, {Albelo-Corchado}, {Tamburri},
  and {Welsch}}]{Kazachenko2022}
{Kazachenko} MD, {Albelo-Corchado} MF, {Tamburri} CA, et~al (2022) {Invited
  Review: Short-term Variability with the Observations from the Helioseismic
  and Magnetic Imager (HMI) Onboard the Solar Dynamics Observatory (SDO):
  Insights into Flare Magnetism}. \solphys 297(5):59.
  \doi{10.1007/s11207-022-01987-6}

\bibitem[{{Kerr}(2022)}]{Kerr2022}
{Kerr} GS (2022) {Interrogating Solar Flare Loop Models with IRIS Observations
  1: Overview of the Models, and Mass flows}. Frontiers in Astronomy and Space
  Sciences 9:1060856. \doi{10.3389/fspas.2022.1060856}

\bibitem[{{Kerr}(2023)}]{Kerr2023}
{Kerr} GS (2023) {Interrogating Solar Flare Loop Models with IRIS Observations
  2: Plasma Properties, Energy Transport, and Future Directions}. Frontiers in
  Astronomy and Space Sciences 9:425. \doi{10.3389/fspas.2022.1060862}

\bibitem[{{Kerr} et~al(2020){Kerr}, {Allred}, and {Polito}}]{Kerr2020}
{Kerr} GS, {Allred} JC, {Polito} V (2020) {Solar Flare Arcade Modeling:
  Bridging the Gap from 1D to 3D Simulations of Optically Thin Radiation}. \apj
  900(1):18. \doi{10.3847/1538-4357/abaa46},
  {\href{https://arxiv.org/abs/2007.13856}{{arXiv:2007.13856}}} {[astro-ph.SR]}

\bibitem[{{K{\H{o}}v{\'a}ri} et~al(2020){K{\H{o}}v{\'a}ri}, {Ol{\'a}h},
  {G{\"u}nther}, {Vida}, {Kriskovics}, {Seli}, {Bakos}, {Hartman}, {Csubry},
  and {Bhatti}}]{Kovari2020}
{K{\H{o}}v{\'a}ri} Z, {Ol{\'a}h} K, {G{\"u}nther} MN, et~al (2020) {Superflares
  on the late-type giant KIC 2852961. Scaling effect behind flaring at
  different energy levels}. \aap 641:A83. \doi{10.1051/0004-6361/202038397},
  {\href{https://arxiv.org/abs/2005.05397}{{arXiv:2005.05397}}} {[astro-ph.SR]}

\bibitem[{{Kiman} et~al(2021){Kiman}, {Faherty}, {Cruz}, {Gagn{\'e}}, {Angus},
  {Schmidt}, {Mann}, {Bardalez Gagliuffi}, and {Rice}}]{Kiman2021}
{Kiman} R, {Faherty} JK, {Cruz} KL, et~al (2021) {Calibration of the
  H{\ensuremath{\alpha}} Age-Activity Relation for M Dwarfs}. \aj 161(6):277.
  \doi{10.3847/1538-3881/abf561},
  {\href{https://arxiv.org/abs/2104.01232}{{arXiv:2104.01232}}} {[astro-ph.SR]}

\bibitem[{{Kleint} et~al(2016){Kleint}, {Heinzel}, {Judge}, and
  {Krucker}}]{Kleint2016}
{Kleint} L, {Heinzel} P, {Judge} P, et~al (2016) {Continuum Enhancements in the
  Ultraviolet, the Visible and the Infrared during the X1 Flare on 2014 March
  29}. \apj 816:88. \doi{10.3847/0004-637X/816/2/88},
  {\href{https://arxiv.org/abs/1511.04161}{{arXiv:1511.04161}}} {[astro-ph.SR]}

\bibitem[{{Kleppner} et~al(1981){Kleppner}, {Littman}, and
  {Zimmerman}}]{Kleppner1981}
{Kleppner} D, {Littman} MG, {Zimmerman} ML (1981) {Highly excited atoms}.
  Scientific American 244:130--138. \doi{10.1038/scientificamerican0581-130}

\bibitem[{{Kneer} and {Nakagawa}(1976)}]{Kneer1976}
{Kneer} F, {Nakagawa} Y (1976) {Radiative hydrodynamics of chromospheric
  transients.} \aap 47(1):65--76

\bibitem[{{Knuth} and {Glesener}(2020)}]{Knuth2020}
{Knuth} T, {Glesener} L (2020) {Subsecond Spikes in Fermi GBM X-Ray Flux as a
  Probe for Solar Flare Particle Acceleration}. \apj 903(1):63.
  \doi{10.3847/1538-4357/abb779},
  {\href{https://arxiv.org/abs/2003.05007}{{arXiv:2003.05007}}} {[astro-ph.SR]}

\bibitem[{{Koch} et~al(2010){Koch}, {Borucki}, {Basri}, {Batalha}, {Brown},
  {Caldwell}, {Christensen-Dalsgaard}, {Cochran}, {DeVore}, {Dunham},
  {Gautier}, {Geary}, {Gilliland}, {Gould}, {Jenkins}, {Kondo}, {Latham},
  {Lissauer}, {Marcy}, {Monet}, {Sasselov}, {Boss}, {Brownlee}, {Caldwell},
  {Dupree}, {Howell}, {Kjeldsen}, {Meibom}, {Morrison}, {Owen}, {Reitsema},
  {Tarter}, {Bryson}, {Dotson}, {Gazis}, {Haas}, {Kolodziejczak}, {Rowe}, {Van
  Cleve}, {Allen}, {Chandrasekaran}, {Clarke}, {Li}, {Quintana}, {Tenenbaum},
  {Twicken}, and {Wu}}]{Koch2010}
{Koch} DG, {Borucki} WJ, {Basri} G, et~al (2010) {Kepler Mission Design,
  Realized Photometric Performance, and Early Science}. \apjl 713(2):L79--L86.
  \doi{10.1088/2041-8205/713/2/L79},
  {\href{https://arxiv.org/abs/1001.0268}{{arXiv:1001.0268}}} {[astro-ph.EP]}

\bibitem[{{Kochukhov} and {Shulyak}(2019)}]{YYGem19}
{Kochukhov} O, {Shulyak} D (2019) {Magnetic Field of the Eclipsing M-dwarf
  Binary YY Gem}. \apj 873(1):69. \doi{10.3847/1538-4357/ab06c5},
  {\href{https://arxiv.org/abs/1902.04157}{{arXiv:1902.04157}}} {[astro-ph.SR]}

\bibitem[{{Koller} et~al(2021){Koller}, {Leitzinger}, {Temmer}, {Odert},
  {Beck}, and {Veronig}}]{Koller2021}
{Koller} F, {Leitzinger} M, {Temmer} M, et~al (2021) {Search for flares and
  associated CMEs on late-type main-sequence stars in optical SDSS spectra}.
  \aap 646:A34. \doi{10.1051/0004-6361/202039003},
  {\href{https://arxiv.org/abs/2012.00786}{{arXiv:2012.00786}}} {[astro-ph.SR]}

\bibitem[{{K{\"o}nig} et~al(2006){K{\"o}nig}, {Guenther}, {Esposito}, and
  {Hatzes}}]{Koning2006}
{K{\"o}nig} B, {Guenther} EW, {Esposito} M, et~al (2006) {Spectral synthesis
  analysis and radial velocity study of the northern F-, G- and K-type flare
  stars}. \mnras 365(3):1050--1056. \doi{10.1111/j.1365-2966.2005.09796.x},
  {\href{https://arxiv.org/abs/astro-ph/0511232}{{arXiv:astro-ph/0511232}}}
  {[astro-ph]}

\bibitem[{{Kontar} et~al(2011){Kontar}, {Brown}, {Emslie}, {Hajdas}, {Holman},
  {Hurford}, {Ka{\v{s}}parov{\'a}}, {Mallik}, {Massone}, {McConnell}, {Piana},
  {Prato}, {Schmahl}, and {Suarez-Garcia}}]{Kontar2011}
{Kontar} EP, {Brown} JC, {Emslie} AG, et~al (2011) {Deducing Electron
  Properties from Hard X-ray Observations}. \ssr 159(1-4):301--355.
  \doi{10.1007/s11214-011-9804-x},
  {\href{https://arxiv.org/abs/1110.1755}{{arXiv:1110.1755}}} {[astro-ph.SR]}

\bibitem[{{Kontar} et~al(2012){Kontar}, {Ratcliffe}, and {Bian}}]{Kontar2012}
{Kontar} EP, {Ratcliffe} H, {Bian} NH (2012) {Wave-particle interactions in
  non-uniform plasma and the interpretation of hard X-ray spectra in solar
  flares}. \aap 539:A43. \doi{10.1051/0004-6361/201118216},
  {\href{https://arxiv.org/abs/1112.4448}{{arXiv:1112.4448}}} {[astro-ph.SR]}

\bibitem[{{Kontar} et~al(2023){Kontar}, {Emslie}, {Motorina}, and
  {Dennis}}]{Kontar2023}
{Kontar} EP, {Emslie} AG, {Motorina} GG, et~al (2023) {The Efficiency of
  Electron Acceleration during the Impulsive Phase of a Solar Flare}. \apjl
  947(1):L13. \doi{10.3847/2041-8213/acc9b7},
  {\href{https://arxiv.org/abs/2304.01088}{{arXiv:2304.01088}}} {[astro-ph.SR]}

\bibitem[{{Kopp} and {Pneuman}(1976)}]{Kopp1976}
{Kopp} RA, {Pneuman} GW (1976) {Magnetic reconnection in the corona and the
  loop prominence phenomenon.} \solphys 50(1):85--98. \doi{10.1007/BF00206193}

\bibitem[{{Kopp} and {Poletto}(1984)}]{Kopp1984}
{Kopp} RA, {Poletto} G (1984) {Extension of the Reconnection Theory of
  Two-Ribbon Solar Flares}. \solphys 93(2):351--361. \doi{10.1007/BF02270846}

\bibitem[{{Kosovichev} and {Zharkova}(2001)}]{Kosovichev2001}
{Kosovichev} AG, {Zharkova} VV (2001) {Magnetic Energy Release and Transients
  in the Solar Flare of 2000 July 14}. \apjl 550(1):L105--L108.
  \doi{10.1086/319484}

\bibitem[{{Kowalski}(2012)}]{Kowalski2012}
{Kowalski} AF (2012) {Time-Resolved Properties and Global Trends in dMe Flares
  from Simultaneous Photometry and Spectra}. PhD thesis, University of
  Washington

\bibitem[{{Kowalski}(2016)}]{Kowalski2016B}
{Kowalski} AF (2016) {White-light continuum in stellar flares}. In:
  {Kosovichev} AG, {Hawley} SL, {Heinzel} P (eds) Solar and Stellar Flares and
  their Effects on Planets, pp 259--267, \doi{10.1017/S1743921316000028}

\bibitem[{{Kowalski}(2022)}]{Kowalski2022Frontiers}
{Kowalski} AF (2022) {Near-ultraviolet continuum modeling of the 1985 April 12
  great flare of AD Leo}. Frontiers in Astronomy and Space Sciences 9:1034458.
  \doi{10.3389/fspas.2022.1034458},
  {\href{https://arxiv.org/abs/2210.16980}{{arXiv:2210.16980}}} {[astro-ph.SR]}

\bibitem[{{Kowalski}(2023)}]{Kowalski2023}
{Kowalski} AF (2023) {Bridging High-density Electron Beam Coronal Transport and
  Deep Chromospheric Heating in Stellar Flares}. \apjl 943(2):L23.
  \doi{10.3847/2041-8213/acb144},
  {\href{https://arxiv.org/abs/2301.03477}{{arXiv:2301.03477}}} {[astro-ph.SR]}

\bibitem[{{Kowalski} and {Allred}(2018)}]{KA18}
{Kowalski} AF, {Allred} JC (2018) {Parameterizations of Chromospheric
  Condensations in dG and dMe Model Flare Atmospheres}. \apj 852(1):61.
  \doi{10.3847/1538-4357/aa9d91},
  {\href{https://arxiv.org/abs/1711.09488}{{arXiv:1711.09488}}} {[astro-ph.SR]}

\bibitem[{{Kowalski} et~al(2009){Kowalski}, {Hawley}, {Hilton}, {Becker},
  {West}, {Bochanski}, and {Sesar}}]{Kowalski2009}
{Kowalski} AF, {Hawley} SL, {Hilton} EJ, et~al (2009) {M Dwarfs in Sloan
  Digital Sky Survey Stripe 82: Photometric Light Curves and Flare Rate
  Analysis}. \aj 138(2):633--648. \doi{10.1088/0004-6256/138/2/633},
  {\href{https://arxiv.org/abs/0906.2030}{{arXiv:0906.2030}}} {[astro-ph.SR]}

\bibitem[{{Kowalski} et~al(2010){Kowalski}, {Hawley}, {Holtzman}, {Wisniewski},
  and {Hilton}}]{Kowalski2010}
{Kowalski} AF, {Hawley} SL, {Holtzman} JA, et~al (2010) {A White Light
  Megaflare on the dM4.5e Star YZ CMi}. \apjl 714:L98--L102.
  \doi{10.1088/2041-8205/714/1/L98},
  {\href{https://arxiv.org/abs/1003.3057}{{arXiv:1003.3057}}} {[astro-ph.SR]}

\bibitem[{{Kowalski} et~al(2011){Kowalski}, {Mathioudakis}, {Hawley}, {Hilton},
  {Dhillon}, {Marsh}, and {Copperwheat}}]{Kowalski2011ASPC}
{Kowalski} AF, {Mathioudakis} M, {Hawley} SL, et~al (2011) {White Light Flare
  Continuum Observations with ULTRACAM}. In: {Johns-Krull} C, {Browning} MK,
  {West} AA (eds) 16th Cambridge Workshop on Cool Stars, Stellar Systems, and
  the Sun, p 1157, \doi{10.48550/arXiv.1103.0822}, \eprint{1103.0822}

\bibitem[{{Kowalski} et~al(2013){Kowalski}, {Hawley}, {Wisniewski}, {Osten},
  {Hilton}, {Holtzman}, {Schmidt}, and {Davenport}}]{Kowalski2013}
{Kowalski} AF, {Hawley} SL, {Wisniewski} JP, et~al (2013) {Time-resolved
  Properties and Global Trends in dMe Flares from Simultaneous Photometry and
  Spectra}. \apjs 207:15. \doi{10.1088/0067-0049/207/1/15},
  {\href{https://arxiv.org/abs/1307.2099}{{arXiv:1307.2099}}} {[astro-ph.SR]}

\bibitem[{{Kowalski} et~al(2015){Kowalski}, {Hawley}, {Carlsson}, {Allred},
  {Uitenbroek}, {Osten}, and {Holman}}]{Kowalski2015}
{Kowalski} AF, {Hawley} SL, {Carlsson} M, et~al (2015) {New Insights into
  White-Light Flare Emission from Radiative-Hydrodynamic Modeling of a
  Chromospheric Condensation}. \solphys 290:3487--3523.
  \doi{10.1007/s11207-015-0708-x},
  {\href{https://arxiv.org/abs/1503.07057}{{arXiv:1503.07057}}} {[astro-ph.SR]}

\bibitem[{{Kowalski} et~al(2016){Kowalski}, {Mathioudakis}, {Hawley},
  {Wisniewski}, {Dhillon}, {Marsh}, {Hilton}, and {Brown}}]{Kowalski2016}
{Kowalski} AF, {Mathioudakis} M, {Hawley} SL, et~al (2016) {M Dwarf Flare
  Continuum Variations on One-second Timescales: Calibrating and Modeling of
  ULTRACAM Flare Color Indices}. \apj 820:95. \doi{10.3847/0004-637X/820/2/95},
  {\href{https://arxiv.org/abs/1602.04879}{{arXiv:1602.04879}}} {[astro-ph.SR]}

\bibitem[{{Kowalski} et~al(2017{\natexlab{a}}){Kowalski}, {Allred}, {Daw},
  {Cauzzi}, and {Carlsson}}]{Kowalski2017Mar29}
{Kowalski} AF, {Allred} JC, {Daw} A, et~al (2017{\natexlab{a}}) {The
  Atmospheric Response to High Nonthermal Electron Beam Fluxes in Solar Flares.
  I. Modeling the Brightest NUV Footpoints in the X1 Solar Flare of 2014 March
  29}. \apj 836:12. \doi{10.3847/1538-4357/836/1/12},
  {\href{https://arxiv.org/abs/1609.07390}{{arXiv:1609.07390}}} {[astro-ph.SR]}

\bibitem[{{Kowalski} et~al(2017{\natexlab{b}}){Kowalski}, {Allred},
  {Uitenbroek}, {Tremblay}, {Brown}, {Carlsson}, {Osten}, {Wisniewski}, and
  {Hawley}}]{Kowalski2017Broadening}
{Kowalski} AF, {Allred} JC, {Uitenbroek} H, et~al (2017{\natexlab{b}})
  {Hydrogen Balmer Line Broadening in Solar and Stellar Flares}. \apj 837:125.
  \doi{10.3847/1538-4357/aa603e},
  {\href{https://arxiv.org/abs/1702.03321}{{arXiv:1702.03321}}} {[astro-ph.SR]}

\bibitem[{{Kowalski} et~al(2019{\natexlab{a}}){Kowalski}, {Butler}, {Daw},
  {Fletcher}, {Allred}, {De Pontieu}, {Kerr}, and {Cauzzi}}]{Kowalski2019IRIS}
{Kowalski} AF, {Butler} E, {Daw} AN, et~al (2019{\natexlab{a}}) {Spectral
  Evidence for Heating at Large Column Mass in Umbral Solar Flare Kernels. I.
  IRIS Near-UV Spectra of the X1 Solar Flare of 2014 October 25}. \apj
  878(2):135. \doi{10.3847/1538-4357/ab1f8b},
  {\href{https://arxiv.org/abs/1905.02111}{{arXiv:1905.02111}}} {[astro-ph.SR]}

\bibitem[{{Kowalski} et~al(2019{\natexlab{b}}){Kowalski}, {Wisniewski},
  {Hawley}, {Osten}, {Brown}, {Fari{\~n}a}, {Valenti}, {Brown}, {Xilouris},
  {Schmidt}, and {Johns-Krull}}]{Kowalski2019HST}
{Kowalski} AF, {Wisniewski} JP, {Hawley} SL, et~al (2019{\natexlab{b}}) {The
  Near-ultraviolet Continuum Radiation in the Impulsive Phase of HF/GF-type dMe
  Flares. I. Data}. \apj 871:167. \doi{10.3847/1538-4357/aaf058},
  {\href{https://arxiv.org/abs/1811.04021}{{arXiv:1811.04021}}} {[astro-ph.SR]}

\bibitem[{{Kowalski} et~al(2022){Kowalski}, {Allred}, {Carlsson}, {Kerr},
  {Tremblay}, {Namekata}, {Kuridze}, and {Uitenbroek}}]{Kowalski2022}
{Kowalski} AF, {Allred} JC, {Carlsson} M, et~al (2022) {The Atmospheric
  Response to High Nonthermal Electron-beam Fluxes in Solar Flares. II.
  Hydrogen-broadening Predictions for Solar Flare Observations with the Daniel
  K. Inouye Solar Telescope}. \apj 928(2):190. \doi{10.3847/1538-4357/ac5174},
  {\href{https://arxiv.org/abs/2201.13349}{{arXiv:2201.13349}}} {[astro-ph.SR]}

\bibitem[{{Kretzschmar}(2011)}]{Kretzschmar2011}
{Kretzschmar} M (2011) {The Sun as a star: observations of white-light flares}.
  \aap 530:A84. \doi{10.1051/0004-6361/201015930},
  {\href{https://arxiv.org/abs/1103.3125}{{arXiv:1103.3125}}} {[astro-ph.SR]}

\bibitem[{{Krucker} and {Battaglia}(2014)}]{Krucker2014}
{Krucker} S, {Battaglia} M (2014) {Particle Densities within the Acceleration
  Region of a Solar Flare}. \apj 780(1):107. \doi{10.1088/0004-637X/780/1/107}

\bibitem[{{Krucker} et~al(2010){Krucker}, {Hudson}, {Glesener}, {White},
  {Masuda}, {Wuelser}, and {Lin}}]{Krucker2010}
{Krucker} S, {Hudson} HS, {Glesener} L, et~al (2010) {Measurements of the
  Coronal Acceleration Region of a Solar Flare}. \apj 714(2):1108--1119.
  \doi{10.1088/0004-637X/714/2/1108}

\bibitem[{{Krucker} et~al(2011){Krucker}, {Hudson}, {Jeffrey}, {Battaglia},
  {Kontar}, {Benz}, {Csillaghy}, and {Lin}}]{Krucker2011}
{Krucker} S, {Hudson} HS, {Jeffrey} NLS, et~al (2011) {High-resolution Imaging
  of Solar Flare Ribbons and Its Implication on the Thick-target Beam Model}.
  \apj 739:96. \doi{10.1088/0004-637X/739/2/96}

\bibitem[{{Krucker} et~al(2015){Krucker}, {Saint-Hilaire}, {Hudson},
  {Haberreiter}, {Martinez-Oliveros}, {Fivian}, {Hurford}, {Kleint},
  {Battaglia}, {Kuhar}, and {Arnold}}]{Krucker2015}
{Krucker} S, {Saint-Hilaire} P, {Hudson} HS, et~al (2015) {Co-Spatial White
  Light and Hard X-Ray Flare Footpoints Seen Above the Solar Limb}. \apj
  802(1):19. \doi{10.1088/0004-637X/802/1/19}

\bibitem[{{Kundu} et~al(2009){Kundu}, {Grechnev}, {White}, {Schmahl},
  {Meshalkina}, and {Kashapova}}]{Kundu2009}
{Kundu} MR, {Grechnev} VV, {White} SM, et~al (2009) {High-Energy Emission from
  a Solar Flare in Hard X-rays and Microwaves}. \solphys 260(1):135--156.
  \doi{10.1007/s11207-009-9437-3},
  {\href{https://arxiv.org/abs/0908.0385}{{arXiv:0908.0385}}} {[astro-ph.SR]}

\bibitem[{{Kunkel}(1970)}]{Kunkel1970}
{Kunkel} WE (1970) {On the Spectra of Stellar Flares}. \apj 161:503.
  \doi{10.1086/150555}

\bibitem[{{Kunze}(2009)}]{Kunze2009}
{Kunze} HJ (2009) {Introduction to Plasma Spectroscopy}, vol~56.
  \doi{10.1007/978-3-642-02233-3}

\bibitem[{{Kurochka} and {Maslennikova}(1970)}]{Kurochka1970}
{Kurochka} LN, {Maslennikova} LB (1970) {Determination of Electron Density in
  Plasma by the Number of the Extreme Resolved Line}. \solphys 11:33--41.
  \doi{10.1007/BF00156548}

\bibitem[{{Kuznetsov} and {Kolotkov}(2021)}]{Kuznetsov2021}
{Kuznetsov} AA, {Kolotkov} DY (2021) {Stellar Superflares Observed
  Simultaneously with Kepler and XMM-Newton}. \apj 912(1):81.
  \doi{10.3847/1538-4357/abf569},
  {\href{https://arxiv.org/abs/2103.10866}{{arXiv:2103.10866}}} {[astro-ph.SR]}

\bibitem[{{Lacy}(1977)}]{Lacy1977}
{Lacy} CH (1977) {Absolute dimensions and masses of the remarkable spotted dM4e
  eclipsing binary flare star CM Draconis.} \apj 218:444--460.
  \doi{10.1086/155698}

\bibitem[{{Lacy} et~al(1976){Lacy}, {Moffett}, and {Evans}}]{Lacy1976}
{Lacy} CH, {Moffett} TJ, {Evans} DS (1976) {UV Ceti stars - Statistical
  analysis of observational data}. \apjs 30:85--96. \doi{10.1086/190358}

\bibitem[{{Lalitha} et~al(2013){Lalitha}, {Fuhrmeister}, {Wolter}, {Schmitt},
  {Engels}, and {Wieringa}}]{Lalitha2013}
{Lalitha} S, {Fuhrmeister} B, {Wolter} U, et~al (2013) {A multi-wavelength view
  of AB Doradus outer atmosphere . Simultaneous X-ray and optical spectroscopy
  at high cadence}. \aap 560:A69. \doi{10.1051/0004-6361/201321419}

\bibitem[{{Laming}(2004)}]{Laming2004}
{Laming} JM (2004) {A Unified Picture of the First Ionization Potential and
  Inverse First Ionization Potential Effects}. \apj 614(2):1063--1072.
  \doi{10.1086/423780},
  {\href{https://arxiv.org/abs/astro-ph/0405230}{{arXiv:astro-ph/0405230}}}
  {[astro-ph]}

\bibitem[{{Laming}(2021)}]{Laming2021}
{Laming} JM (2021) {The FIP and Inverse-FIP Effects in Solar Flares}. \apj
  909(1):17. \doi{10.3847/1538-4357/abd9c3},
  {\href{https://arxiv.org/abs/2101.03038}{{arXiv:2101.03038}}} {[astro-ph.SR]}

\bibitem[{{Lang} et~al(1983){Lang}, {Bookbinder}, {Golub}, and
  {Davis}}]{Lang1983}
{Lang} KR, {Bookbinder} J, {Golub} L, et~al (1983) {Bright, rapid, highly
  circularly polarized radio spikes from the M dwarf AD Leonis.} \apjl
  272:L15--L18. \doi{10.1086/184108}

\bibitem[{{Lawson} et~al(2019){Lawson}, {Wisniewski}, {Bellm}, {Kowalski}, and
  {Shupe}}]{Lawson2019}
{Lawson} KD, {Wisniewski} JP, {Bellm} EC, et~al (2019) {Identification of
  Stellar Flares Using Differential Evolution Template Optimization}. \aj
  158(3):119. \doi{10.3847/1538-3881/ab3461},
  {\href{https://arxiv.org/abs/1903.03240}{{arXiv:1903.03240}}} {[astro-ph.SR]}

\bibitem[{{Lazarian} et~al(2020){Lazarian}, {Eyink}, {Jafari}, {Kowal}, {Li},
  {Xu}, and {Vishniac}}]{Lazarian2020}
{Lazarian} A, {Eyink} GL, {Jafari} A, et~al (2020) {3D turbulent reconnection:
  Theory, tests, and astrophysical implications}. Physics of Plasmas
  27(1):012305. \doi{10.1063/1.5110603},
  {\href{https://arxiv.org/abs/2001.00868}{{arXiv:2001.00868}}} {[astro-ph.HE]}

\bibitem[{{Leach} and {Petrosian}(1981)}]{Leach1981}
{Leach} J, {Petrosian} V (1981) {Impulsive phase of solar flares. I -
  Characteristics of high energy electrons}. \apj 251:781--791.
  \doi{10.1086/159521}

\bibitem[{{Leggett}(1992)}]{Leggett1992}
{Leggett} SK (1992) {Infrared Colors of Low-Mass Stars}. \apjs 82:351.
  \doi{10.1086/191720}

\bibitem[{{Lemen} et~al(2012){Lemen}, {Title}, {Akin}, {Boerner}, {Chou},
  {Drake}, {Duncan}, {Edwards}, {Friedlaender}, {Heyman}, {Hurlburt}, {Katz},
  {Kushner}, {Levay}, {Lindgren}, {Mathur}, {McFeaters}, {Mitchell}, {Rehse},
  {Schrijver}, {Springer}, {Stern}, {Tarbell}, {Wuelser}, {Wolfson}, {Yanari},
  {Bookbinder}, {Cheimets}, {Caldwell}, {Deluca}, {Gates}, {Golub}, {Park},
  {Podgorski}, {Bush}, {Scherrer}, {Gummin}, {Smith}, {Auker}, {Jerram},
  {Pool}, {Soufli}, {Windt}, {Beardsley}, {Clapp}, {Lang}, and
  {Waltham}}]{Lemen2012}
{Lemen} JR, {Title} AM, {Akin} DJ, et~al (2012) {The Atmospheric Imaging
  Assembly (AIA) on the Solar Dynamics Observatory (SDO)}. \solphys 275:17--40.
  \doi{10.1007/s11207-011-9776-8}

\bibitem[{{Lemke}(1997)}]{Lemke1997}
{Lemke} M (1997) {Extended VCS Stark broadening tables for hydrogen -- Lyman to
  Brackett series}. \aaps 122:285--292. \doi{10.1051/aas:1997134}

\bibitem[{{Leone} et~al(2004){Leone}, {Paoletti}, and {Robotti}}]{Leone2004}
{Leone} M, {Paoletti} A, {Robotti} N (2004) {A Simultaneous Discovery: The Case
  of Johannes Stark and Antonino Lo Surdo}. Physics in Perspective
  6(3):271--294. \doi{10.1007/s00016-003-0170-2}

\bibitem[{{Lestrade} et~al(1993){Lestrade}, {Phillips}, {Hodges}, and
  {Preston}}]{Lestrade1993}
{Lestrade} JF, {Phillips} RB, {Hodges} MW, et~al (1993) {VLBI Astrometric
  Identification of the Radio-emitting Region in Algol and Determination of the
  Orientation of the Close Binary}. \apj 410:808. \doi{10.1086/172798}

\bibitem[{{Liebert} et~al(1999){Liebert}, {Kirkpatrick}, {Reid}, and
  {Fisher}}]{Liebert1999}
{Liebert} J, {Kirkpatrick} JD, {Reid} IN, et~al (1999) {A 2MASS Ultracool M
  Dwarf Observed in a Spectacular Flare}. \apj 519(1):345--353.
  \doi{10.1086/307349}

\bibitem[{{Liefke} et~al(2008){Liefke}, {Ness}, {Schmitt}, and
  {Maggio}}]{Liefke2008}
{Liefke} C, {Ness} JU, {Schmitt} JHMM, et~al (2008) {Coronal properties of the
  EQ Pegasi binary system}. \aap 491(3):859--872.
  \doi{10.1051/0004-6361:200810054},
  {\href{https://arxiv.org/abs/0810.0150}{{arXiv:0810.0150}}} {[astro-ph]}

\bibitem[{{Liefke} et~al(2010){Liefke}, {Fuhrmeister}, and
  {Schmitt}}]{Liefke2010}
{Liefke} C, {Fuhrmeister} B, {Schmitt} JHMM (2010) {Multiwavelength
  observations of a giant flare on CN Leonis. III. Temporal evolution of
  coronal properties}. \aap 514:A94. \doi{10.1051/0004-6361/201014012},
  {\href{https://arxiv.org/abs/1003.4128}{{arXiv:1003.4128}}} {[astro-ph.SR]}

\bibitem[{{Lin} et~al(2001){Lin}, {Forbes}, and {Isenberg}}]{Lin2001}
{Lin} J, {Forbes} TG, {Isenberg} PA (2001) {Prominence eruptions and coronal
  mass ejections triggered by newly emerging flux}. \jgr 106(A11):25053--25074.
  \doi{10.1029/2001JA000046}

\bibitem[{{Lin} and {Hudson}(1976)}]{Lin1976}
{Lin} RP, {Hudson} HS (1976) {Non-thermal processes in large solar flares.}
  \solphys 50(1):153--178. \doi{10.1007/BF00206199}

\bibitem[{{Lin} et~al(2002){Lin}, {Dennis}, {Hurford}, {Smith}, {Zehnder},
  {Harvey}, {Curtis}, {Pankow}, {Turin}, {Bester}, {Csillaghy}, {Lewis},
  {Madden}, {van Beek}, {Appleby}, {Raudorf}, {McTiernan}, {Ramaty}, {Schmahl},
  {Schwartz}, {Krucker}, {Abiad}, {Quinn}, {Berg}, {Hashii}, {Sterling},
  {Jackson}, {Pratt}, {Campbell}, {Malone}, {Landis}, {Barrington-Leigh},
  {Slassi-Sennou}, {Cork}, {Clark}, {Amato}, {Orwig}, {Boyle}, {Banks},
  {Shirey}, {Tolbert}, {Zarro}, {Snow}, {Thomsen}, {Henneck}, {McHedlishvili},
  {Ming}, {Fivian}, {Jordan}, {Wanner}, {Crubb}, {Preble}, {Matranga}, {Benz},
  {Hudson}, {Canfield}, {Holman}, {Crannell}, {Kosugi}, {Emslie}, {Vilmer},
  {Brown}, {Johns-Krull}, {Aschwanden}, {Metcalf}, and {Conway}}]{Lin2002}
{Lin} RP, {Dennis} BR, {Hurford} GJ, et~al (2002) {The Reuven Ramaty
  High-Energy Solar Spectroscopic Imager (RHESSI)}. \solphys 210(1):3--32.
  \doi{10.1023/A:1022428818870}

\bibitem[{{Lin} et~al(2003){Lin}, {Krucker}, {Hurford}, {Smith}, {Hudson},
  {Holman}, {Schwartz}, {Dennis}, {Share}, {Murphy}, {Emslie}, {Johns-Krull},
  and {Vilmer}}]{Lin2003}
{Lin} RP, {Krucker} S, {Hurford} GJ, et~al (2003) {RHESSI Observations of
  Particle Acceleration and Energy Release in an Intense Solar Gamma-Ray Line
  Flare}. \apjl 595(2):L69--L76. \doi{10.1086/378932}

\bibitem[{{Linsky} and {Wood}(1994)}]{Linsky1994}
{Linsky} JL, {Wood} BE (1994) {High-Velocity Plasma in the Transition Region of
  AU Microscopii: Evidence for Magnetic Reconnection and Saturated Heating
  during Quiescent and Flaring Conditions}. \apj 430:342. \doi{10.1086/174409}

\bibitem[{{Linsky} et~al(1995){Linsky}, {Wood}, {Brown}, {Giampapa}, and
  {Ambruster}}]{Linsky1995}
{Linsky} JL, {Wood} BE, {Brown} A, et~al (1995) {Stellar Activity at the End of
  the Main Sequence: GHRS Observations of the M8 Ve Star VB 10}. \apj 455:670.
  \doi{10.1086/176614}

\bibitem[{{Littman} et~al(1978){Littman}, {Kash}, and {Kleppner}}]{Littman1978}
{Littman} MG, {Kash} MM, {Kleppner} D (1978) {Field-ionization processes in
  excited atoms}. \prl 41(2):103--107. \doi{10.1103/PhysRevLett.41.103}

\bibitem[{{Liu} et~al(2013{\natexlab{a}}){Liu}, {Chen}, and
  {Petrosian}}]{Liu2013}
{Liu} W, {Chen} Q, {Petrosian} V (2013{\natexlab{a}}) {Plasmoid Ejections and
  Loop Contractions in an Eruptive M7.7 Solar Flare: Evidence of Particle
  Acceleration and Heating in Magnetic Reconnection Outflows}. \apj 767(2):168.
  \doi{10.1088/0004-637X/767/2/168},
  {\href{https://arxiv.org/abs/1303.3321}{{arXiv:1303.3321}}} {[astro-ph.SR]}

\bibitem[{{Liu} et~al(2013{\natexlab{b}}){Liu}, {Qiu}, {Longcope}, and
  {Caspi}}]{LiuQiu2013}
{Liu} WJ, {Qiu} J, {Longcope} DW, et~al (2013{\natexlab{b}}) {Determining
  Heating Rates in Reconnection Formed Flare Loops of the M8.0 Flare on 2005
  May 13}. \apj 770(2):111. \doi{10.1088/0004-637X/770/2/111},
  {\href{https://arxiv.org/abs/1304.4521}{{arXiv:1304.4521}}} {[astro-ph.SR]}

\bibitem[{{Livshits} et~al(1981){Livshits}, {Badalian}, {Kosovichev}, and
  {Katsova}}]{Livshits1981}
{Livshits} MA, {Badalian} OG, {Kosovichev} AG, et~al (1981) {The Optical
  Continuum of Solar and Stellar Flares}. \solphys 73(2):269--288.
  \doi{10.1007/BF00151682}

\bibitem[{{Loh} et~al(2017){Loh}, {Corbel}, and {Dubus}}]{Loh2017}
{Loh} A, {Corbel} S, {Dubus} G (2017) {Fermi/LAT detection of a transient
  gamma-ray flare in the vicinity of the binary star DG CVn}. \mnras
  467(4):4462--4466. \doi{10.1093/mnras/stx396},
  {\href{https://arxiv.org/abs/1702.03754}{{arXiv:1702.03754}}} {[astro-ph.HE]}

\bibitem[{{Longair}(2013)}]{Longair2013}
{Longair} M (2013) {Quantum Concepts in Physics}

\bibitem[{{Longcope} et~al(2016){Longcope}, {Qiu}, and {Brewer}}]{Longcope2016}
{Longcope} D, {Qiu} J, {Brewer} J (2016) {A Reconnection-driven Model of the
  Hard X-Ray Loop-top Source from Flare 2004-Feb-26}. \apj 833:211.
  \doi{10.3847/1538-4357/833/2/211},
  {\href{https://arxiv.org/abs/1610.07953}{{arXiv:1610.07953}}} {[astro-ph.SR]}

\bibitem[{{Longcope} et~al(2018){Longcope}, {Unverferth}, {Klein}, {McCarthy},
  and {Priest}}]{Longcope2018}
{Longcope} D, {Unverferth} J, {Klein} C, et~al (2018) {Evidence for Downflows
  in the Narrow Plasma Sheet of 2017 September 10 and Their Significance for
  Flare Reconnection}. \apj 868(2):148. \doi{10.3847/1538-4357/aaeac4}

\bibitem[{{Loyd} and {France}(2014)}]{Loyd2014}
{Loyd} ROP, {France} K (2014) {Fluctuations and Flares in the Ultraviolet Line
  Emission of Cool Stars: Implications for Exoplanet Transit Observations}.
  \apjs 211(1):9. \doi{10.1088/0067-0049/211/1/9},
  {\href{https://arxiv.org/abs/1402.0073}{{arXiv:1402.0073}}} {[astro-ph.SR]}

\bibitem[{{Loyd} et~al(2018{\natexlab{a}}){Loyd}, {France}, {Youngblood},
  {Schneider}, {Brown}, {Hu}, {Segura}, {Linsky}, {Redfield}, {Tian},
  {Rugheimer}, {Miguel}, and {Froning}}]{Loyd2018}
{Loyd} ROP, {France} K, {Youngblood} A, et~al (2018{\natexlab{a}}) {The MUSCLES
  Treasury Survey. V. FUV Flares on Active and Inactive M Dwarfs}. \apj
  867(1):71. \doi{10.3847/1538-4357/aae2bd},
  {\href{https://arxiv.org/abs/1809.07322}{{arXiv:1809.07322}}} {[astro-ph.SR]}

\bibitem[{{Loyd} et~al(2018{\natexlab{b}}){Loyd}, {Shkolnik}, {Schneider},
  {Barman}, {Meadows}, {Pagano}, and {Peacock}}]{Loyd2018Hazmat}
{Loyd} ROP, {Shkolnik} EL, {Schneider} AC, et~al (2018{\natexlab{b}}) {HAZMAT.
  IV. Flares and Superflares on Young M Stars in the Far Ultraviolet}. \apj
  867(1):70. \doi{10.3847/1538-4357/aae2ae},
  {\href{https://arxiv.org/abs/1810.03277}{{arXiv:1810.03277}}} {[astro-ph.SR]}

\bibitem[{{Lurie} et~al(2015){Lurie}, {Davenport}, {Hawley}, {Wilkinson},
  {Wisniewski}, {Kowalski}, and {Hebb}}]{Lurie2015}
{Lurie} JC, {Davenport} JRA, {Hawley} SL, et~al (2015) {Kepler Flares III:
  Stellar Activity on GJ 1245A and B}. \apj 800(2):95.
  \doi{10.1088/0004-637X/800/2/95},
  {\href{https://arxiv.org/abs/1412.6109}{{arXiv:1412.6109}}} {[astro-ph.SR]}

\bibitem[{{Lynch} et~al(2016){Lynch}, {Edmondson}, {Kazachenko}, and
  {Guidoni}}]{Lynch2016}
{Lynch} BJ, {Edmondson} JK, {Kazachenko} MD, et~al (2016) {Reconnection
  Properties of Large-scale Current Sheets During Coronal Mass Ejection
  Eruptions}. \apj 826(1):43. \doi{10.3847/0004-637X/826/1/43},
  {\href{https://arxiv.org/abs/1410.1089}{{arXiv:1410.1089}}} {[astro-ph.SR]}

\bibitem[{{Maas} et~al(2022){Maas}, {Ilin}, {Oshagh}, {Pall{\'e}},
  {Parviainen}, {Molaverdikhani}, {Quirrenbach}, {Esparza-Borges}, {Murgas},
  {B{\'e}jar}, {Narita}, {Fukui}, {Lin}, {Mori}, and {Klagyivik}}]{Maas2022}
{Maas} AJ, {Ilin} E, {Oshagh} M, et~al (2022) {Lower-than-expected flare
  temperatures for TRAPPIST-1}. \aap 668:A111.
  \doi{10.1051/0004-6361/202243869},
  {\href{https://arxiv.org/abs/2210.11103}{{arXiv:2210.11103}}} {[astro-ph.SR]}

\bibitem[{{MacGregor} et~al(2020){MacGregor}, {Osten}, and
  {Hughes}}]{MacGregor2020}
{MacGregor} AM, {Osten} RA, {Hughes} AM (2020) {Properties of M Dwarf Flares at
  Millimeter Wavelengths}. \apj 891(1):80. \doi{10.3847/1538-4357/ab711d}

\bibitem[{{MacGregor} et~al(2018){MacGregor}, {Weinberger}, {Wilner},
  {Kowalski}, and {Cranmer}}]{MacGregor2018}
{MacGregor} MA, {Weinberger} AJ, {Wilner} DJ, et~al (2018) {Detection of a
  Millimeter Flare from Proxima Centauri}. \apjl 855:L2.
  \doi{10.3847/2041-8213/aaad6b},
  {\href{https://arxiv.org/abs/1802.08257}{{arXiv:1802.08257}}} {[astro-ph.EP]}

\bibitem[{{MacGregor} et~al(2021){MacGregor}, {Weinberger}, {Loyd}, {Shkolnik},
  {Barclay}, {Howard}, {Zic}, {Osten}, {Cranmer}, {Kowalski}, {Lenc},
  {Youngblood}, {Estes}, {Wilner}, {Forbrich}, {Hughes}, {Law}, {Murphy},
  {Boley}, and {Matthews}}]{MacGregor2021}
{MacGregor} MA, {Weinberger} AJ, {Loyd} ROP, et~al (2021) {Discovery of an
  Extremely Short Duration Flare from Proxima Centauri Using Millimeter through
  Far-ultraviolet Observations}. \apjl 911(2):L25.
  \doi{10.3847/2041-8213/abf14c},
  {\href{https://arxiv.org/abs/2104.09519}{{arXiv:2104.09519}}} {[astro-ph.SR]}

\bibitem[{{Maehara} et~al(2012){Maehara}, {Shibayama}, {Notsu}, {Notsu},
  {Nagao}, {Kusaba}, {Honda}, {Nogami}, and {Shibata}}]{Maehara2012}
{Maehara} H, {Shibayama} T, {Notsu} S, et~al (2012) {Superflares on solar-type
  stars}. \nat 485:478--481. \doi{10.1038/nature11063}

\bibitem[{{Maehara} et~al(2015){Maehara}, {Shibayama}, {Notsu}, {Notsu},
  {Honda}, {Nogami}, and {Shibata}}]{Maehara2015}
{Maehara} H, {Shibayama} T, {Notsu} Y, et~al (2015) {Statistical properties of
  superflares on solar-type stars based on 1-min cadence data}. Earth, Planets,
  and Space 67:59. \doi{10.1186/s40623-015-0217-z},
  {\href{https://arxiv.org/abs/1504.00074}{{arXiv:1504.00074}}} {[astro-ph.SR]}

\bibitem[{{Maehara} et~al(2021){Maehara}, {Notsu}, {Namekata}, {Honda},
  {Kowalski}, {Katoh}, {Ohshima}, {Iida}, {Oeda}, {Murata}, {Yamanaka},
  {Takagi}, {Sasada}, {Akitaya}, {Ikuta}, {Okamoto}, {Nogami}, and
  {Shibata}}]{Maehara2021}
{Maehara} H, {Notsu} Y, {Namekata} K, et~al (2021) {Time-resolved spectroscopy
  and photometry of M dwarf flare star YZ Canis Minoris with OISTER and TESS:
  Blue asymmetry in the H{\ensuremath{\alpha}} line during the non-white light
  flare}. \pasj 73(1):44--65. \doi{10.1093/pasj/psaa098},
  {\href{https://arxiv.org/abs/2009.14412}{{arXiv:2009.14412}}} {[astro-ph.SR]}

\bibitem[{{Magaudda} et~al(2020){Magaudda}, {Stelzer}, {Covey}, {Raetz},
  {Matt}, and {Scholz}}]{Magaudda2020}
{Magaudda} E, {Stelzer} B, {Covey} KR, et~al (2020) {Relation of X-ray activity
  and rotation in M dwarfs and predicted time-evolution of the X-ray
  luminosity}. \aap 638:A20. \doi{10.1051/0004-6361/201937408},
  {\href{https://arxiv.org/abs/2004.02904}{{arXiv:2004.02904}}} {[astro-ph.SR]}

\bibitem[{{Maggio} et~al(2000){Maggio}, {Pallavicini}, {Reale}, and
  {Tagliaferri}}]{Maggio2000}
{Maggio} A, {Pallavicini} R, {Reale} F, et~al (2000) {Twin X-ray flares and the
  active corona of AB Dor observed with BeppoSAX}. \aap 356:627--642

\bibitem[{{Ma{\'\i}z Apell{\'a}niz}(2006)}]{MA2006}
{Ma{\'\i}z Apell{\'a}niz} J (2006) {A Recalibration of Optical Photometry:
  Tycho-2, Str{\"o}mgren, and Johnson Systems}. \aj 131(2):1184--1199.
  \doi{10.1086/499158},
  {\href{https://arxiv.org/abs/astro-ph/0510785}{{arXiv:astro-ph/0510785}}}
  {[astro-ph]}

\bibitem[{{Mamajek} and {Bell}(2014)}]{Mamajek2014}
{Mamajek} EE, {Bell} CPM (2014) {On the age of the {\ensuremath{\beta}}
  Pictoris moving group}. \mnras 445(3):2169--2180. \doi{10.1093/mnras/stu1894}

\bibitem[{{Mamajek} and {Hillenbrand}(2008)}]{Mamajek2008}
{Mamajek} EE, {Hillenbrand} LA (2008) {Improved Age Estimation for Solar-Type
  Dwarfs Using Activity-Rotation Diagnostics}. \apj 687(2):1264--1293.
  \doi{10.1086/591785}

\bibitem[{{Massi} et~al(2006){Massi}, {Forbrich}, {Menten},
  {Torricelli-Ciamponi}, {Neidh{\"o}fer}, {Leurini}, and
  {Bertoldi}}]{Massi2006}
{Massi} M, {Forbrich} J, {Menten} KM, et~al (2006) {Synchrotron emission from
  the T Tauri binary system V773 Tauri A}. \aap 453:959--964.
  \doi{10.1051/0004-6361:20053535},
  {\href{https://arxiv.org/abs/astro-ph/0604124}{{astro-ph/0604124}}}

\bibitem[{{Masuda} et~al(1994){Masuda}, {Kosugi}, {Hara}, {Tsuneta}, and
  {Ogawara}}]{Masuda1994}
{Masuda} S, {Kosugi} T, {Hara} H, et~al (1994) {A loop-top hard X-ray source in
  a compact solar flare as evidence for magnetic reconnection}. \nat
  371(6497):495--497. \doi{10.1038/371495a0}

\bibitem[{{Mathieu} et~al(1997){Mathieu}, {Stassun}, {Basri}, {Jensen},
  {Johns-Krull}, {Valenti}, and {Hartmann}}]{Mathieu1997}
{Mathieu} RD, {Stassun} K, {Basri} G, et~al (1997) {The Classical T Tauri
  Spectroscopic Binary DQ Tau.I.Orbital Elements and Light Curves}. \aj
  113:1841. \doi{10.1086/118395}

\bibitem[{{Mathioudakis} and {Doyle}(1990)}]{Math644}
{Mathioudakis} M, {Doyle} JG (1990) {IUE observations of GL 644 AB (=Wolf 630)
  in the wavelength region 1150-1950 A, in June 1981.} \aap 232:114

\bibitem[{{Mathioudakis} et~al(1992){Mathioudakis}, {Doyle}, {Avgoloupis},
  {Mavridis}, and {Seiradakis}}]{Mathioudakis1992}
{Mathioudakis} M, {Doyle} JG, {Avgoloupis} V, et~al (1992) {Optical flares on
  the RS CVn binary II Peg.} \mnras 255:48--50. \doi{10.1093/mnras/255.1.48}

\bibitem[{{Mathioudakis} et~al(1999){Mathioudakis}, {McKenny}, {Keenan},
  {Williams}, and {Phillips}}]{Mathioudakis1999}
{Mathioudakis} M, {McKenny} J, {Keenan} FP, et~al (1999) {The effects of
  opacity in the transition region of YZ CMi}. \aap 351:L23--L26

\bibitem[{{Mathioudakis} et~al(2003){Mathioudakis}, {Seiradakis}, {Williams},
  {Avgoloupis}, {Bloomfield}, and {McAteer}}]{Mathioudakis2003}
{Mathioudakis} M, {Seiradakis} JH, {Williams} DR, et~al (2003) {White-light
  oscillations during a flare on II Peg.} \aap 403:1101--1104.
  \doi{10.1051/0004-6361:20030394}

\bibitem[{{Mathioudakis} et~al(2006){Mathioudakis}, {Bloomfield}, {Jess},
  {Dhillon}, and {Marsh}}]{Mathioudakis2006}
{Mathioudakis} M, {Bloomfield} DS, {Jess} DB, et~al (2006) {The periodic
  variations of a white-light flare observed with ULTRACAM}. \aap
  456(1):323--327. \doi{10.1051/0004-6361:20054752},
  {\href{https://arxiv.org/abs/astro-ph/0605196}{{arXiv:astro-ph/0605196}}}
  {[astro-ph]}

\bibitem[{{Matt} et~al(2011){Matt}, {Do Cao}, {Brown}, and {Brun}}]{Matt2011}
{Matt} SP, {Do Cao} O, {Brown} BP, et~al (2011) {Convection and differential
  rotation properties of G and K stars computed with the ASH code}.
  Astronomische Nachrichten 332:897. \doi{10.1002/asna.201111624},
  {\href{https://arxiv.org/abs/1111.5585}{{arXiv:1111.5585}}} {[astro-ph.SR]}

\bibitem[{{Mauas} and {Falchi}(1996)}]{Mauas1996}
{Mauas} PJD, {Falchi} A (1996) {Atmospheric models of flare stars: the flaring
  state of AD Leonis.} \aap 310:245--258

\bibitem[{{Mauas} and {G{\'o}mez}(1997)}]{Mauas1997}
{Mauas} PJD, {G{\'o}mez} DO (1997) {Fokker-Planck Description of Electron Beams
  in the Solar Chromosphere}. \apj 483(1):496--506. \doi{10.1086/304203}

\bibitem[{{McClymont} and {Canfield}(1983)}]{McClymont1983}
{McClymont} AN, {Canfield} RC (1983) {Flare loop radiative hydrodynamics. I -
  Basic methods}. \apj 265:483--506. \doi{10.1086/160692}

\bibitem[{{McClymont} and {Canfield}(1986)}]{McClymont1986}
{McClymont} AN, {Canfield} RC (1986) {The Solar Flare Extreme Ultraviolet to
  Hard X-Ray Ratio}. \apj 305:936. \doi{10.1086/164307}

\bibitem[{{McIntosh} et~al(2023){McIntosh}, {Leamon}, and
  {Egeland}}]{McIntosh2023}
{McIntosh} SW, {Leamon} RJ, {Egeland} R (2023) {Deciphering Solar Magnetic
  Activity: The (Solar) Hale Cycle Terminator of 2021}. Frontiers in Astronomy
  and Space Sciences 10:16. \doi{10.3389/fspas.2023.1050523},
  {\href{https://arxiv.org/abs/2209.10577}{{arXiv:2209.10577}}} {[astro-ph.SR]}

\bibitem[{{McTiernan} and {Petrosian}(1990)}]{McTiernan1990}
{McTiernan} JM, {Petrosian} V (1990) {The Behavior of Beams of Relativistic
  Nonthermal Electrons under the Influence of Collisions and Synchrotron
  Losses}. \apj 359:524. \doi{10.1086/169084}

\bibitem[{{McTiernan} et~al(1999){McTiernan}, {Fisher}, and
  {Li}}]{McTiernan1999}
{McTiernan} JM, {Fisher} GH, {Li} P (1999) {The Solar Flare Soft X-Ray
  Differential Emission Measure and the Neupert Effect at Different
  Temperatures}. \apj 514(1):472--483. \doi{10.1086/306924}

\bibitem[{{Medina} et~al(2020){Medina}, {Winters}, {Irwin}, and
  {Charbonneau}}]{Medina2020}
{Medina} AA, {Winters} JG, {Irwin} JM, et~al (2020) {Flare Rates, Rotation
  Periods, and Spectroscopic Activity Indicators of a Volume-complete Sample of
  Mid- to Late-M Dwarfs within 15 pc}. \apj 905(2):107.
  \doi{10.3847/1538-4357/abc686}

\bibitem[{{Meigs} et~al(2013){Meigs}, {Brezinsek}, {Clever}, {Huber}, {Marsen},
  {Nicholas}, {Stamp}, {Zastrow}, and {JET EFDA Contributors}}]{Meigs2013}
{Meigs} AG, {Brezinsek} S, {Clever} M, et~al (2013) {Deuterium Balmer/Stark
  spectroscopy and impurity profiles: First results from mirror-link divertor
  spectroscopy system on the JET ITER-like wall}. Journal of Nuclear Materials
  438:S607--S611. \doi{10.1016/j.jnucmat.2013.01.127},
  {\href{https://arxiv.org/abs/1307.6985}{{arXiv:1307.6985}}}
  {[physics.plasm-ph]}

\bibitem[{{Melrose} and {Brown}(1976)}]{Melrose1976}
{Melrose} DB, {Brown} JC (1976) {Precipitation in trap models for solar hard
  X-ray bursts.} \mnras 176:15--30. \doi{10.1093/mnras/176.1.15}

\bibitem[{{Mihalas}(1978)}]{Mihalas1978}
{Mihalas} D (1978) {Stellar atmospheres /2nd edition/}

\bibitem[{{Miller} et~al(1996){Miller}, {Larosa}, and {Moore}}]{Miller1996}
{Miller} JA, {Larosa} TN, {Moore} RL (1996) {Stochastic Electron Acceleration
  by Cascading Fast Mode Waves in Impulsive Solar Flares}. \apj 461:445.
  \doi{10.1086/177072}

\bibitem[{{Milligan} and {Dennis}(2009)}]{Milligan2009}
{Milligan} RO, {Dennis} BR (2009) {Velocity Characteristics of Evaporated
  Plasma Using Hinode/EUV Imaging Spectrometer}. \apj 699(2):968--975.
  \doi{10.1088/0004-637X/699/2/968},
  {\href{https://arxiv.org/abs/0905.1669}{{arXiv:0905.1669}}} {[astro-ph.SR]}

\bibitem[{{Million} et~al(2016){Million}, {Fleming}, {Shiao}, {Seibert},
  {Loyd}, {Tucker}, {Smith}, {Thompson}, and {White}}]{Million2016}
{Million} C, {Fleming} SW, {Shiao} B, et~al (2016) {gPhoton: The GALEX Photon
  Data Archive}. \apj 833(2):292. \doi{10.3847/1538-4357/833/2/292},
  {\href{https://arxiv.org/abs/1609.09492}{{arXiv:1609.09492}}} {[astro-ph.IM]}

\bibitem[{{Mitra-Kraev} et~al(2005{\natexlab{a}}){Mitra-Kraev}, {Harra},
  {G{\"u}del}, {Audard}, {Branduardi-Raymont}, {Kay}, {Mewe}, {Raassen}, and
  {van Driel-Gesztelyi}}]{Mitra2005}
{Mitra-Kraev} U, {Harra} LK, {G{\"u}del} M, et~al (2005{\natexlab{a}})
  {Relationship between X-ray and ultraviolet emission of flares from dMe stars
  observed by XMM-Newton}. \aap 431:679--686. \doi{10.1051/0004-6361:20041201},
  {\href{https://arxiv.org/abs/astro-ph/0410592}{{astro-ph/0410592}}}

\bibitem[{{Mitra-Kraev} et~al(2005{\natexlab{b}}){Mitra-Kraev}, {Harra},
  {Williams}, and {Kraev}}]{MitraKraev2005B}
{Mitra-Kraev} U, {Harra} LK, {Williams} DR, et~al (2005{\natexlab{b}}) {The
  first observed stellar X-ray flare oscillation: Constraints on the flare loop
  length and the magnetic field}. \aap 436(3):1041--1047.
  \doi{10.1051/0004-6361:20052834},
  {\href{https://arxiv.org/abs/astro-ph/0503384}{{arXiv:astro-ph/0503384}}}
  {[astro-ph]}

\bibitem[{{Mochnacki} and {Zirin}(1980)}]{Mochnacki1980}
{Mochnacki} SW, {Zirin} H (1980) {Multichannel spectrophotometry of stellar
  flares.} \apjl 239:L27--L31. \doi{10.1086/183285}

\bibitem[{{Moffett}(1974)}]{Moffett1974}
{Moffett} TJ (1974) {UV Ceti flare stars: observational data.} \apjs 29:1--42.
  \doi{10.1086/190330}

\bibitem[{{Mondrik} et~al(2019){Mondrik}, {Newton}, {Charbonneau}, and
  {Irwin}}]{Mondrik2019}
{Mondrik} N, {Newton} E, {Charbonneau} D, et~al (2019) {An Increased Rate of
  Large Flares at Intermediate Rotation Periods for Mid-to-late M Dwarfs}. \apj
  870(1):10. \doi{10.3847/1538-4357/aaee64},
  {\href{https://arxiv.org/abs/1809.09177}{{arXiv:1809.09177}}} {[astro-ph.SR]}

\bibitem[{{Montes} et~al(1999){Montes}, {Saar}, {Collier Cameron}, and
  {Unruh}}]{Montes1999}
{Montes} D, {Saar} SH, {Collier Cameron} A, et~al (1999) {Optical and
  ultraviolet observations of a strong flare in the young, single K2 dwarf LQ
  Hya}. \mnras 305(1):45--60. \doi{10.1046/j.1365-8711.1999.02373.x},
  {\href{https://arxiv.org/abs/astro-ph/9811452}{{arXiv:astro-ph/9811452}}}
  {[astro-ph]}

\bibitem[{{Montgomery} et~al(2022){Montgomery}, {Dunlap}, {Cho}, and
  {Gomez}}]{Montgomery2022}
{Montgomery} MH, {Dunlap} BH, {Cho} PB, et~al (2022) {Hydrogen Line Shape
  Uncertainties in White Dwarf Model Atmospheres}. Frontiers in Astronomy and
  Space Sciences 9:830163. \doi{10.3389/fspas.2022.830163}

\bibitem[{{Moore} et~al(2014){Moore}, {Chamberlin}, and {Hock}}]{Moore2014}
{Moore} CS, {Chamberlin} PC, {Hock} R (2014) {Measurements and Modeling of
  Total Solar Irradiance in X-class Solar Flares}. \apj 787(1):32.
  \doi{10.1088/0004-637X/787/1/32},
  {\href{https://arxiv.org/abs/1509.06074}{{arXiv:1509.06074}}} {[astro-ph.IM]}

\bibitem[{{Morales} et~al(2009){Morales}, {Ribas}, {Jordi}, {Torres},
  {Gallardo}, {Guinan}, {Charbonneau}, {Wolf}, {Latham}, {Anglada-Escud{\'e}},
  {Bradstreet}, {Everett}, {O'Donovan}, {Mandushev}, and
  {Mathieu}}]{Morales2009}
{Morales} JC, {Ribas} I, {Jordi} C, et~al (2009) {Absolute Properties of the
  Low-Mass Eclipsing Binary CM Draconis}. \apj 691(2):1400--1411.
  \doi{10.1088/0004-637X/691/2/1400},
  {\href{https://arxiv.org/abs/0810.1541}{{arXiv:0810.1541}}} {[astro-ph]}

\bibitem[{{Morchenko} et~al(2015){Morchenko}, {Bychkov}, and
  {Livshits}}]{Morchenko2015}
{Morchenko} E, {Bychkov} K, {Livshits} M (2015) {Continuum and line emission of
  flares on red dwarf stars}. \apss 357(2):119.
  \doi{10.1007/s10509-015-2347-y},
  {\href{https://arxiv.org/abs/1504.02749}{{arXiv:1504.02749}}} {[astro-ph.SR]}

\bibitem[{{Morgan} et~al(2016){Morgan}, {West}, and {Becker}}]{Morgan2016}
{Morgan} DP, {West} AA, {Becker} AC (2016) {Using Close White Dwarf + M Dwarf
  Stellar Pairs to Constrain the Flare Rates in Close Stellar Binaries}. \aj
  151(5):114. \doi{10.3847/0004-6256/151/5/114},
  {\href{https://arxiv.org/abs/1602.06296}{{arXiv:1602.06296}}} {[astro-ph.SR]}

\bibitem[{{Morin} et~al(2008){Morin}, {Donati}, {Petit}, {Delfosse},
  {Forveille}, {Albert}, {Auri{\`e}re}, {Cabanac}, {Dintrans}, {Fares},
  {Gastine}, {Jardine}, {Ligni{\`e}res}, {Paletou}, {Ramirez Velez}, and
  {Th{\'e}ado}}]{Morin2008}
{Morin} J, {Donati} JF, {Petit} P, et~al (2008) {Large-scale magnetic
  topologies of mid M dwarfs}. \mnras 390(2):567--581.
  \doi{10.1111/j.1365-2966.2008.13809.x},
  {\href{https://arxiv.org/abs/0808.1423}{{arXiv:0808.1423}}} {[astro-ph]}

\bibitem[{{Morris} et~al(2018){Morris}, {Agol}, {Davenport}, and
  {Hawley}}]{Morris2018}
{Morris} BM, {Agol} E, {Davenport} JRA, et~al (2018) {Possible Bright Starspots
  on TRAPPIST-1}. \apj 857(1):39. \doi{10.3847/1538-4357/aab6a5},
  {\href{https://arxiv.org/abs/1803.04543}{{arXiv:1803.04543}}} {[astro-ph.SR]}

\bibitem[{{Morrissey} et~al(2007){Morrissey}, {Conrow}, {Barlow}, {Small},
  {Seibert}, {Wyder}, {Budav{\'a}ri}, {Arnouts}, {Friedman}, {Forster},
  {Martin}, {Neff}, {Schiminovich}, {Bianchi}, {Donas}, {Heckman}, {Lee},
  {Madore}, {Milliard}, {Rich}, {Szalay}, {Welsh}, and {Yi}}]{Morrissey2007}
{Morrissey} P, {Conrow} T, {Barlow} TA, et~al (2007) {The Calibration and Data
  Products of GALEX}. \apjs 173(2):682--697. \doi{10.1086/520512},
  {\href{https://arxiv.org/abs/0706.0755}{{arXiv:0706.0755}}} {[astro-ph]}

\bibitem[{{Mott} and {Massey}(1949)}]{Mott1949}
{Mott} NF, {Massey} HSW (1949) {The theory of atomic collisions}

\bibitem[{{Mullan}(1976)}]{Mullan1976}
{Mullan} DJ (1976) {Mean colors of stellar flare continuum.} \apj 210:702--712.
  \doi{10.1086/154877}

\bibitem[{{Mullan}(2009)}]{Mullan2009}
{Mullan} DJ (2009) {Flares on a Bp Star}. \apj 702(1):759--766.
  \doi{10.1088/0004-637X/702/1/759}

\bibitem[{Mullan and Mathioudakis(2000)}]{Mullan2000}
Mullan DJ, Mathioudakis M (2000) Extreme-ultraviolet flares in an f2 star. The
  Astrophysical Journal 544(1):475. \doi{10.1086/317202},
  \urlprefix\url{https://dx.doi.org/10.1086/317202}

\bibitem[{{Mullan} and {Paudel}(2019)}]{Mullan2019}
{Mullan} DJ, {Paudel} RR (2019) {Origin of Radio-quiet Coronal Mass Ejections
  in Flare Stars}. \apj 873(1):1. \doi{10.3847/1538-4357/ab041b},
  {\href{https://arxiv.org/abs/1902.00810}{{arXiv:1902.00810}}} {[astro-ph.SR]}

\bibitem[{{Mullan} and {Tarter}(1977)}]{MullanTarter1977}
{Mullan} DJ, {Tarter} CB (1977) {Influence of stellar flare X-rays on the
  optical light curve.} \apj 212:179--185. \doi{10.1086/155033}

\bibitem[{{Mullan} et~al(2006){Mullan}, {Mathioudakis}, {Bloomfield}, and
  {Christian}}]{Mullan2006}
{Mullan} DJ, {Mathioudakis} M, {Bloomfield} DS, et~al (2006) {A Comparative
  Study of Flaring Loops in Active Stars}. \apjs 164(1):173--201.
  \doi{10.1086/502629}

\bibitem[{{Murphy} et~al(1997){Murphy}, {Share}, {Grove}, {Johnson}, {Kinzer},
  {Kurfess}, {Strickman}, and {Jung}}]{Murphy1997}
{Murphy} RJ, {Share} GH, {Grove} JE, et~al (1997) {Accelerated Particle
  Composition and Energetics and Ambient Abundances from Gamma-Ray Spectroscopy
  of the 1991 June 4 Solar Flare}. \apj 490(2):883--900. \doi{10.1086/304902}

\bibitem[{{Nakariakov} et~al(2006){Nakariakov}, {Foullon}, {Verwichte}, and
  {Young}}]{Nakariakov2006}
{Nakariakov} VM, {Foullon} C, {Verwichte} E, et~al (2006) {Quasi-periodic
  modulation of solar and stellar flaring emission by magnetohydrodynamic
  oscillations in a nearby loop}. \aap 452(1):343--346.
  \doi{10.1051/0004-6361:20054608}

\bibitem[{{Namekata} et~al(2017){Namekata}, {Sakaue}, {Watanabe}, {Asai},
  {Maehara}, {Notsu}, {Notsu}, {Honda}, {Ishii}, {Ikuta}, {Nogami}, and
  {Shibata}}]{Namekata2017}
{Namekata} K, {Sakaue} T, {Watanabe} K, et~al (2017) {Statistical Studies of
  Solar White-light Flares and Comparisons with Superflares on Solar-type
  Stars}. \apj 851:91. \doi{10.3847/1538-4357/aa9b34},
  {\href{https://arxiv.org/abs/1710.11325}{{arXiv:1710.11325}}} {[astro-ph.SR]}

\bibitem[{{Namekata} et~al(2020){Namekata}, {Maehara}, {Sasaki}, {Kawai},
  {Notsu}, {Kowalski}, {Allred}, {Iwakiri}, {Tsuboi}, {Murata}, {Niwano},
  {Shiraishi}, {Adachi}, {Iida}, {Oeda}, {Honda}, {Tozuka}, {Katoh}, {Onozato},
  {Okamoto}, {Isogai}, {Kimura}, {Kojiguchi}, {Wakamatsu}, {Tampo}, {Nogami},
  and {Shibata}}]{Namekata2020}
{Namekata} K, {Maehara} H, {Sasaki} R, et~al (2020) {Optical and X-ray
  observations of stellar flares on an active M dwarf AD Leonis with the Seimei
  Telescope, SCAT, NICER, and OISTER}. \pasj \doi{10.1093/pasj/psaa051},
  {\href{https://arxiv.org/abs/2005.04336}{{arXiv:2005.04336}}} {[astro-ph.SR]}

\bibitem[{{Namekata} et~al(2022{\natexlab{a}}){Namekata}, {Ichimoto}, {Ishii},
  and {Shibata}}]{Namekata2022}
{Namekata} K, {Ichimoto} K, {Ishii} TT, et~al (2022{\natexlab{a}})
  {Sun-as-a-star Analysis of H{\ensuremath{\alpha}} Spectra of a Solar Flare
  Observed by SMART/SDDI: Time Evolution of Red Asymmetry and Line Broadening}.
  \apj 933(2):209. \doi{10.3847/1538-4357/ac75cd},
  {\href{https://arxiv.org/abs/2206.01395}{{arXiv:2206.01395}}} {[astro-ph.SR]}

\bibitem[{{Namekata} et~al(2022{\natexlab{b}}){Namekata}, {Maehara}, {Honda},
  {Notsu}, {Okamoto}, {Takahashi}, {Takayama}, {Ohshima}, {Saito}, {Katoh},
  {Tozuka}, {Murata}, {Ogawa}, {Niwano}, {Adachi}, {Oeda}, {Shiraishi},
  {Isogai}, {Nogami}, and {Shibata}}]{Namekata2022B}
{Namekata} K, {Maehara} H, {Honda} S, et~al (2022{\natexlab{b}}) {Discovery of
  a Long-duration Superflare on a Young Solar-type Star EK Draconis with Nearly
  Similar Time Evolution for H{\ensuremath{\alpha}} and White-light Emissions}.
  \apjl 926(1):L5. \doi{10.3847/2041-8213/ac4df0},
  {\href{https://arxiv.org/abs/2201.09416}{{arXiv:2201.09416}}} {[astro-ph.SR]}

\bibitem[{{Namekata} et~al(2022{\natexlab{c}}){Namekata}, {Maehara}, {Honda},
  {Notsu}, {Okamoto}, {Takahashi}, {Takayama}, {Ohshima}, {Saito}, {Katoh},
  {Tozuka}, {Murata}, {Ogawa}, {Niwano}, {Adachi}, {Oeda}, {Shiraishi},
  {Isogai}, {Seki}, {Ishii}, {Ichimoto}, {Nogami}, and
  {Shibata}}]{Namekata2021}
{Namekata} K, {Maehara} H, {Honda} S, et~al (2022{\natexlab{c}}) {Probable
  detection of an eruptive filament from a superflare on a solar-type star}.
  Nature Astronomy 6:241--248. \doi{10.1038/s41550-021-01532-8},
  {\href{https://arxiv.org/abs/2112.04808}{{arXiv:2112.04808}}} {[astro-ph.SR]}

\bibitem[{{Namizaki} et~al(2023){Namizaki}, {Namekata}, {Maehara}, {Notsu},
  {Honda}, {Nogami}, and {Shibata}}]{Namizaki2023}
{Namizaki} K, {Namekata} K, {Maehara} H, et~al (2023) {A Superflare on YZ Canis
  Minoris Observed by the Seimei Telescope and TESS: Red Asymmetry of
  H{\ensuremath{\alpha}} Emission Associated with White-light Emission}. \apj
  945(1):61. \doi{10.3847/1538-4357/acb928},
  {\href{https://arxiv.org/abs/2302.03007}{{arXiv:2302.03007}}} {[astro-ph.SR]}

\bibitem[{{Nayfonov} et~al(1999){Nayfonov}, {D{\"a}ppen}, {Hummer}, and
  {Mihalas}}]{Nayfonov1999}
{Nayfonov} A, {D{\"a}ppen} W, {Hummer} DG, et~al (1999) {The MHD Equation of
  State with Post-Holtsmark Microfield Distributions}. \apj 526(1):451--464.
  \doi{10.1086/307972},
  {\href{https://arxiv.org/abs/astro-ph/9901360}{{arXiv:astro-ph/9901360}}}
  {[astro-ph]}

\bibitem[{{Neidig}(1983)}]{Neidig1983}
{Neidig} DF (1983) {Spectral analysis of the optical continuum in the 24 April
  1981 flare}. \solphys 85(2):285--302. \doi{10.1007/BF00148655}

\bibitem[{{Neidig}(1989)}]{Neidig1989}
{Neidig} DF (1989) {The Importance of Solar White-Light Flares}. \solphys
  121(1-2):261--269. \doi{10.1007/BF00161699}

\bibitem[{{Neidig} and {Kane}(1993)}]{NeidigKane1993}
{Neidig} DF, {Kane} SR (1993) {Energetics and Timing of the Hard and Soft X-Ray
  Emissions in White-Light Flares}. \solphys 143(1):201--204.
  \doi{10.1007/BF00619106}

\bibitem[{{Neidig} and {Wiborg}(1984)}]{Neidig1984}
{Neidig} DF, {Wiborg} JP.~H. (1984) {The Hydrogen Emission Spectrum in Three
  White Light Flares}. \solphys 92(1-2):217--225. \doi{10.1007/BF00157247}

\bibitem[{{Neidig} et~al(1993{\natexlab{a}}){Neidig}, {Kiplinger}, {Cohl}, and
  {Wiborg}}]{Neidig1993}
{Neidig} DF, {Kiplinger} AL, {Cohl} HS, et~al (1993{\natexlab{a}}) {The Solar
  White-Light Flare of 1989 March 7: Simultaneous Multiwavelength Observations
  at High Time Resolution}. \apj 406:306. \doi{10.1086/172442}

\bibitem[{{Neidig} et~al(1993{\natexlab{b}}){Neidig}, {Wiborg}, and
  {Gilliam}}]{Neidig1993Limb}
{Neidig} DF, {Wiborg} PH, {Gilliam} LB (1993{\natexlab{b}}) {Physical
  properties of white-light flares derived from their center-to-limb
  distribution}. \solphys 144(1):169--194. \doi{10.1007/BF00667990}

\bibitem[{{Neidig} et~al(1994){Neidig}, {Grosser}, and {Hrovat}}]{Neidig1994}
{Neidig} DF, {Grosser} H, {Hrovat} M (1994) {Optical output of the 24 April
  1984 white-light flare}. \solphys 155(1):199--202. \doi{10.1007/BF00670740}

\bibitem[{{Neupert}(1968)}]{Neupert1968}
{Neupert} WM (1968) {Comparison of Solar X-Ray Line Emission with Microwave
  Emission during Flares}. \apjl 153:L59. \doi{10.1086/180220}

\bibitem[{{Newton} et~al(2017){Newton}, {Irwin}, {Charbonneau}, {Berlind},
  {Calkins}, and {Mink}}]{Newton2017}
{Newton} ER, {Irwin} J, {Charbonneau} D, et~al (2017) {The
  H{\ensuremath{\alpha}} Emission of Nearby M Dwarfs and its Relation to
  Stellar Rotation}. \apj 834(1):85. \doi{10.3847/1538-4357/834/1/85},
  {\href{https://arxiv.org/abs/1611.03509}{{arXiv:1611.03509}}} {[astro-ph.SR]}

\bibitem[{{Nishikawa} et~al(2021){Nishikawa}, {Du{\c{t}}an}, {K{\"o}hn}, and
  {Mizuno}}]{Nishikawa2021}
{Nishikawa} K, {Du{\c{t}}an} I, {K{\"o}hn} C, et~al (2021) {PIC methods in
  astrophysics: simulations of relativistic jets and kinetic physics in
  astrophysical systems}. Living Reviews in Computational Astrophysics 7(1):1.
  \doi{10.1007/s41115-021-00012-0},
  {\href{https://arxiv.org/abs/2008.02105}{{arXiv:2008.02105}}} {[astro-ph.HE]}

\bibitem[{{Nizamov}(2019)}]{Nizamov2019}
{Nizamov} BA (2019) {Soft X-ray heating as a mechanism of optical continuum
  generation in solar-type star superflares}. \mnras 489(3):4338--4345.
  \doi{10.1093/mnras/stz2478},
  {\href{https://arxiv.org/abs/1905.05054}{{arXiv:1905.05054}}} {[astro-ph.SR]}

\bibitem[{{Nogami} et~al(2014){Nogami}, {Notsu}, {Honda}, {Maehara}, {Notsu},
  {Shibayama}, and {Shibata}}]{Nogami2014}
{Nogami} D, {Notsu} Y, {Honda} S, et~al (2014) {Two sun-like superflare stars
  rotating as slow as the Sun*}. \pasj 66(2):L4. \doi{10.1093/pasj/psu012},
  {\href{https://arxiv.org/abs/1402.3772}{{arXiv:1402.3772}}} {[astro-ph.SR]}

\bibitem[{{Nordon} and {Behar}(2007)}]{Nordon2007}
{Nordon} R, {Behar} E (2007) {Six large coronal X-ray flares observed with
  Chandra}. \aap 464(1):309--321. \doi{10.1051/0004-6361:20066449},
  {\href{https://arxiv.org/abs/astro-ph/0611386}{{arXiv:astro-ph/0611386}}}
  {[astro-ph]}

\bibitem[{{Nordon} and {Behar}(2008)}]{Nordon2008}
{Nordon} R, {Behar} E (2008) {Abundance variations and first ionization
  potential trends during large stellar flares}. \aap 482(2):639--651.
  \doi{10.1051/0004-6361:20078848},
  {\href{https://arxiv.org/abs/0712.0482}{{arXiv:0712.0482}}} {[astro-ph]}

\bibitem[{{Notsu} et~al(2013{\natexlab{a}}){Notsu}, {Honda}, {Notsu}, {Nagao},
  {Shibayama}, {Maehara}, {Nogami}, and {Shibata}}]{Notsu2013}
{Notsu} S, {Honda} S, {Notsu} Y, et~al (2013{\natexlab{a}}) {High-Dispersion
  Spectroscopy of the Superflare Star KIC 6934317}. \pasj 65:112.
  \doi{10.1093/pasj/65.5.112}

\bibitem[{{Notsu} et~al(2013{\natexlab{b}}){Notsu}, {Shibayama}, {Maehara},
  {Notsu}, {Nagao}, {Honda}, {Ishii}, {Nogami}, and {Shibata}}]{Notsu2013B}
{Notsu} Y, {Shibayama} T, {Maehara} H, et~al (2013{\natexlab{b}}) {Superflares
  on Solar-type Stars Observed with Kepler II. Photometric Variability of
  Superflare-generating Stars: A Signature of Stellar Rotation and Starspots}.
  \apj 771(2):127. \doi{10.1088/0004-637X/771/2/127},
  {\href{https://arxiv.org/abs/1304.7361}{{arXiv:1304.7361}}} {[astro-ph.SR]}

\bibitem[{{Notsu} et~al(2015{\natexlab{a}}){Notsu}, {Honda}, {Maehara},
  {Notsu}, {Shibayama}, {Nogami}, and {Shibata}}]{Notsu2015A}
{Notsu} Y, {Honda} S, {Maehara} H, et~al (2015{\natexlab{a}}) {High dispersion
  spectroscopy of solar-type superflare stars. I. Temperature, surface gravity,
  metallicity, and vsin i}. \pasj 67(3):32. \doi{10.1093/pasj/psv001}

\bibitem[{{Notsu} et~al(2015{\natexlab{b}}){Notsu}, {Honda}, {Maehara},
  {Notsu}, {Shibayama}, {Nogami}, and {Shibata}}]{Notsu2015B}
{Notsu} Y, {Honda} S, {Maehara} H, et~al (2015{\natexlab{b}}) {High dispersion
  spectroscopy of solar-type superflare stars. II. Stellar rotation, starspots,
  and chromospheric activities}. \pasj 67(3):33. \doi{10.1093/pasj/psv002}

\bibitem[{{Notsu} et~al(2019){Notsu}, {Maehara}, {Honda}, {Hawley},
  {Davenport}, {Namekata}, {Notsu}, {Ikuta}, {Nogami}, and
  {Shibata}}]{Notsu2019}
{Notsu} Y, {Maehara} H, {Honda} S, et~al (2019) {Do Kepler Superflare Stars
  Really Include Slowly Rotating Sun-like Stars?{\textemdash}Results Using APO
  3.5 m Telescope Spectroscopic Observations and Gaia-DR2 Data}. \apj
  876(1):58. \doi{10.3847/1538-4357/ab14e6},
  {\href{https://arxiv.org/abs/1904.00142}{{arXiv:1904.00142}}} {[astro-ph.SR]}

\bibitem[{{Notsu} et~al(2023){Notsu}, {Kowalski}, {Maehara}, {Namekata},
  {Hamaguchi}, {Enoto}, {Tristan}, {Hawley}, {Davenport}, {Honda}, {Ikuta},
  {Inoue}, {Namizaki}, {Nogami}, and {Shibata}}]{Notsu2023}
{Notsu} Y, {Kowalski} AF, {Maehara} H, et~al (2023) {APO \& SMARTS flare star
  campaign observations I. Blue wing asymmetries in chromospheric lines during
  mid M dwarf flares from simultaneous spectroscopic and photometric
  observation data}. arXiv e-prints arXiv:2310.02450.
  \doi{10.48550/arXiv.2310.02450},
  {\href{https://arxiv.org/abs/2310.02450}{{arXiv:2310.02450}}} {[astro-ph.SR]}

\bibitem[{{Oka} et~al(2010){Oka}, {Phan}, {Krucker}, {Fujimoto}, and
  {Shinohara}}]{Oka2010}
{Oka} M, {Phan} TD, {Krucker} S, et~al (2010) {Electron Acceleration by
  Multi-Island Coalescence}. \apj 714(1):915--926.
  \doi{10.1088/0004-637X/714/1/915},
  {\href{https://arxiv.org/abs/1004.1154}{{arXiv:1004.1154}}} {[astro-ph.SR]}

\bibitem[{{Oka} et~al(2013){Oka}, {Ishikawa}, {Saint-Hilaire}, {Krucker}, and
  {Lin}}]{Oka2013}
{Oka} M, {Ishikawa} S, {Saint-Hilaire} P, et~al (2013) {Kappa Distribution
  Model for Hard X-Ray Coronal Sources of Solar Flares}. \apj 764(1):6.
  \doi{10.1088/0004-637X/764/1/6},
  {\href{https://arxiv.org/abs/1212.2579}{{arXiv:1212.2579}}} {[astro-ph.SR]}

\bibitem[{{Okamoto} et~al(2021){Okamoto}, {Notsu}, {Maehara}, {Namekata},
  {Honda}, {Ikuta}, {Nogami}, and {Shibata}}]{Okamoto2021}
{Okamoto} S, {Notsu} Y, {Maehara} H, et~al (2021) {Statistical Properties of
  Superflares on Solar-type Stars: Results Using All of the Kepler Primary
  Mission Data}. \apj 906(2):72. \doi{10.3847/1538-4357/abc8f5},
  {\href{https://arxiv.org/abs/2011.02117}{{arXiv:2011.02117}}} {[astro-ph.SR]}

\bibitem[{{Oks} and {Gershberg}(2016)}]{Oks2016}
{Oks} E, {Gershberg} RE (2016) {Flare Stars{\textemdash}A Favorable Object for
  Studying Mechanisms of Nonthermal Astrophysical Phenomena}. \apj 819(1):16.
  \doi{10.3847/0004-637X/819/1/16}

\bibitem[{{Ol{\'a}h} et~al(2022){Ol{\'a}h}, {Seli}, {K{\H{o}}v{\'a}ri},
  {Kriskovics}, and {Vida}}]{Olah2022}
{Ol{\'a}h} K, {Seli} B, {K{\H{o}}v{\'a}ri} Z, et~al (2022) {Characteristics of
  flares on giant stars}. arXiv e-prints arXiv:2210.09710.
  {\href{https://arxiv.org/abs/2210.09710}{{arXiv:2210.09710}}} {[astro-ph.SR]}

\bibitem[{{Orrall} and {Zirker}(1976)}]{OZ}
{Orrall} FQ, {Zirker} JB (1976) {Lyman-alpha emission from nonthermal proton
  beams.} \apj 208:618--632. \doi{10.1086/154642}

\bibitem[{{Osten} and {Bastian}(2006)}]{Osten2006}
{Osten} RA, {Bastian} TS (2006) {Wide-Band Spectroscopy of Two Radio Bursts on
  AD Leonis}. \apj 637(2):1016--1024. \doi{10.1086/498410},
  {\href{https://arxiv.org/abs/astro-ph/0509815}{{arXiv:astro-ph/0509815}}}
  {[astro-ph]}

\bibitem[{{Osten} and {Bastian}(2008)}]{Osten2008}
{Osten} RA, {Bastian} TS (2008) {Ultrahigh Time Resolution Observations of
  Radio Bursts on AD Leonis}. \apj 674(2):1078--1085. \doi{10.1086/525013},
  {\href{https://arxiv.org/abs/0710.5881}{{arXiv:0710.5881}}} {[astro-ph]}

\bibitem[{{Osten} and {Brown}(1999)}]{Osten1999}
{Osten} RA, {Brown} A (1999) {Extreme Ultraviolet Explorer Photometry of RS
  Canum Venaticorum Systems: Four Flaring Megaseconds}. \apj 515(2):746--761.
  \doi{10.1086/307034}

\bibitem[{{Osten} and {Wolk}(2015)}]{Osten2015}
{Osten} RA, {Wolk} SJ (2015) {Connecting Flares and Transient Mass-loss Events
  in Magnetically Active Stars}. \apj 809:79. \doi{10.1088/0004-637X/809/1/79},
  {\href{https://arxiv.org/abs/1506.04994}{{arXiv:1506.04994}}} {[astro-ph.SR]}

\bibitem[{{Osten} et~al(2000){Osten}, {Brown}, {Ayres}, {Linsky}, {Drake},
  {Gagn{\'e}}, and {Stern}}]{Osten2000}
{Osten} RA, {Brown} A, {Ayres} TR, et~al (2000) {Radio, X-Ray, and
  Extreme-Ultraviolet Coronal Variability of the Short-Period RS Canum
  Venaticorum Binary {\ensuremath{\sigma}}$^{2}$ Coronae Borealis}. \apj
  544(2):953--976. \doi{10.1086/317249}

\bibitem[{{Osten} et~al(2002){Osten}, {Brown}, {Wood}, and {Brady}}]{Osten2002}
{Osten} RA, {Brown} A, {Wood} BE, et~al (2002) {Multiwavelength Observations of
  Three Short-Period Active Binary Systems: ER Vulpeculae, CC Eridani, and EI
  Eridani}. \apjs 138(1):99--120. \doi{10.1086/323666}

\bibitem[{{Osten} et~al(2003){Osten}, {Ayres}, {Brown}, {Linsky}, and
  {Krishnamurthi}}]{Osten2003}
{Osten} RA, {Ayres} TR, {Brown} A, et~al (2003) {Chandra, Extreme Ultraviolet
  Explorer, and Very Large Array Observations of the Active Binary System
  {\ensuremath{\sigma}}$^{2}$ Coronae Borealis}. \apj 582(2):1073--1101.
  \doi{10.1086/344797}

\bibitem[{{Osten} et~al(2004){Osten}, {Brown}, {Ayres}, {Drake}, {Franciosini},
  {Pallavicini}, {Tagliaferri}, {Stewart}, {Skinner}, and {Linsky}}]{Osten2004}
{Osten} RA, {Brown} A, {Ayres} TR, et~al (2004) {A Multiwavelength Perspective
  of Flares on HR 1099: 4 Years of Coordinated Campaigns}. \apjs
  153(1):317--362. \doi{10.1086/420770},
  {\href{https://arxiv.org/abs/astro-ph/0402613}{{arXiv:astro-ph/0402613}}}
  {[astro-ph]}

\bibitem[{{Osten} et~al(2005){Osten}, {Hawley}, {Allred}, {Johns-Krull}, and
  {Roark}}]{Osten2005}
{Osten} RA, {Hawley} SL, {Allred} JC, et~al (2005) {From Radio to X-Ray: Flares
  on the dMe Flare Star EV Lacertae}. \apj 621:398--416. \doi{10.1086/427275},
  {\href{https://arxiv.org/abs/astro-ph/0411236}{{astro-ph/0411236}}}

\bibitem[{{Osten} et~al(2007){Osten}, {Drake}, {Tueller}, {Cummings}, {Perri},
  {Moretti}, and {Covino}}]{Osten2007}
{Osten} RA, {Drake} S, {Tueller} J, et~al (2007) {Nonthermal Hard X-Ray
  Emission and Iron K{\ensuremath{\alpha}} Emission from a Superflare on II
  Pegasi}. \apj 654(2):1052--1067. \doi{10.1086/509252},
  {\href{https://arxiv.org/abs/astro-ph/0609205}{{arXiv:astro-ph/0609205}}}
  {[astro-ph]}

\bibitem[{{Osten} et~al(2010){Osten}, {Godet}, {Drake}, {Tueller}, {Cummings},
  {Krimm}, {Pye}, {Pal'shin}, {Golenetskii}, {Reale}, {Oates}, {Page}, and
  {Melandri}}]{Osten2010}
{Osten} RA, {Godet} O, {Drake} S, et~al (2010) {The Mouse That Roared: A
  Superflare from the dMe Flare Star EV Lac Detected by Swift and Konus-Wind}.
  \apj 721(1):785--801. \doi{10.1088/0004-637X/721/1/785},
  {\href{https://arxiv.org/abs/1007.5300}{{arXiv:1007.5300}}} {[astro-ph.SR]}

\bibitem[{{Osten} et~al(2012){Osten}, {Kowalski}, {Sahu}, and
  {Hawley}}]{Osten2012}
{Osten} RA, {Kowalski} A, {Sahu} K, et~al (2012) {DRAFTS: A Deep, Rapid
  Archival Flare Transient Search in the Galactic Bulge}. \apj 754(1):4.
  \doi{10.1088/0004-637X/754/1/4},
  {\href{https://arxiv.org/abs/1205.1485}{{arXiv:1205.1485}}} {[astro-ph.SR]}

\bibitem[{{Osten} et~al(2016){Osten}, {Kowalski}, {Drake}, {Krimm}, {Page},
  {Gazeas}, {Kennea}, {Oates}, {Page}, {de Miguel}, {Nov{\'a}k}, {Apeltauer},
  and {Gehrels}}]{Osten2016}
{Osten} RA, {Kowalski} A, {Drake} SA, et~al (2016) {A Very Bright, Very Hot,
  and Very Long Flaring Event from the M Dwarf Binary System DG CVn}. \apj
  832:174. \doi{10.3847/0004-637X/832/2/174},
  {\href{https://arxiv.org/abs/1609.04674}{{arXiv:1609.04674}}} {[astro-ph.SR]}

\bibitem[{{Pagani} et~al(2011){Pagani}, {Beardmore}, {Abbey}, {Mountford},
  {Osborne}, {Capalbi}, {Perri}, {Angelini}, {Burrows}, {Campana}, {Cusumano},
  {Evans}, {Kennea}, {Moretti}, {Page}, and {Starling}}]{Pagani2011}
{Pagani} C, {Beardmore} AP, {Abbey} AF, et~al (2011) {Recovering Swift-XRT
  energy resolution through CCD charge trap mapping}. \aap 534:A20.
  \doi{10.1051/0004-6361/201117660},
  {\href{https://arxiv.org/abs/1108.5049}{{arXiv:1108.5049}}} {[astro-ph.IM]}

\bibitem[{{Pandey} and {Singh}(2008)}]{Pandey2008}
{Pandey} JC, {Singh} KP (2008) {A study of X-ray flares - I. Active late-type
  dwarfs}. \mnras 387(4):1627--1648. \doi{10.1111/j.1365-2966.2008.13342.x},
  {\href{https://arxiv.org/abs/0805.3882}{{arXiv:0805.3882}}} {[astro-ph]}

\bibitem[{{Pandey} and {Singh}(2012)}]{Pandey2012}
{Pandey} JC, {Singh} KP (2012) {A study of X-ray flares - II. RS CVn-type
  binaries}. \mnras 419(2):1219--1237. \doi{10.1111/j.1365-2966.2011.19776.x},
  {\href{https://arxiv.org/abs/1110.2008}{{arXiv:1110.2008}}} {[astro-ph.SR]}

\bibitem[{{Paudel} et~al(2018){Paudel}, {Gizis}, {Mullan}, {Schmidt},
  {Burgasser}, {Williams}, and {Berger}}]{Paudel2018}
{Paudel} RR, {Gizis} JE, {Mullan} DJ, et~al (2018) {K2 Ultracool Dwarfs Survey.
  III. White Light Flares Are Ubiquitous in M6-L0 Dwarfs}. \apj 858(1):55.
  \doi{10.3847/1538-4357/aab8fe},
  {\href{https://arxiv.org/abs/1803.07708}{{arXiv:1803.07708}}} {[astro-ph.SR]}

\bibitem[{{Paudel} et~al(2019){Paudel}, {Gizis}, {Mullan}, {Schmidt},
  {Burgasser}, {Williams}, {Youngblood}, and {Stassun}}]{Paudel2019}
{Paudel} RR, {Gizis} JE, {Mullan} DJ, et~al (2019) {K2 Ultracool Dwarfs Survey
  - V. High superflare rates on rapidly rotating late-M dwarfs}. \mnras
  486(1):1438--1447. \doi{10.1093/mnras/stz886},
  {\href{https://arxiv.org/abs/1812.07631}{{arXiv:1812.07631}}} {[astro-ph.SR]}

\bibitem[{{Paudel} et~al(2020){Paudel}, {Gizis}, {Mullan}, {Schmidt},
  {Burgasser}, and {Williams}}]{Paudel2020}
{Paudel} RR, {Gizis} JE, {Mullan} DJ, et~al (2020) {K2 Ultracool Dwarfs Survey
  - VI. White light superflares observed on an L5 dwarf and flare rates of L
  dwarfs}. \mnras 494(4):5751--5760. \doi{10.1093/mnras/staa1137},
  {\href{https://arxiv.org/abs/2004.10579}{{arXiv:2004.10579}}} {[astro-ph.SR]}

\bibitem[{{Paudel} et~al(2021){Paudel}, {Barclay}, {Schlieder}, {Quintana},
  {Gilbert}, {Vega}, {Youngblood}, {Silverstein}, {Osten}, {Tucker}, {Huber},
  {Do}, {Hamaguchi}, {Mullan}, {Gizis}, {Monsue}, {Col{\'o}n}, {Boyd},
  {Davenport}, and {Walkowicz}}]{Paudel2021}
{Paudel} RR, {Barclay} T, {Schlieder} JE, et~al (2021) {Simultaneous
  Multiwavelength Flare Observations of EV Lacertae}. \apj 922(1):31.
  \doi{10.3847/1538-4357/ac1946},
  {\href{https://arxiv.org/abs/2108.04753}{{arXiv:2108.04753}}} {[astro-ph.SR]}

\bibitem[{{Paulson} et~al(2006){Paulson}, {Allred}, {Anderson}, {Hawley},
  {Cochran}, and {Yelda}}]{Paulson2006}
{Paulson} DB, {Allred} JC, {Anderson} RB, et~al (2006) {Optical Spectroscopy of
  a Flare on Barnard's Star}. \pasp 118:227--235. \doi{10.1086/499497},
  {\href{https://arxiv.org/abs/astro-ph/0511281}{{astro-ph/0511281}}}

\bibitem[{{Paxton} et~al(2011){Paxton}, {Bildsten}, {Dotter}, {Herwig},
  {Lesaffre}, and {Timmes}}]{Paxton2011}
{Paxton} B, {Bildsten} L, {Dotter} A, et~al (2011) {Modules for Experiments in
  Stellar Astrophysics (MESA)}. \apjs 192(1):3.
  \doi{10.1088/0067-0049/192/1/3},
  {\href{https://arxiv.org/abs/1009.1622}{{arXiv:1009.1622}}} {[astro-ph.SR]}

\bibitem[{{Pedersen} et~al(2017){Pedersen}, {Antoci}, {Korhonen}, {White},
  {Jessen-Hansen}, {Lehtinen}, {Nikbakhsh}, and {Viuho}}]{Pedersen2017}
{Pedersen} MG, {Antoci} V, {Korhonen} H, et~al (2017) {Do A-type stars flare?}
  \mnras 466(3):3060--3076. \doi{10.1093/mnras/stw3226},
  {\href{https://arxiv.org/abs/1612.04575}{{arXiv:1612.04575}}} {[astro-ph.SR]}

\bibitem[{{Peres} et~al(1982){Peres}, {Serio}, {Vaiana}, and {Rosner}}]{PM1}
{Peres} G, {Serio} S, {Vaiana} GS, et~al (1982) {Coronal closed structures. IV
  - Hydrodynamical stability and response to heating perturbations}. \apj
  252:791--799. \doi{10.1086/159601}

\bibitem[{{Peterson} et~al(2010){Peterson}, {Mutel}, {G{\"u}del}, and
  {Goss}}]{Petersen2010}
{Peterson} WM, {Mutel} RL, {G{\"u}del} M, et~al (2010) {A large coronal loop in
  the Algol system}. \nat 463(7278):207--209. \doi{10.1038/nature08643}

\bibitem[{{Petit} et~al(2014){Petit}, {Louge}, {Th{\'e}ado}, {Paletou},
  {Manset}, {Morin}, {Marsden}, and {Jeffers}}]{PolarBase2}
{Petit} P, {Louge} T, {Th{\'e}ado} S, et~al (2014) {PolarBase: A Database of
  High-Resolution Spectropolarimetric Stellar Observations}. \pasp
  126(939):469. \doi{10.1086/676976},
  {\href{https://arxiv.org/abs/1401.1082}{{arXiv:1401.1082}}} {[astro-ph.SR]}

\bibitem[{{Petrosian}(2012)}]{Petrosian2012}
{Petrosian} V (2012) {Stochastic Acceleration by Turbulence}. \ssr
  173(1-4):535--556. \doi{10.1007/s11214-012-9900-6},
  {\href{https://arxiv.org/abs/1205.2136}{{arXiv:1205.2136}}} {[astro-ph.HE]}

\bibitem[{{Petrosian} and {Liu}(2004)}]{Petrosian2004}
{Petrosian} V, {Liu} S (2004) {Stochastic Acceleration of Electrons and
  Protons. I. Acceleration by Parallel-Propagating Waves}. \apj 610:550--571.
  \doi{10.1086/421486},
  {\href{https://arxiv.org/abs/astro-ph/0401585}{{astro-ph/0401585}}}

\bibitem[{{Pettersen}(1983)}]{V780Tau}
{Pettersen} BR (1983) {The flare activity of V780 Tau.} \aap 120:192--196

\bibitem[{{Pettersen}(1989)}]{Pettersen1989}
{Pettersen} BR (1989) {A Review of Stellar Flares and Their Characteristics}.
  \solphys 121(1-2):299--312. \doi{10.1007/BF00161702}

\bibitem[{{Pettersen}(2016)}]{Pettersen2016}
{Pettersen} BR (2016) {An Optical Megaflare On EV Lac}. In: 19th Cambridge
  Workshop on Cool Stars, Stellar Systems, and the Sun (CS19), Cambridge
  Workshop on Cool Stars, Stellar Systems, and the Sun, p 117,
  \doi{10.5281/zenodo.59128}

\bibitem[{{Pettersen} et~al(1984){Pettersen}, {Coleman}, and
  {Evans}}]{Pettersen1984}
{Pettersen} BR, {Coleman} LA, {Evans} DS (1984) {The flare activity of AD
  Leonis.} \apjs 54:375--386. \doi{10.1086/190934}

\bibitem[{{Phillips} et~al(1992){Phillips}, {Bromage}, and
  {Doyle}}]{Phillips1992}
{Phillips} KJH, {Bromage} GE, {Doyle} JG (1992) {The Origin of the
  Far-Ultraviolet Continuum in Solar and Stellar Flares}. \apj 385:731.
  \doi{10.1086/170979}

\bibitem[{{Phillips} et~al(1996){Phillips}, {Lonsdale}, {Feigelson}, and
  {Deeney}}]{Phillips1996}
{Phillips} RB, {Lonsdale} CJ, {Feigelson} ED, et~al (1996) {Polarized Radio
  Emission From the Multiple T Tauri System HD 283447}. \aj 111:918.
  \doi{10.1086/117839}

\bibitem[{{Pillet} et~al(1984){Pillet}, {van Linden van den Heuvell}, {Smith},
  {Kachru}, {Tran}, and {Gallagher}}]{Pillet1984}
{Pillet} P, {van Linden van den Heuvell} HB, {Smith} WW, et~al (1984)
  {Microwave ionization of Na Rydberg atoms}. \pra 30:280--294.
  \doi{10.1103/PhysRevA.30.280}

\bibitem[{{Pizzolato} et~al(2003){Pizzolato}, {Maggio}, {Micela}, {Sciortino},
  and {Ventura}}]{Pizzolato2003}
{Pizzolato} N, {Maggio} A, {Micela} G, et~al (2003) {The stellar
  activity-rotation relationship revisited: Dependence of saturated and
  non-saturated X-ray emission regimes on stellar mass for late-type dwarfs}.
  \aap 397:147--157. \doi{10.1051/0004-6361:20021560}

\bibitem[{{Poletto} et~al(1988){Poletto}, {Pallavicini}, and
  {Kopp}}]{Poletto1988}
{Poletto} G, {Pallavicini} R, {Kopp} RA (1988) {Modeling of long-duration
  two-ribbon flares on M dwarf stars.} \aap 201:93--99

\bibitem[{{Pontin} and {Priest}(2022)}]{Pontin2022}
{Pontin} DI, {Priest} ER (2022) {Magnetic reconnection: MHD theory and
  modelling}. Living Reviews in Solar Physics 19(1):1.
  \doi{10.1007/s41116-022-00032-9}

\bibitem[{{Priest}(2014)}]{Priest2014}
{Priest} E (2014) {Magnetohydrodynamics of the Sun}.
  \doi{10.1017/CBO9781139020732}

\bibitem[{{Proch{\'a}zka} et~al(2018){Proch{\'a}zka}, {Reid}, {Milligan},
  {Sim{\~o}es}, {Allred}, and {Mathioudakis}}]{Ondrej2}
{Proch{\'a}zka} O, {Reid} A, {Milligan} RO, et~al (2018) {Reproducing Type II
  White-light Solar Flare Observations with Electron and Proton Beam
  Simulations}. \apj 862(1):76. \doi{10.3847/1538-4357/aaca37},
  {\href{https://arxiv.org/abs/1806.00249}{{arXiv:1806.00249}}} {[astro-ph.SR]}

\bibitem[{{Pye} et~al(2015){Pye}, {Rosen}, {Fyfe}, and
  {Schr{\"o}der}}]{Pye2015}
{Pye} JP, {Rosen} S, {Fyfe} D, et~al (2015) {A survey of stellar X-ray flares
  from the XMM-Newton serendipitous source catalogue: HIPPARCOS-Tycho cool
  stars}. \aap 581:A28. \doi{10.1051/0004-6361/201526217},
  {\href{https://arxiv.org/abs/1506.05289}{{arXiv:1506.05289}}} {[astro-ph.SR]}

\bibitem[{{Qiu} and {Longcope}(2016)}]{Qiu2016}
{Qiu} J, {Longcope} DW (2016) {Long Duration Flare Emission: Impulsive Heating
  or Gradual Heating?} \apj 820(1):14. \doi{10.3847/0004-637X/820/1/14},
  {\href{https://arxiv.org/abs/1604.05342}{{arXiv:1604.05342}}} {[astro-ph.SR]}

\bibitem[{{Qiu} et~al(2010){Qiu}, {Liu}, {Hill}, and {Kazachenko}}]{Qiu2010}
{Qiu} J, {Liu} W, {Hill} N, et~al (2010) {Reconnection and Energetics in
  Two-ribbon Flares: A Revisit of the Bastille-day Flare}. \apj 725:319--330.
  \doi{10.1088/0004-637X/725/1/319}

\bibitem[{{Qiu} et~al(2012){Qiu}, {Cheng}, {Hurford}, {Xu}, and
  {Wang}}]{Qiu2012}
{Qiu} J, {Cheng} JX, {Hurford} GJ, et~al (2012) {Solar flare hard X-ray spikes
  observed by RHESSI: a case study}. \aap 547:A72.
  \doi{10.1051/0004-6361/201118609},
  {\href{https://arxiv.org/abs/1210.7040}{{arXiv:1210.7040}}} {[astro-ph.SR]}

\bibitem[{{Qiu} et~al(2017){Qiu}, {Longcope}, {Cassak}, and {Priest}}]{Qiu2017}
{Qiu} J, {Longcope} DW, {Cassak} PA, et~al (2017) {Elongation of Flare
  Ribbons}. \apj 838(1):17. \doi{10.3847/1538-4357/aa6341},
  {\href{https://arxiv.org/abs/1707.02478}{{arXiv:1707.02478}}} {[astro-ph.SR]}

\bibitem[{{Ramsay} et~al(2013){Ramsay}, {Doyle}, {Hakala}, {Garcia-Alvarez},
  {Brooks}, {Barclay}, and {Still}}]{Ramsay2013}
{Ramsay} G, {Doyle} JG, {Hakala} P, et~al (2013) {Short-duration high-amplitude
  flares detected on the M dwarf star KIC 5474065}. \mnras 434(3):2451--2457.
  \doi{10.1093/mnras/stt1182},
  {\href{https://arxiv.org/abs/1306.5938}{{arXiv:1306.5938}}} {[astro-ph.SR]}

\bibitem[{{Rathore} and {Carlsson}(2015)}]{Rathore2015A}
{Rathore} B, {Carlsson} M (2015) {The Formation of IRIS Diagnostics. V. A
  Quintessential Model Atom of C II and General Formation Properties of the C
  II Lines at 133.5 nm}. \apj 811(2):80. \doi{10.1088/0004-637X/811/2/80},
  {\href{https://arxiv.org/abs/1508.04365}{{arXiv:1508.04365}}} {[astro-ph.SR]}

\bibitem[{{Rausch v. Traubenberg} et~al(1930){Rausch v. Traubenberg},
  {Gebauer}, and {Lewin}}]{Traubenberg1930}
{Rausch v. Traubenberg} H, {Gebauer} R, {Lewin} G (1930) {{\"U}ber die
  Existenzgrenzen von Anregungszust{\"a}nden des Wasserstoffatoms in starken
  elektrischen Feldern}. Naturwissenschaften 18(19):417--418.
  \doi{10.1007/BF01501125}

\bibitem[{{Reale} et~al(1997){Reale}, {Betta}, {Peres}, {Serio}, and
  {McTiernan}}]{Reale1997}
{Reale} F, {Betta} R, {Peres} G, et~al (1997) {Determination of the length of
  coronal loops from the decay of X-ray flares I. Solar flares observed with
  YOHKOH SXT.} \aap 325:782--790

\bibitem[{{Reale} et~al(2004){Reale}, {G{\"u}del}, {Peres}, and
  {Audard}}]{Reale2004}
{Reale} F, {G{\"u}del} M, {Peres} G, et~al (2004) {Modeling an X-ray flare on
  Proxima Centauri: Evidence of two flaring loop components and of two heating
  mechanisms at work}. \aap 416:733--747. \doi{10.1051/0004-6361:20034027},
  {\href{https://arxiv.org/abs/astro-ph/0312267}{{arXiv:astro-ph/0312267}}}
  {[astro-ph]}

\bibitem[{{Reames}(2021)}]{Reames2021}
{Reames} DV (2021) {Solar Energetic Particles. A Modern Primer on Understanding
  Sources, Acceleration and Propagation}, vol 978.
  \doi{10.1007/978-3-030-66402-2}

\bibitem[{{Redfield} et~al(2002){Redfield}, {Linsky}, {Ake}, {Ayres}, {Dupree},
  {Robinson}, {Wood}, and {Young}}]{Redfield2002}
{Redfield} S, {Linsky} JL, {Ake} TB, et~al (2002) {A Far Ultraviolet
  Spectroscopic Explorer Survey of Late-Type Dwarf Stars}. \apj
  581(1):626--653. \doi{10.1086/344153}

\bibitem[{{Reep} and {Toriumi}(2017)}]{Reep2017}
{Reep} JW, {Toriumi} S (2017) {The Direct Relation between the Duration of
  Magnetic Reconnection and the Evolution of GOES Light Curves in Solar
  Flares}. \apj 851(1):4. \doi{10.3847/1538-4357/aa96fe},
  {\href{https://arxiv.org/abs/1711.00422}{{arXiv:1711.00422}}} {[astro-ph.SR]}

\bibitem[{{Reep} et~al(2022){Reep}, {Ugarte-Urra}, {Warren}, and
  {Barnes}}]{Reep2022}
{Reep} JW, {Ugarte-Urra} I, {Warren} HP, et~al (2022) {Geometric Assumptions in
  Hydrodynamic Modeling of Coronal and Flaring Loops}. \apj 933(1):106.
  \doi{10.3847/1538-4357/ac7398}

\bibitem[{{Reeves}(2022)}]{Reeves2022}
{Reeves} KK (2022) {A window into magnetic reconnection: IRIS observations of
  the consequences of reconnection during solar flares}. Frontiers in Astronomy
  and Space Sciences 9:392. \doi{10.3389/fspas.2022.1041951}

\bibitem[{{Reeves} and {Forbes}(2005)}]{Reeves2005}
{Reeves} KK, {Forbes} TG (2005) {Predicted Light Curves for a Model of Solar
  Eruptions}. \apj 630(2):1133--1147. \doi{10.1086/432047}

\bibitem[{{Reeves} and {Warren}(2002)}]{Reeves2002}
{Reeves} KK, {Warren} HP (2002) {Modeling the Cooling of Postflare Loops}. \apj
  578(1):590--597. \doi{10.1086/342310}

\bibitem[{{Reeves} et~al(2007){Reeves}, {Warren}, and {Forbes}}]{Reeves2007}
{Reeves} KK, {Warren} HP, {Forbes} TG (2007) {Theoretical Predictions of X-Ray
  and Extreme-UV Flare Emissions Using a Loss-of-Equilibrium Model of Solar
  Eruptions}. \apj 668(2):1210--1220. \doi{10.1086/521291}

\bibitem[{{Reid} and {Hawley}(2005)}]{NLDS}
{Reid} IN, {Hawley} SL (2005) {New light on dark stars : red dwarfs, low-mass
  stars, brown dwarfs}. \doi{10.1007/3-540-27610-6}

\bibitem[{{Reid} et~al(1995){Reid}, {Hawley}, and {Gizis}}]{PMSU1}
{Reid} IN, {Hawley} SL, {Gizis} JE (1995) {The Palomar/MSU Nearby-Star
  Spectroscopic Survey. I. The Northern M Dwarfs -Bandstrengths and
  Kinematics}. \aj 110:1838. \doi{10.1086/117655}

\bibitem[{{Reid} et~al(2002){Reid}, {Gizis}, and {Hawley}}]{PMSU4}
{Reid} IN, {Gizis} JE, {Hawley} SL (2002) {The Palomar/MSU Nearby Star
  Spectroscopic Survey. IV. The Luminosity Function in the Solar Neighborhood
  and M Dwarf Kinematics}. \aj 124(5):2721--2738. \doi{10.1086/343777}

\bibitem[{{Reiners} and {Basri}(2008)}]{Reiners2008}
{Reiners} A, {Basri} G (2008) {The moderate magnetic field of the flare star
  Proxima Centauri}. \aap 489(3):L45--L48. \doi{10.1051/0004-6361:200810491}

\bibitem[{{Reiners} et~al(2014){Reiners}, {Sch{\"u}ssler}, and
  {Passegger}}]{Reiners2014}
{Reiners} A, {Sch{\"u}ssler} M, {Passegger} VM (2014) {Generalized
  Investigation of the Rotation-Activity Relation: Favoring Rotation Period
  instead of Rossby Number}. \apj 794(2):144.
  \doi{10.1088/0004-637X/794/2/144},
  {\href{https://arxiv.org/abs/1408.6175}{{arXiv:1408.6175}}} {[astro-ph.SR]}

\bibitem[{{Reiners} et~al(2022){Reiners}, {Shulyak}, {K{\"a}pyl{\"a}}, {Ribas},
  {Nagel}, {Zechmeister}, {Caballero}, {Shan}, {Fuhrmeister}, {Quirrenbach},
  {Amado}, {Montes}, {Jeffers}, {Azzaro}, {B{\'e}jar}, {Chaturvedi}, {Henning},
  {K{\"u}rster}, and {Pall{\'e}}}]{Reiners2022}
{Reiners} A, {Shulyak} D, {K{\"a}pyl{\"a}} PJ, et~al (2022) {Magnetism,
  rotation, and nonthermal emission in cool stars. Average magnetic field
  measurements in 292 M dwarfs}. \aap 662:A41.
  \doi{10.1051/0004-6361/202243251},
  {\href{https://arxiv.org/abs/2204.00342}{{arXiv:2204.00342}}} {[astro-ph.SR]}

\bibitem[{{Rempel} et~al(2023){Rempel}, {Chintzoglou}, {Cheung}, {Fan}, and
  {Kleint}}]{Rempel2023}
{Rempel} M, {Chintzoglou} G, {Cheung} MCM, et~al (2023) {Comprehensive
  radiative MHD simulations of flares above collisional polarity inversion
  lines}. arXiv e-prints arXiv:2303.05299. \doi{10.48550/arXiv.2303.05299},
  {\href{https://arxiv.org/abs/2303.05299}{{arXiv:2303.05299}}} {[astro-ph.SR]}

\bibitem[{{Reyl{\'e}} et~al(2021){Reyl{\'e}}, {Jardine}, {Fouqu{\'e}},
  {Caballero}, {Smart}, and {Sozzetti}}]{Gaia10pc}
{Reyl{\'e}} C, {Jardine} K, {Fouqu{\'e}} P, et~al (2021) {The 10 parsec sample
  in the Gaia era}. \aap 650:A201. \doi{10.1051/0004-6361/202140985},
  {\href{https://arxiv.org/abs/2104.14972}{{arXiv:2104.14972}}} {[astro-ph.SR]}

\bibitem[{{Ribas} et~al(2005){Ribas}, {Guinan}, {G{\"u}del}, and
  {Audard}}]{Ribas2005}
{Ribas} I, {Guinan} EF, {G{\"u}del} M, et~al (2005) {Evolution of the Solar
  Activity over Time and Effects on Planetary Atmospheres. I. High-Energy
  Irradiances (1-1700 {\r{A}})}. \apj 622(1):680--694. \doi{10.1086/427977},
  {\href{https://arxiv.org/abs/astro-ph/0412253}{{arXiv:astro-ph/0412253}}}
  {[astro-ph]}

\bibitem[{{Ricchiazzi} and {Canfield}(1983)}]{RC83}
{Ricchiazzi} PJ, {Canfield} RC (1983) {A static model of chromospheric heating
  in solar flares}. \apj 272:739--755. \doi{10.1086/161336}

\bibitem[{{Ricker} et~al(2015){Ricker}, {Winn}, {Vanderspek}, {Latham},
  {Bakos}, {Bean}, {Berta-Thompson}, {Brown}, {Buchhave}, {Butler}, {Butler},
  {Chaplin}, {Charbonneau}, {Christensen-Dalsgaard}, {Clampin}, {Deming},
  {Doty}, {De Lee}, {Dressing}, {Dunham}, {Endl}, {Fressin}, {Ge}, {Henning},
  {Holman}, {Howard}, {Ida}, {Jenkins}, {Jernigan}, {Johnson}, {Kaltenegger},
  {Kawai}, {Kjeldsen}, {Laughlin}, {Levine}, {Lin}, {Lissauer}, {MacQueen},
  {Marcy}, {McCullough}, {Morton}, {Narita}, {Paegert}, {Palle}, {Pepe},
  {Pepper}, {Quirrenbach}, {Rinehart}, {Sasselov}, {Sato}, {Seager},
  {Sozzetti}, {Stassun}, {Sullivan}, {Szentgyorgyi}, {Torres}, {Udry}, and
  {Villasenor}}]{Ricker2015}
{Ricker} GR, {Winn} JN, {Vanderspek} R, et~al (2015) {Transiting Exoplanet
  Survey Satellite (TESS)}. Journal of Astronomical Telescopes, Instruments,
  and Systems 1:014003. \doi{10.1117/1.JATIS.1.1.014003}

\bibitem[{{Rimmele} et~al(2020){Rimmele}, {Warner}, {Keil}, {Goode},
  {Kn{\"o}lker}, {Kuhn}, {Rosner}, {McMullin}, {Casini}, {Lin}, {W{\"o}ger},
  {von der L{\"u}he}, {Tritschler}, {Davey}, {de Wijn}, {Elmore}, {Fehlmann},
  {Harrington}, {Jaeggli}, {Rast}, {Schad}, {Schmidt}, {Mathioudakis},
  {Mickey}, {Anan}, {Beck}, {Marshall}, {Jeffers}, {Oschmann}, {Beard},
  {Berst}, {Cowan}, {Craig}, {Cross}, {Cummings}, {Donnelly}, {de Vanssay},
  {Eigenbrot}, {Ferayorni}, {Foster}, {Galapon}, {Gedrites}, {Gonzales},
  {Goodrich}, {Gregory}, {Guzman}, {Guzzo}, {Hegwer}, {Hubbard}, {Hubbard},
  {Johansson}, {Johnson}, {Liang}, {Liang}, {McQuillen}, {Mayer}, {Newman},
  {Onodera}, {Phelps}, {Puentes}, {Richards}, {Rimmele}, {Sekulic}, {Shimko},
  {Simison}, {Smith}, {Starman}, {Sueoka}, {Summers}, {Szabo}, {Szabo},
  {Wampler}, {Williams}, and {White}}]{Rimmele2020}
{Rimmele} TR, {Warner} M, {Keil} SL, et~al (2020) {The Daniel K. Inouye Solar
  Telescope - Observatory Overview}. \solphys 295(12):172.
  \doi{10.1007/s11207-020-01736-7}

\bibitem[{{Robinson} et~al(1993){Robinson}, {Carpenter}, {Woodgate}, and
  {Maran}}]{Robinson1993}
{Robinson} RD, {Carpenter} KG, {Woodgate} BE, et~al (1993) {A search for proton
  beams during flares on AU Microscopii}. \apj 414:872--876.
  \doi{10.1086/173129}

\bibitem[{{Robinson} et~al(1995){Robinson}, {Carpenter}, {Percival}, and
  {Bookbinder}}]{Robinson1995}
{Robinson} RD, {Carpenter} KG, {Percival} JW, et~al (1995) {A Search for
  Microflaring Activity on dMe Flare Stars. I. Observations of the dM8e Star CN
  Leonis}. \apj 451:795. \doi{10.1086/176266}

\bibitem[{{Robinson} et~al(2001){Robinson}, {Linsky}, {Woodgate}, and
  {Timothy}}]{Robinson2001}
{Robinson} RD, {Linsky} JL, {Woodgate} BE, et~al (2001) {Far-Ultraviolet
  Observations of Flares on the dM0e Star AU Microscopii}. \apj
  554(1):368--382. \doi{10.1086/321379}

\bibitem[{{Robinson} et~al(2005){Robinson}, {Wheatley}, {Welsh}, {Forster},
  {Morrissey}, {Seibert}, {Rich}, {Salim}, {Barlow}, {Bianchi}, {Byun},
  {Donas}, {Friedman}, {Heckman}, {Jelinsky}, {Lee}, {Madore}, {Malina},
  {Martin}, {Milliard}, {Neff}, {Schiminovich}, {Siegmund}, {Small}, {Szalay},
  and {Wyder}}]{Robinson2005}
{Robinson} RD, {Wheatley} JM, {Welsh} BY, et~al (2005) {GALEX Observations of
  an Energetic Ultraviolet Flare on the dM4e Star GJ 3685A}. \apj 633:447--451.
  \doi{10.1086/444608},
  {\href{https://arxiv.org/abs/astro-ph/0507396}{{astro-ph/0507396}}}

\bibitem[{{Robrade} and {Schmitt}(2005)}]{Robrade2005}
{Robrade} J, {Schmitt} JHMM (2005) {X-ray properties of active M dwarfs as
  observed by XMM-Newton}. \aap 435(3):1073--1085.
  \doi{10.1051/0004-6361:20041941},
  {\href{https://arxiv.org/abs/astro-ph/0504145}{{arXiv:astro-ph/0504145}}}
  {[astro-ph]}

\bibitem[{{Robrade} and {Schmitt}(2009)}]{Altair}
{Robrade} J, {Schmitt} JHMM (2009) {Altair - the ``hottest'' magnetically
  active star in X-rays}. \aap 497(2):511--520.
  \doi{10.1051/0004-6361/200811348},
  {\href{https://arxiv.org/abs/0903.0966}{{arXiv:0903.0966}}} {[astro-ph.SR]}

\bibitem[{{Robrade} et~al(2004){Robrade}, {Ness}, and {Schmitt}}]{Robrade2004}
{Robrade} J, {Ness} JU, {Schmitt} JHMM (2004) {Spatially resolved X-ray
  emission of EQ Pegasi}. \aap 413:317--321. \doi{10.1051/0004-6361:20034084},
  {\href{https://arxiv.org/abs/astro-ph/0310600}{{arXiv:astro-ph/0310600}}}
  {[astro-ph]}

\bibitem[{{Rodr{\'\i}guez Mart{\'\i}nez} et~al(2020){Rodr{\'\i}guez
  Mart{\'\i}nez}, {Lopez}, {Shappee}, {Schmidt}, {Jayasinghe}, {Kochanek},
  {Auchettl}, and {Holoien}}]{Rodriguez2020}
{Rodr{\'\i}guez Mart{\'\i}nez} R, {Lopez} LA, {Shappee} BJ, et~al (2020) {A
  Catalog of M-dwarf Flares with ASAS-SN}. \apj 892(2):144.
  \doi{10.3847/1538-4357/ab793a},
  {\href{https://arxiv.org/abs/1912.05549}{{arXiv:1912.05549}}} {[astro-ph.SR]}

\bibitem[{{Rosenbluth} et~al(1957){Rosenbluth}, {MacDonald}, and
  {Judd}}]{Rosenbluth1957}
{Rosenbluth} MN, {MacDonald} WM, {Judd} DL (1957) {Fokker-Planck Equation for
  an Inverse-Square Force}. Physical Review 107(1):1--6.
  \doi{10.1103/PhysRev.107.1}

\bibitem[{{Rosner} and {Vaiana}(1978)}]{Rosner1978}
{Rosner} R, {Vaiana} GS (1978) {Cosmic flare transients: constraints upon
  models for energy storage and release derived from the event frequency
  distribution.} \apj 222:1104--1108. \doi{10.1086/156227}

\bibitem[{{Rosner} et~al(1978){Rosner}, {Tucker}, and {Vaiana}}]{RTV78}
{Rosner} R, {Tucker} WH, {Vaiana} GS (1978) {Dynamics of the quiescent solar
  corona.} \apj 220:643--645. \doi{10.1086/155949}

\bibitem[{{Rubbmark} et~al(1981){Rubbmark}, {Kash}, {Littman}, and
  {Kleppner}}]{Rubbmark1981}
{Rubbmark} JR, {Kash} MM, {Littman} MG, et~al (1981) {Dynamical effects at
  avoided level crossings: A study of the Landau-Zener effect using Rydberg
  atoms}. \pra 23(6):3107--3117. \doi{10.1103/PhysRevA.23.3107}

\bibitem[{{Rubio da Costa} and {Kleint}(2017)}]{Rubio2017}
{Rubio da Costa} F, {Kleint} L (2017) {A Parameter Study for Modeling Mg II h
  and k Emission during Solar Flares}. \apj 842(2):82.
  \doi{10.3847/1538-4357/aa6eaf},
  {\href{https://arxiv.org/abs/1704.05874}{{arXiv:1704.05874}}} {[astro-ph.SR]}

\bibitem[{{Rubio da Costa} et~al(2016){Rubio da Costa}, {Kleint}, {Petrosian},
  {Liu}, and {Allred}}]{Rubio2016}
{Rubio da Costa} F, {Kleint} L, {Petrosian} V, et~al (2016) {Data-driven
  Radiative Hydrodynamic Modeling of the 2014 March 29 X1.0 Solar Flare}. \apj
  827:38. \doi{10.3847/0004-637X/827/1/38},
  {\href{https://arxiv.org/abs/1603.04951}{{arXiv:1603.04951}}} {[astro-ph.SR]}

\bibitem[{{Rutledge} et~al(2000){Rutledge}, {Basri}, {Mart{\'\i}n}, and
  {Bildsten}}]{Rutledge2000}
{Rutledge} RE, {Basri} G, {Mart{\'\i}n} EL, et~al (2000) {Chandra Detection of
  an X-Ray Flare from the Brown Dwarf LP 944-20}. \apjl 538(2):L141--L144.
  \doi{10.1086/312817},
  {\href{https://arxiv.org/abs/astro-ph/0005559}{{arXiv:astro-ph/0005559}}}
  {[astro-ph]}

\bibitem[{{Rutten}(2003)}]{Rutten2003}
{Rutten} RJ (2003) {Radiative Transfer in Stellar Atmospheres}

\bibitem[{{Ryan} et~al(2013){Ryan}, {Chamberlin}, {Milligan}, and
  {Gallagher}}]{Ryan2013}
{Ryan} DF, {Chamberlin} PC, {Milligan} RO, et~al (2013) {Decay-phase Cooling
  and Inferred Heating of M- and X-class Solar Flares}. \apj 778(1):68.
  \doi{10.1088/0004-637X/778/1/68},
  {\href{https://arxiv.org/abs/1401.4079}{{arXiv:1401.4079}}} {[astro-ph.SR]}

\bibitem[{{Sakurai}(2022)}]{Sakurai2022}
{Sakurai} T (2022) {Probability Distribution Functions of Solar and Stellar
  Flares}. Physics 5(1):11--23. \doi{10.3390/physics5010002},
  {\href{https://arxiv.org/abs/2212.02678}{{arXiv:2212.02678}}} {[astro-ph.SR]}

\bibitem[{{Salter} et~al(2008){Salter}, {Hogerheijde}, and
  {Blake}}]{Salter2008}
{Salter} DM, {Hogerheijde} MR, {Blake} GA (2008) {Captured at millimeter
  wavelengths: a flare from the classical T Tauri star DQ Tauri}. \aap
  492(1):L21--L24. \doi{10.1051/0004-6361:200810807},
  {\href{https://arxiv.org/abs/0810.4162}{{arXiv:0810.4162}}} {[astro-ph]}

\bibitem[{{Salter} et~al(2010){Salter}, {K{\'o}sp{\'a}l}, {Getman},
  {Hogerheijde}, {van Kempen}, {Carpenter}, {Blake}, and {Wilner}}]{Salter2010}
{Salter} DM, {K{\'o}sp{\'a}l} {\'A}, {Getman} KV, et~al (2010) {Recurring
  millimeter flares as evidence for star-star magnetic reconnection events in
  the DQ Tauri PMS binary system}. \aap 521:A32.
  \doi{10.1051/0004-6361/201015197},
  {\href{https://arxiv.org/abs/1008.0981}{{arXiv:1008.0981}}} {[astro-ph.SR]}

\bibitem[{{Samus'} et~al(2017){Samus'}, {Kazarovets}, {Durlevich}, {Kireeva},
  and {Pastukhova}}]{VarCat}
{Samus'} NN, {Kazarovets} EV, {Durlevich} OV, et~al (2017) {General catalogue
  of variable stars: Version GCVS 5.1}. Astronomy Reports 61(1):80--88.
  \doi{10.1134/S1063772917010085}

\bibitem[{{Sarna} et~al(1998){Sarna}, {Yerli}, and {Muslimov}}]{Sarna1998}
{Sarna} MJ, {Yerli} SK, {Muslimov} AG (1998) {Magnetic activity and evolution
  of Algol-type stars - II}. \mnras 297(3):760--768.
  \doi{10.1046/j.1365-8711.1998.01539.x}

\bibitem[{{Schaefer} et~al(2000){Schaefer}, {King}, and
  {Deliyannis}}]{Schaeffer2000}
{Schaefer} BE, {King} JR, {Deliyannis} CP (2000) {Superflares on Ordinary
  Solar-Type Stars}. \apj 529(2):1026--1030. \doi{10.1086/308325},
  {\href{https://arxiv.org/abs/astro-ph/9909188}{{arXiv:astro-ph/9909188}}}
  {[astro-ph]}

\bibitem[{{Scherrer} et~al(2012){Scherrer}, {Schou}, {Bush}, {Kosovichev},
  {Bogart}, {Hoeksema}, {Liu}, {Duvall}, {Zhao}, {Title}, {Schrijver},
  {Tarbell}, and {Tomczyk}}]{Scherrer2012}
{Scherrer} PH, {Schou} J, {Bush} RI, et~al (2012) {The Helioseismic and
  Magnetic Imager (HMI) Investigation for the Solar Dynamics Observatory
  (SDO)}. \solphys 275(1-2):207--227. \doi{10.1007/s11207-011-9834-2}

\bibitem[{{Schmidt} et~al(2007){Schmidt}, {Cruz}, {Bongiorno}, {Liebert}, and
  {Reid}}]{Schmidt2007}
{Schmidt} SJ, {Cruz} KL, {Bongiorno} BJ, et~al (2007) {Activity and Kinematics
  of Ultracool Dwarfs, Including an Amazing Flare Observation}. \aj
  133(5):2258--2273. \doi{10.1086/512158},
  {\href{https://arxiv.org/abs/astro-ph/0701055}{{arXiv:astro-ph/0701055}}}
  {[astro-ph]}

\bibitem[{{Schmidt} et~al(2012){Schmidt}, {Kowalski}, {Hawley}, {Hilton},
  {Wisniewski}, and {Tofflemire}}]{Schmidt2012}
{Schmidt} SJ, {Kowalski} AF, {Hawley} SL, et~al (2012) {Probing the Flare
  Atmospheres of M Dwarfs Using Infrared Emission Lines}. \apj 745(1):14.
  \doi{10.1088/0004-637X/745/1/14},
  {\href{https://arxiv.org/abs/1111.7072}{{arXiv:1111.7072}}} {[astro-ph.SR]}

\bibitem[{{Schmidt} et~al(2014){Schmidt}, {Prieto}, {Stanek}, {Shappee},
  {Morrell}, {Bardalez Gagliuffi}, {Kochanek}, {Jencson}, {Holoien}, {Basu},
  {Beacom}, {Szczygie{\l}}, {Pojmanski}, {Brimacombe}, {Dubberley}, {Elphick},
  {Foale}, {Hawkins}, {Mullins}, {Rosing}, {Ross}, and {Walker}}]{Schmidt2014}
{Schmidt} SJ, {Prieto} JL, {Stanek} KZ, et~al (2014) {Characterizing a Dramatic
  {\ensuremath{\Delta}}V \raisebox{-0.5ex}\textasciitilde -9 Flare on an
  Ultracool Dwarf Found by the ASAS-SN Survey}. \apjl 781(2):L24.
  \doi{10.1088/2041-8205/781/2/L24},
  {\href{https://arxiv.org/abs/1310.4515}{{arXiv:1310.4515}}} {[astro-ph.SR]}

\bibitem[{{Schmidt} et~al(2016){Schmidt}, {Shappee}, {Gagn{\'e}}, {Stanek},
  {Prieto}, {Holoien}, {Kochanek}, {Chomiuk}, {Dong}, {Seibert}, and
  {Strader}}]{Schmidt2016}
{Schmidt} SJ, {Shappee} BJ, {Gagn{\'e}} J, et~al (2016) {ASASSN-16ae: A
  Powerful White-light Flare on an Early-L Dwarf}. \apjl 828(2):L22.
  \doi{10.3847/2041-8205/828/2/L22},
  {\href{https://arxiv.org/abs/1605.04313}{{arXiv:1605.04313}}} {[astro-ph.SR]}

\bibitem[{{Schmidt} et~al(2019){Schmidt}, {Shappee}, {van Saders}, {Stanek},
  {Brown}, {Kochanek}, {Dong}, {Drout}, {Frank}, {Holoien}, {Johnson},
  {Madore}, {Prieto}, {Seibert}, {Seidel}, and {Simonian}}]{Schmidt2019}
{Schmidt} SJ, {Shappee} BJ, {van Saders} JL, et~al (2019) {The Largest M Dwarf
  Flares from ASAS-SN}. \apj 876(2):115. \doi{10.3847/1538-4357/ab148d},
  {\href{https://arxiv.org/abs/1809.04510}{{arXiv:1809.04510}}} {[astro-ph.SR]}

\bibitem[{{Schmitt} and {Favata}(1999)}]{Algol1}
{Schmitt} JHMM, {Favata} F (1999) {Continuous heating of a giant X-ray flare on
  Algol}. \nat 401(6748):44--46. \doi{10.1038/43389},
  {\href{https://arxiv.org/abs/astro-ph/9909040}{{arXiv:astro-ph/9909040}}}
  {[astro-ph]}

\bibitem[{{Schmitt} and {Liefke}(2004)}]{Schmitt2004}
{Schmitt} JHMM, {Liefke} C (2004) {NEXXUS: A comprehensive ROSAT survey of
  coronal X-ray emission among nearby solar-like stars}. \aap 417:651--665.
  \doi{10.1051/0004-6361:20030495},
  {\href{https://arxiv.org/abs/astro-ph/0308510}{{arXiv:astro-ph/0308510}}}
  {[astro-ph]}

\bibitem[{{Schmitt} et~al(2003){Schmitt}, {Ness}, and {Franco}}]{Algol2}
{Schmitt} JHMM, {Ness} JU, {Franco} G (2003) {A spatially resolved limb flare
  on Algol B observed with XMM-Newton}. \aap 412:849--855.
  \doi{10.1051/0004-6361:20034057},
  {\href{https://arxiv.org/abs/astro-ph/0308394}{{arXiv:astro-ph/0308394}}}
  {[astro-ph]}

\bibitem[{{Schmitt} et~al(2008){Schmitt}, {Reale}, {Liefke}, {Wolter},
  {Fuhrmeister}, {Reiners}, and {Peres}}]{Schmitt2008}
{Schmitt} JHMM, {Reale} F, {Liefke} C, et~al (2008) {A coronal explosion on the
  flare star CN Leonis}. \aap 481:799--805. \doi{10.1051/0004-6361:20079017},
  {\href{https://arxiv.org/abs/0801.3752}{{arXiv:0801.3752}}}

\bibitem[{{Schmitt} et~al(2019){Schmitt}, {Ioannidis}, {Robrade}, {Czesla}, and
  {Schneider}}]{Schmitt2019}
{Schmitt} JHMM, {Ioannidis} P, {Robrade} J, et~al (2019) {Superflares on AB
  Doradus observed with TESS}. \aap 628:A79. \doi{10.1051/0004-6361/201935374}

\bibitem[{{Schrijver} et~al(2012){Schrijver}, {Beer}, {Baltensperger},
  {Cliver}, {G{\"u}del}, {Hudson}, {McCracken}, {Osten}, {Peter}, {Soderblom},
  {Usoskin}, and {Wolff}}]{Schrijver2012}
{Schrijver} CJ, {Beer} J, {Baltensperger} U, et~al (2012) {Estimating the
  frequency of extremely energetic solar events, based on solar, stellar,
  lunar, and terrestrial records}. Journal of Geophysical Research (Space
  Physics) 117(A8):A08103. \doi{10.1029/2012JA017706},
  {\href{https://arxiv.org/abs/1206.4889}{{arXiv:1206.4889}}} {[astro-ph.SR]}

\bibitem[{{Schrijver} et~al(2019){Schrijver}, {Bagenal}, {Bastian}, {Beer},
  {Bisi}, {Bogdan}, {Bougher}, {Boteler}, {Brain}, {Brasseur}, {Brownlee},
  {Charbonneau}, {Cohen}, {Christensen}, {Crowley}, {Fischer}, {Forbes},
  {Fuller-Rowell}, {Galand}, {Giacalone}, {Gloeckler}, {Gosling}, {Green},
  {Gross}, {Guetersloh}, {Hansteen}, {Hartmann}, {Horanyi}, {Hudson},
  {Jakowski}, {Jokipii}, {Kivelson}, {Krauss-Varban}, {Krupp}, {Lean},
  {Linsky}, {Longcope}, {Marsh}, {Miesch}, {Moldwin}, {Moore}, {Odenwald},
  {Opher}, {Osten}, {Rempel}, {Schmidt}, {Siscoe}, {Siskind}, {Smith},
  {Solomon}, {Stallard}, {Stanley}, {Sojka}, {Tobiska}, {Toffoletto},
  {Tribble}, {Vasyliunas}, {Walterscheid}, {Wang}, {Wood}, {Woods}, and
  {Zapp}}]{Schrijver2019}
{Schrijver} K, {Bagenal} F, {Bastian} T, et~al (2019) {Principles Of
  Heliophysics: a textbook on the universal processes behind planetary
  habitability}. arXiv e-prints arXiv:1910.14022.
  \doi{10.48550/arXiv.1910.14022},
  {\href{https://arxiv.org/abs/1910.14022}{{arXiv:1910.14022}}} {[astro-ph.SR]}

\bibitem[{{Seaton}(1990)}]{Seaton1990}
{Seaton} MJ (1990) {Atomic data for opacity calculations. XIII - Line profiles
  for transitions in hydrogenic ions}. Journal of Physics B Atomic Molecular
  Physics 23:3255--3296. \doi{10.1088/0953-4075/23/19/012}

\bibitem[{{Segura}(2018)}]{Segura2018}
{Segura} A (2018) {Star-Planet Interactions and Habitability: Radiative
  Effects}. In: {Deeg} HJ, {Belmonte} JA (eds) Handbook of Exoplanets. p~73,
  \doi{10.1007/978-3-319-55333-7_73}

\bibitem[{{Seidel} et~al(1995){Seidel}, {Arndt}, and {Kraeft}}]{Seidel1995}
{Seidel} J, {Arndt} S, {Kraeft} WD (1995) {Energy spectrum of hydrogen atoms in
  dense plasmas}. \pre 52(5):5387--5400. \doi{10.1103/PhysRevE.52.5387}

\bibitem[{{Shakhovskaia}(1989)}]{Shakh1989}
{Shakhovskaia} NI (1989) {Stellar flare statistics {\textemdash} Physical
  consequences}. \solphys 121(1-2):375--386. \doi{10.1007/BF00161707}

\bibitem[{{Shappee} et~al(2017){Shappee}, {Simon}, {Drout}, {Piro}, {Morrell},
  {Prieto}, {Kasen}, {Holoien}, {Kollmeier}, {Kelson}, {Coulter}, {Foley},
  {Kilpatrick}, {Siebert}, {Madore}, {Murguia-Berthier}, {Pan}, {Prochaska},
  {Ramirez-Ruiz}, {Rest}, {Adams}, {Alatalo}, {Ba{\~n}ados}, {Baughman},
  {Bernstein}, {Bitsakis}, {Boutsia}, {Bravo}, {Di Mille}, {Higgs}, {Ji},
  {Maravelias}, {Marshall}, {Placco}, {Prieto}, and {Wan}}]{Shappee2017}
{Shappee} BJ, {Simon} JD, {Drout} MR, et~al (2017) {Early spectra of the
  gravitational wave source GW170817: Evolution of a neutron star merger}.
  Science 358(6370):1574--1578. \doi{10.1126/science.aaq0186},
  {\href{https://arxiv.org/abs/1710.05432}{{arXiv:1710.05432}}} {[astro-ph.HE]}

\bibitem[{{Shibata} and {Magara}(2011)}]{Shibata2011}
{Shibata} K, {Magara} T (2011) {Solar Flares: Magnetohydrodynamic Processes}.
  Living Reviews in Solar Physics 8(1):6. \doi{10.12942/lrsp-2011-6}

\bibitem[{{Shibata} and {Tanuma}(2001)}]{Shibata2001}
{Shibata} K, {Tanuma} S (2001) {Plasmoid-induced-reconnection and fractal
  reconnection}. Earth, Planets and Space 53:473--482.
  \doi{10.1186/BF03353258},
  {\href{https://arxiv.org/abs/astro-ph/0101008}{{arXiv:astro-ph/0101008}}}
  {[astro-ph]}

\bibitem[{{Shibata} and {Yokoyama}(1999)}]{Shibata1999}
{Shibata} K, {Yokoyama} T (1999) {Origin of the Universal Correlation between
  the Flare Temperature and the Emission Measure for Solar and Stellar Flares}.
  \apjl 526(1):L49--L52. \doi{10.1086/312354}

\bibitem[{{Shibata} et~al(1995){Shibata}, {Masuda}, {Shimojo}, {Hara},
  {Yokoyama}, {Tsuneta}, {Kosugi}, and {Ogawara}}]{Shibata1995}
{Shibata} K, {Masuda} S, {Shimojo} M, et~al (1995) {Hot-Plasma Ejections
  Associated with Compact-Loop Solar Flares}. \apjl 451:L83.
  \doi{10.1086/309688}

\bibitem[{{Shibata} et~al(2013){Shibata}, {Isobe}, {Hillier}, {Choudhuri},
  {Maehara}, {Ishii}, {Shibayama}, {Notsu}, {Notsu}, {Nagao}, {Honda}, and
  {Nogami}}]{Shibata2013}
{Shibata} K, {Isobe} H, {Hillier} A, et~al (2013) {Can Superflares Occur on Our
  Sun?} \pasj 65:49. \doi{10.1093/pasj/65.3.49},
  {\href{https://arxiv.org/abs/1212.1361}{{arXiv:1212.1361}}} {[astro-ph.SR]}

\bibitem[{{Shibayama} et~al(2013){Shibayama}, {Maehara}, {Notsu}, {Notsu},
  {Nagao}, {Honda}, {Ishii}, {Nogami}, and {Shibata}}]{Shibayama2013}
{Shibayama} T, {Maehara} H, {Notsu} S, et~al (2013) {Superflares on Solar-type
  Stars Observed with Kepler. I. Statistical Properties of Superflares}. \apjs
  209:5. \doi{10.1088/0067-0049/209/1/5},
  {\href{https://arxiv.org/abs/1308.1480}{{arXiv:1308.1480}}} {[astro-ph.SR]}

\bibitem[{{Shields} et~al(2016){Shields}, {Ballard}, and
  {Johnson}}]{Shields2016}
{Shields} AL, {Ballard} S, {Johnson} JA (2016) {The habitability of planets
  orbiting M-dwarf stars}. \physrep 663:1. \doi{10.1016/j.physrep.2016.10.003},
  {\href{https://arxiv.org/abs/1610.05765}{{arXiv:1610.05765}}} {[astro-ph.EP]}

\bibitem[{{Shull} et~al(2012){Shull}, {Stevans}, and {Danforth}}]{Shull2012}
{Shull} JM, {Stevans} M, {Danforth} CW (2012) {HST-COS Observations of AGNs. I.
  Ultraviolet Composite Spectra of the Ionizing Continuum and Emission Lines}.
  \apj 752(2):162. \doi{10.1088/0004-637X/752/2/162},
  {\href{https://arxiv.org/abs/1204.3908}{{arXiv:1204.3908}}} {[astro-ph.CO]}

\bibitem[{{Siess} and {Livio}(1999)}]{Siess1999}
{Siess} L, {Livio} M (1999) {The accretion of brown dwarfs and planets by giant
  stars - II. Solar-mass stars on the red giant branch}. \mnras
  308(4):1133--1149. \doi{10.1046/j.1365-8711.1999.02784.x},
  {\href{https://arxiv.org/abs/astro-ph/9905235}{{arXiv:astro-ph/9905235}}}
  {[astro-ph]}

\bibitem[{{Silverberg} et~al(2016){Silverberg}, {Kowalski}, {Davenport},
  {Wisniewski}, {Hawley}, and {Hilton}}]{Silverberg2016}
{Silverberg} SM, {Kowalski} AF, {Davenport} JRA, et~al (2016) {Kepler Flares.
  IV. A Comprehensive Analysis of the Activity of the dM4e Star GJ 1243}. \apj
  829:129. \doi{10.3847/0004-637X/829/2/129},
  {\href{https://arxiv.org/abs/1607.03886}{{arXiv:1607.03886}}} {[astro-ph.SR]}

\bibitem[{{Simnett} and {Haines}(1990)}]{Simnett1990}
{Simnett} GM, {Haines} MG (1990) {On the Production of Hard X-Rays in Solar
  Flares}. \solphys 130(1-2):253--263. \doi{10.1007/BF00156793}

\bibitem[{{Simon} and {Drake}(1989)}]{Simon1989}
{Simon} T, {Drake} SA (1989) {The Evolution of Chromospheric Activity of Cool
  Giant and Subgiant Stars}. \apj 346:303. \doi{10.1086/168012}

\bibitem[{{Singh} et~al(1996){Singh}, {Drake}, and {White}}]{Singh1996}
{Singh} KP, {Drake} SA, {White} NE (1996) {RS CVn Versus Algol-Type Binaries: A
  Comparative Study of Their X-Ray Emission}. \aj 111:2415.
  \doi{10.1086/117975}

\bibitem[{{Sirianni} et~al(2005){Sirianni}, {Jee}, {Ben{\'{\i}}tez},
  {Blakeslee}, {Martel}, {Meurer}, {Clampin}, {De Marchi}, {Ford}, {Gilliland},
  {Hartig}, {Illingworth}, {Mack}, and {McCann}}]{Sirianni2005}
{Sirianni} M, {Jee} MJ, {Ben{\'{\i}}tez} N, et~al (2005) {The Photometric
  Performance and Calibration of the Hubble Space Telescope Advanced Camera for
  Surveys}. \pasp 117:1049--1112. \doi{10.1086/444553},
  {\href{https://arxiv.org/abs/astro-ph/0507614}{{astro-ph/0507614}}}

\bibitem[{{Siversky} and {Zharkova}(2009)}]{Siversky2009}
{Siversky} TV, {Zharkova} VV (2009) {Stationary and impulsive injection of
  electron beams in converging magnetic field}. \aap 504(3):1057--1070.
  \doi{10.1051/0004-6361/200912341},
  {\href{https://arxiv.org/abs/0907.1911}{{arXiv:0907.1911}}} {[astro-ph.SR]}

\bibitem[{{Smith} and {Auer}(1980)}]{Smith1980}
{Smith} DF, {Auer} LH (1980) {Thermal models for solar hard X-ray bursts}. \apj
  238:1126--1133. \doi{10.1086/158078}

\bibitem[{{Smith} et~al(1969){Smith}, {Vidal}, and {Cooper}}]{Smith1969}
{Smith} EW, {Vidal} CR, {Cooper} J (1969) {Classical Path Methods in Line
  Broadening. I. The Classical Path Approximation}. J Res Natl Bur Stand A Phys
  Chem 73A(4):389--404. \doi{10.6028/jres.073A.030}

\bibitem[{{Smith} et~al(2005){Smith}, {G{\"u}del}, and {Audard}}]{Smith2005}
{Smith} K, {G{\"u}del} M, {Audard} M (2005) {Flares observed with XMM-Newton
  and the VLA}. \aap 436:241--251. \doi{10.1051/0004-6361:20042054},
  {\href{https://arxiv.org/abs/astro-ph/0503022}{{astro-ph/0503022}}}

\bibitem[{{Snyder} and {Scott}(1949)}]{Snyder1949}
{Snyder} HS, {Scott} WT (1949) {Multiple Scattering of Fast Charged Particles}.
  Physical Review 76(2):220--225. \doi{10.1103/PhysRev.76.220}

\bibitem[{{Sobel'Man} et~al(1995){Sobel'Man}, {Vainshtein}, and
  {Yukov}}]{Sobelman1995}
{Sobel'Man} II, {Vainshtein} LA, {Yukov} EA (1995) {Excitation of Atoms and
  Broadening of Spectral Lines}

\bibitem[{{Somov} and {Kosugi}(1997)}]{Somov1997}
{Somov} BV, {Kosugi} T (1997) {Collisionless Reconnection and High-Energy
  Particle Acceleration in Solar Flares}. \apj 485(2):859--868.
  \doi{10.1086/304449}

\bibitem[{{Song} and {Paglione}(2020)}]{Song2020}
{Song} Y, {Paglione} TAD (2020) {A Stacking Search for Gamma-Ray Emission from
  Nearby Flare Stars and the Periodic Source TVLM 513-46546}. \apj 900(2):185.
  \doi{10.3847/1538-4357/abac5f},
  {\href{https://arxiv.org/abs/2008.01143}{{arXiv:2008.01143}}} {[astro-ph.HE]}

\bibitem[{{Stehle}(1994)}]{Stehle1994}
{Stehle} C (1994) {Stark broadening of hydrogen Lyman and Balmer in the
  conditions of stellar envelopes}. \aaps 104:509--527

\bibitem[{{Stehl{\'e}} and {Hutcheon}(1999)}]{Stehle1999}
{Stehl{\'e}} C, {Hutcheon} R (1999) {Extensive tabulations of Stark broadened
  hydrogen line profiles}. \aaps 140:93--97. \doi{10.1051/aas:1999118}

\bibitem[{{Stehle} and {Jacquemot}(1993)}]{Stehle1993}
{Stehle} C, {Jacquemot} S (1993) {Line shapes in hydrogen opacities}. \aap
  271:348

\bibitem[{{Stelzer} et~al(2000){Stelzer}, {Neuh{\"a}user}, and
  {Hambaryan}}]{Stelzer2000}
{Stelzer} B, {Neuh{\"a}user} R, {Hambaryan} V (2000) {X-ray flares on zero-age-
  and pre-main sequence stars in Taurus-Auriga-Perseus}. \aap 356:949--971.
  \doi{10.48550/arXiv.astro-ph/0002354},
  {\href{https://arxiv.org/abs/astro-ph/0002354}{{arXiv:astro-ph/0002354}}}
  {[astro-ph]}

\bibitem[{{Stelzer} et~al(2002){Stelzer}, {Burwitz}, {Audard}, {G{\"u}del},
  {Ness}, {Grosso}, {Neuh{\"a}user}, {Schmitt}, {Predehl}, and
  {Aschenbach}}]{Stelzer2002}
{Stelzer} B, {Burwitz} V, {Audard} M, et~al (2002) {Simultaneous X-ray
  spectroscopy of YY Gem with Chandra and XMM-Newton}. \aap 392:585--598.
  \doi{10.1051/0004-6361:20021188},
  {\href{https://arxiv.org/abs/astro-ph/0206429}{{arXiv:astro-ph/0206429}}}
  {[astro-ph]}

\bibitem[{{Stelzer} et~al(2006){Stelzer}, {Schmitt}, {Micela}, and
  {Liefke}}]{Stelzer2006}
{Stelzer} B, {Schmitt} JHMM, {Micela} G, et~al (2006) {Simultaneous optical and
  X-ray observations of a giant flare on the ultracool dwarf LP 412-31}. \aap
  460(2):L35--L38. \doi{10.1051/0004-6361:20066488},
  {\href{https://arxiv.org/abs/astro-ph/0610582}{{arXiv:astro-ph/0610582}}}
  {[astro-ph]}

\bibitem[{{Stepien}(1993)}]{Stepien1993}
{Stepien} K (1993) {HR 1362: The Evolved 53 Camelopardalis}. \apj 416:368.
  \doi{10.1086/173240}

\bibitem[{{St{\"o}kl} and {Dorfi}(2007)}]{Stokl2007}
{St{\"o}kl} A, {Dorfi} EA (2007) {2-dimensional implicit hydrodynamics on
  adaptive grids}. Computer Physics Communications 177(11):815--831.
  \doi{10.1016/j.cpc.2007.06.012}

\bibitem[{{Strassmeier} et~al(1993){Strassmeier}, {Hall}, {Fekel}, and
  {Scheck}}]{Strass1993}
{Strassmeier} KG, {Hall} DS, {Fekel} FC, et~al (1993) {A catalog of
  chromospherically active binary stars (second edition).} \aaps 100:173--225

\bibitem[{{Sturrock}(1966)}]{Sturrock1966}
{Sturrock} PA (1966) {Model of the High-Energy Phase of Solar Flares}. \nat
  211(5050):695--697. \doi{10.1038/211695a0}

\bibitem[{{Sun} et~al(2012){Sun}, {Hoeksema}, {Liu}, {Wiegelmann}, {Hayashi},
  {Chen}, and {Thalmann}}]{Sun2012}
{Sun} X, {Hoeksema} JT, {Liu} Y, et~al (2012) {Evolution of Magnetic Field and
  Energy in a Major Eruptive Active Region Based on SDO/HMI Observation}. \apj
  748(2):77. \doi{10.1088/0004-637X/748/2/77},
  {\href{https://arxiv.org/abs/1201.3404}{{arXiv:1201.3404}}} {[astro-ph.SR]}

\bibitem[{{Sutton}(1978)}]{Sutton1978}
{Sutton} K (1978) {Approximate line shapes for hydrogen}. \jqsrt 20:333--343.
  \doi{10.1016/0022-4073(78)90102-4}

\bibitem[{{Svestka}(1976)}]{Svestka1976}
{Svestka} Z (1976) {Solar Flares}

\bibitem[{{Tamazian} et~al(2008){Tamazian}, {Docobo}, {Balega}, {Melikian},
  {Maximov}, and {Malogolovets}}]{Tamazian2008}
{Tamazian} VS, {Docobo} JA, {Balega} YY, et~al (2008) {Preliminary Orbit and
  Differential Photometry of the Nearby Flare Star CR Dra}. \aj
  136(3):974--979. \doi{10.1088/0004-6256/136/3/974}

\bibitem[{{Tandberg-Hanssen} and {Emslie}(2009)}]{Tandberg2009}
{Tandberg-Hanssen} E, {Emslie} AG (2009) {The Physics of Solar Flares}

\bibitem[{{Tei} et~al(2018){Tei}, {Sakaue}, {Okamoto}, {Kawate}, {Heinzel},
  {UeNo}, {Asai}, {Ichimoto}, and {Shibata}}]{Tei2018}
{Tei} A, {Sakaue} T, {Okamoto} TJ, et~al (2018) {Blue-wing enhancement of the
  chromospheric Mg II h and k lines in a solar flare}. \pasj 70(6):100.
  \doi{10.1093/pasj/psy047},
  {\href{https://arxiv.org/abs/1803.05237}{{arXiv:1803.05237}}} {[astro-ph.SR]}

\bibitem[{{Testa} et~al(2004){Testa}, {Drake}, and {Peres}}]{Testa2004}
{Testa} P, {Drake} JJ, {Peres} G (2004) {The Density of Coronal Plasma in
  Active Stellar Coronae}. \apj 617(1):508--530. \doi{10.1086/422355},
  {\href{https://arxiv.org/abs/astro-ph/0405019}{{arXiv:astro-ph/0405019}}}
  {[astro-ph]}

\bibitem[{{Testa} et~al(2007){Testa}, {Reale}, {Garcia-Alvarez}, and
  {Huenemoerder}}]{Testa2007}
{Testa} P, {Reale} F, {Garcia-Alvarez} D, et~al (2007) {Detailed Diagnostics of
  an X-Ray Flare in the Single Giant HR 9024}. \apj 663(2):1232--1243.
  \doi{10.1086/518241},
  {\href{https://arxiv.org/abs/astro-ph/0703422}{{arXiv:astro-ph/0703422}}}
  {[astro-ph]}

\bibitem[{{Testa} et~al(2008){Testa}, {Drake}, {Ercolano}, {Reale},
  {Huenemoerder}, {Affer}, {Micela}, and {Garcia-Alvarez}}]{Testa2008}
{Testa} P, {Drake} JJ, {Ercolano} B, et~al (2008) {Geometry Diagnostics of a
  Stellar Flare from Fluorescent X-Rays}. \apjl 675(2):L97.
  \doi{10.1086/533461},
  {\href{https://arxiv.org/abs/0801.3857}{{arXiv:0801.3857}}} {[astro-ph]}

\bibitem[{{Thalmann} et~al(2019){Thalmann}, {Moraitis}, {Linan}, {Pariat},
  {Valori}, and {Dalmasse}}]{Thalmann2019}
{Thalmann} JK, {Moraitis} K, {Linan} L, et~al (2019) {Magnetic Helicity Budget
  of Solar Active Regions Prolific of Eruptive and Confined Flares}. \apj
  887(1):64. \doi{10.3847/1538-4357/ab4e15},
  {\href{https://arxiv.org/abs/1910.06563}{{arXiv:1910.06563}}} {[astro-ph.SR]}

\bibitem[{{Thorne} and {Blandford}(2017)}]{Thorne2017}
{Thorne} KS, {Blandford} RD (2017) {Modern Classical Physics: Optics, Fluids,
  Plasmas, Elasticity, Relativity, and Statistical Physics}

\bibitem[{{Tilley} et~al(2019){Tilley}, {Segura}, {Meadows}, {Hawley}, and
  {Davenport}}]{Tilley2019}
{Tilley} MA, {Segura} A, {Meadows} V, et~al (2019) {Modeling Repeated M Dwarf
  Flaring at an Earth-like Planet in the Habitable Zone: Atmospheric Effects
  for an Unmagnetized Planet}. Astrobiology 19(1):64--86.
  \doi{10.1089/ast.2017.1794}

\bibitem[{{Tofflemire} et~al(2012){Tofflemire}, {Wisniewski}, {Kowalski},
  {Schmidt}, {Kundurthy}, {Hilton}, {Holtzman}, and {Hawley}}]{Tofflemire2012}
{Tofflemire} BM, {Wisniewski} JP, {Kowalski} AF, et~al (2012) {The Implications
  of M Dwarf Flares on the Detection and Characterization of Exoplanets at
  Infrared Wavelengths}. \aj 143(1):12. \doi{10.1088/0004-6256/143/1/12},
  {\href{https://arxiv.org/abs/1111.1793}{{arXiv:1111.1793}}} {[astro-ph.SR]}

\bibitem[{{Tofflemire} et~al(2017){Tofflemire}, {Mathieu}, {Ardila}, {Akeson},
  {Ciardi}, {Johns-Krull}, {Herczeg}, and {Quijano-Vodniza}}]{Tofflemire2017}
{Tofflemire} BM, {Mathieu} RD, {Ardila} DR, et~al (2017) {Accretion and
  Magnetic Reconnection in the Classical T Tauri Binary DQ Tau}. \apj 835(1):8.
  \doi{10.3847/1538-4357/835/1/8},
  {\href{https://arxiv.org/abs/1612.02431}{{arXiv:1612.02431}}} {[astro-ph.SR]}

\bibitem[{{Toledo-Padr{\'o}n} et~al(2019){Toledo-Padr{\'o}n}, {Gonz{\'a}lez
  Hern{\'a}ndez}, {Rodr{\'\i}guez-L{\'o}pez}, {Su{\'a}rez Mascare{\~n}o},
  {Rebolo}, {Butler}, {Ribas}, {Anglada-Escud{\'e}}, {Johnson}, {Reiners},
  {Caballero}, {Quirrenbach}, {Amado}, {B{\'e}jar}, {Morales}, {Perger},
  {Jeffers}, {Vogt}, {Teske}, {Shectman}, {Crane}, {D{\'\i}az}, {Arriagada},
  {Holden}, {Burt}, {Rodr{\'\i}guez}, {Herrero}, {Murgas}, {Pall{\'e}},
  {Morales}, {L{\'o}pez-Gonz{\'a}lez}, {D{\'\i}ez Alonso}, {Tuomi}, {Kiraga},
  {Engle}, {Guinan}, {Strachan}, {Aceituno}, {Aceituno}, {Casanova},
  {Mart{\'\i}n-Ruiz}, {Montes}, {Ortiz}, {Sota}, {Briol}, {Barbieri},
  {Cervini}, {Deldem}, {Dubois}, {Hambsch}, {Harris}, {Kotnik}, {Logie},
  {Lopez}, {McNeely}, {Ogmen}, {P{\'e}rez}, {Rau}, {Rodr{\'\i}guez}, {Urquijo},
  and {Vanaverbeke}}]{Toledo2019}
{Toledo-Padr{\'o}n} B, {Gonz{\'a}lez Hern{\'a}ndez} JI,
  {Rodr{\'\i}guez-L{\'o}pez} C, et~al (2019) {Stellar activity analysis of
  Barnard's Star: very slow rotation and evidence for long-term activity
  cycle}. \mnras 488(4):5145--5161. \doi{10.1093/mnras/stz1975},
  {\href{https://arxiv.org/abs/1812.06712}{{arXiv:1812.06712}}} {[astro-ph.SR]}

\bibitem[{{Topka} and {Marsh}(1982)}]{TopkaMarsh1982}
{Topka} K, {Marsh} KA (1982) {Detection of microwave emission from both
  components of the red dwarfbinary EQ Peg.} \apj 254:641--645.
  \doi{10.1086/159773}

\bibitem[{{Toriumi} and {Wang}(2019)}]{Toriumi2019}
{Toriumi} S, {Wang} H (2019) {Flare-productive active regions}. Living Reviews
  in Solar Physics 16(1):3. \doi{10.1007/s41116-019-0019-7},
  {\href{https://arxiv.org/abs/1904.12027}{{arXiv:1904.12027}}} {[astro-ph.SR]}

\bibitem[{{Torrence} and {Compo}(1998)}]{Torrence1998}
{Torrence} C, {Compo} GP (1998) {A Practical Guide to Wavelet Analysis.}
  Bulletin of the American Meteorological Society 79(1):61--78.
  \doi{10.1175/1520-0477(1998)079<0061:APGTWA>2.0.CO;2}

\bibitem[{{Torres} and {Ribas}(2002)}]{Torres2002}
{Torres} G, {Ribas} I (2002) {Absolute Dimensions of the M-Type Eclipsing
  Binary YY Geminorum (Castor C): A Challenge to Evolutionary Models in the
  Lower Main Sequence}. \apj 567(2):1140--1165. \doi{10.1086/338587},
  {\href{https://arxiv.org/abs/astro-ph/0111167}{{arXiv:astro-ph/0111167}}}
  {[astro-ph]}

\bibitem[{{Tremblay} and {Bergeron}(2009)}]{Tremblay2009}
{Tremblay} PE, {Bergeron} P (2009) {Spectroscopic Analysis of DA White Dwarfs:
  Stark Broadening of Hydrogen Lines Including Nonideal Effects}. \apj
  696:1755--1770. \doi{10.1088/0004-637X/696/2/1755},
  {\href{https://arxiv.org/abs/0902.4182}{{arXiv:0902.4182}}} {[astro-ph.SR]}

\bibitem[{{Tristan} et~al(2023){Tristan}, {Notsu}, {Kowalski}, {Brown},
  {Wisniewski}, {Osten}, {Vrijmoet}, {White}, {Carter}, {Grady}, {Henry},
  {Hinojosa}, {Lomax}, {Neff}, {Paredes}, and {Soutter}}]{Tristan2023}
{Tristan} II, {Notsu} Y, {Kowalski} AF, et~al (2023) {A Seven-Day
  Multi-Wavelength Flare Campaign on AU Mic I: High-Time Resolution Light
  Curves and the Thermal Empirical Neupert Effect}. arXiv e-prints
  arXiv:2304.05692. \doi{10.48550/arXiv.2304.05692},
  {\href{https://arxiv.org/abs/2304.05692}{{arXiv:2304.05692}}} {[astro-ph.SR]}

\bibitem[{{Trubnikov}(1965)}]{Trubnikov1965}
{Trubnikov} BA (1965) {Particle Interactions in a Fully Ionized Plasma}.
  Reviews of Plasma Physics 1:105

\bibitem[{{Tsiklauri}(2017)}]{Tsiklauri2017}
{Tsiklauri} D (2017) {Electron plasma wake field acceleration in solar coronal
  and chromospheric plasmas}. Physics of Plasmas 24(7):072902.
  \doi{10.1063/1.4990560},
  {\href{https://arxiv.org/abs/1706.05265}{{arXiv:1706.05265}}} {[astro-ph.SR]}

\bibitem[{{Tsvetkova} et~al(2013){Tsvetkova}, {Petit}, {Auri{\`e}re},
  {Konstantinova-Antova}, {Wade}, {Charbonnel}, {Decressin}, and
  {Bogdanovski}}]{Tsvetkova2013}
{Tsvetkova} S, {Petit} P, {Auri{\`e}re} M, et~al (2013) {Magnetic field
  structure in single late-type giants: {\ensuremath{\beta}} Ceti in
  2010-2012}. \aap 556:A43. \doi{10.1051/0004-6361/201321051},
  {\href{https://arxiv.org/abs/1301.1592}{{arXiv:1301.1592}}} {[astro-ph.SR]}

\bibitem[{{Uitenbroek}(2001)}]{Uitenbroek2001}
{Uitenbroek} H (2001) {Multilevel Radiative Transfer with Partial Frequency
  Redistribution}. \apj 557:389--398. \doi{10.1086/321659}

\bibitem[{{Upton} and {Hathaway}(2023)}]{Upton2023}
{Upton} LA, {Hathaway} DH (2023) {Solar Cycle Precursors and the Outlook for
  Cycle 25}. Journal of Geophysical Research (Space Physics)
  128(10):e2023JA031681. \doi{10.1029/2023JA031681},
  {\href{https://arxiv.org/abs/2305.06516}{{arXiv:2305.06516}}} {[astro-ph.SR]}

\bibitem[{{Usoskin}(2023)}]{Usoskin2023LRSP}
{Usoskin} IG (2023) {A history of solar activity over millennia}. Living
  Reviews in Solar Physics 20(1):2. \doi{10.1007/s41116-023-00036-z}

\bibitem[{{Uzdensky}(2007)}]{Uzdensky2007}
{Uzdensky} DA (2007) {The Fast Collisionless Reconnection Condition and the
  Self-Organization of Solar Coronal Heating}. \apj 671(2):2139--2153.
  \doi{10.1086/522915},
  {\href{https://arxiv.org/abs/0707.1316}{{arXiv:0707.1316}}} {[astro-ph]}

\bibitem[{{van den Oord}(1990)}]{Oord1990}
{van den Oord} GHJ (1990) {The electrodynamics of beam/return current systems
  in the solar corona}. \aap 234(1-2):496--518

\bibitem[{{van den Oord} and {Mewe}(1989)}]{Oord1989A}
{van den Oord} GHJ, {Mewe} R (1989) {The X-ray flare and the quiescent emission
  from Algol as detected by EXOSAT.} \aap 213:245--260

\bibitem[{{van den Oord} et~al(1989){van den Oord}, {Kuijpers}, {White}, {van
  der Hulst}, and {Culhane}}]{Oord1989B}
{van den Oord} GHJ, {Kuijpers} J, {White} NE, et~al (1989) {A combined radio
  and X-ray observation of Algol.} \aap 209:296--304

\bibitem[{{van Dien}(1949)}]{vanDien1949}
{van Dien} E (1949) {The Stark Effect of the Higher Balmer Lines in Stars of
  Spectral Types a and B.} \apj 109:452. \doi{10.1086/145150}

\bibitem[{{Van Doorsselaere} et~al(2017){Van Doorsselaere}, {Shariati}, and
  {Debosscher}}]{VanD}
{Van Doorsselaere} T, {Shariati} H, {Debosscher} J (2017) {Stellar Flares
  Observed in Long-cadence Data from the Kepler Mission}. \apjs 232:26.
  \doi{10.3847/1538-4365/aa8f9a},
  {\href{https://arxiv.org/abs/1711.02587}{{arXiv:1711.02587}}} {[astro-ph.SR]}

\bibitem[{{van Leer}(1977)}]{vanLeer1977}
{van Leer} B (1977) {Towards the Ultimate Conservative Difference Scheme. IV. A
  New Approach to Numerical Convection}. Journal of Computational Physics
  23:276. \doi{10.1016/0021-9991(77)90095-X}

\bibitem[{{van Leer}(1979)}]{vanLeer1979}
{van Leer} B (1979) {Towards the Ultimate Conservative Difference Scheme. V. A
  Second-Order Sequel to Godunov's Method}. Journal of Computational Physics
  32(1):101--136. \doi{10.1016/0021-9991(79)90145-1}

\bibitem[{{van Leeuwen}(2007)}]{Hipparcos2007}
{van Leeuwen} F (2007) {Validation of the new Hipparcos reduction}. \aap
  474(2):653--664. \doi{10.1051/0004-6361:20078357},
  {\href{https://arxiv.org/abs/0708.1752}{{arXiv:0708.1752}}} {[astro-ph]}

\bibitem[{{Veronig} et~al(2002{\natexlab{a}}){Veronig}, {Temmer}, {Hanslmeier},
  {Otruba}, and {Messerotti}}]{Veronig2002Xrays}
{Veronig} A, {Temmer} M, {Hanslmeier} A, et~al (2002{\natexlab{a}}) {Temporal
  aspects and frequency distributions of solar soft X-ray flares}. \aap
  382:1070--1080. \doi{10.1051/0004-6361:20011694},
  {\href{https://arxiv.org/abs/astro-ph/0207234}{{arXiv:astro-ph/0207234}}}
  {[astro-ph]}

\bibitem[{{Veronig} et~al(2002{\natexlab{b}}){Veronig}, {Vr{\v s}nak},
  {Dennis}, {Temmer}, {Hanslmeier}, and {Magdaleni{\'c}}}]{Veronig2002}
{Veronig} A, {Vr{\v s}nak} B, {Dennis} BR, et~al (2002{\natexlab{b}})
  {Investigation of the Neupert effect in solar flares. I. Statistical
  properties and the evaporation model}. \aap 392:699--712.
  \doi{10.1051/0004-6361:20020947},
  {\href{https://arxiv.org/abs/astro-ph/0207217}{{astro-ph/0207217}}}

\bibitem[{{Veronig} et~al(2005){Veronig}, {Brown}, {Dennis}, {Schwartz}, {Sui},
  and {Tolbert}}]{Veronig2005}
{Veronig} AM, {Brown} JC, {Dennis} BR, et~al (2005) {Physics of the Neupert
  Effect: Estimates of the Effects of Source Energy, Mass Transport, and
  Geometry Using RHESSI and GOES Data}. \apj 621:482--497. \doi{10.1086/427274}

\bibitem[{{Vida} et~al(2016){Vida}, {Kriskovics}, {Ol{\'a}h}, {Leitzinger},
  {Odert}, {K{\H{o}}v{\'a}ri}, {Korhonen}, {Greimel}, {Robb}, {Cs{\'a}k}, and
  {Kov{\'a}cs}}]{Vida2016}
{Vida} K, {Kriskovics} L, {Ol{\'a}h} K, et~al (2016) {Investigating magnetic
  activity in very stable stellar magnetic fields. Long-term photometric and
  spectroscopic study of the fully convective M4 dwarf V374 Pegasi}. \aap
  590:A11. \doi{10.1051/0004-6361/201527925},
  {\href{https://arxiv.org/abs/1603.00867}{{arXiv:1603.00867}}} {[astro-ph.SR]}

\bibitem[{{Vida} et~al(2017){Vida}, {K{\H{o}}v{\'a}ri}, {P{\'a}l}, {Ol{\'a}h},
  and {Kriskovics}}]{Vida2017}
{Vida} K, {K{\H{o}}v{\'a}ri} Z, {P{\'a}l} A, et~al (2017) {Frequent Flaring in
  the TRAPPIST-1 System{\textemdash}Unsuited for Life?} \apj 841(2):124.
  \doi{10.3847/1538-4357/aa6f05},
  {\href{https://arxiv.org/abs/1703.10130}{{arXiv:1703.10130}}} {[astro-ph.SR]}

\bibitem[{{Vida} et~al(2019){Vida}, {Leitzinger}, {Kriskovics}, {Seli},
  {Odert}, {Kov{\'a}cs}, {Korhonen}, and {van Driel-Gesztelyi}}]{Vida2019}
{Vida} K, {Leitzinger} M, {Kriskovics} L, et~al (2019) {The quest for stellar
  coronal mass ejections in late-type stars. I. Investigating Balmer-line
  asymmetries of single stars in Virtual Observatory data}. \aap 623:A49.
  \doi{10.1051/0004-6361/201834264},
  {\href{https://arxiv.org/abs/1901.04229}{{arXiv:1901.04229}}} {[astro-ph.SR]}

\bibitem[{{Vidal} et~al(1970){Vidal}, {Cooper}, and {Smith}}]{Vidal1970}
{Vidal} CR, {Cooper} J, {Smith} EW (1970) {Hydrogen Stark broadening
  calculations with the unified classical path theory.} \jqsrt 10:1011--1063.
  \doi{10.1016/0022-4073(70)90121-4}

\bibitem[{{Vidal} et~al(1971){Vidal}, {Cooper}, and {Smith}}]{Vidal1971}
{Vidal} CR, {Cooper} J, {Smith} EW (1971) {Unified theory calculations of Stark
  broadened hydrogen lines including lower state interactions.} \jqsrt
  11(3):263--281. \doi{10.1016/0022-4073(71)90013-6}

\bibitem[{{Vidal} et~al(1973){Vidal}, {Cooper}, and {Smith}}]{Vidal1973}
{Vidal} CR, {Cooper} J, {Smith} EW (1973) {Hydrogen Stark-Broadening Tables}.
  \apjs 25:37. \doi{10.1086/190264}

\bibitem[{{Vidotto} et~al(2014){Vidotto}, {Gregory}, {Jardine}, {Donati},
  {Petit}, {Morin}, {Folsom}, {Bouvier}, {Cameron}, {Hussain}, {Marsden},
  {Waite}, {Fares}, {Jeffers}, and {do Nascimento}}]{Vidotto2014}
{Vidotto} AA, {Gregory} SG, {Jardine} M, et~al (2014) {Stellar magnetism:
  empirical trends with age and rotation}. \mnras 441(3):2361--2374.
  \doi{10.1093/mnras/stu728},
  {\href{https://arxiv.org/abs/1404.2733}{{arXiv:1404.2733}}} {[astro-ph.SR]}

\bibitem[{{Vievering} et~al(2019){Vievering}, {Glesener}, {Grefenstette}, and
  {Smith}}]{Vievering2019}
{Vievering} JT, {Glesener} L, {Grefenstette} BW, et~al (2019) {New Star
  Observations with NuSTAR: Flares from Young Stellar Objects in the
  {\ensuremath{\rho}} Ophiuchi Cloud Complex in Hard X-Rays}. \apj 882(1):72.
  \doi{10.3847/1538-4357/ab2e0d}

\bibitem[{{Vievering} et~al(2023){Vievering}, {Glesener}, {Caspi}, {Allred},
  {Chen}, {Lustig-Yaeger}, {Kerr}, {Inglis}, {Alaoui}, and
  {Mayorga}}]{Vivering2023}
{Vievering} JT, {Glesener} L, {Caspi} A, et~al (2023) {High-Energy Aspects of
  the Solar-Stellar Connection}. In: Bulletin of the American Astronomical
  Society, p 419, \doi{10.3847/25c2cfeb.18c000ce}

\bibitem[{{Villadsen} and {Hallinan}(2019)}]{Villadsen2019}
{Villadsen} J, {Hallinan} G (2019) {Ultra-wideband Detection of 22 Coherent
  Radio Bursts on M Dwarfs}. \apj 871(2):214. \doi{10.3847/1538-4357/aaf88e},
  {\href{https://arxiv.org/abs/1810.00855}{{arXiv:1810.00855}}} {[astro-ph.SR]}

\bibitem[{{Vilmer} et~al(2011){Vilmer}, {MacKinnon}, and
  {Hurford}}]{Vilmer2011}
{Vilmer} N, {MacKinnon} AL, {Hurford} GJ (2011) {Properties of Energetic Ions
  in the Solar Atmosphere from {\ensuremath{\gamma}}-Ray and Neutron
  Observations}. \ssr 159(1-4):167--224. \doi{10.1007/s11214-010-9728-x},
  {\href{https://arxiv.org/abs/1110.2432}{{arXiv:1110.2432}}} {[astro-ph.SR]}

\bibitem[{{Vlahos} and {Isliker}(2019)}]{Vlahos2019}
{Vlahos} L, {Isliker} H (2019) {Particle acceleration and heating in a
  turbulent solar corona}. Plasma Physics and Controlled Fusion 61(1):014020.
  \doi{10.1088/1361-6587/aadbe7},
  {\href{https://arxiv.org/abs/1808.07136}{{arXiv:1808.07136}}} {[astro-ph.HE]}

\bibitem[{{Wahlstrom} and {Carlsson}(1994)}]{Whalstrom1994}
{Wahlstrom} C, {Carlsson} M (1994) {The Formation of the Solar He II 1640.4
  Angstrom Emission Line}. \apj 433:417. \doi{10.1086/174654}

\bibitem[{{Walker}(1981)}]{Walker1981}
{Walker} AR (1981) {Flare activity of proxima Cen.} \mnras 195:1029--1035.
  \doi{10.1093/mnras/195.4.1029}

\bibitem[{{Walkowicz} and {Hawley}(2009)}]{WH09}
{Walkowicz} LM, {Hawley} SL (2009) {Tracers of Chromospheric Structure. I.
  Observations of Ca II K and H{\ensuremath{\alpha}} in M Dwarfs}. \aj
  137(2):3297--3313. \doi{10.1088/0004-6256/137/2/3297},
  {\href{https://arxiv.org/abs/0811.1778}{{arXiv:0811.1778}}} {[astro-ph]}

\bibitem[{{Walkowicz} et~al(2011){Walkowicz}, {Basri}, {Batalha}, {Gilliland},
  {Jenkins}, {Borucki}, {Koch}, {Caldwell}, {Dupree}, {Latham}, {Meibom},
  {Howell}, {Brown}, and {Bryson}}]{Walkowicz2011}
{Walkowicz} LM, {Basri} G, {Batalha} N, et~al (2011) {White-light Flares on
  Cool Stars in the Kepler Quarter 1 Data}. \aj 141:50.
  \doi{10.1088/0004-6256/141/2/50},
  {\href{https://arxiv.org/abs/1008.0853}{{arXiv:1008.0853}}} {[astro-ph.SR]}

\bibitem[{{Wall} and {Jenkins}(2003)}]{WJ}
{Wall} JV, {Jenkins} CR (2003) {Practical Statistics for Astronomers}, vol~3

\bibitem[{{Wargelin} et~al(2008){Wargelin}, {Kashyap}, {Drake},
  {Garc{\'\i}a-Alvarez}, and {Ratzlaff}}]{Wargelin2008}
{Wargelin} BJ, {Kashyap} VL, {Drake} JJ, et~al (2008) {X-Ray Flaring on the dMe
  Star, Ross 154}. \apj 676(1):610--627. \doi{10.1086/528702},
  {\href{https://arxiv.org/abs/0712.2791}{{arXiv:0712.2791}}} {[astro-ph]}

\bibitem[{{Warmuth} and {Mann}(2016)}]{WarmuthMann2016}
{Warmuth} A, {Mann} G (2016) {Constraints on energy release in solar flares
  from RHESSI and GOES X-ray observations. II. Energetics and energy
  partition}. \aap 588:A116. \doi{10.1051/0004-6361/201527475}

\bibitem[{{Warmuth} and {Mann}(2020)}]{Warmuth2020}
{Warmuth} A, {Mann} G (2020) {Thermal-nonthermal energy partition in solar
  flares derived from X-ray, EUV, and bolometric observations. Discussion of
  recent studies}. \aap 644:A172. \doi{10.1051/0004-6361/202039529},
  {\href{https://arxiv.org/abs/2011.04442}{{arXiv:2011.04442}}} {[astro-ph.SR]}

\bibitem[{{Warmuth} et~al(2009){Warmuth}, {Holman}, {Dennis}, {Mann}, {Aurass},
  and {Milligan}}]{Warmuth2009}
{Warmuth} A, {Holman} GD, {Dennis} BR, et~al (2009) {Rapid Changes of Electron
  Acceleration Characteristics at the End of the Impulsive Phase of an X-class
  Solar Flare}. \apj 699:917--922. \doi{10.1088/0004-637X/699/1/917}

\bibitem[{{Warren}(2006)}]{Warren2006}
{Warren} HP (2006) {Multithread Hydrodynamic Modeling of a Solar Flare}. \apj
  637:522--530. \doi{10.1086/497904},
  {\href{https://arxiv.org/abs/astro-ph/0507328}{{astro-ph/0507328}}}

\bibitem[{{Warren}(2014)}]{Warren2014}
{Warren} HP (2014) {Measurements of Absolute Abundances in Solar Flares}. \apjl
  786(1):L2. \doi{10.1088/2041-8205/786/1/L2},
  {\href{https://arxiv.org/abs/1310.4765}{{arXiv:1310.4765}}} {[astro-ph.SR]}

\bibitem[{{Webb} et~al(2021){Webb}, {Flynn}, {Cooke}, {Zhang}, {Mahabal},
  {Abbott}, {Allen}, {Andreoni}, {Bird}, {Goode}, {Lochner}, and
  {Pritchard}}]{Webb2021}
{Webb} S, {Flynn} C, {Cooke} J, et~al (2021) {The Deeper, Wider, Faster
  programme: exploring stellar flare activity with deep, fast cadenced DECam
  imaging via machine learning}. \mnras 506(2):2089--2103.
  \doi{10.1093/mnras/stab1798},
  {\href{https://arxiv.org/abs/2106.13026}{{arXiv:2106.13026}}} {[astro-ph.SR]}

\bibitem[{{Weinberg}(2021)}]{Weinberg2021}
{Weinberg} S (2021) {Foundations of Modern Physics}.
  \doi{10.1017/9781108894845}

\bibitem[{{Weisheit} and {Murillo}(2006)}]{Springer2006_Ch86}
{Weisheit} J, {Murillo} M (2006) {Atoms in Dense Plasmas}. In: Springer
  Handbook of Atomic, Molecular, and Optical Physics. p 1303,
  \doi{10.1007/978-0-387-26308-3_86}

\bibitem[{{Welsh} et~al(2006){Welsh}, {Wheatley}, {Browne}, {Siegmund},
  {Doyle}, {O'Shea}, {Antonova}, {Forster}, {Seibert}, {Morrissey}, and
  {Taroyan}}]{Welsh2006}
{Welsh} BY, {Wheatley} J, {Browne} SE, et~al (2006) {GALEX high time-resolution
  ultraviolet observations of dMe flare events}. \aap 458:921--930.
  \doi{10.1051/0004-6361:20065304},
  {\href{https://arxiv.org/abs/astro-ph/0608254}{{astro-ph/0608254}}}

\bibitem[{{Welsh} et~al(2007){Welsh}, {Wheatley}, {Seibert}, {Browne}, {West},
  {Siegmund}, {Barlow}, {Forster}, {Friedman}, {Martin}, {Morrissey}, {Small},
  {Wyder}, {Schiminovich}, {Neff}, and {Rich}}]{Welsh2007}
{Welsh} BY, {Wheatley} JM, {Seibert} M, et~al (2007) {The Detection of M Dwarf
  UV Flare Events in the GALEX Data Archives}. \apjs 173(2):673--681.
  \doi{10.1086/516640},
  {\href{https://arxiv.org/abs/astro-ph/0605328}{{arXiv:astro-ph/0605328}}}
  {[astro-ph]}

\bibitem[{{West} et~al(2008){West}, {Hawley}, {Bochanski}, {Covey}, {Reid},
  {Dhital}, {Hilton}, and {Masuda}}]{West2008}
{West} AA, {Hawley} SL, {Bochanski} JJ, et~al (2008) {Constraining the
  Age-Activity Relation for Cool Stars: The Sloan Digital Sky Survey Data
  Release 5 Low-Mass Star Spectroscopic Sample}. \aj 135(3):785--795.
  \doi{10.1088/0004-6256/135/3/785},
  {\href{https://arxiv.org/abs/0712.1590}{{arXiv:0712.1590}}} {[astro-ph]}

\bibitem[{{West} et~al(2015){West}, {Weisenburger}, {Irwin}, {Berta-Thompson},
  {Charbonneau}, {Dittmann}, and {Pineda}}]{West2015}
{West} AA, {Weisenburger} KL, {Irwin} J, et~al (2015) {An Activity-Rotation
  Relationship and Kinematic Analysis of Nearby Mid-to-Late-Type M Dwarfs}.
  \apj 812(1):3. \doi{10.1088/0004-637X/812/1/3},
  {\href{https://arxiv.org/abs/1509.01590}{{arXiv:1509.01590}}} {[astro-ph.SR]}

\bibitem[{{White} et~al(2003){White}, {Krucker}, {Shibasaki}, {Yokoyama},
  {Shimojo}, and {Kundu}}]{White2003}
{White} SM, {Krucker} S, {Shibasaki} K, et~al (2003) {Radio and Hard X-Ray
  Images of High-Energy Electrons in an X-Class Solar Flare}. \apjl
  595(2):L111--L114. \doi{10.1086/379274}

\bibitem[{{White} et~al(2011){White}, {Benz}, {Christe}, {F{\'a}rn{\'\i}k},
  {Kundu}, {Mann}, {Ning}, {Raulin}, {Silva-V{\'a}lio}, {Saint-Hilaire},
  {Vilmer}, and {Warmuth}}]{White2011}
{White} SM, {Benz} AO, {Christe} S, et~al (2011) {The Relationship Between
  Solar Radio and Hard X-ray Emission}. \ssr 159(1-4):225--261.
  \doi{10.1007/s11214-010-9708-1},
  {\href{https://arxiv.org/abs/1109.6629}{{arXiv:1109.6629}}} {[astro-ph.SR]}

\bibitem[{{Willmer}(2018)}]{Willmer2018}
{Willmer} CNA (2018) {The Absolute Magnitude of the Sun in Several Filters}.
  \apjs 236(2):47. \doi{10.3847/1538-4365/aabfdf}

\bibitem[{{Wilson} and {Mega-Muscles Collaboration}(2021)}]{Wilson2021Mega}
{Wilson} D, {Mega-Muscles Collaboration} (2021) {Mega-MUSCLES}. In: The 20.5th
  Cambridge Workshop on Cool Stars, Stellar Systems, and the Sun (CS20.5),
  Cambridge Workshop on Cool Stars, Stellar Systems, and the Sun, p 253,
  \doi{10.5281/zenodo.4567579}

\bibitem[{{Wood} et~al(1996){Wood}, {Harper}, {Linsky}, and
  {Dempsey}}]{Wood1996}
{Wood} BE, {Harper} GM, {Linsky} JL, et~al (1996) {Goddard High-Resolution
  Spectrograph Observations of Procyon and HR 1099}. \apj 458:761.
  \doi{10.1086/176857}

\bibitem[{{Wood} et~al(2021){Wood}, {M{\"u}ller}, {Redfield}, {Konow},
  {Vannier}, {Linsky}, {Youngblood}, {Vidotto}, {Jardine},
  {Alvarado-G{\'o}mez}, and {Drake}}]{Wood2021}
{Wood} BE, {M{\"u}ller} HR, {Redfield} S, et~al (2021) {New Observational
  Constraints on the Winds of M dwarf Stars}. \apj 915(1):37.
  \doi{10.3847/1538-4357/abfda5},
  {\href{https://arxiv.org/abs/2105.00019}{{arXiv:2105.00019}}} {[astro-ph.SR]}

\bibitem[{{Woodgate} et~al(1992){Woodgate}, {Robinson}, {Carpenter}, {Maran},
  and {Shore}}]{Woodgate1992}
{Woodgate} BE, {Robinson} RD, {Carpenter} KG, et~al (1992) {Detection of a
  proton beam during the impulsive phase of a stellar flare}. \apjl
  397:L95--L98. \doi{10.1086/186553}

\bibitem[{{Woods} et~al(2004){Woods}, {Eparvier}, {Fontenla}, {Harder}, {Kopp},
  {McClintock}, {Rottman}, {Smiley}, and {Snow}}]{Woods2004}
{Woods} TN, {Eparvier} FG, {Fontenla} J, et~al (2004) {Solar irradiance
  variability during the October 2003 solar storm period}. \grl 31:L10802.
  \doi{10.1029/2004GL019571}

\bibitem[{{Woods} et~al(2006){Woods}, {Kopp}, and {Chamberlin}}]{Woods2006}
{Woods} TN, {Kopp} G, {Chamberlin} PC (2006) {Contributions of the solar
  ultraviolet irradiance to the total solar irradiance during large flares}.
  Journal of Geophysical Research (Space Physics) 111(A10):A10S14.
  \doi{10.1029/2005JA011507}

\bibitem[{{Wright} and {Drake}(2016)}]{Wright2016}
{Wright} NJ, {Drake} JJ (2016) {Solar-type dynamo behaviour in fully convective
  stars without a tachocline}. \nat 535(7613):526--528.
  \doi{10.1038/nature18638},
  {\href{https://arxiv.org/abs/1607.07870}{{arXiv:1607.07870}}} {[astro-ph.SR]}

\bibitem[{{Wright} et~al(2011){Wright}, {Drake}, {Mamajek}, and
  {Henry}}]{Wright2011}
{Wright} NJ, {Drake} JJ, {Mamajek} EE, et~al (2011) {The
  Stellar-activity-Rotation Relationship and the Evolution of Stellar Dynamos}.
  \apj 743(1):48. \doi{10.1088/0004-637X/743/1/48},
  {\href{https://arxiv.org/abs/1109.4634}{{arXiv:1109.4634}}} {[astro-ph.SR]}

\bibitem[{{Wright} et~al(2018){Wright}, {Newton}, {Williams}, {Drake}, and
  {Yadav}}]{Wright2018}
{Wright} NJ, {Newton} ER, {Williams} PKG, et~al (2018) {The stellar
  rotation-activity relationship in fully convective M dwarfs}. \mnras
  479(2):2351--2360. \doi{10.1093/mnras/sty1670},
  {\href{https://arxiv.org/abs/1807.03304}{{arXiv:1807.03304}}} {[astro-ph.SR]}

\bibitem[{{Wu} et~al(2015){Wu}, {Ip}, and {Huang}}]{Wu2015}
{Wu} CJ, {Ip} WH, {Huang} LC (2015) {A Study of Variability in the Frequency
  Distributions of the Superflares of G-type Stars Observed by the Kepler
  Mission}. \apj 798(2):92. \doi{10.1088/0004-637X/798/2/92}

\bibitem[{{Wu} et~al(2022){Wu}, {Chen}, {Tian}, {Zhang}, {Shi}, {He}, {Lu},
  {Xu}, and {Wang}}]{Wu2022}
{Wu} Y, {Chen} H, {Tian} H, et~al (2022) {Broadening and Redward Asymmetry of
  H{\ensuremath{\alpha}} Line Profiles Observed by LAMOST during a Stellar
  Flare on an M-type Star}. \apj 928(2):180. \doi{10.3847/1538-4357/ac5897},
  {\href{https://arxiv.org/abs/2203.02292}{{arXiv:2203.02292}}} {[astro-ph.SR]}

\bibitem[{{Wulser} et~al(1992){Wulser}, {Canfield}, and {Zarro}}]{Wulser1992}
{Wulser} JP, {Canfield} RC, {Zarro} DM (1992) {Energetics and Dynamics in a
  Large Solar Flare of 1989 March}. \apj 384:341. \doi{10.1086/170877}

\bibitem[{{Xia} et~al(2021){Xia}, {Su}, {Wang}, {Wang}, {Warmuth}, {Gan}, and
  {Li}}]{Xia2021}
{Xia} F, {Su} Y, {Wang} W, et~al (2021) {Detection of Energy Cutoffs in
  Flare-accelerated Electrons}. \apj 908(1):111. \doi{10.3847/1538-4357/abce5c}

\bibitem[{{Yaakobi} et~al(1977){Yaakobi}, {Thorsos}, {Hauer}, {Perry}, and
  {Steel}}]{Yaakobi1977}
{Yaakobi} B, {Thorsos} E, {Hauer} A, et~al (1977) {Direct Measurement of
  Compression of Laser-Imploded Targets Using X-Ray Spectroscopy}. \prl
  39(24):1526--1529. \doi{10.1103/PhysRevLett.39.1526}

\bibitem[{{Yang} and {Liu}(2019)}]{Yang2019}
{Yang} H, {Liu} J (2019) {The Flare Catalog and the Flare Activity in the
  Kepler Mission}. \apjs 241(2):29. \doi{10.3847/1538-4365/ab0d28},
  {\href{https://arxiv.org/abs/1903.01056}{{arXiv:1903.01056}}} {[astro-ph.SR]}

\bibitem[{{Yang} et~al(2018){Yang}, {Liu}, {Qiao}, {Zhang}, {Gao}, {Cui}, and
  {Han}}]{Yang2018}
{Yang} H, {Liu} J, {Qiao} E, et~al (2018) {Do Long-cadence Data of the Kepler
  Spacecraft Capture Basic Properties of Flares?} \apj 859(2):87.
  \doi{10.3847/1538-4357/aabd31},
  {\href{https://arxiv.org/abs/1804.02621}{{arXiv:1804.02621}}} {[astro-ph.SR]}

\bibitem[{{Yokoyama} and {Shibata}(1998)}]{Yokoyama1998}
{Yokoyama} T, {Shibata} K (1998) {A Two-dimensional Magnetohydrodynamic
  Simulation of Chromospheric Evaporation in a Solar Flare Based on a Magnetic
  Reconnection Model}. \apjl 494(1):L113--L116. \doi{10.1086/311174}

\bibitem[{{Youngblood} et~al(2017){Youngblood}, {France}, {Loyd}, {Brown},
  {Mason}, {Schneider}, {Tilley}, {Berta-Thompson}, {Buccino}, {Froning},
  {Hawley}, {Linsky}, {Mauas}, {Redfield}, {Kowalski}, {Miguel}, {Newton},
  {Rugheimer}, {Segura}, {Roberge}, and {Vieytes}}]{Youngblood2017}
{Youngblood} A, {France} K, {Loyd} ROP, et~al (2017) {The MUSCLES Treasury
  Survey. IV. Scaling Relations for Ultraviolet, Ca II K, and Energetic
  Particle Fluxes from M Dwarfs}. \apj 843(1):31.
  \doi{10.3847/1538-4357/aa76dd},
  {\href{https://arxiv.org/abs/1705.04361}{{arXiv:1705.04361}}} {[astro-ph.SR]}

\bibitem[{{Zarro} and {Canfield}(1989)}]{Zarro1989}
{Zarro} DM, {Canfield} RC (1989) {H alpha Redshifts as a Diagnostic of Solar
  Flare Heating}. \apjl 338:L33. \doi{10.1086/185394}

\bibitem[{{Zel'dovich} and {Raizer}(1967)}]{ZeldovichRazier}
{Zel'dovich} YB, {Raizer} YP (1967) {Physics of shock waves and
  high-temperature hydrodynamic phenomena}

\bibitem[{{Zharkova} and {Gordovskyy}(2006)}]{Zharkova2006}
{Zharkova} VV, {Gordovskyy} M (2006) {The Effect of the Electric Field Induced
  by Precipitating Electron Beams on Hard X-Ray Photon and Mean Electron
  Spectra}. \apj 651(1):553--565. \doi{10.1086/506423}

\bibitem[{{Zharkova} and {Kobylinskii}(1993)}]{Zharkova1993}
{Zharkova} VV, {Kobylinskii} VA (1993) {The effect of non-thermal excitation
  and ionization on the hydrogen emission in impulsive solar flares}. \solphys
  143(2):259--274. \doi{10.1007/BF00646487}

\bibitem[{{Zharkova} et~al(2011){Zharkova}, {Arzner}, {Benz}, {Browning},
  {Dauphin}, {Emslie}, {Fletcher}, {Kontar}, {Mann}, {Onofri}, {Petrosian},
  {Turkmani}, {Vilmer}, and {Vlahos}}]{Zharkova2011}
{Zharkova} VV, {Arzner} K, {Benz} AO, et~al (2011) {Recent Advances in
  Understanding Particle Acceleration Processes in Solar Flares}. \ssr
  159(1-4):357--420. \doi{10.1007/s11214-011-9803-y},
  {\href{https://arxiv.org/abs/1110.2359}{{arXiv:1110.2359}}} {[astro-ph.SR]}

\bibitem[{{Zhilyaev} et~al(2007){Zhilyaev}, {Romanyuk}, {Svyatogorov},
  {Verlyuk}, {Kaminsky}, {Andreev}, {Sergeev}, {Gershberg}, {Lovkaya},
  {Avgoloupis}, {Seiradakis}, {Contadakis}, {Antov}, {Konstantinova-Antova},
  and {Bogdanovski}}]{Zhilyaev2007}
{Zhilyaev} BE, {Romanyuk} YO, {Svyatogorov} OA, et~al (2007) {Fast colorimetry
  of the flare star EV Lacertae from UBVRI observations in 2004}. \aap
  465:235--240. \doi{10.1051/0004-6361:20065632}

\bibitem[{{Zhu} et~al(2018){Zhu}, {Qiu}, and {Longcope}}]{Zhu2018}
{Zhu} C, {Qiu} J, {Longcope} DW (2018) {Two-phase Heating in Flaring Loops}.
  \apj 856(1):27. \doi{10.3847/1538-4357/aaad10},
  {\href{https://arxiv.org/abs/1802.00871}{{arXiv:1802.00871}}} {[astro-ph.SR]}

\bibitem[{{Zhu} et~al(2019){Zhu}, {Kowalski}, {Tian}, {Uitenbroek}, {Carlsson},
  and {Allred}}]{Zhu2019}
{Zhu} Y, {Kowalski} AF, {Tian} H, et~al (2019) {Modeling Mg II h, k and Triplet
  Lines at Solar Flare Ribbons}. \apj 879(1):19.
  \doi{10.3847/1538-4357/ab2238},
  {\href{https://arxiv.org/abs/1904.12285}{{arXiv:1904.12285}}} {[astro-ph.SR]}

\bibitem[{{Zimmerman} et~al(1979){Zimmerman}, {Littman}, {Kash}, and
  {Kleppner}}]{Zimmerman1979}
{Zimmerman} ML, {Littman} MG, {Kash} MM, et~al (1979) {Stark structure of the
  Rydberg states of alkali-metal atoms}. \pra 20(6):2251--2275.
  \doi{10.1103/PhysRevA.20.2251}

\bibitem[{{Zimovets} et~al(2021){Zimovets}, {McLaughlin}, {Srivastava},
  {Kolotkov}, {Kuznetsov}, {Kupriyanova}, {Cho}, {Inglis}, {Reale}, {Pascoe},
  {Tian}, {Yuan}, {Li}, and {Zhang}}]{Zimovets2021}
{Zimovets} IV, {McLaughlin} JA, {Srivastava} AK, et~al (2021) {Quasi-Periodic
  Pulsations in Solar and Stellar Flares: A Review of Underpinning Physical
  Mechanisms and Their Predicted Observational Signatures}. \ssr 217(5):66.
  \doi{10.1007/s11214-021-00840-9}

\bibitem[{Zwiebach(2022)}]{Zwiebach2022}
Zwiebach B (2022) Mastering quantum mechanics: Essentials, Theory, and
  Applications. The MIT Press, Cambridge

\end{thebibliography}

%\input main_v5.bbl


\end{document}